\documentclass[preprint2]{aastex}
\usepackage{graphicx,lscape}
\usepackage[usenames]{color}
\usepackage{float}
\usepackage{amsmath}
\usepackage{fixltx2e}
\usepackage{dblfloatfix}
\usepackage{epstopdf}
\usepackage{chngpage}
\usepackage{geometry}
\usepackage{pdflscape}
\usepackage{calc}

\newcommand{\msun}{\mbox{${M_\odot}$}}
\newcommand{\msunyr}{$\rm M_\odot\,yr^{-1}$}
\newcommand{\msunyrkpc}{$\rm M_\odot\,yr^{-1}\,kpc^{-2}$}

\newcommand{\hi}{H\,{\sc i}}
\newcommand{\hii}{H\,{\sc ii}}
\newcommand{\halpha}{H${\alpha}$}
\newcommand{\RCT}{${\rm RC_{Th}}$}
\newcommand{\RCNT}{${\rm RC_{Nth}}$}

\newcommand{\et}{et al.}
\newcommand{\Cband}{{\em C-}band}
\newcommand{\CRe}{\mbox{CR$e$}}

\shorttitle{Radio Continuum Study of Dwarf Galaxies} 
\shortauthors{Kitchener, Brinks \& Heesen et al.}

\begin{document}

\title{A Radio Continuum Study of Dwarf Galaxies: 6\,cm imaging of LITTLE THINGS}

\author{Luke Hindson\altaffilmark{1}, Ged Kitchener\altaffilmark{1}, Elias Brinks\altaffilmark{1}, Volker Heesen\altaffilmark{2,6}, Jonathan Westcott\altaffilmark{1}, Deidre Hunter\altaffilmark{3}, Hong-Xin Zhang\altaffilmark{4}, Michael Rupen\altaffilmark{5,7}, Urvashi Rau\altaffilmark{5}}

\altaffiltext{1}{Centre for Astrophysics Research, University of Hertfordshire, Hatfield, AL10 9AB, United Kingdom}
\altaffiltext{2}{School of Physics and Astronomy, University of Southampton, Southampton, SO17 1BJ, United Kingdom}
\altaffiltext{3}{Lowell Observatory, 1400 West Mars Hill Road, Flagstaff, AZ 86001, USA}
\altaffiltext{4}{Department of Astronomy, Peking University, Beijing 100871, China}
\altaffiltext{5}{National Radio Astronomy Observatory, 1003 Lopezville Road, Socorro, NM 87801, USA}
\altaffiltext{6}{Current address: Hamburger Sternwarte, Gojenbergsweg 112, 21029  Hamburg, Germany}
\altaffiltext{7}{Current address:National Research Council of Canada, Herzberg Astronomy and Astrophysics Programs, Dominion Radio Astrophysical Observatory, PO Box 248, Penticton, BC V2A 6J9, Canada}
\email{l.hindson3@herts.ac.uk}

\begin{abstract}

In this paper we examine to what extent the radio continuum can be used as an extinction free probe of star formation in dwarf galaxies. To that aim we observe $40$ nearby dwarf galaxies with the Very Large Array at 6\,cm ($4$--$8$\,GHz) in C--configuration. We obtained images with $3$--$8^{\prime\prime}$ resolution and noise levels of $3$--$15{\rm \,\mu Jy\,beam^{-1}}$. We detected emission associated with $22$ of the $40$ dwarf galaxies, $8$ of which are new detections. The general picture is that of an interstellar medium largely devoid of radio continuum emission, interspersed by isolated pockets of emission associated with star formation. We find an average thermal fraction of $\sim 50$--$70$\% and an average magnetic field strength of $\sim 5$--$8\,{\rm \mu G}$, only slightly lower than that found in larger, spiral galaxies. At 100\,pc scales, we find surprisingly high values for the average magnetic field strength of up to 50$\,{\rm \mu G}$. We find that dwarf galaxies follow the theoretical predictions of the radio continuum--star formation rate relation within regions of significant radio continuum emission but that the non-thermal radio continuum is suppressed relative to the star formation rate when considering the entire optical disk. We examine the far-infrared--star formation rate relation for our sample and find that the far-infrared is suppressed compared to the expected star formation rate. We discuss explanations for these observed relations and the impact of our findings on the radio continuum--far-infrared relation. We conclude that radio continuum emission at centimetre wavelengths has the promise of being a largely extinction--free star formation rate indicator. We find that star formation rates of gas rich, low mass galaxies can be estimated with an uncertainty of $\pm 0.2$\,dex between the values of $2 \times 10^{-4}$ and $0.1 {\rm M_\odot\,yr^{-1}}$.
\end{abstract}

\keywords{galaxies: dwarf --- galaxies: magnetic fields --- galaxies: SF --- radio continuum: galaxies --- radio continuum: ISM}

\section{INTRODUCTION}
\label{section:Paper1Introduction}

The radio continuum -- far-infrared (RC--FIR) relation of galaxies holds over 4 orders of magnitude in luminosity, irrespective of galaxy type \cite[][]{Helou1985,deJong1985,Yun2001}. It displays a mere $0.26$\,dex scatter \cite[][]{Yun2001} and has been observed to hold at least out to a redshift of $3$ \cite[][]{Appleton2004}. The radio continuum \cite[RC; see][for a review]{Condon1992} and the far-infrared (FIR) have long been attributed to the input of energy following a star formation (SF) event. At 6\,cm the RC is comprised of two main contributions: a thermal component (\RCT) from thermal electrons in HII regions, and a non-thermal component (\RCNT) generated by cosmic ray electrons (\CRe) accelerated in supernova shocks. The \RCT\ and \RCNT\ emission both have an unambiguous link to recent SF, while the FIR originates from the modified blackbody radiation of interstellar dust that is heated by the interstellar radiation field \cite[][]{Li2010}.

The classical scenario leading to the RC--FIR relation assumes a galaxy behaves as a calorimeter \cite[][]{Voelk1989}. This model assumes that galaxies are optically thick to dust-heating UV photons which are absorbed by dust within the galaxy that goes on to reradiate the energy gained in the FIR. It also requires that magnetic fields retain all \CRe, which eventually produce synchrotron radiation. Since all the energy contained within the \CRe\ is radiated, the strength of the magnetic field is irrelevant, i.e., whether the energy contained in the \CRe\ is radiated over $1$\,Myr in a relatively strong magnetic field, or over $1$\,Gyr in a weaker magnetic field, the total energy emitted is the same. The calorimeter assumption implies that the mean-free-path of dust-heating photons is less than the galaxy disk scale height, and that the typical energy loss timescale of \CRe\ is less than the diffusion timescale to traverse the scale height.

Clearly, the calorimeter model is not perfect. Dust-heating photons {\em are} observed to be coming from galaxies and \RCNT\ emission {\em is} observed in the haloes of larger spiral galaxies \cite[][]{Heesen2009}. Therefore, for galaxies to follow the RC--FIR relation, the escape of \CRe\ from galactic magnetic fields must be in proportion to the escape of dust-heating photons from the disk \citep{Helou1993}. Some of the earliest evidence of this comes from \cite{Klein1991} who stated that the deficiency of \RCNT\ `happens to be balanced' by a lack of dust in a study of Blue Compact Dwarfs (BCDs). This is known as the `conspiracy' between the emission at RC and FIR wavelengths (e.g., \citealp{Bell2003}; \citealp{Dale2009}; \citealp{Lacki2010}). The first paper with a quantitative model of the RC-FIR correlation for non-calorimeter galaxies and the "conspiracy" between the processes involved was that by \cite{NiklasB1997} who predicted a slightly nonlinear correlation for the synchrotron emission. To complicate the picture yet further, heating of the diffuse cold dust by photons may not be sufficient to make the RC--FIR relation as tight as is observed; \cite{Xu1990} found that a significant fraction of the heating of diffuse cool dust could not be accounted for by UV radiation in their sample of $40$ spiral galaxies. An alternative source to compensate for this insufficient UV--heating could be heating by \CRe\ \cite[as for example in Ultra--Luminous Infrared Galaxies;][]{Papadopoulos2010}.

It is difficult to disentangle the many factors that lead to the RC--FIR relation. This is especially true in large spiral galaxies where within any kpc--size area the \CRe\ population stemming from recent SF can be contaminated by older \CRe\ from neighbouring areas. The interstellar medium (ISM) in spirals likewise is in a constant state of flux with differential rotation and spiral arms transporting material in and out of such a kpc--size region. We argue therefore that dwarf galaxies create a more accessible route to understanding the relationship between the RC and FIR emission and SF. The low mass of dwarf galaxies leads to SF which simulations suggest is episodic \cite[e.g.,][liken the SF history of isolated dwarf galaxies to `breathing']{Stinson2007}. If this is the case then within a set region one is ever only receiving emission from one generation of \CRe. However, observational evidence suggests that the duration of bursts of SF in dwarf galaxies may actually be quite long in some cases ($0.5$--$1.3$\,Gyr; \citealt{McQuinn2010}) which may complicate this interpretation. Dwarf galaxies also contain proportionally less dust than spirals, as confirmed by several authors \cite[e.g.,][]{Lisenfeld1998,Bigiel2008}, and should therefore be fainter in the FIR for a given level of radio emission. Understanding the origin of \RCNT\ emission generated should also be more straightforward in dwarf galaxies. They lack differential rotation \cite[][]{Gallagher1984} and thus lack the associated dynamo action present in larger, grand--design spirals that leads to amplification of the magnetic field and ordered fields of $\sim 5\mathrm{\mu G}$ between spiral arms \citep{Beck2009}. Observations suggest that dwarf galaxies differ markedly from spirals in terms of their magnetic field strength \cite[e.g.,][]{Chyzy2011,Roychowdhury2012}. These features make dwarf galaxies ideal laboratories in which to study the RC--FIR relationship. 

Historically, spatially resolved studies of the RC in dwarf galaxies have been limited by their intrinsically low surface brightness. To date, resolved observations of dwarf galaxies have been restricted to the few brightest: the near and bright IC\,10 \cite[][]{Heesen2011}, IC\,1613 \cite[][]{Chyzy2011}, NGC\,4214 \cite[][]{Kepley2011}, NGC\,1569 (\citealp{Lisenfeld2004}; \citealp{Kepley2010}; \citealp{Westcott2017}), and the Magellanic Clouds (e.g., \citealp{Filipovic1995}; \citealp{Filipovic1998}; \citealp{Leverenz2013}). The NRAO\footnote{The National Radio Astronomy Observatory is a facility of the National Science Foundation operated under cooperative agreement by Associated Universities, Inc.} Karl G.\ Jansky Very Large Array (VLA), after recently benefiting from a major upgrade, provided the prospect of routinely delivering quality, good signal-to-noise (S/N) observations of nearby dwarf galaxies. This opportunity is exploited here to revisit the relationship between the RC, FIR and star formation rate (SFR) in the dwarf galaxy regime.

The study presented here is based on VLA \Cband\ (4--8\,GHz) images of the 40 dwarf galaxies corresponding to the LITTLE THINGS sample \cite[][]{Hunter2012}, mainly focussing on the relation of RC emission with SFR indicators. The ultimate aim is to increase our understanding of the RC--SFR relation of low mass, low metallicity systems. With the development of the Square Kilometre Array (SKA; \citealt{Dewdney2015}), calibrating the RC--SFR relation in quiescent/low SFR galaxies will become more important than ever. The benefit of the RC is that observations can be carried out with ground based instruments rather than expensive (cryogenically cooled) IR satellites. Our calibration of the RC--SFR relation may provide a better understanding of how this indicator may work at higher redshift, in the domain that will be accessible to the SKA. 

This paper is organised as follows: in Section~\ref{section:Paper1sample_and_data_reduction} we describe the observations, calibration, and imaging of our sample. We present our results (images and integrated emission) in Section~\ref{section:Paper1Results}. We then discuss our results, including the RC--SFR, FIR--SFR, and RC--FIR relations in Sections~\ref{section:Paper1rc--sfr_relation}, \ref{section:Paper1fir--sfr_relation}, and \ref{section:Paper1rc--fir_relation}, respectively. We summarise our results in Section~\ref{section:Paper1Summary}.

\begin{landscape}
\notetoeditor{Table~\ref{Table:tab-sample} should be able to fit on a landscape-orientated page with tabletypesize=tiny. There is something wrong with the formatting that pushes the table off the page. Can this be sorted? I have copied this table to the end of the document to show how it should be viewed.}
\begin{deluxetable}{lccccccccccc}
\tabletypesize{\tiny}
\tablecolumns{12}
\tablewidth{689pt}
\tablecaption{The Galaxy Sample\label{Table:tab-sample} }
\tablehead{ &  & \colhead{D} &  & \colhead{M$_V$} & \colhead{$R_H$\tablenotemark{c}} & \colhead{$R_D$\tablenotemark{c}} &  
& \colhead{log$_{10}$ $\Sigma_\mathrm{SFR}\mathrm{(H\alpha)}$} & \colhead{log$_{10}$ $\Sigma_\mathrm{SFR}\mathrm{(FUV)}$} &  &  \\ \colhead{Galaxy}  &  \colhead{Other names\tablenotemark{a}} & \colhead{(Mpc)} & \colhead{Ref\tablenotemark{b}} & \colhead{(mag)} & \colhead{(arcmin)} & \colhead{(kpc)} & \colhead{E($B-V$)\tablenotemark{d}} & \colhead{(\msun yr$^{-1}$ kpc$^{-2}$)\tablenotemark{e}} & \colhead{(\msun yr$^{-1}$ kpc$^{-2}$)\tablenotemark{e}} & \colhead{$12+\log_{10} {\rm O/H}$\tablenotemark{f}}  & \colhead{Ref\tablenotemark{g}}  }
\startdata
\cutinhead{Im Galaxies}
\object[CVn I dwA]{CVnIdwA}   & UGCA 292                                                                          &  3.6 & 1             & -12.4 & 0.87 & $0.57 \pm 0.12$ & 0.01 & $-2.58 \pm 0.01$ & $-2.48 \pm 0.01$ & $7.3 \pm 0.06$ & 24 \\
\object{DDO 43}     &  PGC 21073, UGC 3860                                                  &  7.8 & 2              & -15.1 & 0.89 & $0.41\pm 0.03$ & 0.05 & $-1.78 \pm 0.01$ & $-1.55 \pm 0.01$ & $8.3 \pm 0.09$ & 25 \\
\object{DDO 46}     &  PGC 21585, UGC 3966                                                  &  6.1 & \nodata  & -14.7 & \nodata & $1.14 \pm 0.06$ & 0.05 & $-2.89 \pm 0.01$ & $-2.46 \pm 0.01$ & $8.1 \pm 0.1$  & 25 \\
\object{DDO 47}     &  PGC 21600, UGC 3974                                                  &  5.2  & 3             & -15.5 & 2.24 & $1.37 \pm 0.06$ & 0.02 & $-2.70 \pm 0.01$ & $-2.40 \pm 0.01$ & $7.8 \pm 0.2$ & 26 \\
\object{DDO 50}     & PGC 23324, UGC 4305, Holmberg II, VIIZw 223        &  3.4 & 1             & -16.6 & 3.97 & $1.10 \pm 0.05$ & 0.02 & $-1.67 \pm 0.01$ & $-1.55 \pm 0.01$ & $7.7 \pm 0.14$ & 27 \\
\object{DDO 52}     &  PGC 23769, UGC 4426                                                  & 10.3 & 4            & -15.4 & 1.08 & $1.30 \pm 0.13$ & 0.03 & $-3.20 \pm 0.01$ & $-2.43 \pm 0.01$ & (7.7)                   & 28 \\
\object{DDO 53}     &  PGC 24050, UGC 4459, VIIZw 238                              &  3.6 & 1             & -13.8 & 1.37 & $0.72 \pm 0.06$ & 0.03 & $-2.42 \pm 0.01$ &$-2.41 \pm 0.01$ & $7.6 \pm 0.11$ & 27 \\
\object{DDO 63}     &  PGC 27605, Holmberg I, UGC 5139,  Mailyan 044   &  3.9 & 1             & -14.8 & 2.17 & $0.68 \pm 0.01$ & 0.01 & $-2.32 \pm 0.01$ & $-1.95 \pm 0.01$ & $7.6 \pm 0.11$ & 27 \\
\object{DDO 69}     &  PGC 28868, UGC 5364, Leo A                                     &  0.8 & 5             & -11.7 & 2.40 & $0.19 \pm  0.01$ & 0.00 & $-2.83 \pm 0.01$ & $-2.22 \pm 0.01$ & $7.4 \pm 0.10$ & 29 \\
\object{DDO 70}     &  PGC 28913, UGC 5373, Sextans B                             &  1.3 & 6             & -14.1 & 3.71 & $0.48 \pm 0.01$ & 0.01 & $-2.85 \pm 0.01$ & $-2.16 \pm 0.01$ & $7.5 \pm 0.06$ & 30 \\
\object{DDO 75}     &  PGC 29653, UGCA 205, Sextans A                             &  1.3 & 7             & -13.9 & 3.09 & $0.22 \pm 0.01$ & 0.02 & $-1.28 \pm 0.01$ & $-1.07 \pm 0.01$ & $7.5 \pm 0.06$ & 30 \\
\object{DDO 87}     &  PGC 32405, UGC 5918, VIIZw 347                              &  7.7 & \nodata & -15.0 & 1.15 & $1.31 \pm 0.12$ & 0.00 & $-1.36 \pm 0.01$ & $-1.00 \pm 0.01$ & $7.8 \pm 0.04$ & 31 \\
\object{DDO 101}   &  PGC 37449, UGC 6900                                                 &  6.4 & \nodata & -15.0 & 1.05 & $0.94 \pm 0.03$ & 0.01 & $-2.85 \pm 0.01$ & $-2.81 \pm 0.01$ & $8.7 \pm 0.03$ & 25 \\
\object{DDO 126}   &  PGC 40791, UGC 7559                                                 &  4.9 & 8             & -14.9 & 1.76 & $0.87 \pm 0.03$ & 0.00 & $-2.37 \pm 0.01$ & $-2.10 \pm 0.01$ & (7.8)                    & 28 \\
\object{DDO 133}   &  PGC 41636, UGC 7698                                                 &  3.5 & \nodata  & -14.8 & 2.33 & $1.24 \pm 0.09$ & 0.00 & $-2.88 \pm 0.01$ & $-2.62 \pm 0.01$ & $8.2 \pm 0.09$  & 25 \\
\object{DDO 154}   &  PGC 43869, UGC 8024, NGC 4789A                         &  3.7 & \nodata  & -14.2 & 1.55 & $0.59 \pm 0.03$ & 0.01 & $-2.50 \pm 0.01$ & $-1.93 \pm 0.01$ & $7.5 \pm 0.09$ & 27 \\
\object{DDO 155}   &  PGC 44491, UGC 8091, GR 8, LSBC D646-07        &  2.2 & 9             & -12.5 & 0.95 & $0.15 \pm 0.01$ & 0.01 & $-1.44 \pm 0.01$ &\nodata & $7.7 \pm 0.06$ & 29 \\
\object{DDO 165}   &  PGC 45372, UGC 8201, IIZw 499, Mailyan 82          &  4.6 & 10           & -15.6 & 2.14 & $2.26 \pm 0.08$ & 0.01 & $-3.67 \pm 0.01$ & \nodata & $7.6 \pm0.08$ & 27 \\
\object{DDO 167}   &  PGC 45939, UGC 8308                                                 &  4.2 & 8             & -13.0 & 0.75 & $0.33 \pm 0.05$ & 0.00 & $-2.36 \pm 0.01$ & $-1.83 \pm 0.01$ & $7.7 \pm 0.2$ & 26 \\
\object{DDO 168}   &  PGC 46039, UGC 8320                                                 &  4.3 & 8             & -15.7 & 2.32 & $0.82 \pm 0.01$ & 0.00 & $-2.27 \pm 0.01$ &$-2.04 \pm 0.01$ & $8.3 \pm 0.07$ & 25 \\
\object{DDO 187}   &  PGC 50961, UGC 9128                                                  &  2.2 & 1            & -12.7 & 1.06 & $0.18 \pm 0.01$ & 0.00 & $-2.52 \pm 0.01$ & $-1.98 \pm 0.01$ & $7.7 \pm 0.09$ & 32 \\
\object{DDO 210}   &  PGC 65367, Aquarius Dwarf                                         &  0.9 & 10          & -10.9 & 1.31 & $0.17 \pm 0.01$ & 0.03 &  \nodata   & $-2.71 \pm 0.06$ & (7.2)                                      & 28 \\
\object{DDO 216}   &  PGC 71538, UGC 12613, Peg DIG, Pegasus Dwarf & 1.1  & 11         & -13.7 & 4.00 & $0.54 \pm 0.01$ & 0.02 & $-4.10 \pm 0.07$ & $-3.21 \pm 0.01$ & $7.9 \pm 0.15$ & 33 \\
\object[F564 V3]{F564-V3}    &  LSBC D564-08                                                                  &  8.7 & 4           & -14.0 & \nodata  & $0.53 \pm 0.03$ & 0.02 &  \nodata  & $-2.79 \pm 0.02$ & (7.6)                   & 28 \\
\object{IC 10}          &  PGC 1305, UGC 192                                                        &  0.7 & 12        & -16.3 & \nodata  & $0.40 \pm 0.01$ & 0.75 & $-1.11 \pm 0.01$ & \nodata & $8.2 \pm 0.12$ & 34 \\
\object{IC 1613}     &  PGC 3844, UGC 668, DDO 8                                          &  0.7 & 13         & -14.6 & 9.10 & $0.58 \pm 0.02$ & 0.00 & $-2.56 \pm 0.01$ & $-1.99 \pm 0.01$ & $7.6 \pm 0.05$ & 35 \\
\object{LGS 3}        &  PGC 3792, Pisces dwarf                                                  &  0.7 & 14         &    -9.7 & 0.96 & $0.23 \pm 0.02$ & 0.04 &  \nodata   & $-3.88 \pm 0.06$ & (7.0)                   & 28 \\
\object[M81 DwA]{M81dwA}    & PGC 23521                                                                         &  3.5 & 15         & -11.7 & \nodata  & $0.26 \pm 0.01$ & 0.02 &   \nodata   & $-2.26 \pm 0.01$ & (7.3)                   & 28 \\
\object{NGC 1569} & PGC 15345, UGC 3056, Arp 210, VIIZw 16                 &  3.4 & 16         & -18.2 & \nodata  & $0.38 \pm 0.02$& 0.51 & $0.19 \pm 0.01$ & $-0.01 \pm 0.01$ & $ 8.2 \pm 0.05$ & 36 \\
\object{NGC 2366} & PGC 21102, UGC 3851, DDO 42                                  &  3.4 & 17          & -16.8 & 4.72 & $1.36 \pm 0.04$ & 0.04 & $-1.67 \pm 0.01$ & $-1.66 \pm 0.01$ & $7.9 \pm 0.01$  & 37 \\
\object{NGC 3738} & PGC 35856, UGC 6565, Arp 234                                   &  4.9 & 3            & -17.1 & 2.40 & $0.78 \pm 0.01$ & 0.00 & $-1.66 \pm 0.01$ & $-1.53 \pm 0.01$ & $8.4 \pm 0.01$  & 25 \\
\object{NGC 4163} & PGC 38881, NGC 4167, UGC 7199                              &  2.9 & 1            & -14.4 & 1.47 & $0.27 \pm 0.03$ & 0.00 & $-2.28 \pm 0.13$ & $-1.74 \pm 0.01$ & $7.9 \pm 0.2$ & 38 \\
\object{NGC 4214} & PGC 39225, UGC 7278                                                   &  3.0 & 1            & -17.6 & 4.67 & $0.75 \pm 0.01$ & 0.00 & $-1.03 \pm 0.01$ & $-1.08 \pm 0.01$ & $8.2 \pm 0.06$ & 39 \\
\object[Sag DIG]{SagDIG}      & PGC 63287, Lowal's Object                                            &  1.1 & 19          & -12.5 & \nodata  & $0.23 \pm 0.03$ & 0.14 & $-2.97 \pm 0.04$ & $-2.11 \pm 0.01$ & $7.3 \pm 0.1$ & 35 \\
\object{UGC 8508} & PGC 47495, IZw 60                                                           &  2.6  & 1           & -13.6 & 1.28 & $0.27 \pm 0.01$ & 0.00 & $-2.03 \pm 0.01$ & \nodata & $7.9 \pm 0.2$ & 38 \\
\object{WLM}           & PGC 143, UGCA 444, DDO 221, Wolf-Lundmark-Melott & 1.0 & 20     & -14.4 & 5.81 & $0.57 \pm 0.03$ & 0.02 & $-2.77 \pm 0.01$ & $-2.05 \pm 0.01$ & $7.8 \pm 0.06$ & 40 \\
\cutinhead{BCD Galaxies}
\object{Haro 29}     & PGC 40665, UGCA 281, Mrk 209, I Zw 36                    &  5.8 & 21         & -14.6 & 0.84 & $0.29 \pm 0.01$ & 0.00 & $-0.77 \pm 0.01$ & $-1.07 \pm 0.01$ & $7.9 \pm 0.07$ & 41 \\
\object{Haro 36}     & PGC 43124, UGC 7950                                                  &  9.3 & \nodata & -15.9 & \nodata  & $0.69 \pm 0.01$ & 0.00 & $-1.86 \pm 0.01$ & $-1.55 \pm 0.01$ & $8.4 \pm 0.08$ & 25 \\
\object{Mrk 178}     & PGC 35684, UGC 6541                                                   &  3.9 & 8            & -14.1 & 1.01 & $0.33 \pm 0.01$ & 0.00 & $-1.60 \pm 0.01$ & $-1.66 \pm 0.01$ & $7.7 \pm 0.02 $ & 42 \\
\object[VII Zw 403]{VIIZw 403}  & PGC 35286, UGC 6456                                                   &  4.4 & 22,23    & -14.3 & 1.11 & $0.52 \pm 0.02$ & 0.02 & $-1.71 \pm 0.01$ & $-1.67 \pm 0.01$ & $7.7 \pm 0.01$ & 25 \\
\enddata
\tablenotetext{a}{Selected alternate identifications obtained from NED.}
\tablenotetext{b}{Reference for the distance to the galaxy. If no reference is given, the distance was determined from the galaxy's radial velocity, given by de Vaucouleurs et al. (1991), corrected for infall to the Virgo Cluster (Mould \et\ 2000) and a Hubble constant of 73 km s$^{-1}$ Mpc$^{-1}$.}
\tablenotetext{c}{$R_H$ is the Holmberg radius, the radius of the galaxy at a $B$-band isophote, corrected for reddening, of 26.7 mag arcsec$^{-2}$. $R_D$ is the disk scale length measured from $V$-band images. (Table from Hunter \& Elmegreen 2006).}
\tablenotetext{d}{Foreground reddening from Burstein \& Heiles (1984).}
\tablenotetext{e}{$\Sigma_\mathrm{SFR}\mathrm{(H\alpha)}$ is the Star Formation Rate Density (SFRD) measured from H$\alpha$, calculated over the area $\pi R_D^2$, where $R_D$ is the disk scale length (Hunter \& Elmegreen 2004). $\Sigma_\mathrm{SFR}\mathrm{(FUV)}$ is the SFR density determined from {\it GALEX} FUV fluxes (Hunter et al.\ 2010, with an update of the  {\it GALEX} FUV photometry to the GR4/GR5 pipeline reduction).}
\tablenotetext{f}{Values in parentheses were determined from the empirical relationship between oxygen abundance and $M_B$ given by Richer \& McCall (1995) and are particularly uncertain.}
\tablenotetext{g}{Reference for the oxygen abundance.}
\tablerefs{
(1) Dalcanton et al. 2009;
(2) Karachentsev et al. 2004;
(3) Karachentsev et al. 2003a;
(4) Karachentsev et al. 2006;
(5) Dolphin et al. 2002;
(6) Sakai et al. 2004;
(7) Dolphin et al. 2003;
(8) Karachentsev et al. 2003b;
(9) Tolstoy et al. 1995a;
(10) Karachentsev et al. 2002;
(11) Meschin et al. 2009;
(12) Sakai et al. 1999;
(13) Pietrzynski et al. 2006;
(14) Miller et al. 2001;
(15) Freedman et al. 2001;
(16) Grocholski et al. 2008;
(17) Tolstoy et al. 1995b;
(18) Gieren et al. 2006;
(19) Momany et al. 2002;
(20) Gieren et al. 2008;
(21) Schulte-Ladbeck et al. 2001;
(22) Lynds et al. 1998;
(23) M\'endez et al. 2002;
(24) van Zee \& Haynes 2006;
(25) Hunter \& Hoffman 1999;
(26) Skillman, Kennicutt, \& Hodge 1989;
(27) Moustakas et al.\ 2010;
(28) Richer \& McCall 1995;
(29) van Zee et al. 2006;
(30) Kniazev et al. 2005;
(31) Croxall et al. 2009;
(32) Lee et al. 2003b;
(33) Skillman et al. 1997;
(34) Lequex et al.\ 1979;
(35) Lee et al. 2003a;
(36) Kobulnicky \& Skillman 1997;
(37) Gonz\'alez-Delgado et al.\ 1994;
(38) Moustakas \& Kennicutt (2006);
(39) Masegosa et al. 1991;
(40) Lee et al. 2005;
(41) Viallefond \& Thuan 1983;
(42) Gon\'zalez-Riestra \et\ 1988.
}
\end{deluxetable}
\end{landscape}
\restoregeometry
\section{OBSERVATIONS AND DATA REDUCTION}
\label{section:Paper1sample_and_data_reduction}

\subsection{Observations}

The LITTLE THINGS sample consists of 40 gas-rich dwarf galaxies within 11\,Mpc \cite[cf.][for sample details]{Hunter2012} and is listed in table~\ref{Table:tab-sample}. The sample spans 4\,dex in both SFR and gas mass, and a factor of 50 in metallicity.

Observations of the LITTLE THINGS sample were obtained (project ID: 12A-234) with the VLA at C-band (6\,cm: 4--8\,GHz) and in its C-configuration in 9 observing runs between March and May of 2012. All observing runs included one of four NRAO primary calibrators to calibrate the flux scale, and a calibrator within $10^{\circ}$ of each dwarf galaxy to correct the complex gain on timescales of around ${\rm 10\,minutes}$. For details of the various calibrators used see table~\ref{table:12A-234_Observations}. One of the primary goals of these observations is to resolve the faint low surface brightness emission associated with dwarf galaxies. C-configuration provided the best compromise between resolution and surface brightness sensitivity. We note that IC\,1613 is 0.7\,Mpc away and so has a large angular size. We utilised archival observations taken in D-configuration (project ID: AH1006) to minimise the effect of missing large scale emission for this galaxy. At C-band we expect a roughly equal mix of \RCT\ and \RCNT\ emission and sensitivity to spatial scales up to $\sim 4^\prime$. Given that most galaxies have angular sizes smaller than this we do not expect significant loss of large scale flux.

\begin{deluxetable}{ccccccccccc}
\tablewidth{0pt}
\tabletypesize{\tiny}
\tablecolumns{11}
\tablecaption{\Cband\,\,Observations and Imaging Properties of LITTLE THINGS\label{table:12A-234_Observations}}
\tablehead{  & \multicolumn{3}{c}{Observation}  & & \multicolumn{6}{c}{Imaging} \\ \\ \cline{2-4} \cline{6-11} \\ \colhead{Galaxy} & \colhead{Date} & \colhead{Flux Cal.} & \colhead{Gain Cal.} & & \multicolumn{2}{c}{Phase Centre} & \colhead{Scale} & \colhead{Res.} & \colhead{Noise} & \colhead{Notes} \\ \\ \cline{6-7} \\ \colhead{Name} &  & \colhead{Name} & \colhead{Name} & & \colhead{R.A} & \colhead{Dec.} & \colhead{pc arcsec$^{-1}$} & \colhead{arcsec} & \colhead{$\mu$Jy\,beam$^{-1}$} & \\ (1)                            &         (2)                 &        (3)                     & (4)                            &  & (5)                        & (6)                         & (7)                                                 & (8)                            & (9)           & (10)      }
\startdata
 CVn I dwA  & 2012 Mar 17 & 3C286 & J1310+3220& & $12\,38\,40.2$ & $+32\,45\,40$ & 6.3  & $3.0 \times 3.0$ & $4.3$ & N\\
 DDO 43 & 2012 Mar 22 & 3C286 & J0818+4222    & & $07\,28\,17.8$ & $+40\,46\,13$ & 8.5  & $2.5 \times 2.3$ & $6.9$ & R,S\\
 DDO 46 & 2012 Mar 22 & 3C286 & J0818+4222    & & $07\,41\,26.6$ & $+40\,06\,39$ & 8.5  & $3.0 \times 2.8$ & $5.1$ & N\\
 DDO 47 & 2012 Mar 20 & 3C286 & J0738+1742    & & $07\,41\,55.3$ & $+16\,48\,08$ & 8.0  & $3.2 \times 3.0$ & $5.0$ & N\\
 DDO 50 & 2012 Mar 17 & 3C147 & J0841+7053    & & $08\,19\,08.7$ & $+70\,43\,25$ & 5.2  & $4.4 \times 3.5$ & $5.5$ & N,S\\
 DDO 52 & 2012 Mar 22 & 3C286 & J0818+4222    & & $08\,28\,28.5$ & $+41\,51\,21$ & 9.3  & $2.2 \times 2.0$ & $8.3$ & R,S\\
 DDO 53 & 2012 Mar 16 & 3C147 & J0841+7053    & & $08\,34\,08.0$ & $+66\,10\,37$ & 5.6  & $4.9 \times 4.0$ & $5.4$ & N\\
 DDO 63 & 2012 Mar 25 & 3C286 & J0841+7053    & & $09\,40\,30.4$ & $+71\,11\,02$ & 5.9  & $6.1 \times 3.4$ & $4.6$ & N\\
 DDO 69 & 2012 Mar 20 & 3C286 & J0956+2515    & & $09\,59\,25.0$ & $+30\,44\,42$ & 1.2  & $4.1 \times 3.6$ & $4.0$ & N\\
 DDO 70 & 2012 Mar 20 & 3C286 & J0925+0019    & & $10\,00\,00.9$ & $+05\,19\,50$ & 2.0  & $4.5 \times 3.4$ & $5.8$ & N\\
 DDO 75 & 2012 Mar 20 & 3C286 & J1024$-$0052    & & $10\,10\,59.2$ & $-04\,41\,56$ & 2.0  & $3.3 \times 2.4$ & $9.7$ & N,S\\
 DDO 87 & 2012 Mar 25 & 3C286 & J1048+7143    & & $10\,49\,34.7$ & $+65\,31\,46$ & 10.3 & $3.8 \times 2.2$ & $6.2$ & R\\
 DDO 101  & 2012 Mar 17 & 3C286 & J1221+2813  & & $11\,55\,39.4$ & $+31\,31\,08$ & 13.9 & $3.1 \times 3.0$ & $15.1$ & R,S,P\\
 DDO 126  & 2012 Apr 05 & 3C286 & J1215+3448  & & $12\,27\,06.5$ & $+37\,08\,23$ & 7.6  & $4.6 \times 4.0$ & $5.4$ & N,S\\
 DDO 133  & 2012 Mar 17 & 3C286 & J1310+3220  & & $12\,32\,55.4$ & $+31\,32\,14$ & 9.4  & $3.8 \times 3.7$ & $4.4$ & N,S\\
 DDO 154  & 2012 Mar 17 & 3C286 & J1310+3220  & & $12\,54\,06.2$ & $+27\,09\,02$ & 6.6  & $2.2 \times 2.2$ & $7.3$ & R,\\
 DDO 155  & 2012 Mar 17 & 3C286 & J1309+1154  & & $12\,58\,39.8$ & $+14\,13\,10$ & 3.4  & $3.8 \times 3.5$ & $4.7$ & N\\
 DDO 165  & 2012 Mar 25 & 3C286 & J1313+6735  & & $13\,06\,25.3$ & $+67\,42\,25$ & 7.4  & $3.7 \times 2.8$ & $4.5$ & R\\
 DDO 167  & 2012 Apr 20 & 3C286 & J1327+4326  & & $13\,13\,22.9$ & $+46\,19\,11$ & 6.5  & $3.3 \times 3.0$ & $5.1$ & N\\
 DDO 168  & 2012 Apr 20 & 3C286 & J1327+4326  & & $13\,14\,27.2$ & $+45\,55\,46$ & 5.4  & $3.9 \times 3.5$ & $4.4$ & N\\
 DDO 187  & 2012 Mar 17 & 3C286 & J1407+2827  & & $14\,15\,56.7$ & $+23\,03\,19$ & 3.9  & $2.7 \times 2.5$ & $6.9$ & R,S\\
 DDO 210  & 2012 May 19 & 3C48  & J2047$-$1639  & & $20\,46\,52.0$ & $-12\,50\,51$ & 1.4  & $3.1 \times 1.7$ & $4.6$ & R\\
 DDO 216  & 2012 Mar 31 & 3C48  & J2253+1608  & & $23\,28\,35.0$ & $+14\,44\,30$ & 1.4  & $3.1 \times 2.9$ & $5.1$ & R\\
 F564-V03 & 2012 Mar 20 & 3C286 & J0854+2006  & & $09\,02\,53.9$ & $+20\,04\,29$ & 9.6  & $3.3 \times 3.0$ & $5.4$ & N\\
 Haro 29  & 2012 Apr 20 & 3C286 & 1219+484    & & $12\,26\,16.7$ & $+48\,29\,38$ & 8.3  & $3.9 \times 3.6$ & $5.1$ & N\\
 Haro 36  & 2012 Apr 20 & 3C286 & 1219+484    & & $12\,46\,56.3$ & $+51\,36\,48$ & 13.9 & $3.9 \times 3.6$ & $5.2$ & N\\
 IC 1613  & 2010 Aug 19 & 3C48  & J0108+0135  & & $01\,04\,49.2$ & $+02\,07\,48$ & 1.1  & $9.3 \times 7.8$ & $5.1$ & R\\
 IC 10  & 2012 Apr 28 & 3C84  & J0102+5824    & & $00\,20\,17.3$ & $+59\,18\,14$ & 1.5  & $2.6 \times 2.3$ & $7.8$ & R\\
 LGS 3  & 2012 Mar 31 & 3C48  & J0112+2244    & & $01\,03\,55.2$ & $+21\,52\,39$ & 0.9  & $3.0 \times 2.8$ & $5.5$ & R\\
 M81 dwA  & 2012 Mar 17 & 3C147 & J0841+7053  & & $08\,23\,57.2$ & $+71\,01\,51$ & 5.6  & $2.7 \times 1.9$ & $10.8$ & R,S,P\\
 Mrk 178  & 2012 Apr 20 & 3C286 & 1219+484    & & $11\,33\,29.0$ & $+49\,14\,24$ & 6.0  & $4.4 \times 4.0$ & $9.3$ & N\\
 NGC 1569 & 2012 Mar 16 & 3C147 & J0449+6332  & & $04\,30\,49.8$ & $+64\,50\,51$ & 3.9  & $2.7 \times 2.3$ & $6.8$ & R\\
 NGC 2366 & 2012 Mar 16 & 3C147 & J0841+7053  & & $07\,28\,48.8$ & $+69\,12\,22$ & 4.9  & $4.2 \times 3.4$ & $5.1$ & N\\
 NGC 3738 & 2012 Apr 20 & 3C286 & J1146+5356  & & $11\,35\,49.0$ & $+54\,31\,23$ & 7.6  & $2.5 \times 2.5$ & $7.6$ & N,S\\
 NGC 4163 & 2012 Apr 05 & 3C286 & J1215+3448  & & $12\,12\,09.2$ & $+36\,10\,13$ & 4.3  & $3.3 \times 2.9$ & $4.5$ & N\\
 NGC 4214 & 2012 Apr 05 & 3C286 & J1215+3448  & & $12\,15\,39.2$ & $+36\,19\,38$ & 4.5  & $4.5 \times 4.0$ & $6.3$ & N,S\\
 Sag DIG  & 2012 May 19 & 3C48  & J1911$-$2006  & & $19\,30\,00.6$ & $-17\,40\,56$ & 1.7  & $3.5 \times 1.4$ & $8.2$ & R\\
 UGC 8508 & 2012 Apr 20 & 3C286 & J1349+5341  & & $13\,30\,44.9$ & $+54\,54\,29$ & 4.0  & $2.6 \times 2.5$ & $6.0$ & N\\
 VIIZw 403  & 2012 Mar 25 & 3C286 & J1153+8058& & $11\,27\,58.2$ & $+78\,59\,39$ & 6.8  & $5.8 \times 3.7$ & $5.8$ & N\\
 WLM  & 2012 May 19 & 3C48  & J2348$-$1631      & & $00\,01\,59.2$ & $-15\,27\,41$ & 1.5  & $5.0 \times 1.5$ & $5.3$ & R\\
\enddata
\tablecomments{ (Column 1) Name of dwarf galaxy observed; 
(Column 2) Date of observation; 
(Column 3) Name of primary calibrator; 
(Column 4) Name of secondary calibrator; 
(Columns 5 \& 6) J$2000$ equatorial coordinate of observation (dwarf galaxy) phase centre; 
(Column 7) Physical scale at distance of galaxy; 
(Column 8) Resolution of image. Note that some images were made using {\sc robust=0.0} and others using {\sc robust=+2.0} where CASA robust values range between $-2.0$ (uniform weighting) and $+2.0$ (natural weighting); 
(Column 9) rms noise; 
(Column 10) Comments regarding deviations from the typical imaging process: R signifies that the {\sc clean} algorithm was performed using {\sc robust=0.0} weighting, whereas N signifies an approach closer to natural weighting; S means that the generated image benefited from self-calibration; P refers to those images that were strongly affected by a bright, nearby background source of $\sim 0.1{\rm \,Jy}$ which was located such that it entered the sidelobes of the primary beam. Because of the Alt--Az mounting of the VLA antennas, the primary beam rotates on the sky making the detected signal time varying; self-calibration failed as a result. To minimise the effect of the offending source, only about a quarter of the bandwidth was used, using those spectral windows in which the first null of the primary beam coincides as near as possible to the offending source.}
\end{deluxetable}

\subsection{Radio Continuum Calibration \& Imaging}
We calibrated the data using the Common Astronomy Software Applications ({\sc casa}\footnote{http://casa.nrao.edu/}; \citealt{McMullin2007}) package following standard procedures that we present in the following subsections.

\subsubsection{Flagging}
Before calibration we used the {\sc tflagdata} task to apply two automatic flagging algorithms: {\sc tfcrop} \citep{Rau2003} and {\sc rflag} (based on {\sc aips}; \citealt{Greisen2011}). The {\sc tfcrop} algorithm identifies outliers by splitting each baseline into `chunks' along the frequency--domain (each channel) and time--domain (every $50$ seconds). The amplitude of all visibilities within a given chunk were averaged and then any chunks with an amplitude greater than $4\,\sigma_{\rm pre}$ from the mean were flagged. Here, $\sigma_{\rm pre}$ refers to the pre-calibration dispersion of amplitudes around the mean. We opted for a conservative threshold value as, at this point, we were only concerned with removing extremely high--amplitude data such that subsequent steps in the calibration would not be affected. The {\sc rflag} algorithm detects outliers by using a sliding window in the time and then spectral window domain to determine local statistics and identify data that exceed $4\,\sigma_{\rm pre}$. The algorithm first calculates the local rms within each sliding window. It then calculates the median rms across the time windows, deviations of the local rms from this median, and the median deviation. Data is flagged if the local rms is larger than $4\times(\rm medianRMS + medianDev)$. For a more in depth description of these algorithms see \cite{Rau2003} and \cite{Greisen2011}. Bad baselines, scans and channels, as well as wide--band radio frequency interference (RFI) were generally caught by the algorithms although the measurement sets were manually checked to identify any discrepant visibilities that were missed. This approach typically resulted in the removal of $15$--$20$\% of the observed visibilities. 

\subsubsection{Calibration}
The flux scale in our images was set using one of recommended VLA primary flux calibrators given in column\,3 of table~\ref{table:12A-234_Observations} using the task {\sc setjy}. This flux calibrator was also bright enough to be used to correct for the bandpass shape using the task {\sc bandpass}. Calibration of the time-dependant complex gain was achieved by regular observations of a nearby gain calibrator (table~\ref{table:12A-234_Observations}; column\,4) using {\sc gaincal}.

Once calibration was completed, each measurement set was inspected a final time for low--level RFI. First, a manual check was performed to flag baselines, scans, or channels that exhibited deviant amplitudes or phases. In addition to this, a second round of automated flagging was performed (this time designed to catch outliers greater than $3.5\,\sigma_{\rm post}$ from the mean). Here, $\sigma_{\rm post}$ refers to the post--calibration dispersion of amplitudes around the mean. This flagging on the calibrated data often reduced the rms noise in subsequent imaging by a further $\sim 10$\% (compared to when this second round of flagging was omitted).

\subsubsection{Imaging}
\label{Section:paper1_Imaging}

We generated images of our targets using the {\sc casa} {\sc clean} task, using the Multi-Scale, Multi-Frequency Synthesis (MS-MFS) algorithm developed by \cite{Rau2011}. Due to the various angular scales of emission observed in the galaxies, the cleaning scales chosen were unique to each observation to give the optimum {\sc clean} map. At least two scales of $1$ and $3$ times the synthesised beam width were used. In a few cases larger angular scales were added to deal with large--scale emission in the brighter, more extended galaxies such as DDO\,50 and NGC\,1569.

Due to the faint nature of the dwarfs, observations were generally Fourier-transformed using natural weighting ({\sc robust=+2.0}). This ensured we optimised our images for S/N. Some observations were mapped using Brigg's robust imaging method ({\sc robust=0.0}) because either 1) the galaxy was sufficiently bright that a high enough S/N was reached using {\sc robust=0.0} weighting, or 2) the natural weighting {\sc clean} left significant image artefacts throughout the image due to the rather sparse sampling of the ($u,v$) plane. Using the Brigg's {\sc robust=0.0} method ensures that the image is not dominated by visibilities representing the more numerous short baselines. This method increases the resolution, results in a synthesised beam that more closely resembles a Gaussian shape, and improves the image quality but at the expense of a slight (${\rm \sim 20\%}$) increase of the rms noise. Typical rms noise values in these cleaned images fell between $4$ and $8{\rm \,\mu Jy\,beam^{-1}}$ in close agreement with expected values. Table~\ref{table:12A-234_Observations} states whether the image of the galaxy was generated using {\sc robust=0.0} weighting (R) or an approach closer to natural weighting (N).

Self-calibration (phase only) was performed on 11 of our 40 observations to improve the dynamic range across the image; these are marked in table~\ref{table:12A-234_Observations} (S symbol in column 10). In only one case (NGC\,4214) the emission originating from the galaxy itself produced strong enough artefacts to warrant self-calibration; in all other cases, the offending source was an unresolved background object.

Observations of DDO\,101 and M81\,DwA (marked in table~\ref{table:12A-234_Observations}) harboured the strongest background sources in our survey. These sources have a flux density of $> 0.1{\rm \,Jy}$ and are located approximately $9^\prime$ and $6^\prime$ from the observation's phase centre, respectively. Self-calibration was not successful in sufficiently improving the dynamic range for these images. This is due to a combination of both offending sources residing near the edge of the primary beam combined with the VLA antennas operating on an Alt-Az mount. This causes the offending sources to have a time-varying signal due to the source passing through the sidelobes of the primary beam. The result is that the MS--MFS {\sc clean} algorithm cannot successfully remove the sidelobes of the confusing source. Since these sources are not of interest to our project --- they lie beyond the FWHM of the primary beam anyway --- we decided to select solely the spectral windows least affected by the offending background source, i.e., by choosing $2$ or $3$ spectral windows for which the first null of the primary beam fell close to the offending source. In doing this, the rms noise was approximately doubled to $15{\rm \,\mu Jy\,beam^{-1}}$ while the side lobes of the confusing source were considerably suppressed. We note that in an earlier study, \cite{Stil2002} do not list a flux density for DDO\,101 for the same technical reason.

We maintained as much consistency as possible by using the same calibration and imaging pipeline for all observations. Our images prior to primary beam correction had a flat noise background lacking in significant structure. Very few images had artefacts from nearby strong ($>0.5{\rm \,mJy}$) sources. Those that did had the offending regions masked manually. Our residual maps comprise a Gaussian intensity distribution consistent with pure noise, having an average of $0$ and variance of $\sigma$ suggesting that the MS-MFS algorithm successfully modelled all genuine emission present in the ($u$,$v$) data. Only NGC\,1569 and NGC\,4214 showed any indication of sitting in a negative bowl, suggesting they suffer from missing flux on the largest scales (see Section~\ref{Section:MissingStructures} for further discussion). The observations and imaging properties of all LITTLE THINGS targets are summarised in table~\ref{table:12A-234_Observations}. Notes on the data reduction of individual galaxies can be found in Appendix\label{section:Appendix_IndividualNotes}.

\subsection{Ancillary Data}
\label{section:Paper1AncillaryData}

The LITTLE THINGS project has acquired a large collection of multi-wavelength and spatially resolved data on each of the $40$ dwarf galaxies \cite[see][for details]{Hunter2012,Zhang2012}. We make use of the following ancillary data in this study:
\begin{itemize}
\item \halpha\ line emission: the FWHM of the filter used for the \halpha\ observations was $30$\,\AA\ centred on $6562.8$\,\AA\ \cite[][]{Hunter2004}, while the FWHM of the point spread function (PSF) was $\sim2^{\prime\prime}$. The maps were continuum subtracted and the fluxes corrected for [NII] contribution. \cite{Hunter2012} used \cite{Burstein1982} values to correct \halpha\ and FUV maps for foreground reddening. Internal extinction in dwarf galaxies can generally be ignored because they have low--metallicity and consequently a low dust-to-gas ratio with respect to spirals \citep{Ficut2016}. However, internal extinction may be important in some of the more actively star forming dwarfs. We discuss this further in Section~\ref{sect:rccomp};
\item Far-Ultraviolet broadband emission: the FUV data were taken with {\em GALEX} in the $1350$--$1750$ \AA\ band (effective wavelength of $1516$ \AA) with a resolution of $4^{\prime\prime}$ at the FWHM. The data were calibrated with the GR4/5 pipeline except DDO\,165 and NGC\,4214 which were processed with the GR6 pipeline \cite[][]{Zhang2012}. The resulting images have been sky subtracted and geometrically transformed to match the optical {\em V}-band orientation. UGC\,8508 was not observed due to bright foreground stars, and IC\,10 was not observed due to its low Galactic latitude placing it in a region of high extinction. For surface brightness measurements, and hence for extended emission, the estimated uncertainty for the {\em GALEX} FUV maps is $0.15$\,mag \cite[][]{GildePaz2007};
\item Infrared (IR) broadband emission: the IR data were taken with the {\em Spitzer} space telescope using the Multiband Imaging Photometer for Spitzer (MIPS). The two bands used were mid-infrared (MIR), with an effective wavelength of 24\micron\ with a resolution of $6^{\prime\prime}$ at FWHM, and FIR with an effective wavelength of 70\micron\ and a resolution of $17^{\prime\prime}.5$ at FWHM. The {\em Spitzer} $24$ and 70\micron\ maps were taken from either the Local Volume Legacy (LVL) survey \cite[see][for details]{Dale2009} or the {\em Spitzer} Infrared Nearby Galaxies Survey (SINGS). A pixel-dependent background subtraction was performed and images were convolved with a custom kernel to make a near Gaussian PSF. For the Spitzer 24\micron\ maps, the photometric uncertainty is 2\% for both unresolved sources and extended emission \cite[][]{Engelbracht2007}.
\end{itemize}

\section{RESULTS}
\label{section:Paper1Results}

\label{section:Paper1Images}
\label{section:Paper1radio_continuum_images}
\begin{figure*}
  \begin{tabular}{ccc}
    \includegraphics[width=0.31\linewidth,clip]{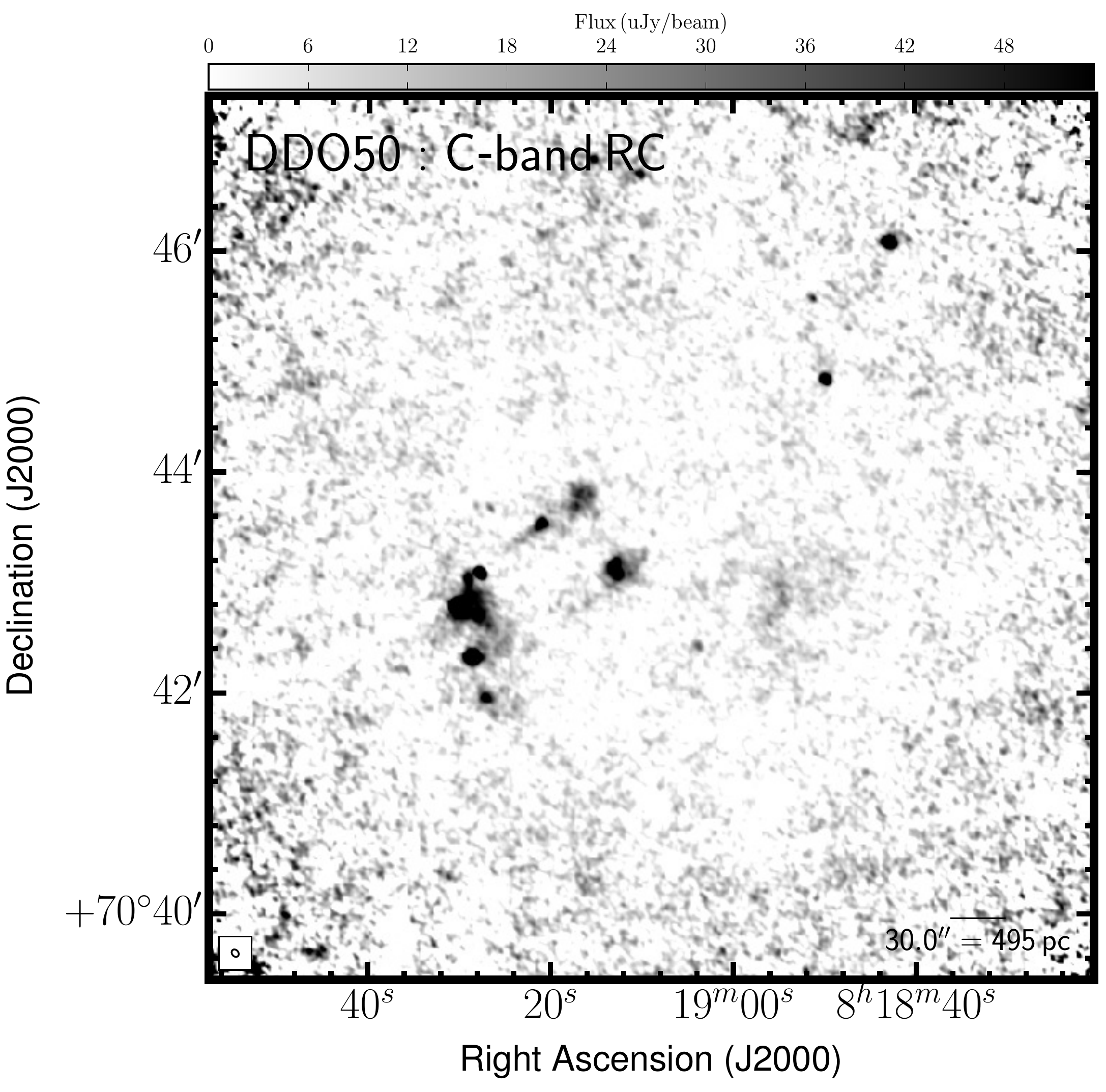} & \ 
    \includegraphics[width=0.31\linewidth,clip]{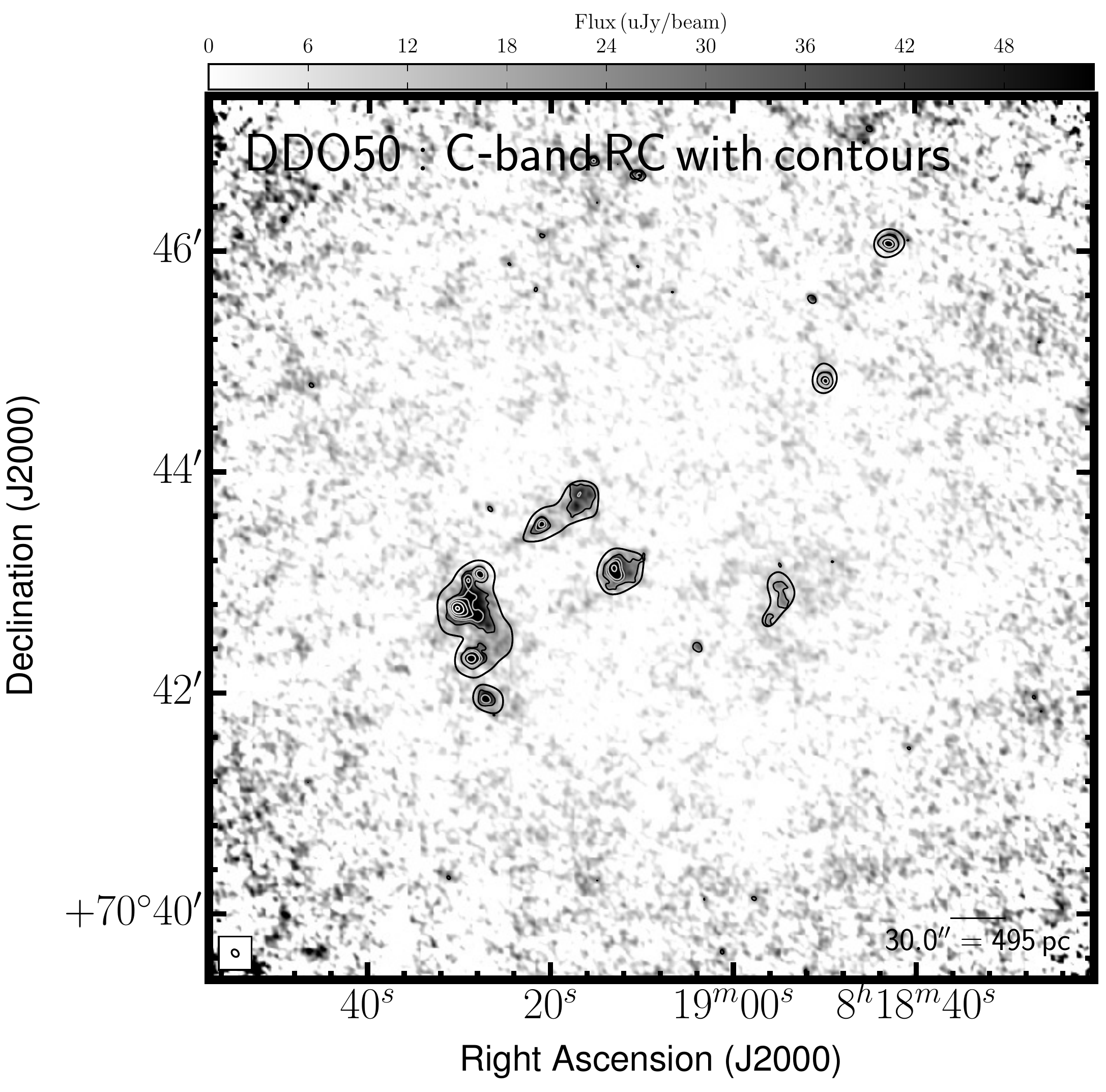} & \ 
    \includegraphics[width=0.31\linewidth,clip]{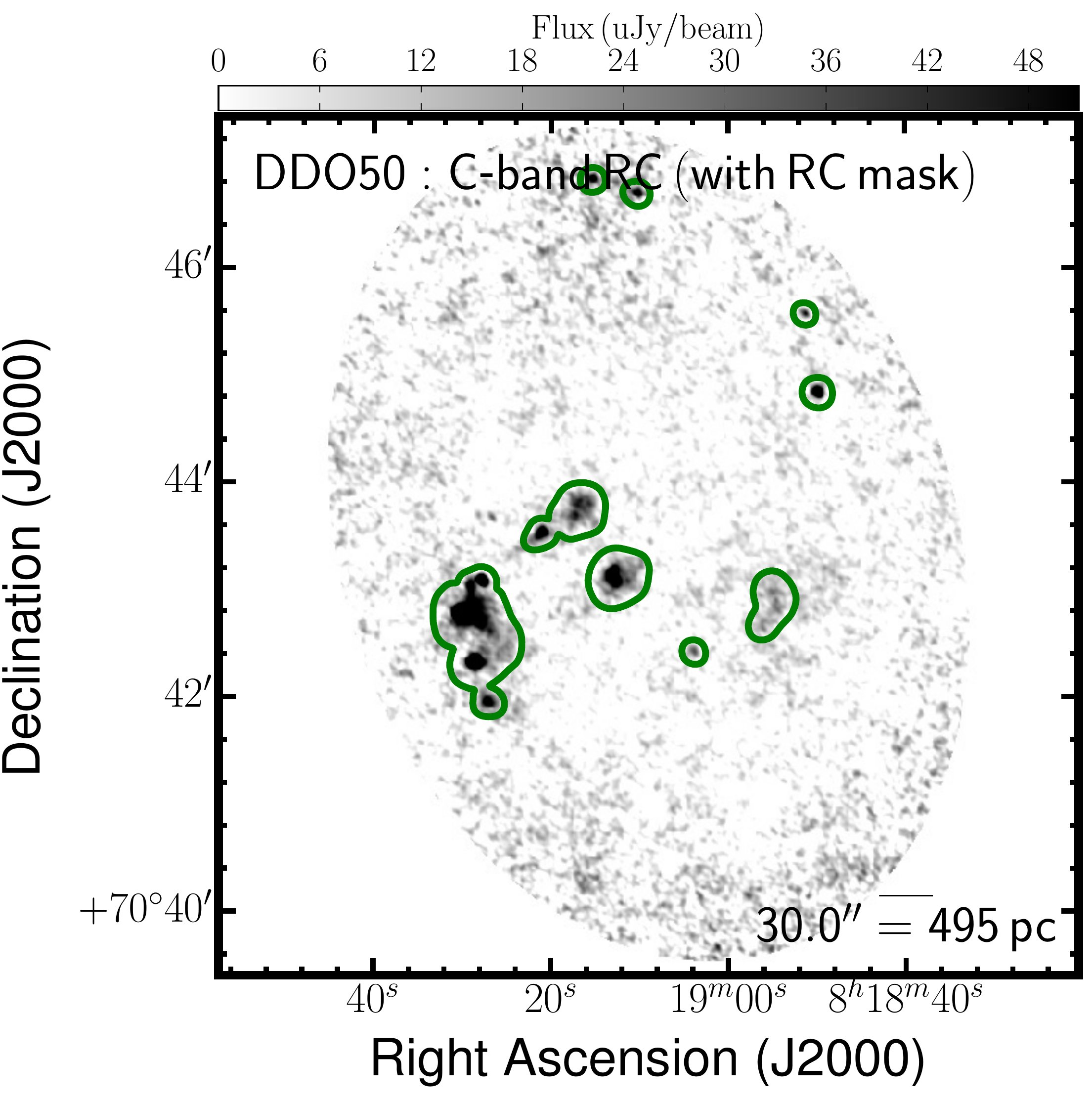} \\
    \includegraphics[width=0.31\linewidth,clip]{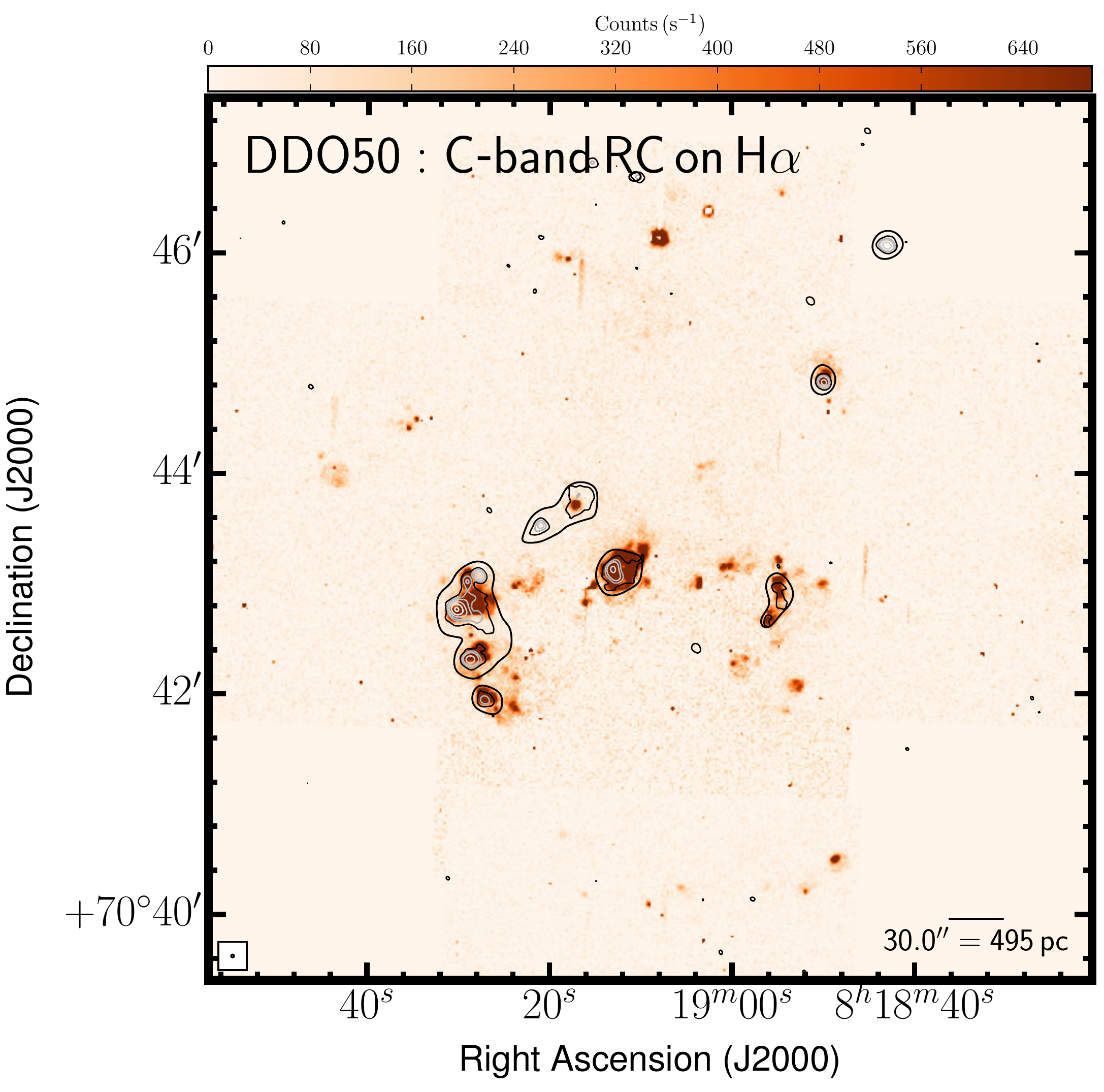} & \ 
    \includegraphics[width=0.31\linewidth,clip]{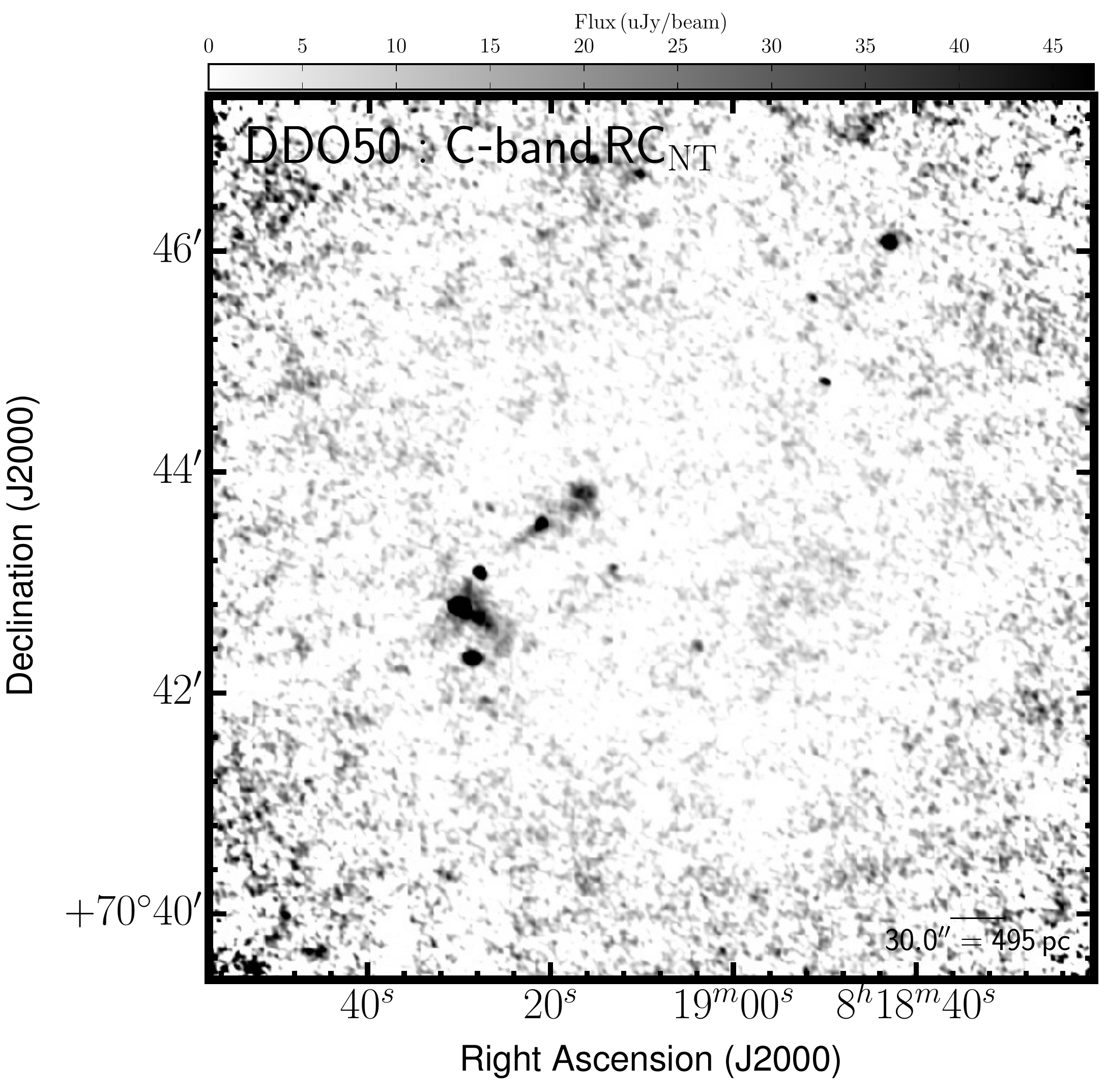} & \ 
    \includegraphics[width=0.31\linewidth,clip]{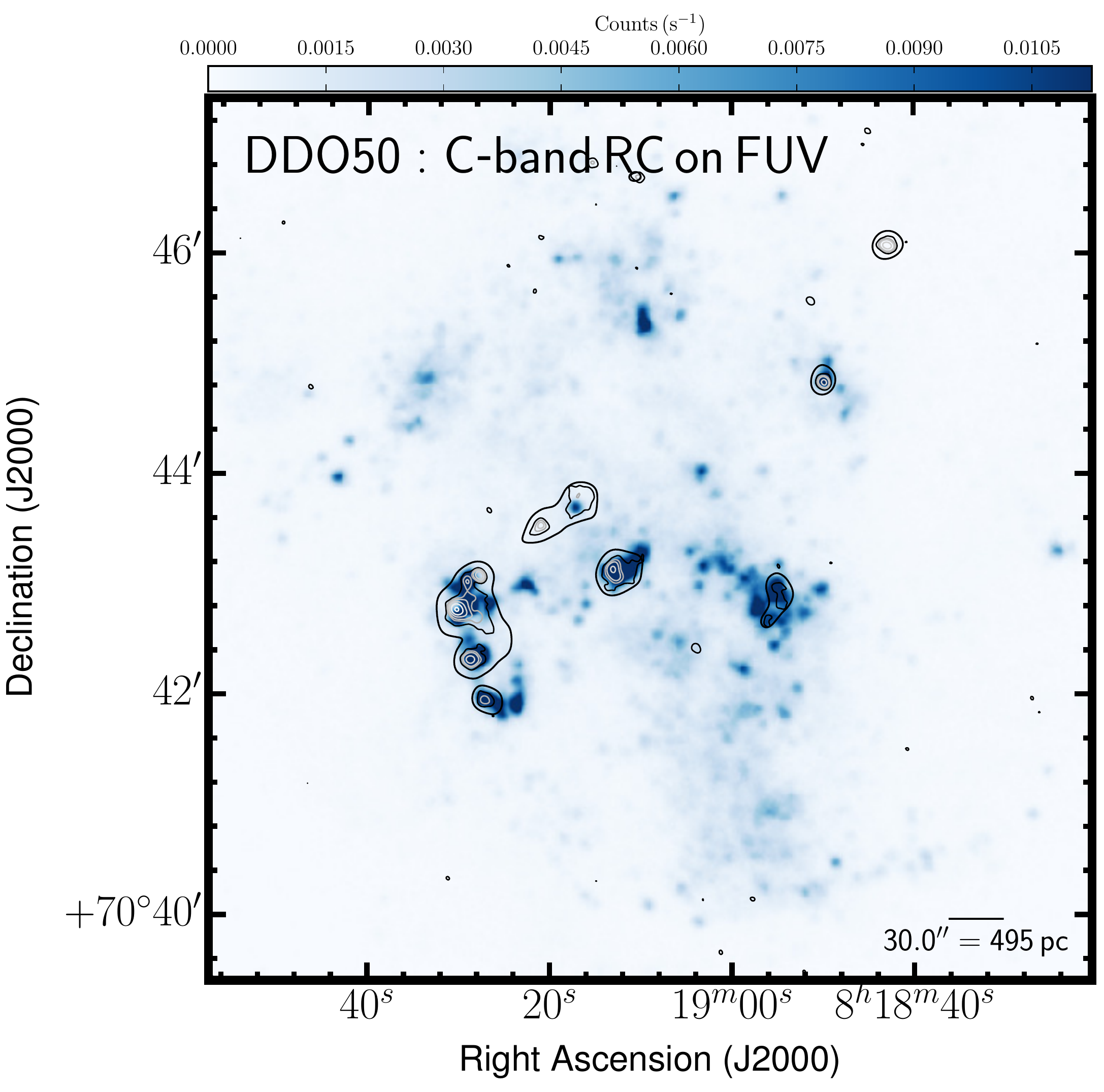} \\
    \includegraphics[width=0.31\linewidth,clip]{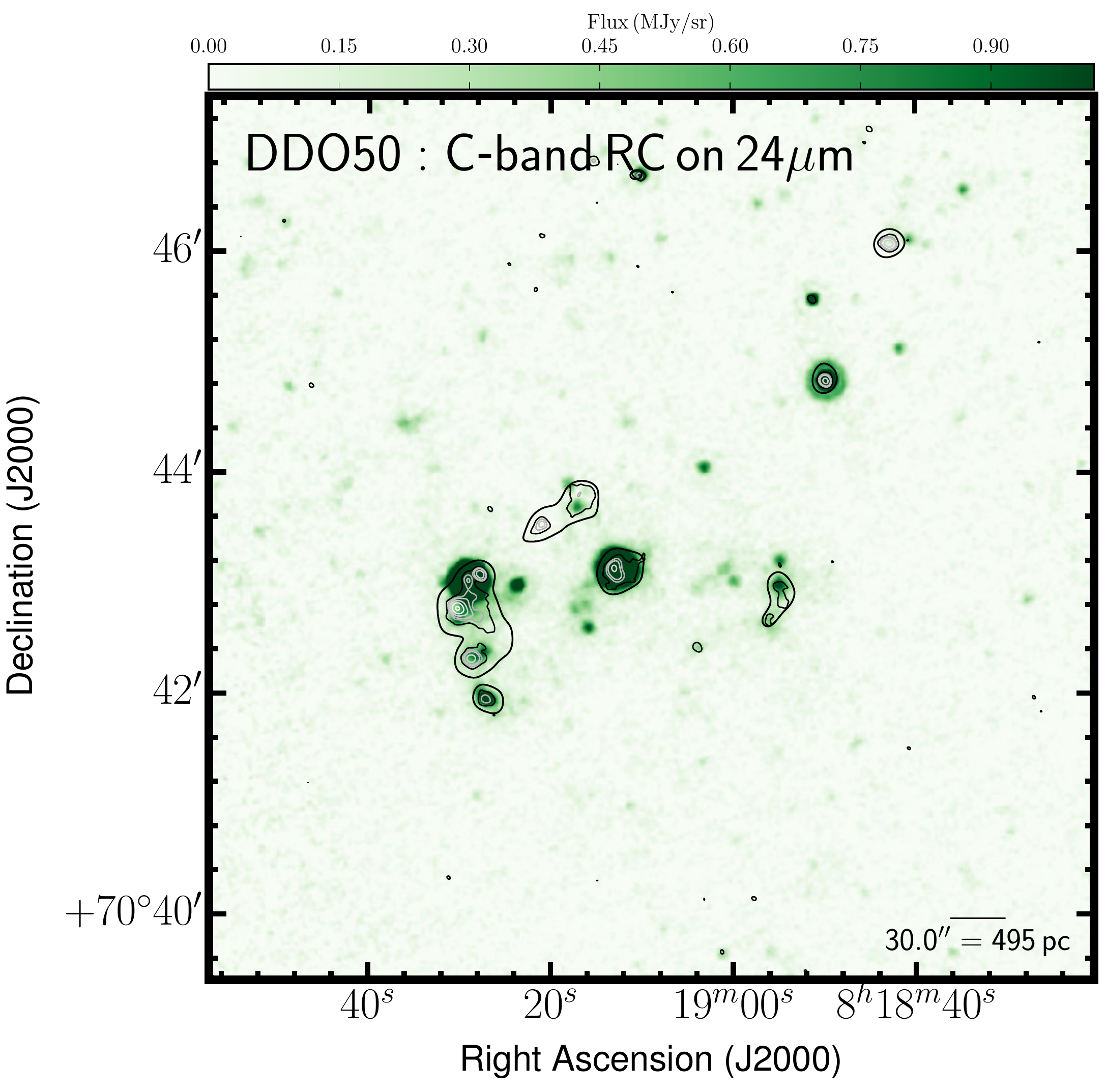} & \ 
    \includegraphics[width=0.31\linewidth,clip]{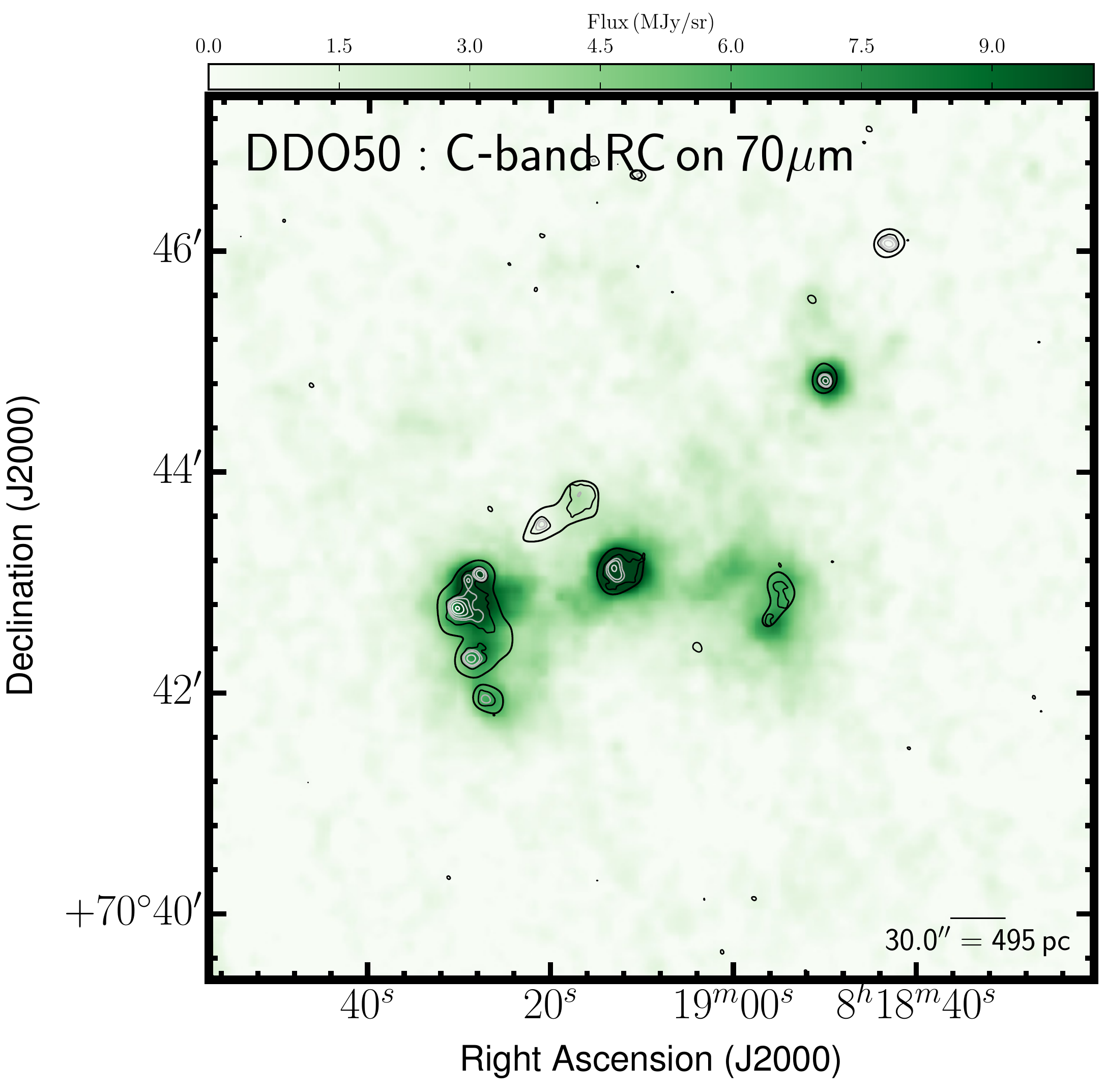} & \ 
    \includegraphics[width=0.31\linewidth,clip]{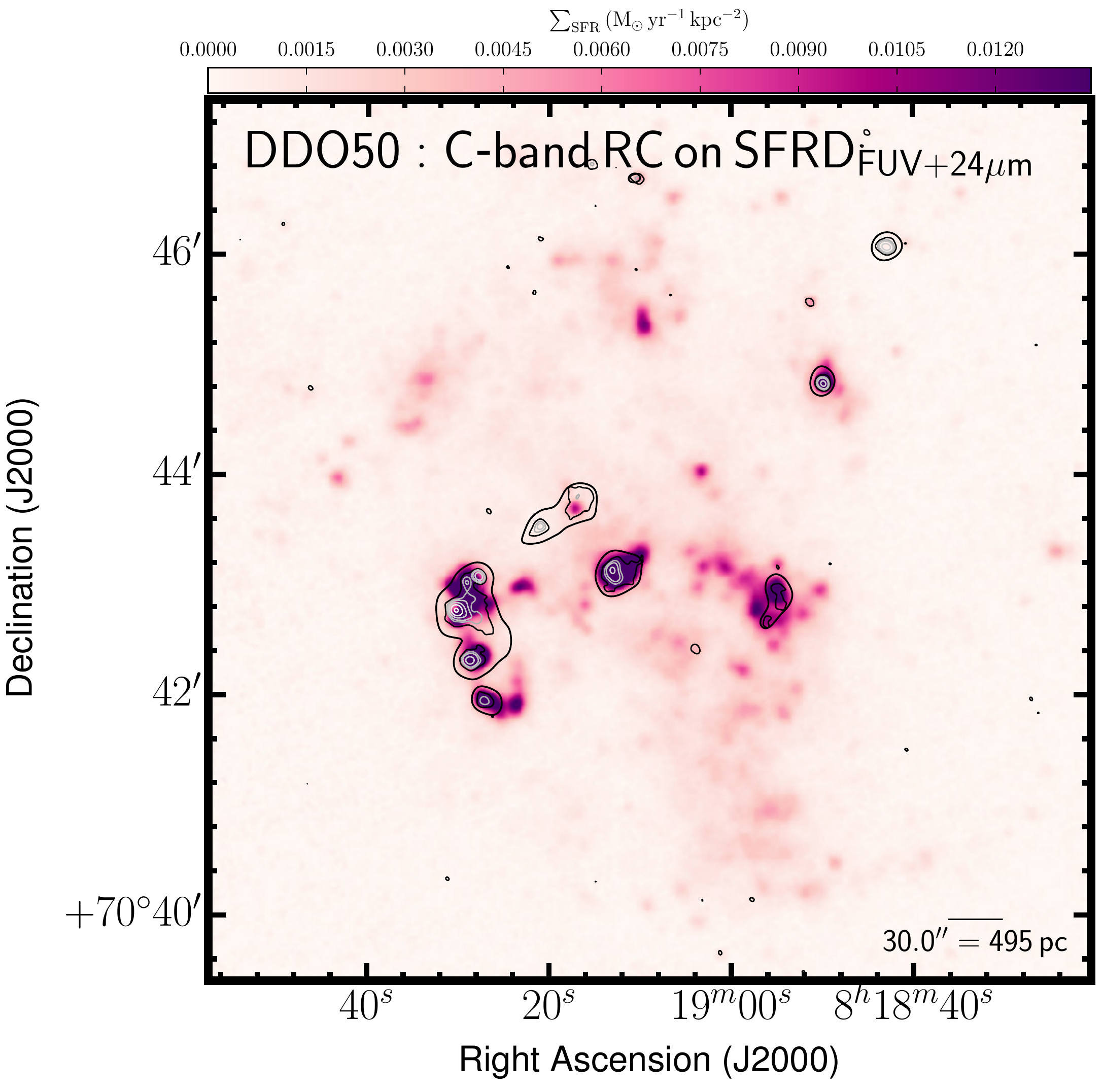} \\
  \end{tabular}
\caption{Multi-wavelength coverage of DDO\,50 displaying an $8.0^\prime \times 8.0^\prime$ area. We show total RC flux density at the native resolution (top-left) and again with contours (top-centre). The lowest contour highlights low--surface brightness emission at a S/N level of 3 in the image smoothed to $10''$. The remaining contours are at S/N levels of 3, 6, 9, and then multiples of twice the previous contour level from our native resolution images. These contours are also superposed on ancillary LITTLE THINGS images where possible: \halpha\ (middle-left); \RCNT\ (middle-centre); {\em GALEX} FUV (middle-right); {\em Spitzer} 24\micron\ (bottom-left); {\em Spitzer} 70\micron\ (bottom-centre); FUV$+24{\rm \mu m}$--inferred  SFRD (bottom-right). We also show the RC that is isolated by the RC--based and disk masking technique (top-right). In this panel the green contours outline the RC mask and includes background and ambiguous sources. The elliptical outline corresponds to the area hence forth referred to as the disk mask.}
  \label{figure:example_maps}
\end{figure*}

We present an example of our multi-wavelength data set in Fig.~\ref{figure:example_maps}, which shows our data for DDO\,50. This includes the results of our RC observations and contours overlaid onto the \halpha, FUV, and 24\micron\ images. Multi-wavelength images for our entire sample can be found in the online electronic version in Appendix~\ref{Section:Appendix_Images}. 


\subsection{Identifying Emission unrelated to the Target Object}
\label{section:Paper1background_source_removal}

Contamination by background sources in the RC is an issue since their emission is often brighter than, or similar to, the emission originating from the dwarf galaxy \citep{Padovani2011}. Low resolution observations reported in the literature are predominantly from single dish observations and will have suffered from contamination to varying degrees. Our resolved maps make it possible to remove the effects of contamination by identifying emission unrelated to our galaxies. 

We inspected each of our RC images and classified features in a manner similar to \cite{Chomiuk2009}. Flux was attributed as originating from either: %
\begin{itemize}
\item the dwarf galaxy (exactly coincident with a SF tracer); %
\item a background galaxy; %
\item ambiguous emission of unknown origin (i.e., unable to discern between a) background origin, or b) non-thermal emission from unresolved SNRs or diffuse non--thermal emission). 
\end{itemize}

\noindent We applied a two step process to classify the RC emission in our images into these three catagories. First, we cross-matched our RC sources with the literature. Following this we applied a procedure designed to isolate RC emission features from background galaxies based on their proximity to \halpha\ emission. We describe these two steps in more detail below. 

\subsubsection{Cross-matching with line-of-sight Optical Counterparts} 
We manually cross matched unresolved sources of RC emission with the NASA/IPAC Extragalactic Database\footnote{\scriptsize{\url{http://ned.ipac.caltech.edu/forms/nearposn.html}}} (NED). If an archived galaxy was found within $2^{\prime\prime}$ (approximately half the FWHM of the synthesised beam at the native resolution) of the unresolved RC source, we characterised that source as a background galaxy.

\subsubsection{Isolating obvious Background Galaxies}
RC emission coming from the same line-of-sight as the \halpha\ emission from \hii\ regions was assumed to originate from the dwarf galaxy. All galaxies in our sample have heliocentric velocities and rotational speeds \cite[][]{Hunter2012} that ensure all \halpha\ emission falls within the FWHM of the filter used, which is $30$\,\AA\ wide and centred on $6562.8$\,\AA\ \cite[][]{Hunter2004}. Unresolved background galaxies and SNRs look similar and share broadly similar values for their non--thermal spectral index, with values of $-0.85\pm0.13$ \citep{Niklas1996} and $-0.5 \pm 0.2$ \citep{Green2014}, for background galaxies and SNRs, respectively. SNRs from core--collapse supernovae are expected to be associated with SF regions in our dwarf galaxies. This is because the stellar velocity dispersion in dwarf galaxies is low \citep{Walker2007}, which implies that over the lifetime of a SNR, it will not have strayed very far from its host massive star cluster. Studies of dwarf galaxies have measured velocity dispersions of $\lesssim 10{\rm \,km\,s^{-1}}$ \citep{Walker2007,Mateo1998,Martin2007}, but the stellar velocity dispersion would be yet lower for the sub-population of high mass stars (i.e., the core-collapse supernova progenitors) since these would generally sink to the bottom of the parent cluster's gravitational potential. Based on the above we assume a stellar velocity dispersion of $5{\rm \,km\,s^{-1}}$ for the stars that eventually lead to the injection of \CRe\ (and the associated \RCNT\ emission). Given that a SN progenitor may live up to ${\rm 55\,Myr}$ and assuming a typical distance of 5\,Mpc, a SNR can travel a projected distance of $<250$\,pc or $<10^{\prime\prime}$ (for a face-on galaxy). Any significant RC source, unresolved or extended, that had little to no \halpha\ emission within this projected radius was marked as a background source and was removed by placing a mask over the source. For a Gaussian-like synthesised beam, $99$\% of the power of an unresolved source is contained within $3 \times {\rm FWHM_{\rm native}}$, and so this was the diameter of the mask placed over the background source. Even for a strong background source (e.g., $1{\rm \,m Jy}$), this removal technique leaves at most $10{\rm \,\mu Jy}$ unmasked in the image, whilst not masking out too much of the dwarf galaxy. 

\subsubsection{Ambiguous Sources} 
\begin{figure*}
  \begin{tabular}{cc}
    \includegraphics[width=0.48\linewidth,clip]{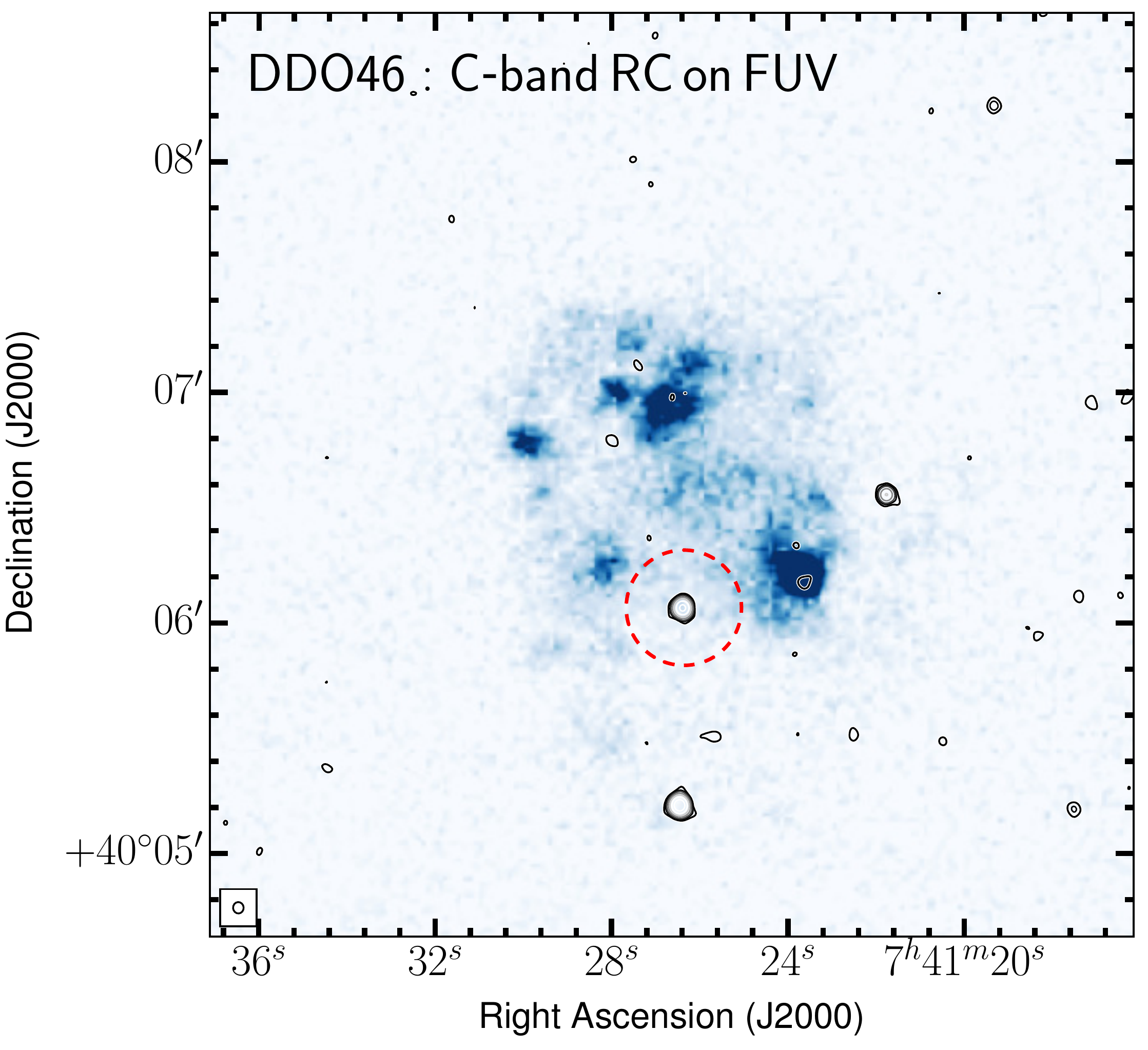} & \
    \includegraphics[width=0.48\linewidth,clip]{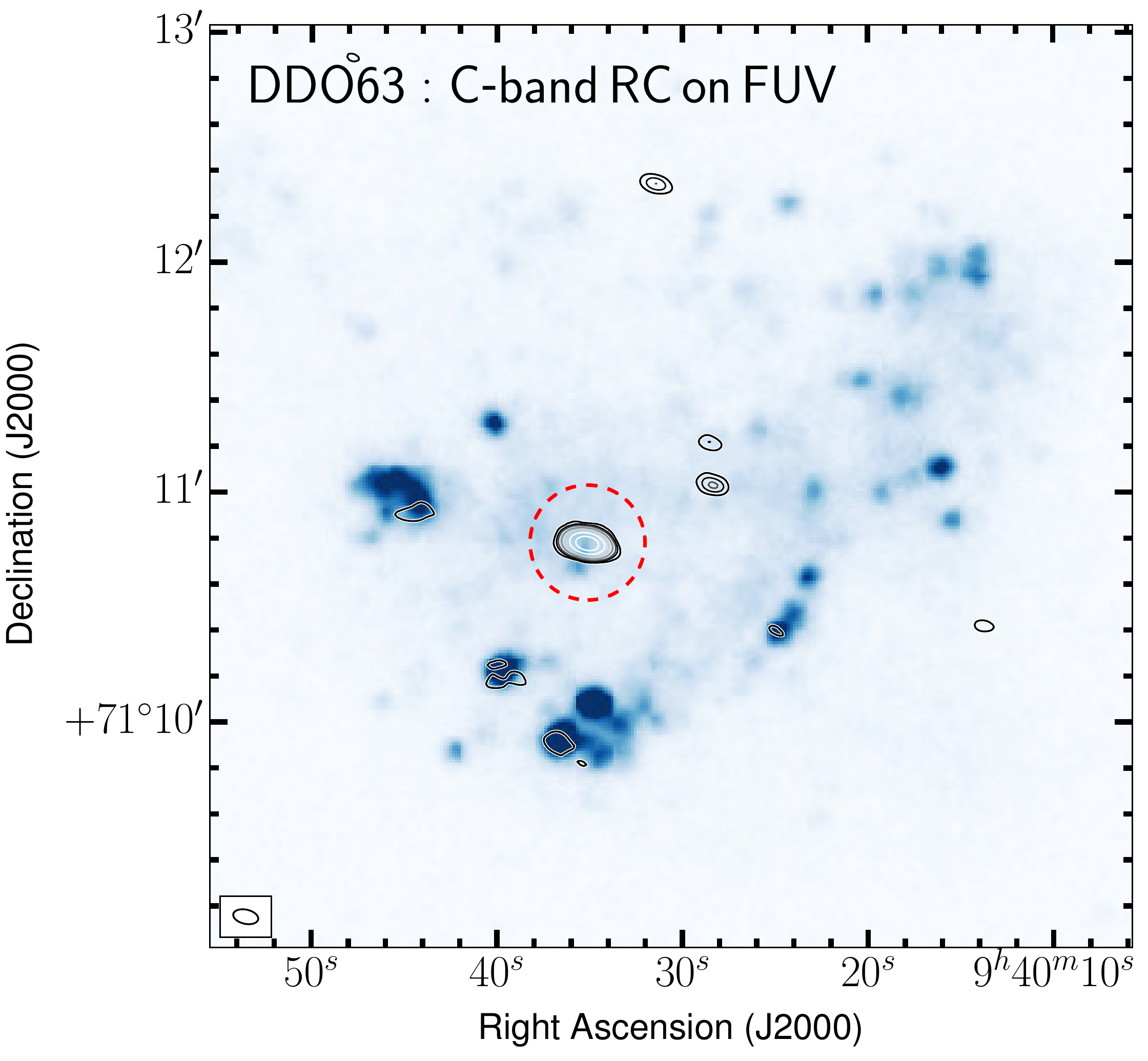} \\
    \includegraphics[width=0.48\linewidth,clip]{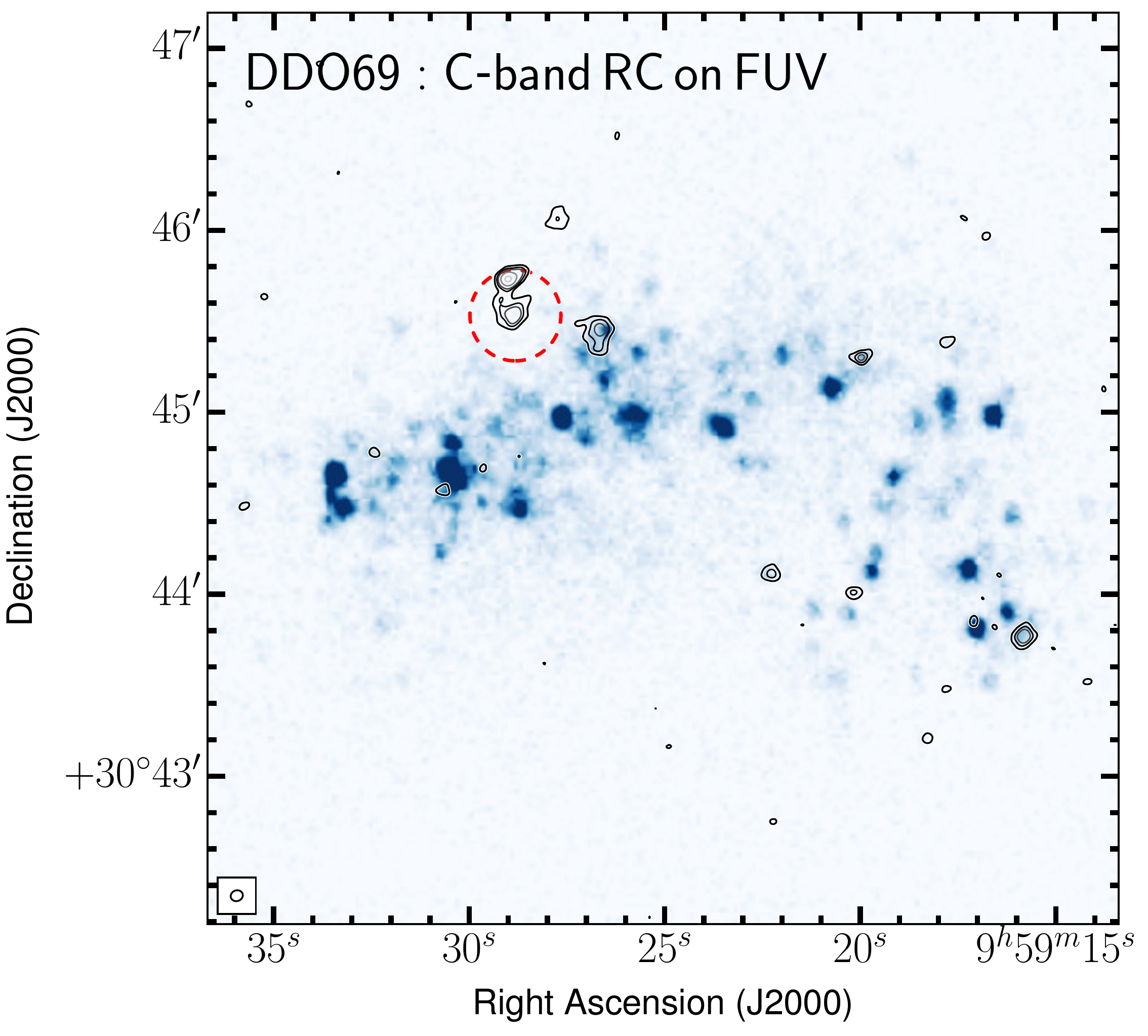} & \
    \includegraphics[width=0.48\linewidth,clip]{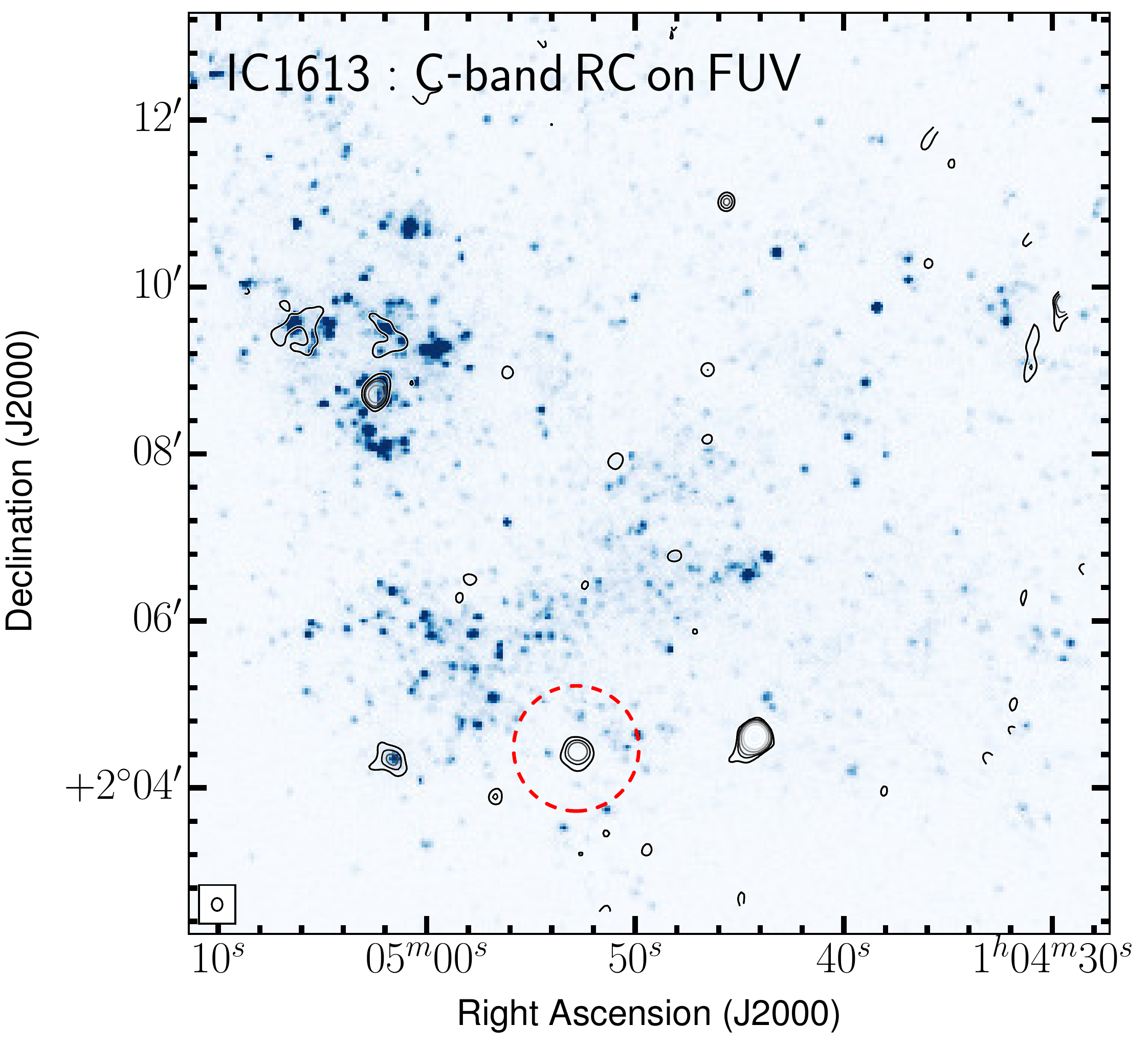}
  \end{tabular}
  \caption{\footnotesize{Examples of our definition of ambiguous emission (red dashed circles). We show DDO\,46 and DDO\,63 each of which containing an unresolved source of $1{\rm \,mJy}$ (top-left) and ${\rm 1.4\,mJy}$ (top-right), respectively. We also show DDO\,69 and IC\,1613 which both contain an extended source (bottom panels). The RC emission could not be attributed as definitely coming from a background galaxy, but at the same time was not close enough to a SF site to be confidently classified as originating from the target galaxy either; accordingly, these sources were designated ambiguous.}}
  \label{figure:ambiguous_examples}
\end{figure*}

After cross matching with NED and isolating ambiguous sources by comparing to \halpha\ there remained sources that we could not attribute as coming from a background galaxy, but at the same time were not close enough to a SF site to be confidently classified as originating from the target galaxy; we refer to these sources of RC emission as ambiguous. To illustrate our definition of ambiguous RC emission, we present four of our observations that contained such a source in Fig.~\ref{figure:ambiguous_examples}. A strong unresolved source can be seen in DDO\,46 and DDO\,63, whilst DDO\,69 and IC\,1613 demonstrate galaxies with significantly extended sources.

Most of our observations contained at least one ambiguous source; none of these had a non-thermal luminosity that exceeded a reference threshold---that of a known bright SNR ($1 \times 10^{19} {\rm \,W}{\rm \,Hz}^{-1}$ or $3.3 {\rm \,mJy}$ at 5\,Mpc at 6\,GHz). This reference luminosity was based on observations of SNR N4449-12, which resides in the dwarf galaxy NGC\,4449 at a distance of $4.2$\,Mpc. In 2002 this SNR had a luminosity of $S_{\rm 6cm}=4.84{\rm \,mJy}$ with a spectral index of $\alpha=-0.7$ between $20$\,cm and $6$\,cm \cite[][]{Chomiuk2009}. For comparison, this is $10$ times the luminosity of Cassiopeia\,A. Since the luminosity terminally declines for the majority of the SNR's lifetime, we treat the observed luminosity of SNR N4449-12 in 2002 as an approximate empirical upper limit to the luminosity of a supernova remnant. We justify our use of SNR N4449-12 as it was the most luminous from a sample of 43 SNRs from 4 irregular galaxies (35 of which are in galaxies that overlap with our sample, namely: NGC\,1569, NGC\,2366, and NGC\,4214). 

\begin{figure*}
\centering
    \includegraphics[width=0.6\linewidth]{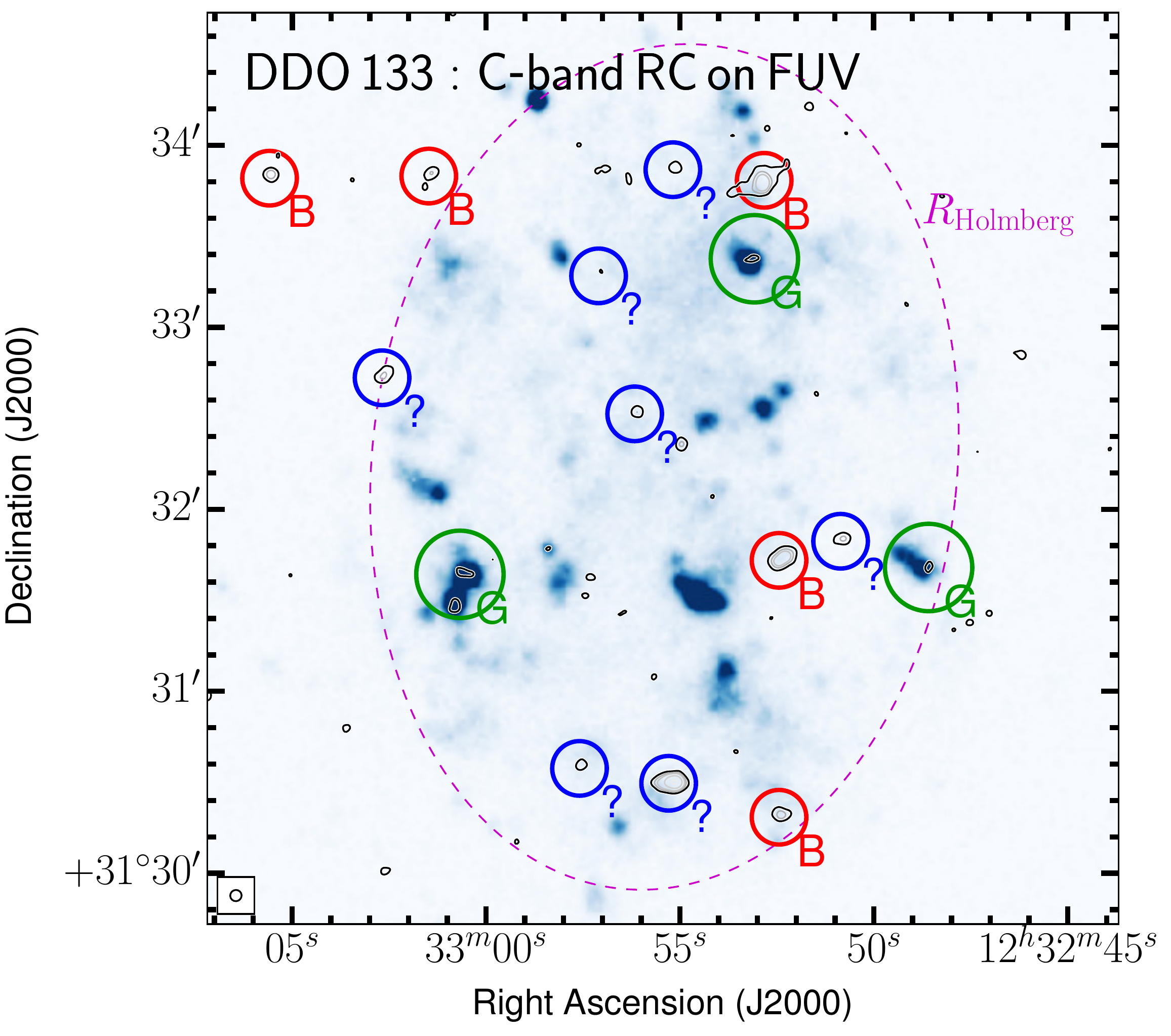}
  \caption{\footnotesize{{\em GALEX} FUV emission of DDO\,133 overlaid with our RC contours. Following the procedure outlined in the text we attribute RC emission as being from either the galaxy itself (G, green), a background galaxy (B, red), or an unknown or ambiguous source (?, blue). We also overlay the optical disk size (defined by the Holmberg radius; purple dashed ellipse). }}
  \label{figure:ddo133_sources}
\end{figure*}

Using the method above we are able to classify all of the observed RC emission in our images. As an example, we show DDO\,133 in Fig.~\ref{figure:ddo133_sources} along with the classification attributed to each source of RC emission.

\subsection{Missing Large-scale Structures}
\label{Section:MissingStructures}

Owing to the way that interferometers function, large angular structures in the sky can be completely missed if their corresponding visibilities are not recorded by the interferometer. The largest angular scale ($\theta_{\rm LAS}$) that the VLA is sensitive to in C-configuration (shortest baseline of 35\,m) at 6\,cm is $\sim 4$ arcminutes. This assumes an observation of 12\,hours that is uniformly weighted and untapered. Observations of a shorter duration will have a slightly lower $\theta_{\rm LAS}$ value and for weighting schemes closer to natural weighting the $\theta_{\rm LAS}$ will be larger. In our observations angular scales of $\sim 4$ arcminutes and above may not be adequately sampled leading to a lower than expected flux density; there are only $7$ galaxies with an angular size greater than $4^\prime$ (see column $4$ of table~\ref{table:12A-234_DiskQuantities}). Under the assumption that RC emission coincides with optical emission, it is only these galaxies that are vulnerable to having large angular structures absent in the ($u$,$v$) data. Even so, SF in dwarf galaxies is intermittent on scales of one to a few Gyr, whereas \CRe\ age over much shorter timescales of tens of Myr; therefore, in the majority of our sample no significant emission is expected from, for example, a \CRe\ halo. We note that NGC\,1569 was found to have an extended radio halo extending beyond the optical emission when observed between 0.6 and 1.4\,GHz \citep{Israel1988}. This is attributed to the post-starburst nature of the galaxy which is not reflected in the majority of targets in our sample. We do not see any evidence of such a halo in our 6\,GHz image. This may be due to spatial filtering or spectral ageing which has shifted the halo emission below our detection threshold.

\subsection{Disk Integrated Quantities}

With background and ambiguous sources removed (see Section~\ref{section:Paper1background_source_removal}), emission from our RC and ancillary images was integrated within each of the dwarf galaxy's optical disks (hereafter the disk mask; see table~\ref{table:12A-234_DiskQuantities} for the disk parameters). We also extract the integrated properties including the ambiguous sources; these can be found in the online material (Appendix~\ref{Section:Appendix_Tables}). The semi-major axis of the disk was based on optical isophotes: using either the Holmberg radius \cite[defined as the isophote where the {\em B}-band surface brightness drops to a magnitude of 26.66;][]{Hunter2006} or $3$ times the {\em V}-band disk scale length \cite[][]{Hunter2006} if the {\em B}-band radius was not defined. All emission outside this radius was masked.

\begin{deluxetable}{lccccccccccc}
\tablewidth{0pt}
\tabletypesize{\tiny}
\tablecolumns{13}
\rotate
\tablecaption{Integrated emission over the disk of the LITTLE THINGS Galaxies\label{table:12A-234_DiskQuantities}}
\tablehead{
\colhead{Galaxy}  & \colhead{Size} & \colhead{P.A.} & \colhead{6\,cm RC} & \colhead{\halpha}        & \colhead{FUV} & \colhead{24\micron\ MIR} & \colhead{70\micron\ FIR} & \colhead{6\,cm \RCNT} & \colhead{$B_{\rm eq}$} \\
  & ($^{\prime}$) & ($^\circ$)  & (mJy)  & ($10^{-13} $\,ergs\,s$^{-1}$\,cm$^{-2}$)   & (mJy) & ($10^{-2}$\,Jy) & ($10^{-2}$\,Jy) & (mJy) & (${\rm \mu G}$) \\
(1) & (2) & (3) & (4) & (5) & (6) & (7) & (8) & (9) & (10)}
\startdata
CVn I dwA	& $1.7 \times 1.4$ & $80$ &  $>0.29$ & $1.95\pm 0.03$ & $1.04\pm 0.11$ & $0.15\pm 0.06$ & $2.46\pm 0.04$ & $>0.29$ & $<2$ \\
DDO 43	& $1.8 \times 1.2$ & $6$ &  $>0.99$ & $1.28\pm 0.03$ & $1.07\pm 0.11$ & \nodata & \nodata & $>0.99$ & $<2$ \\
DDO 46$^\mathrm{V}$	& $3.8 \times 3.4$ & $84$ &  $>1.16$ & $1.08\pm 0.02$ & $1.75\pm 0.17$ & \nodata & \nodata & $>1.16$ & $<2$ \\
DDO 47	& $4.5 \times 2.3$ & $-79$ &  $>0.61$ & $3.00\pm 0.03$ & $3.00\pm 0.30$ & \nodata & \nodata & $>0.61$ & $<2$ \\
DDO 50	& $7.9 \times 5.7$ & $18$ &  $6.70 \pm 0.60$ & $60.10\pm 0.49$ & $41.95\pm 4.20$ & $17.27\pm 0.01$ & $319.90\pm 0.28$ & $0.99\pm 0.60$ & $<2$ \\
DDO 52	& $2.2 \times 1.4$ & $4$ &  $>1.28$ & $0.29\pm 0.01$ & $0.61\pm 0.06$ & $-0.04\pm 0.02$ & $1.81\pm 0.05$ & $>1.28$ & $<1$ \\
DDO 53	& $2.7 \times 1.4$ & $81$ &  $0.65 \pm 0.13$ & $4.32\pm 0.04$ & $2.55\pm 0.26$ & $2.32\pm 0.02$ & $24.05\pm 0.03$ & $0.24\pm 0.13$ & $<1$ \\
DDO 63	& $4.3 \times 4.3$ & $0$ &  $>0.71$ & $4.39\pm 0.04$ & $5.03\pm 0.50$ & $1.77\pm 0.01$ & $3.74\pm 0.13$ & $>0.71$ & $<1$ \\
DDO 69	& $4.8 \times 2.7$ & $-64$ &  $>0.89$ & $1.66\pm 0.01$ & $4.67\pm 0.47$ & $-0.65\pm 0.01$ & $11.08\pm 0.07$ & $>0.89$ & $<1$ \\
DDO 70	& $7.4 \times 4.4$ & $88$ &  $>1.48$ & $6.27\pm 0.04$ & $11.53\pm 1.15$ & $0.59\pm 0.01$ & $63.09\pm 0.13$ & $>1.48$ & $<1$ \\
DDO 75	& $6.2 \times 5.2$ & $42$ &  $>2.04$ & $40.44\pm 0.10$ & $29.46\pm 2.95$ & $0.20\pm 0.01$ & $77.89\pm 0.20$ & $>2.04$ & $<1$ \\
DDO 87	& $2.3 \times 1.3$ & $76$ &  $>0.70$ & $0.68\pm 0.01$ & $0.65\pm 0.06$ & $0.07\pm 0.02$ & $7.03\pm 0.03$ & $>0.70$ & $<2$ \\
DDO 101	& $2.1 \times 1.5$ & $-69$ &  $>1.79$ & $0.82\pm 0.01$ & $0.39\pm 0.04$ & $0.24\pm 0.02$ & $-0.54\pm 0.04$ & $>1.79$ & $<2$ \\
DDO 126	& $3.5 \times 1.7$ & $-41$ &  $>0.57$ & $3.66\pm 0.08$ & $2.91\pm 0.29$ & $0.32\pm 0.03$ & $14.92\pm 0.10$ & $>0.57$ & $<2$ \\
DDO 133	& $4.7 \times 3.2$ & $-6$ &  $>1.17$ & $4.55\pm 0.03$ & $4.09\pm 0.41$ & $0.53\pm 0.01$ & $33.04\pm 0.13$ & $>1.17$ & $<2$ \\
DDO 154	& $3.1 \times 1.6$ & $46$ &  $>1.73$ & $2.21\pm 0.02$ & $3.77\pm 0.38$ & $0.28\pm 0.03$ & $3.66\pm 0.04$ & $>1.73$ & $<1$ \\
DDO 155	& $1.9 \times 1.3$ & $51$ &  $>0.47$ & $4.85\pm 0.05$ & \nodata & $0.22\pm 0.03$ & $16.15\pm 0.05$ & $>0.47$ & $<2$ \\
DDO 165	& $4.3 \times 2.3$ & $89$ &  $>1.19$ & $1.53\pm 0.01$ & \nodata & $0.04\pm 0.01$ & $10.64\pm 0.06$ & $>1.19$ & $<2$ \\
DDO 167	& $1.5 \times 1.0$ & $-24$ &  $>0.51$ & $0.80\pm 0.01$ & $1.05\pm 0.11$ & \nodata & \nodata & $>0.51$ & $<3$ \\
DDO 168	& $4.6 \times 2.9$ & $-25$ &  $>0.94$ & $5.91\pm 0.03$ & $5.55\pm 0.56$ & $0.67\pm 0.01$ & $41.94\pm 0.10$ & $>0.94$ & $<1$ \\
DDO 187	& $2.1 \times 1.7$ & $37$ &  $>1.17$ & $0.57\pm 0.01$ & $1.15\pm 0.12$ & $-0.02\pm 0.03$ & $-1.94\pm 0.09$ & $>1.17$ & $<1$ \\
DDO 210	& $2.6 \times 1.3$ & $-85$ &  $>0.87$ & \nodata & $0.80\pm 0.08$ & $-0.16\pm 0.02$ & $5.26\pm 0.04$ & $>0.87$ & $<2$ \\
DDO 216	& $8.0 \times 3.6$ & $-58$ &  $>1.28$ & $0.09\pm 0.01$ & $2.00\pm 0.20$ & $-0.12\pm 0.01$ & $9.87\pm 0.08$ & $>1.28$ & $<1$ \\
F564-V03$^\mathrm{V}$	& $1.3 \times 1.0$ & $7$ &  $>0.35$ & \nodata & $0.10\pm 0.01$ & \nodata & \nodata & $>0.35$ & $<3$ \\
Haro 29	& $1.7 \times 1.0$ & $85$ &  $2.14 \pm 0.11$ & $13.02\pm 0.45$ & $3.02\pm 0.32$ & $5.83\pm 0.05$ & $39.00\pm 0.05$ & $0.91\pm 0.12$ & $6$ \\
Haro 36$^\mathrm{V}$	& $1.5 \times 1.2$ & $90$ &  $0.94 \pm 0.09$ & $2.41\pm 0.03$ & $2.84\pm 0.29$ & $0.94\pm 0.04$ & $23.66\pm 0.06$ & $0.71\pm 0.09$ & $5$ \\
IC 1613	& $18.2 \times 14.7$ & $71$ &  $4.49 \pm 0.55$ & $55.81\pm 0.87$ & $91.86\pm 9.24$ & $6.85\pm 0.02$ & $408.70\pm 1.73$ & $-0.77\pm 0.55$ & $<1$ \\
IC 10$^\mathrm{V}$	& $11.6 \times 9.1$ & $-38$ &  $96.38 \pm 0.81$ & $1191.00\pm 5.73$ & \nodata & $3741.00\pm 4.83$ & $9547.00\pm 12.08$ & $-16.78\pm 0.97$ & $<1$ \\
LGS 3	& $1.9 \times 1.0$ & $-3$ &  $>0.57$ & \nodata & $0.08\pm 0.01$ & \nodata & \nodata & $>0.57$ & $<2$ \\
M81 dwA$^\mathrm{V}$	& $1.5 \times 1.1$ & $86$ &  $>1.28$ & \nodata & $0.38\pm 0.04$ & \nodata & \nodata & $>1.28$ & $<2$ \\
Mrk 178	& $2.0 \times 0.9$ & $-51$ &  $1.01 \pm 0.14$ & $5.38\pm 0.09$ & $2.56\pm 0.27$ & $0.45\pm 0.03$ & $0.45\pm 0.01$ & $0.50\pm 0.14$ & $5$ \\
NGC 1569$^\mathrm{V}$	& $2.3 \times 1.3$ & $-59$ &  $149.60 \pm 0.31$ & $486.70\pm 3.02$ & $746.50\pm 75.63$ & $705.50\pm 13.61$ & $3543.00\pm 2.66$ & $71.38\pm 0.57$ & $17$ \\
NGC 2366	& $9.4 \times 4.0$ & $33$ &  $9.66 \pm 0.59$ & $96.38\pm 1.11$ & $37.26\pm 3.73$ & $65.47\pm 0.01$ & $506.20\pm 0.30$ & $0.51\pm 0.60$ & $17$ \\
NGC 3738	& $4.8 \times 4.8$ & $0$ &  $2.62 \pm 0.48$ & $16.26\pm 0.17$ & $11.22\pm 1.13$ & $11.65\pm 0.03$ & $248.10\pm 0.41$ & $1.07\pm 0.48$ & $17$ \\
NGC 4163	& $2.9 \times 1.9$ & $18$ &  $>0.69$ & $1.48\pm 0.02$ & $2.68\pm 0.27$ & $0.43\pm 0.03$ & $10.20\pm 0.11$ & $>0.69$ & $<2$ \\
NGC 4214	& $9.3 \times 8.5$ & $16$ &  $27.78 \pm 0.57$ & $178.60\pm 0.92$ & $80.72\pm 8.08$ & $199.60\pm 0.01$ & $2393.00\pm 1.13$ & $10.81\pm 0.57$ & $6$ \\
Sag DIG$^\mathrm{V}$	& $4.3 \times 2.3$ & $88$ &  $>2.47$ & $1.28\pm 0.01$ & $4.52\pm 0.45$ & \nodata & \nodata & $>2.47$ & $<1$ \\
UGC 8508	& $2.5 \times 1.4$ & $-60$ &  $0.38 \pm 0.13$ & $2.75\pm 0.04$ & \nodata & $0.37\pm 0.03$ & $12.52\pm 0.04$ & $0.12\pm 0.13$ & $<1$ \\
VIIZw 403	& $2.2 \times 1.1$ & $-11$ &  $1.37 \pm 0.10$ & $7.48\pm 0.15$ & $3.67\pm 0.38$ & $1.87\pm 15.06$ & $57.05\pm 1.36$ & $0.66\pm 0.10$ & $5$ \\
WLM	& $11.6 \times 5.1$ & $-2$ &  $>2.51$ & $16.81\pm 0.06$ & $29.53\pm 2.96$ & $4.61\pm 0.01$ & $117.70\pm 0.18$ & $>2.51$ & $<1$ \\

\enddata
\tablecomments{ (Column 1) Name of dwarf galaxy. The superscript V means that disk properties (columns $2$--$5$) are taken from {\em V}--band data;for all others, properties are taken from {\em B}--band; 
(Columns 2 \& 3) Size (major and minor axes) and position angle (P.A.) of the optical disk \cite[][]{Hunter2006}; 
(Column 4) 6\,cm ($\sim 6$\,GHz) radio continuum flux density. This value and those following have ambiguous sources removed. For values where we retain ambiguous sources see Appendix~\ref{Section:Appendix_Tables}; 
(Column 5) \halpha\ flux; 
(Column 6) {\em GALEX} FUV flux density; 
(Column 7) {\em Spitzer} 24\micron\ MIR flux density; 
(Column 8) {\em Spitzer} 70\micron\ FIR flux density; 
(Column 9) 6\,cm ($\sim 6$\,GHz) radio continuum non-thermal (synchrotron) flux density. All \RCNT\ emission is assumed to be synchrotron and is inferred by subtracting the \RCT\ component from the total RC following \cite{Deeg1997}. The quantity in parentheses is the amount that was regarded as ambiguous; 
(Column 13) Equipartition magnetic field strength in the plane of the sky \cite[see Equation 3 in][]{Beck2005}. }
\end{deluxetable}

\subsection{Isolating Target RC emission}
\label{section:Paper1RCbasedMask}

The majority of our dwarf galaxy sample only exhibits significant RC emission in isolated regions, which is attributed to both the episodic nature of SF in dwarf galaxies \cite[e.g.,][]{Stinson2007} and the surface brightness sensitivity of our RC observations which limits our RC maps to to detecting SFRDs greater than $\sim5\times10^{-3}$\msunyrkpc. When integrated over the disk, the signal from most galaxies is dominated by the contribution of noise from the individual beams within the integration area. The uncertainty, $\delta N$, is given by $\sigma_{\rm rms} \sqrt{N}$, where $\sigma_{\rm rms}$ is the rms noise level and $N$ is the number of individual beams. This motivates the use of masks to isolate genuine emission from background noise (i.e., reduce the integration area which is proportional to $N$) in order to improve the RC S/N.

\subsubsection{Radio Continuum--based Mask}
\label{section:Paper1masking_techniques}

To characterise the RC emission within our images we first estimate the spatially varying background noise across each image using the {\sc bane} algorithm \citep{Hancock2012}. {\sc bane} works by selecting each pixel in the image on a specified grid and then defines a boxed region. This region is first clipped at $3\sigma$ to remove the contribution of source pixels. The median of the remaining pixels in the box is calculated and used as the background estimate. Linear interpolation is then used to smooth the background across the image. We found that the default options for {\sc bane}, which uses a grid size of four times the beam area and a box size of five times the grid size, produced good estimates of the background noise for the majority of our images. In cases where there is large scale extended emission such as NGC\,1569 and IC\,10 the grid size was increased to the approximate size of the most extended feature in the image, six and nine, respectively, and the default box size was applied. Estimating the background noise allows us to create S/N images that account for local variations in the image background caused by the primary beam sensitivity pattern and any residual low-level artefacts. This results in a robust threshold for our source detection. The average noise towards our galaxies is presented in table~\ref{table:12A-234_Observations}, column\,9.

We apply an automated approach to source identification using the {\sc fellwalker} source finding algorithm \citep{Berry2015} available in the {\sc starlink} distribution {\sc cupid}. {\sc fellwalker} is a thresholding approach to source detection that identifies contiguous features in an image by finding the steepest gradient for each pixel. Starting with the first pixel in an image, above a user defined threshold, each of the surrounding pixels is inspected to locate the pixel with the highest ascending gradient; this process continues until a peak is located (i.e. a pixel surrounded by flat or descending gradients). The pixels along this path are assigned an arbitrary integer to represent their connection along a path. All pixels in the image are inspected in a similar process and the image is segmented into clumps by grouping together all paths that lead to the same peak value. The pixels belonging to paths that lead to the same peak are then defined as belonging to that particular clump. For a full description of this process see \cite{Berry2007}.

Using {\sc fellwalker} we create two masks for each S/N image: the first is at full resolution whilst the second is smoothed to an angular resolution of 10$''$. The former image is used to characterise unresolved point sources whilst the latter is used to define regions of extended emission. We assign a threshold level corresponding to a S/N level of 3 in both cases where the noise levels are derived independently for each image. Fluctuations that are smaller than the beam are excluded; they are identified as noise spikes. We verify the robustness of this approach by comparing our mask to those produced by the {\sc clumpfind} algorithm, which is also available in {\sc cupid}, and by checking each mask by eye to ensure that no spurious emission is included in the maps. An example of the results of this approach can be seen in the top-right panel of Fig.~\ref{section:Paper1Images} and Appendix~\ref{Section:Appendix_Images}.

Using our RC based mask we extract the integrated properties towards our sample of dwarf galaxies excluding background and ambiguous sources and present the results in table~\ref{table:12A-234_MaskQuantities}. A table containing the integrated properties including ambiguous sources can be found in Appendix~\ref{Section:Appendix_Tables}.

\changetext{0.5cm}{}{}{1cm}{} 
\begin{deluxetable}{lcccccccccc}
\tablewidth{0pt}
\tabletypesize{\tiny}
\tablecolumns{13}
\rotate
\tablecaption{Integrated emission over the RC mask of the LITTLE THINGS Galaxies\label{table:12A-234_MaskQuantities}}
\tablehead{
\colhead{Galaxy}  & \colhead{R.A} & \colhead{Dec.} & \colhead{$f_{\rm disk}$} & \colhead{6\,cm RC} & \colhead{\halpha}        & \colhead{FUV} & \colhead{24\micron\ MIR} & \colhead{70\micron\ FIR} & \colhead{6\,cm \RCNT} & \colhead{$B_{\rm eq}$} \\
  &  hh\,mm\,ss.s    & dd\,mm\,ss.s     &  (\%) & (mJy)  & ($10^{-13} $\,ergs\,s$^{-1}$\,cm$^{-2}$)   & (mJy) & ($10^{-2}$\,Jy) & ($10^{-2}$\,Jy) & (mJy) & (${\rm \mu G}$) \\ 
(1) & (2) & (3) & (4) & (5) & (6) & (7) & (8) & (9) & (10) & (11)}
\startdata
DDO 46	& $07\,41\,26.6$ & $+40\,06\,39$ & $0.1$ 	& $0.02 \pm 0.01$ & $0.16\pm 0.01$ & $0.02\pm 0.01$ & \nodata & \nodata & $0.01\pm 0.01$ & $<1$ \\
DDO 47	& $07\,41\,55.3$ & $+16\,48\,08$ & $0.1$ 	& $0.03 \pm 0.01$ & $0.15\pm 0.01$ & $0.05\pm 0.01$ & \nodata & \nodata & $0.02\pm 0.01$ & $<1$ \\
DDO 50	& $08\,19\,08.7$ & $+70\,43\,25$ & $2.2$ 	& $6.27 \pm 0.09$ & $25.28\pm 0.42$ & $7.45\pm 0.76$ & $7.69\pm 0.06$ & $53.17\pm 0.04$ & $3.95\pm 0.10$ & $4$ \\
DDO 53	& $08\,34\,08.0$ & $+66\,10\,37$ & $3.2$ 	& $0.33 \pm 0.02$ & $1.90\pm 0.04$ & $0.71\pm 0.07$ & $1.32\pm 0.09$ & $5.30\pm 0.01$ & $0.16\pm 0.02$ & $4$ \\
DDO 63	& $09\,40\,30.4$ & $+71\,11\,02$ & $0.1$ 	& $0.06 \pm 0.01$ & $0.24\pm 0.01$ & $0.14\pm 0.02$ & $0.04\pm 0.34$ & $0.33\pm 0.01$ & $0.04\pm 0.01$ & $2$ \\
DDO 70	& $10\,00\,00.9$ & $+05\,19\,50$ & $0.1$ 	& $0.07 \pm 0.02$ & $0.29\pm 0.02$ & $0.21\pm 0.03$ & $0.03\pm 0.34$ & $0.34\pm 0.01$ & $0.04\pm 0.02$ & $<1$ \\
DDO 75	& $10\,10\,59.2$ & $-04\,41\,56$ & $0.3$ 	& $0.24 \pm 0.03$ & $2.87\pm 0.06$ & $0.85\pm 0.09$ & $0.07\pm 0.23$ & $1.61\pm 0.01$ & $0.01\pm 0.01$ & $<1$ \\
DDO 126	& $12\,27\,06.5$ & $+37\,08\,23$ & $2.3$ 	& $0.35 \pm 0.03$ & $1.46\pm 0.07$ & $0.54\pm 0.06$ & $0.15\pm 0.20$ & $2.11\pm 0.02$ & $0.23\pm 0.03$ & $3$ \\
DDO 155	& $12\,58\,39.8$ & $+14\,13\,10$ & $5.2$ 	& $0.28 \pm 0.04$ & $2.23\pm 0.04$ & \nodata & $0.15\pm 0.12$ & $2.42\pm 0.01$ & $0.08\pm 0.04$ & $<1$ \\
DDO 168	& $13\,14\,27.2$ & $+45\,55\,46$ & $0.2$ 	& $0.11 \pm 0.01$ & $0.24\pm 0.01$ & $0.12\pm 0.01$ & $0.04\pm 0.18$ & $0.67\pm 0.01$ & $0.09\pm 0.01$ & $2$ \\
Haro 29	& $12\,26\,16.7$ & $+48\,29\,38$ & $13.4$ 	& $2.01 \pm 0.04$ & $12.54\pm 0.45$ & $2.65\pm 0.29$ & $4.96\pm 0.14$ & $21.21\pm 0.02$ & $0.82\pm 0.06$ & $6$ \\
Haro 36	& $12\,46\,56.3$ & $+51\,36\,48$ & $9.3$ 	& $0.37 \pm 0.03$ & $1.17\pm 0.03$ & $1.94\pm 0.21$ & $0.41\pm 0.13$ & $6.88\pm 0.02$ & $0.26\pm 0.03$ & $4$ \\
IC 1613	& $01\,04\,49.2$ & $+02\,07\,48$ & $0.7$ 	& $2.51 \pm 0.05$ & $10.26\pm 0.43$ & $5.08\pm 0.71$ & $1.69\pm 0.23$ & $23.10\pm 0.12$ & $1.63\pm 0.06$ & $3$ \\
IC 10	& $00\,20\,17.5$ & $+59\,18\,14$ & $22.9$ 	& $99.33 \pm 0.39$ & $887.90\pm 5.68$ & \nodata & $1369.00\pm 10.10$ & $5482.00\pm 6.68$ & $14.96\pm 0.66$ & $8$ \\
Mrk 178	& $11\,33\,29.0$ & $+49\,14\,24$ & $3.8$ 	& $0.46 \pm 0.03$ & $2.33\pm 0.08$ & $0.97\pm 0.12$ & $0.16\pm 0.17$ & $0.16\pm 0.01$ & $0.25\pm 0.03$ & $4$ \\
NGC 1569	& $04\,30\,49.8$ & $+64\,50\,51$ & $126.3$ 	& $155.40 \pm 0.34$ & $503.90\pm 3.03$ & $755.60\pm 76.53$ & $716.20\pm 12.11$ & $3758.00\pm 2.99$ & $74.41\pm 0.60$ & $17$ \\
NGC 2366	& $07\,28\,48.8$ & $+69\,12\,22$ & $2.2$ 	& $11.98 \pm 0.09$ & $66.97\pm 1.10$ & $12.64\pm 1.28$ & $52.01\pm 0.04$ & $179.50\pm 0.05$ & $5.65\pm 0.14$ & $5$ \\
NGC 3738	& $11\,35\,49.0$ & $+54\,31\,23$ & $6.2$ 	& $2.98 \pm 0.12$ & $11.83\pm 0.17$ & $7.29\pm 0.75$ & $7.58\pm 0.13$ & $91.12\pm 0.10$ & $1.85\pm 0.12$ & $7$ \\
NGC 4214	& $12\,15\,39.2$ & $+36\,19\,38$ & $2.2$ 	& $22.58 \pm 0.08$ & $117.20\pm 0.91$ & $32.63\pm 3.29$ & $140.40\pm 0.09$ & $941.10\pm 0.17$ & $11.55\pm 0.12$ & $6$ \\
UGC 8508	& $13\,30\,44.9$ & $+54\,54\,29$ & $3.2$ 	& $0.16 \pm 0.02$ & $0.65\pm 0.02$ & \nodata & $0.06\pm 0.15$ & $1.18\pm 0.01$ & $0.10\pm 0.02$ & $2$ \\
VIIZw 403	& $11\,27\,58.2$ & $+78\,59\,39$ & $15.8$ 	& $1.29 \pm 0.04$ & $6.49\pm 0.15$ & $3.21\pm 0.33$ & $2.10\pm 37.85$ & $33.77\pm 0.54$ & $0.68\pm 0.04$ & $5$ \\
WLM	& $00\,01\,59.2$ & $-15\,27\,41$ & $0.1$ 	& $0.16 \pm 0.02$ & $0.79\pm 0.05$ & $0.10\pm 0.01$ & $0.27\pm 0.30$ & $0.31\pm 0.01$ & $0.01\pm 0.01$ & $<1$ \\
\enddata
\tablecomments{ (Column 1) Name of dwarf galaxy;
(Columns 2 \& 3) Equatorial coordinates (J2000) of centre of the galaxy defined by the optical disk; 
(Column 4) Fraction of the disk (see table\,\ref{table:12A-234_DiskQuantities}) that has significant RC emission; 
(Column 5) 6\,cm ($\sim 6$\,GHz) radio continuum flux density. This value and those following have ambiguous sources removed. For values where we retain ambiguous sources see Appendix~\ref{Section:Appendix_Tables}; 
(Column 6) \halpha\ flux; 
(Column 7) {\em GALEX} FUV flux density; 
(Column 8) {\em Spitzer} 24\micron\ MIR flux density; 
(Column 9) {\em Spitzer} 70\micron\ FIR flux density; 
(Column 10) 6\,cm ($\sim 6$\,GHz) radio continuum non-thermal (synchrotron) flux density. All \RCNT\ emission is assumed to be synchrotron and is inferred by subtracting the \RCT\ component from the total RC following \cite{Deeg1997}. The quantity in parentheses is the amount that was regarded as ambiguous; 
(Column 11) Equipartition magnetic field strength in the plane of the sky \cite[see Equation 3 in][]{Beck2005}. }
\end{deluxetable}


In order to compare the RC emission to our ancillary data we first investigate which masks best represent the global emission in our dwarf galaxies. Ideally, we would like to compare the various quantities over the same optically derived disk mask as our ancillary data in general present emission over a large fraction of the disk leading. However, if we integrate the RC emission over the disk we find that only 11 of our 40 observations have significant integrated RC flux density measurements. Using instead our RC mask we identify RC emission associated with 22 out of the 40 LITTLE THINGS galaxies (excluding ambiguous sources); $8$ are new RC detections. It is for this reason that in the course of the analysis of our data we will present results integrated over both the RC and disk based masks.



\subsection{Radio Continuum Source Counts}

\begin{figure}[t!]
  \centering
    \includegraphics[width=1.0\linewidth]{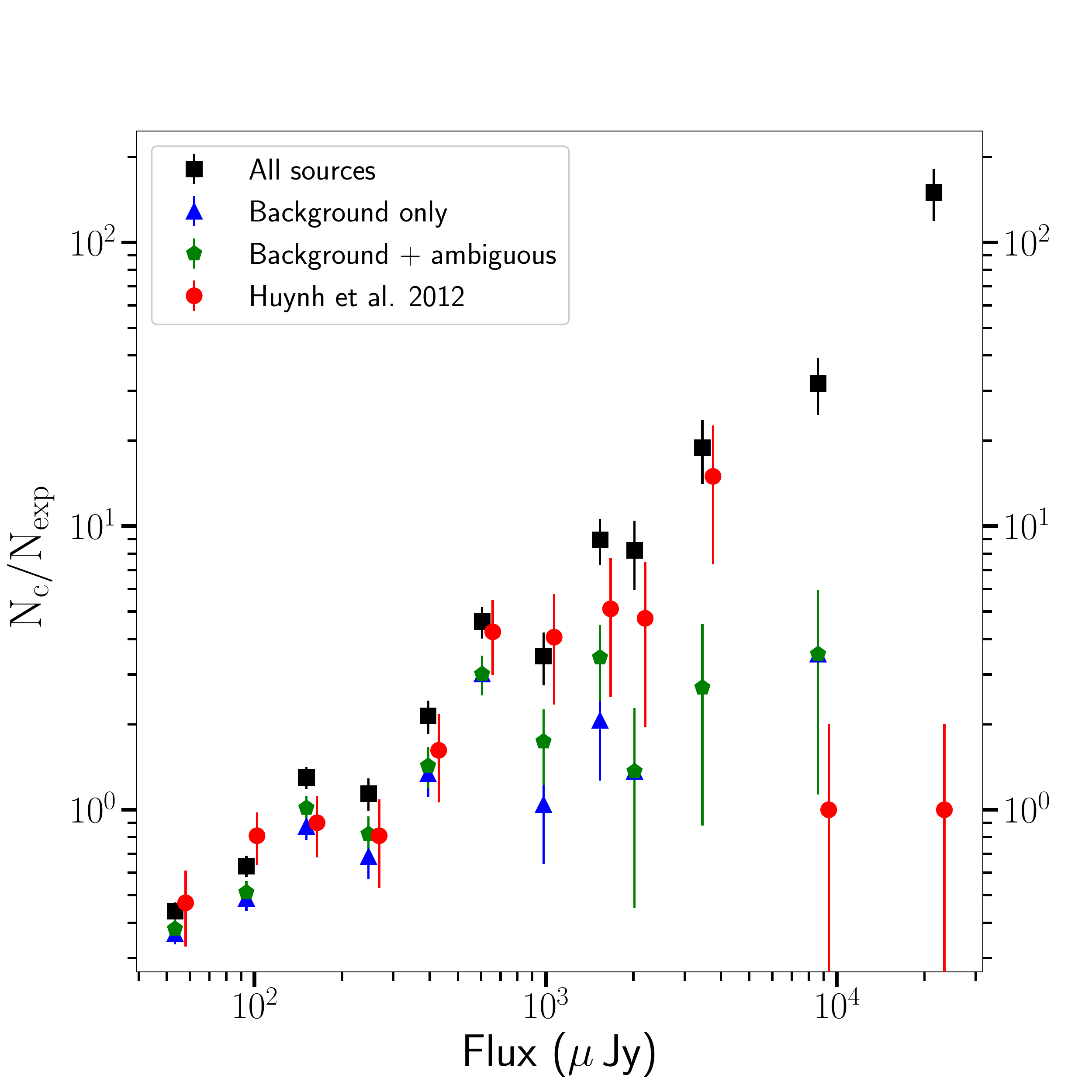}
  \caption{\footnotesize{Corrected and normalised source counts recovered from our images. Sources are separated based on our source identification approach into: all sources (black squares), background sources (blue triangles), background and ambiguous sources (green pentagon). We compare these results to those of \cite{Huynh2012} (red circles).}}
  \label{figure:sourcecounts}
\end{figure}

\begin{deluxetable}{c|ccc|ccc|ccc|ccc}
\tablewidth{0pt}
\tabletypesize{\tiny}
\tablecolumns{5}
\tablecaption{6\,cm Source Counts\label{Table:SourceCounts}}
\tablehead{
\colhead{$\Delta S$} & \multicolumn{3}{|c}{$N$}  & \multicolumn{3}{|c}{$N_{\rm c}$}  & \multicolumn{3}{|c}{${\rm d}N_{\rm c}/{\rm d}S$} & \multicolumn{3}{|c}{$N_{\rm c}$ / $N_{\rm exp}$} \\ \
(${\rm \mu Jy}$) & & &  & & &  & &(${\rm sr^{-1}\,Jy^{-1}}$) & & &&  \\ \
  & all & bg & amb & all & bg & amb & all & bg & amb & all & bg & amb }
\startdata
$ 46 $ -- $ 73 $ & $ 60 $ & $ 50 $ & $ 52 $ & $ 180.18 $ & $ 150.15 $ & $ 156.15 $ & $ 1983.58 $ & $ 1652.99 $ & $ 1719.11 $ & $ 0.44 \pm 0.03 $ & $ 0.37 \pm 0.03 $ & $ 0.38 \pm 0.03 $ \\
$ 73 $ -- $ 116 $ & $ 52 $ & $ 40 $ & $ 42 $ & $ 125.43 $ & $ 96.48 $ & $ 101.31 $ & $ 785.19 $ & $ 603.99 $ & $ 634.19 $ & $ 0.60 \pm 0.05 $ & $ 0.46 \pm 0.05 $ & $ 0.49 \pm 0.05 $ \\
$ 116 $ -- $ 183 $ & $ 55 $ & $ 37 $ & $ 43 $ & $ 124.68 $ & $ 83.87 $ & $ 97.47 $ & $ 485.42 $ & $ 326.55 $ & $ 379.51 $ & $ 1.23 \pm 0.11 $ & $ 0.83 \pm 0.09 $ & $ 0.96 \pm 0.10 $ \\
$ 183 $ -- $ 290 $ & $ 25 $ & $ 15 $ & $ 18 $ & $ 55.52 $ & $ 33.31 $ & $ 39.98 $ & $ 132.29 $ & $ 79.37 $ & $ 95.25 $ & $ 1.08 \pm 0.15 $ & $ 0.65 \pm 0.11 $ & $ 0.78 \pm 0.12 $ \\
$ 290 $ -- $ 460 $ & $ 24 $ & $ 15 $ & $ 16 $ & $ 52.78 $ & $ 32.99 $ & $ 35.19 $ & $ 78.56 $ & $ 49.10 $ & $ 52.37 $ & $ 2.04 \pm 0.28 $ & $ 1.27 \pm 0.22 $ & $ 1.36 \pm 0.23 $ \\
$ 460 $ -- $ 728 $ & $ 26 $ & $ 17 $ & $ 17 $ & $ 56.74 $ & $ 37.10 $ & $ 37.10 $ & $ 55.06 $ & $ 36.00 $ & $ 36.00 $ & $ 4.39 \pm 0.58 $ & $ 2.87 \pm 0.47 $ & $ 2.87 \pm 0.47 $ \\
$ 728 $ -- $ 1155 $ & $ 10 $ & $ 3 $ & $ 5 $ & $ 21.65 $ & $ 6.50 $ & $ 10.83 $ & $ 12.93 $ & $ 3.88 $ & $ 6.47 $ & $ 3.34 \pm 0.72 $ & $ 1.00 \pm 0.39 $ & $ 1.67 \pm 0.51 $ \\
$ 1155 $ -- $ 1831 $ & $ 13 $ & $ 3 $ & $ 5 $ & $ 27.96 $ & $ 6.45 $ & $ 10.75 $ & $ 10.68 $ & $ 2.47 $ & $ 4.11 $ & $ 8.61 \pm 1.63 $ & $ 1.99 \pm 0.78 $ & $ 3.31 \pm 1.01 $ \\
$ 1831 $ -- $ 2901 $ & $ 6 $ & $ 1 $ & $ 1 $ & $ 12.86 $ & $ 2.14 $ & $ 2.14 $ & $ 3.74 $ & $ 0.62 $ & $ 0.62 $ & $ 7.91 \pm 2.20 $ & $ 1.32 \pm 0.90 $ & $ 1.32 \pm 0.90 $ \\
$ 2901 $ -- $ 4598 $ & $ 7 $ & $ 0 $ & $ 1 $ & $ 14.90 $ & $ 0.00 $ & $ 2.13 $ & $ 2.53 $ & $ 0.00 $ & $ 0.36 $ & $ 18.27 \pm 4.73 $ & $ 0.00 \pm 0.00 $ & $ 2.61 \pm 1.79 $ \\
$ 4598 $ -- $ 11478 $ & $ 9 $ & $ 1 $ & $ 1 $ & $ 18.98 $ & $ 2.11 $ & $ 2.11 $ & $ 1.29 $ & $ 0.14 $ & $ 0.14 $ & $ 31.03 \pm 7.12 $ & $ 3.45 \pm 2.37 $ & $ 3.45 \pm 2.37 $ \\
$ 11478 $ -- $ 28653 $ & $ 11 $ & $ 0 $ & $ 0 $ & $ 23.03 $ & $ 0.00 $ & $ 0.00 $ & $ 0.63 $ & $ 0.00 $ & $ 0.00 $ & $ 148.48 \pm 30.94 $ & $ 0.00 \pm 0.00 $ & $ 0.00 \pm 0.00 $ \\
\enddata
\tablecomments{6\,cm ($6.2$\,GHz) source counts. 
(Column 1) flux density bins taken from \cite{Huynh2012} converted to $6.2$\,GHz  assuming a spectral index of $-0.7$; 
(Column 2) number of $>5\,\sigma_{\rm rms}$ RC source counts. We count all sources in the images (all), sources identified as background (bg), sources identified as background or ambiguous (amb); 
(Column 3) the completeness and resolution corrected RC source counts; 
(Column 4) the corrected RC source count rate---the number of sources found per steradian normalised to the mid–point of the flux density bin.
(Column 5) corrected source counts normalised by the expected number from a non evolving Euclidean model.}
\end{deluxetable}

To test the robustness of our source identification and extraction approach we determine the radio continuum source counts from our images. We compare these to \cite{Huynh2012} who performed $5.5$\,GHz observations with the Australia Telescope Compact Array of a $900$\,arcmin$^2$ region with a restoring beam of $4.9^{\prime\prime}\times2.0^{\prime\prime}$ and an rms noise of $12 {\rm \,\mu Jy\,beam^{-1}}$. After correcting for incompleteness and resolution bias, they present normalised source counts in $10$ flux density bins ranging between $50$ and $5000 {\rm \,\mu Jy}$ (see their table\,2).

Our images were generated using a restoring beam of approximately $3^{\prime\prime}$ and attained an rms noise of $\sim6 {\rm \,\mu Jy\,beam^{-1}}$. Therefore, the sensitivity per beam in our data is comparable to that from \cite{Huynh2012}. We scale the \cite{Huynh2012} bins to $6.2$\,GHz, the effective frequency for most of our images, assuming a spectral index of $-0.7 \pm 0.2$. This assumption is supported by various studies that show the average spectral index of star forming galaxies is narrowly concentrated around $\sim -0.7$ with a small dispersion of $\lesssim 0.2$ \citep{Condon1992,Lisenfeld2000,Niklas1997}. For each bin, we cycled through our images counting all sources with flux densities in the range $\Delta S$. We count sources only from within a $4^{\prime}$ circular aperture centred on the image pointing reference to avoid regions where the primary beam response leads to higher noise levels. Sources are assigned to three different groups following our source classification approach described in Section~\ref{section:Paper1background_source_removal}. The first group includes all sources in the field including the galaxy emission, the second counts only sources we are confident are background sources, and the final group consists of both background and ambiguous sources. Sources were not counted if, in the given bin, the low end of the bin was less than $5$ times the rms noise from the image (this only affected the two lowest bins because of a few high rms images). No attempt was made to count resolved sources as originating from the same source (e.g., radio lobes, multiple SF regions from a dwarf, etc.).

To estimate the completeness of our source catalogue we follow a similar approach to \cite{Huynh2012} and perform a Monte-Carlo simulation. We inject a synthetic Gaussian source with a randomly generated position and brightness from 30 to 3000\,${\rm \,\mu Jy}$ into our image and then apply the {\sc fellwalker} source detection algorithm following the same approach as described in Section~\ref{section:Paper1RCbasedMask} to see if the source is recovered. We do this 8000 times and find that sources with flux densities of $5\,\sigma_{\rm rms}$ ($\sim50\,{\rm \mu\,Jy}$) have a detection rate of 50\%, where $\sigma_{\rm rms}$ is the rms noise in the image. The detection rate rises steeply to 90\% at 120$\rm \mu\,Jy$. We also correct for the resolution bias following the same approach as \cite{Huynh2012}. This correction accounts for sources with weak extended emission and large total integrated flux densities that have peaks which fall below the detection threshold. Given our slightly higher sensitivity and resolution we find lower resolution correction factors than \cite{Huynh2012} with values of $1.24$ in our lowest bin and $1.03$ in our highest bin.

We present the results of our source counts in table~\ref{Table:SourceCounts}. For each bin we present the raw source counts ($N$) and the counts corrected for completeness and resolution bias ($N_{\rm c}$). We determine the RC source count rate (${\rm d}N_{\rm c}/{\rm d}S$), which corresponds to the number of sources found per steradian normalised to the mid point of the flux bin. Finally, we normalise our corrected source counts by dividing by the expected number of sources ($N_{\rm exp}$) derived from a non evolving Euclidean model using the relation $N(>S_{\rm 6\,cm}) = 60 * S_{\rm 6\,cm}^{-1.5}$. The Poissonian errors are presented for the normalised and corrected counts with the resolution and completeness correction uncertainties ($10\%$ and $2$--$5\%$, respectively) added in quadrature.

In Fig.~\ref{figure:sourcecounts} we present a comparison of our source counts using all sources (black squares), only background sources (blue triangles), both background and ambiguous sources (green pentagons). We compare our results to the corrected and normalised source counts of \cite{Huynh2012} (red circles). This plot clearly shows that our counts are consistent with \cite{Huynh2012} until $\sim 10^3\,{\rm \mu\,Jy}$. Beyond this flux we see that including galaxy emission in our source counts leads to higher counts than those found in \cite{Huynh2012}, particularly at flux densities above 8.6\,mJy. Ideally, we would like to use the source counts to test the reliability of our source identification approach, in particular we would like to test whether sources we define as ambiguous are background sources or associated with the galaxy emission. If we assume that our source identification approach has reliably identified the galaxy emission and background sources and that the bulk of our ambiguous sources are associated with one of these groups then we should see a signature of this in our source counts. If the ambiguous sources belong to the background sources group we would expect that including them in the source counts whilst excluding the galaxy emission would lead to source counts that are similar to \cite{Huynh2012}. Conversely, if the ambiguous sources are background sources and we do not count them whilst also excluding the galaxy emission we would expect to see lower source counts than expected. In Fig.~\ref{figure:sourcecounts} we do see some tentative evidence that suggests the ambiguous sources are background sources with the background only source counts (blue triangles) tending to be lower than the source counts including both the background and ambiguous sources (green pentagons). However, due to the small number of sources in each bin and the associated errors we are prevented from stating that, statistically, the ambiguous sources belong to the population of background sources.

\section{DISCUSSION}
\label{section:discussion}

\subsection{The Radio Continuum}

\subsubsection{Comparison with Literature Flux Densities}
\label{section:Paper1Discussion_ComparisonFluxDensity}

There are few significant RC detections of dwarf galaxies in the literature. Of the galaxies that overlap with our sample, the literature is dominated by non-detections \cite[e.g.,][]{Altschuler1984,Wynn-Williams1986,Klein1992,Hoeppe1994}. On closer inspection, the seemingly high detection rate of $40$\% in \cite{Klein1986} is actually dominated by $1$--$3\,\sigma$ detections which are likely influenced by the inclusion of background galaxies in the large Effelsberg 100-m single dish beam. We are therefore limited by the number of dwarf galaxies with flux densities in the literature which we can confidently compare our results against\footnote{We note that the flux densities for sources found in the literature may be derived from a range of absolute flux density scales. Commonly used absolute flux scales include \citet{Baars1977}, \citet{Perley2013}, and \citet{Scaife2012}. Variations of the absolute flux scale between these different standards are on the order of 5\% \citep{Perley2016}.}. Reliable RC detections in the literature mostly come from deeper case studies of individual dwarf galaxies. Below we compare our RC flux density integrated over the RC mask that includes ambiguous sources (table~\ref{table:12A-234_MaskQuantities_ambig}) to the few studies available in the literature: 

\paragraph{NGC\,1569:} %
\cite{Lisenfeld2004} find a VLA $8.415$\,GHz flux density of $125\pm12$\,mJy and spectral index of $-0.47$. The same spectral index was found by \cite{Kepley2010} (see their Fig.~\,3). Scaling the $8.415$\,GHz flux density we find an equivalent $6.2$\,GHz flux density of $144\pm14$\,mJy which agrees with our measurement of $157.30 \pm 0.35$\,mJy. Single dish observations performed by the Green Bank telescope at 4.85\,GHz \cite{Gregory1991} found a flux density of 202\,mJy. If we scale this to $6.2$\,GHz, assuming a spectral index of $-0.47$, we find a flux density of 180.0\,mJy. This suggests that we may be missing approximately 12.8\,mJy ($\sim 9$\%) of the flux in our image.

\paragraph{NGC\,4214:} %
\cite{Kepley2011} find a VLA 4.86\,GHz flux density of $34.0\pm6.8$\,mJy (D-array) and spectral index of $-0.43$. The equivalent $6.2$\,GHz flux density is $30\pm6$\,mJy whilst we find $23.16 \pm 0.09$\,mJy. We compare our flux density to that of \cite{Gregory1991} and find that our measured flux density is 3.8\,mJy ($\sim 14\%$) lower. We note that  this suggests that we have missed large scale emission. 

\paragraph{DDO\,50:} %
\cite{Tongue1995} find a VLA 6\,cm flux density of $11.7 \pm 0.1$\,mJy (D-array) which is higher than the $6.81\pm0.09$\,mJy at $6.2$\,GHz that we measured. Again, we note that there is the possibility that we have missed large scale emission.

\paragraph{NGC\,2366:} %
In the absence of a literature flux density at 6\,cm, we resort to a comparison with an {\em L}--band value. \cite{Condon2002} find a 1.4\,GHz flux density of $19.9$\,mJy whilst we report a $6.2$\,GHz flux density of $12.05 \pm 0.09$\,mJy. This implies a spectral index of $-0.34 \pm 0.10$ which is plausible. In light of this, it is unlikely that we have missed large scale emission which would flatten the spectral index and would imply emission even more dominated by \RCT\ emission than derived here.

\paragraph{NGC\,3738:} %
\cite{Stil2002} find a 1.4\,GHz flux density of $13\pm2$\,mJy and we find a $6.2$\,GHz flux density of $2.62\pm0.0.48$\,mJy. This implies a spectral index of $-1.08 \pm 0.04$ which is quite steep. Our image is affected by artefacts from a nearby bright source which may be influencing our flux density measurement.

\paragraph{Haro\,29:} %
\cite{Condon1998} find a 1.4\,GHz flux density of $4.5\pm0.5$\,mJy wheras we find a $6.2$\,GHz flux density of $2.18 \pm 0.11$\,mJy. This implies a spectral index of $-0.49 \pm 0.08$ which is plausible.

\paragraph{Others:} %
\cite{Klein1986} find a number of $\sim 4\,\sigma$ detections at $4.75$\,GHz: $3.5\pm1.0$\,mJy for DDO\,126; $4\pm1$\,mJy for DDO\,133; $9\pm2$\,mJy for DDO\,52. However, we observe less than a mJy for each of these. In all cases, we find nearby background galaxies that will have entered their $2^{\prime}30^{\prime\prime}$ single dish beam and contributed to their flux density to some degree.

\subsubsection{Composition of the Radio Continuum: Thermal and Non--thermal contributions}
\label{sect:rccomp}

The total RC emission is comprised of two contributions: \RCT\ and \RCNT. Since \halpha\ and the \RCT\ both have their origins in hot ($\sim 10^4$\,K) plasma associated with HII regions, a tight spatial and temporal correlation between the two is expected \cite[e.g.,][]{Deeg1997,Murphy2011}. The \halpha--\RCT\ relation taken from \cite{Deeg1997} assumes the form:
\begin{eqnarray}
\frac{{\rm RC_{Th}}}{{\rm W\,m^{-2}}} & = & 1.14 \times 10^{-25} \, \Big( \frac{\nu}{{\rm GHz}} \Big)^{-0.1}\nonumber \\
& & \times\ \Big(\frac{T_{\rm e}}{{\rm 10^4\,K}} \Big)^{0.34} \, \frac{{F_{{\rm H\alpha}}}}{{\rm ergs\,s^{-1}\,cm^{-2}}}.
\label{Equation:Deeg1997}
\end{eqnarray}%
\noindent where $\nu$ is the observed frequency in GHz, ${T_{\rm e}}$ is the electron temperature, which is assumed to be $10^4$\,K, and $F_{\rm H\alpha}$ is the \halpha\  luminosity. On a spatially resolved basis, the \RCT\ flux density can be subtracted from the total RC, yielding the \RCNT\ flux density distribution. 

We do not correct our \halpha\ estimates for internal extinction, following the same approach as \cite{Heesen2014}. As our later analysis utilises the SFR derived by combining the 24\micron\ and FUV emission, we wish to avoid using the 24\micron\ to correct for internal extinction so as not to introduce a spurious correlation. Dwarf galaxies are expected to have low internal extinction due to their low metallicity and therefore this is thought to generally not have a significant impact on our results. To verify this assertion we estimate the internal extinction in our \halpha\ maps following the method of \cite{Kennicutt2009}:
\begin{equation}
I_{\rm H\alpha,corr} = I_{\rm H\alpha,obs}+0.02\nu_{\rm 24\mu m}I_{\rm 24\mu m}
\label{Equation:extinction}
\end{equation}
\noindent where $I_{\rm H\alpha,obs}$ is the observed \halpha\ intensity, which has been corrected for foreground reddening, $I_{\rm H\alpha,corr}$ is the \halpha\ intensity corrected for internal extinction, and $I_{\rm 24\mu m}$ is the 24$\mu$m intensity. Our most intensely star forming galaxies are IC\,10 and NGC\,1569. We calculate the average internal extinction towards these galaxies and find values of 38\% and 35\%, respectively. We have explored the extinction towards NGC\,1569 in \cite{Westcott2017} using a Bayesian approach to separate the RC emission. We were able to estimate an average internal extinction of $\sim 20\%$, slightly lower than our estimate above. Galaxies with lower SFR, $<0.1$\msunyr\, that make up the bulk of our sample have much lower internal extinctions of $<10\%$ as derived from the 24$\mu$m intensity. For example, VIIZw\,403 and DDO\,50 both have an internal extinction of $\sim 8\%$. In light of these results we caution that in our subsequent analysis the \RCT\ flux estimates in galaxies with higher SFRs may be underestimated.

The uncertainty in our estimate of the \RCT\ emission is dominated by the foreground Galactic extinction correction and $T_{\rm e}$. The uncertainty in the Galactic extinction correction for our sample is $\pm 0.015$\,mag for values of $E(B-V)\leq 0.015$ and $\sim 10$\% for $E(B-V)> 0.015$ \cite[][]{Burstein1982}. We assume a single value for the foreground extinction across each galaxy. The foreground extinction may vary considerably across each galaxy, particular for those galaxies in close proximity to the Milky Way such as IC\,10 where the foreground extinction has been shown to vary across the face of the galaxy from $-60\%$ to $+25\%$ of our assumed value \citep{Basu2017}.

The value of $T_{\rm e}$ is assumed to be the standard value of $10^4$\,K but the electron temperature in HII regions has been shown to vary considerably. For example a sample of 61 Galactic HII regions where found to have values of $T_{\rm e}$ ranging from $0.25\times10^{4}$\,K to $1.16\times10^{4}$\,K \citep{Hindson2016}. In a study by \cite{Nicholls2014} the mean electron temperature of 17 HII regions in 14 dwarf irregular galaxies was $T_{\rm e} = 1.4\times 10^4$\,K. Variations in the electron temperature from our assumed value could give rise to up to $\sim 20$\% error in the estimated thermal emission.

After the removal of known background galaxies and ambiguous sources, we apply our RC and disk masks to isolate the \RCT\ (scaled \halpha) emission. When integrating over the RC mask we find that the average thermal fraction for our sample is $\sim 50 \pm 10$\% (upper limit). When integrating over the entire disk we find a higher thermal fraction of $70 \pm 10$\%. For comparison we scale thermal fractions reported for dwarf galaxies in the literature to $6.2$\,GHz assuming a spectral index of $-0.1$ and $-0.7$ for thermal and non-thermal components, respectively. The scaled thermal fractions in dwarf galaxies have been quoted as $51$\% for a sample of stacked faint dwarfs \cite[][]{Roychowdhury2012}, $53$\% in IC\,10 \cite[][]{Heesen2011}, $41$\% in NGC\,1569 \cite[][]{Lisenfeld2004}, and $41$\% in NGC\,4449\cite[][]{Niklas1997}. Our estimate of the thermal fraction integrated over the RC mask are consistent with these literature values. The thermal fraction integrated over the disk mask is significantly greater. We neglect internal extinction in our estimate of the \RCT\ which may lead to slightly lower values of the thermal fraction in the high SFR galaxies such as NGC\,1569 and IC\,10. It is also possible that on the scale of the disk we are missing some flux associated with large-scale RC emission which would lead to higher thermal fractions in the most extended galaxies. A more robust measure of the \RCT\ emission may be obtained using a Bayesian approach \citep{Tabatabaei2017, Westcott2017}, however this requires a large number of observations across the radio SED. 

To estimate the \RCNT\ emission we subtract the \RCT\ emission from the total RC. We caution that the \RCNT\ emission may in some cases turn out to be rather an upper limit because of the previous points.

\subsection{The RC--SFR Relation}
\label{section:Paper1rc--sfr_relation}

We estimate the SFR following the approach of \cite{Leroy2012}. This corrects the FUV-inferred SFR for internal extinction, which is only relevant for our more actively star forming dwarfs. The FUV has been proven to be a reliable SF indicator at low SFR in comparison to \halpha--inferred SFRs (e.g., \citealp{Lee2009};  \citealp{Ficut2016}), and the timescale of \RCNT\ emission is closer to the FUV--inferred SF timescales than to, e.g., \halpha--inferred SF timescales. Galactic foreground extinction is taken into account separately \cite[see][for details]{Hunter2012}. To correct for internal extinction, \cite{Bigiel2008} and \cite{Leroy2012} use {\em Spitzer} 24\micron\ dust emission to empirically correct {\em GALEX} FUV fluxes for the fraction of dust-obscured SF on the assumption that a proportion of energy absorbed by internal dust is reradiated at 24\micron\ \cite[this is based on the original idea by][who use \halpha\ instead of FUV]{Calzetti2007}. We use: 

\begin{align}
\frac{\Sigma_{{\rm SFR}}}{{\rm M_\odot\,yr^{-1}\,kpc^{-2}}} & = 0.081\frac{I_{{\rm FUV}}}{{\rm MJy\,sr^{-1}}} \nonumber\\
 &\qquad + \ 0.0032\frac{I_{{\rm 24\mu m}}}{{\rm MJy\,sr^{-1}}}
\end{align}

\noindent where the FUV and 24\micron\ intensity are in units of $\rm MJy\,sr^{-1}$ and $\Sigma_{{\rm SFR}}$ represents the Star Formation Rate Density (SFRD). We show a map of the SFRD for DDO\,50 in Fig.~\ref{figure:example_maps}. For those galaxies where {\em Spitzer} 24\micron\ data was not available (see table~\ref{table:12A-234_DiskQuantities} and \ref{table:12A-234_MaskQuantities}, column\,7), we used the FUV--inferred SFR without any correction. Due to the low dust content of the majority of our sample the FUV dominates the SFR estimates. The error associated with our SFR estimates is $\sim 20$\%. When compared to other methods of deriving the SFR this approach was found the have a scatter of $\sim 50$\% down to a $\Sigma_{{\rm SFR}}$ of $10^{-4}$\msunyr\ \citep{Leroy2012}.

\begin{figure*}[t!]
  \centering
  \begin{tabular}{cc}
\includegraphics[width=0.48\linewidth,clip]{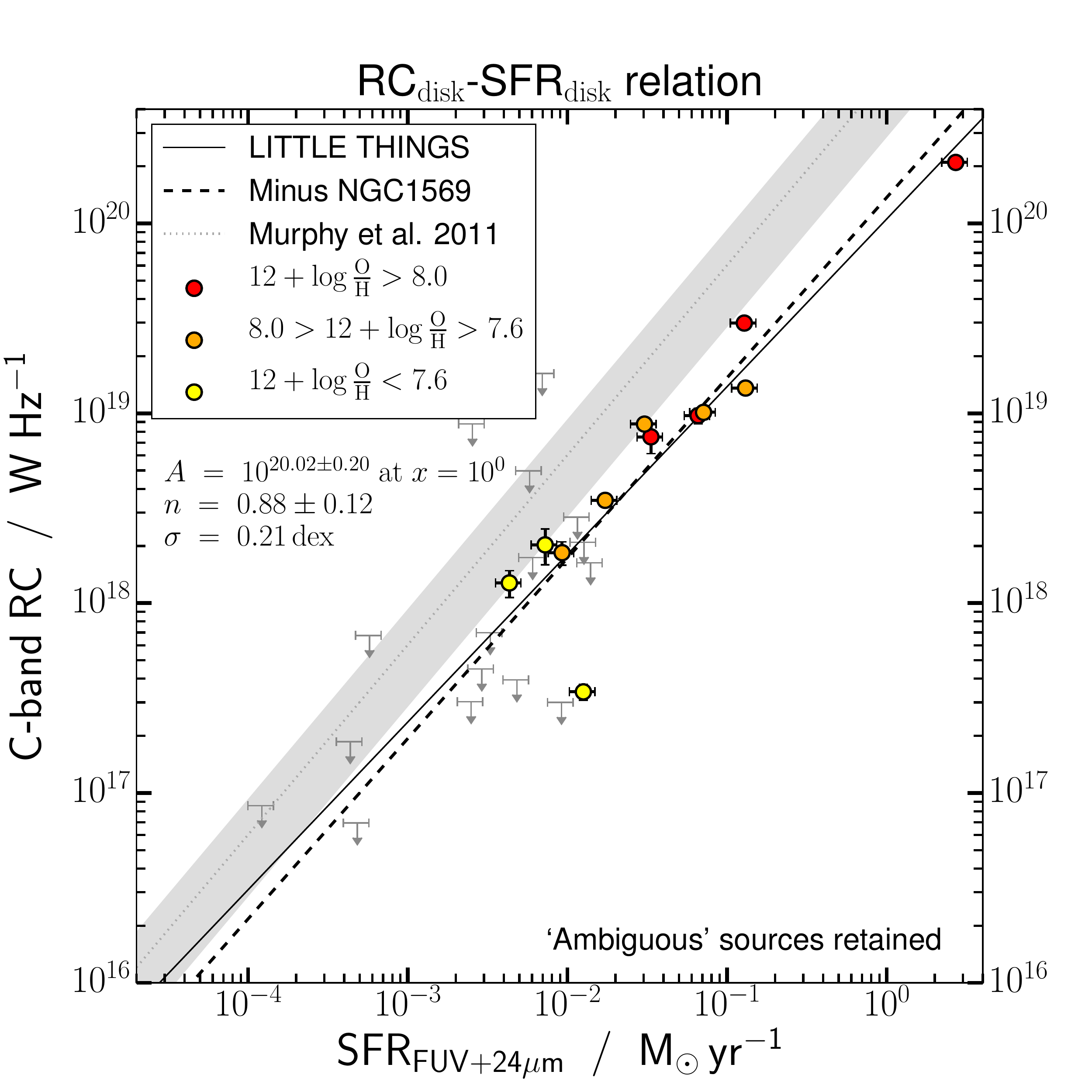}  &\
\includegraphics[width=0.48\linewidth,clip]{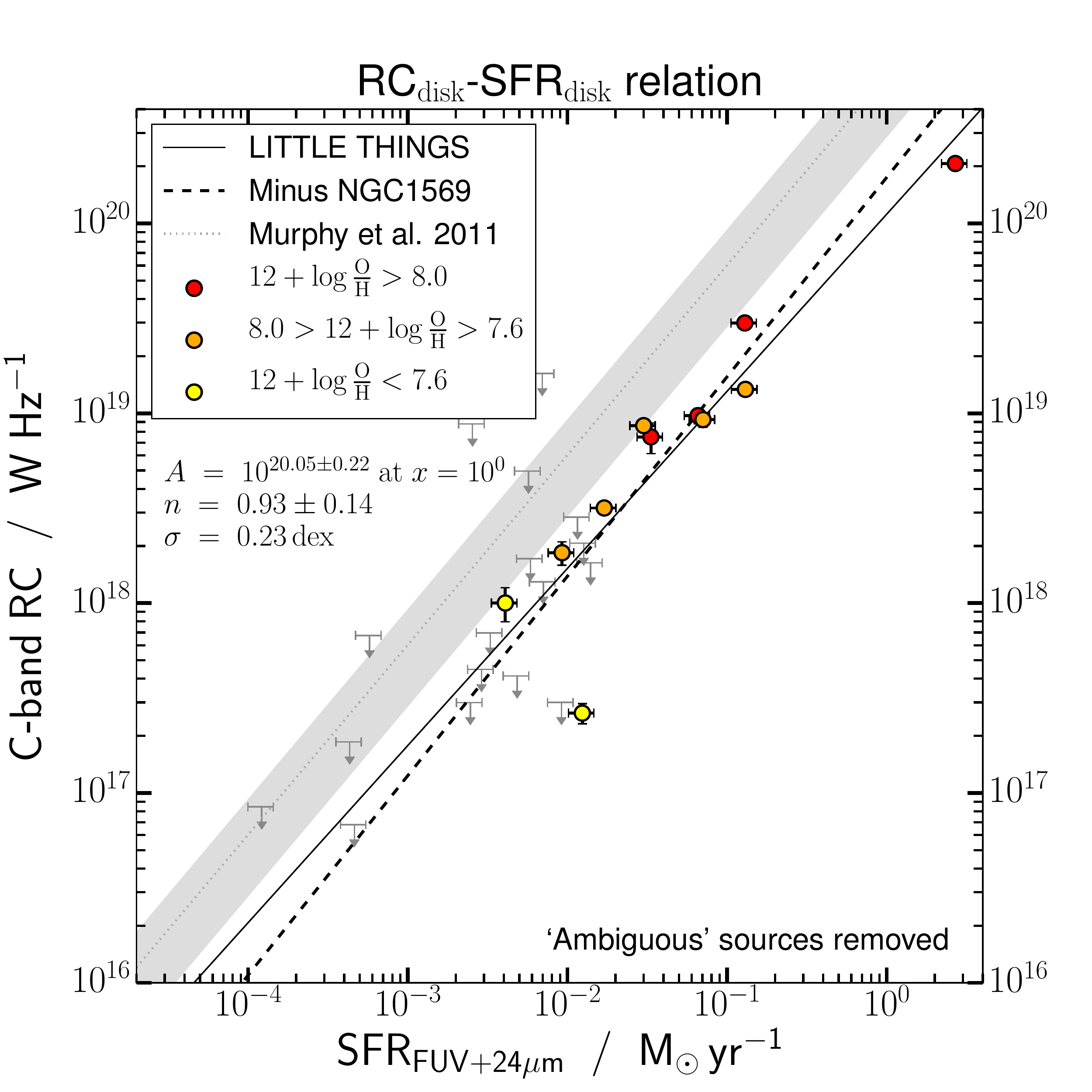}\\
\includegraphics[width=0.48\linewidth,clip]{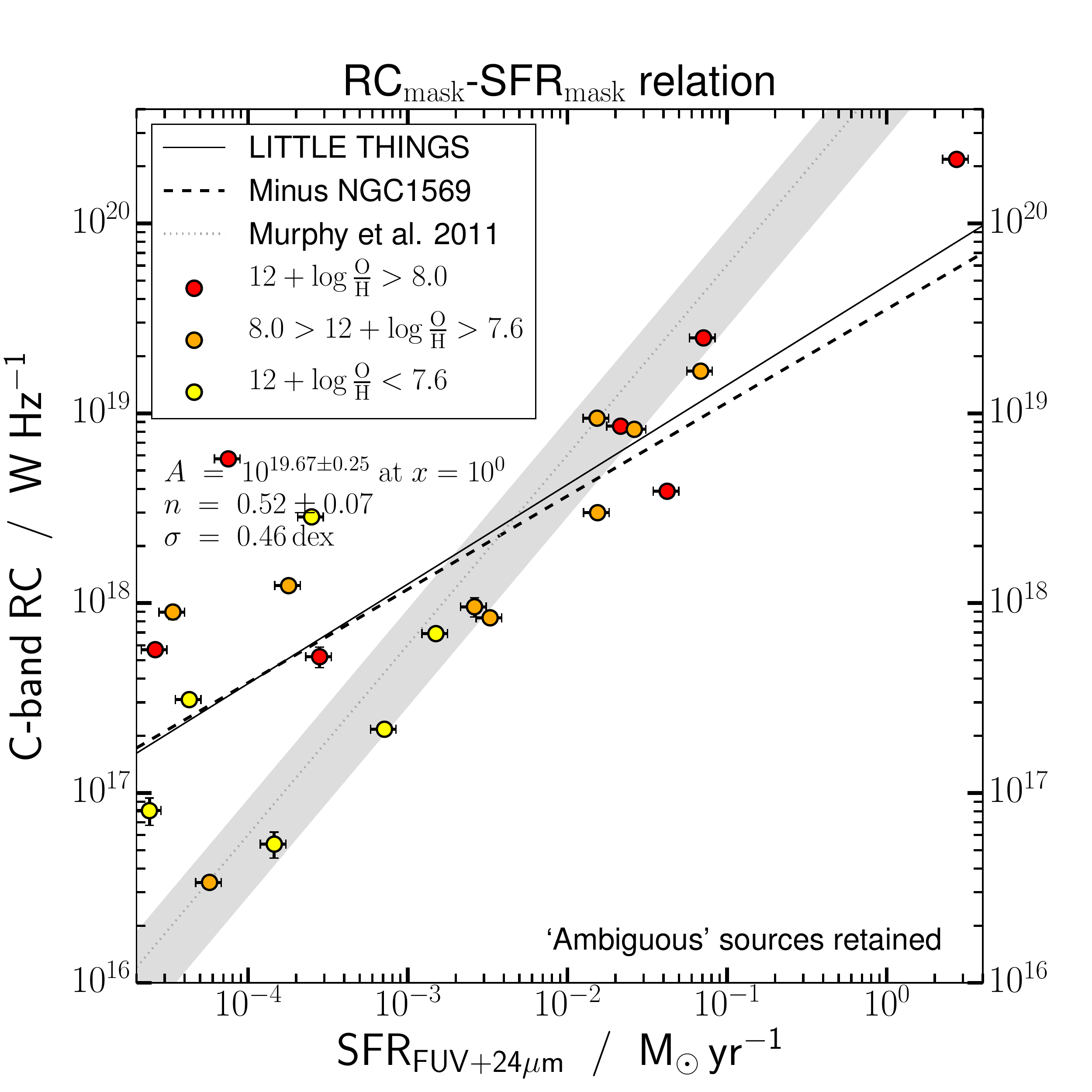} &\
\includegraphics[width=0.48\linewidth,clip]{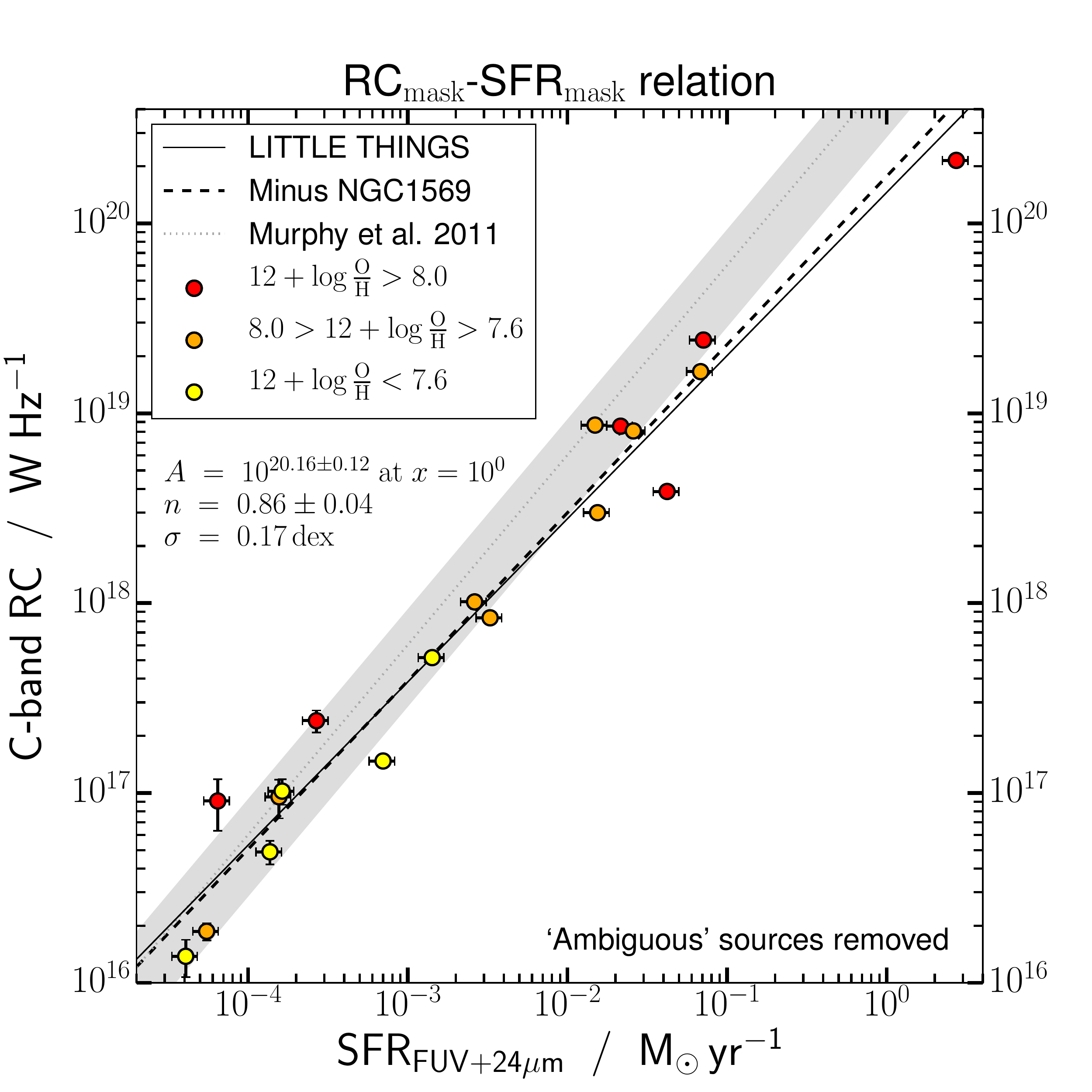}
  \end{tabular}
  \caption{\footnotesize{Total 6\,cm luminosity as a function of SFR over our disk (top-panels) and RC (bottom-panels) masks. Definite background sources have been removed, whilst the ambiguous sources have been retained (left) and removed (right). The solid line is the best-fit power law to our sample. For reference we show the \cite{Murphy2011} RC--SFR relation as a shaded grey band. The errors introduced by our conversion are reflected by the grey shaded band, and the $3\,\sigma$ upper limits of RC emission are shown by grey arrows.}}
  \label{figure:sfr_v_Clumin}
\end{figure*}

\begin{figure}[t!]
  \centering
  \includegraphics[width=0.99\linewidth,clip]{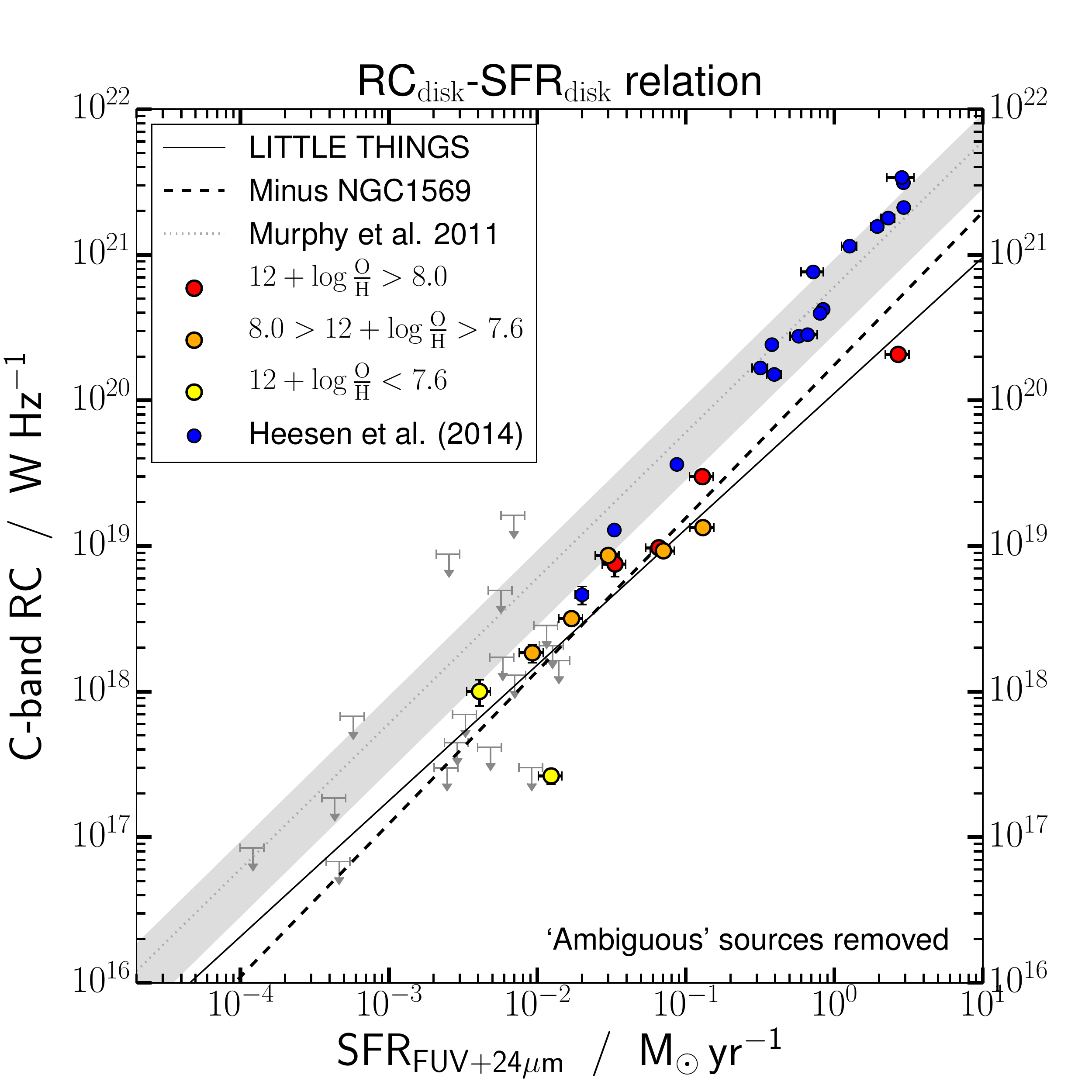}
\includegraphics[width=0.99\linewidth,clip]{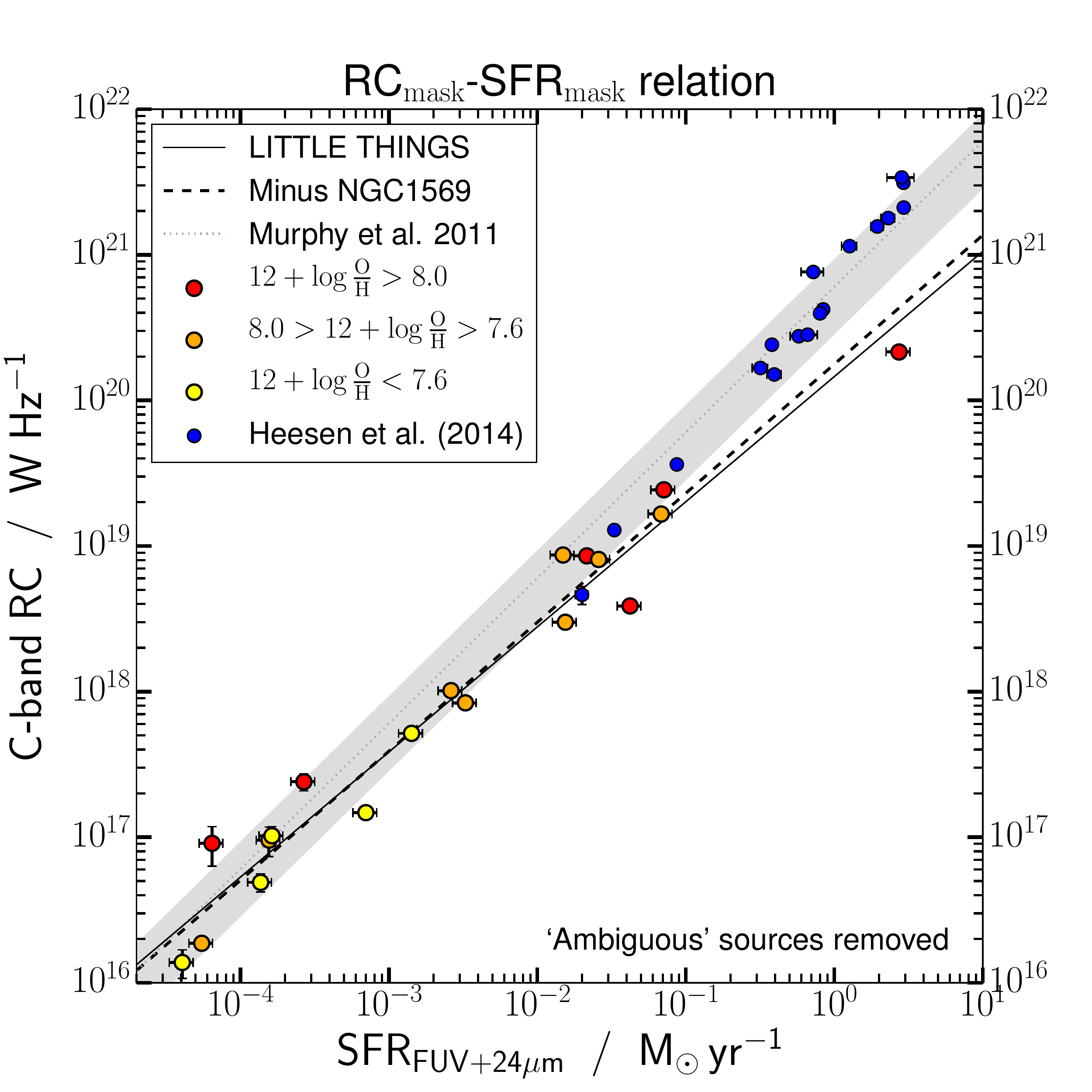}
  \caption{\footnotesize{Comparison of our disk (top) and mask (bottom) integrated RC and SFR properties with that of \cite{Heesen2014} (blue points). Obvious background and ambiguous sources were removed. The WSRT $22$\,cm data were corrected to $6$\,GHz assuming a spectral index of $-0.7$. The LITTLE THINGS galaxies have been coloured according to their metallicity. The solid black line is the best-fit power law to the LITTLE THINGS sample whilst the shaded band shows the \cite{Murphy2011} relation. The dashed black line shows the fit excluding NGC\,1569.}}
  \label{Figure:Heesen2014Comparison}
\end{figure}

\begin{figure*}[h!]
  \centering
  \begin{tabular}{cc}    
\includegraphics[width=0.48\linewidth,clip]{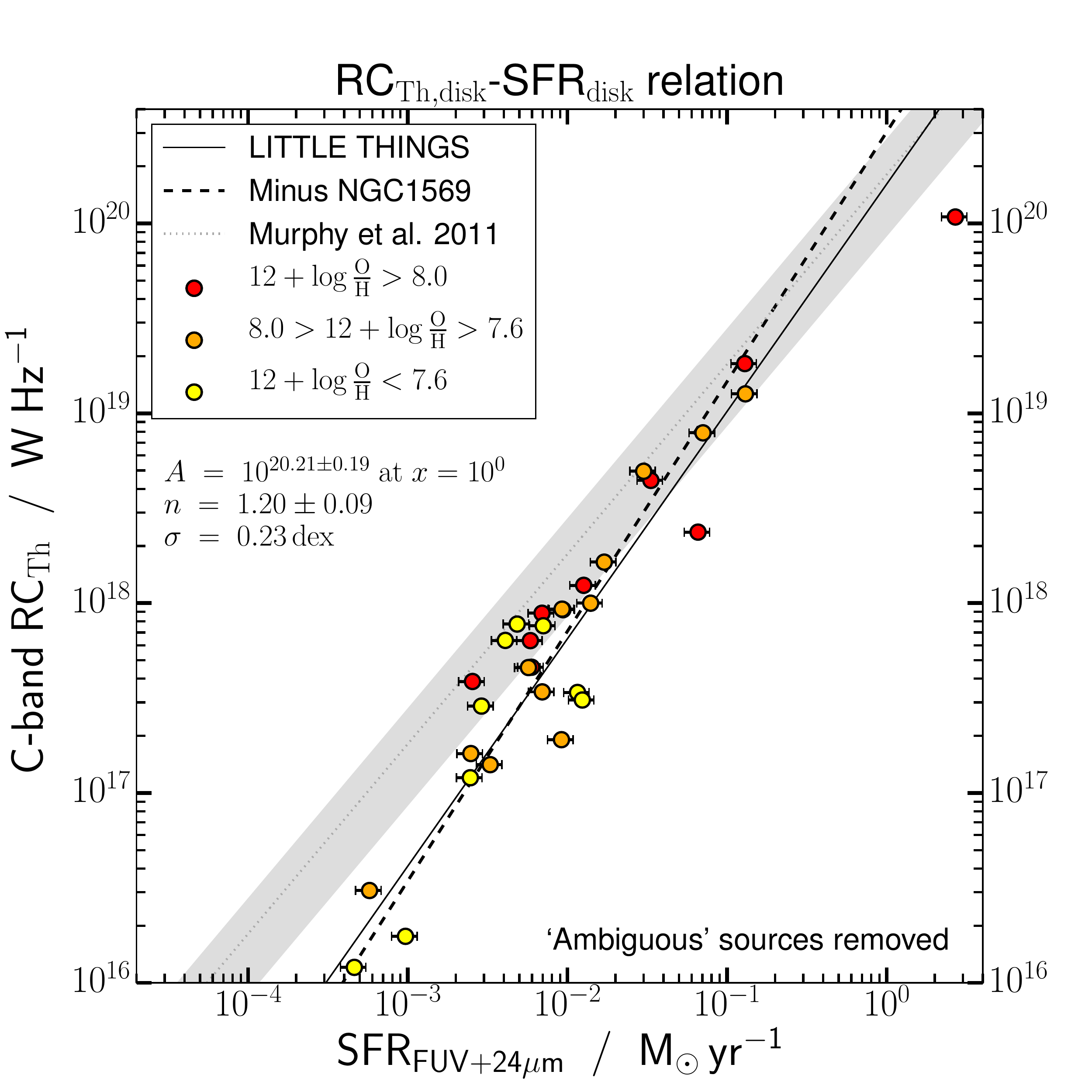} &\
\includegraphics[width=0.48\linewidth,clip]{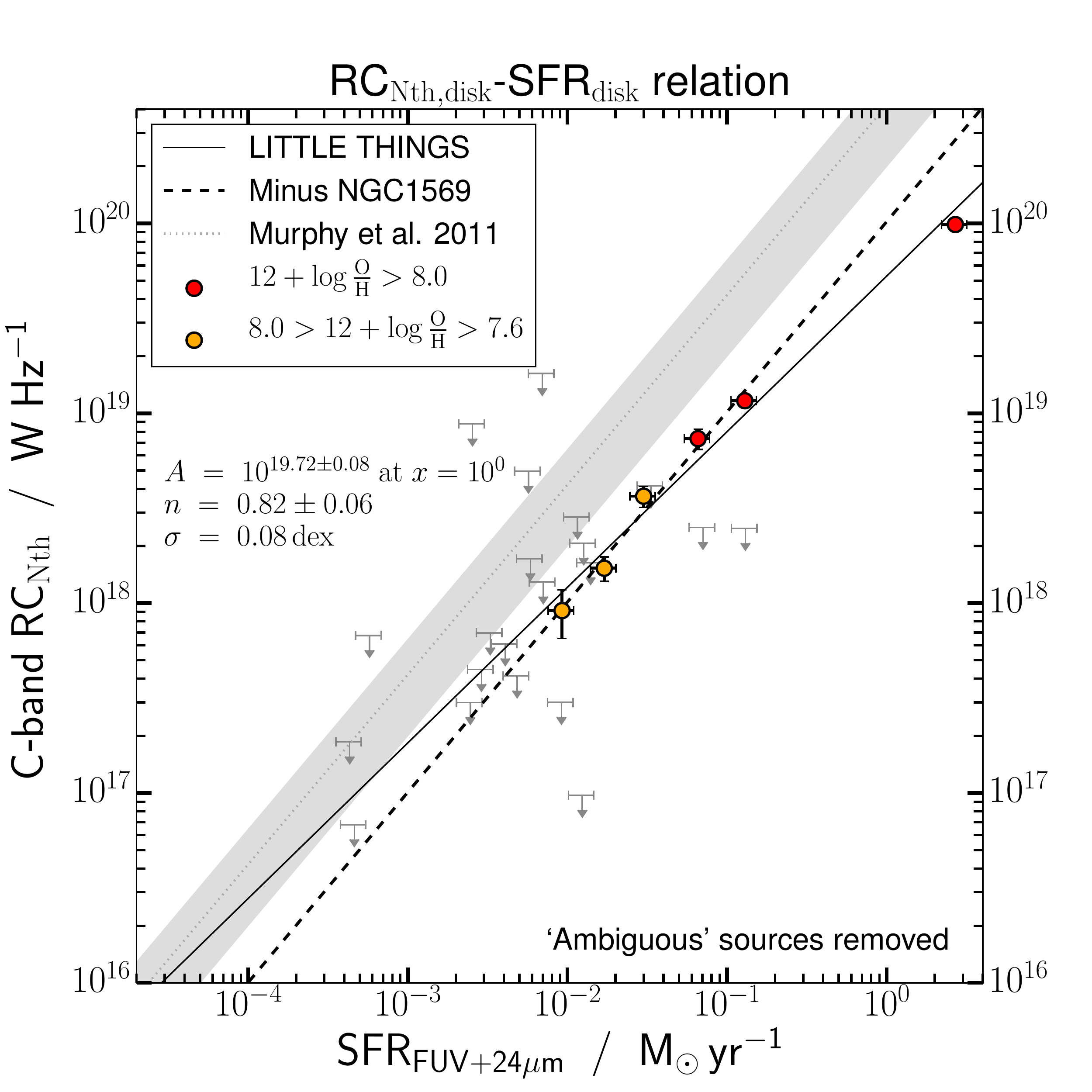} \\
\includegraphics[width=0.48\linewidth,clip]{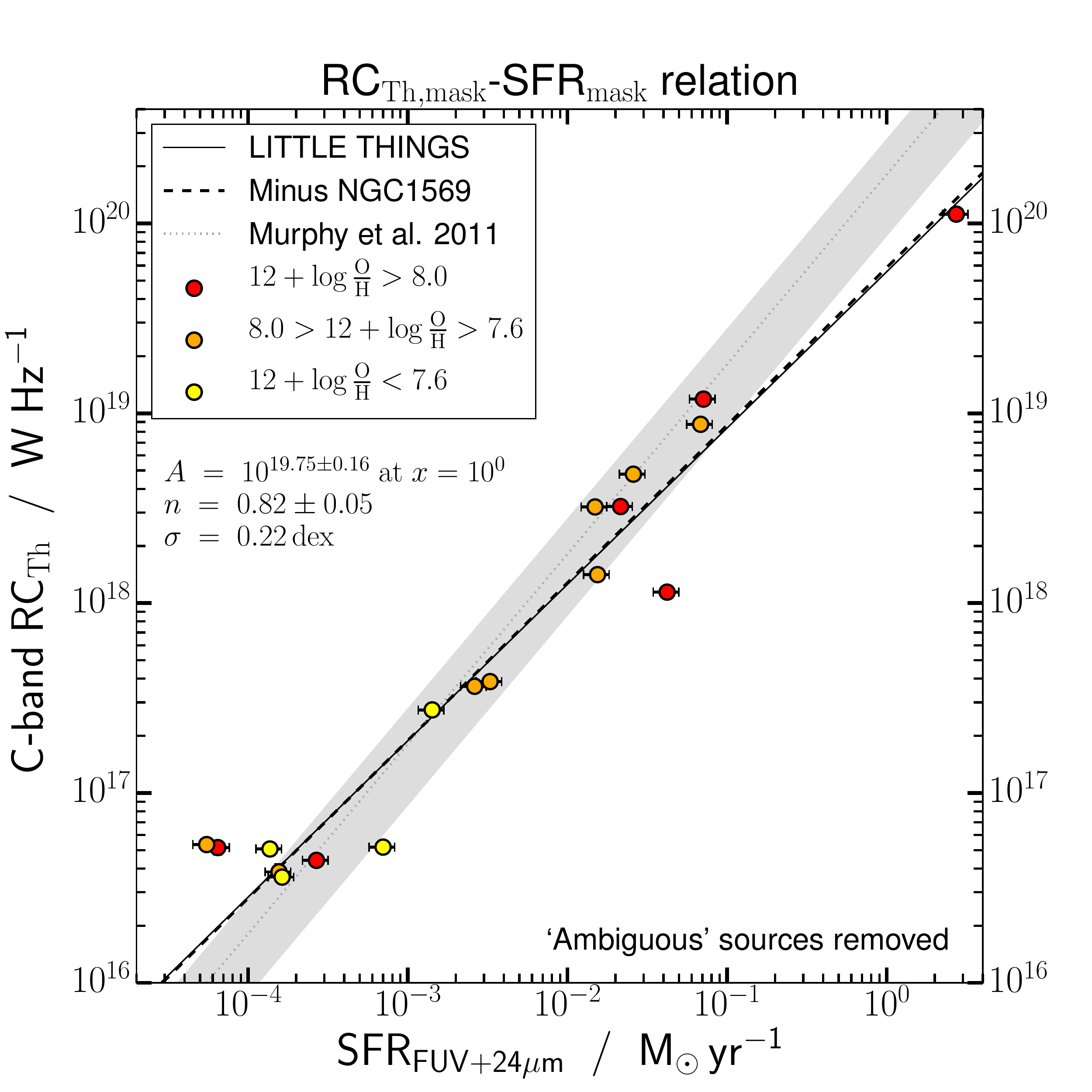} &\
\includegraphics[width=0.48\linewidth,clip]{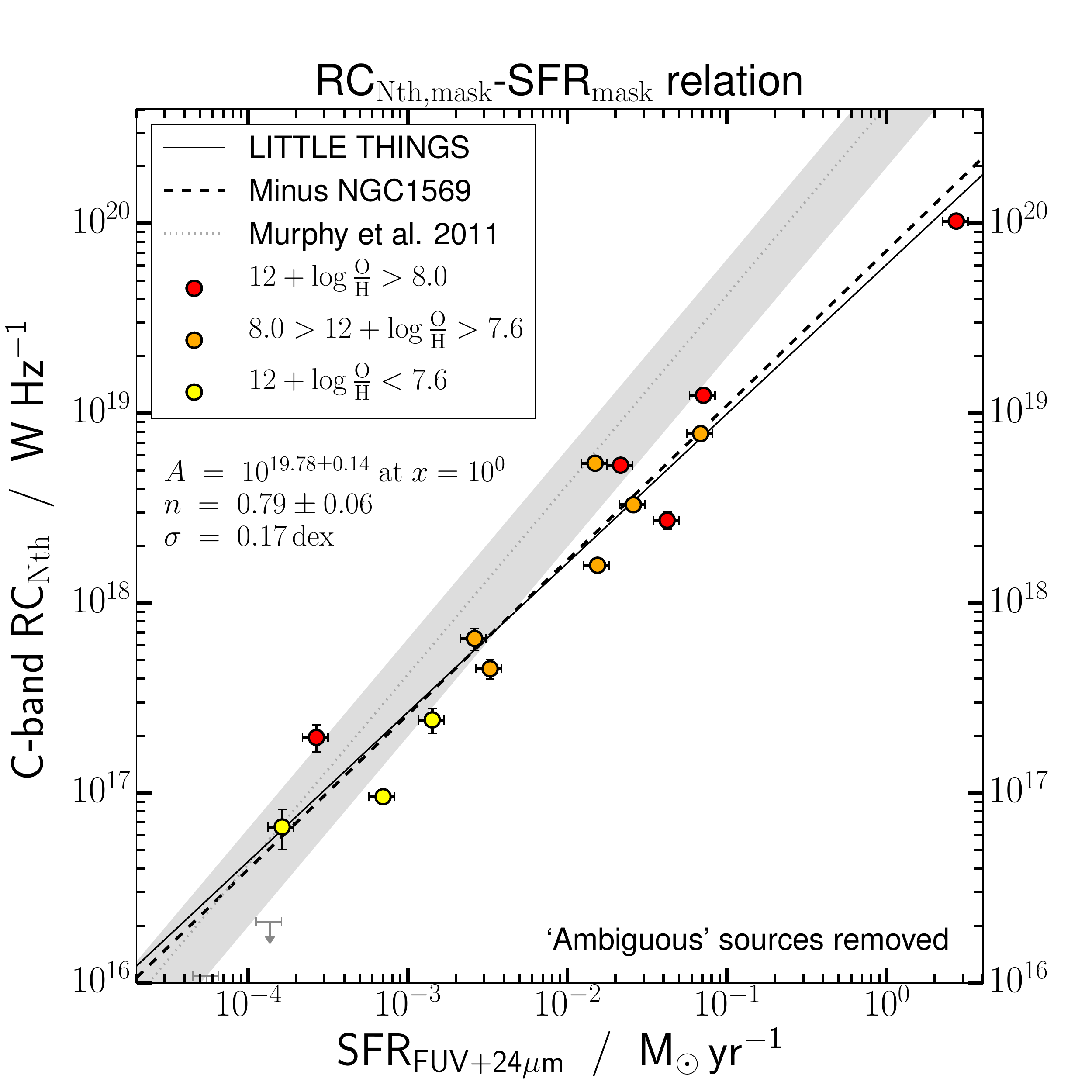}
   \end{tabular}
  \caption{\footnotesize{Total 6\,cm \RCT\ (left) and \RCNT\ (right) luminosity as a function of SFR integrated over our disk (top) and RC mask (bottom). Both definite background sources and ambiguous sources have been removed. The solid line is the best-fit power law to our sample. We show the \cite{Murphy2011} \RCT--SFR and \RCNT--SFR relations for reference as a shaded, grey band; $3\,\sigma$ upper limits of RC emission are shown by the grey symbols with downward arrows.}}
  \label{Figure:rcnt_sfr}
\end{figure*}

In Fig.~\ref{figure:sfr_v_Clumin} we present the RC--SFR relation of our sample when considering the optical disk mask (top) and RC based mask (bottom). We are able to determine the RC and SFR for 11 and 19 galaxies in the disk and RC based masks, respectively. The left panels of Fig.~\ref{figure:sfr_v_Clumin} shows the relation when we include the ambiguous RC sources, whereas the right panel shows the relation with ambiguous sources removed. If we do not remove the ambiguous sources we find a significant flattening and increase in scatter of the fit to the data particularly in the case of the RC based mask results. The most likely cause for this is that these ambiguous sources are background radio galaxies. We therefore continue our analysis focusing only on the results where ambiguous sources are removed. In doing so we may remove at most $10$\% of genuine RC emission in the form of SNRs as according to \citet{Chomiuk2009}, RC emission from SNRs contribute $< 10$\% of the total RC in dwarf galaxies

\begin{figure}[t!]
  \centering
    \includegraphics[width=1.0\linewidth,clip]{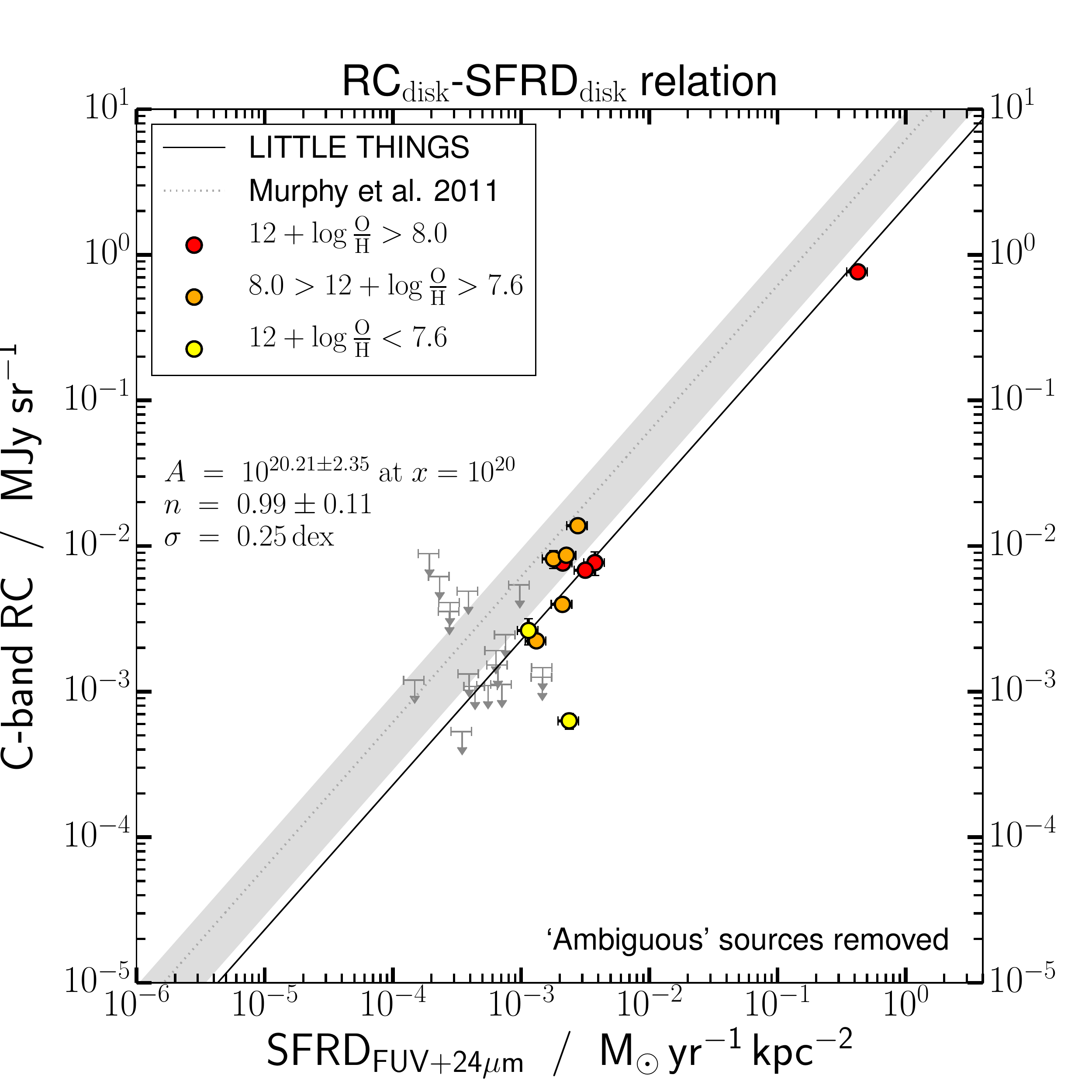} 
    \includegraphics[width=1.0\linewidth,clip]{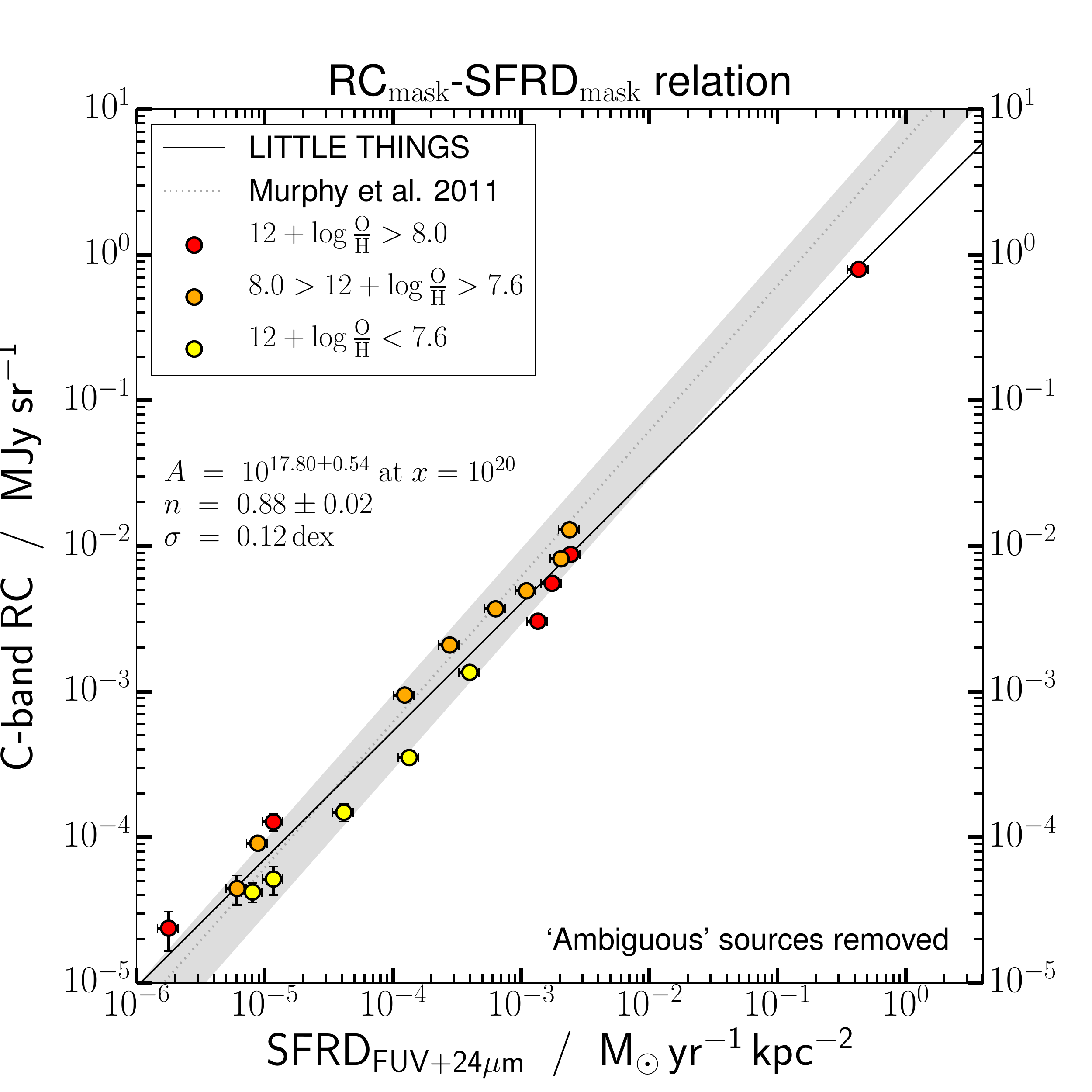}
  \caption{\footnotesize{Galaxy-wide total 6\,cm surface brightness as a function of the FUV-inferred SFRD corrected for internal extinction from dust based on their 24\micron\ emission. Background and ambiguous sources have been removed. The solid line is the best-fit power law to our sample. We show the predicted \cite{Murphy2011} RC--SFR relation for reference as a shaded, grey band; $3\,\sigma$ upper limits of RC emission are shown by the grey symbols with downward arrows.}}
  \label{Figure:SFRD_v_RCSurfaceBrightness}
\end{figure}

The data points of our sample of galaxies in Figs.~\ref{figure:sfr_v_Clumin} -- \ref{Figure:FIR_v_RC} are colour coded based on the galaxy's metallicity. This was done to investigate if there are any trends with metallicity. We find that in general the lowest metallicity objects congregate toward the low--radio continuum, low--SFR end of the plot whereas the high end is populated by the higher metallicity galaxies. This is a direct consequence of the metallicity--luminosity relation \citep{Skillman1989} and the fact that more luminous, hence more massive galaxies tend to have a higher SFR.

We compare our data points with the RC--SFR relation presented by \cite{Murphy2011}. They derive an expression for the \RCT\ and \RCNT\ emission and combine these to determine the total RC in a galaxy. The thermal component is derived from the ionising photon rate, which is directly proportional to the thermal spectral luminosity assuming an optically thin plasma giving:

\begin{align}
\left ( \frac{{\rm SFR}_{\nu}^{\rm Th}}{\rm {M_{\odot} yr^{-1} }} \right ) &= 4.6\times10^{-21}\left ( \frac{T_{\rm e}}{10^4{\rm K}} \right )^{-0.45} \nonumber\\
 &\qquad . \left ( \frac{\nu}{{\rm GHz}} \right )^{0.1}\left ( \frac{L_{\nu}^{\rm Th}}{{\rm W\,Hz^{-1}}} \right )
\end{align}

\noindent Where $T_{\rm e}$ is the electron temperature and $L_{\nu}^{\rm Th}$ is the thermal radio luminosity. This equation assumes solar metallicity, continuous SF, and a Kroupa IMF. Using a Kroupa IMF results in SFR estimates that are $\sim 2.5$ times larger than those found by \cite{Condon1992}. We assume an electron temperature of $10^4$\,K. As mentioned previously this value may vary considerably. A value of  $T_{\rm e} = 1.4\times 10^4$\,K \cite{Nicholls2014} would lead to a 14\% decrease in the SFR. The expected \RCNT\ is derived using:

\begin{align}
 \left ( \frac{{\rm SFR}_{\nu}^{\rm Nth}}{\rm {M_{\odot} yr^{-1} }} \right ) &= 6.64\times10^{-22}  \left ( \frac{\nu}{{\rm GHz}} \right )^{\alpha_{\rm Nth}} \nonumber\\
 &\qquad .  \left ( \frac{L_{\nu}^{\rm Nth}}{{\rm W\,Hz^{-1}}} \right )
\end{align}

\noindent This relationship is derived using the {\sc starburst99} population synthesis code \citep{Leitherer1999} and the empirical relationship between the supernova rate and non-thermal spectral luminosity of the Milky Way. We assume a value for the non-thermal spectral index of $\alpha_{\rm Nth} = -0.7 \pm 0.2$. Finally, the total RC is the combination of the \RCT\ and \RCNT\ leading to:

\begin{align}
 &\left ( \frac{{\rm SFR}_{\nu}}{\rm {M_{\odot} yr^{-1} }} \right ) = 10^{-20} \left [ 2.18 \left ( \frac{T_{\rm e}}{10^4{\rm K}} \right )^{-0.45}  \right.\nonumber\\
 &\qquad \left. {} \left ( \frac{\nu}{{\rm GHz}} \right )^{0.1} + 15.1 \left ( \frac{\nu}{{\rm GHz}} \right )^{\alpha_{\rm Nth}} \right ] \left ( \frac{L_{\nu}}{{\rm W\,Hz^{-1}}} \right )
\end{align}

\noindent where we use a frequency $\nu$ of 6.2\,GHz. These expected relations are plotted as a grey shaded area in Fig.~\ref{figure:sfr_v_Clumin}. The width of the band reflects the overall uncertainty based on a typical error in the spectral index of $0.2$ and a canonical factor of 2 uncertainty in the SFR. 

We performed a bivariate linear regression to quantify the relation between the RC luminosity and SFR, assuming the data follow a power law function of the form $y=Ax^n$ or $\log(y) = n \log(x) + c$, where $c = \log(A)$. We used the {\sc odr}\footnote{\scriptsize{\url{www.scipy.org/doc/api\_docs/SciPy.odr.odrpack.html}}} module from {\sc scipy}, which accepts four arrays of data points ($\log{x}$ and $\log{y}$, and the $1\,\sigma$ errors in log--space: $\frac{\delta x}{x}$ and $\frac{\delta y}{y}$) and the model function, and works to minimise the squares of the orthogonal distance between data points and the model, ultimately returning best-fit values and their standard deviations.

We find that the disk mask RC--SFR relation (Fig.~\ref{figure:sfr_v_Clumin} top-right panel) results are consistent with a linear relationship with $n = 0.93 \pm 0.14$ but the RC luminosity is lower than expected based on the observed SFR by approximately a factor of $\sim 5$. We note that IC\,1613 falls below the relation we find. If we exclude this galaxy we find that the average offset is a factor of $\sim 3$. We find that the RC mask integrated RC--SFR relation in Fig.~\ref{figure:sfr_v_Clumin} (bottom-right panel), where the RC mask is applied to both the RC as well as the SFR map, is marginally shallower than the \cite{Murphy2011} relation with a gradient of $n = 0.86 \pm 0.04$ with a scatter of $0.17$\,dex. If we perform the fit excluding NGC\,1569 (Fig.~\ref{figure:sfr_v_Clumin}, dashed black line) we find a value of $n = 0.91 \pm 0.04$. In Fig.~\ref{Figure:Heesen2014Comparison} we compare the results of our disk integrated and mask integrated RC--SFR to the study of 18 spiral galaxies at 20\,cm by \cite{Heesen2014}. We extend their parameter space by $2$\,dex, down to SFRs of $10^{-4}{\rm \,M_\odot\,yr^{-1}}$. At 20\,cm the \cite{Heesen2014} study found a slope of to the RC--SFR relation of $n = 1.24 \pm 0.04$ which is significantly steeper than our results. 

The relationship between the \RCT\ and \RCNT\ emission and SFR integrated over the disk and RC mask is shown in Fig.~\ref{Figure:rcnt_sfr}. When integrating the two components over the optical disk mask we find slopes of $n = 1.20 \pm 0.09$ and $n = 0.82 \pm 0.06$ for \RCT\ and \RCNT, respectively. We find a slope of $n = 0.82 \pm 0.05$ and $n = 0.79 \pm 0.06$ for \RCT\ and \RCNT\ emission integrated over the RC mask, respectively. If we exclude NGC\,1569 from the fit we find marginally steeper slopes. Our results for the \RCT\ agree with those of \cite{Murphy2011} when integrating over the RC mask but appear to diverge at the low SFR levels ($<10^{-2}$ \,\msunyr) in the disk mask. This may be due to the stochastic nature of SF particularly in the faintest galaxies. It is important to note that the \RCT\ plot is essentially an \halpha-FUV plot. The \RCT\ values are based on the \halpha\ emission and are thus not independently determined, in turn, the SFR relies heavily on the FUV. The scatter, especially for the least active dwarf galaxies (SFR$\,\lesssim 10^{-2} \, \mathrm{\msun \,yr^{-1}}$), is likely due to the \halpha\ emission underestimating the SFR in comparison to that from FUV by a factor of up to $10$ \cite[][]{Lee2009}. These authors argue that as only the highest mass stars ($M \gtrsim 18\,M_{\odot}$) produce a significant number of photons to ionise the surrounding \hi, having a deficit of these stars significantly reduces the amount of \halpha\ emission, while the FUV emission is not affected as much since a larger fraction of the stellar population contributes to the FUV. On the other hand, \cite{Koda2012} find O stars in stellar clusters as small as $100-1000\,{\rm M_\odot}$ coming to the conclusion that the stellar IMF is not necessarily truncated; it could be stochastically populated at the high mass end, accounting for the observed variations in Fig.~\ref{Figure:rcnt_sfr}. We discuss this further in Section~\ref{sect:interplay}.

The \RCNT\ results are shallower than expected based on the predictions of \cite{Murphy2011} in both masking cases. Not only is the slope more shallow, we also see that when using the disk mask the \RCNT\ emission falls below the expected SFR by a factor of 2--4. This agrees with \cite{Bell2003} who finds that the radio emission of low-luminosity galaxies must be suppressed by at least a factor of two to account for the RC--FIR relation at low luminosity. Our results also agree with the findings of \cite{Price1992} who find that the power-law dependence of the synchrotron luminosities and SFR has a slope of $n=0.8$. Using the same method applied here but observing at $20$\,cm, \cite{Heesen2014} found a slope of the \RCNT\ component for spiral galaxies to be $n = 1.16 \pm 0.08$, significantly steeper than our results (Fig.~\ref{Figure:Heesen2014Comparison}). We note that the \RCNT\ may be underestimated particularly for large-scale galaxies that have high SFRs such as NGC\,1569. This would lead to the \RCNT--SFR relation being steeper than we see in Fig.~\ref{Figure:rcnt_sfr}. However, when we remove NGC\,1569 from our fitting our results remain consistent with those with NGC\,1569 included.

The RC surface brightness--SFRD relation is presented in Fig.~\ref{Figure:SFRD_v_RCSurfaceBrightness} where the SFRD is derived over the extent of the galaxy. We find a tight, linear RC--SFRD relation with a slope of $n = 0.99 \pm 0.11$ and $n = 0.88 \pm 0.02$ for the disk and RC based masks, respectively. Within the errors, these slopes are the same as those found for the relations plotted in Fig.~\ref{figure:sfr_v_Clumin}. Unlike the luminosity plots in Fig.~\ref{figure:sfr_v_Clumin}, though, this is independent of distance and so errors introduced by distance uncertainties forcing a linear relation due to flux-to-luminosity scaling are avoided. Figure~\ref{Figure:SFRD_v_RCSurfaceBrightness} could thus be used as a baseline for future studies of normal star forming galaxies---especially those studies that do not have reliable distance measurements (e.g., only photometric redshifts of optical counterparts).

\subsection{The FIR--SFR Relation}
\label{section:Paper1fir--sfr_relation}

\begin{figure*}
  \centering
  \begin{tabular}{cc}
\includegraphics[width=0.48\linewidth,clip]{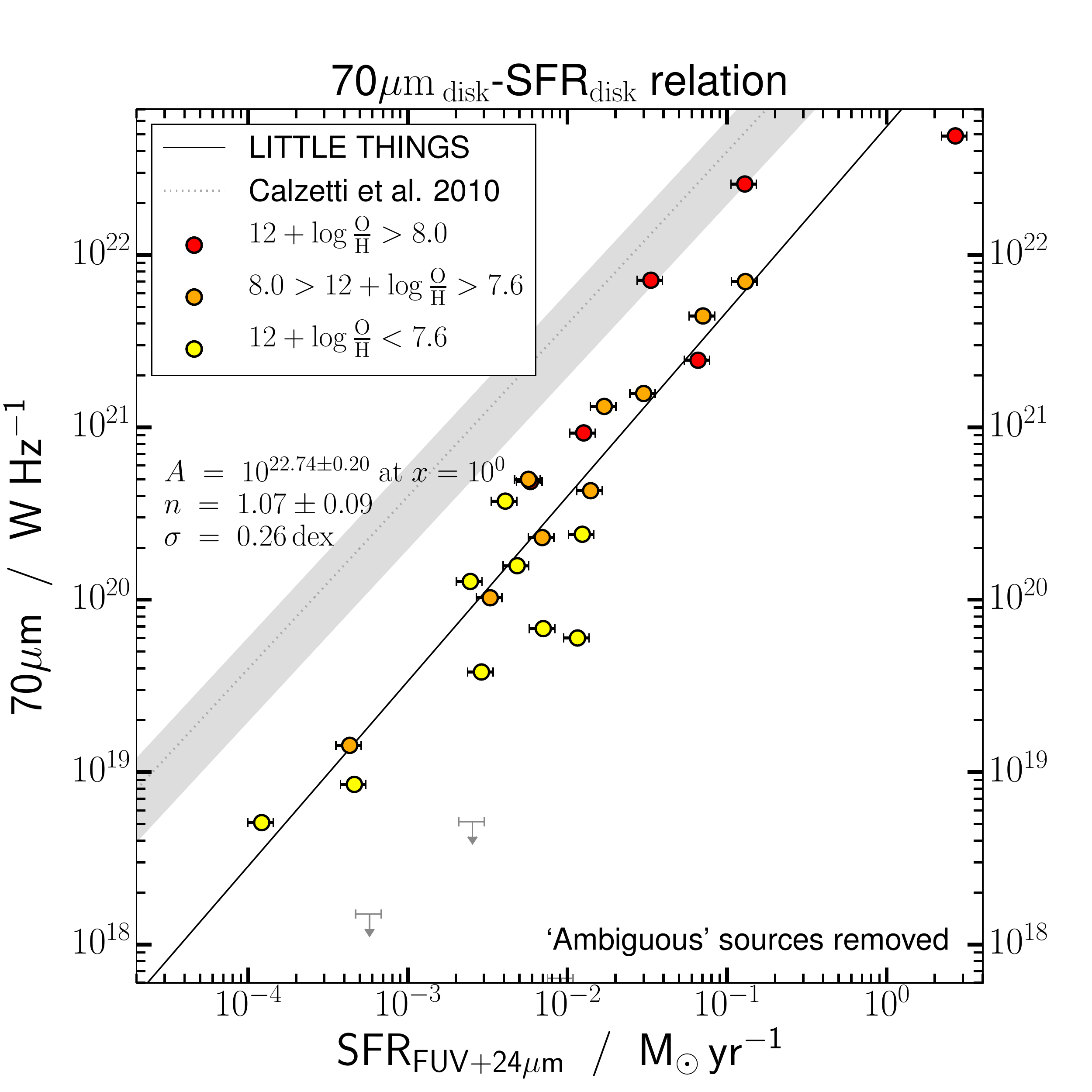}  &\
\includegraphics[width=0.48\linewidth,clip]{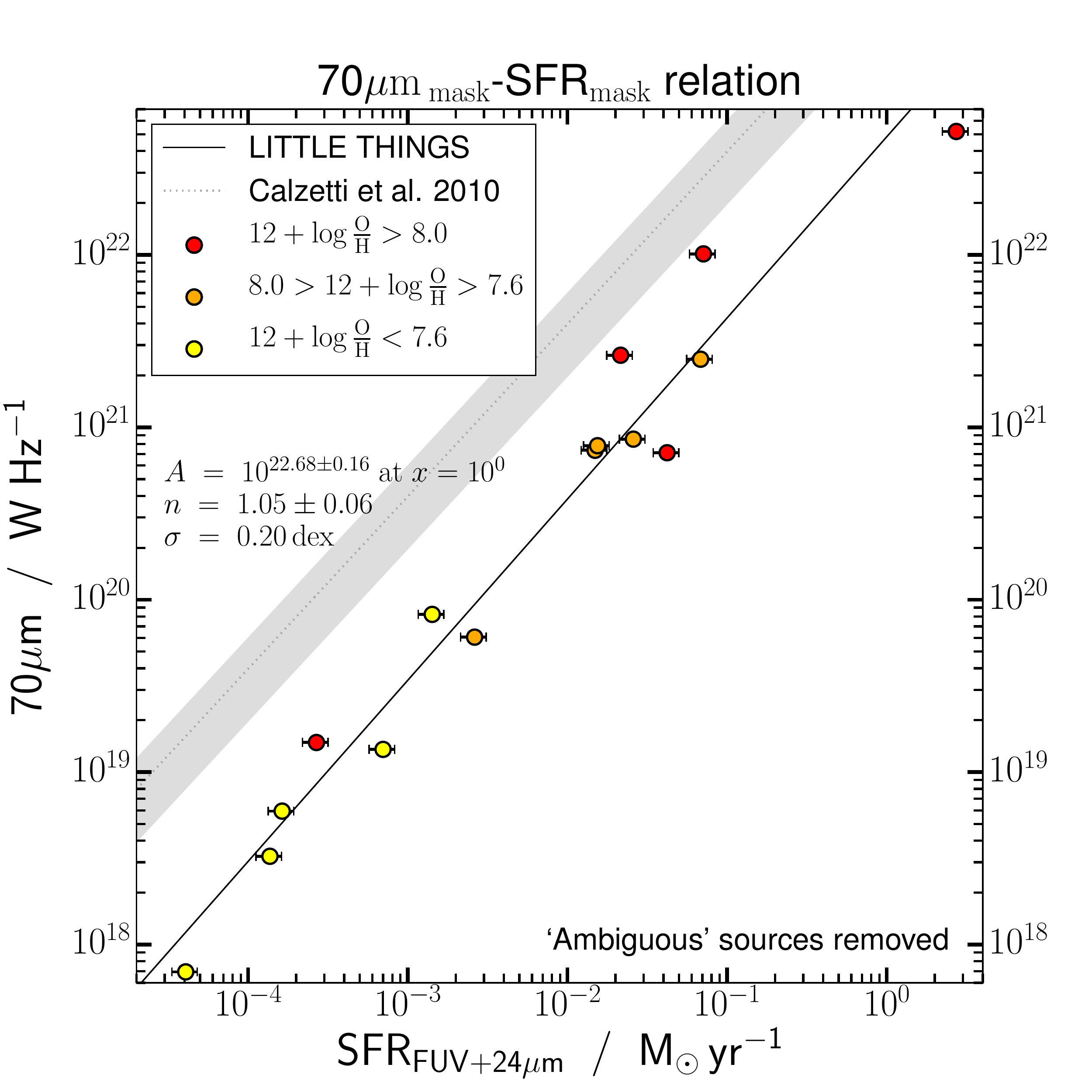}\\
  \end{tabular}
  \caption{\footnotesize{Total 70\micron\ luminosity as a function of SFR over our disk and RC masks. The solid line is the best-fit power law to our sample. For reference we show the \cite{Calzetti2010} 70\micron--SFR relation assuming a factor of two uncertainty (shaded grey band). Upper limits of 70\micron\ emission are shown by grey symbols with downward arrows.}}
  \label{figure:sfr_v_dust}
\end{figure*}

The FIR is often used as a proxy for SFR in studies of unresolved galaxies it is therefore instructive to examine the relationship between the FIR and SFR integrated within the disk and RC masks in Fig.~\ref{figure:sfr_v_dust} of our sample. We compare our estimate to the monochromatic 70\micron\ calibration of \cite{Calzetti2010} using:

\begin{align}
\left ( \frac{{\rm SFR}_{\rm 70\mu m}}{\rm {M_{\odot} yr^{-1} }} \right ) = \left (\frac{L_{\rm 70\mu m}}{{\rm W\,Hz^{-1}}} \right ) \cdot 2.52\times10^{-24}
\end{align}

\noindent where ${L_{\rm 70\mu m}}$ is the $70$\micron\ luminosity in ${\rm W\,Hz^{-1}}$. We find that our best-fit line is the same within the errors for both the disk and RC based masks ($n=1.07\pm0.09$ and $1.05\pm0.06$, respectively) and runs parallel to the \cite{Calzetti2010} relationship. However, for any given SFR we find that our measurement of the integrated 70\,\micron\ emissions is underestimated compared to the expected 70\,\micron\ luminosity by a factor of $\sim 10$. Given the fact that dwarf galaxies have low metallicity this is not surprising. The metallicity of all our galaxies falls below a value of $12+\rm{log(O/H)}\sim 8.1$, below which \cite{Calzetti2010} found the FIR to be an unreliable tracer of the SFR. At these low metallicities the galaxies become basically optically thin and FUV photons can escape before being reprocessed by dust and reemitted in the FIR. This was also suggested as the cause of the ratio between total IR and FUV being  $<1$ in low luminosity galaxies in a study by \cite{Bell2003}.

\subsection{The RC--FIR Relation}
\label{section:Paper1rc--fir_relation}

\begin{figure*}
  \centering
  \begin{tabular}{cc}
\includegraphics[width=0.48\linewidth,clip]{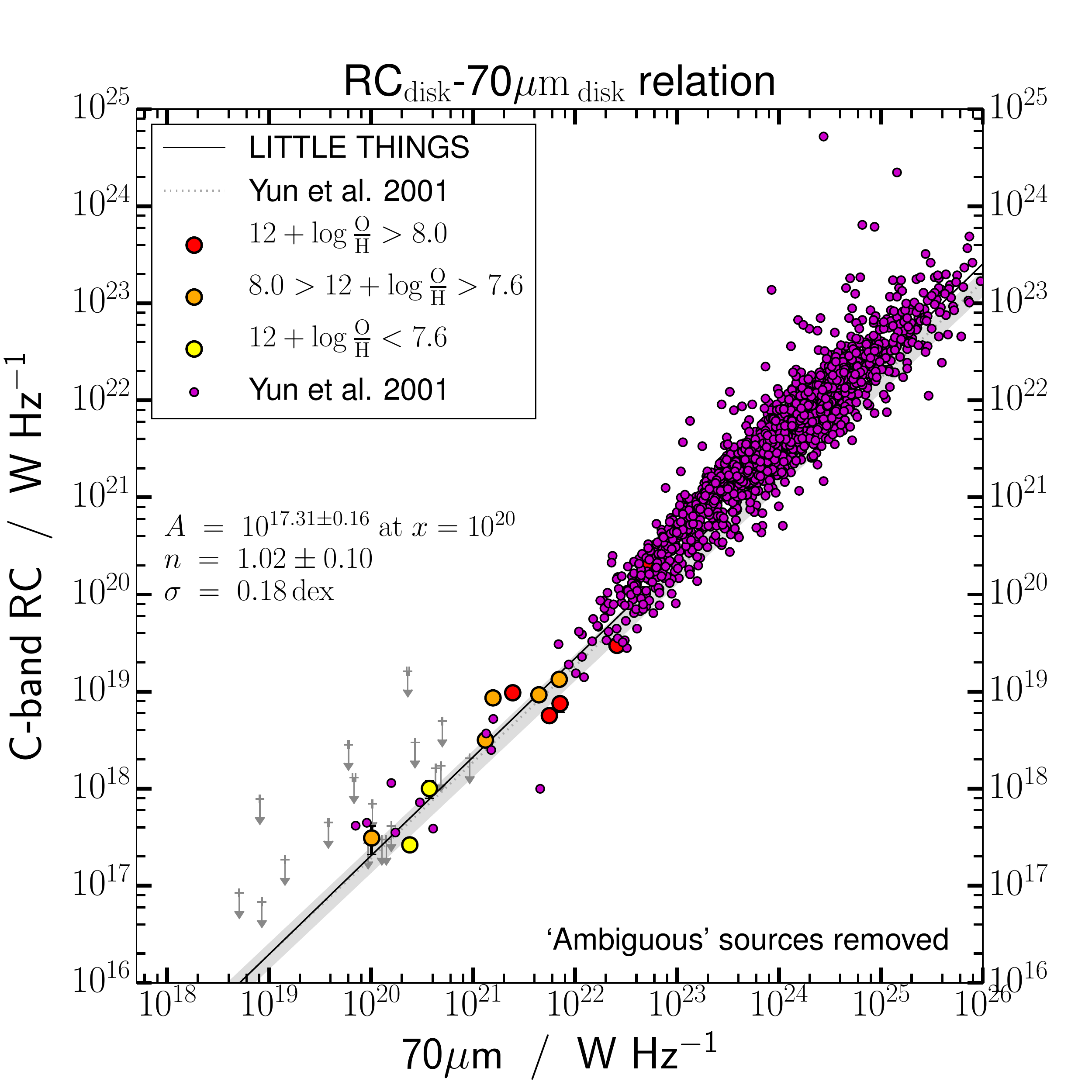} &\    \includegraphics[width=0.48\linewidth,clip]{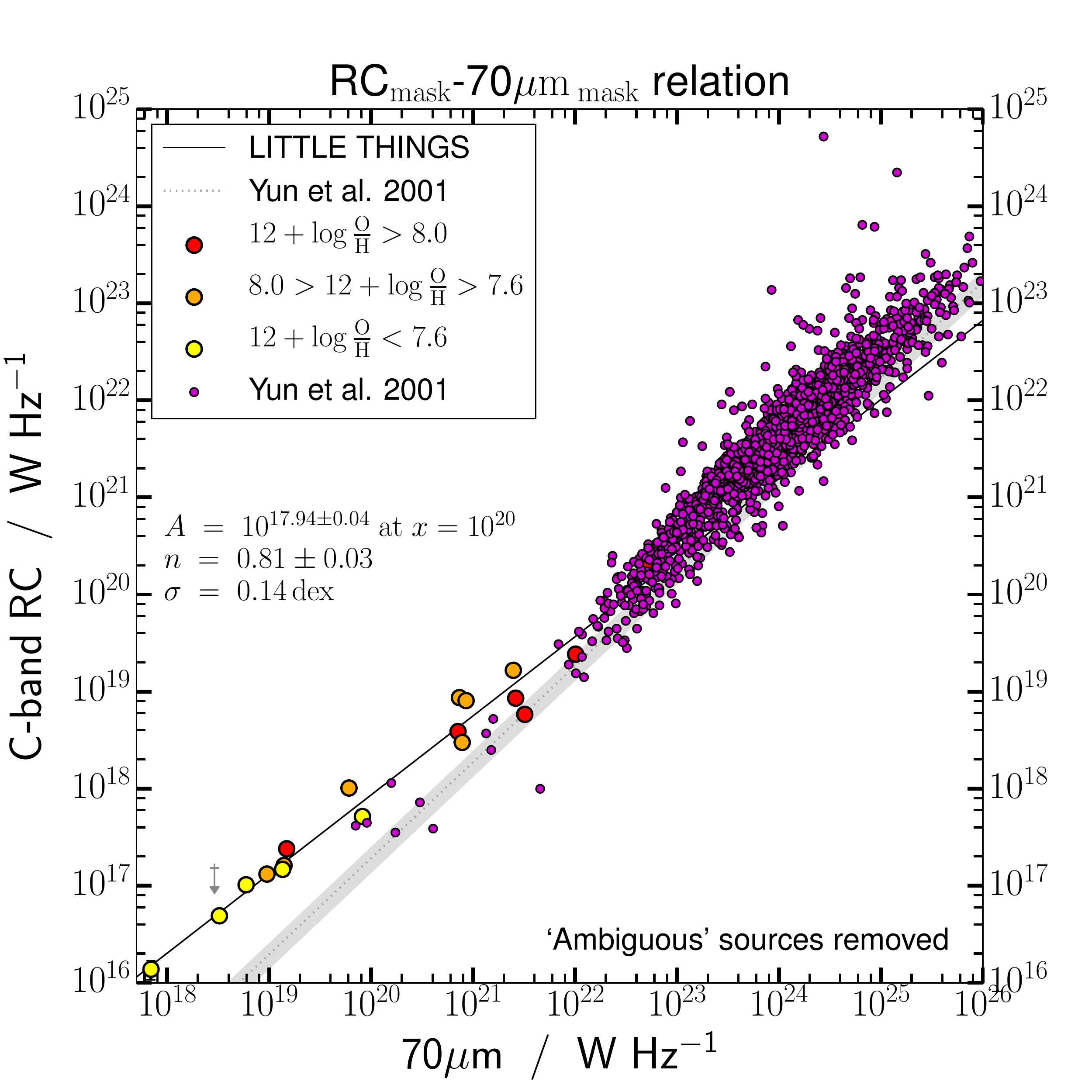}
  \end{tabular}
  \caption{\footnotesize{Comparison of RC and FIR luminosities from this study to those from \cite{Yun2001} (their VLA $1.4$\,GHz data have been extrapolated to $6$\,GHz and the {\em IRAS} 60\micron\ data corrected to {\em Spitzer} 70\micron). Both definite background and ambiguous RC sources have been removed from the LITTLE THINGS galaxies. Integrated quantities were taken from regions of significant RC emission (i.e., the RC-based mask; right-panel) and from over the whole optically defined disk (left). The LITTLE THINGS galaxies that remain undetected are represented by their $3\,\sigma$ upper limits (grey plus symbols with downward arrow). The uncertainties introduced by our conversion of the relation found by \citet{Yun2001} are reflected by the grey shaded band (see text for details). }}
  \label{Figure:Yun2001Comparison}
\end{figure*}

\begin{figure*}
  \centering
  \begin{tabular}{cc}
    \includegraphics[width=0.48\linewidth,clip]{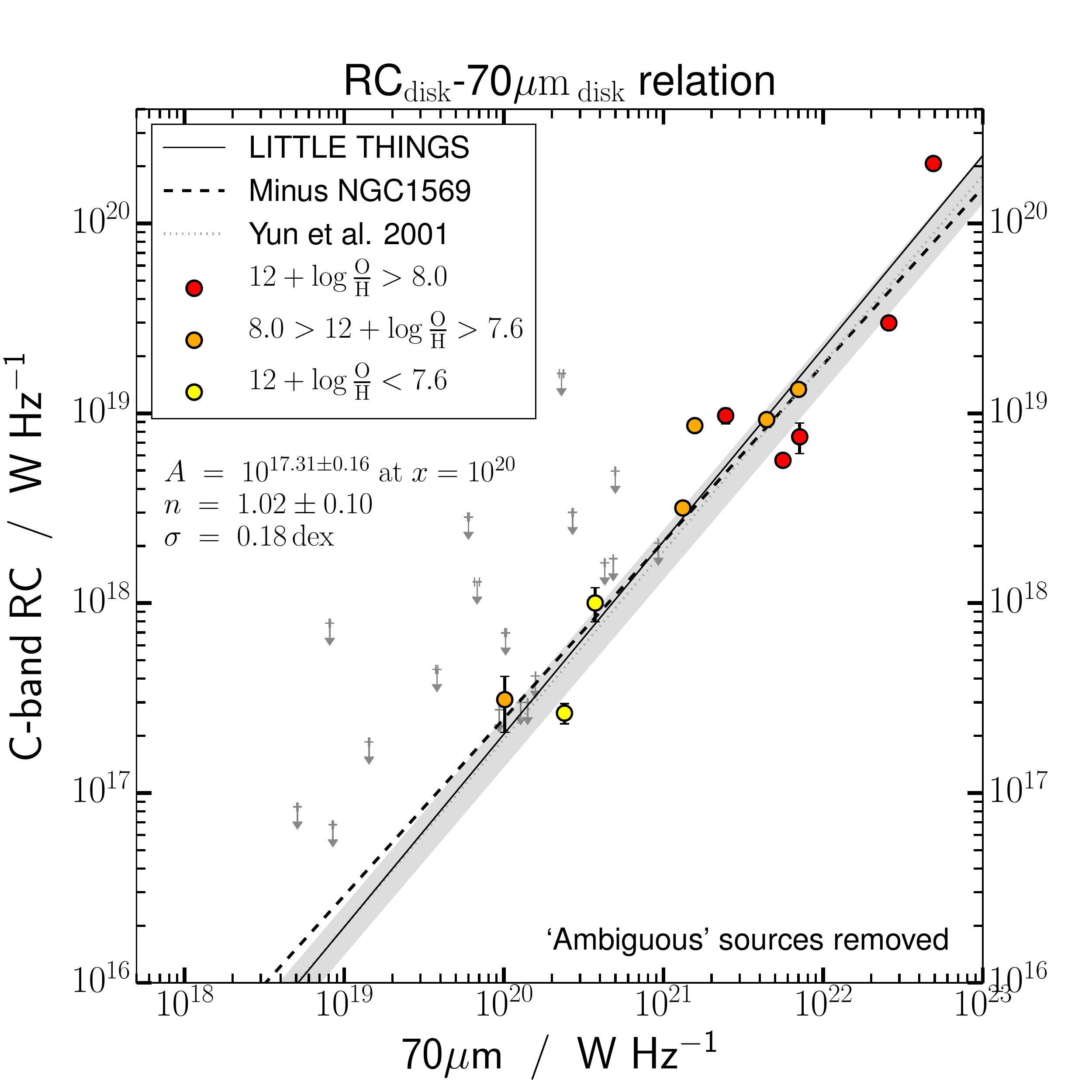} &\
    \includegraphics[width=0.48\linewidth,clip]{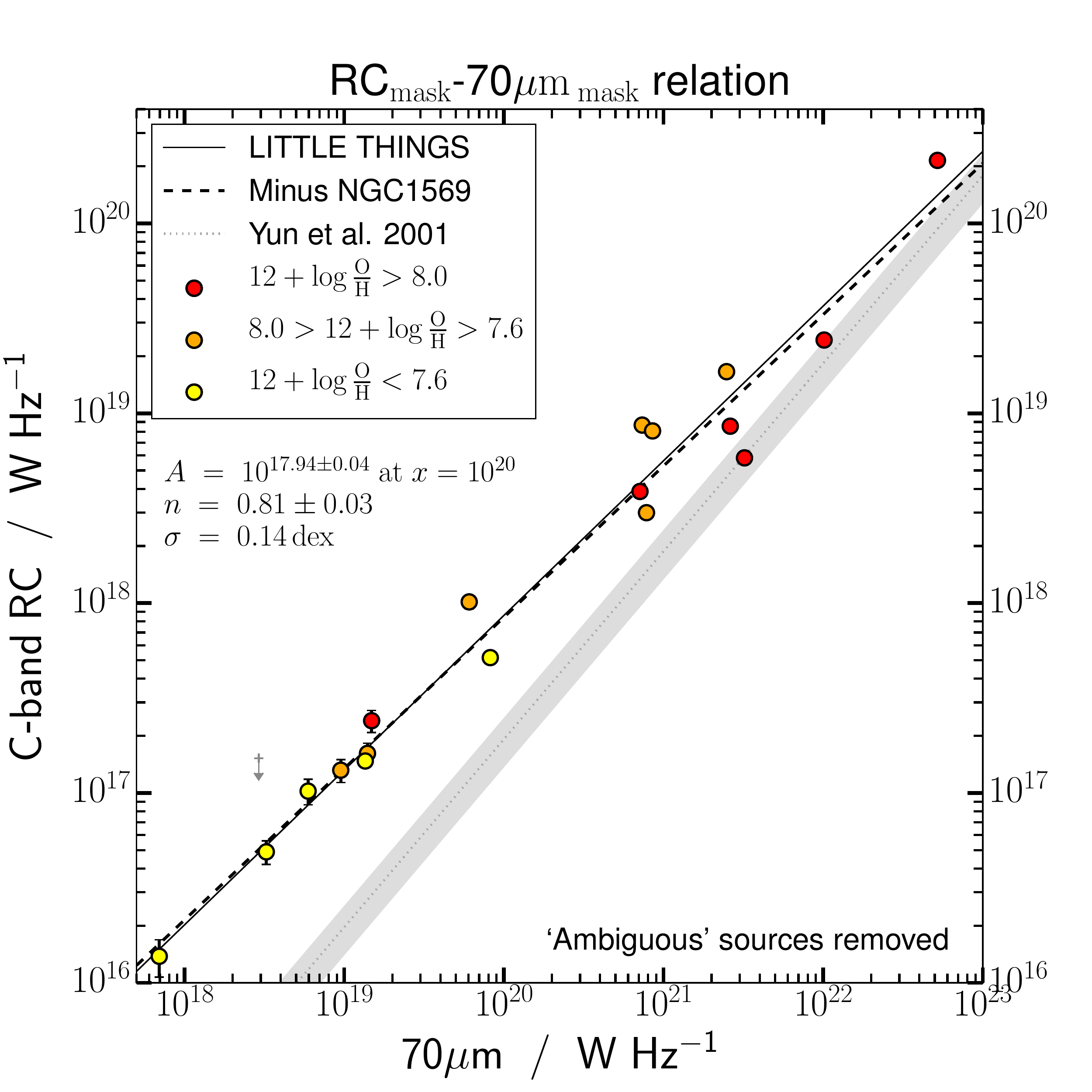} \\
        \includegraphics[width=0.48\linewidth,clip]{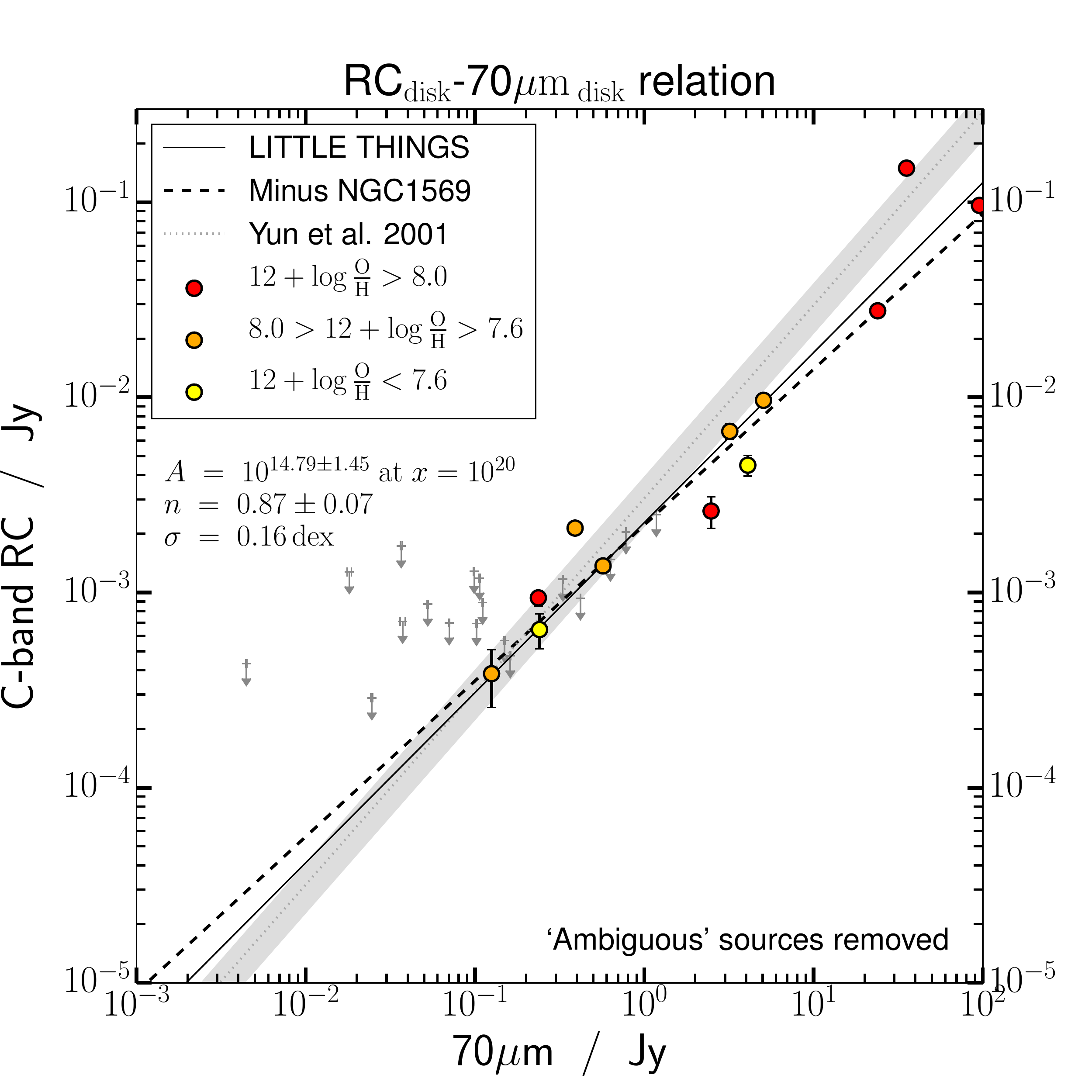} &\
    \includegraphics[width=0.48\linewidth,clip]{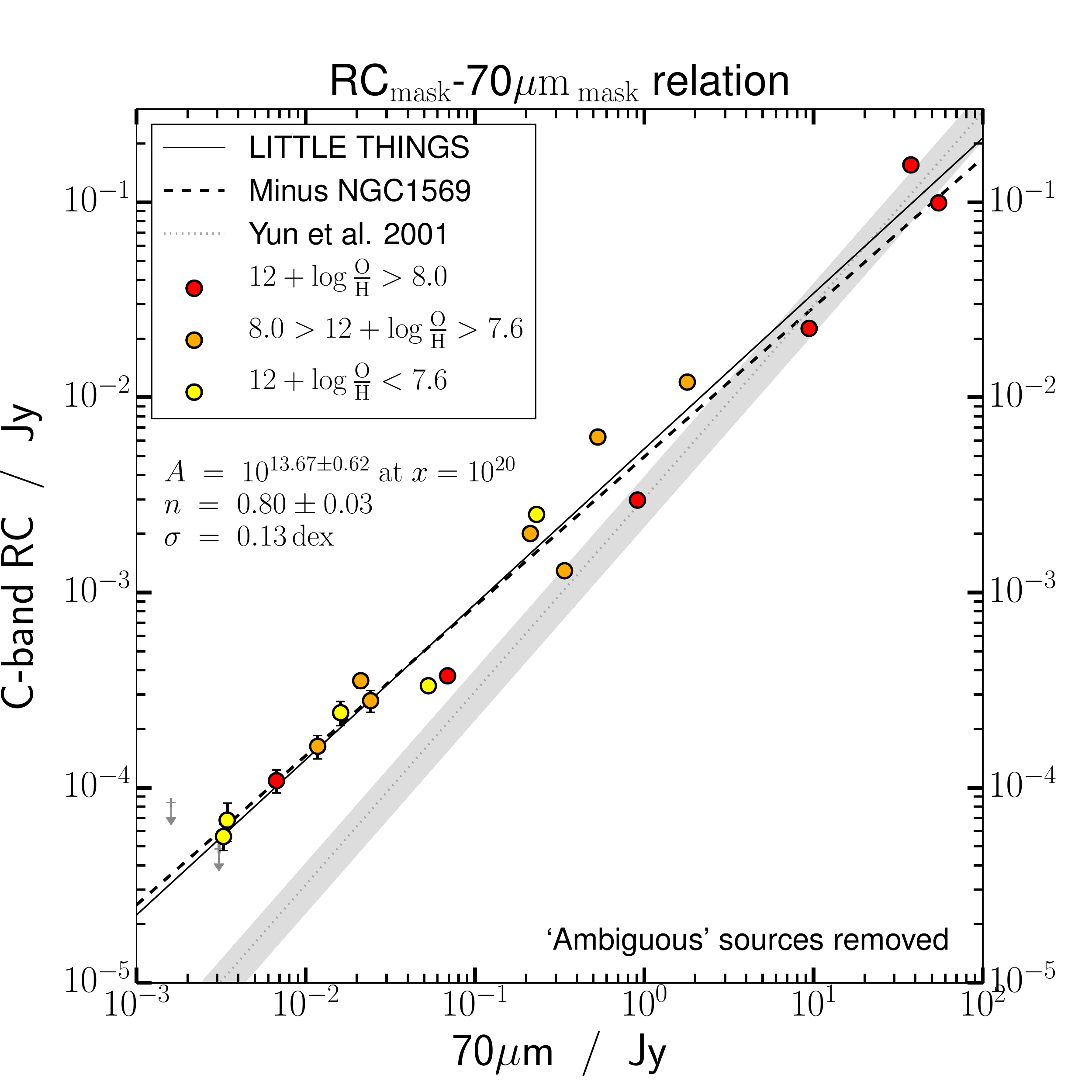}

  \end{tabular}
  \caption{\footnotesize{6\,cm luminosity as a function of {\em Spitzer} 70\micron\ FIR integrated over the disk mask (left) and RC mask (right). The top panels show the luminosity relation whilst the bottom panels have not been corrected for the distance and show the flux-density. Definite and ambiguous background sources have been removed. The solid line is the best-fit power law to our sample. We draw the \cite{Yun2001} RC--FIR relation as described in Equation~\ref{Equation:Yun2001_RC-FIR} (dotted line). The uncertainties introduced by our conversion are reflected by the grey shaded band (see text for details). The LITTLE THINGS galaxies that remain undetected are represented by their $3\,\sigma$ upper limits (grey plus symbols with downward arrow) }}
  \label{Figure:FIR_v_RC}
\end{figure*}

The RC--FIR relation based on 1809 galaxies culled from the NRAO VLA Sky Survey (NVSS: \citealt{Condon1998}) catalog and the 1.2\,Jy IRAS Redshift Survey catalog \citep{Strauss1992} was investigated by \cite{Yun2001}. They related the integrated $1.4$\,GHz RC of an unresolved galaxy to the {\em IRAS} $60{\rm \mu m}$ luminosity and found:
\begin{equation}
{\rm \frac{L_{6GHz}}{W\,Hz^{-1}}}  =  (2.24 \pm 0.67) \times 10^{-3} \Bigg( {\rm \frac{L_{70\mu m}}{W\,Hz^{-1}}} \Bigg)^{0.99}
\label{Equation:Yun2001_RC-FIR}
\end{equation}
\noindent where we converted the {\em IRAS} $60{\rm \mu m}$ luminosity to a `luminosity density' (i.e., from ${\rm W}$ to ${\rm W\,Hz^{-1}}$) by noting that the response from the {\em IRAS} $60{\rm \mu m}$ filter is equivalent to a perfectly transmitting filter with a bandwidth of $2.6 \times 10^{12}$\,Hz. The {\em IRAS} $60{\rm \mu m}$ `luminosity density' was further converted to the equivalent {\em Spitzer} 70\micron\ luminosity by scaling it up by a factor of $1.27$. This value is based on a grey body model for dust emission with $\beta=1.82$ and $T_{\rm dust}=35$\,K; this assumes the \citet{Yun2001} galaxies are in a quiescent mode of star formation, and that there is no significant emission from warm dust. The \cite{Yun2001} VLA $1.4$\,GHz RC data were reduced by a factor of $2.83$, to derive predicted equivalent VLA $6$\,GHz flux densities assuming a constant spectral index of $-0.7\pm0.2$ between 20 and 6\,cm for the galaxies in their sample. 

In Fig.~\ref{Figure:Yun2001Comparison} we show the RC--FIR relation for our dwarf galaxies and compare this to the results of \cite{Yun2001}. The RC--FIR relation traditionally samples the parameter space above FIR luminosities of $\sim 10^{22} {\rm \,W Hz^{-1}}$; we extend this to lower luminosities by $3$\,dex. The uncertainty presented in Fig.~\ref{Figure:Yun2001Comparison} takes into account an uncertainty in the spectral index of $0.2$ and $15$\,K in dust temperature. We show the RC--FIR relation for our dwarf sample where emission is integrated over the entire disk (Fig.~\ref{Figure:Yun2001Comparison}: left) and from the significant regions of RC emission only (i.e., the RC-based mask, Fig.~\ref{Figure:Yun2001Comparison}: right). In Fig.~\ref{Figure:FIR_v_RC} we show the RC--FIR relation for just our dwarf galaxy sample integrated over the disk mask (left) and RC mask (right). The top-panels of this figure show the luminosity and the bottom-panels the flux density, to illustrate any dependence on distance. We find that when integrated over the disk our results for the luminosity match those found by \cite{Yun2001} with a slope if $1.02 \pm 0.10$. The flux density derived slope is slightly shallower at $0.87\pm0.07$. However, when we integrate the RC and 70\micron\ emission using our RC mask we find that our results diverge from the \cite{Yun2001} relation in both the luminosity and flux density plots with a flatter slope of $0.81 \pm 0.03$ and $0.80\pm0.03$ for the luminosity and flux density, respectively. We discuss the possible reasons behind this in Section~\ref{sect:interplay}.

\subsection{$q$-parameter}
\begin{figure*}
  \centering
	\begin{tabular}{cc}
  	\includegraphics[width=0.49\linewidth,clip]{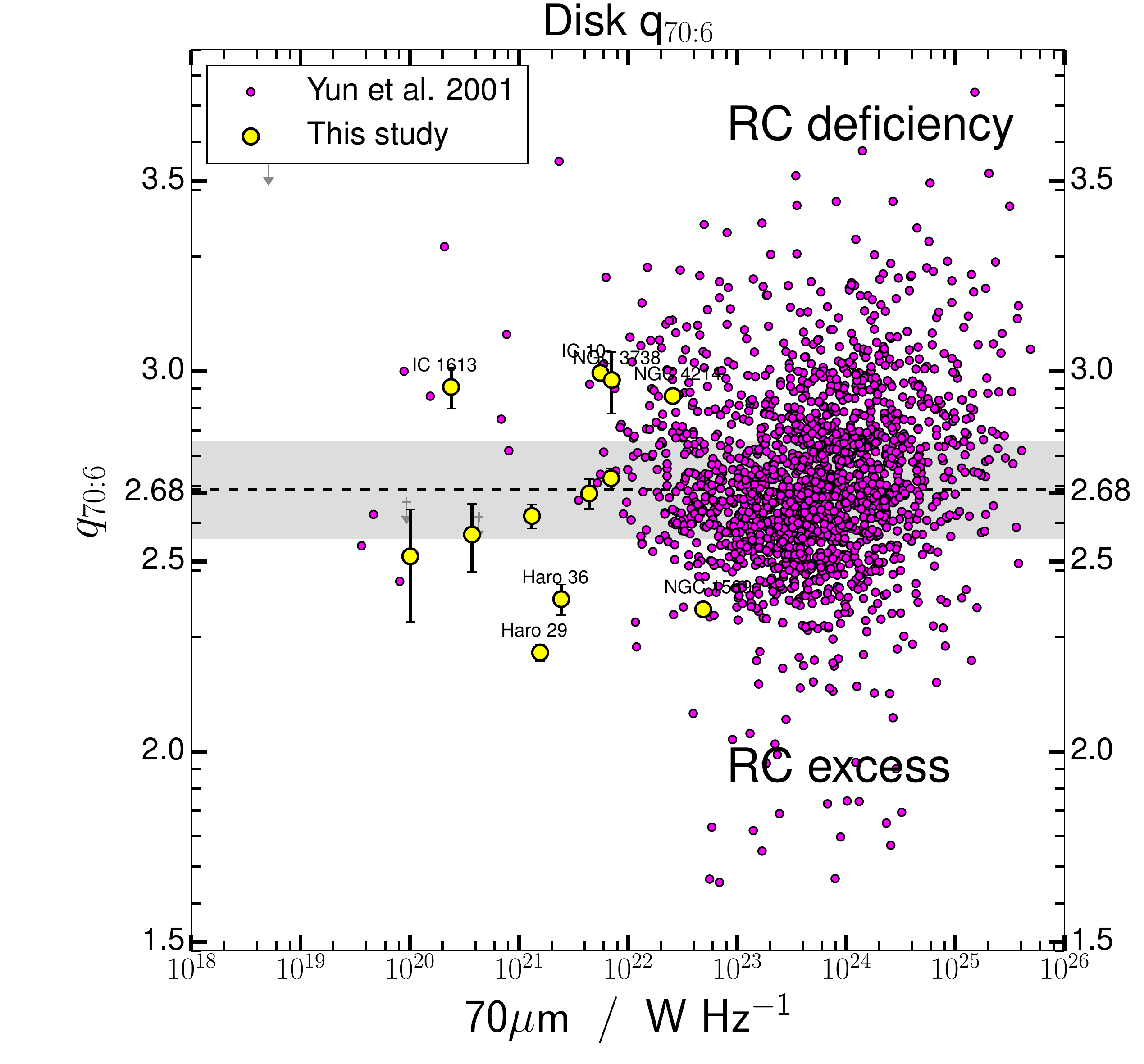} &\
  	\includegraphics[width=0.49\linewidth,clip]{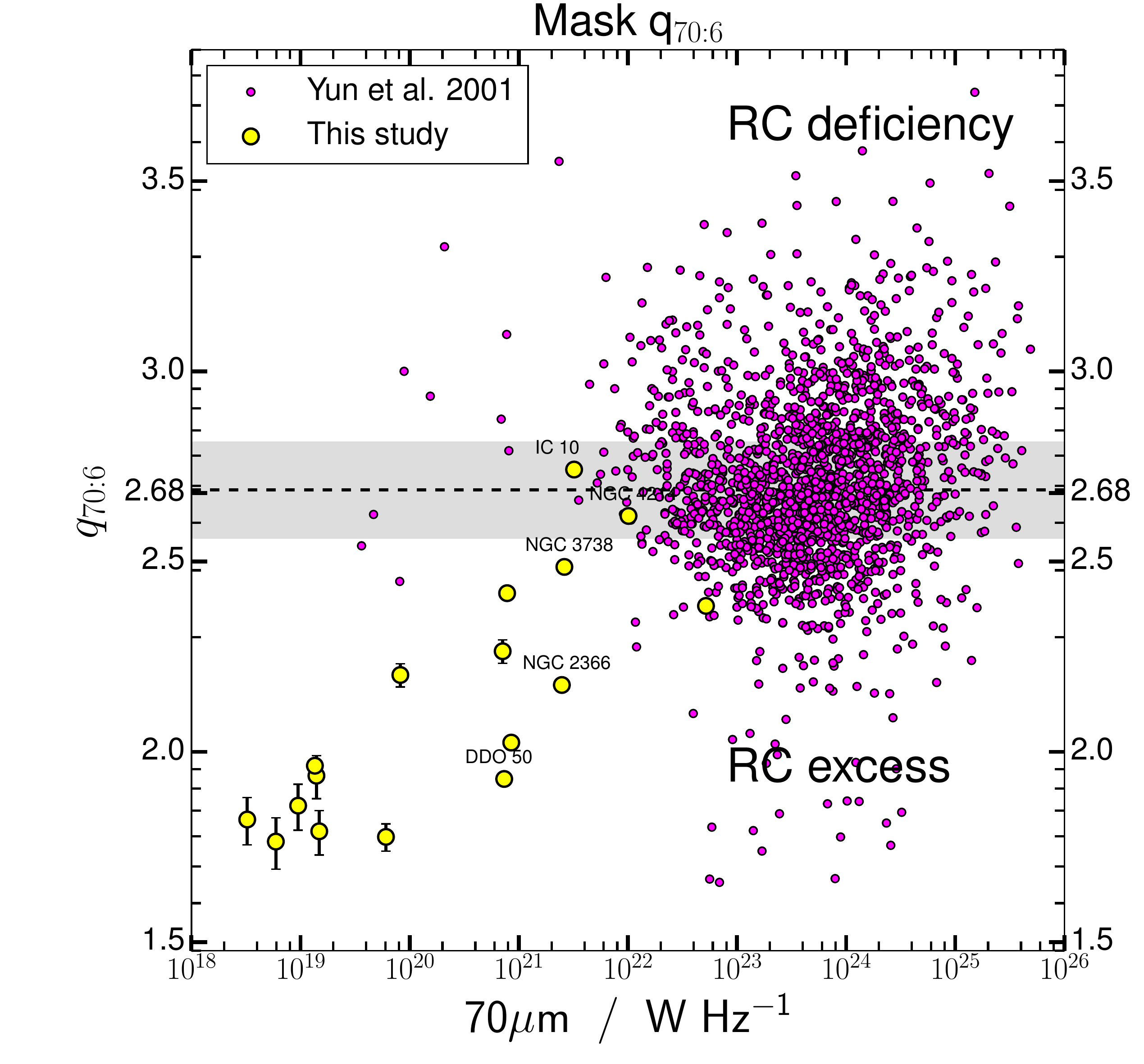}
   \end{tabular}
  \caption{\footnotesize{$q_{70:6}$ parameter as a function of 70\micron\ luminosity. Both obvious and ambiguous background sources have been removed. We also show the \cite{Yun2001} data points (purple) and their $q$-parameter appropriately corrected to our observing bands (dashed line). The errors introduced by our conversion are reflected by the grey shaded band.}}
  \label{Figure:q70}
\end{figure*}

An alternative way of exploring the RC--FIR relation described by \cite{Yun2001} is the $q$ parameter. This is the natural logarithm of the ratio of the {\em IRAS} FIR (a weighted combination of $60$ and $100{\rm \mu m}$ flux) to VLA $1.4$\,GHz flux densities of the \cite{Yun2001} sample and is described by:
\begin{equation}
q_{\rm FIR:1.4} = \log  \Bigg( {\frac{{\rm FIR\ [Jy]}}{{\rm RC\ [Jy]}}} \Bigg) ,
\label{Equation:Yun2001_RCFlux-FIR}
\end{equation}
\noindent the average $q_{\rm FIR:1.4}$ parameter was found to be $2.34 \pm 0.01$. Since our RC and FIR measurements were made in different bands to those used by \cite{Yun2001}, we convert their $q_{\rm FIR:1.4}$ to $q_{\rm 70:6} = 2.68 \pm 0.12$, where the subscript $70$ and $6$ refer to the 70\micron\ FIR and $6$\,GHz RC, respectively. As before, the uncertainty is calculated by assuming an $0.2$ uncertainty in the spectral index between 20 and 6\,cm, and $15$\,K uncertainty in dust temperature.

We plot $q_{\rm 70:6}$ values as a function of 70\micron\ FIR luminosity in Fig.~\ref{Figure:q70}. The LITTLE THINGS dwarfs are consistent with the \cite{Yun2001} sample when integrated over the disk (Fig.~\ref{Figure:q70}; left). This reveals that the RC--FIR `conspiracy' continues in our dwarf galaxy sample. However, we see in the right panel of Fig.~\ref{Figure:q70} that when we integrate the emission over the RC mask there is a considerable excess of RC emission compared to the 70\micron\ for sources that have a low radio luminosity, as was already evident in Figs.~\ref{Figure:Yun2001Comparison} and \ref{Figure:FIR_v_RC} (right hand panels). 

\subsection{The Interplay Between SF, RC and FIR}
\label{sect:interplay}
The relationship between the RC, FIR, and SFR can be summarised by three equations. The RC--SFR and FUV-- SFR relations can be expressed as:
\begin{equation}
{\rm \frac{L_{RC}}{W\,Hz^{-1}}} = 10^{A} \Bigg( {\rm \frac{SFR}{M_\odot\,yr^{-1}}} \Bigg)^{n}
\label{Equation:RC-SFR}
\end{equation}

\noindent and:

\begin{equation}
{\rm \frac{L_{FIR}}{W\,Hz^{-1}}} = 10^{A} \Bigg( {\rm \frac{SFR}{M_\odot\,yr^{-1}}} \Bigg)^{n}
\label{Equation:RC-SFR}
\end{equation}

\noindent whilst the RC--FIR relation can be described by:
\begin{equation}
{\rm \frac{RC}{W\,Hz^{-1}}} = 10^{A} \Bigg( {\rm \frac{70\mu m\ FIR}{10^{20} W\,Hz^{-1}}} \Bigg)^{n}
\label{Equation:RC-FIR}
\end{equation}
\noindent We summarise the results for the various fits for these relations over the disk and RC masks in table~\ref{tab:fits}.

We use the Pearson (P) and Spearman (S) coefficients to describe the strength and direction of the correlation in our relationships, where a value of $-1.0$ indicates a strong anti-correlation, $0.0$ indicates the relationship is random or non-existent, and $+1.0$ indicates a strong, positive correlation. The Pearson coefficient assumes a purely linear relationship and so will approach $0.0$ if there is an inconsistent relationship whilst the Spearman coefficient evaluates the monotonic relationship between two variables and so will remain high. Thus a strong relationship that deviates from linearity will have a lower Pearson than Spearman score. Both the Pearson and Spearman coefficients indicate a strong correlation between all components of the RC and SFR, with values ranging from 0.77 to 1.00 depending on the relation and type of mask used.  

The fitted parameters for $A$ and $n$ for each relation in table~\ref{tab:fits} vary significantly based on the type of mask used except in the case of the FIR--SFR relation. The FIR--SFR relation has lower Pearson and Spearman scores compared to the RC based relations suggesting a weaker relationship and deviation from linearity and shows a larger scatter. Since the RC based mask is able to probe significantly lower SFR galaxies this suggests that the physical processes responsible for the FIR--SFR relation operate in the same way regardless of the level of SFR in our sample. The varying parameters that fit the various RC--SFR relations in the disk and RC masks indicate that there may be some change in the physical processes operating within regions on the scale of the entire disk and resolved scales traced by the RC based masks.

One of the most striking results of this study is the divergence we see in the RC--FIR relation at low luminosities when integrating over the RC mask (Fig.~\ref{Figure:FIR_v_RC}: right). This appears to be caused by the FIR emission being underestimated in the RC mask relative to the expected \cite{Calzetti2010} SFR (Fig.~\ref{figure:sfr_v_dust}) whereas the RC emission continues to follow the SFR down to low values (Fig.~\ref{figure:sfr_v_Clumin}). Our disk masked results on the other hand, both underestimate the RC and FIR luminosity compared to the predicted SFR leading to the linear slope that continues the RC--FIR `conspiracy' seen in the left panel of Fig.~\ref{Figure:FIR_v_RC}. In the case of the RC emission excess we find evidence that it is the \RCNT\ component that is responsible for the suppression of the RC on the scale of the disk (Fig.~\ref{Figure:rcnt_sfr}). 

We propose two possible scenarios that may be responsible for the relations we observe. The first is that dwarf galaxies do not act as calorimeters and \CRe\ and dust heating photons are able to escape the galaxy before they are able to generate the total \RCNT\ and FIR emission associated with their host SF. The second possibility is that we are witnessing the effect of stochastic SF resulting in a partially sampled and/or truncated IMF. This may lead to the \RCNT\ and FIR underestimating the SFR compared to studies of larger galaxies. It is possible that some combination of these scenarios is responsible for our results.
\begin{table*}
\centering
\caption{The best-fit parameters for the RC--SFR, FIR--SFR, and RC--FUV relation for the total RC, \RCT, and \RCNT\ components integrated over the entire disk and only over the RC masks. (1) RC component and mask type; (2 \& 3) Fit parameters; (4) Scatter of the data; (5) Number of galaxies used in the fit; (6 \& 7) The Pearson (P) and Spearman (S) coefficients.}
\label{tab:fits}
\begin{tabular}{lccccccc}
\hline
 Mask & \multicolumn{3}{c}{RC Luminosity--SFR Relation} & N & P & S\\
 & $A$ & $n$ & $\sigma$ & & &  \\
\hline 
RC$_{\rm Disk}$--SFR	&	$20.05 \pm 0.22$	&	$0.93 \pm 0.14$	&	$0.23$	&	$11$	&	$0.87$	&	$0.93$	\\
RC$_{\rm Mask}$--SFR 	&	$20.16 \pm 0.12$	&	$0.86 \pm 0.04$	&	$0.17$	&	$19$	&	$0.92$	&	$0.96$	\\			 
\hline
\RCT$_{\rm , Disk}$--SFR	&	$20.21 \pm 0.19$	&	$1.20 \pm 0.09$	&	$0.23$	&	$32$	&	$0.95$	&	$0.88$		\\
\RCT$_{\rm , Mask}$--SFR	&	$19.75 \pm 0.16$	&	$0.82 \pm 0.05$	&	$0.22$	&	$19$	&	$0.92$	&	$0.90$		\\			 
\hline
\RCNT$_{\rm , Disk}$--SFR	&	$19.72 \pm 0.08$	&	$0.82 \pm 0.06$	&	$0.08$	&	$6$	&	$0.99$	&	$1.00$		\\
\RCNT$_{\rm , Mask}$--SFR	&	$19.78 \pm 0.14$	&	$0.79 \pm 0.06$	&	$0.17$	&	$14$	&	$0.89$	&	$0.91$		\\			 
\hline 
FIR$_{\rm Disk}$--SFR	&	$22.74 \pm 0.20$	&	$1.07 \pm 0.09$	&	$0.26$	&	$24$	&	$0.80$	&	$0.88$		\\
FIR$_{\rm Mask}$--SFR	&	$22.68 \pm 0.16$	&	$1.05 \pm 0.06$	&	$0.20$	&	$15$	&	$0.77$	&	$0.93$		\\			 
\hline 
RC$_{\rm Disk}$--FIR	&	$17.31 \pm 0.16$	&	$1.02 \pm 0.10$	&	$0.18$	&	$12$	&	$0.93$	&	$0.80$		\\
RC$_{\rm Mask}$--FIR	&	$17.94 \pm 0.04$	&	$0.81 \pm 0.03$	&	$0.14$	&	$18$	&	$0.89$	&	$0.94$		\\			 
\end{tabular}
\end{table*}

\subsubsection{Calorimeter breakdown}
Given their low metallicity and dust content it is possible that the mean free path of dust heating photons is less than the galaxy scale height. This would mean that dust heating photons are not completely absorbed and reradiated into the FIR and instead escape the galaxy. This would result in the observed suppression of the FIR relative to the expected SFR from \cite{Calzetti2010} in both the RC and disk mask. Given the low optical depth of dwarf galaxies \citep{Bell2003} this is one plausible explanation for our observed FIR--SFR relation. 

The low gravitational potential wells of dwarf galaxies make them particularly susceptible to outflows where material including \CRe\ can escape the galaxy. In addition \CRe\ may diffuse out of the galaxy before they are able to emit all their radiation if the magnetic field strength is low. Both these processes would lead to a suppression of the \RCNT\ emission relative to the SFR (Fig.~\ref{Figure:rcnt_sfr}: right). The \RCT\ emission is expected to be consistent with theoretical predictions because it is directly associated with ongoing massive star formation \citep{Murphy2011}. This indeed seems to be the case in our study, except  at low SFRs of $<10^{-2}$\msunyr\ where we see evidence of suppression and an increased scatter of \RCT\ (and therefore \halpha) relative to the SFR. The suppression of the \RCNT\ emission relative to expected SFR is only observed in our results when integrated over the disk mask. This may suggest that on the scales of concentrated SF traced by our RC mask magnetic fields are strong enough to act as a local calorimeter. Therefore, \CRe\ expend their energy and age to lower frequencies before they have a chance to diffuse into the wider ISM. We explore this possibility further in Section~\ref{Section:Paper1MagneticFields}.  

\subsubsection{Stochastic star formation}
An alternative explanation for our observed trends in the RC, FIR, and SFR could be due to the effects of stochastic SF within our dwarf galaxy sample. Synthesis models used to calibrate SFR indicators assume a sufficient number of stars to fully populate the IMF. In the case of dwarf galaxies with low SFRs the short lived high-mass stars may be under represented invalidating this assumption. Simulations of dwarf galaxies undergoing stochastic SF have found that FUV inferred SFRs may be biased at the $\sim 0.5$\,dex level for SFRs of $\sim 10^{-4}$\msunyr\ \citep{daSilva2012,daSilva2014}. Moderate variations in the SFR have also been observed in the star formation histories of nearby dwarf galaxies \cite{Dohm1998, Weisz2008}. We would also expect stochasiticity to impact the generation of the FIR, RC, and \halpha\ emission because these are all sensitive to the high-mass end of the IMF. These effects will have complicated spatial and temporal dependancies that vary for each type of emission mechanism. If the high-mass end of the IMF is underpopulated we would expect \halpha\ emission to be suppressed or even absent \citep{Lee2009} resulting in an increase in scatter around SFRs of $10^{-2}$\msunyr\ and a suppression of the \halpha\ emission below $10^{-3}$\msunyr. We see this behaviour in our \RCT--SFR relation in the disk mask results (Fig.~\ref{Figure:rcnt_sfr}: top left). This agrees with the simple calculation of \cite{Lee2009} who find that the \halpha\ flux is only a robust tracer of SFR above $1.4\times10^{-3}$\msunyr. However, \cite{Lee2009} argue that the stochastic SF alone is not sufficient to explain the suppression of \halpha\ emission. In any case the suppression of the RC and FIR we see in our disk masks are both evident at SFRs greater than $10^{-2}$\msunyr.

From our current data set it is unclear which of the scenarios we discuss above is responsible for our observed relations. Further work is required to investigate the impact of stochastic SF and a possible breakdown of the calorimeter model in these galaxies and the impact on our RC, FIR, and SFR measurements. In order to establish the origin of the observed suppression of the RC relative to SFR we require further information on the spectral and spatial variation of the RC--SFR relation in these galaxies so we may explore the effects of cosmic ray ageing and transport \citep{Heesen2016}. In order to explore the impact of stochastic SF we require detailed Monte-Carlo simulations of the underlying stellar populations or observations of the resolved stellar populations.

\subsection{Cosmic Ray Electrons and Magnetic Fields}
\label{Section:Paper1MagneticFields}

Our results suggest that it is the suppression of the \RCNT\ emission that is responsible for the RC--FIR relation remaining consistent at low SFRs. To explore the source of the \RCNT\ emission, we investigate the synchrotron emissivity in an optically thin region:
\begin{equation}
\epsilon_{\rm NTh} \propto n_{\rm CR} B_{\perp}^{\frac{\gamma + 1}{2}},
\end{equation}
\noindent where $n_{\rm CR}$ is the number density of \CRe\ present in the dwarf's galactic magnetic field, $B_{\perp}$ is the strength of the transverse magnetic field, and $\gamma$ is the power--law slope of the \CRe\ injection spectrum.

The \RCNT\ emission depends both on the energy density contained within the magnetic field and that of the population of \CRe. The combined energy density associated with the magnetic field and \CRe\ is usually assumed to be at a minimum \cite[see Section 16.5 of][]{Longair1981}. In galaxies, this is a reasonable assumption. If the energy densities are not equal, they will tend to balance: for example, if the energy density is dominated by the \CRe, then they will rise out of the galactic disk in Parker lobes due to their buoyancy, expand, and escape thus reducing their energy density until it is balanced with that in the magnetic field. 

It is, however, conceivable that dwarf galaxies in particular deviate from equipartition. This would lead to a reduction in synchrotron emission (see Fig.~\ref{Figure:rcnt_sfr}) in two different ways:
\begin{itemize}
\item[1)] a low number density of \CRe\ ($n_{\rm CR}$) present in the dwarf's galactic magnetic field. Dwarf galaxies in particular are prone to galactic outflows since they have low masses and correspondingly shallow gravitational potentials, and winds can advect plasma and resident \CRe\ away from the galaxy;
\item[2)] the magnetic field strength ($B$) being lower than the equipartition value at which the energy density of the magnetic field is equal to that of the cosmic rays (electrons and protons combined). In the standard paradigm of a mean field $\alpha$--$\Omega$ dynamo, the key ingredients are turbulence and shear. Dwarf galaxies may be sites of weak, large-scale, ordered magnetic fields, so magnetic field amplification may be less efficient. Studies by \cite{Chyzy2011,Roychowdhury2012} found global magnetic field strengths on the order of $<5{\rm \mu G}$ towards dwarf galaxies. However, the turbulent field in and around the SF regions may not necessarily be weaker than that found in spirals \cite[e.g.,][]{Tabatabaei2013b} as $30{\rm \mu G}$ is observed across some $100$\,pc regions.
\end{itemize}

In the following, the magnetic field strength in our sample of dwarf galaxies is estimated under the assumption of equipartition; this is the only practical way of estimating the field strengths given our current data set. We apply the equipartition formula for the total magnetic field following equation\,3 from \cite{Beck2005}. We made the standard assumptions of a spectral index of $-0.7\pm0.2$, proton--to--electron number density ratio {\textbf K} is $100 \pm 50$ \cite[][]{Beck2005, Murphy2011}, and that the dwarf galaxy has a scale height of $400 \pm 200$\,pc independent of distance from the galaxy centre \cite[][]{Banerjee2011,Elmegreen2015}. We note that both these assumptions are prone to significant uncertainty. The value of {\textbf K} depends on the acceleration process, propagation and the energy losses of the protons and electrons. As \CRe\ propagate away from their sites of acceleration they rapidly lose energy leading to values of {\textbf K}$>100$. If this is the case in our dwarf galaxies and if \CRe\ are escaping the galaxy altogether this will lead to an underestimate of the equipartition magnetic field. The scale height of dwarf galaxies is also prone to large variation due to their low mass and the potential for outflows. Typical scale heights have values ranging from 200-400\,pc in the inner regions and to 600-1000\,pc in the outer regions.

The average transverse magnetic field strength of our sample is $5.2 \pm 2.6{\rm \,\mu G}$ and $7.5 \pm 3.3{\rm \,\mu G}$ when integrated over the RC and disk masks, respectively (table~\ref{table:12A-234_DiskQuantities} and \ref{table:12A-234_MaskQuantities} for galaxy specific values). These values are greater than the $\sim 2 {\rm \mu G}$ found in $\sim 50$ faint dwarf galaxies from the NVSS catalogue \cite[][]{Roychowdhury2012} and within the errors they are consistent with the $4.2{\rm \mu G}$ found towards 12 local group dwarf galaxies reported in \cite{Chyzy2011}.The transverse field strength we measure in both masks is lower than the $9.7{\rm \,\mu G}$ found in larger spiral galaxies in the WSRT SINGS sample \cite[][]{Heesen2014}. The disk integrated magnetic field strength is greater than the mask integrated value because only the brightest galaxies can be integrated over the entire optical disk. 

Our data allow the magnetic field strength to be measured on a resolved basis. In a few dwarf galaxies (e.g., NGC 1569, NGC\,2366, NGC\, 4214), we find numerous regions where the magnetic field strength was measured to be as high as $30$--$50{\rm \,\mu G}$ in localised $100$\,pc regions (approximate area of the synthesised beam). In fact, the brightest \RCNT\ flux density in our sample comes from a $\sim 100$\,pc region in NGC\,1569---the flux density from this unresolved region implied a magnetic field strength of $\sim 50{\rm \,\mu G}$. \citet{Heesen2015} analysed in detail the \RCNT\ spectrum of the non--thermal super bubble in IC\,10, deriving a magnetic field strength of $44{\rm \,\mu G}$. These are all strong magnetic fields akin to those found in the SF regions of larger spirals \cite[e.g., the turbulent magnetic fields in NGC\,6946's SF regions;][]{Tabatabaei2013b}. With such high magnetic field strengths, \CRe\ could lose all their energy before diffusing into the ISM (rendering the region a local calorimeter). In IC\,10, for example, \citet{Heesen2015} find that at 6.3\,GHz the \CRe\ lifetime in the non--thermal super bubble is only 0.9\,Myr, comparable with the age of 1\,Myr derived from the observed curvature of the spectrum. For less intense SF regions, such as DDO\,168, DDO\,47, and DDO\,53 we find peak local magnetic field strengths of 10--15${\rm \,\mu G}$ where \CRe\ may have sufficient time to escape into the ISM. Once there:
\begin{itemize}
\item 1) the \CRe, now losing energy through synchrotron radiation at a much slower rate, diffuse or are advected into the intergalactic medium (IGM) before they have the time to radiate all their energy---this is the `non-calorimetric' situation that leads to the RC--FIR `conspiracy' \cite[e.g.,][]{Bell2003,Dale2009,Lacki2010}, or
\item 2) the \CRe\ continue to diffuse to $1$\,kpc but, because they continue to radiate and lose energy, the frequency of synchrotron emission shifts gradually to lower frequencies to the extent that emission falls outside of the 6\,cm window.
\end{itemize}

Exploring these possibilities and their impact on the \RCNT--SFR relation requires further information regarding the spectral index of the RC emission so that we can explore the \CRe\ transport and ageing timescales. 

\section{SUMMARY}
\label{section:Paper1Summary}

We used the VLA in C--configuration to make 6\,cm ($\nu = 6.2$\,GHz) observations of 40 dwarf galaxies taken from LITTLE THINGS \cite[][]{Hunter2012}. Our images have a resolution of $3$--$8^{\prime \prime}$ and an rms noise of $3$--$15 {\rm \mu Jy\,beam^{-1}}$. We summarise our findings as follows:
\begin{itemize}
\item Contamination from background sources is a prominent issue in earlier, low resolution observations. Even at the high resolution of the current survey, it is not trivial to reliably remove the contribution from all unrelated background sources;
\item  After removing background and ambiguous sources, a total of $22$ out of the $40$ LITTLE THINGS galaxies are associated with significant RC emission; $8$ are new detections. Where reliable flux densities of our sample  exist in the literature, we find that our observations are in general good agreement;
\item We find that the average thermal fraction in our sample is $50 \pm 10$\%  and $70 \pm 10$\% at $6.2$\,GHz when integrating over the RC and disk based mask, respectively;
\item We present fits for the RC--SFR and FIR--SFR between SFRs of $\sim 10^{-4}$ and 1\,M$_\odot$\,yr$^{-1}$ integrated over the RC mask and disk mask; 
\item We find that the RC--SFR is broadly consistent with theoretical predictions when considering the RC mask but we find that the \RCNT\ is suppressed when integrating over the disk;
\item The FIR emission in our sample is suppressed in both the RC and disk masks given the measured SFR;
\item When integrating the galaxy properties within the optical disk we find that the dwarf galaxies are consistent with the linear \cite{Yun2001} RC--FIR relation. The `conspiracy' invoked to explain the relation continues to hold for our sample of dwarf galaxies (see Fig.~\ref{Figure:FIR_v_RC}). The RC--FIR relation based on our RC mask integrated results shows that our dwarf galaxies diverge from the the linear \cite{Yun2001} relation with a RC excess at low luminosity;
\item In a few dwarf galaxies, the equipartition magnetic field strength reaches as high as $30{\rm \,\mu G}$ in several $100$\,pc regions, and in one case, $50{\rm \,\mu G}$;
\item The average strength of the transverse magnetic field, based on equipartition, is $\sim 5.2{\rm \mu G}$ (RC mask) and $\sim 7.5{\rm \mu G}$ (disk averaged). This value is slightly lower than larger galaxies \cite[e.g., $9.7{\rm \mu G}$ in WSRT SINGS;][]{Heesen2014} but consistent with other studies of dwarf galaxies. 
\end{itemize}

\acknowledgments
We thank Dan Smith for valuable discussions on the RC--FIR relation, and his help with FIR--related work. We also appreciate Min Yun's readiness to provide us with his {\em IRAS} and VLA data from the \cite{Yun2001} study. The manuscript has benefitted from constructive comments on an earlier version by the referee. GK acknowledges support from the UK Science and Technology Facilities Council [grant number ST/J501050/1]. Likewise EB and LH acknowledge support from the UK Science and Technology Facilities Council [grant number ST/M001008/1]. This research has made use of the NASA/IPAC Extragalactic Database (NED) which is operated by the Jet Propulsion Laboratory, California Institute of Technology, under contract with the National Aeronautics and Space Administration.

Facilities: \facility{VLA}; \facility{\em GALEX}; \facility{\em Spitzer}.

\appendix
\section{Appendix}
\label{section:Appendix_IndividualNotes}

The appendix details any notes on our individual RC observations. Here, we focus on prominent features, and also on notable deviations from our normal line of calibration and image generation. Where no frequency is mentioned,  flux densities were determined from the maps presented here.  All other flux density values   were taken from the NASA/IPAC Extragalactic Database (NED).

\paragraph{ DDO\,43: }
A bright source (87GB[BWE91] 0724+4053: flux density $37\,{\rm mJy}$, located $2.5^\prime$ from the image phase centre) gave rise to prominent sidelobes across the FOV. DDO\,43 was directly affected by, in particular, an E--W artefact.

\paragraph{ DDO\,50: }
A bright source (NVSS J081920+704907: flux density  $18\,{\rm mJy}$ located $5.5^\prime$ from the image phase centre) exhibited weak sidelobes across the FOV. Selected parts of DDO\,50 were directly affected by low--level artefacts. A single round of self--calibration was performed which was successful in reducing the prominent sidelobes originating from NVSS\,J081920+704907. 

\paragraph{ DDO\,52: }
Two bright sources (NVSS J082842+415056: flux density of $19\,{\rm mJy}$ located $2.5^\prime$ from the image phase centre, and NVSS\,J082814+415353: flux density of $39\,{\rm mJy}$ located $4^\prime$ from the image phase centre) generated weak sidelobes across the FOV. DDO\,52 was not substantially affected by the artefacts. Self--calibration was not deemed necessary. 

\paragraph{ DDO\,75: }
A bright source (NVSS J101030--044006) with a 1.4\,GHz flux density of 305\,mJy located $7^\prime$ from the image phase centre gave rise to weak sidelobes across the FOV. Parts of DDO\,75 were directly affected by low--level artefacts. A single round of self--calibration was performed which was successful in reducing the prominent sidelobes originating from NVSS\,J101030--044006.

\paragraph{ DDO\,101: }
A bright source (NVSS J115618+312805), with a 4.85\,GHz flux density of 1.03\,Jy located $9^\prime$ from the image phase centre caused prominent sidelobes across the FOV. DDO\,101 was directly affected by the artefacts. 
Self--calibration was not successful in reducing the effects of the sidelobes, which was attributed to the fact that NVSS\,J115618+312805 enters the first sidelobe of the primary beam which results in it being seemingly time--variable. We decided to use just 3 spectral windows for which NVSS\,J115618+312805 fell near the first null of the  primary beam. This was successful in reducing its prominent sidelobes.

\paragraph{ DDO\,126: } 
A bright double--source (NVSS J122658+370719: flux density of $4.6{\rm m Jy}$ located $1.5^\prime$ from the image phase centre) exhibited prominent sidelobes crossing the FOV. DDO\,126 was directly affected by the artefacts. A single round of self--calibration was performed which was successful in reducing the prominent sidelobes originating from NVSS\,J122658+370719. 

\paragraph{ DDO\,154: }
Two bright sources (NVSS J125401+270357: flux density of $18{\rm m Jy}$ located $5.5^\prime$ from the image phase centre and an uncatalogued source of unknown flux density located $5.5^\prime$ from the image phase centre) led to weak sidelobes which crossed through part of the FOV. DDO\,154 was not directly affected. 
A single round of self--calibration was performed which was successful in reducing the prominent sidelobes from both sources. Maps created with robust weighting ({\sc robust=0.0}) did not reveal any significant regions of emission and instead another {\sc clean} was run using a robust weighting that was closer to natural weighting ({\sc robust=0.5}).

\paragraph{ DDO\,187: }
A bright double--source (NVSS J141556+230730:  flux density of $55{\rm m Jy}$ located $4.5^\prime$ from the image phase centre) caused prominent sidelobes across the FOV. DDO\,187 was directly affected by the artefacts. A single round of self--calibration was performed which was successful in reducing the prominent sidelobes originating from NVSS\,J141556+230730.

\paragraph{ M81\,dwA: }
A bright source (NVSS J082451+705808: 4.85\,GHz flux density of 63\,mJy, located $5.5^\prime$ from the image phase centre) gave rise to prominent sidelobes across the FOV. M81\,dwA was directly affected by the artefacts. Self--calibration was not successful in reducing the effects of the sidelobes, which was attributed to the fact that NVSS\,J082451+705808 enters the first sidelobe of the primary beam which results in it being seemingly time--variable. We decided to use just 3 spectral windows for which NVSS\,J082451+705808 fell near the primary beam null. This was successful in reducing the prominent sidelobes originating from NVSS\,J082451+705808. 

\paragraph{ Mrk\,178: }
The {\em GALEX} FUV image was dropped from the analysis due to being of poor quality.

\paragraph{ NGC\,1569: }
The {\em Spitzer} 24\micron\ and 70\micron\ images were dropped from the analysis due to being of poor quality.

\paragraph{ NGC\,3738: }
A bright triple--source (NVSS J113545+543319: combined flux density of $63{\rm m Jy}$ located $2^\prime$ from the image phase centre) exhibited prominent sidelobes across the FOV. NGC\,3738 was directly affected by the artefacts. A single round of self--calibration was performed which was successful in reducing the prominent sidelobes originating from NVSS\,J113545+543319.

\paragraph{ NGC\,4214:} 
NGC\,4214 (especially the \hii\, region centred on 12${\rm ^h}$15${\rm ^m}$41${\rm ^s}$.2, +36${\rm ^\circ}$19${\rm ^\prime}$04${\rm ^{\prime\prime}}$.6) was bright enough that prominent sidelobes were produced throughout the FOV. A single round of self--calibration was performed which was successful in reducing the prominent sidelobes originating from NGC\,4214.

\paragraph{ UGC\,8508:} 
Two sources (not coincident with \halpha\, emission) from the $4^\prime$ square aperture were judged as {\em not} originating from UGC\,8508 and were accordingly masked out. 

\paragraph{ VII\,Zw\,403:} 
The {\em Spitzer} 24\micron\ and 70\micron\ images were dropped from the analysis due to being of poor quality.

\paragraph{ WLM:} 
A bright source (NVSS\,J000141--154040: 4.85\,GHz flux density of 145\,mJy, located $13^\prime$ from the image phase centre) caused prominent sidelobes across the FOV. WLM was directly affected by the artefacts. A single round of self--calibration was performed which was successful in reducing the prominent sidelobes originating from NVSS\,J000141--154040.

\section{$4$--$8$\,GHz Radio Continuum Images of the LITTLE THINGS sample}
\label{Section:Appendix_Images}

In this section, we provide images of the LITTLE THINGS sample. We show RC flux density contours superposed on ancillary LITTLE THINGS images \cite[][]{Hunter2012}. As the dwarf galaxies are faint, extended RC emission does not show well when plotting contours at the native resolution. Therefore, for the lowest surface brightness contour, we smooth the RC image with a Gaussian kernel of $10''$, and use a contour level corresponding to a S/N ratio of 3. The remaining contours, at full spatial resolution, are drawn at S/N levels of $3,6,9,18,36,72,144$. We present the results of our RC based masking technique and disk mask which includes ambiguous and background sources. Where the ancillary data were available, we also show the following images: \halpha; {\em GALEX} FUV; {\em Spitzer} 24 and 70\micron\ images; FUV-inferred $\Sigma_{\rm SFR}$ from \cite{Leroy2012}.

\begin{figure}
  \begin{tabular}{ccc}
    \includegraphics[width=0.31\linewidth,clip]{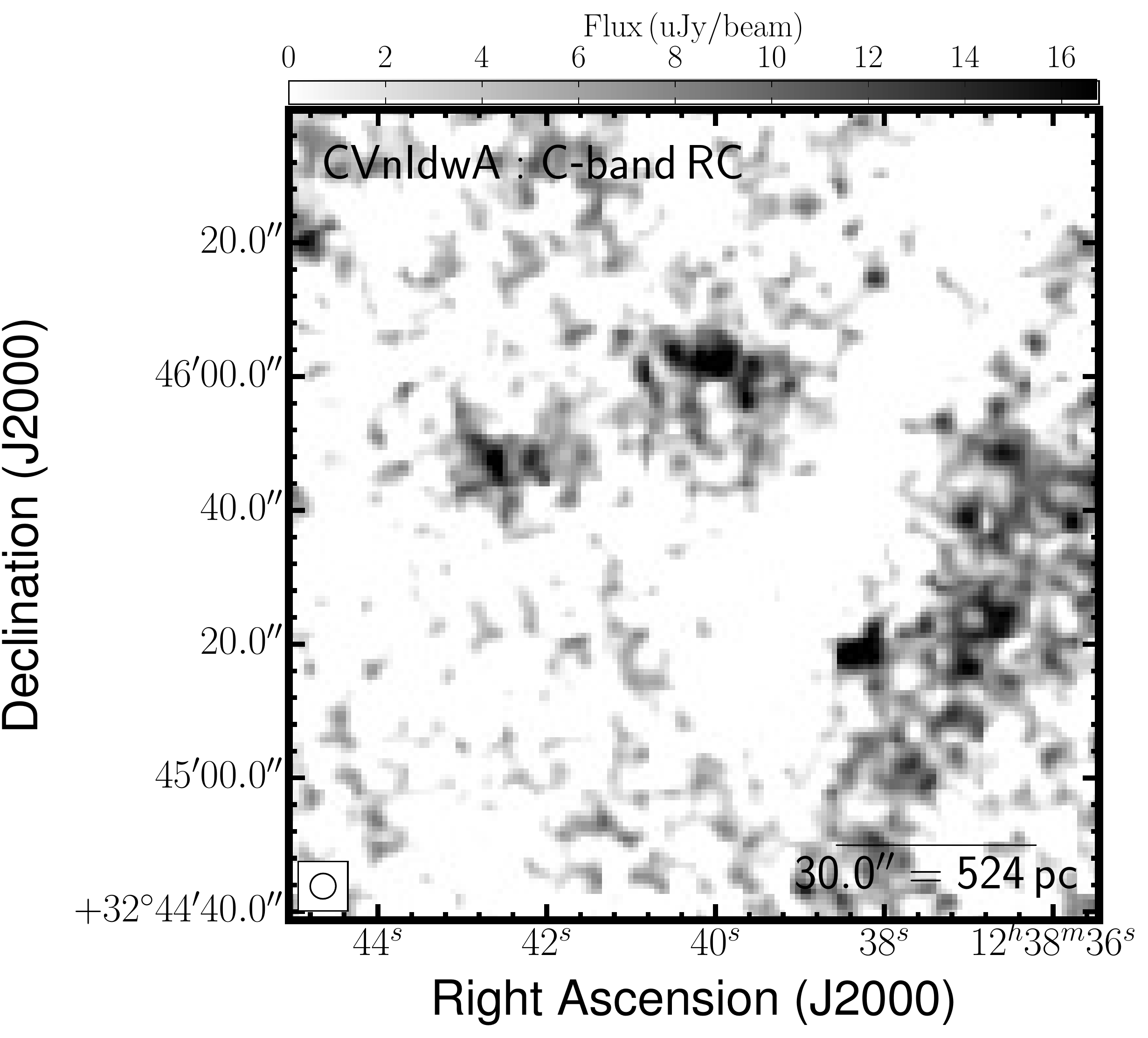} & \ 
    \includegraphics[width=0.31\linewidth,clip]{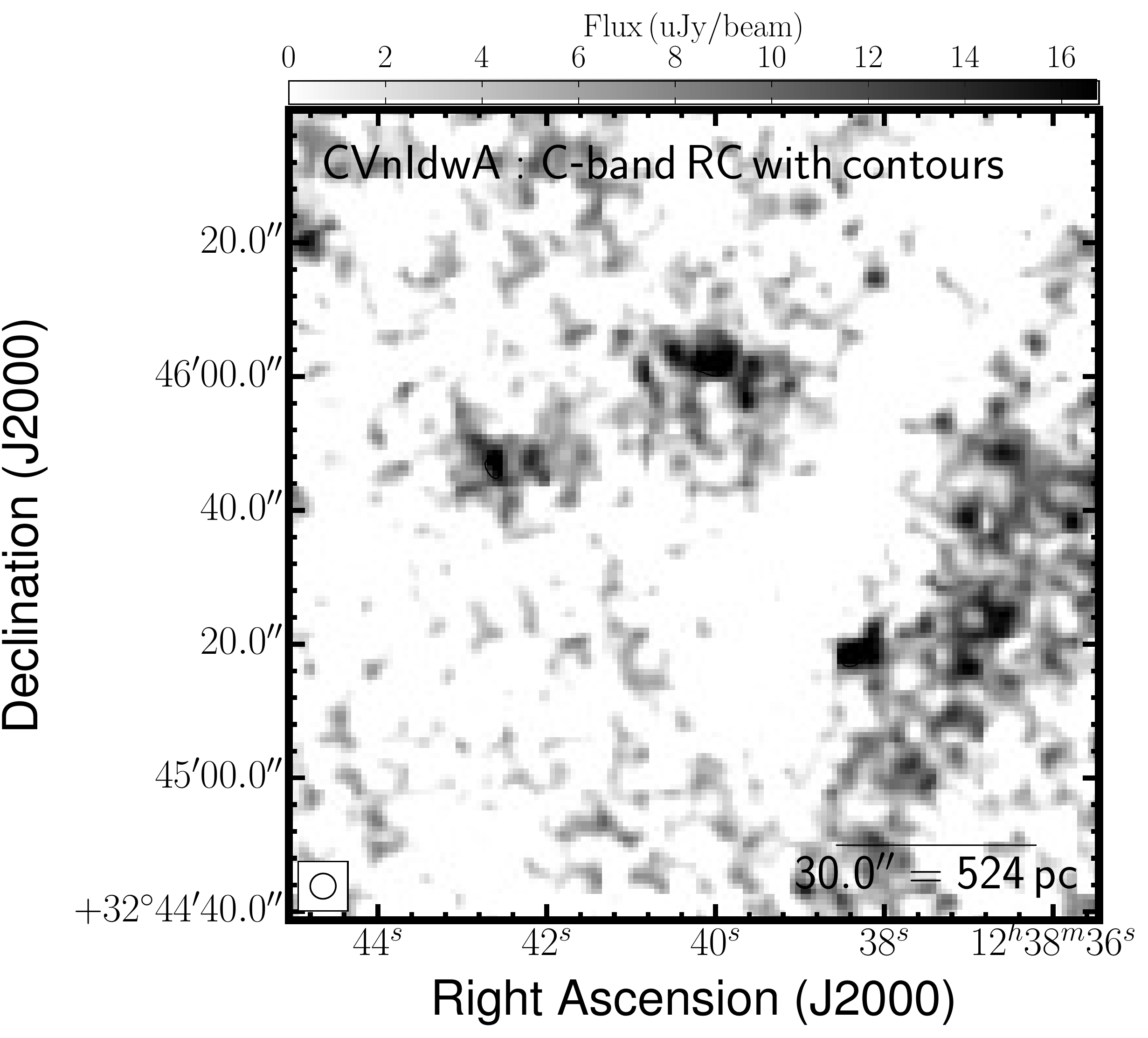} & \ 
    \includegraphics[width=0.31\linewidth,clip]{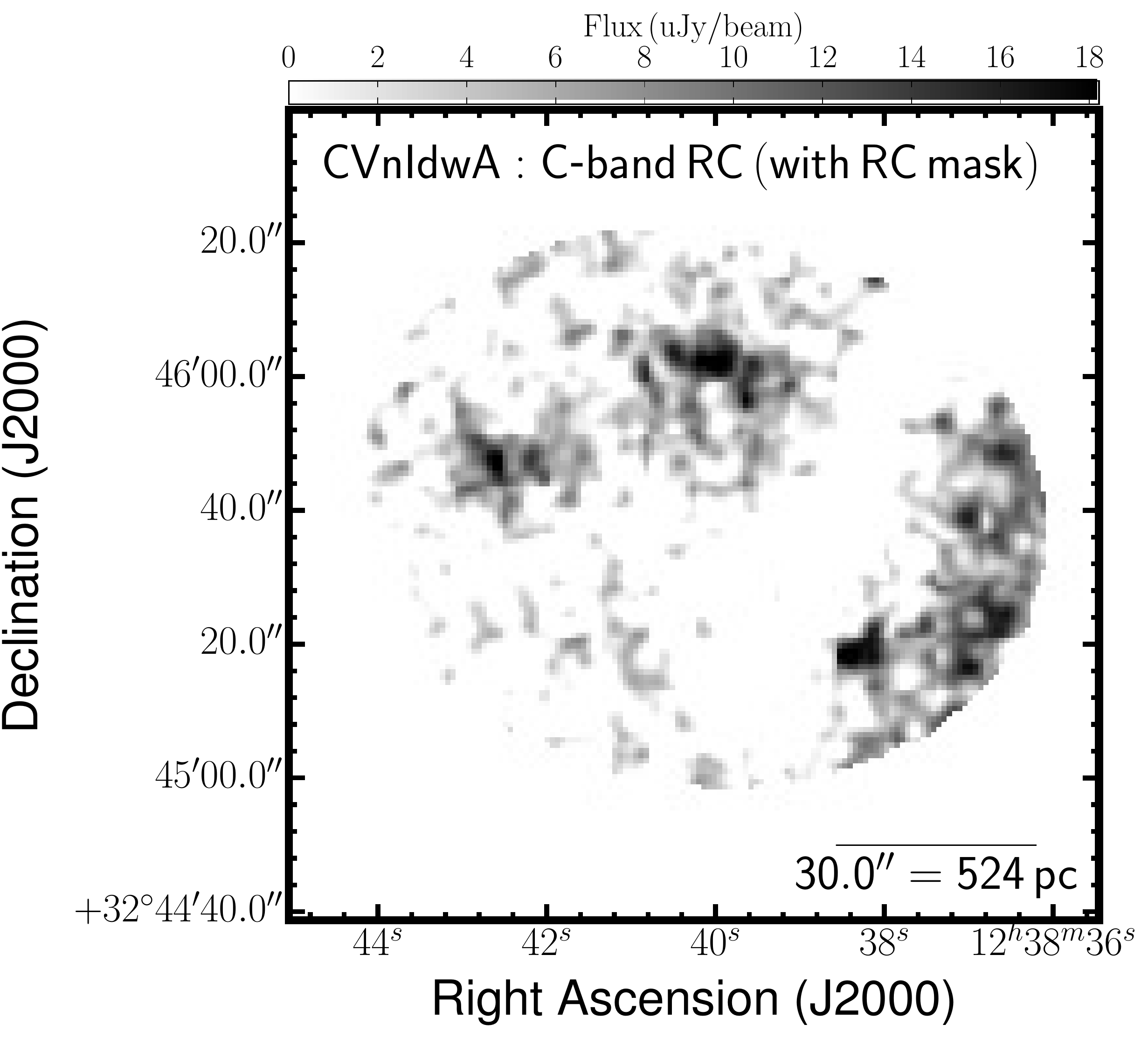} \\
    \includegraphics[width=0.31\linewidth,clip]{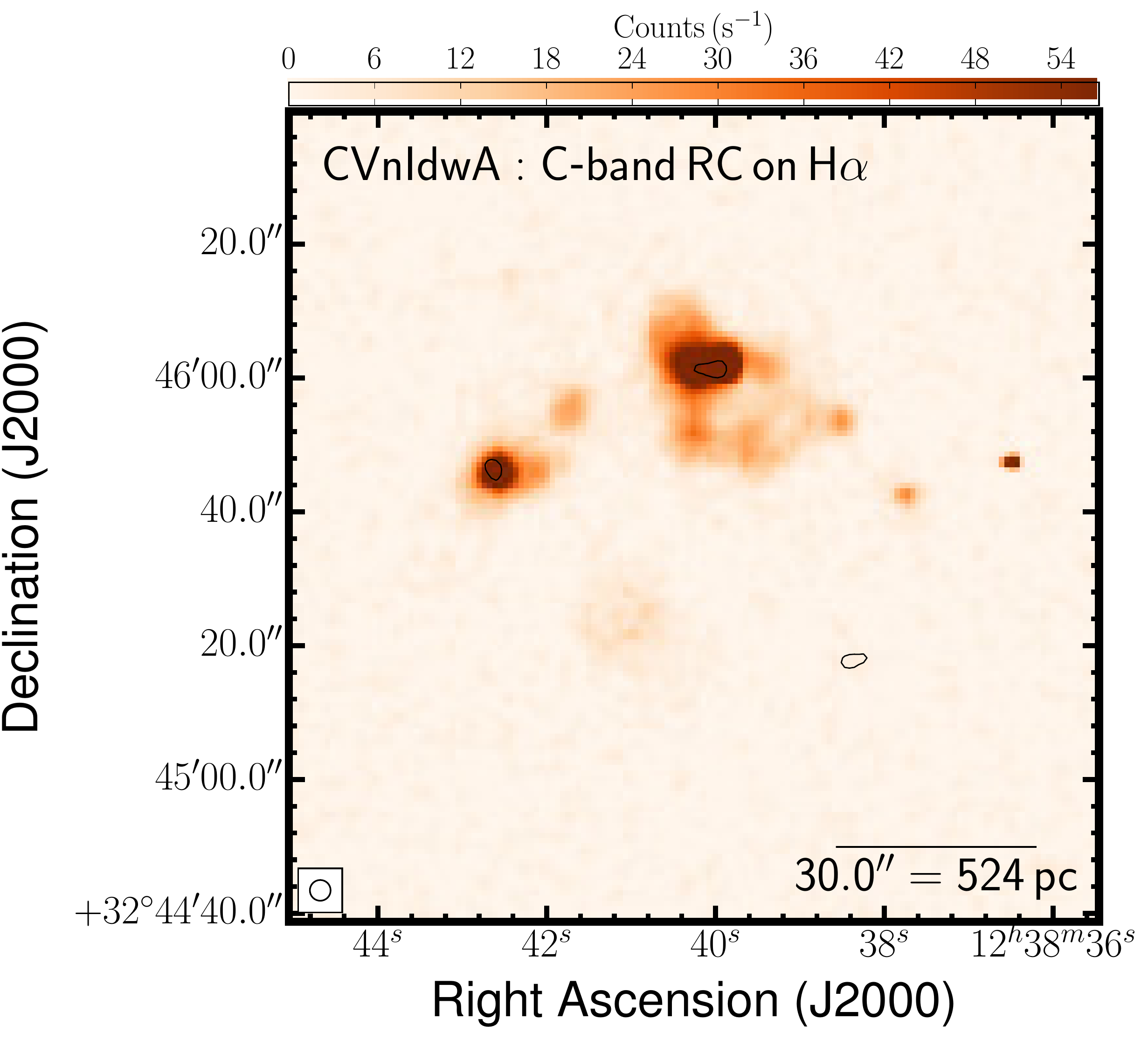} & \ 
    \includegraphics[width=0.31\linewidth,clip]{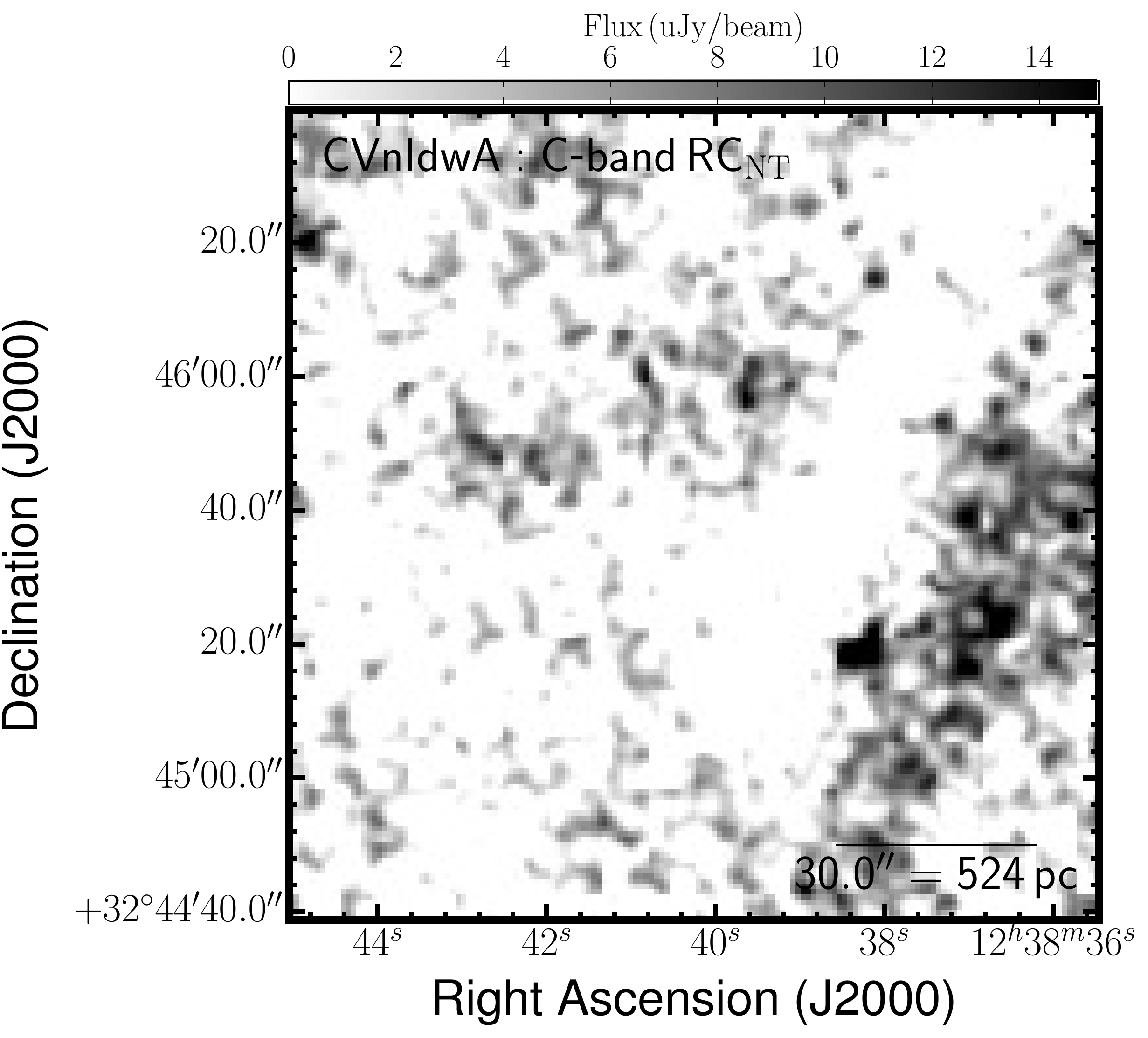} & \ 
    \includegraphics[width=0.31\linewidth,clip]{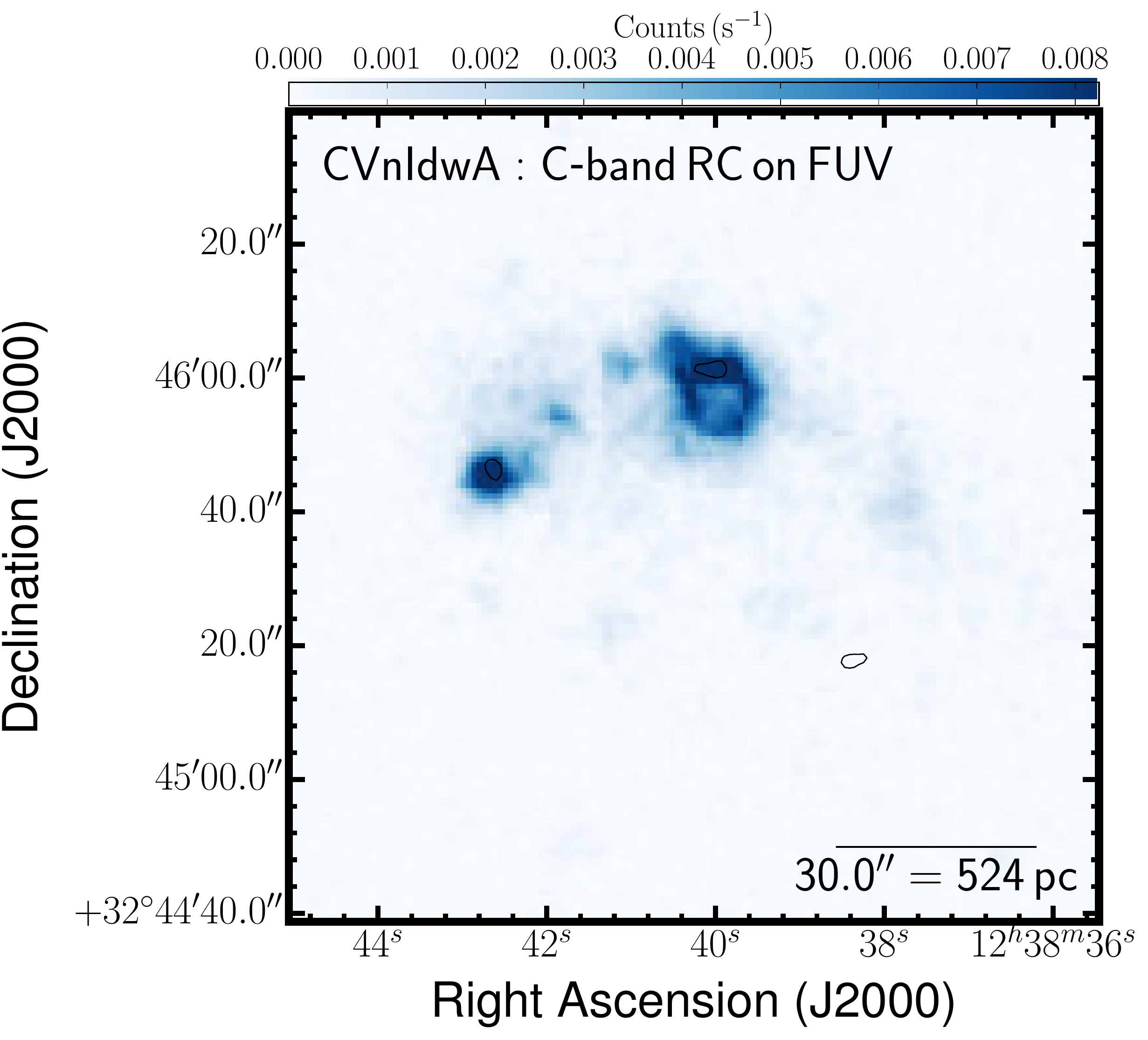} \\
    \includegraphics[width=0.31\linewidth,clip]{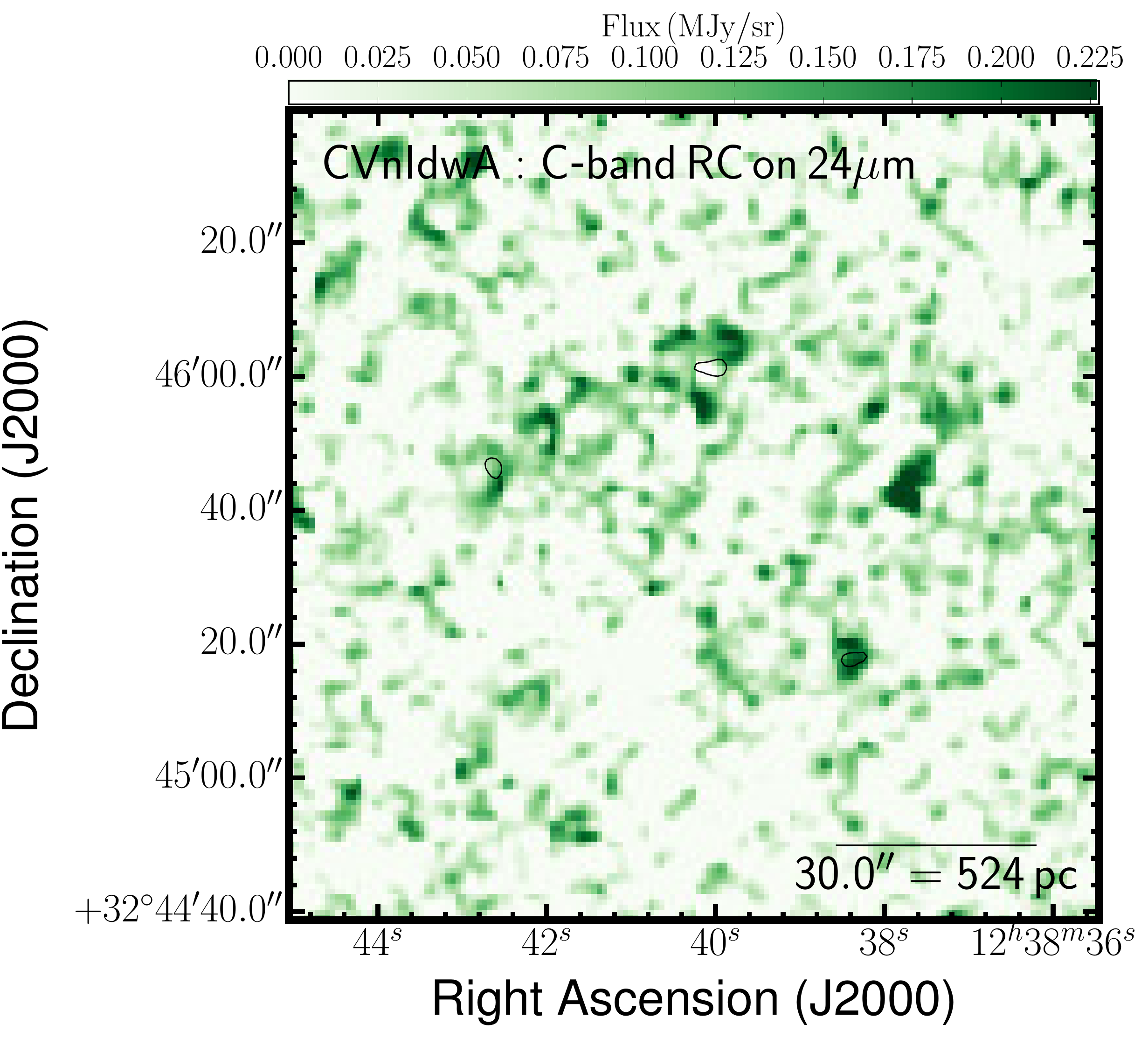} & \ 
    \includegraphics[width=0.31\linewidth,clip]{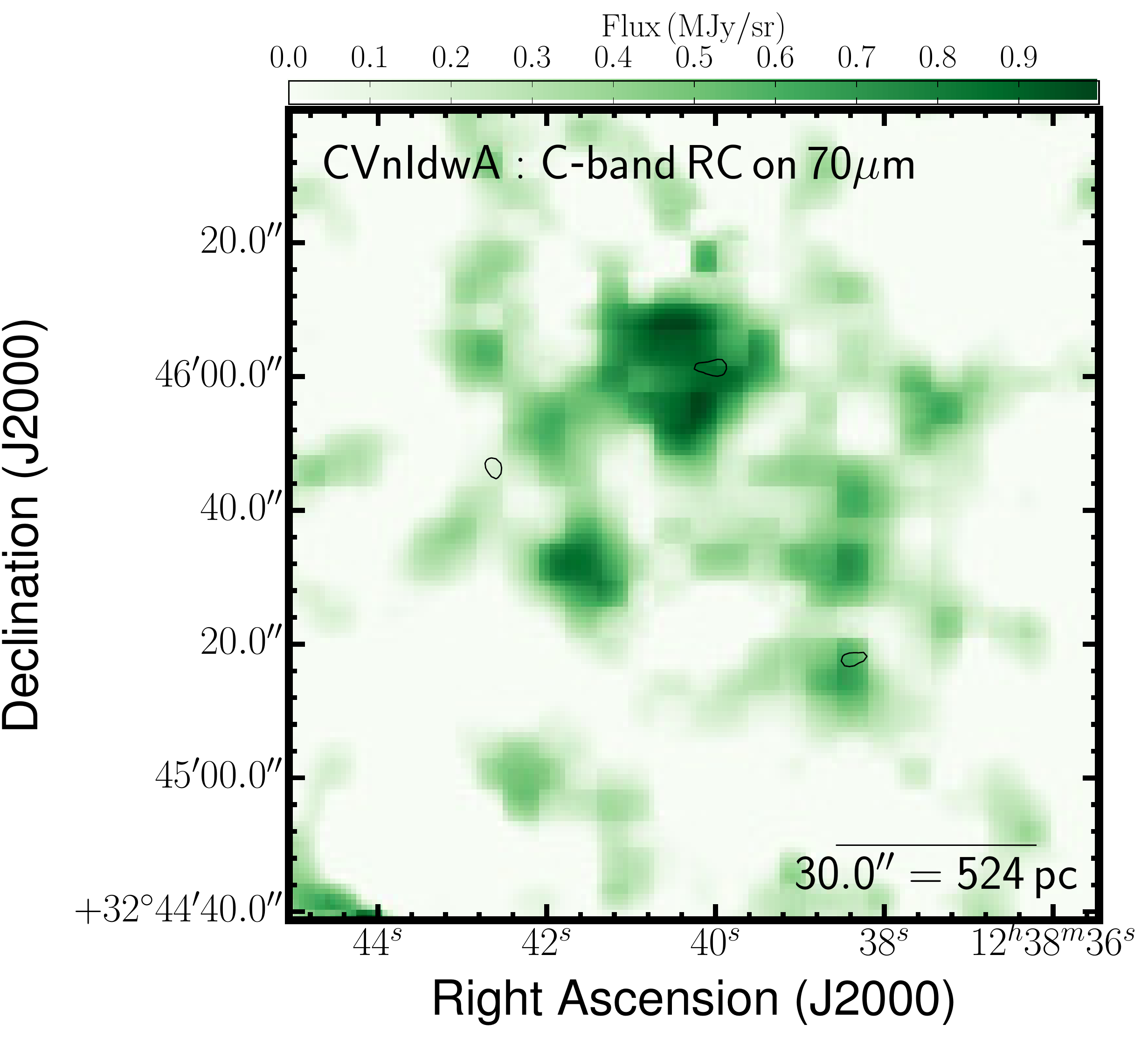} & \ 
    \includegraphics[width=0.31\linewidth,clip]{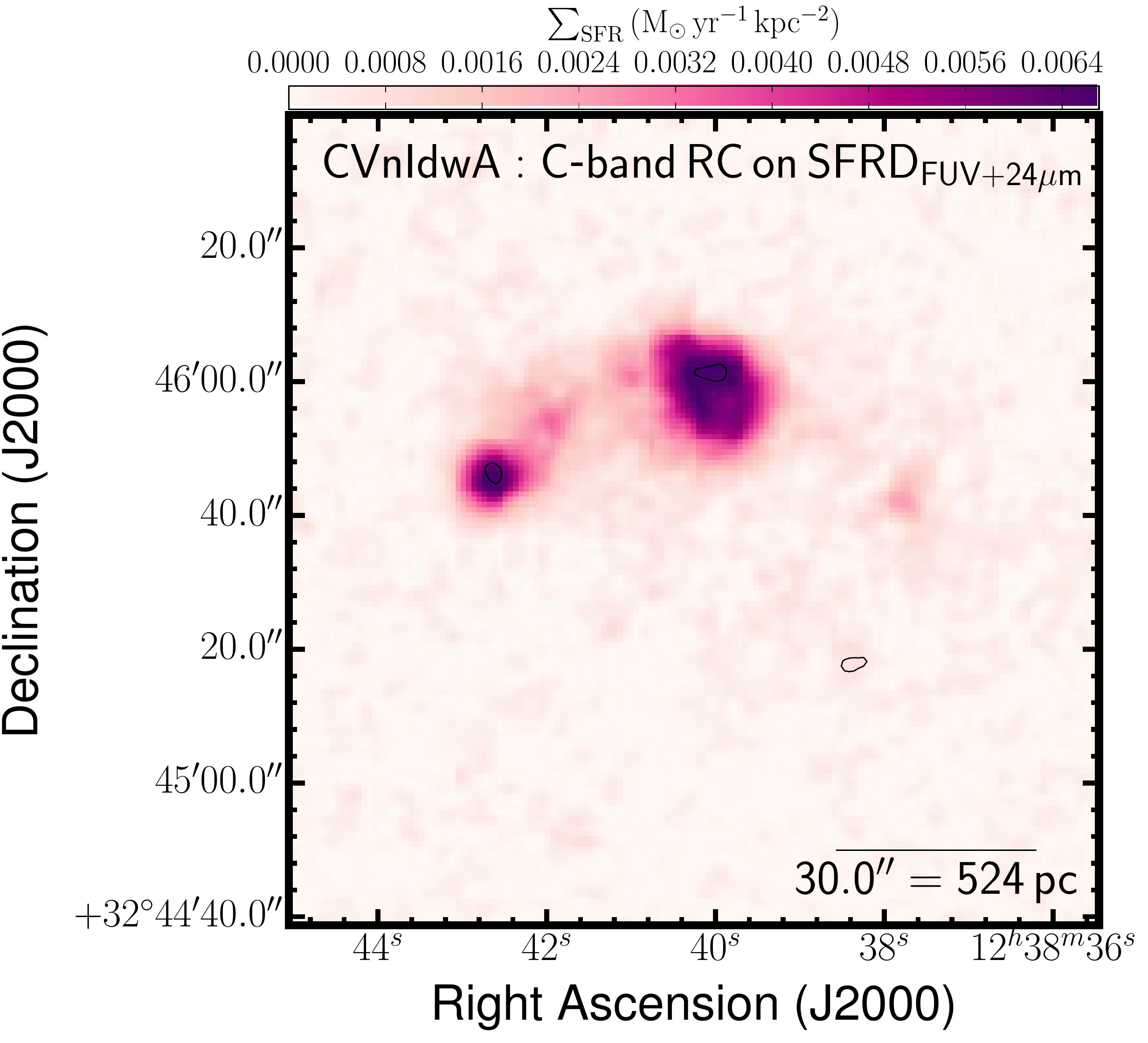} \\
  \end{tabular}
\caption[CVn\,I\,DwA images: RC, IR, optical, and FUV]{Multi-wavelength coverage of CVn I dwA displaying a $2.0^\prime \times 2.0^\prime$ area. We show total RC flux density at the native resolution (top-left) and again with contours (top-centre). The RC contours are superposed on ancillary LITTLE THINGS images where possible: \halpha\ (middle-left); \RCNT\ obtained by subtracting the expected \RCT\ based on the \halpha-\RCT\ scaling factor of \cite{Deeg1997} from the total RC; {\em GALEX} FUV (middle-right); {\em Spitzer} 24\micron\ (bottom-left); {\em Spitzer} 70\micron\ (bottom-centre); FUV$+24{\rm \mu m}$--inferred SFRD from \citealp{Leroy2012} (bottom-right). We also show the RC that was isolated by the RC--based masking technique (top-right).}
  \label{figure:cvnidwaCc_maps}
\end{figure}

\begin{figure}
  \begin{tabular}{ccc}
    \includegraphics[width=0.31\linewidth,clip]{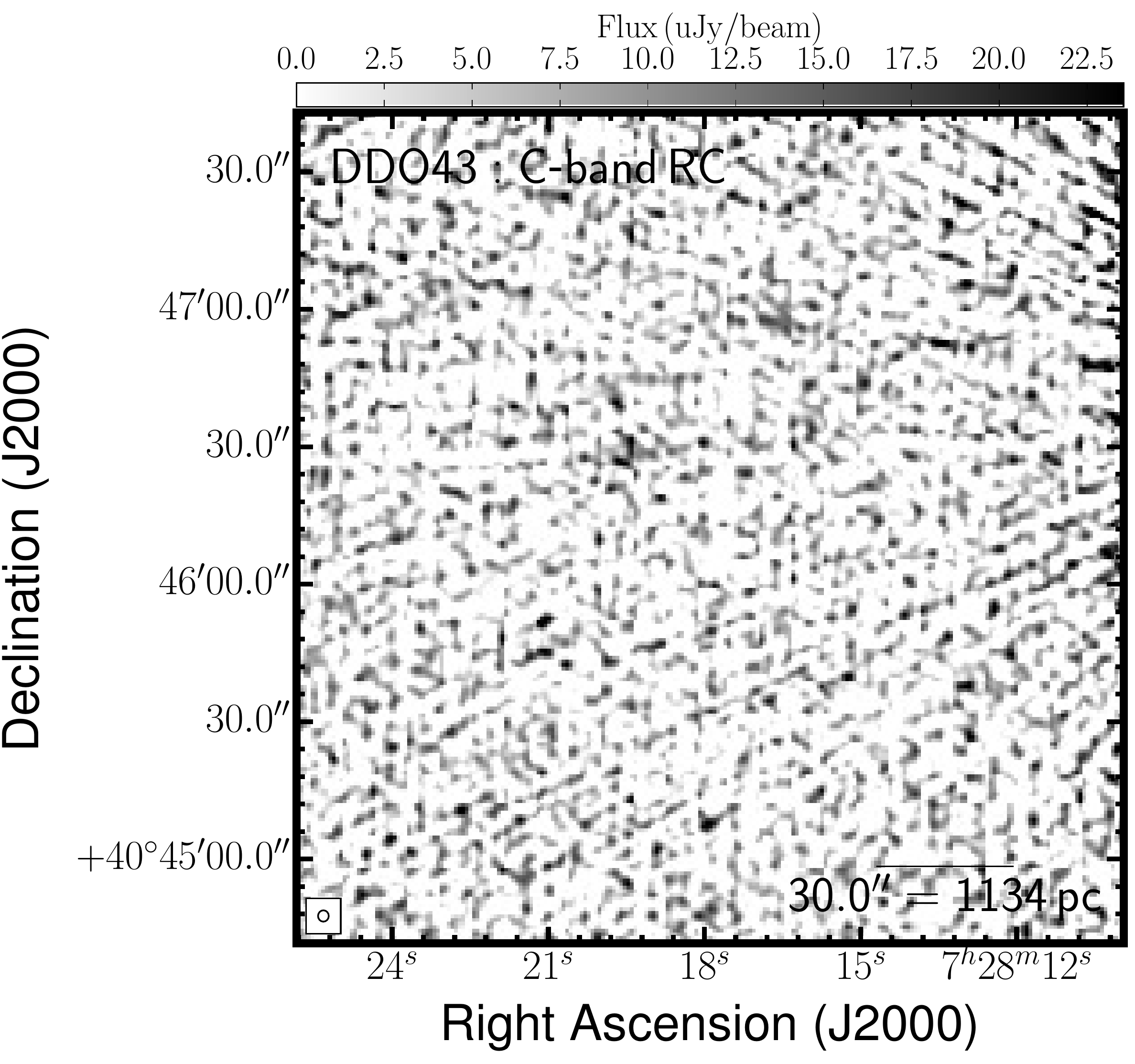} & \ 
    \includegraphics[width=0.31\linewidth,clip]{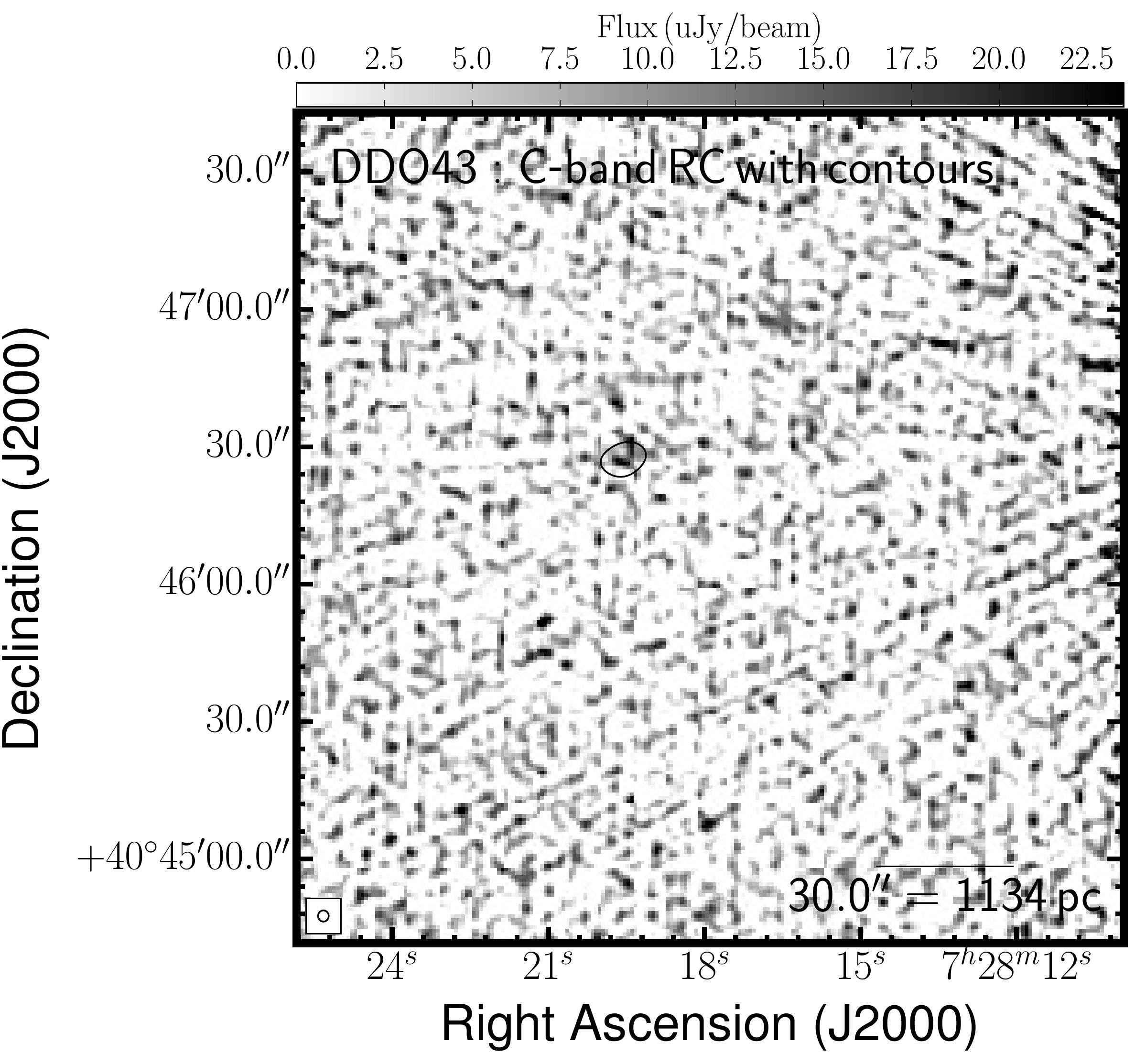} & \ 
    \includegraphics[width=0.31\linewidth,clip]{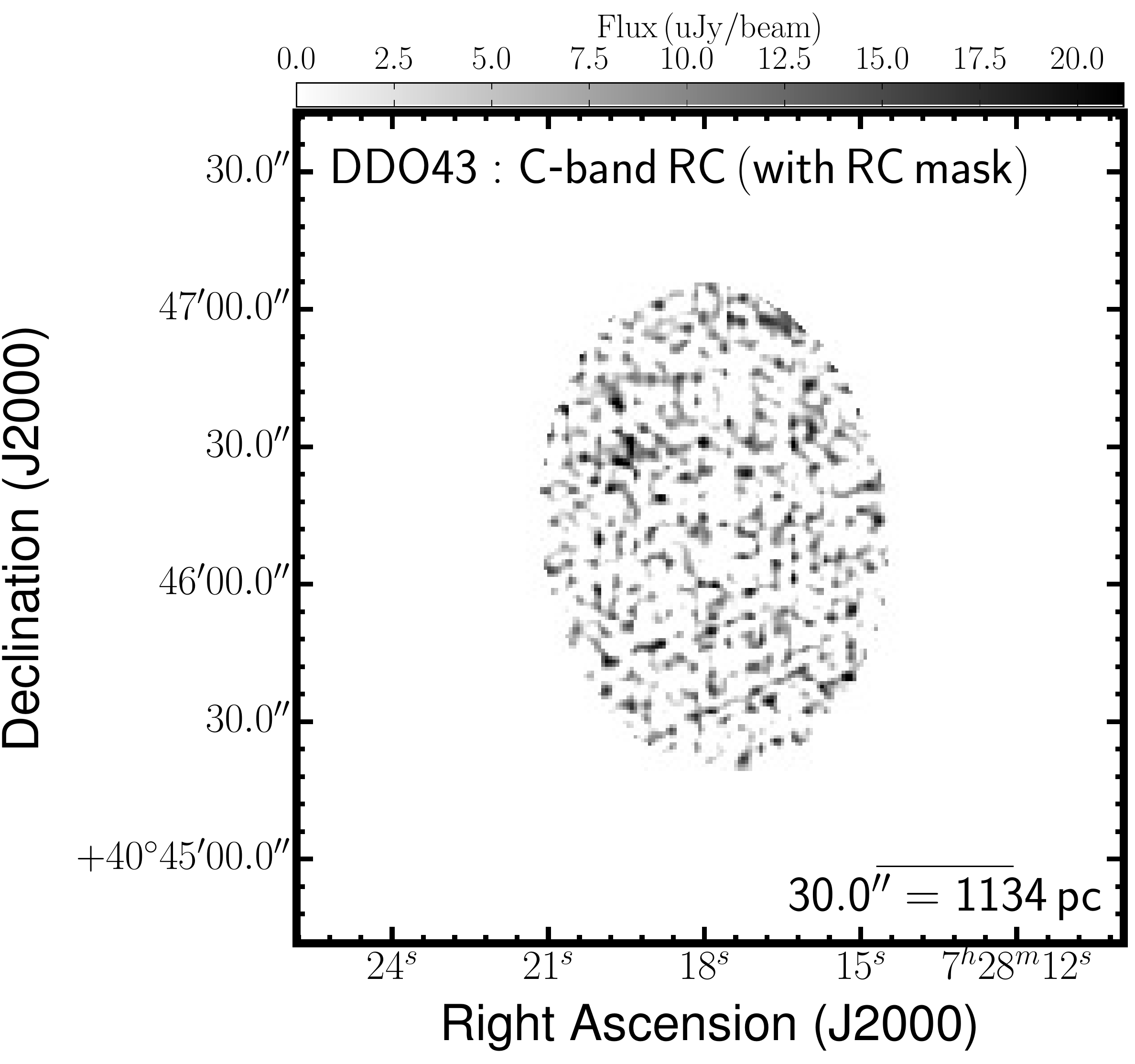} \\
    \includegraphics[width=0.31\linewidth,clip]{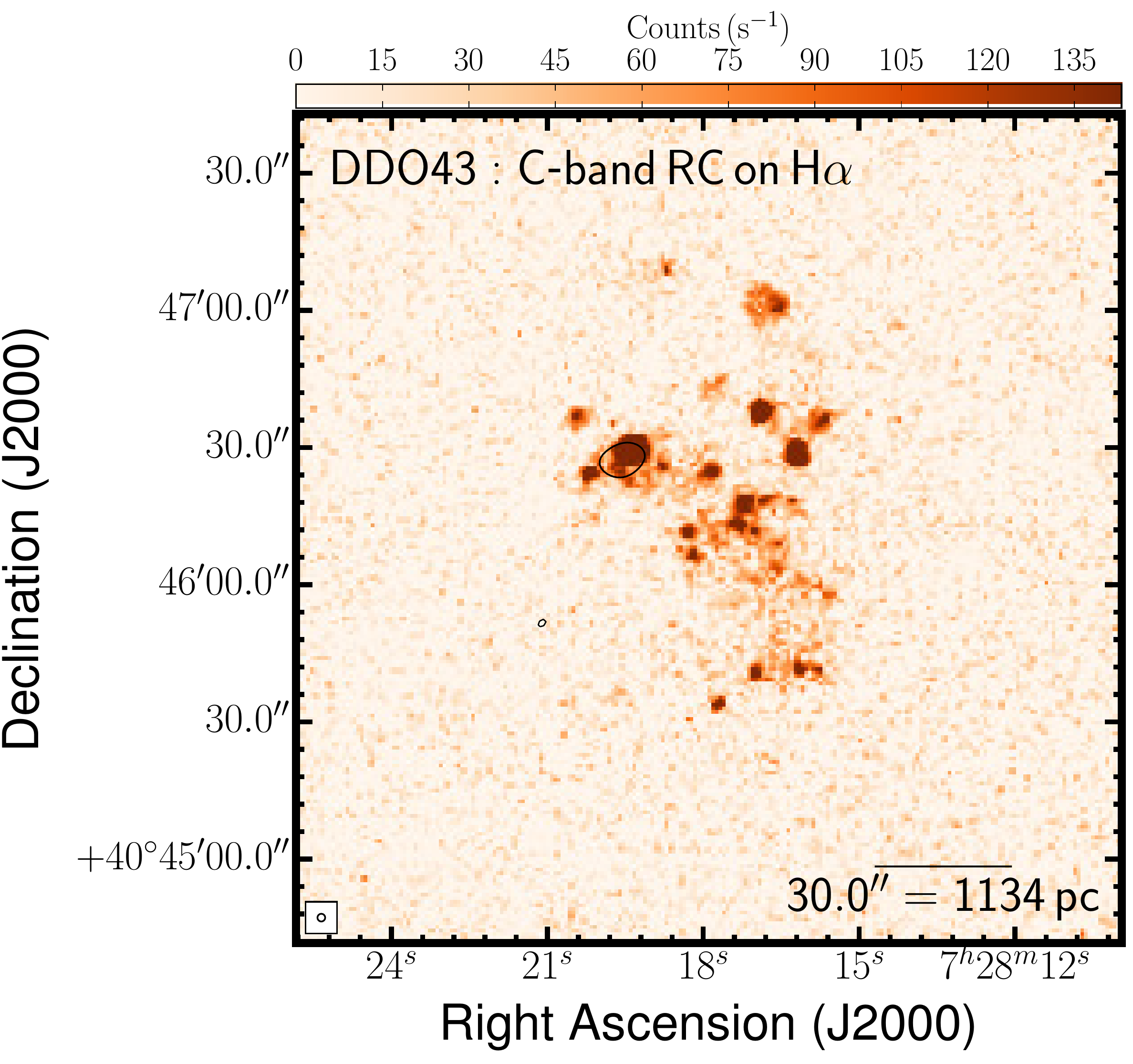} & \ 
    \includegraphics[width=0.31\linewidth,clip]{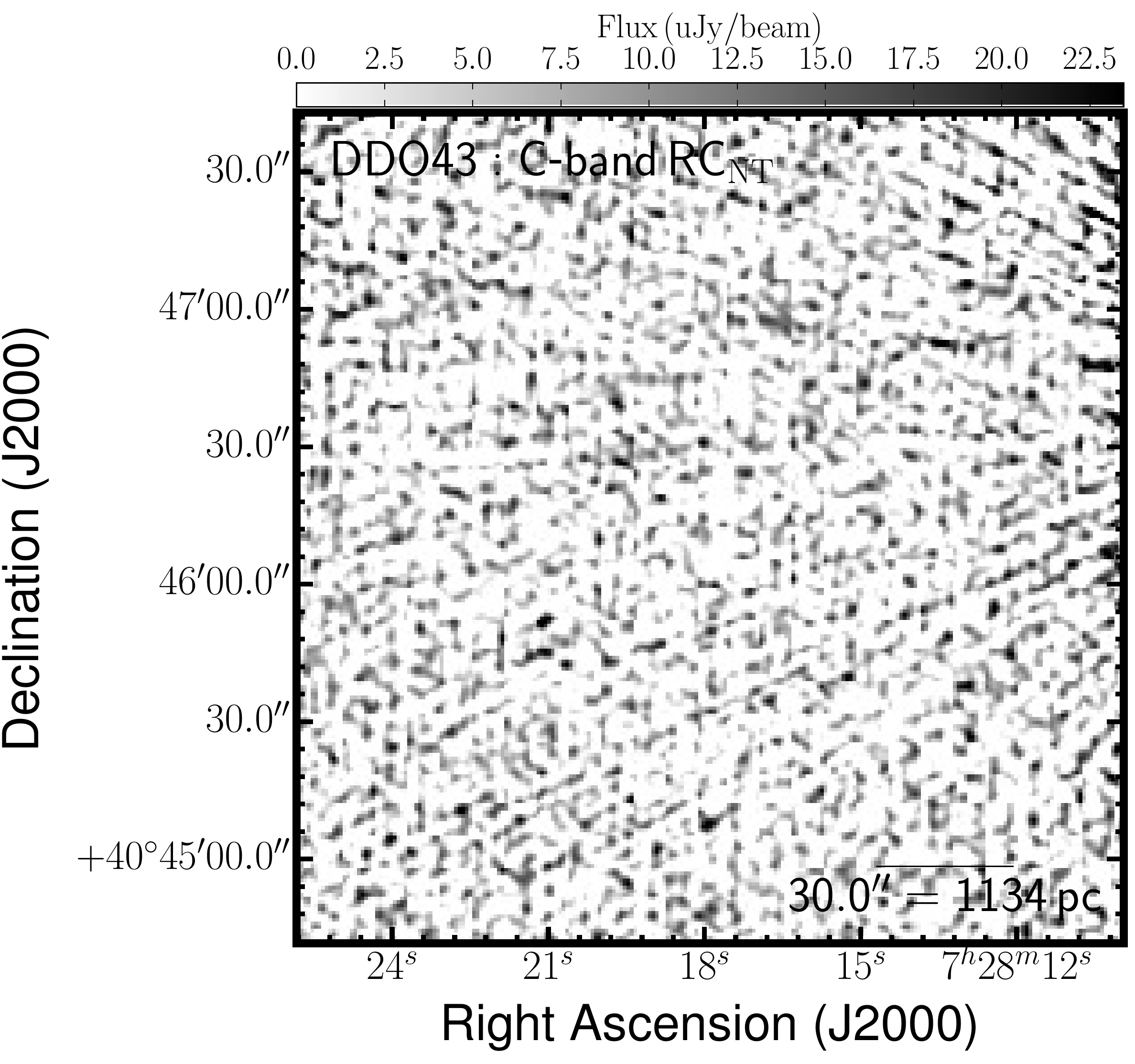} & \ 
    \includegraphics[width=0.31\linewidth,clip]{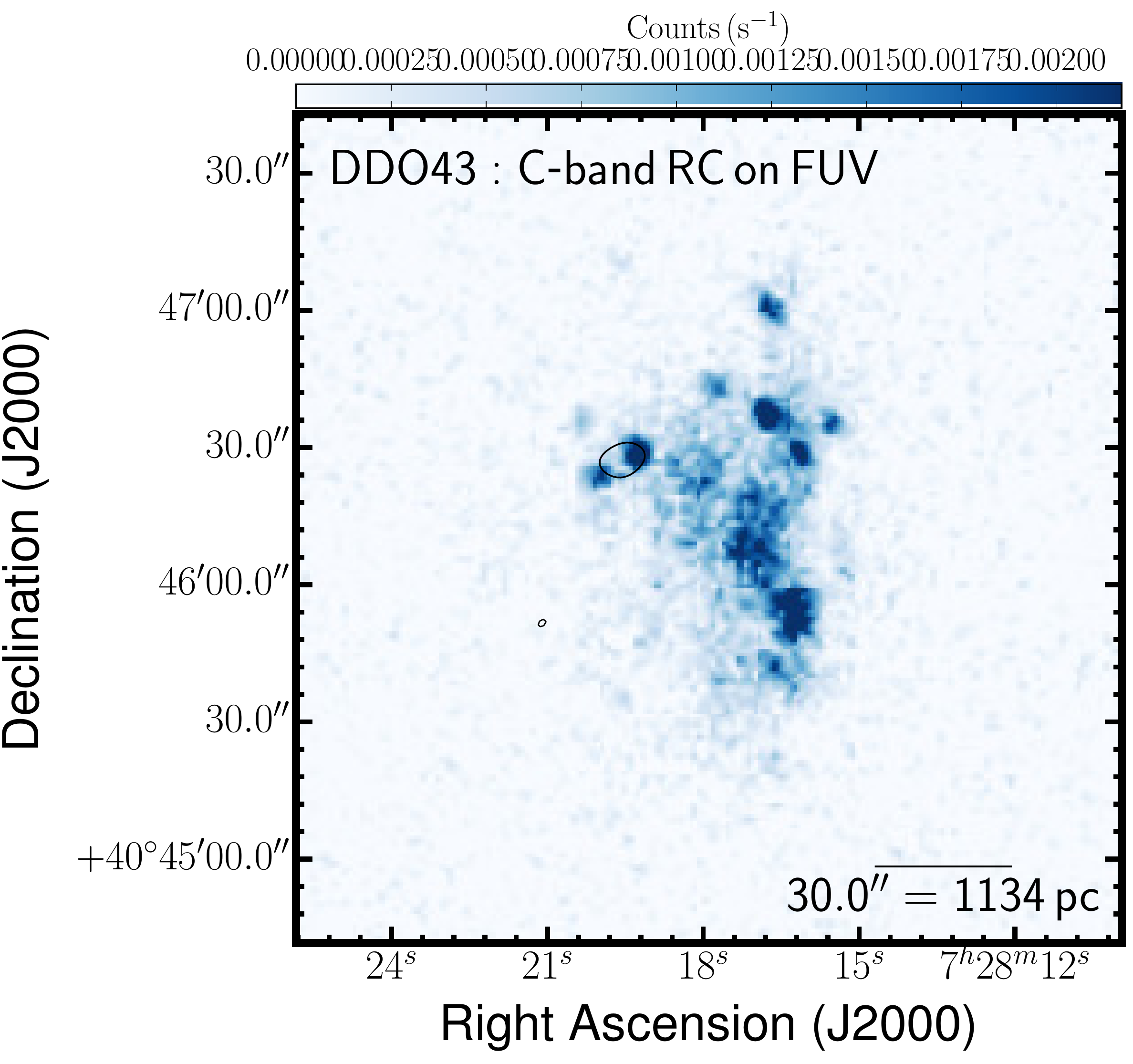} \\
    \includegraphics[width=0.31\linewidth,clip]{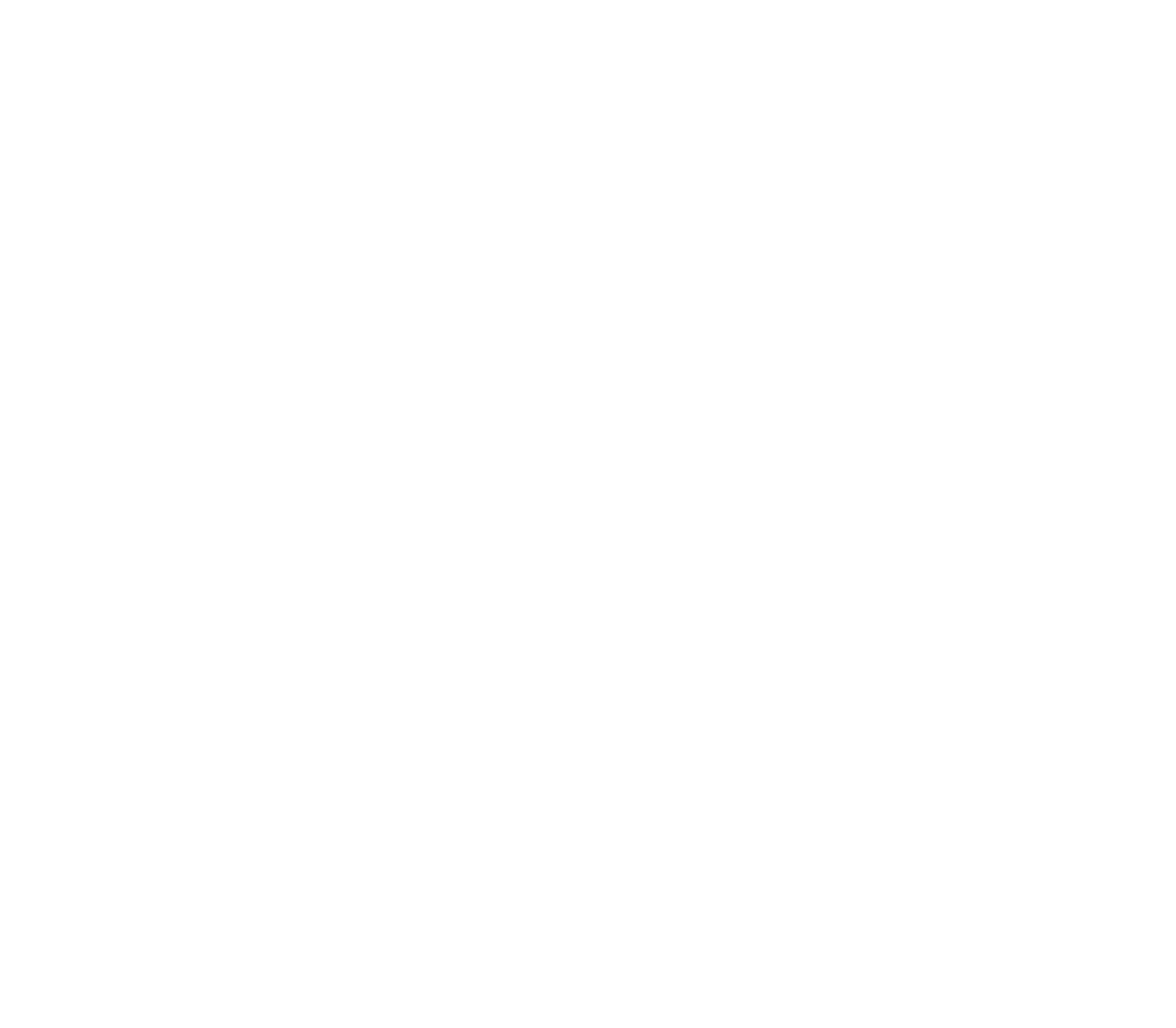} & \ 
    \includegraphics[width=0.31\linewidth,clip]{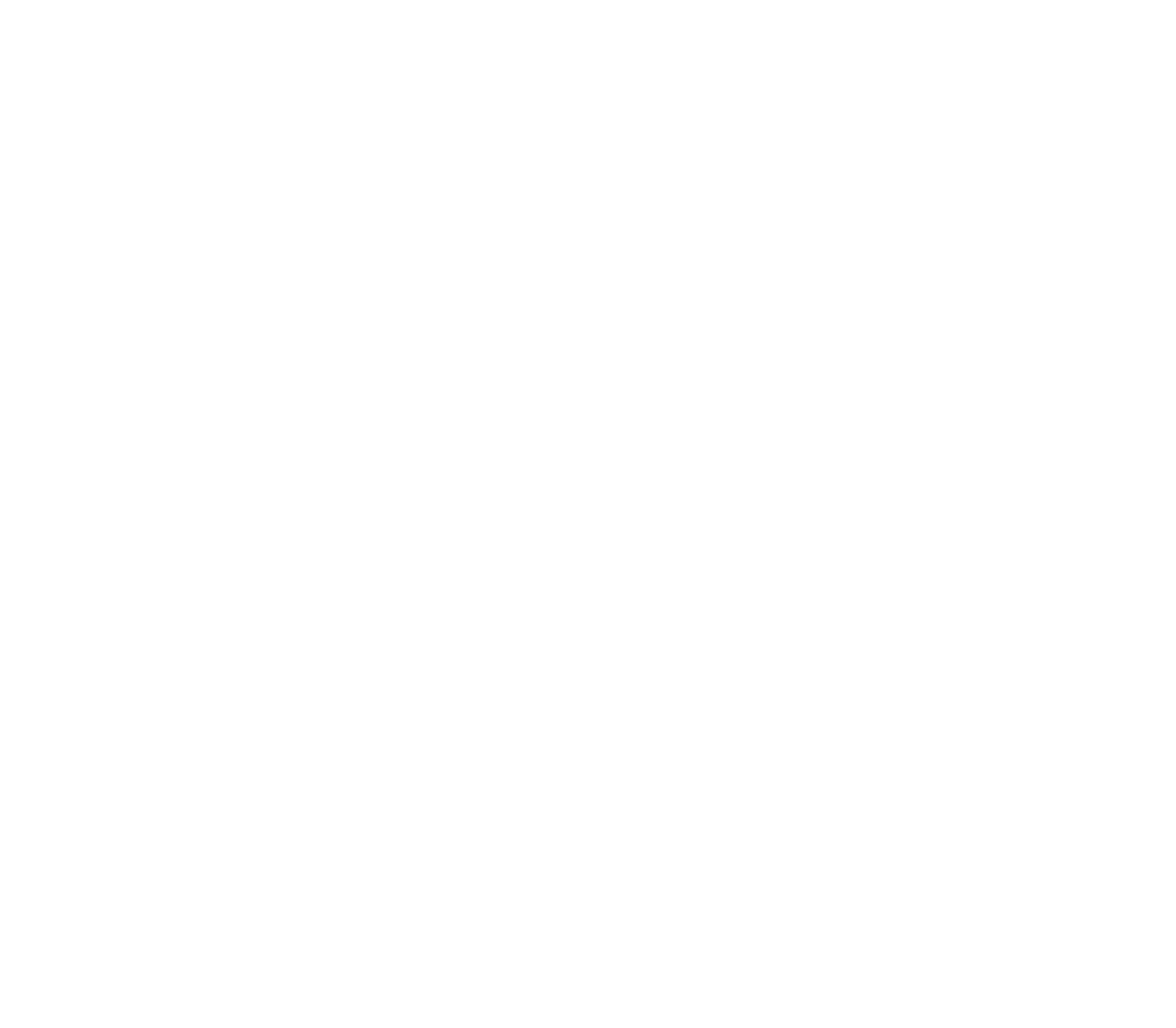} & \ 
    \includegraphics[width=0.31\linewidth,clip]{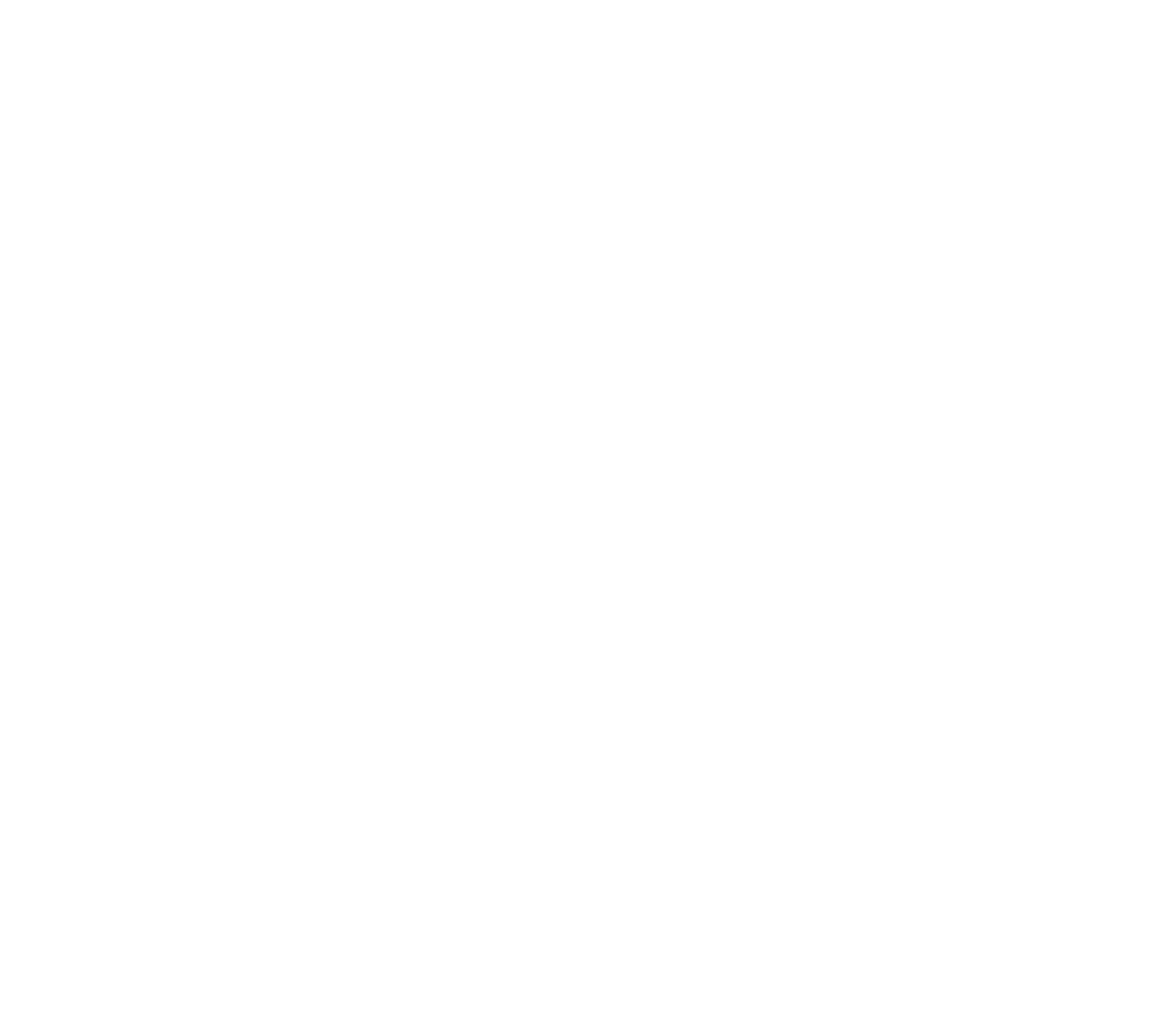} \\
  \end{tabular}
\caption[DDO\,43 images: RC, IR, optical, and FUV]{Multi-wavelength coverage of DDO 43 displaying a $3.0^\prime \times 3.0^\prime$ area. We show total RC flux density at the native resolution (top-left) and again with contours (top-centre). The RC contours are superposed on ancillary LITTLE THINGS images where possible: \halpha\ (middle-left); \RCNT\ obtained by subtracting the expected \RCT\ based on the \halpha-\RCT\ scaling factor of \cite{Deeg1997} from the total RC; {\em GALEX} FUV (middle-right); {\em Spitzer} 24\micron\ (bottom-left); {\em Spitzer} 70\micron\ (bottom-centre); FUV$+24{\rm \mu m}$--inferred SFRD from \citealp{Leroy2012} (bottom-right). We also show the RC that was isolated by the RC--based masking technique (top-right).}
  \label{figure:ddo43Cc_maps}
\end{figure}

\begin{figure}
  \begin{tabular}{ccc}
    \includegraphics[width=0.31\linewidth,clip]{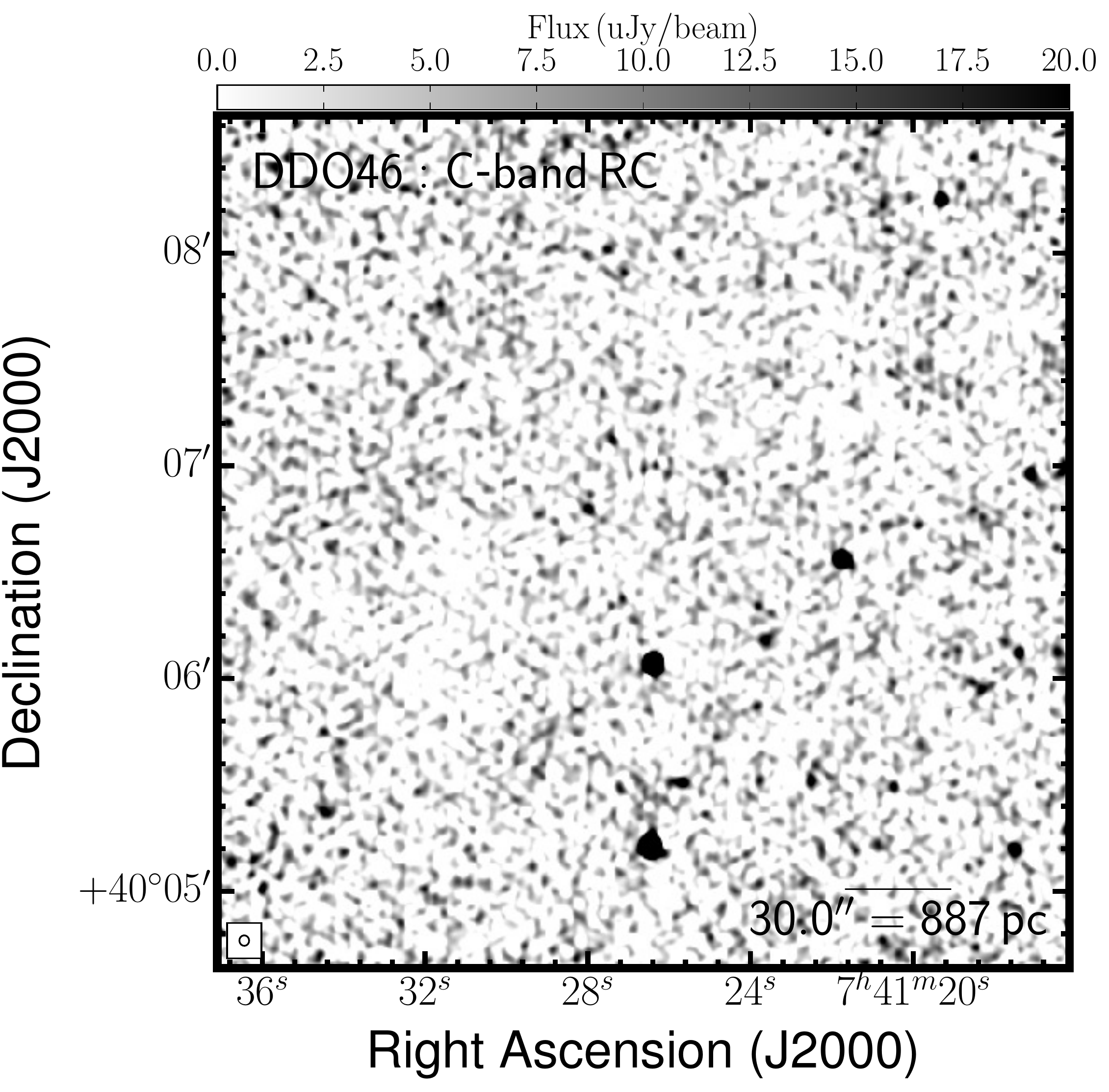} & \ 
    \includegraphics[width=0.31\linewidth,clip]{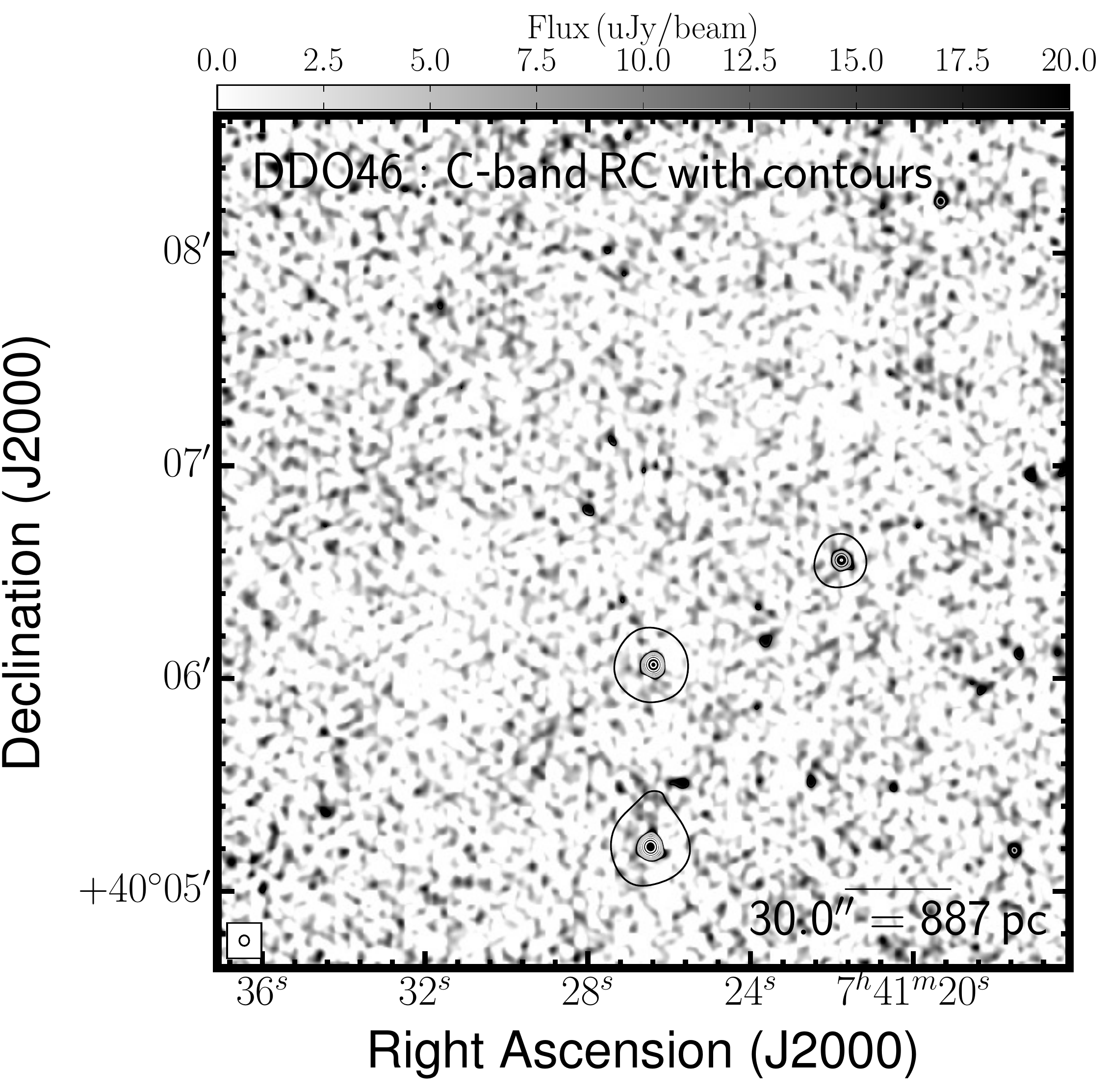} & \ 
    \includegraphics[width=0.31\linewidth,clip]{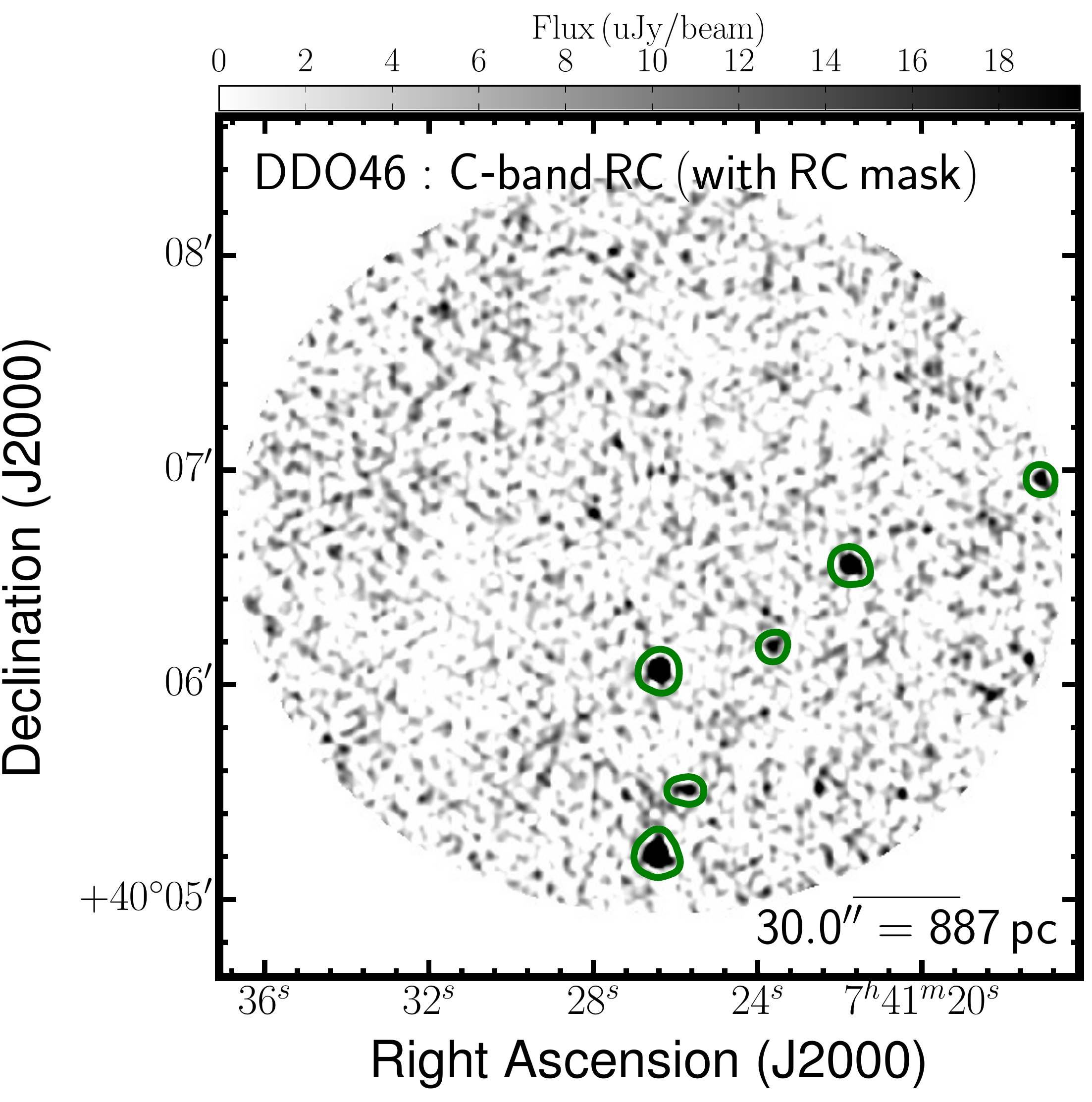} \\
    \includegraphics[width=0.31\linewidth,clip]{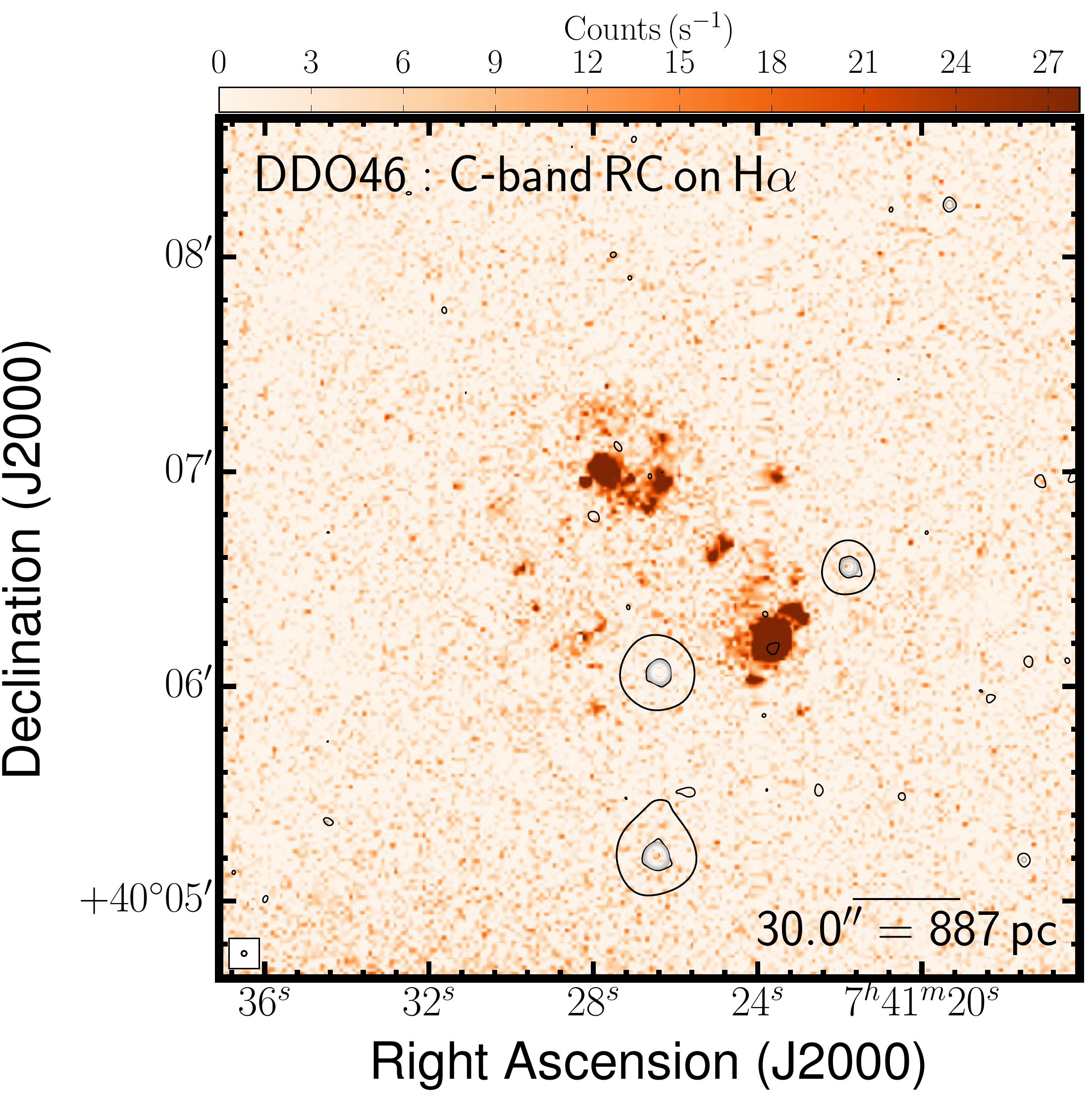} & \ 
    \includegraphics[width=0.31\linewidth,clip]{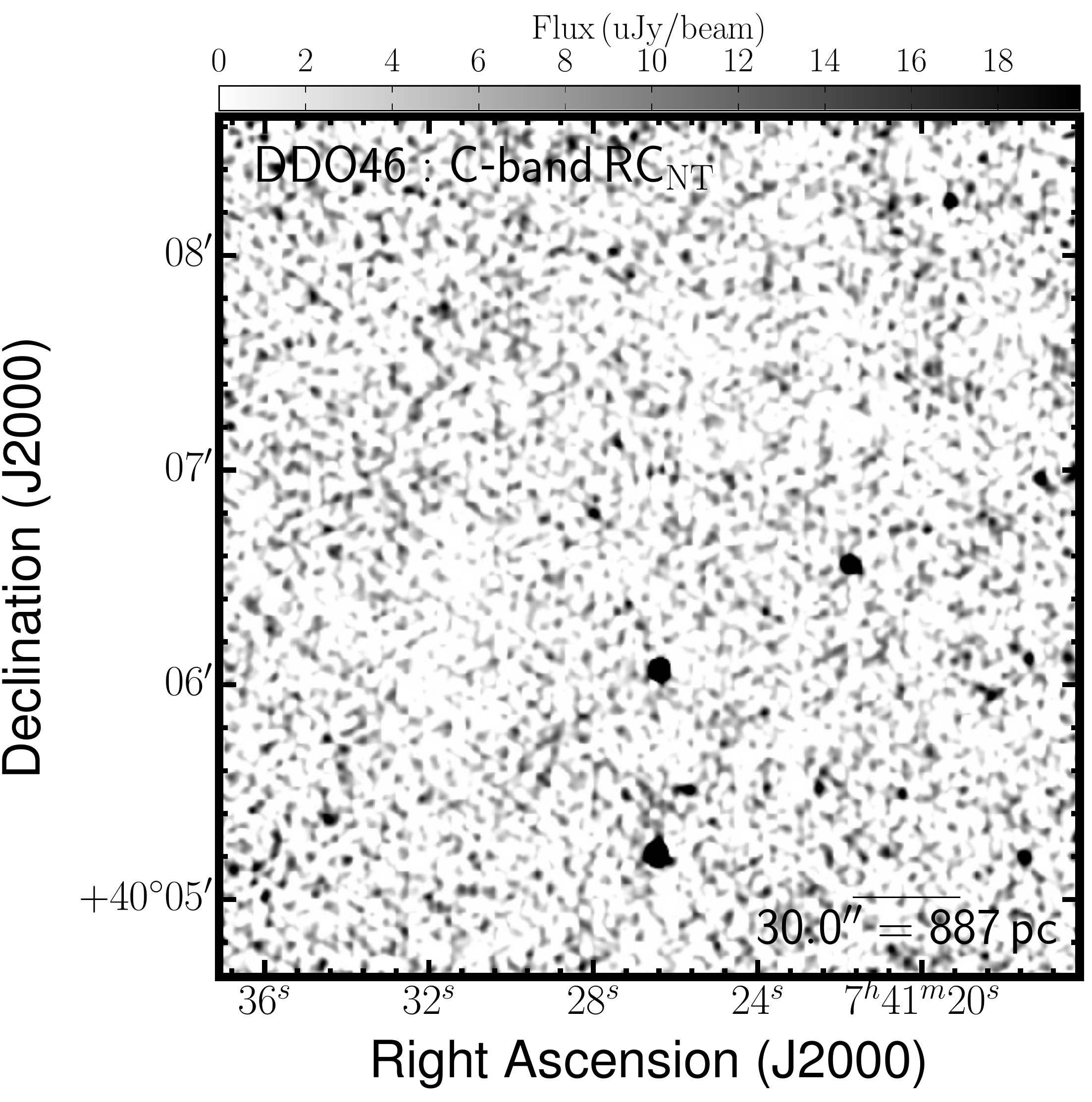} & \ 
    \includegraphics[width=0.31\linewidth,clip]{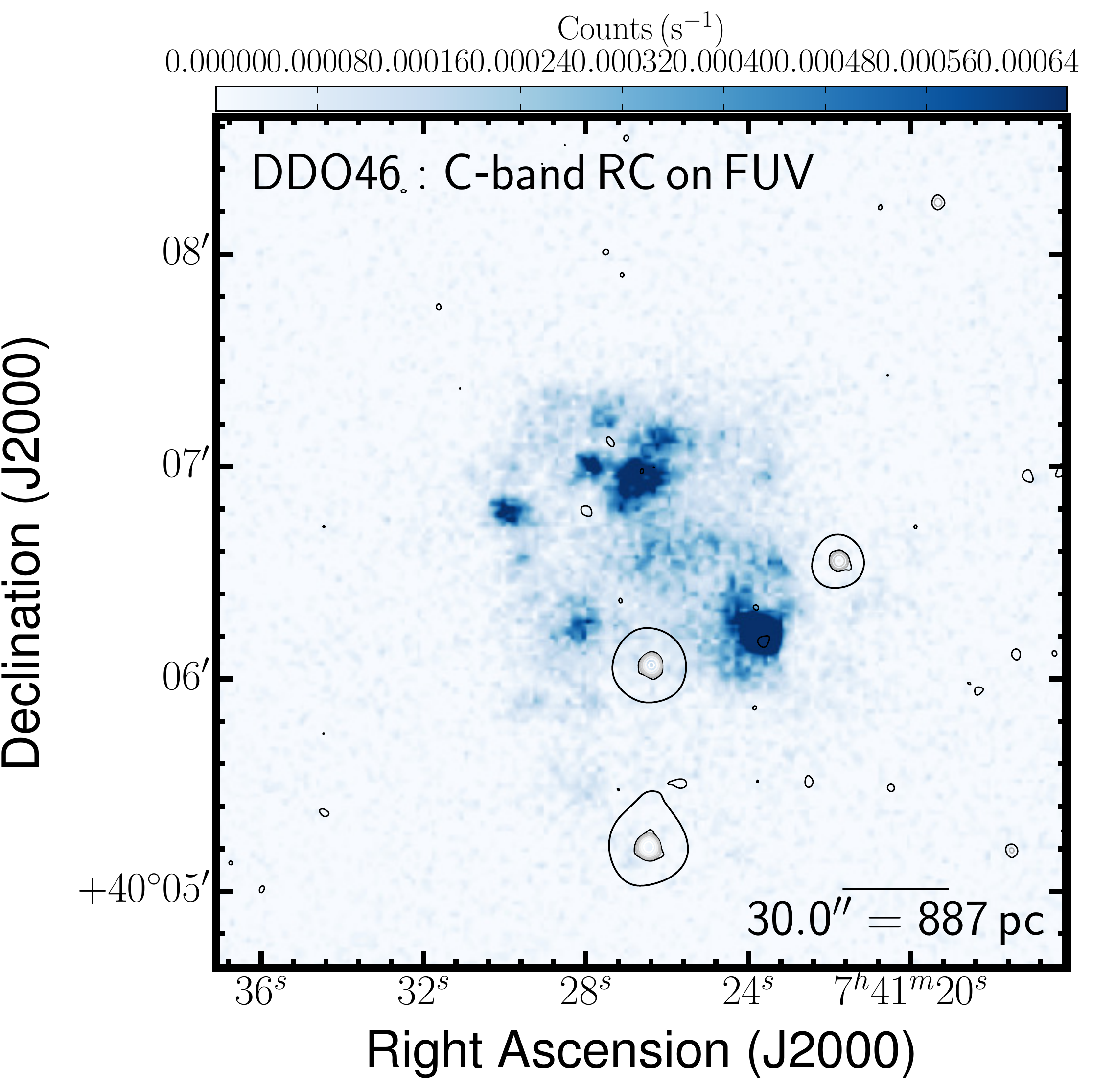} \\
    \includegraphics[width=0.31\linewidth,clip]{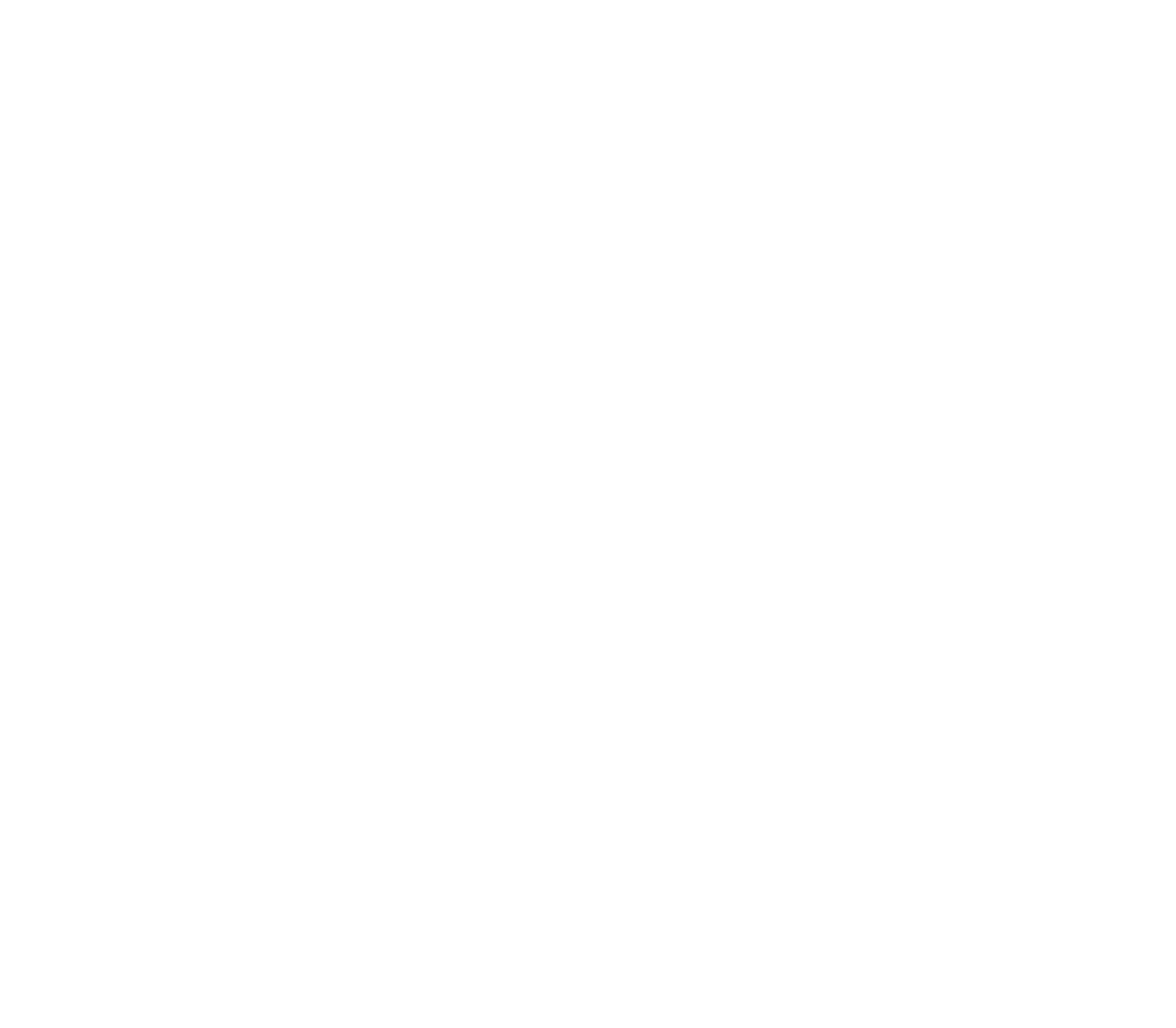} & \ 
    \includegraphics[width=0.31\linewidth,clip]{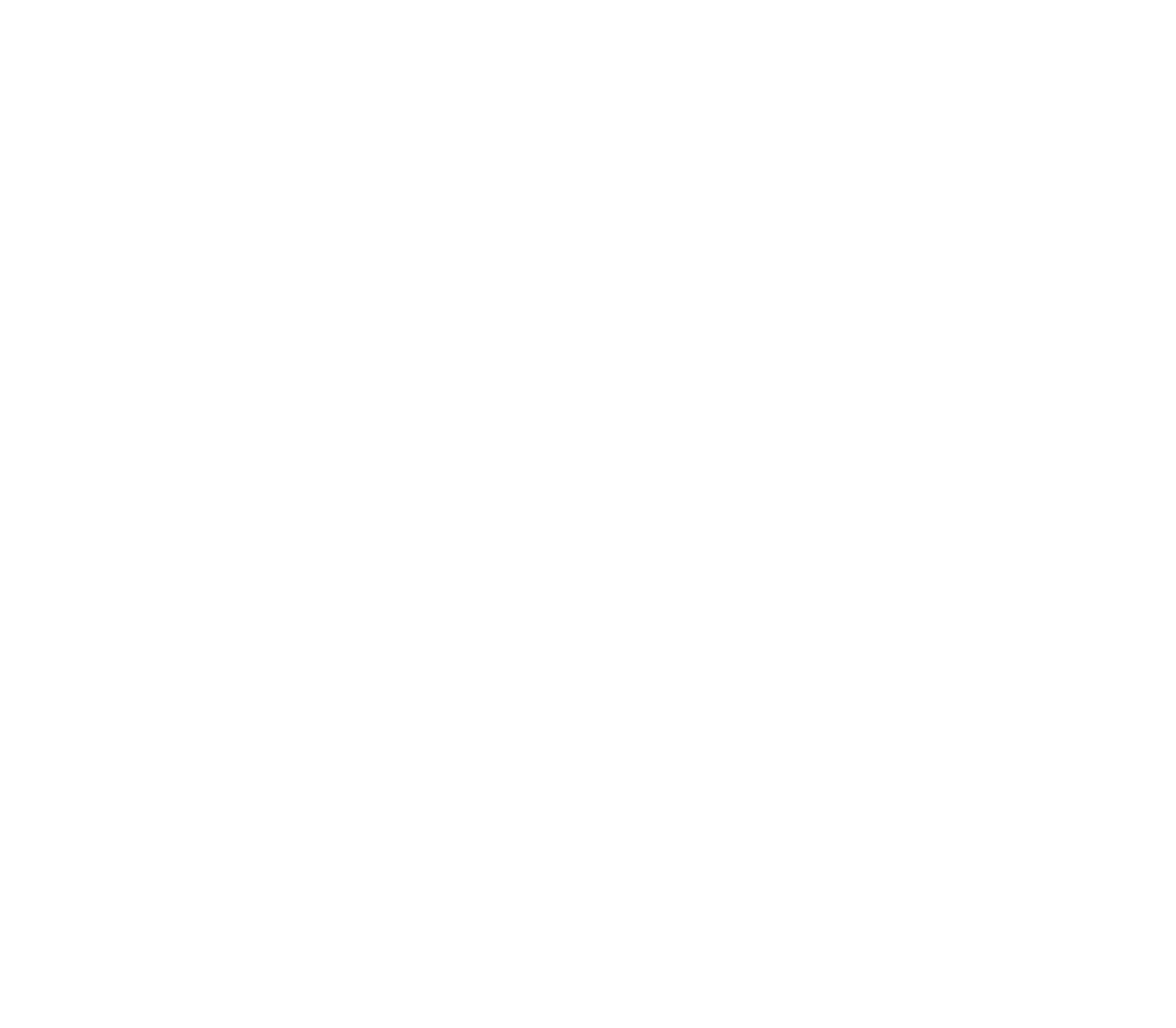} & \ 
    \includegraphics[width=0.31\linewidth,clip]{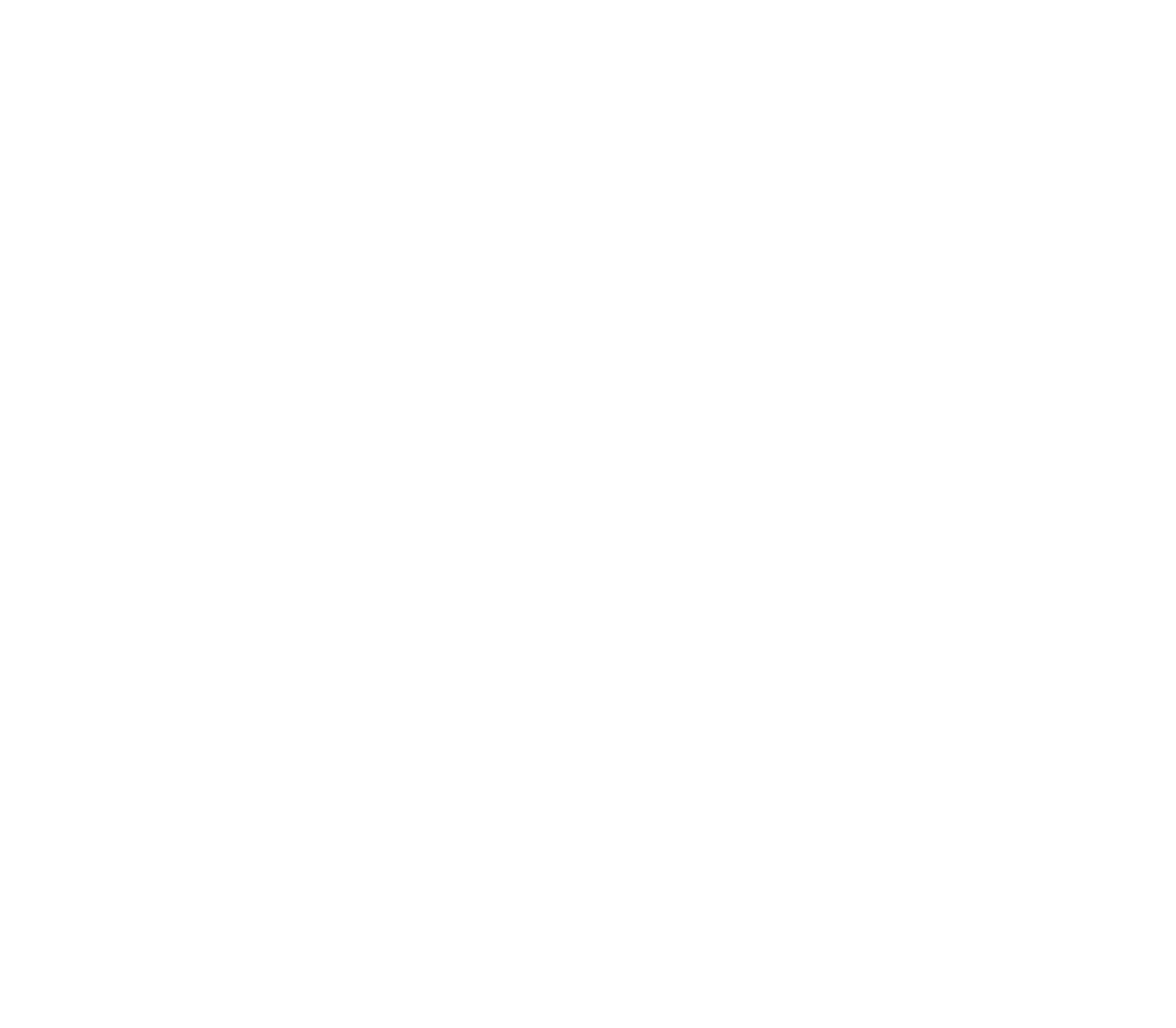} \\
  \end{tabular}
\caption[DDO\,46 images: RC, IR, optical, and FUV]{Multi-wavelength coverage of DDO 46 displaying a $4.0^\prime \times 4.0^\prime$ area. We show total RC flux density at the native resolution (top-left) and again with contours (top-centre). The RC contours are superposed on ancillary LITTLE THINGS images where possible: \halpha\ (middle-left); \RCNT\ obtained by subtracting the expected \RCT\ based on the \halpha-\RCT\ scaling factor of \cite{Deeg1997} from the total RC; {\em GALEX} FUV (middle-right); {\em Spitzer} 24\micron\ (bottom-left); {\em Spitzer} 70\micron\ (bottom-centre); FUV$+24{\rm \mu m}$--inferred SFRD from \citealp{Leroy2012} (bottom-right). We also show the RC that was isolated by the RC--based masking technique (top-right).}
  \label{figure:ddo46Cc_maps}
\end{figure}

\begin{figure}
  \begin{tabular}{ccc}
    \includegraphics[width=0.31\linewidth,clip]{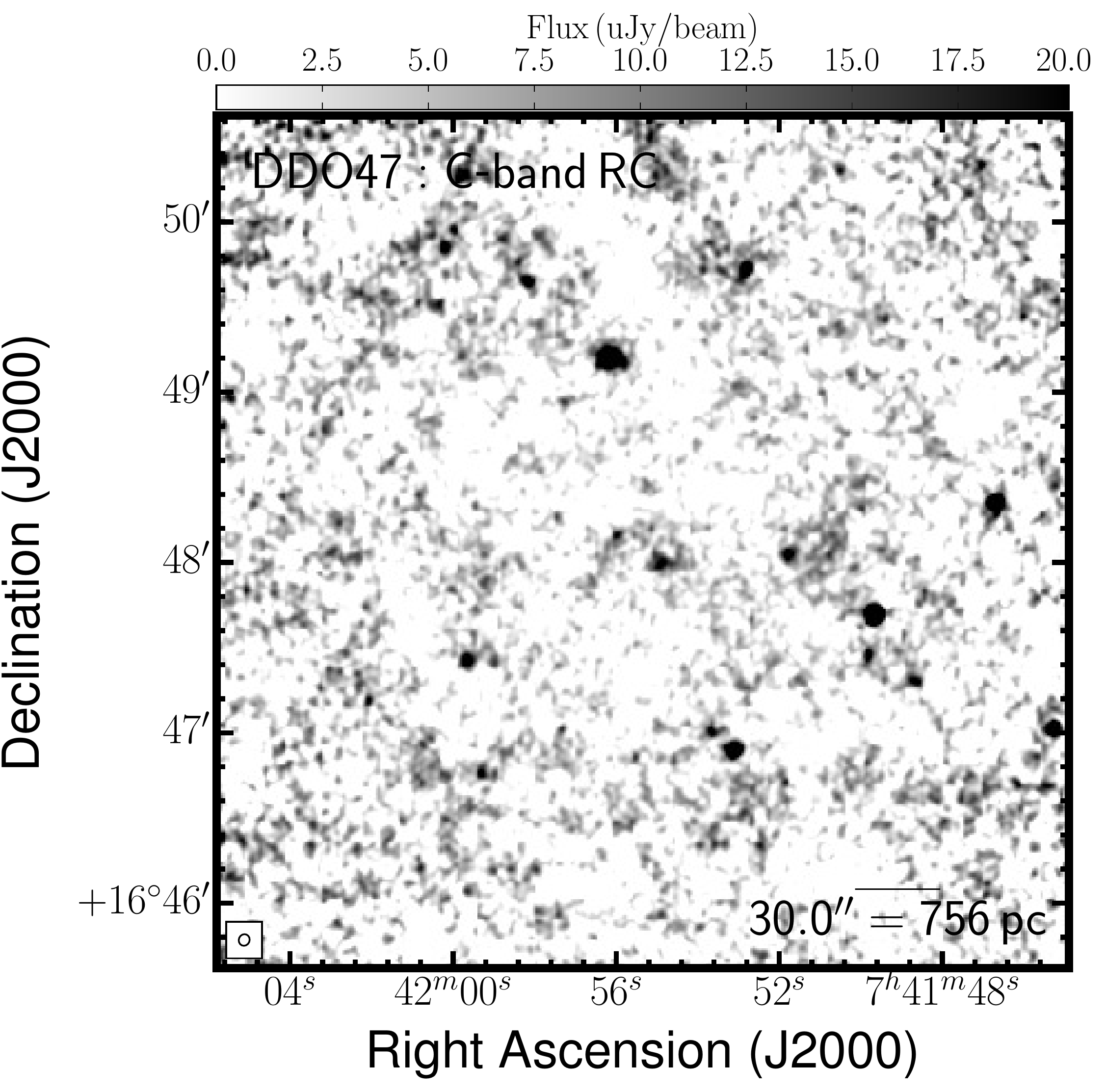} & \ 
    \includegraphics[width=0.31\linewidth,clip]{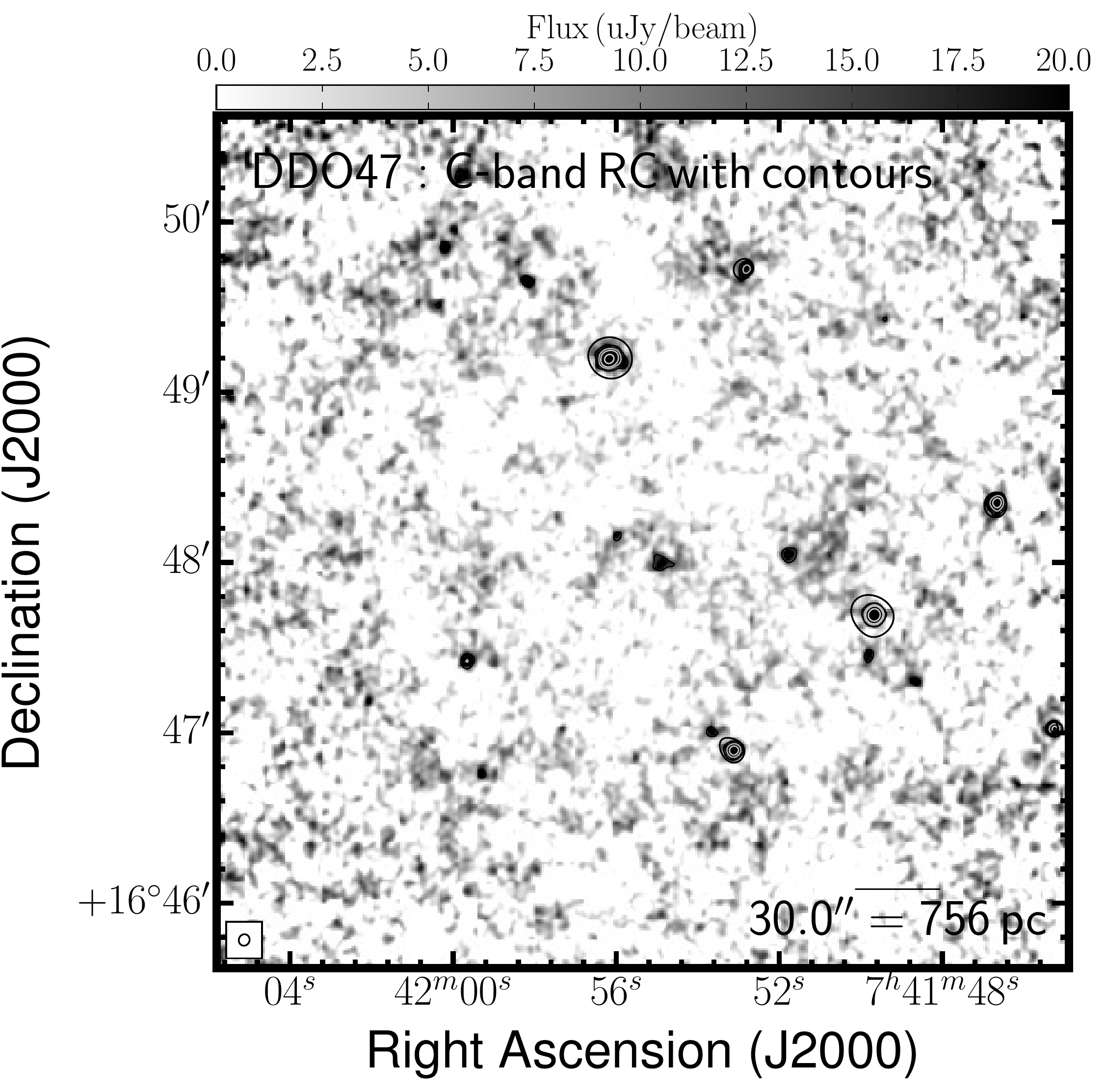} & \ 
    \includegraphics[width=0.31\linewidth,clip]{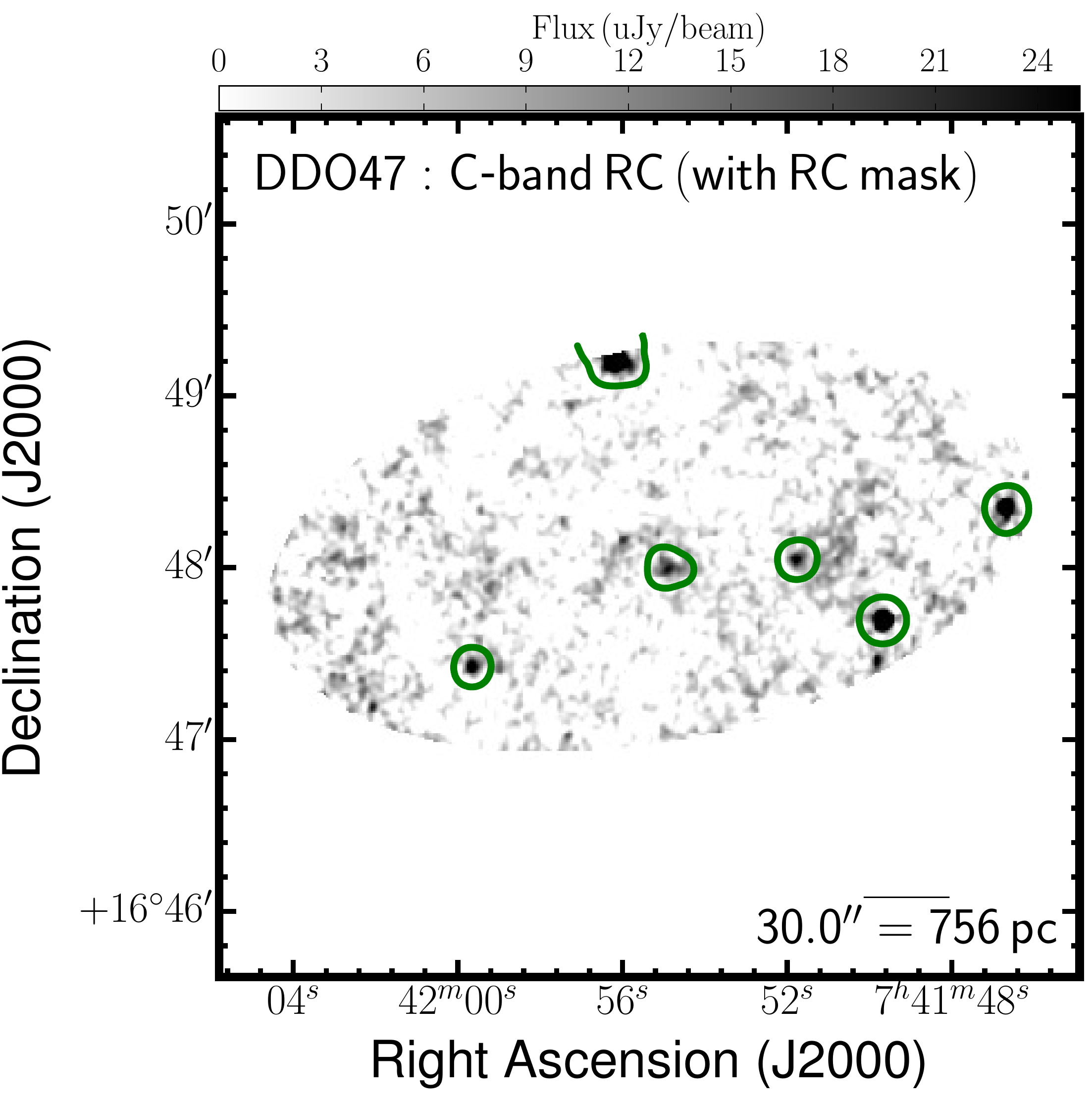} \\
    \includegraphics[width=0.31\linewidth,clip]{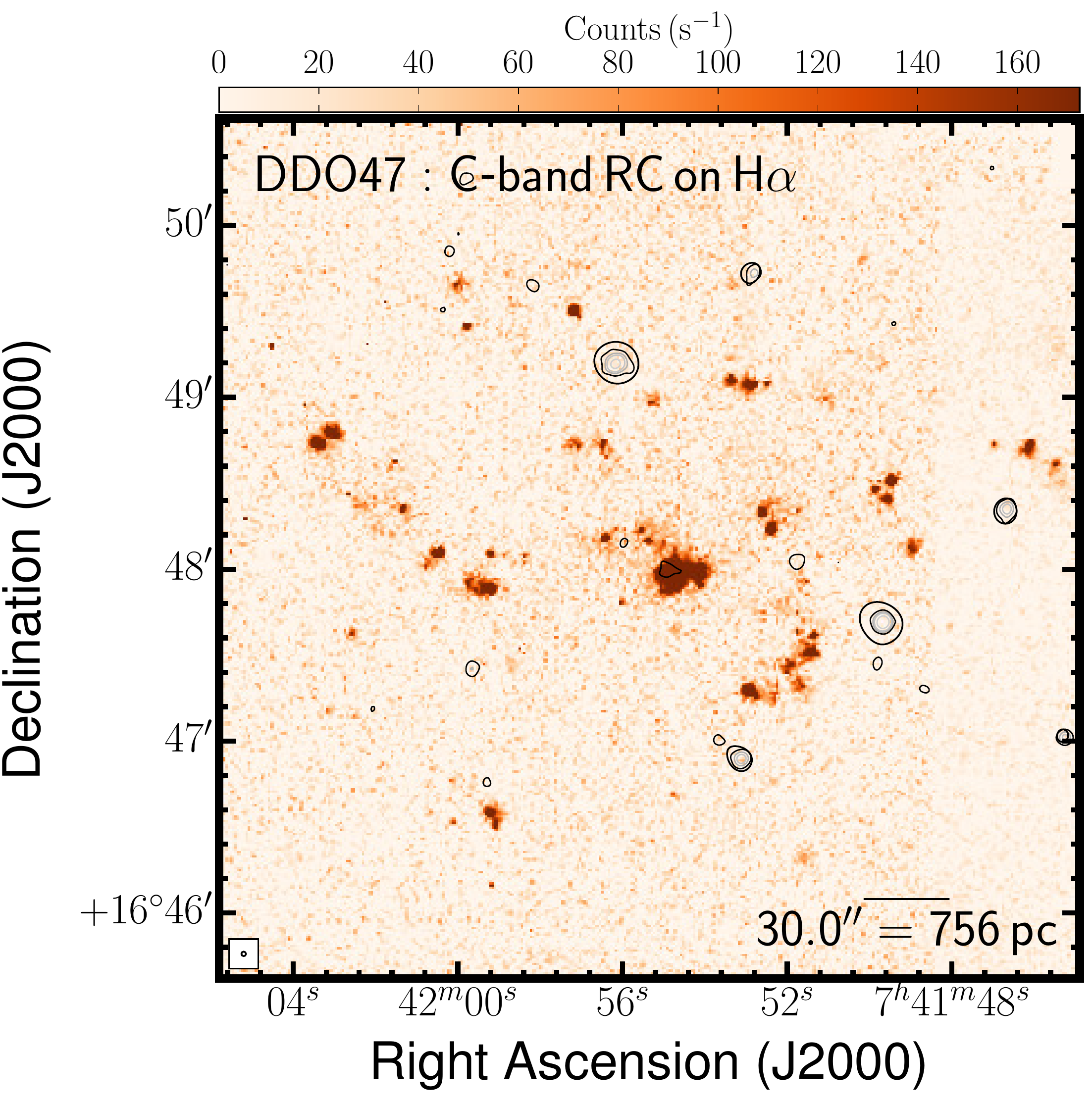} & \ 
    \includegraphics[width=0.31\linewidth,clip]{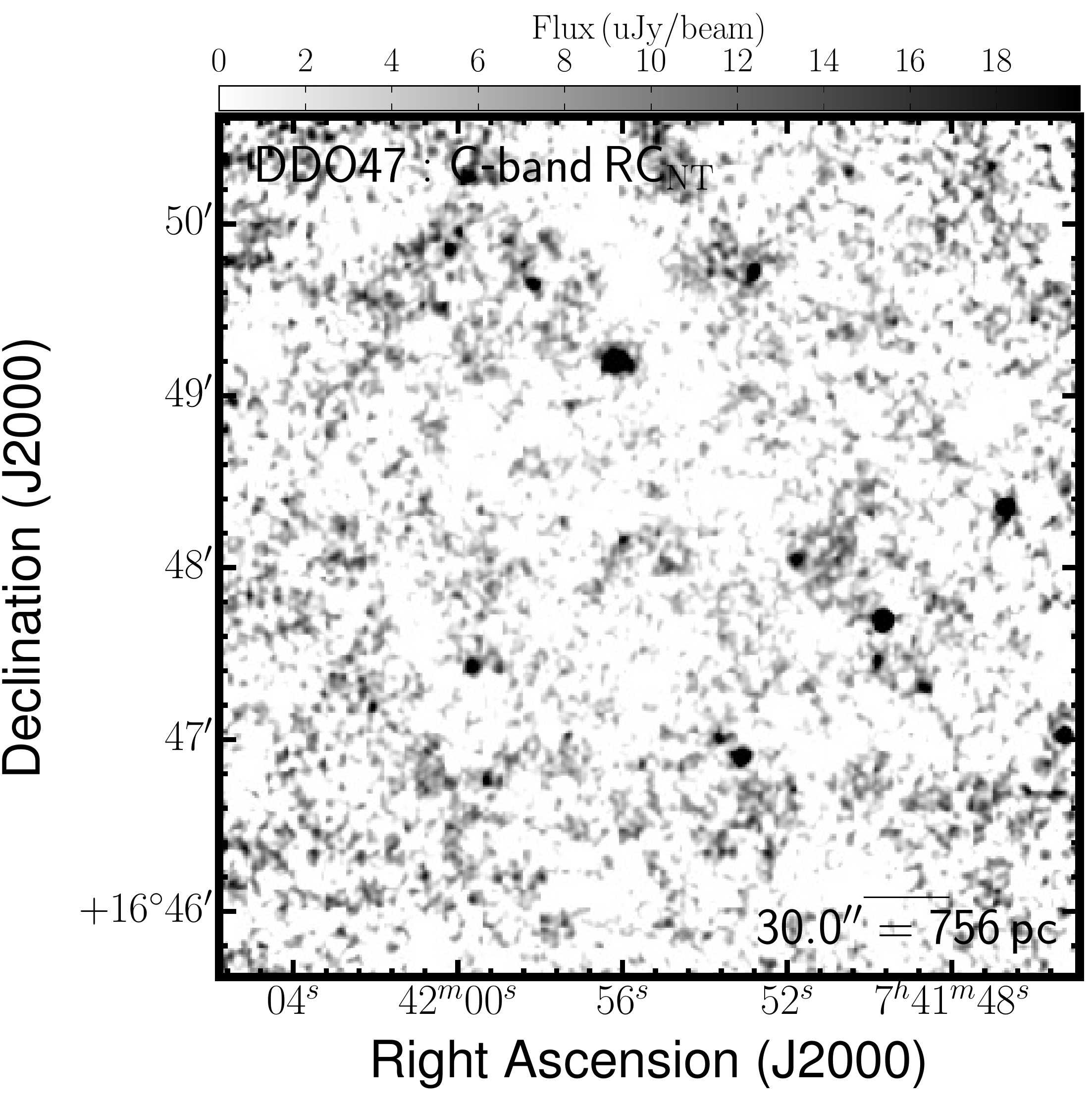} & \ 
    \includegraphics[width=0.31\linewidth,clip]{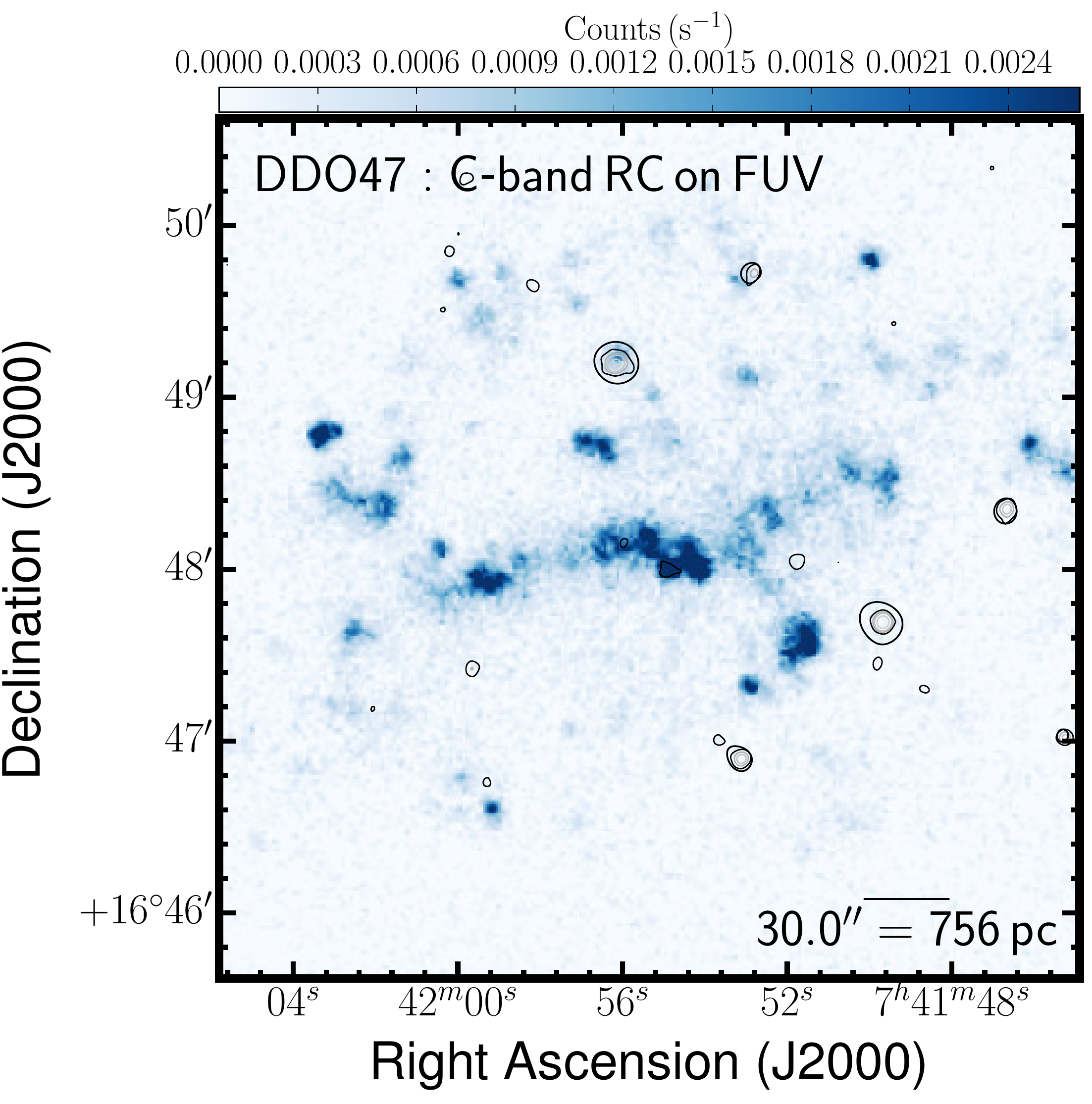} \\
    \includegraphics[width=0.31\linewidth,clip]{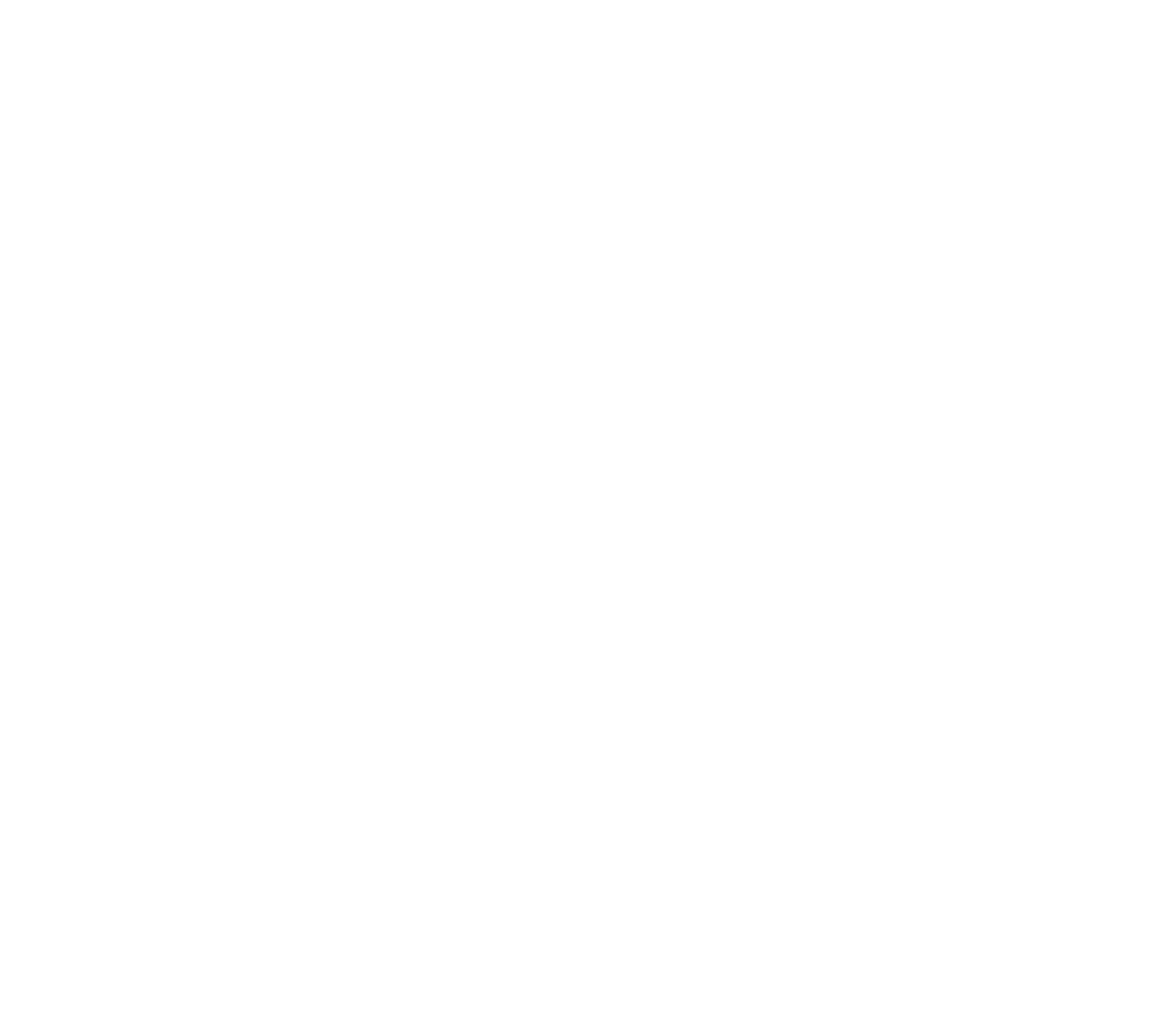} & \ 
    \includegraphics[width=0.31\linewidth,clip]{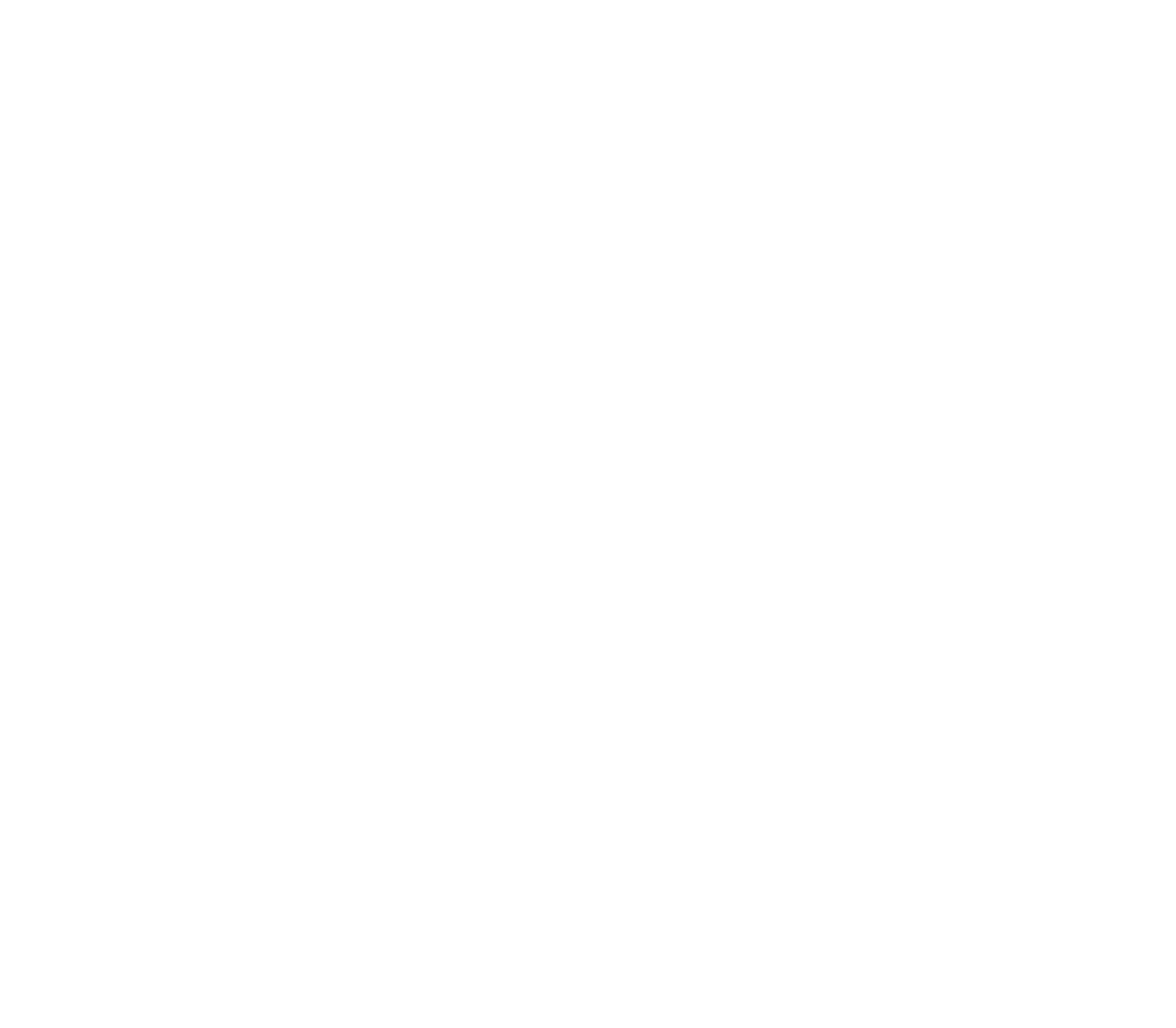} & \ 
    \includegraphics[width=0.31\linewidth,clip]{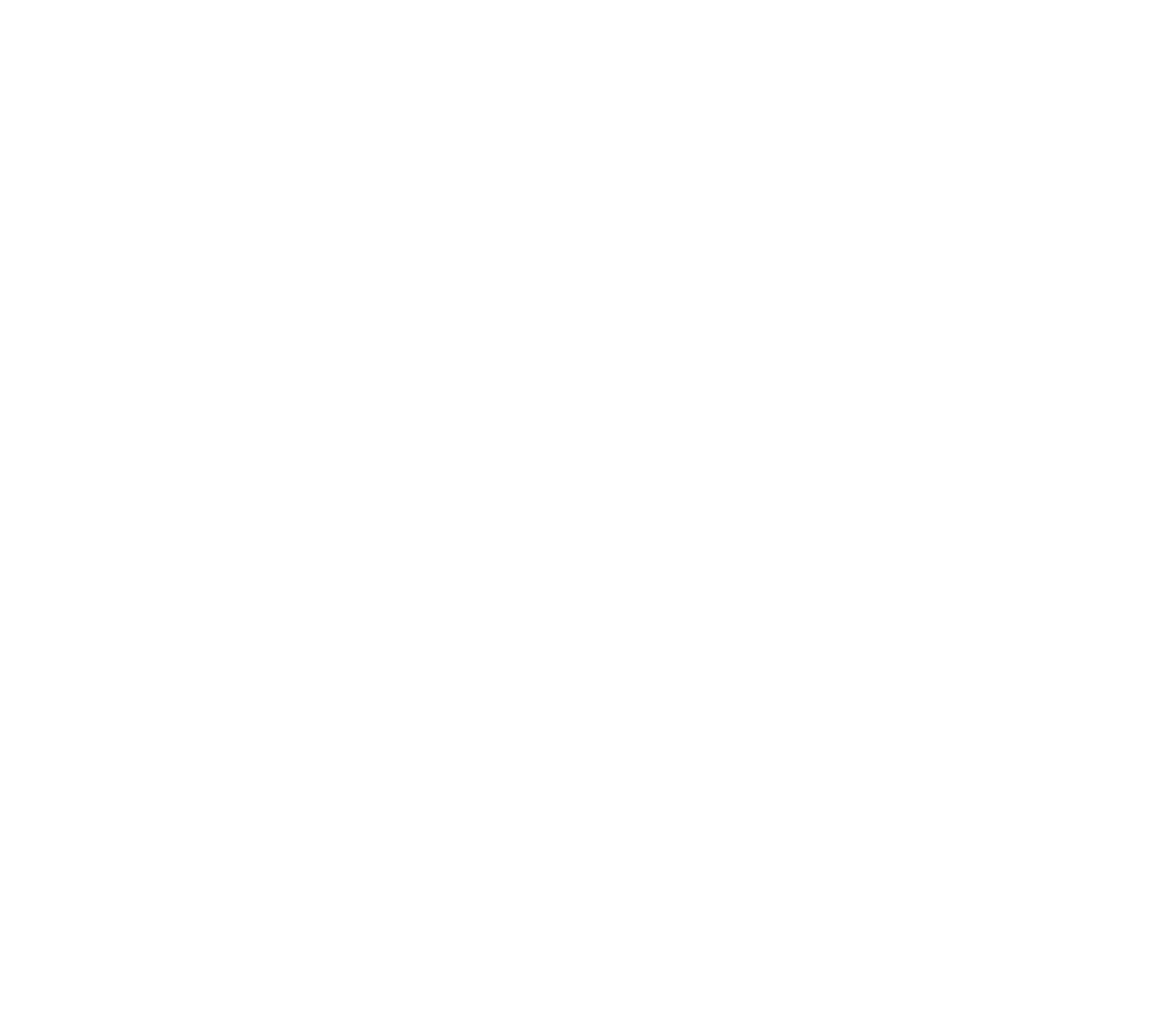} \\
  \end{tabular}
\caption[DDO\,47 images: RC, IR, optical, and FUV]{Multi-wavelength coverage of DDO 47 displaying a $5.0^\prime \times 5.0^\prime$ area. We show total RC flux density at the native resolution (top-left) and again with contours (top-centre). The RC contours are superposed on ancillary LITTLE THINGS images where possible: \halpha\ (middle-left); \RCNT\ obtained by subtracting the expected \RCT\ based on the \halpha-\RCT\ scaling factor of \cite{Deeg1997} from the total RC; {\em GALEX} FUV (middle-right); {\em Spitzer} 24\micron\ (bottom-left); {\em Spitzer} 70\micron\ (bottom-centre); FUV$+24{\rm \mu m}$--inferred SFRD from \citealp{Leroy2012} (bottom-right). We also show the RC that was isolated by the RC--based masking technique (top-right).}
  \label{figure:ddo47Cc_maps}
\end{figure}

\begin{figure}
  \begin{tabular}{ccc}
    \includegraphics[width=0.31\linewidth,clip]{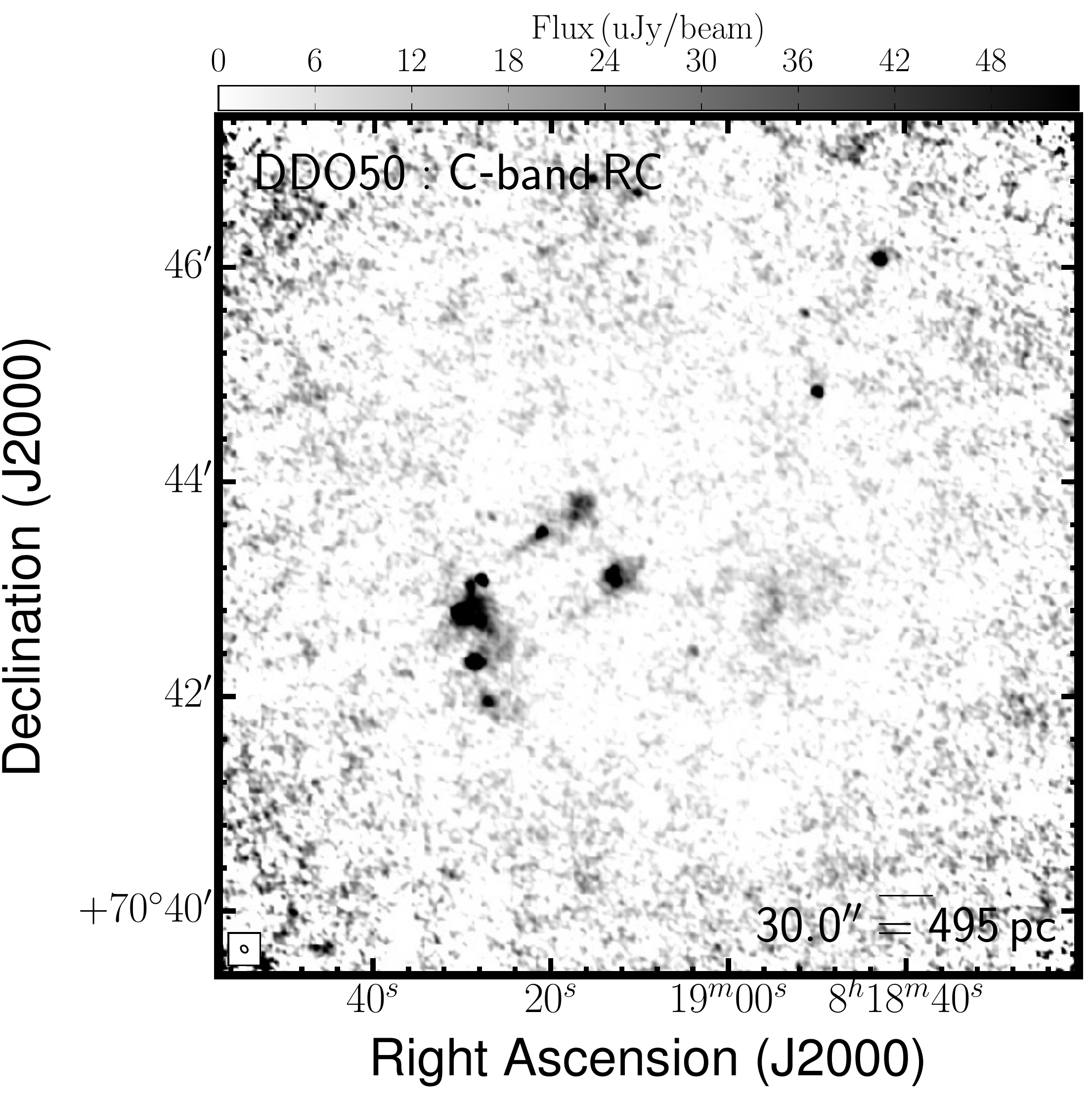} & \ 
    \includegraphics[width=0.31\linewidth,clip]{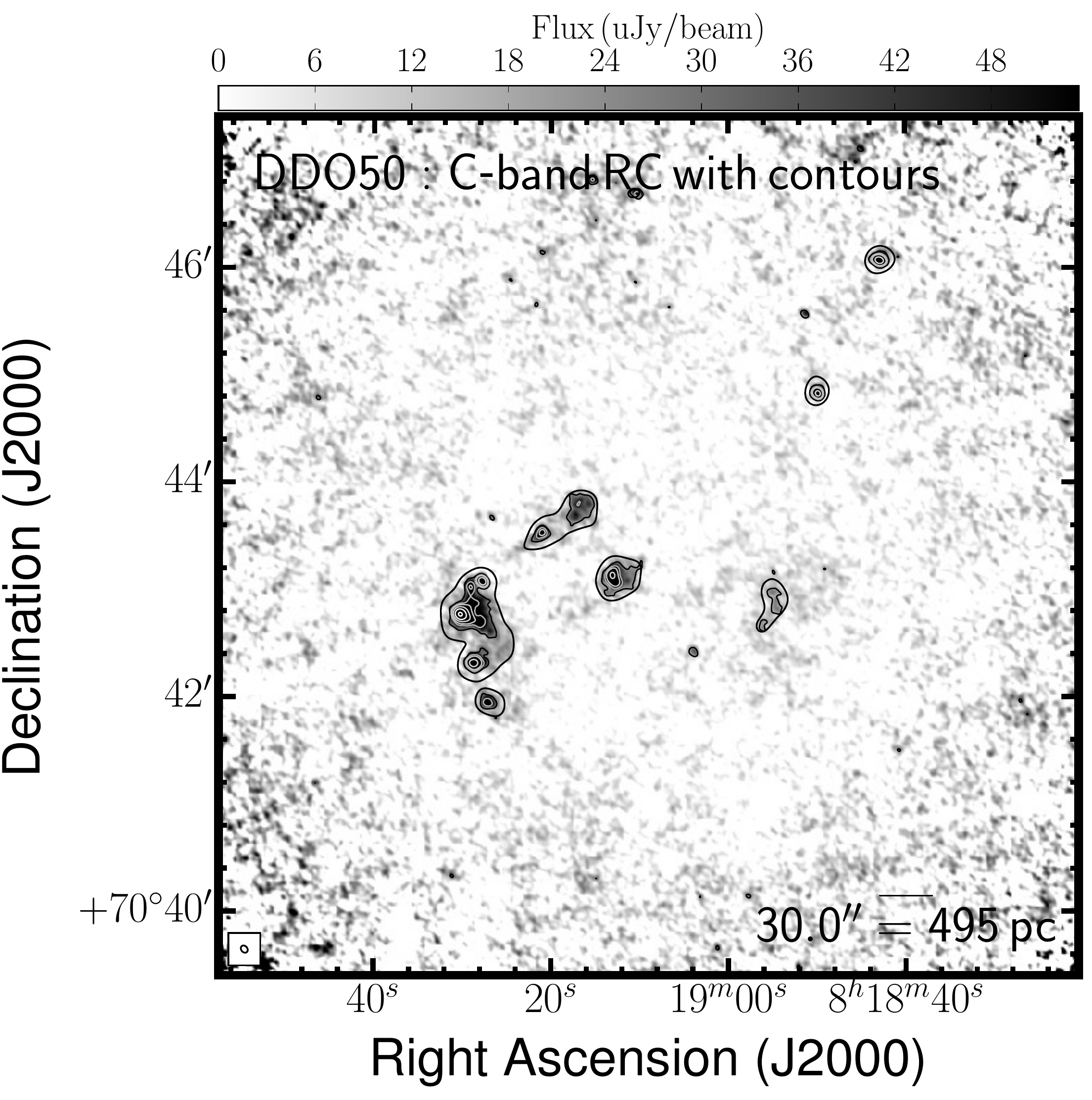} & \ 
    \includegraphics[width=0.31\linewidth,clip]{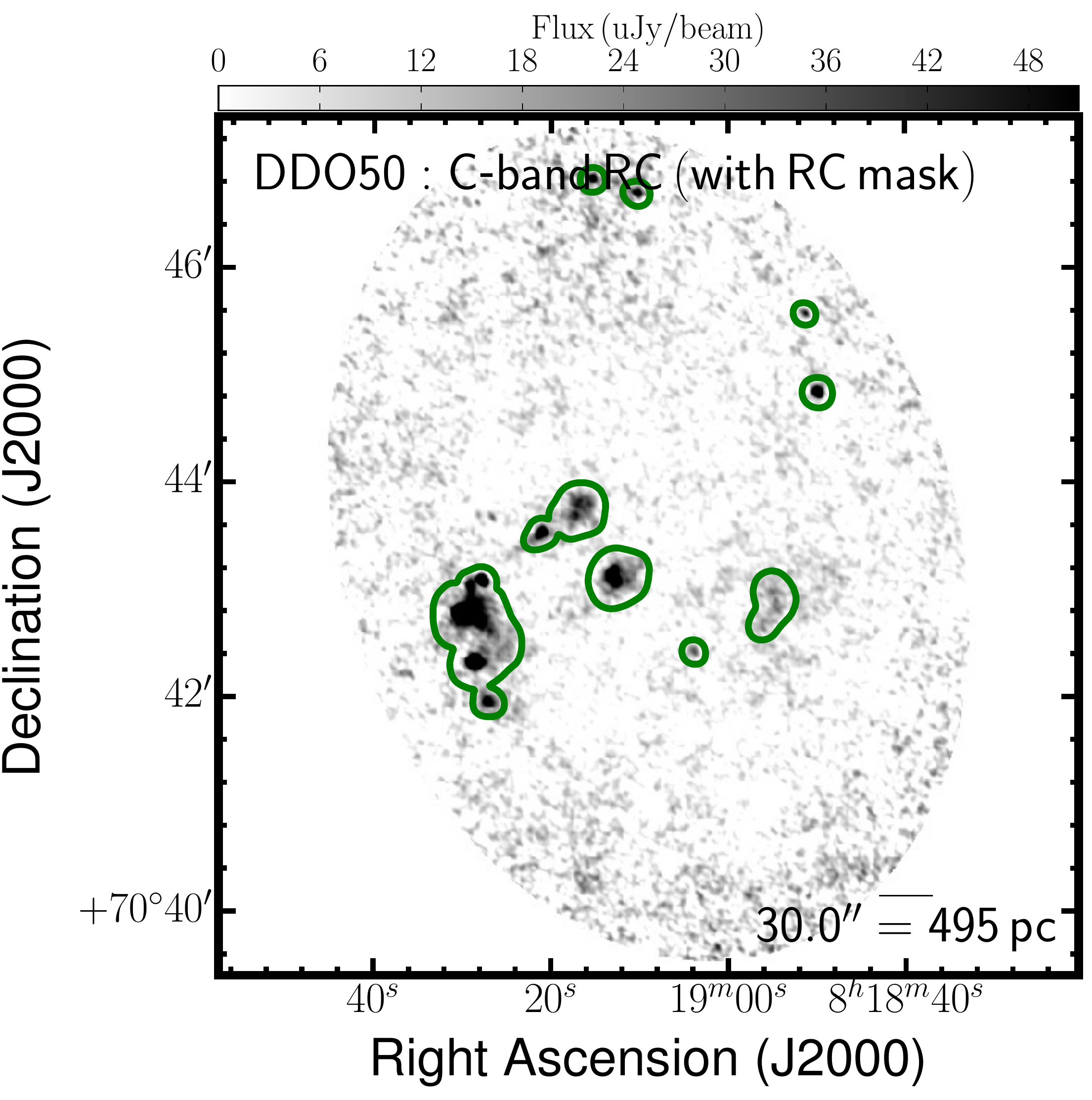} \\
    \includegraphics[width=0.31\linewidth,clip]{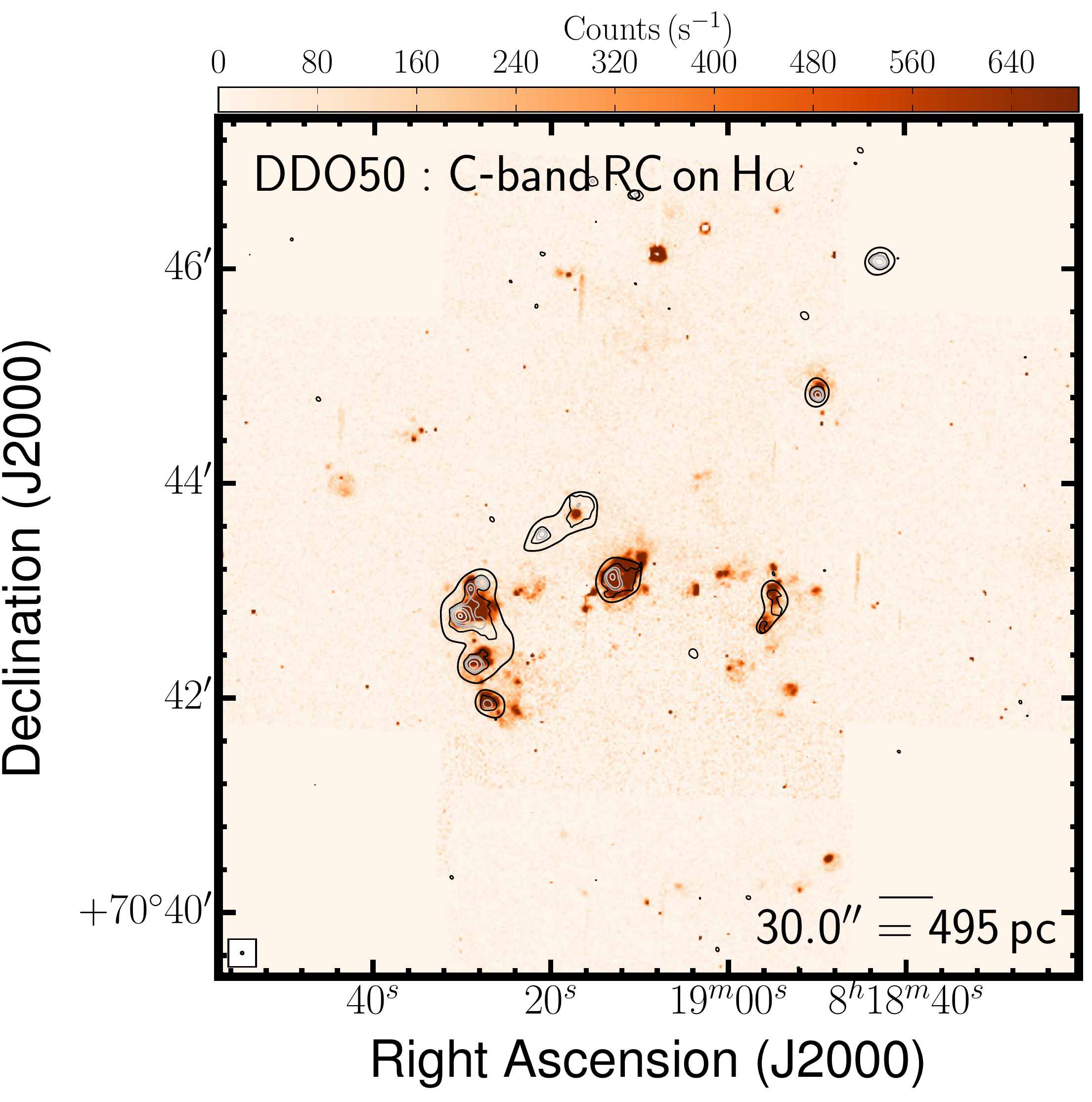} & \ 
    \includegraphics[width=0.31\linewidth,clip]{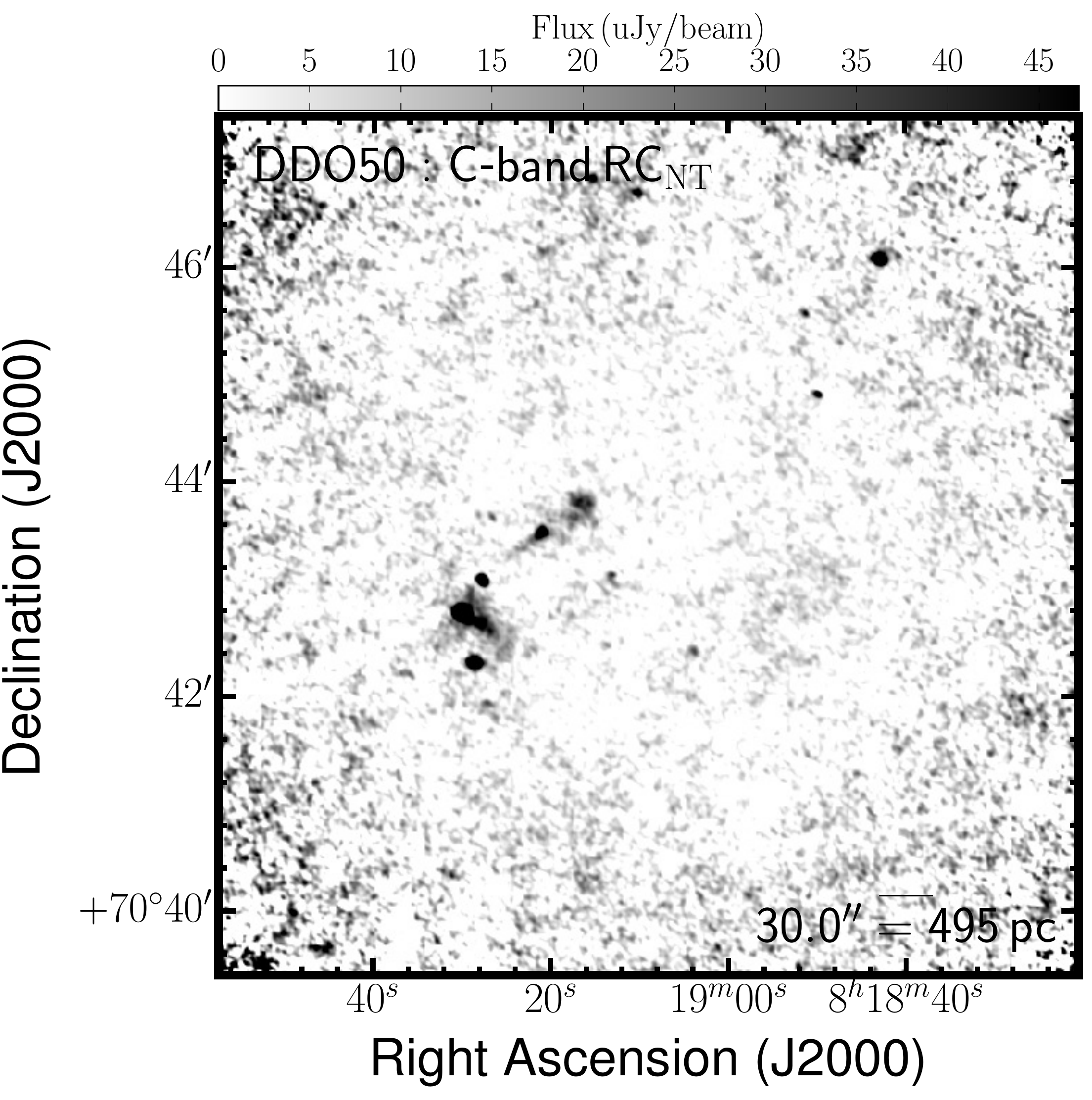} & \ 
    \includegraphics[width=0.31\linewidth,clip]{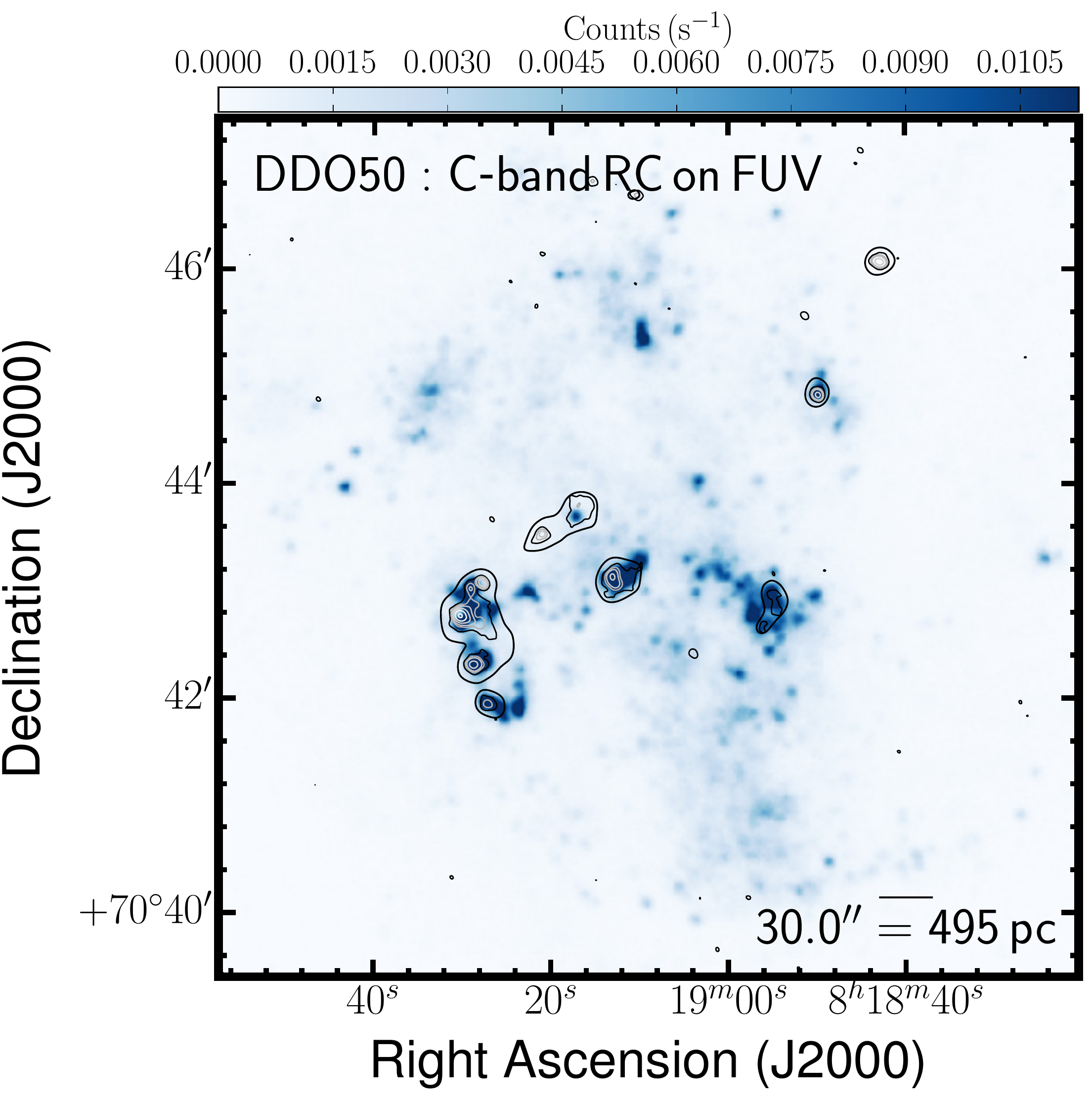} \\
    \includegraphics[width=0.31\linewidth,clip]{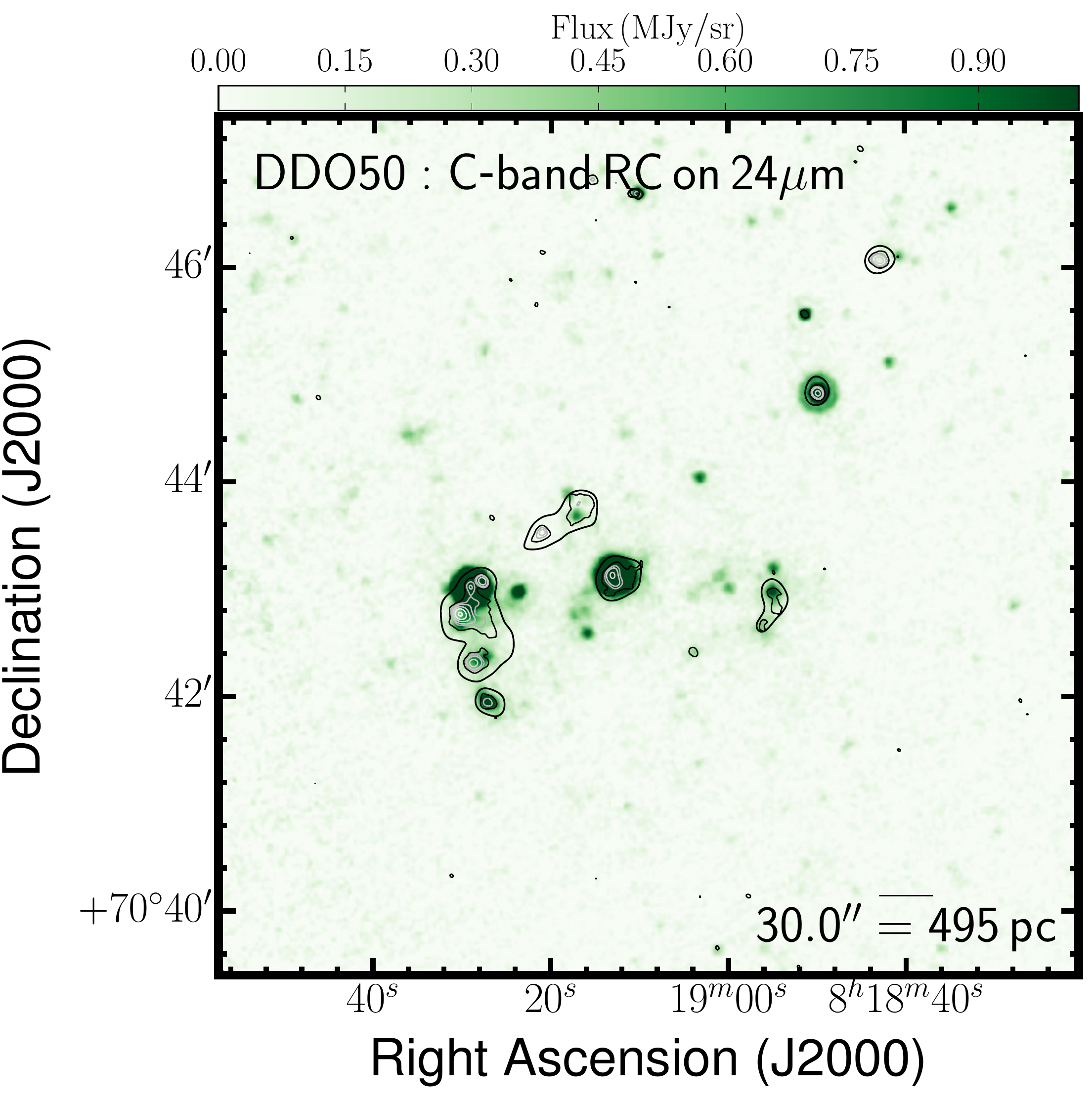} & \ 
    \includegraphics[width=0.31\linewidth,clip]{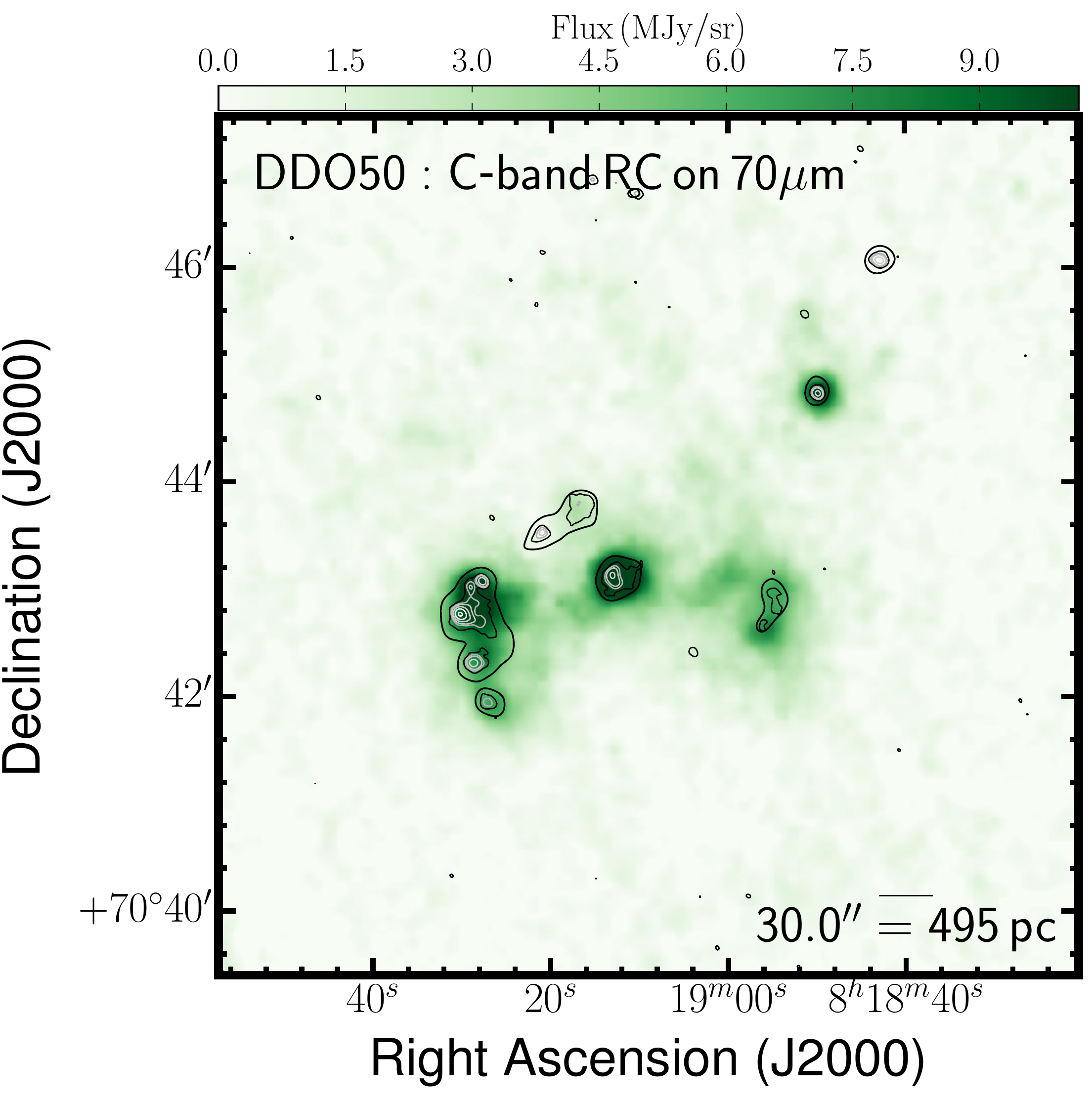} & \ 
    \includegraphics[width=0.31\linewidth,clip]{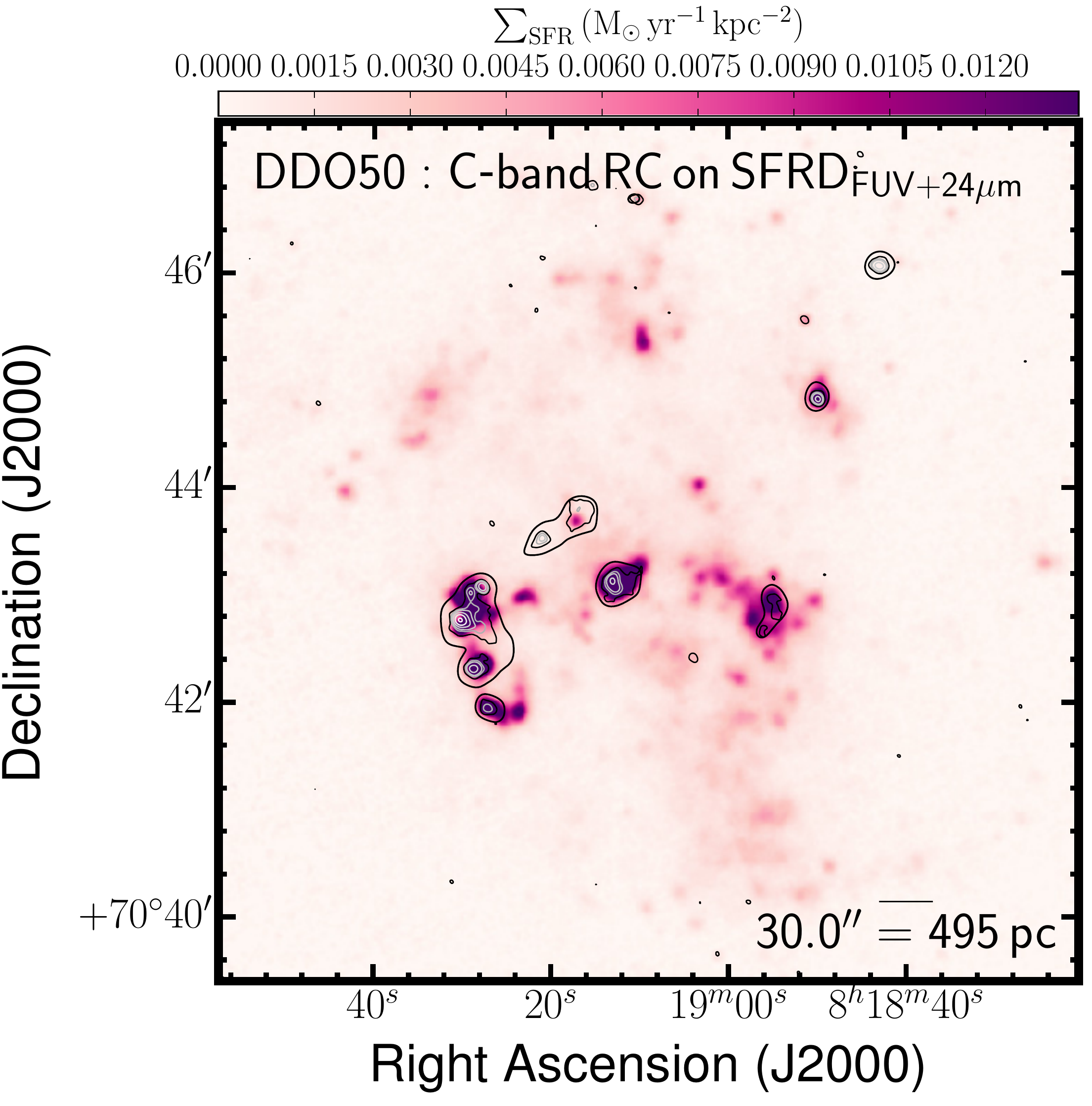} \\
  \end{tabular}
\caption[DDO\,50 images: RC, IR, optical, and FUV]{Multi-wavelength coverage of DDO 50 displaying a $8.0^\prime \times 8.0^\prime$ area. We show total RC flux density at the native resolution (top-left) and again with contours (top-centre). The RC contours are superposed on ancillary LITTLE THINGS images where possible: \halpha\ (middle-left); \RCNT\ obtained by subtracting the expected \RCT\ based on the \halpha-\RCT\ scaling factor of \cite{Deeg1997} from the total RC; {\em GALEX} FUV (middle-right); {\em Spitzer} 24\micron\ (bottom-left); {\em Spitzer} 70\micron\ (bottom-centre); FUV$+24{\rm \mu m}$--inferred SFRD from \citealp{Leroy2012} (bottom-right). We also show the RC that was isolated by the RC--based masking technique (top-right).}
  \label{figure:ddo50Cc_maps}
\end{figure}

\begin{figure}
  \begin{tabular}{ccc}
    \includegraphics[width=0.31\linewidth,clip]{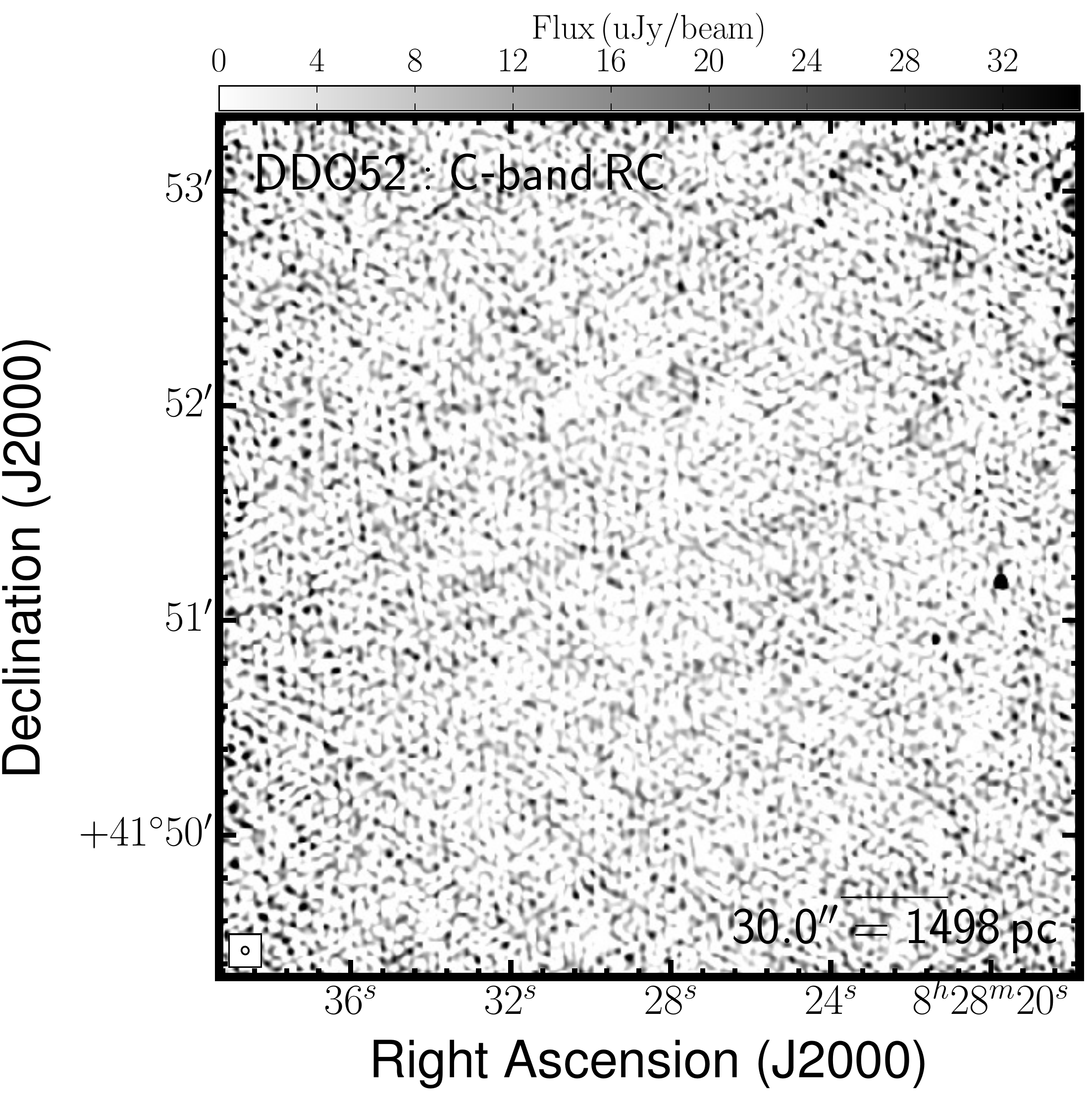} & \ 
    \includegraphics[width=0.31\linewidth,clip]{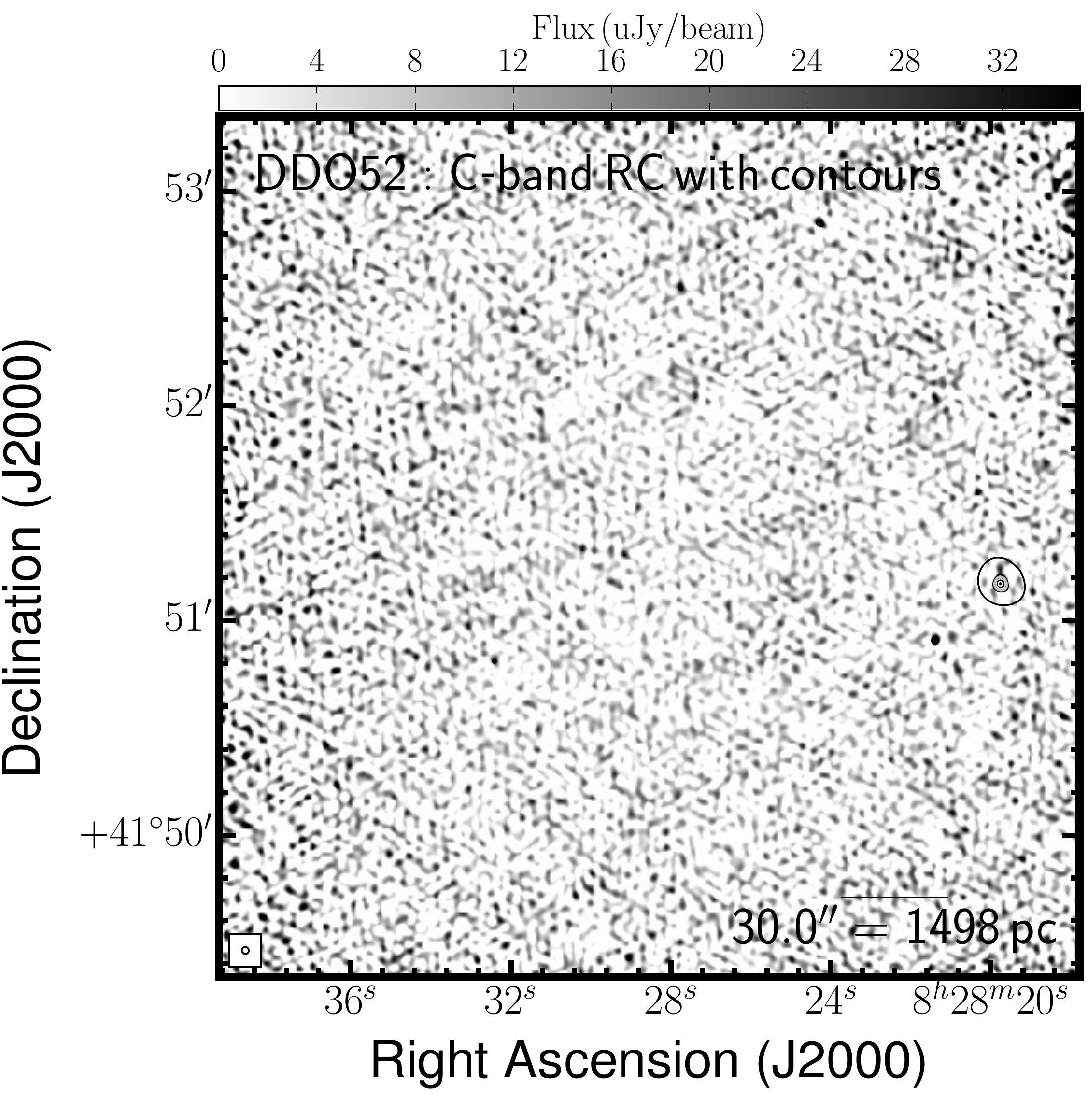} & \ 
    \includegraphics[width=0.31\linewidth,clip]{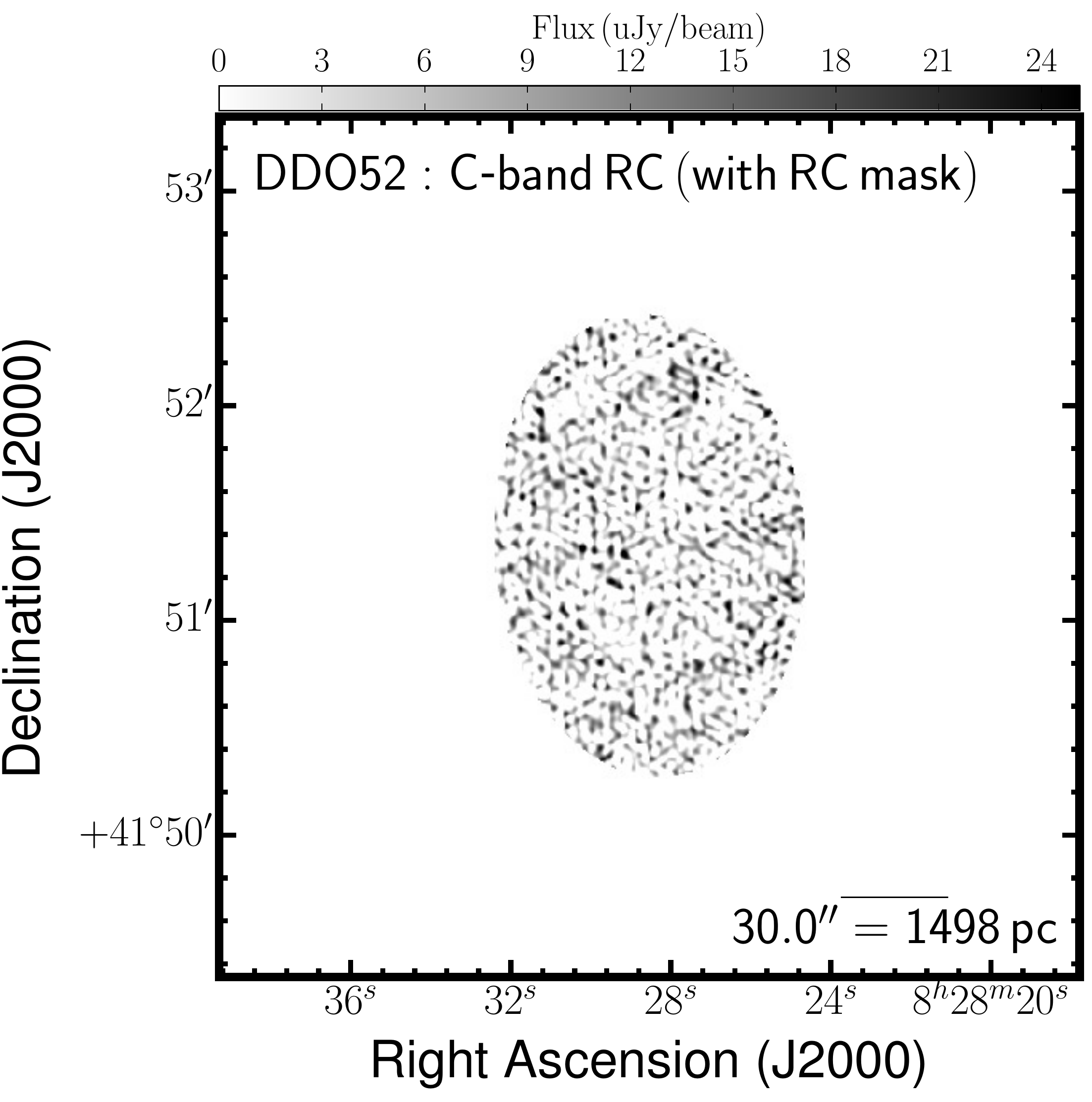} \\
    \includegraphics[width=0.31\linewidth,clip]{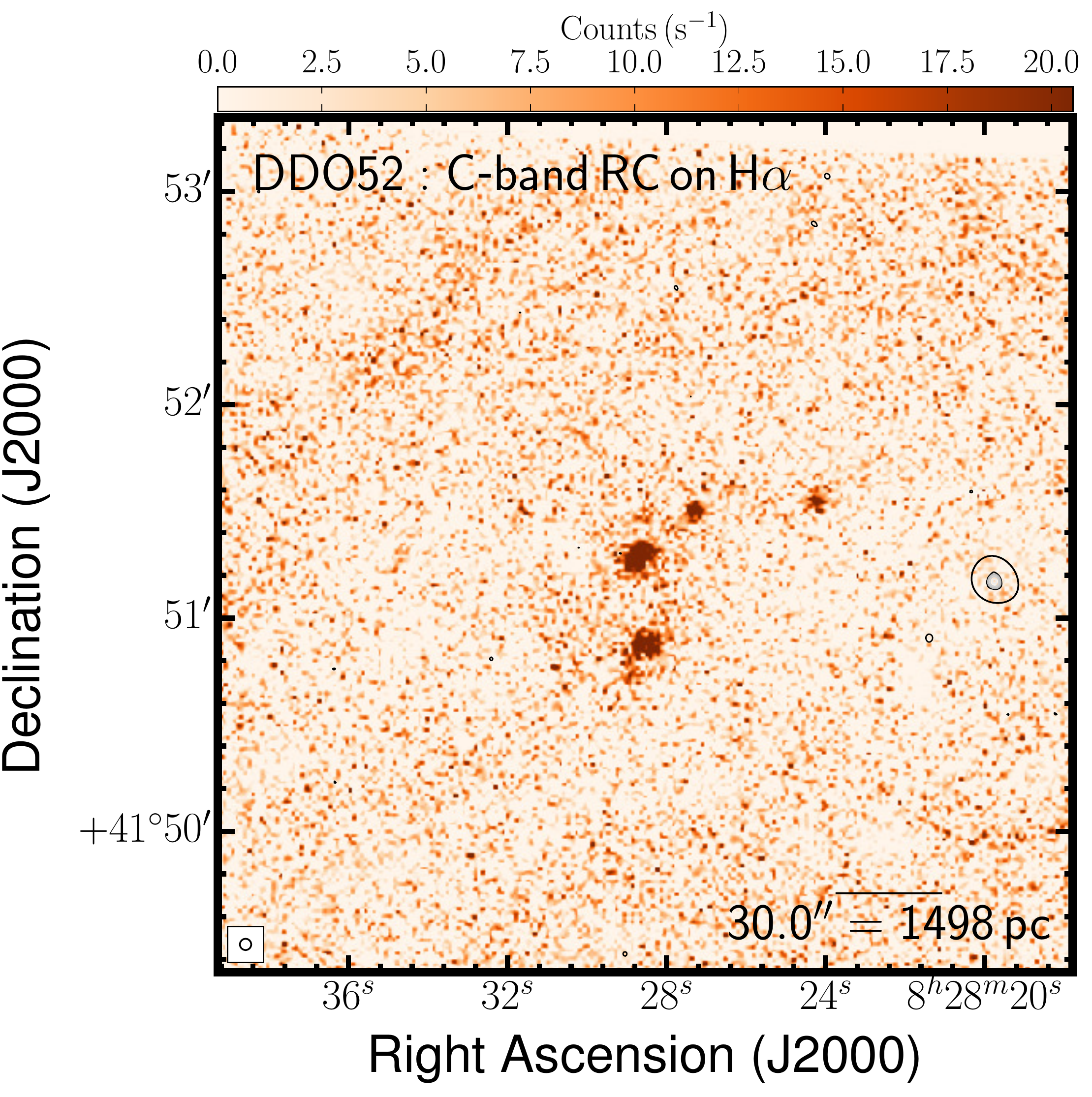} & \ 
    \includegraphics[width=0.31\linewidth,clip]{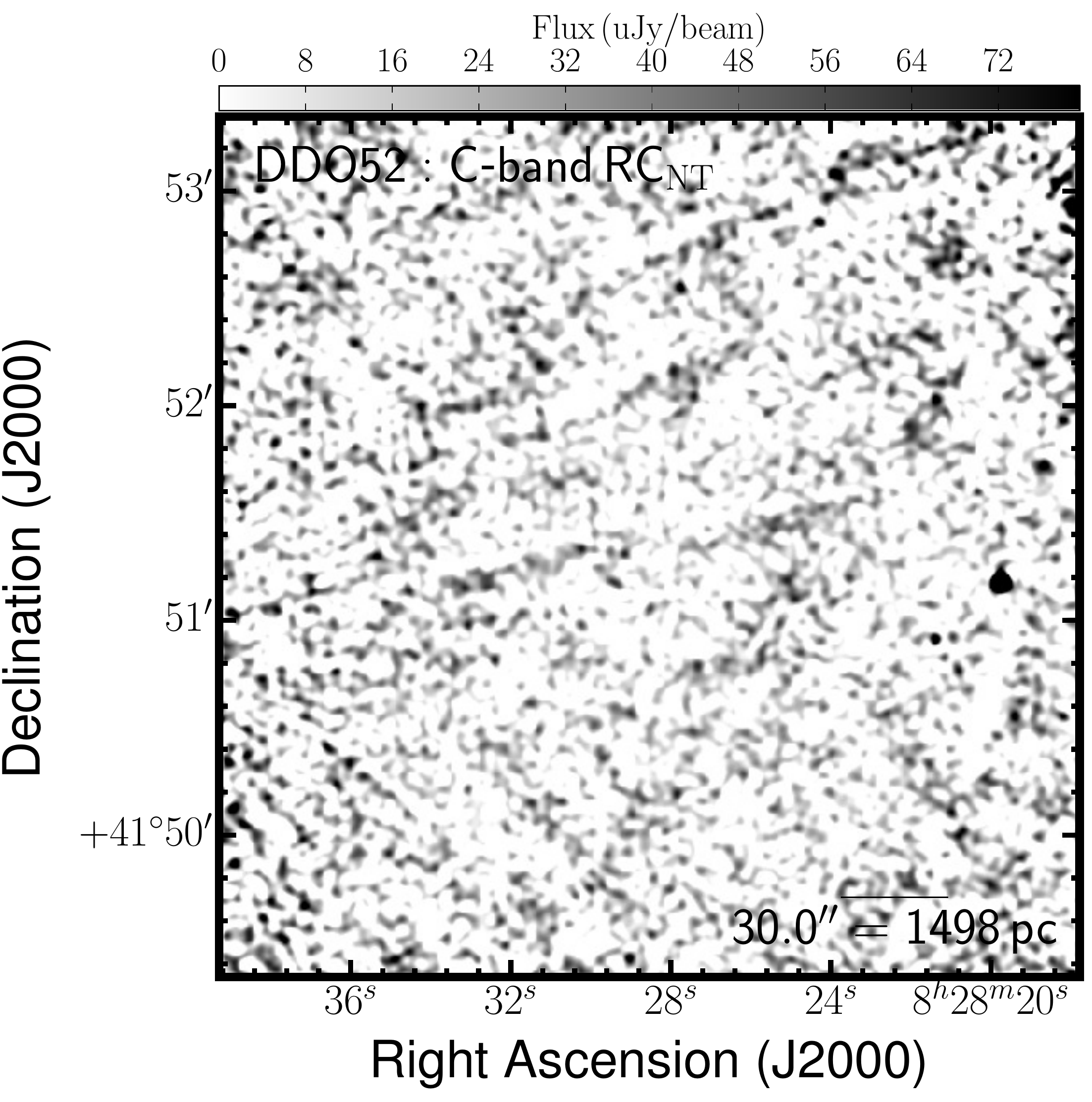} & \ 
    \includegraphics[width=0.31\linewidth,clip]{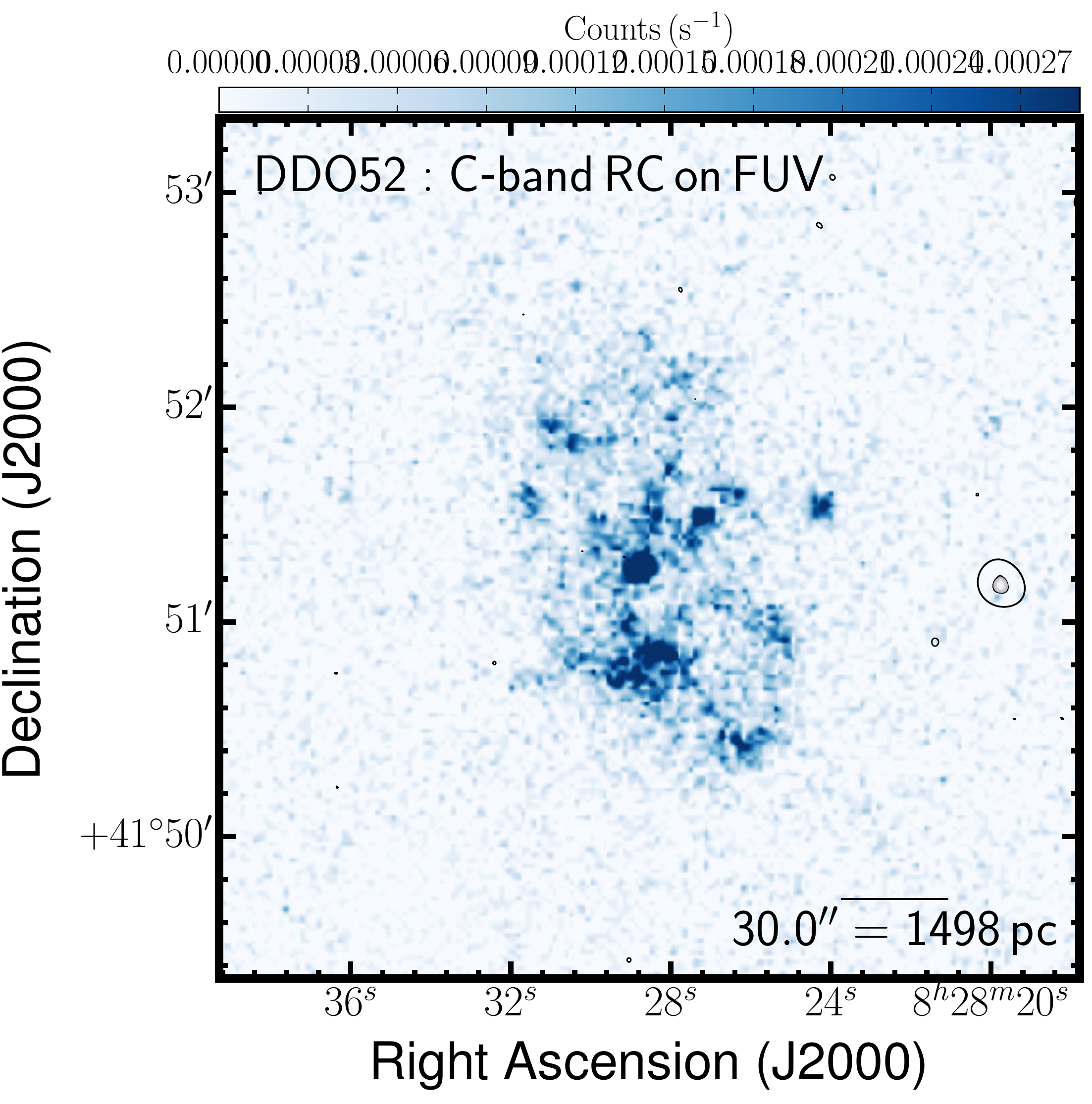} \\
    \includegraphics[width=0.31\linewidth,clip]{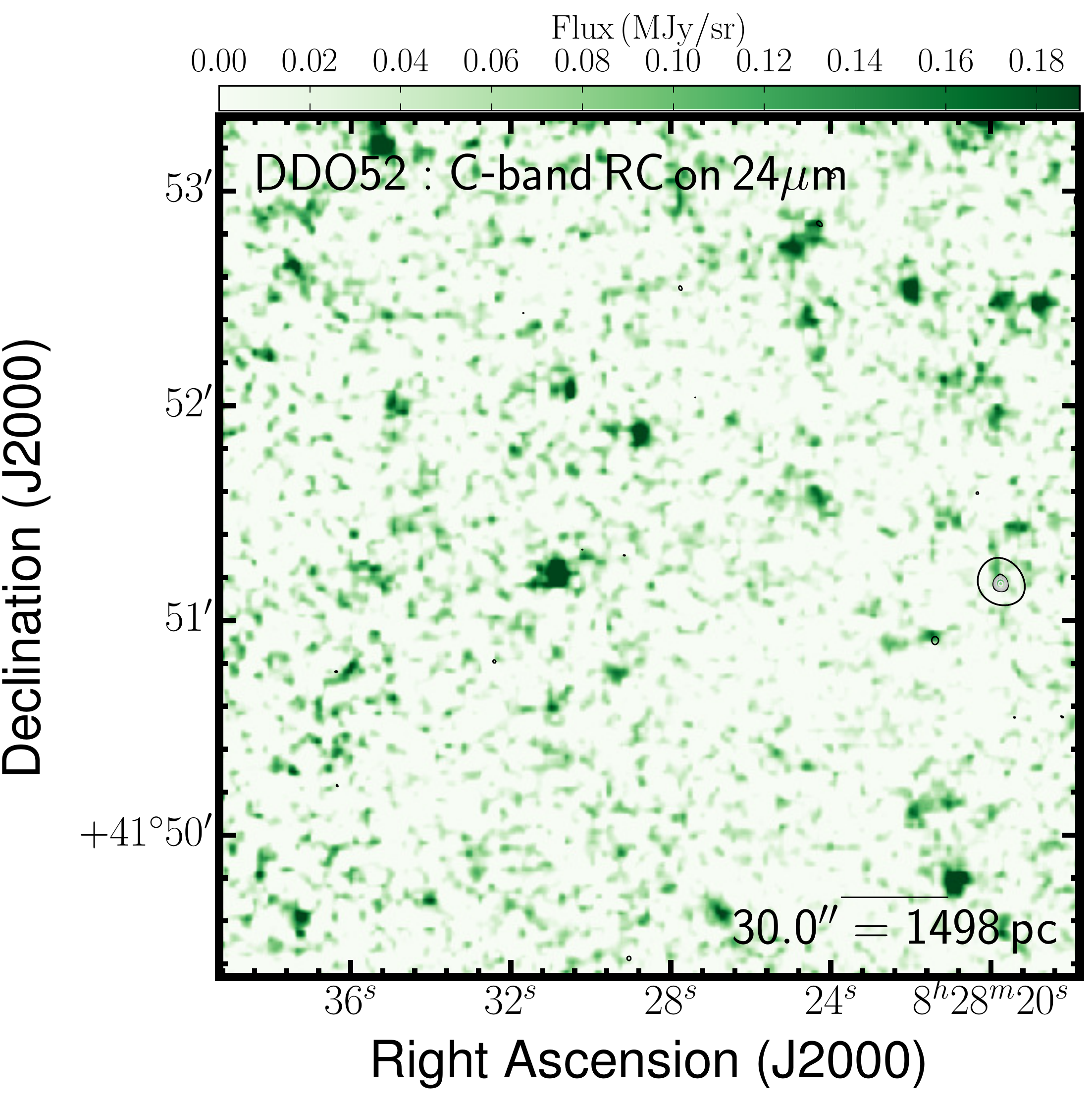} & \ 
    \includegraphics[width=0.31\linewidth,clip]{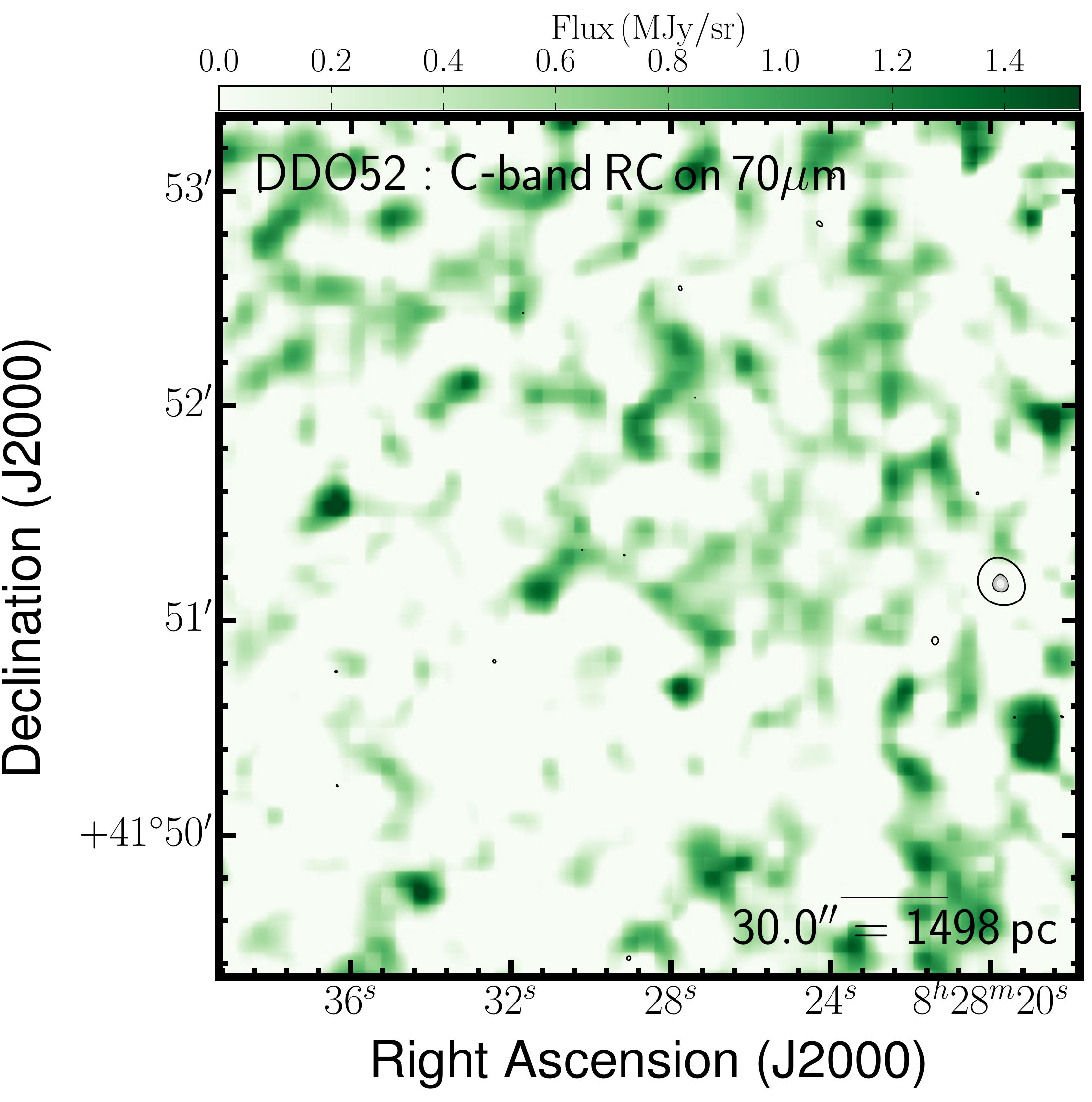} & \ 
    \includegraphics[width=0.31\linewidth,clip]{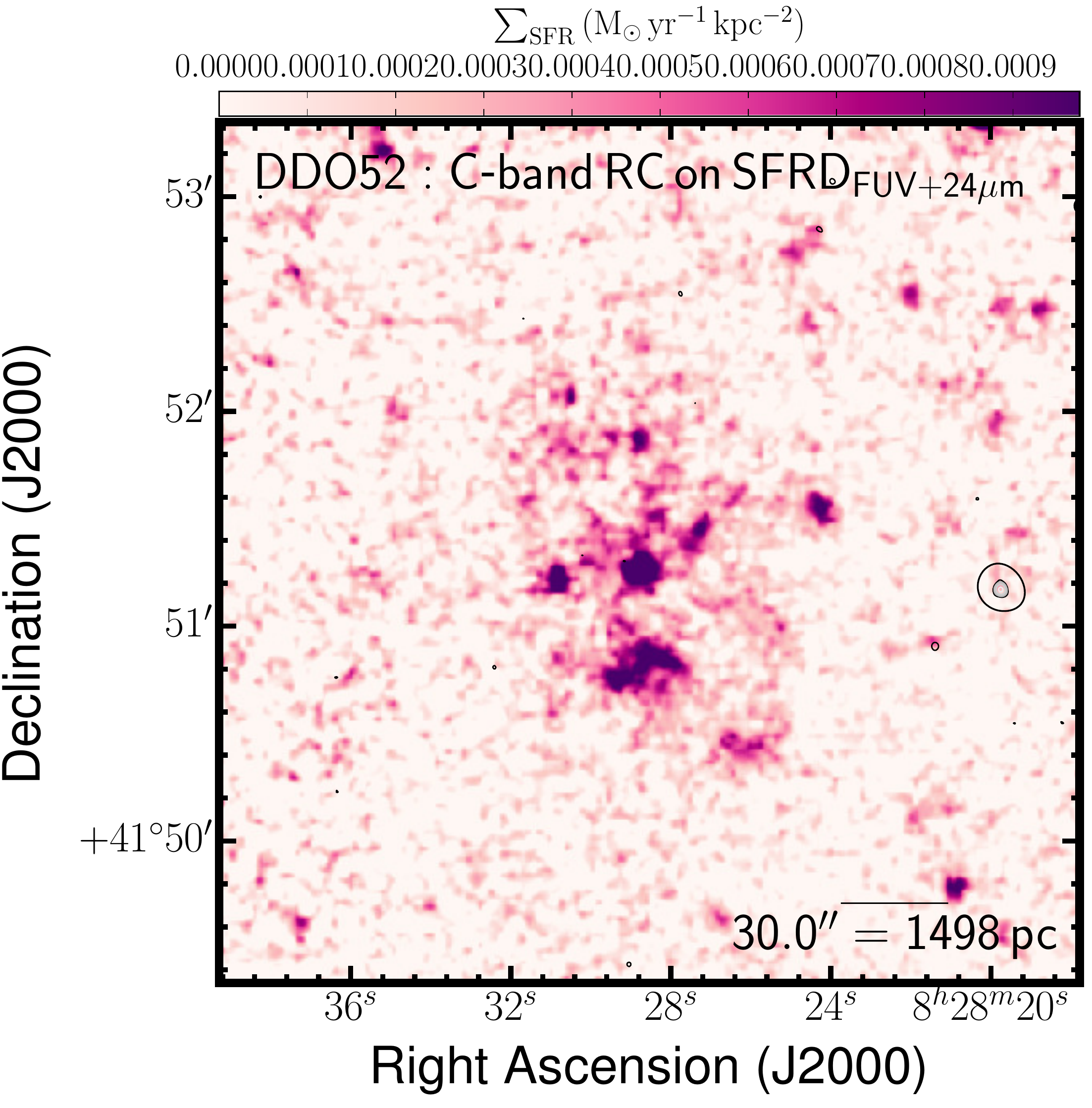} \\
  \end{tabular}
\caption[DDO\,52 images: RC, IR, optical, and FUV]{Multi-wavelength coverage of DDO 52 displaying a $4.0^\prime \times 4.0^\prime$ area. We show total RC flux density at the native resolution (top-left) and again with contours (top-centre). The RC contours are superposed on ancillary LITTLE THINGS images where possible: \halpha\ (middle-left); \RCNT\ obtained by subtracting the expected \RCT\ based on the \halpha-\RCT\ scaling factor of \cite{Deeg1997} from the total RC; {\em GALEX} FUV (middle-right); {\em Spitzer} 24\micron\ (bottom-left); {\em Spitzer} 70\micron\ (bottom-centre); FUV$+24{\rm \mu m}$--inferred SFRD from \citealp{Leroy2012} (bottom-right). We also show the RC that was isolated by the RC--based masking technique (top-right).}
  \label{figure:ddo52Cc_maps}
\end{figure}

\begin{figure}
  \begin{tabular}{ccc}
    \includegraphics[width=0.31\linewidth,clip]{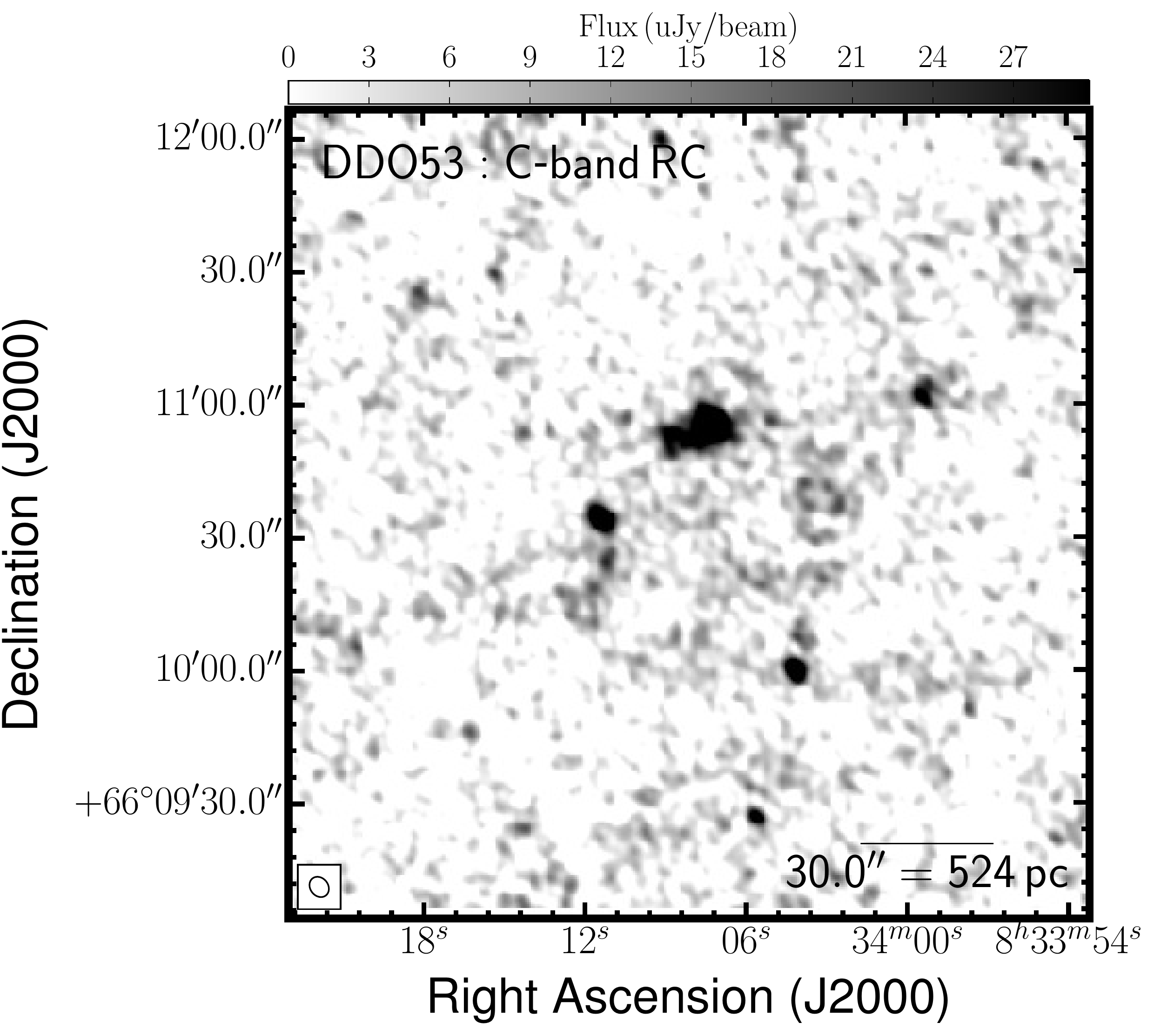} & \ 
    \includegraphics[width=0.31\linewidth,clip]{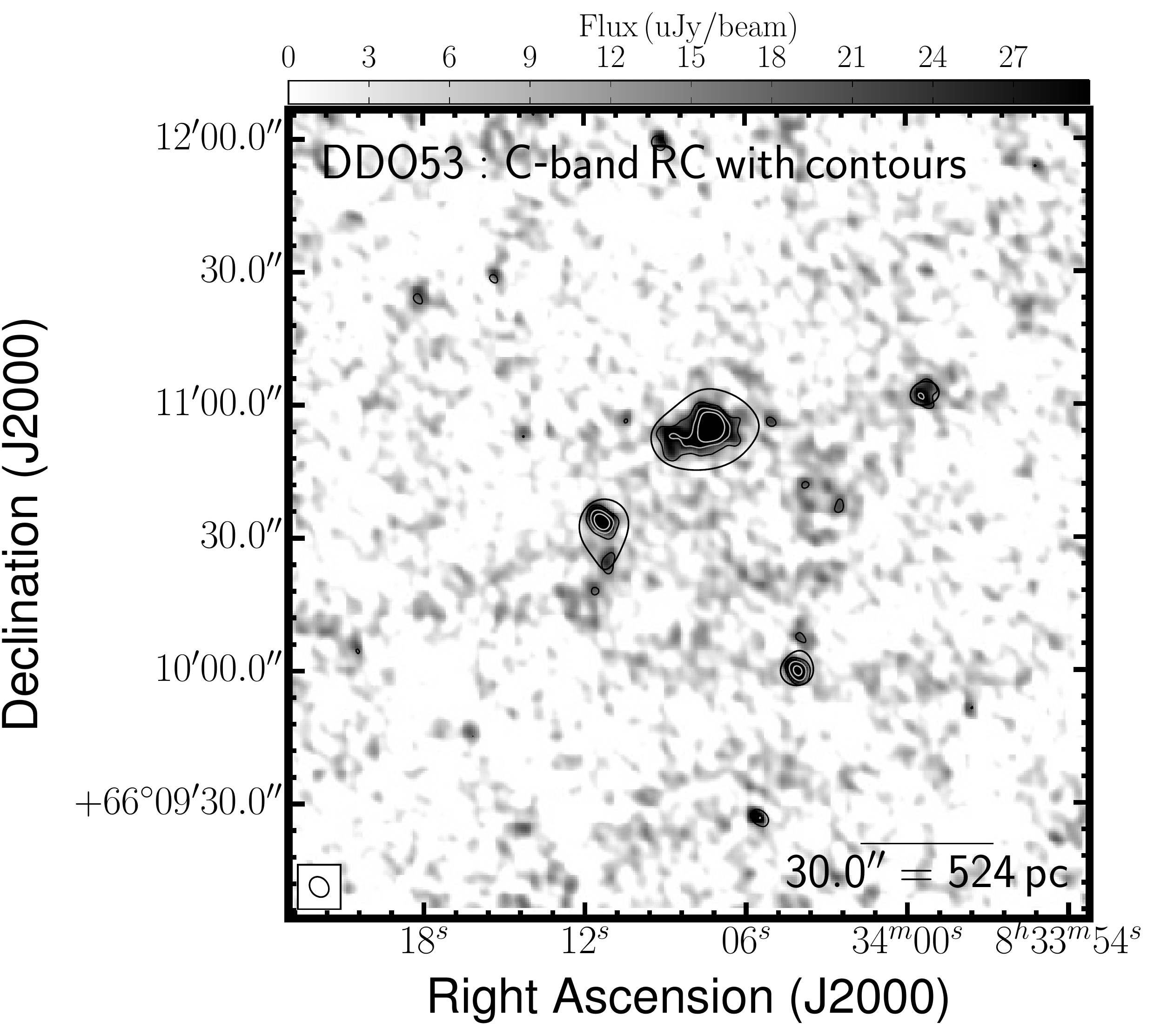} & \ 
    \includegraphics[width=0.31\linewidth,clip]{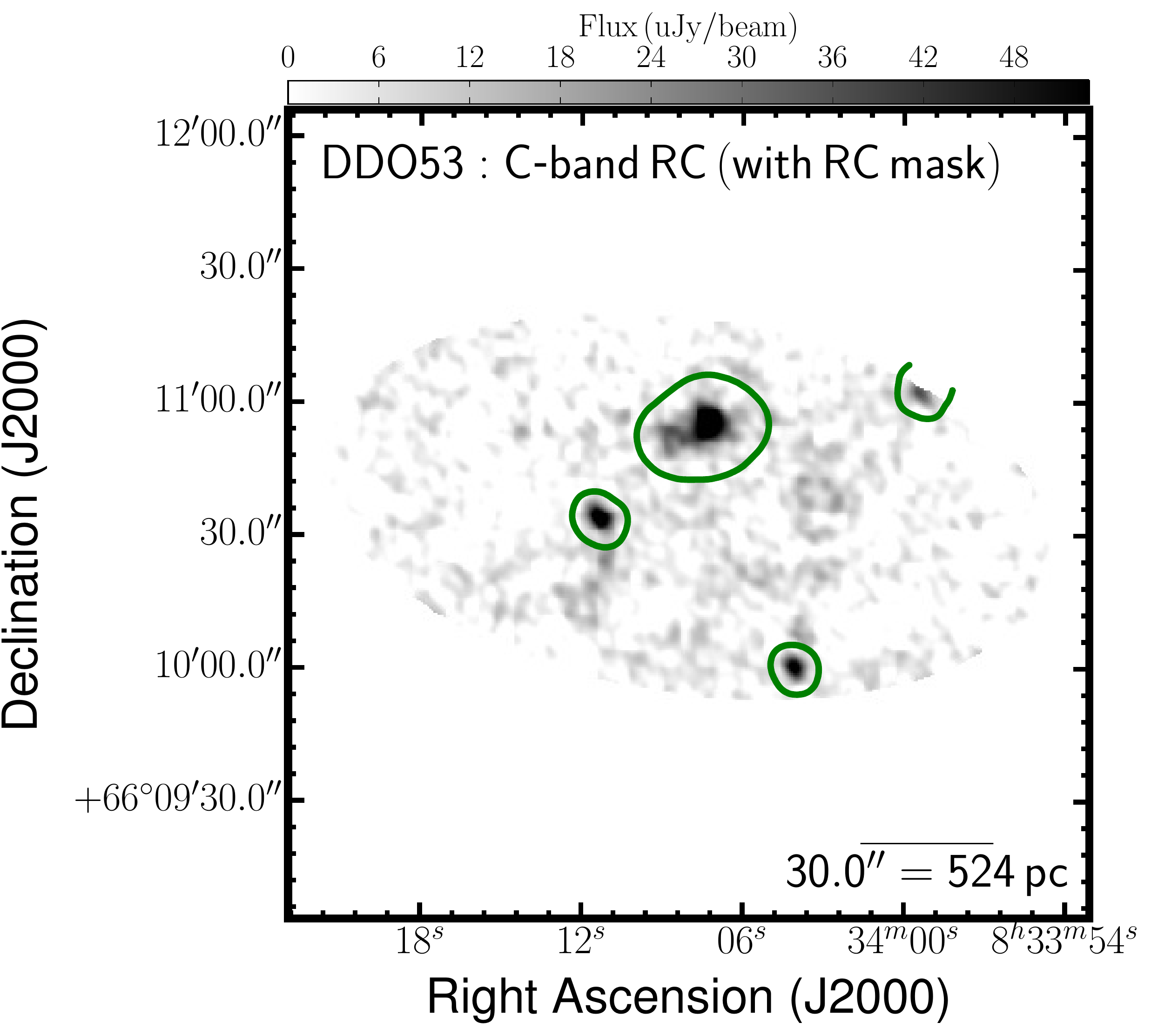} \\
    \includegraphics[width=0.31\linewidth,clip]{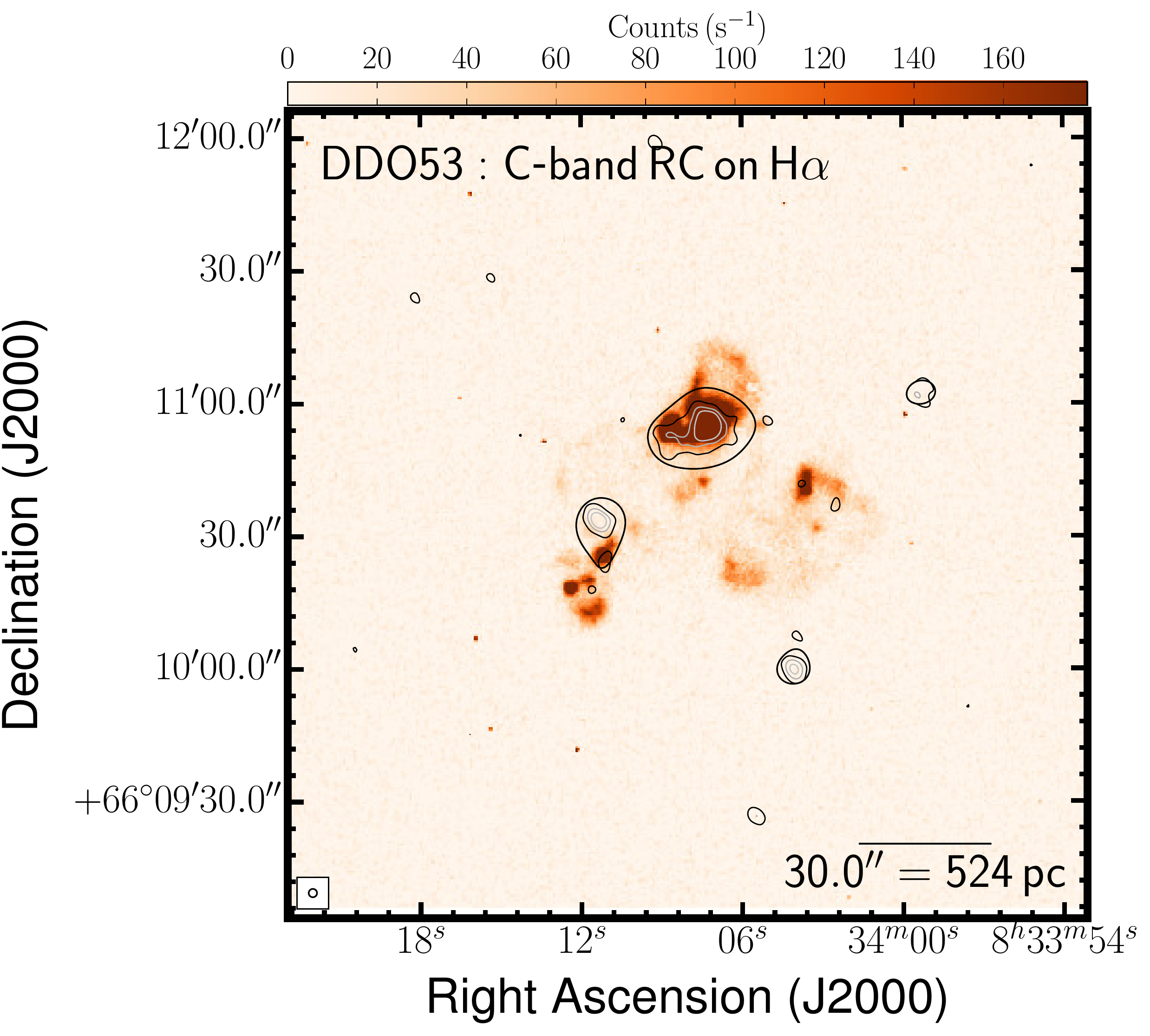} & \ 
    \includegraphics[width=0.31\linewidth,clip]{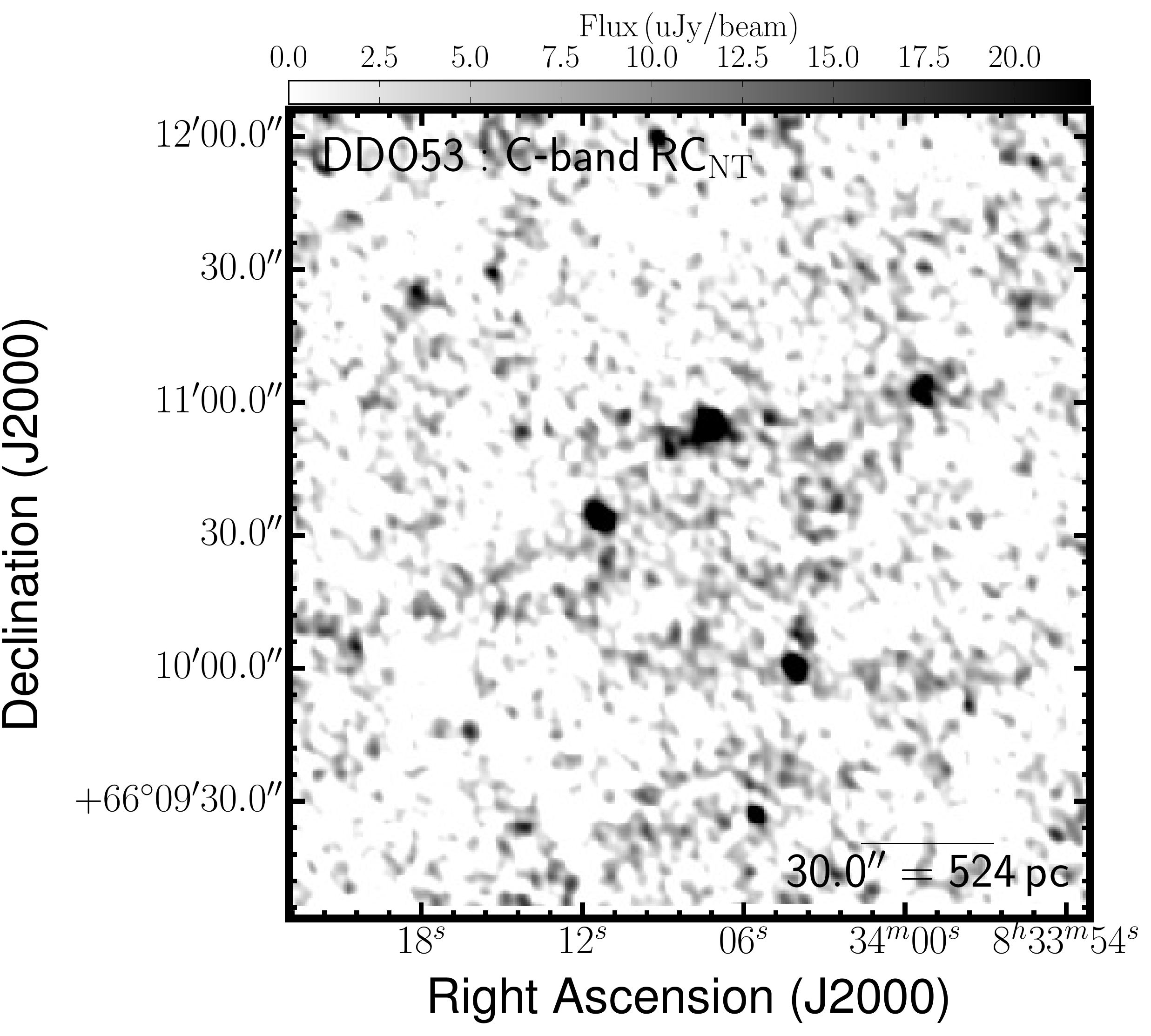} & \ 
    \includegraphics[width=0.31\linewidth,clip]{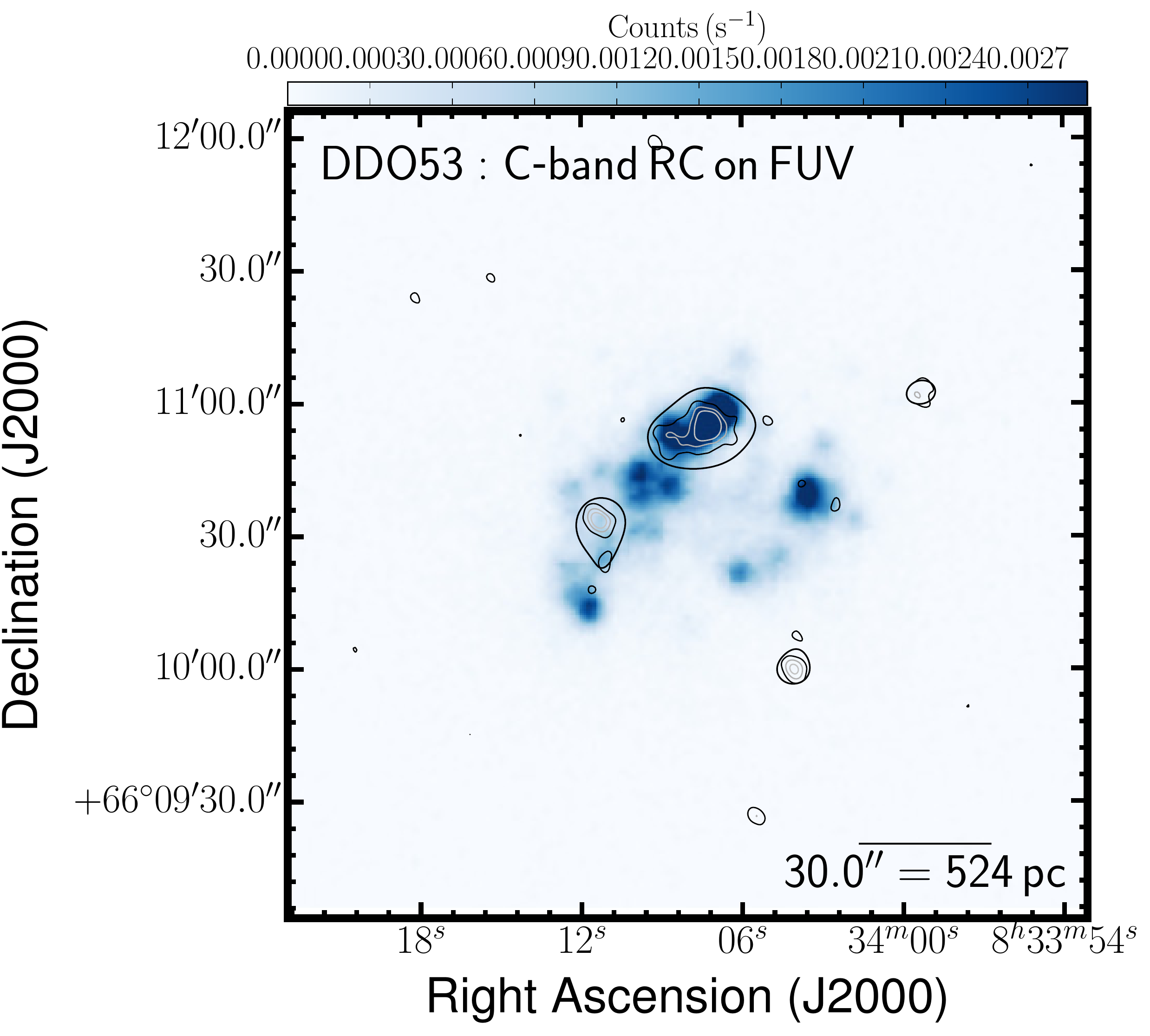} \\
    \includegraphics[width=0.31\linewidth,clip]{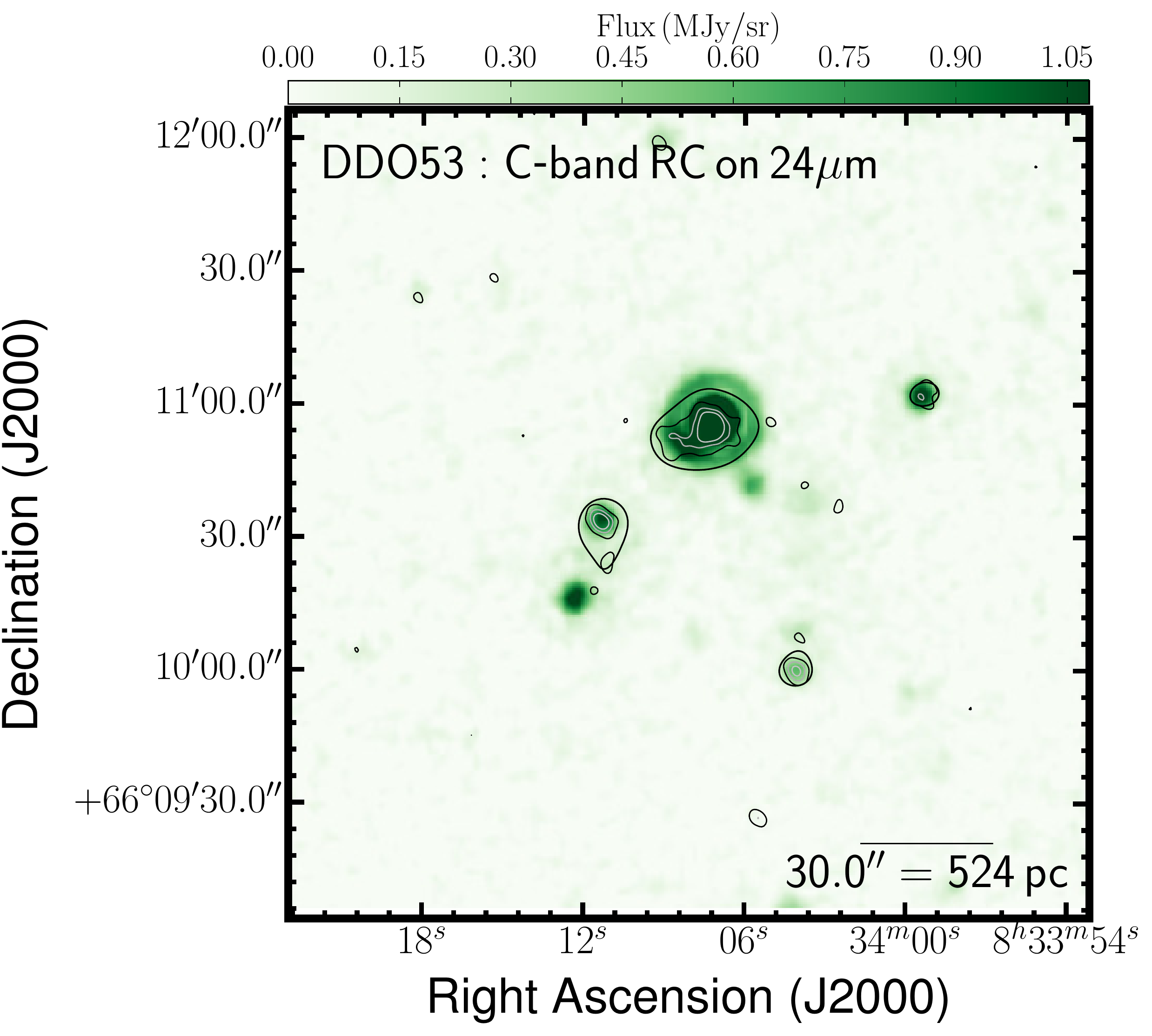} & \ 
    \includegraphics[width=0.31\linewidth,clip]{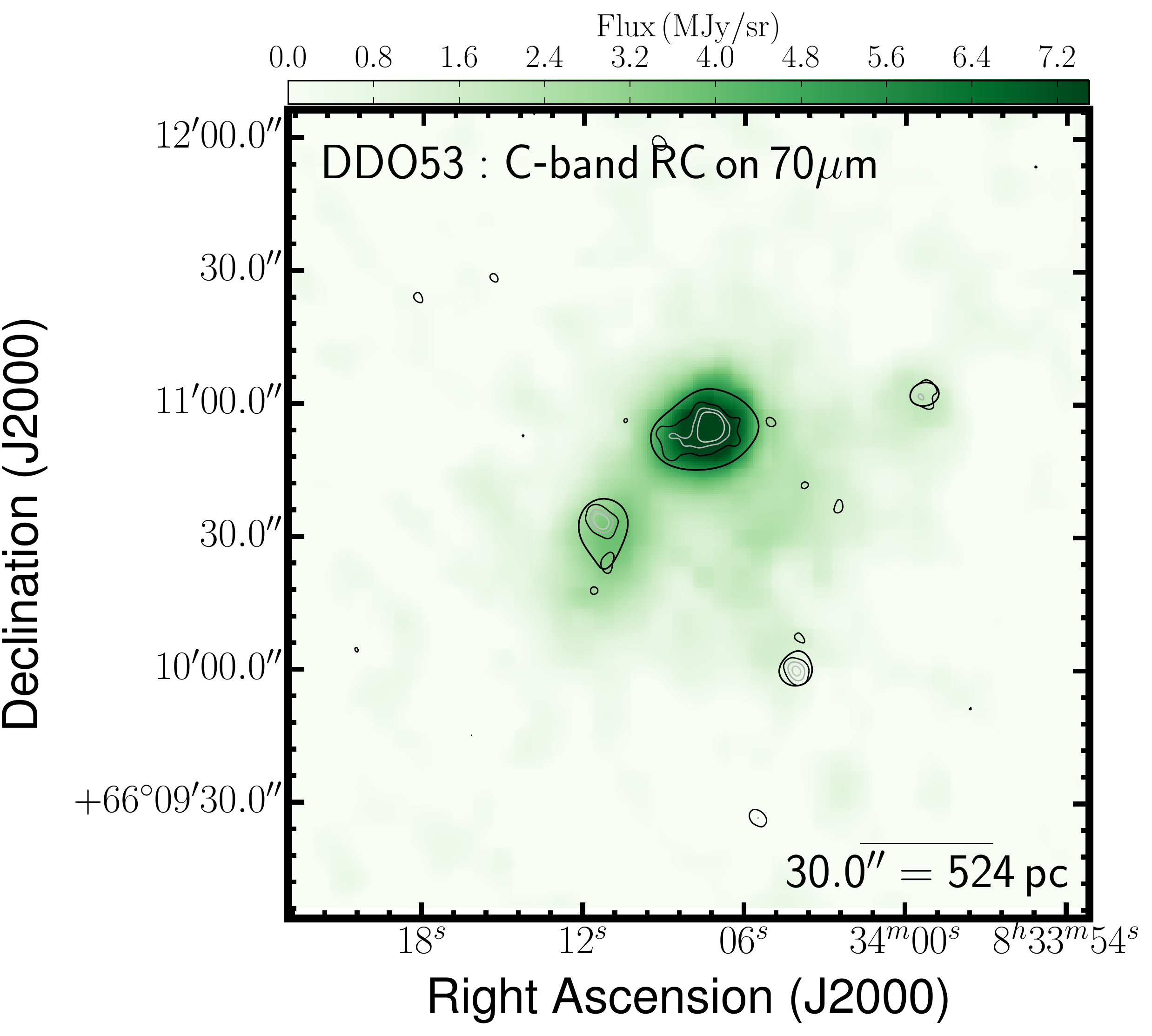} & \ 
    \includegraphics[width=0.31\linewidth,clip]{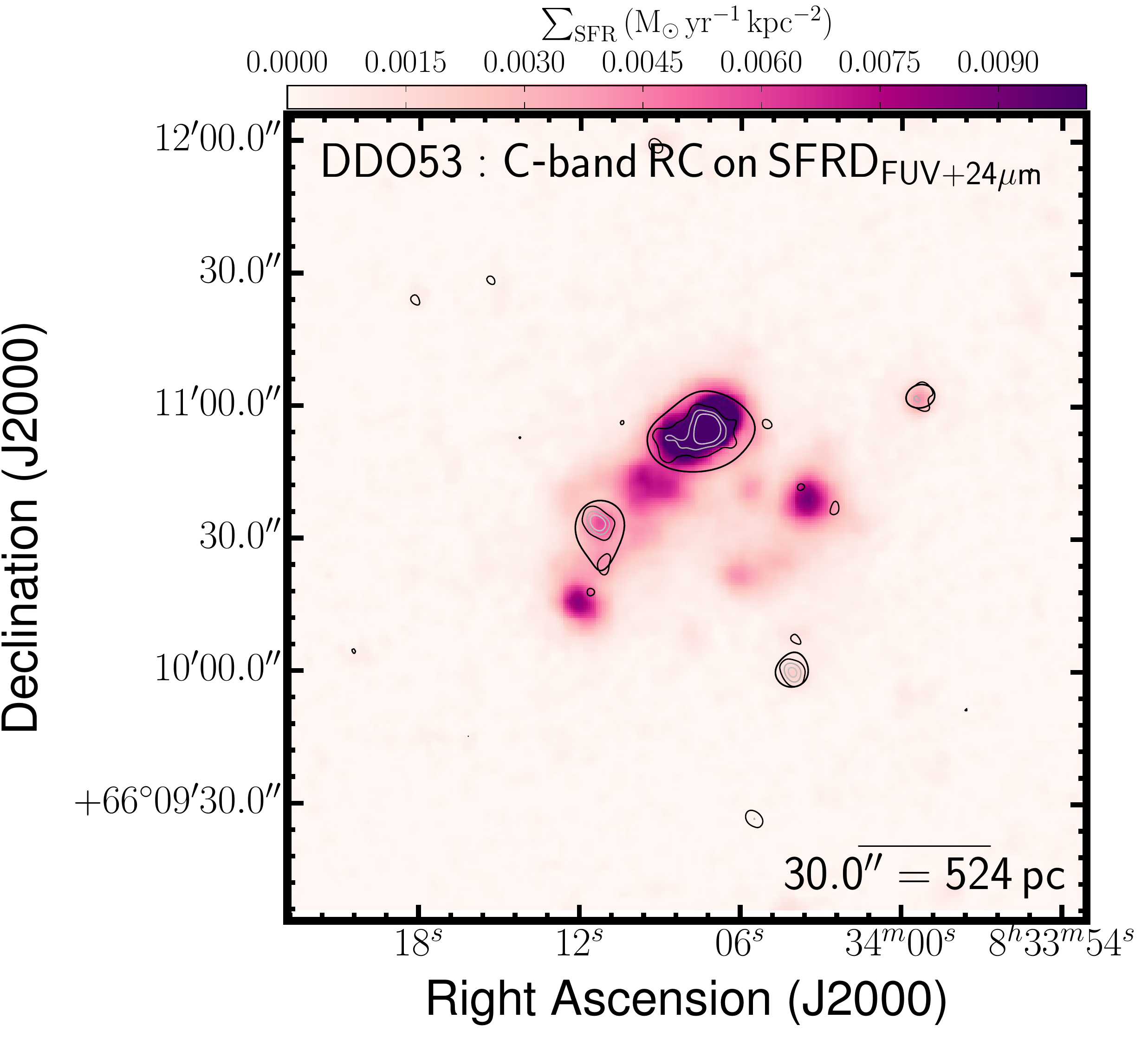} \\
  \end{tabular}
\caption[DDO\,53 images: RC, IR, optical, and FUV]{Multi-wavelength coverage of DDO 53 displaying a $3.6^\prime \times 3.6^\prime$ area. We show total RC flux density at the native resolution (top-left) and again with contours (top-centre). The RC contours are superposed on ancillary LITTLE THINGS images where possible: \halpha\ (middle-left); \RCNT\ obtained by subtracting the expected \RCT\ based on the \halpha-\RCT\ scaling factor of \cite{Deeg1997} from the total RC; {\em GALEX} FUV (middle-right); {\em Spitzer} 24\micron\ (bottom-left); {\em Spitzer} 70\micron\ (bottom-centre); FUV$+24{\rm \mu m}$--inferred SFRD from \citealp{Leroy2012} (bottom-right). We also show the RC that was isolated by the RC--based masking technique (top-right).}
  \label{figure:ddo53Cc_maps}
\end{figure}

\begin{figure}
  \begin{tabular}{ccc}
    \includegraphics[width=0.31\linewidth,clip]{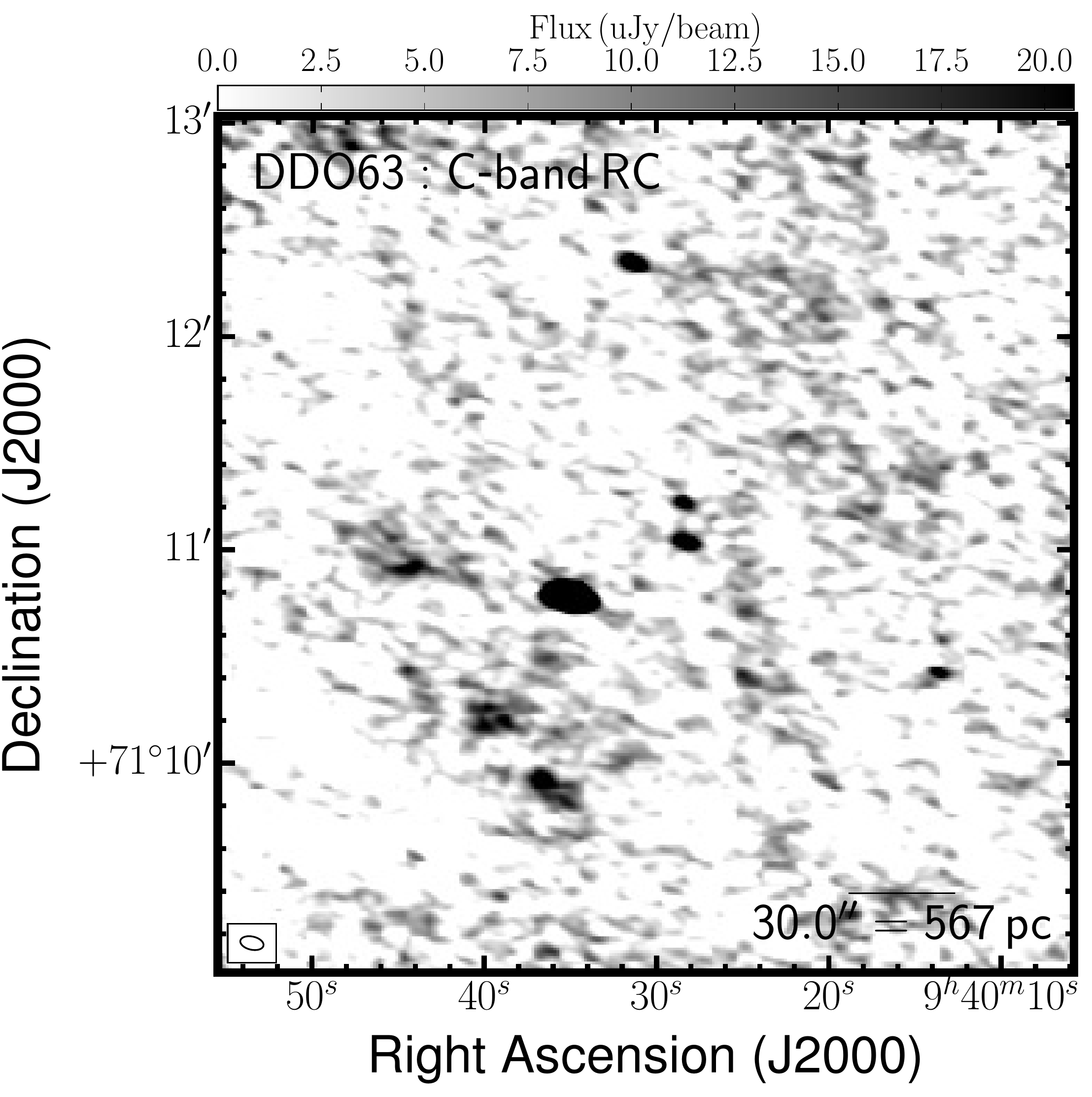} & \ 
    \includegraphics[width=0.31\linewidth,clip]{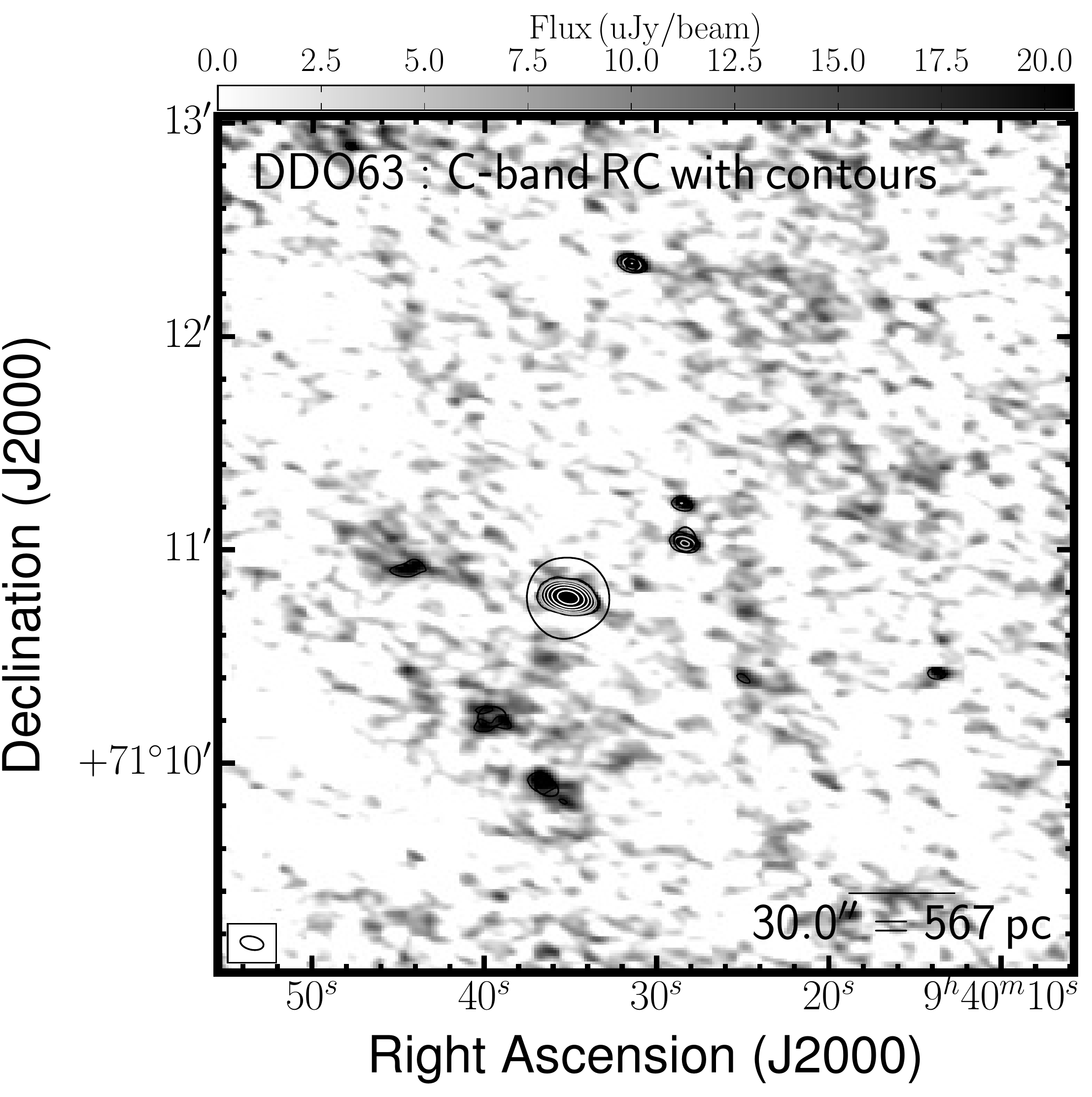} & \ 
    \includegraphics[width=0.31\linewidth,clip]{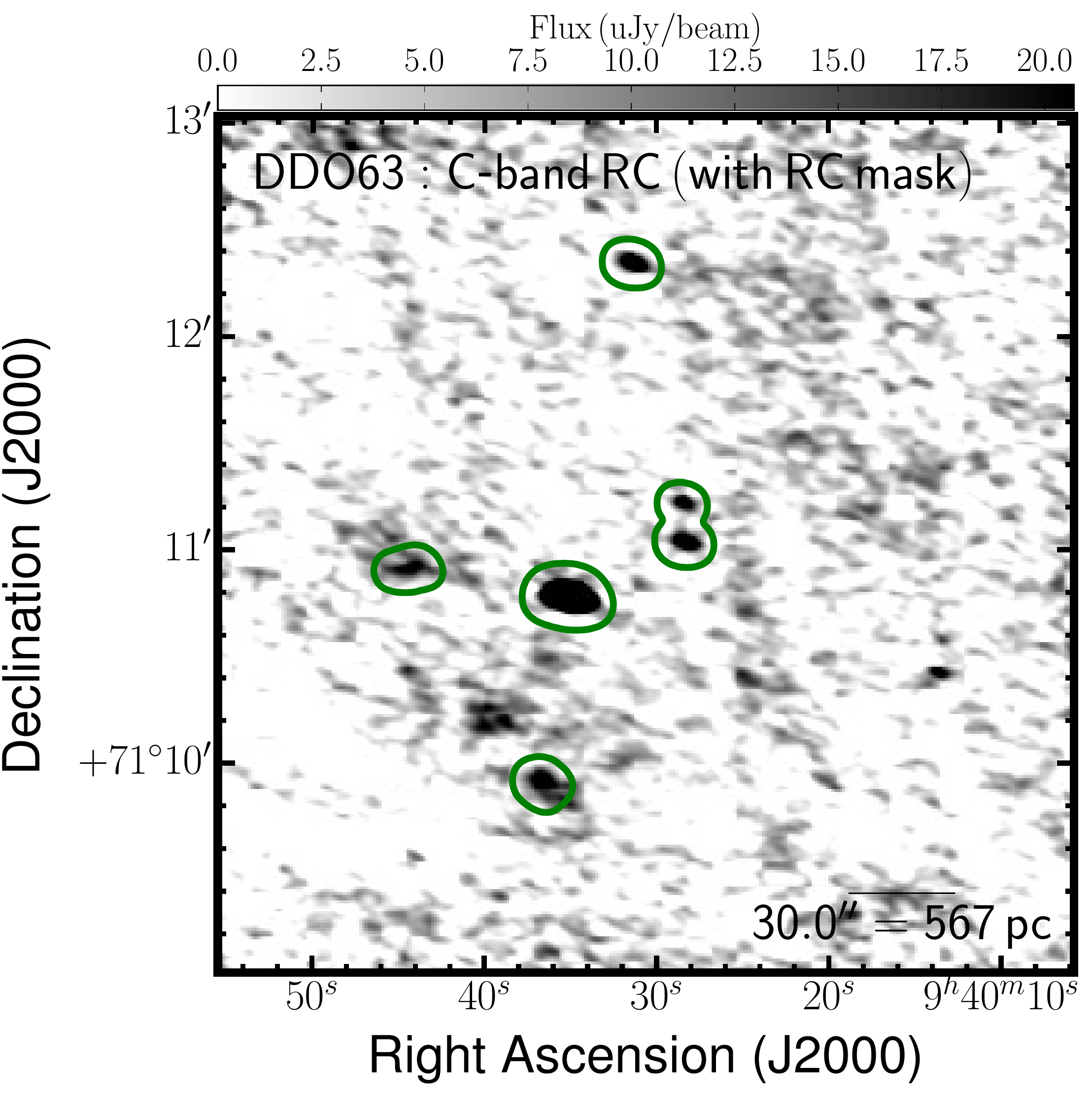} \\
    \includegraphics[width=0.31\linewidth,clip]{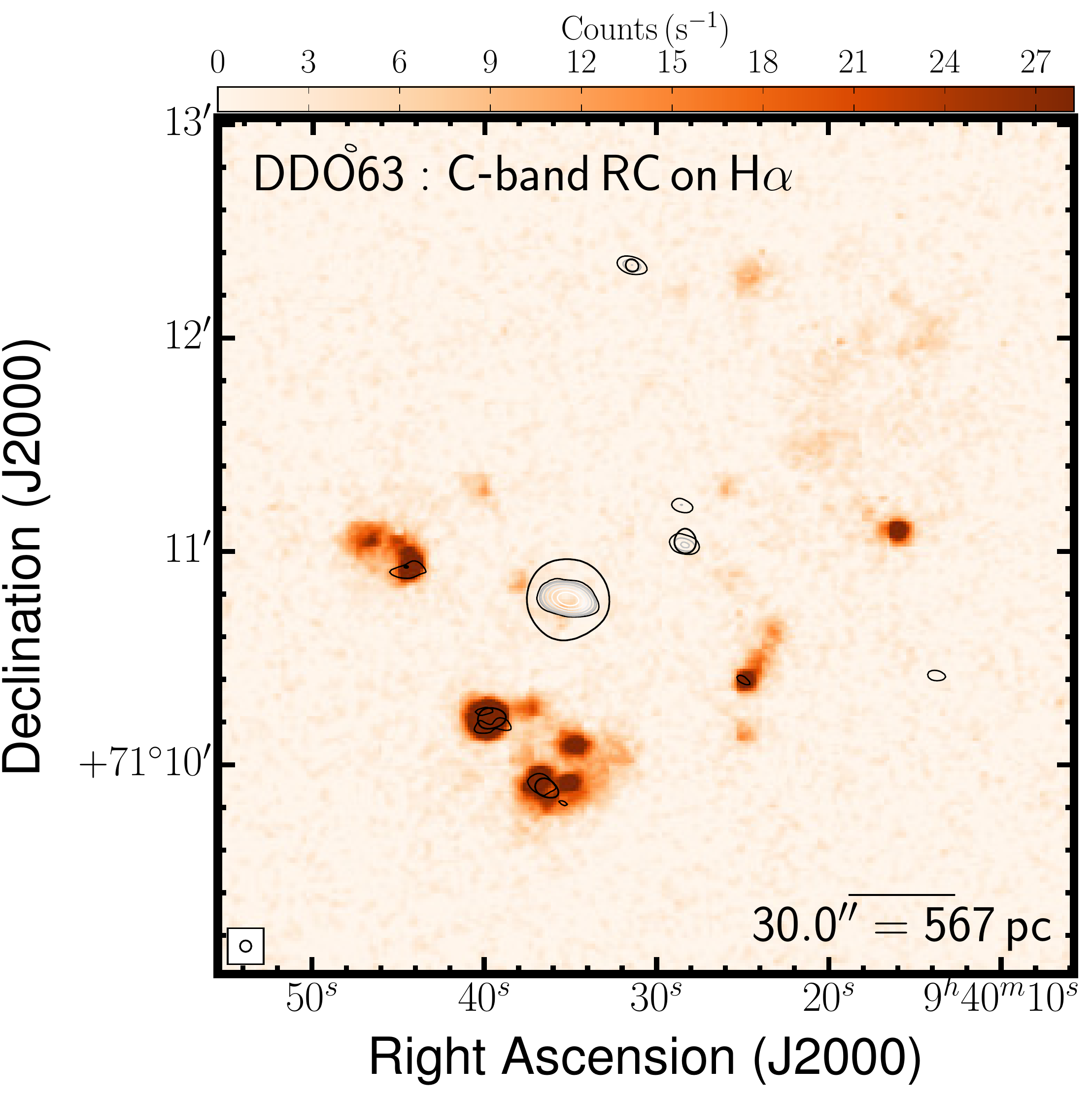} & \ 
    \includegraphics[width=0.31\linewidth,clip]{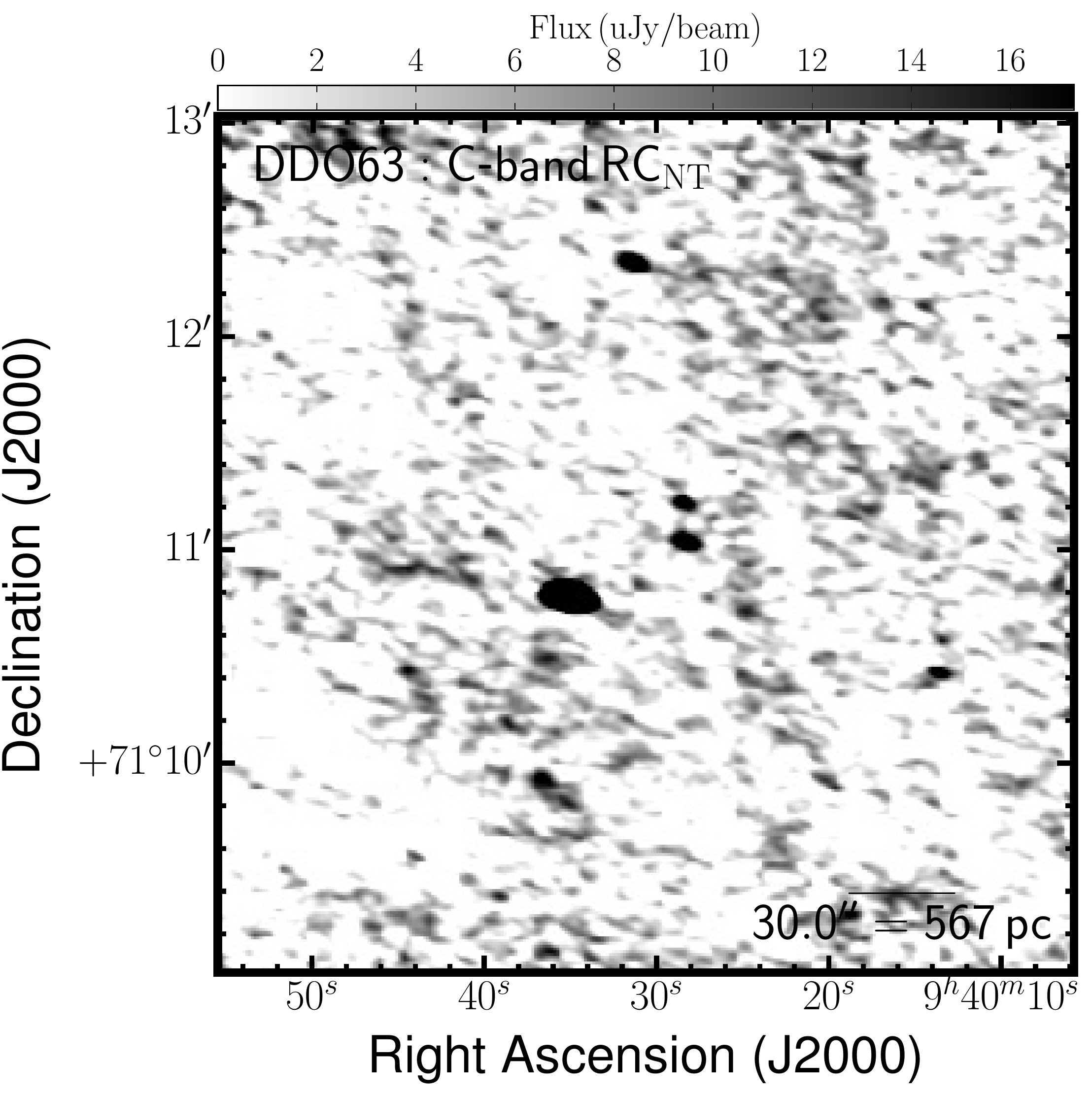} & \ 
    \includegraphics[width=0.31\linewidth,clip]{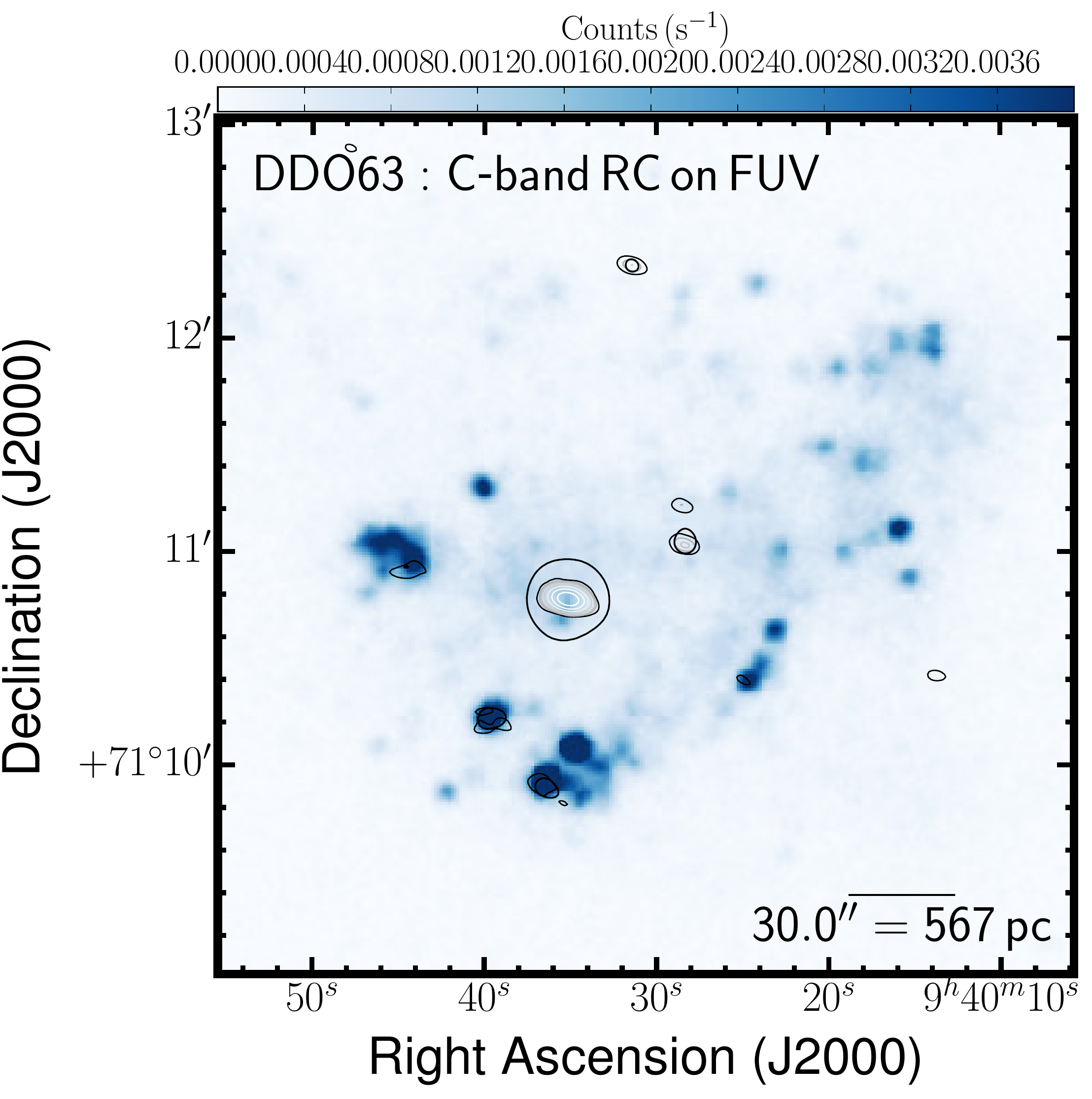} \\
    \includegraphics[width=0.31\linewidth,clip]{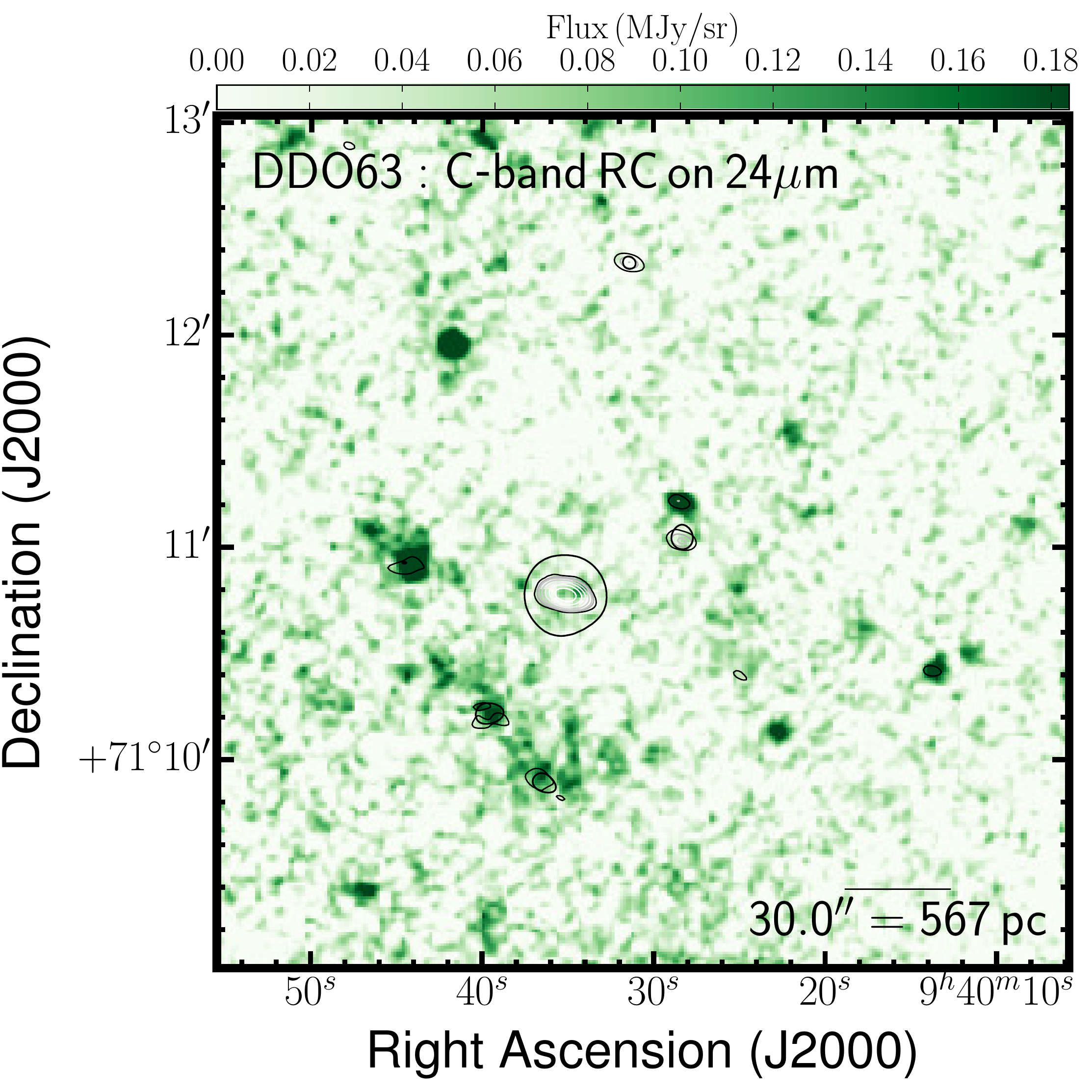} & \ 
    \includegraphics[width=0.31\linewidth,clip]{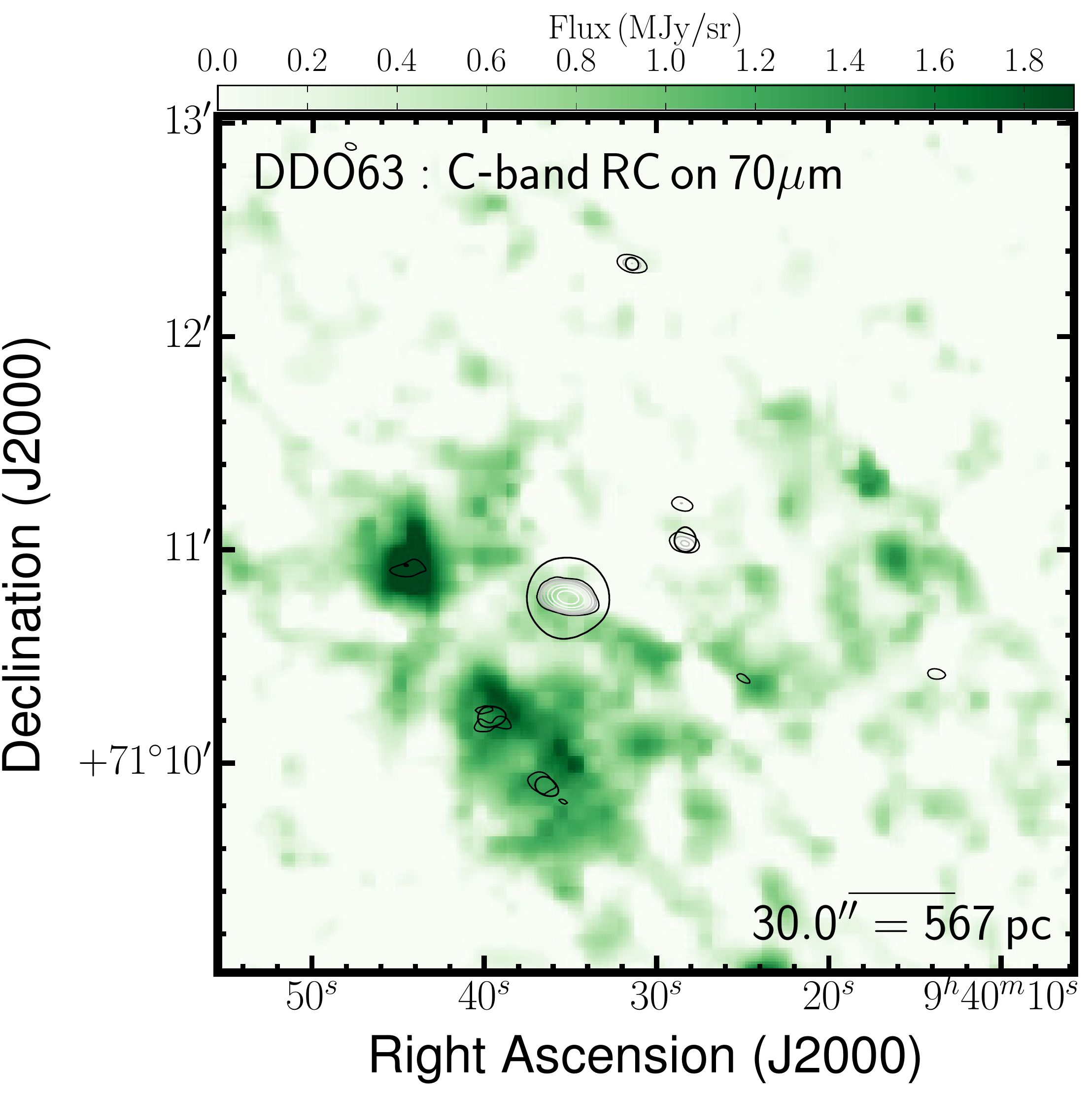} & \ 
    \includegraphics[width=0.31\linewidth,clip]{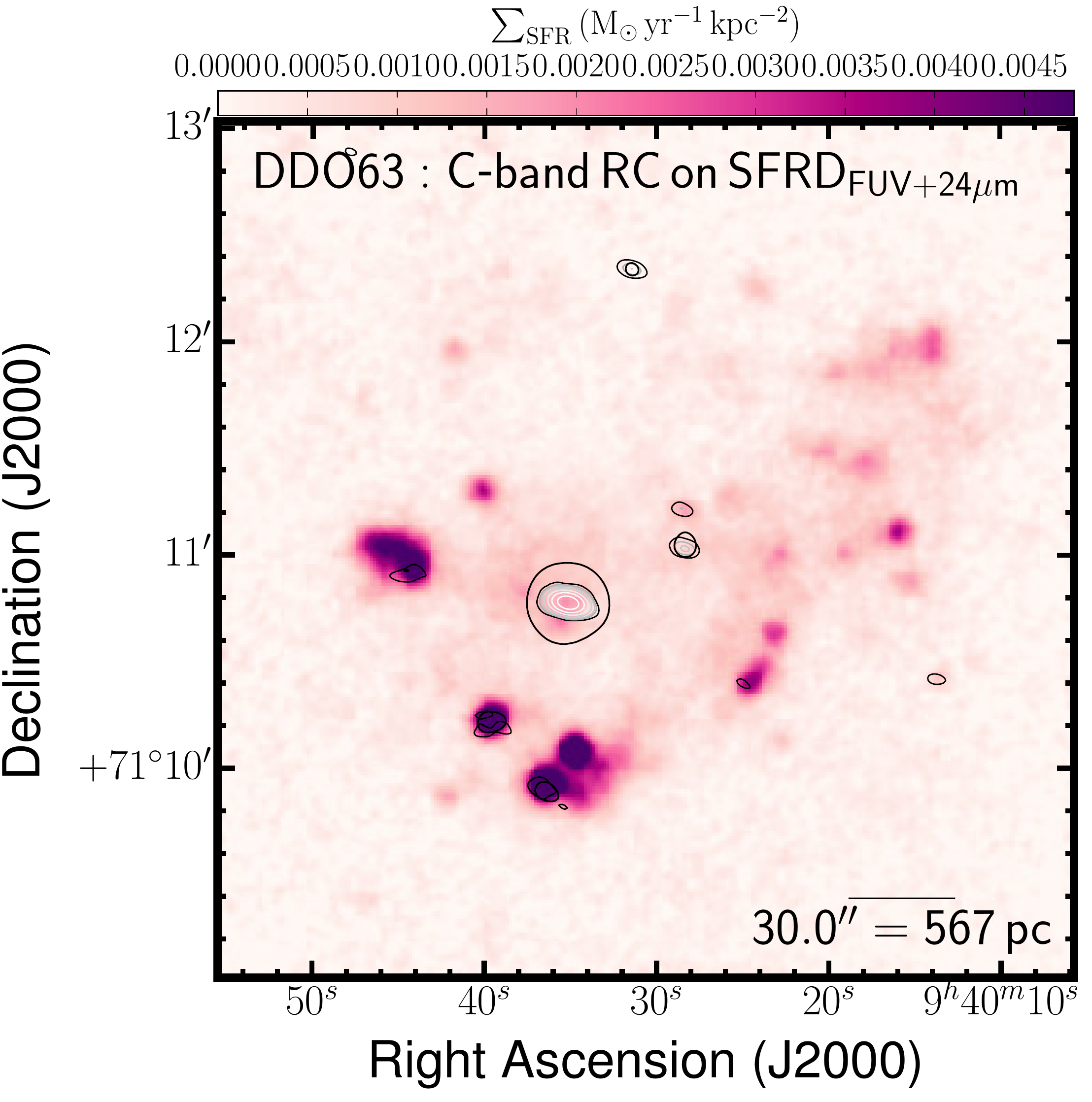} \\
  \end{tabular}
\caption[DDO\,63 images: RC, IR, optical, and FUV]{Multi-wavelength coverage of DDO 63 displaying a $4.0^\prime \times 4.0^\prime$ area. We show total RC flux density at the native resolution (top-left) and again with contours (top-centre). The RC contours are superposed on ancillary LITTLE THINGS images where possible: \halpha\ (middle-left); \RCNT\ obtained by subtracting the expected \RCT\ based on the \halpha-\RCT\ scaling factor of \cite{Deeg1997} from the total RC; {\em GALEX} FUV (middle-right); {\em Spitzer} 24\micron\ (bottom-left); {\em Spitzer} 70\micron\ (bottom-centre); FUV$+24{\rm \mu m}$--inferred SFRD from \citealp{Leroy2012} (bottom-right). We also show the RC that was isolated by the RC--based masking technique (top-right).}
  \label{figure:ddo63Cc_maps}
\end{figure}

\begin{figure}
  \begin{tabular}{ccc}
    \includegraphics[width=0.31\linewidth,clip]{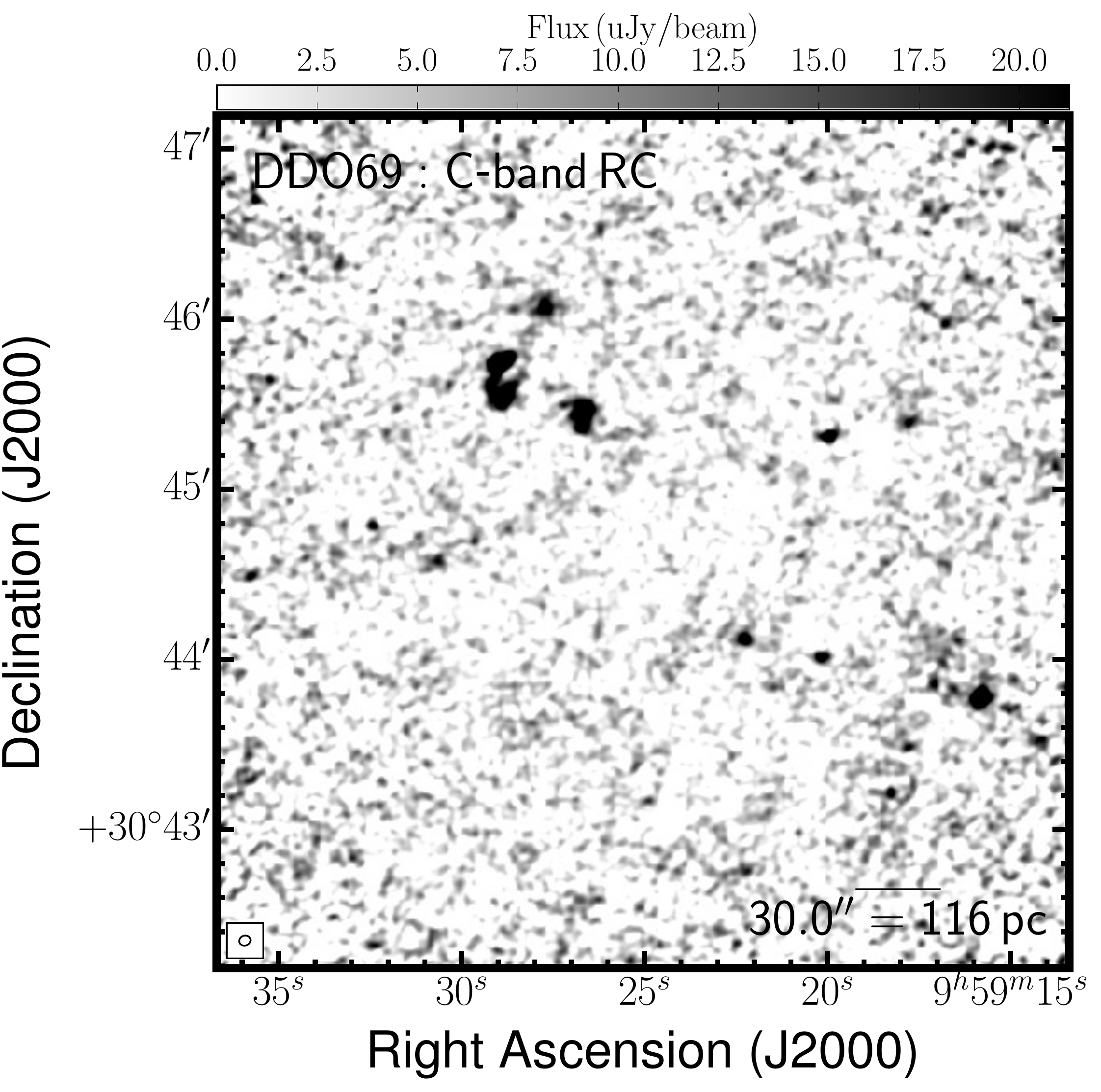} & \ 
    \includegraphics[width=0.31\linewidth,clip]{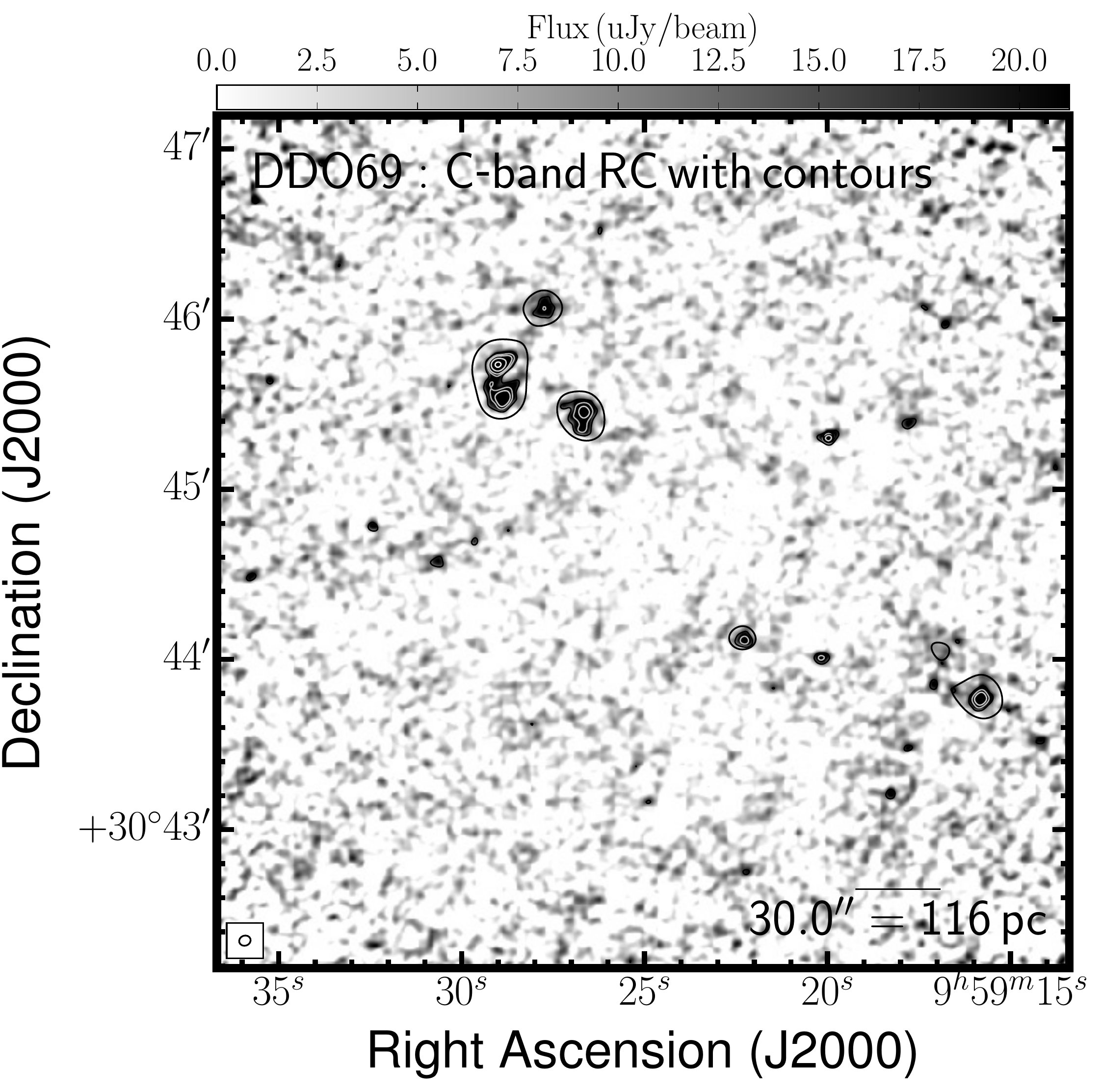} & \ 
    \includegraphics[width=0.31\linewidth,clip]{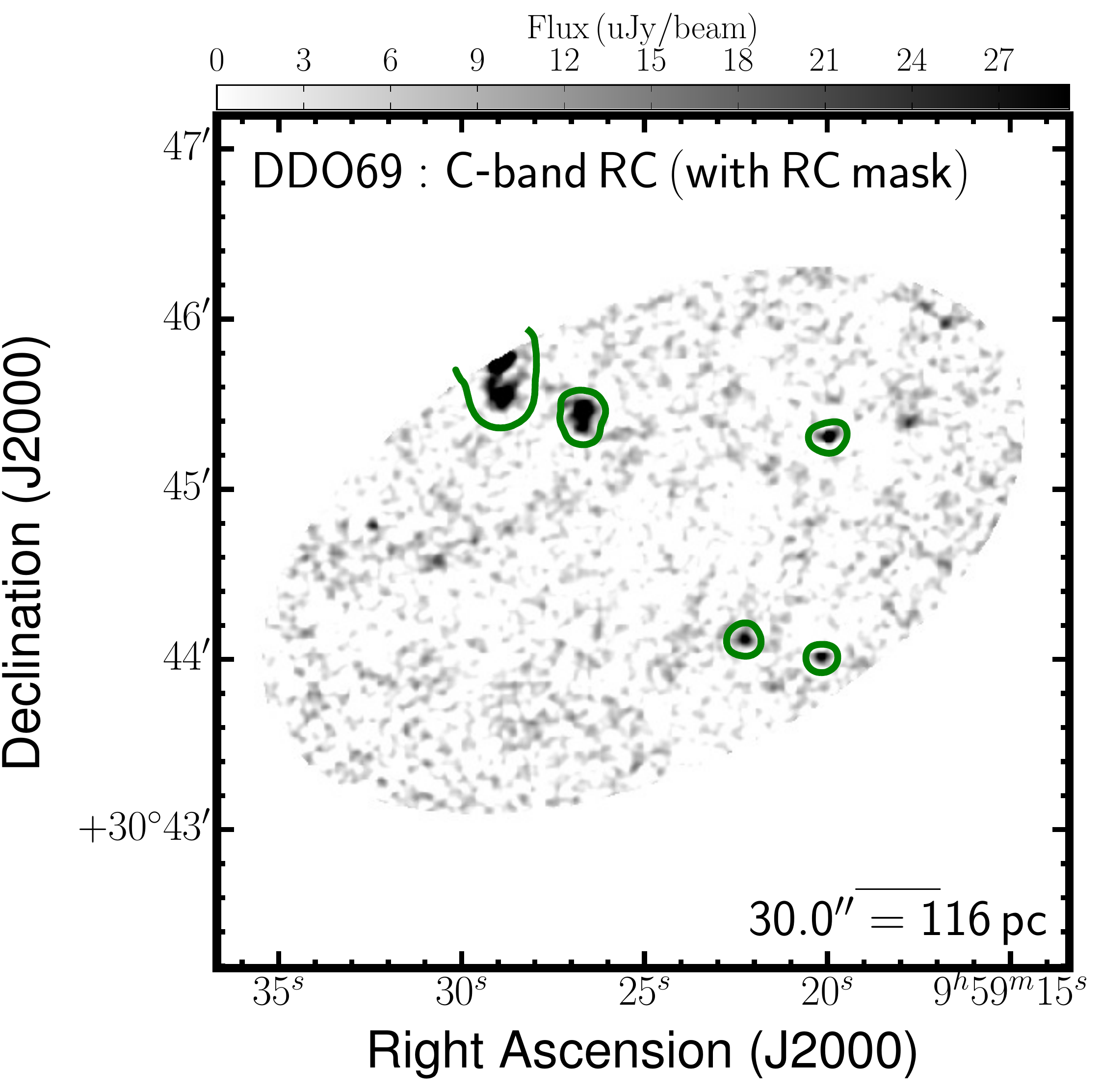} \\
    \includegraphics[width=0.31\linewidth,clip]{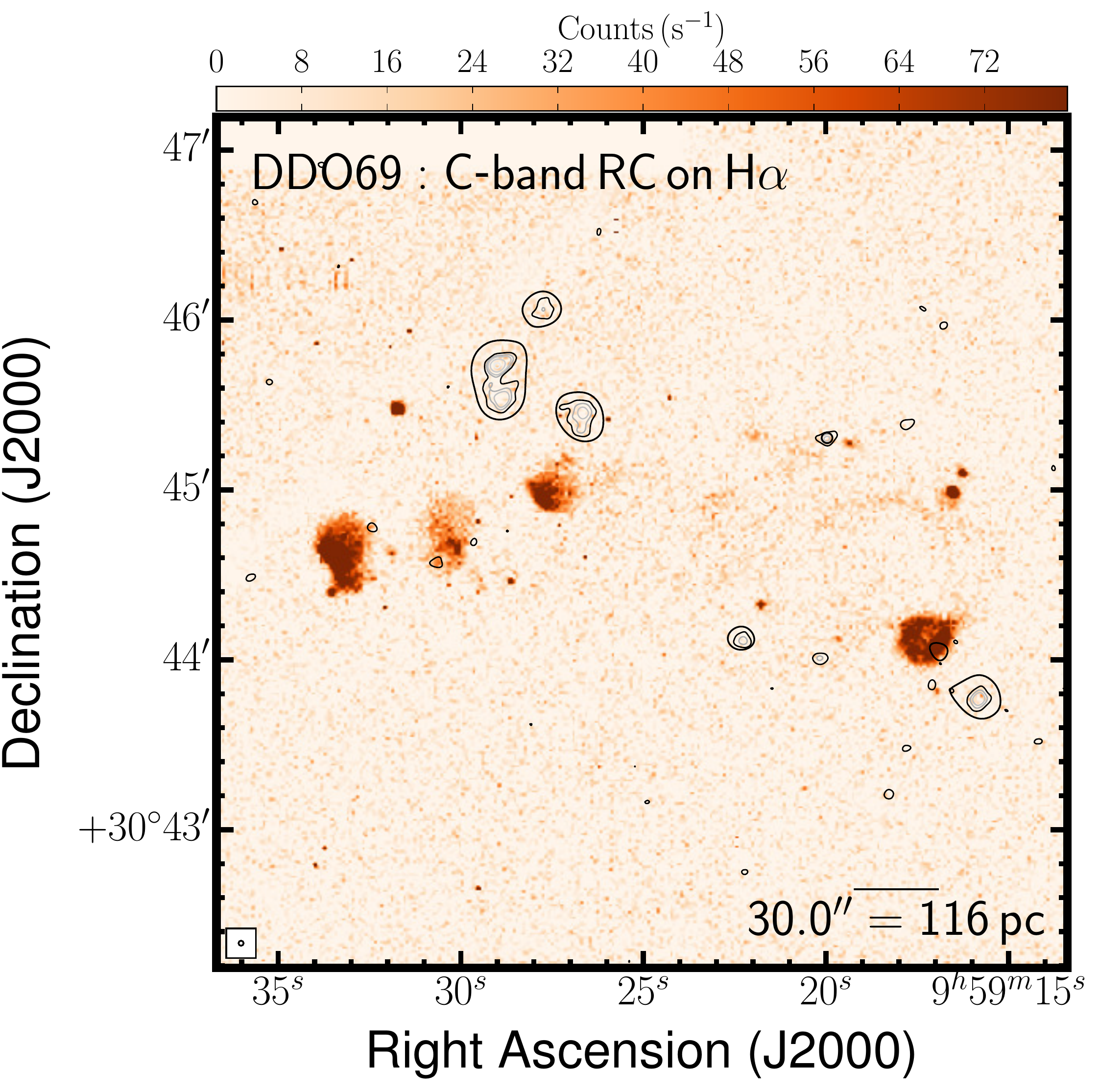} & \ 
    \includegraphics[width=0.31\linewidth,clip]{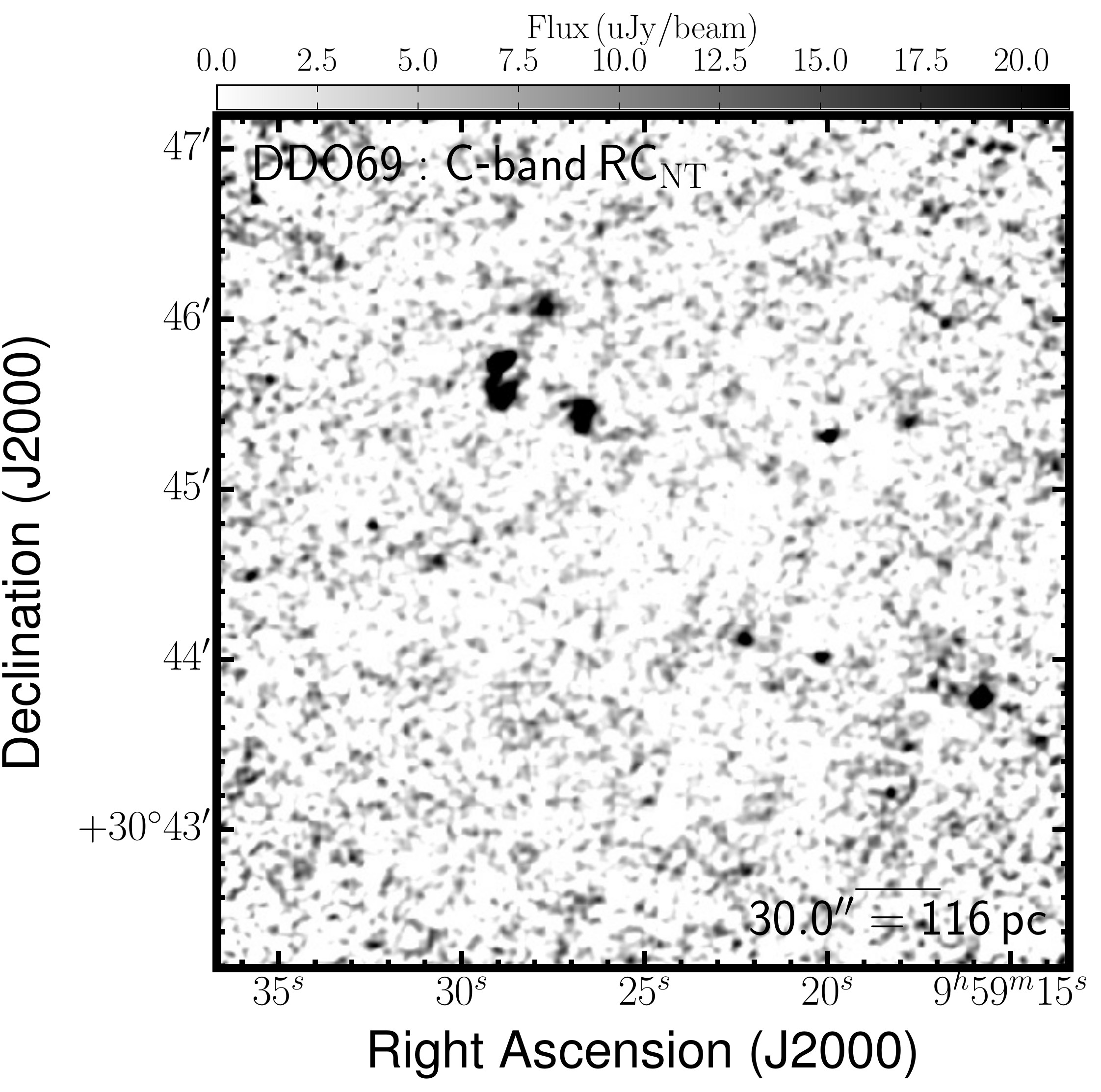} & \ 
    \includegraphics[width=0.31\linewidth,clip]{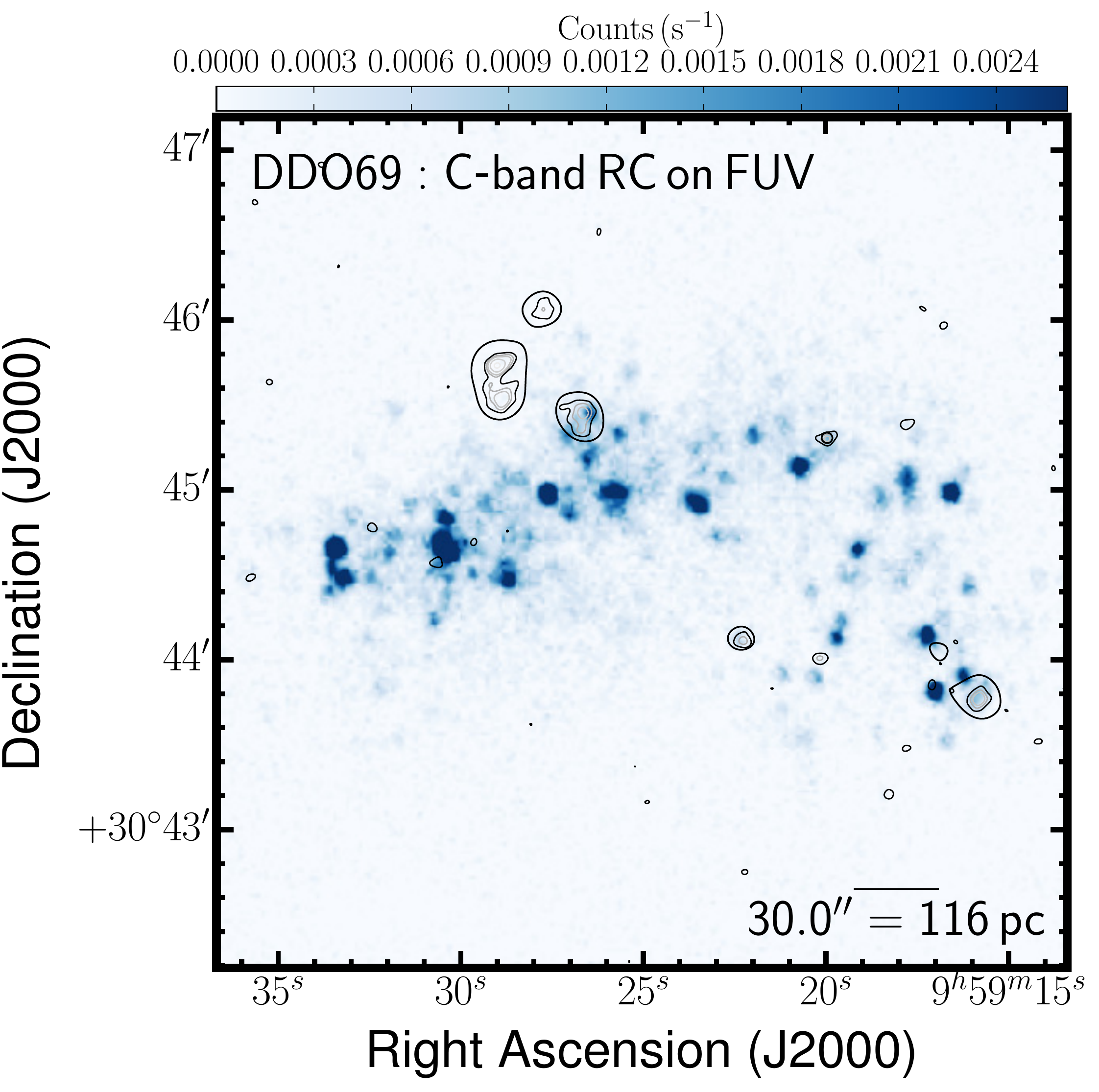} \\
    \includegraphics[width=0.31\linewidth,clip]{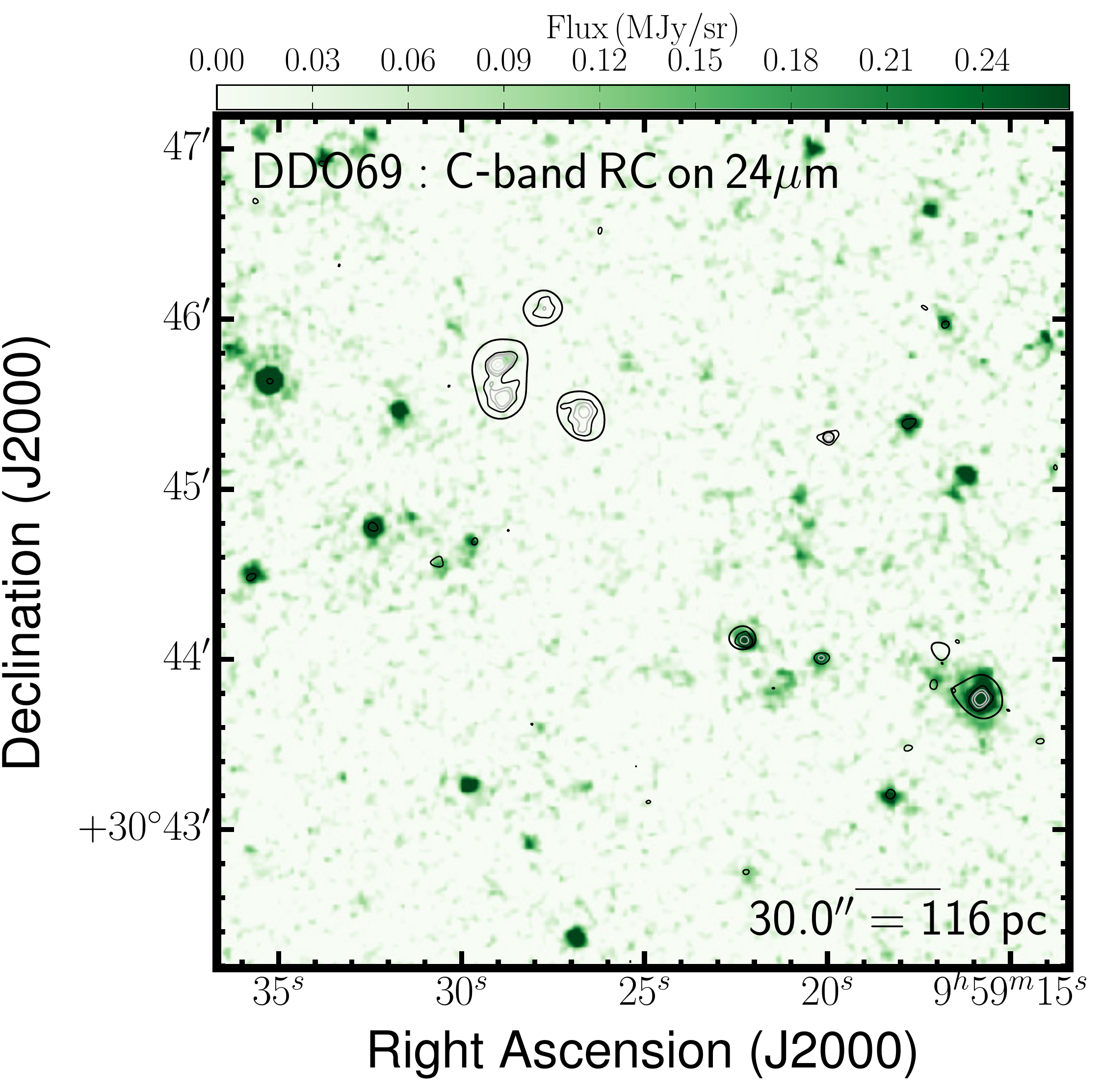} & \ 
    \includegraphics[width=0.31\linewidth,clip]{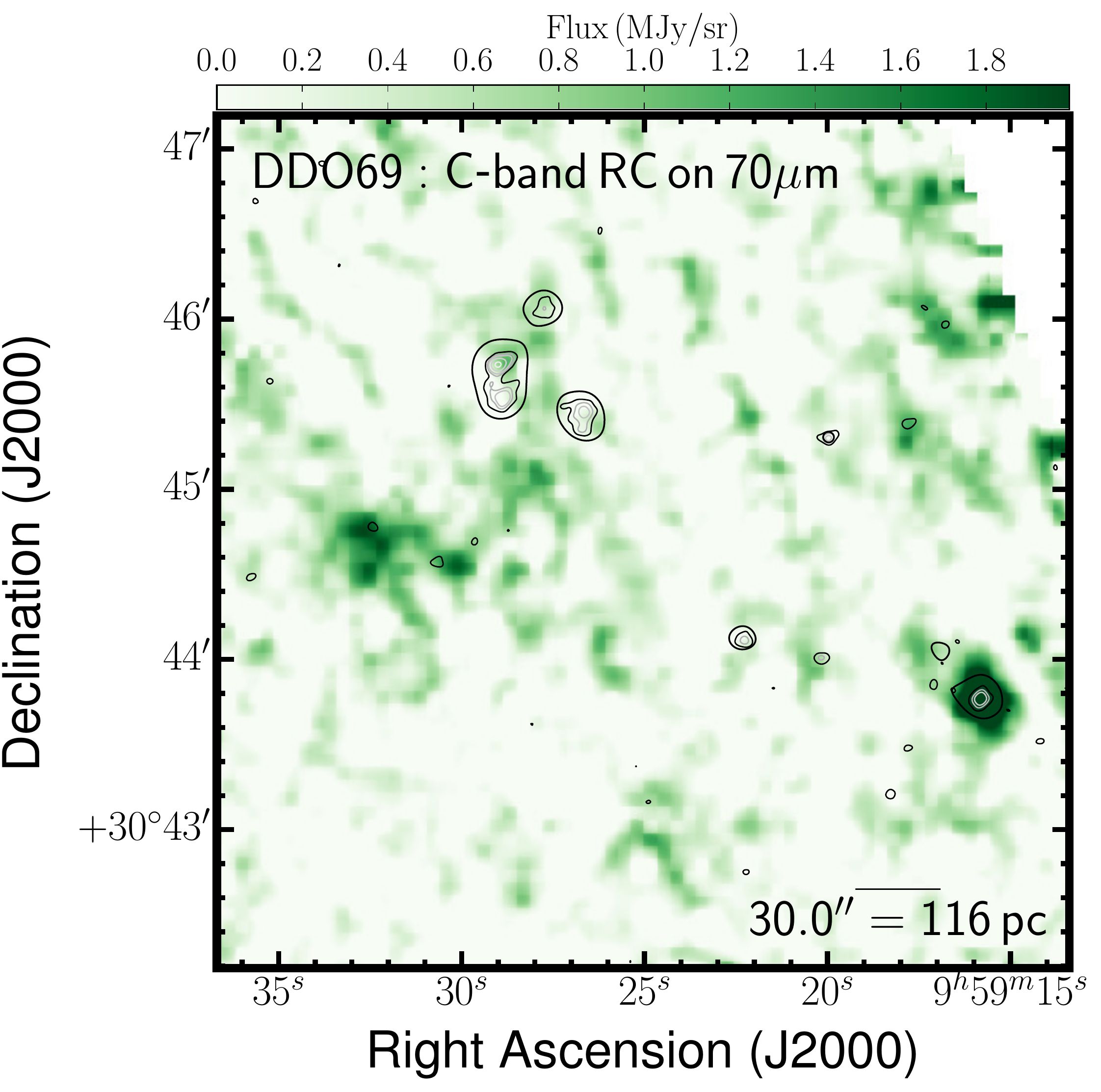} & \ 
    \includegraphics[width=0.31\linewidth,clip]{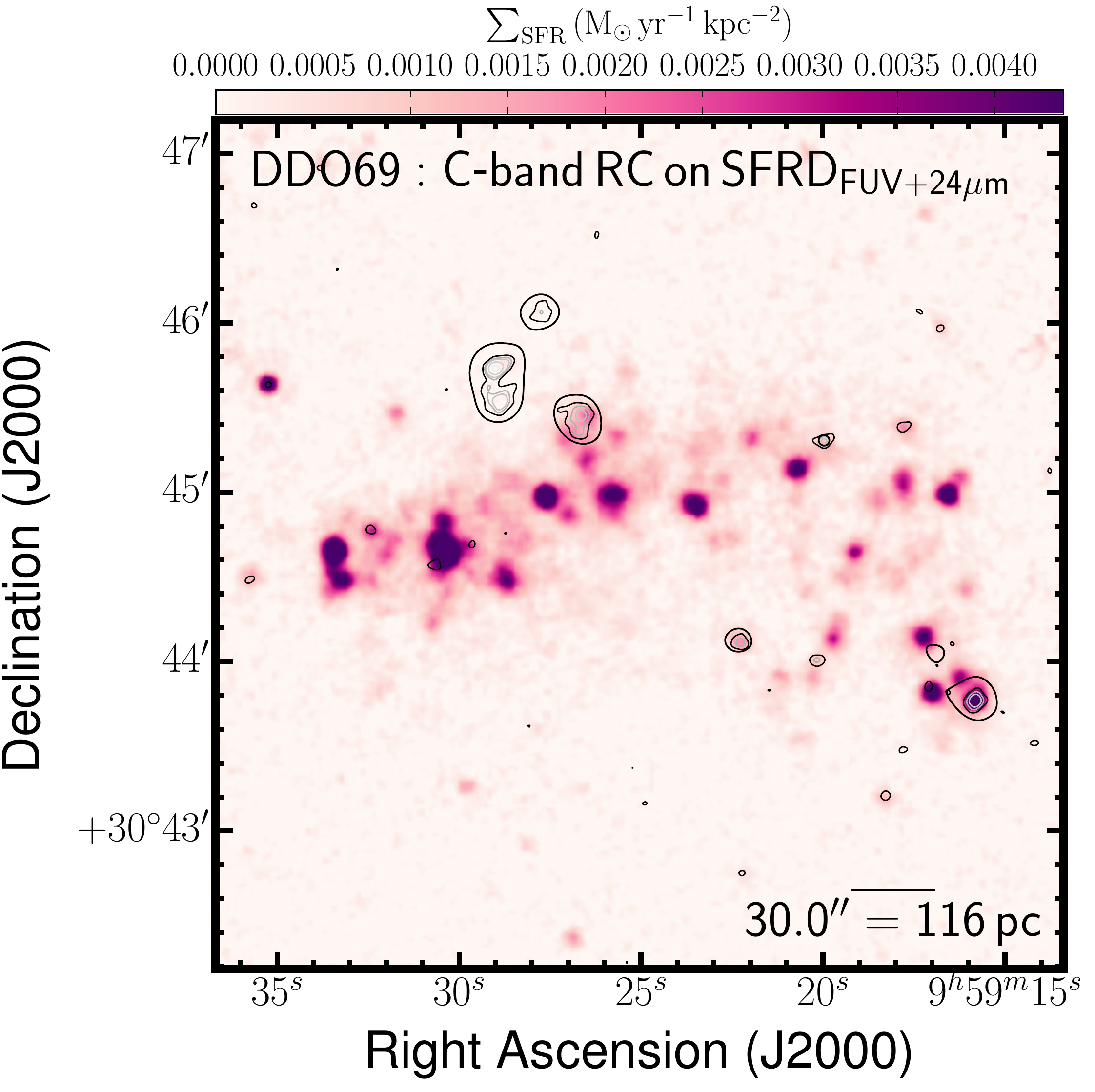} \\
  \end{tabular}
\caption[DDO\,69 images: RC, IR, optical, and FUV]{Multi-wavelength coverage of DDO 69 displaying a $5.0^\prime \times 5.0^\prime$ area. We show total RC flux density at the native resolution (top-left) and again with contours (top-centre). The RC contours are superposed on ancillary LITTLE THINGS images where possible: \halpha\ (middle-left); \RCNT\ obtained by subtracting the expected \RCT\ based on the \halpha-\RCT\ scaling factor of \cite{Deeg1997} from the total RC; {\em GALEX} FUV (middle-right); {\em Spitzer} 24\micron\ (bottom-left); {\em Spitzer} 70\micron\ (bottom-centre); FUV$+24{\rm \mu m}$--inferred SFRD from \citealp{Leroy2012} (bottom-right). We also show the RC that was isolated by the RC--based masking technique (top-right).}
  \label{figure:ddo69Cc_maps}
\end{figure}

\begin{figure}
  \begin{tabular}{ccc}
    \includegraphics[width=0.31\linewidth,clip]{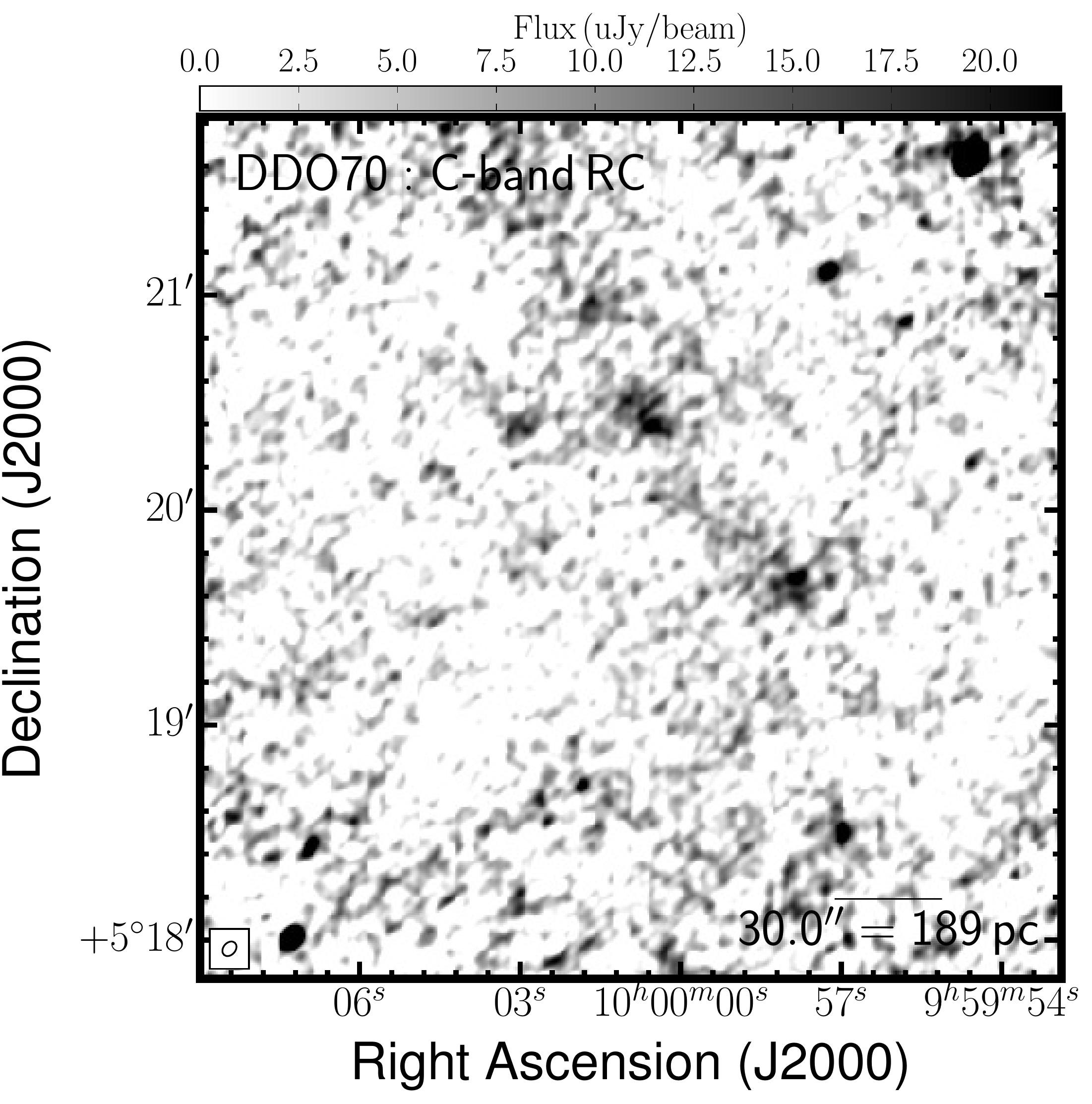} & \ 
    \includegraphics[width=0.31\linewidth,clip]{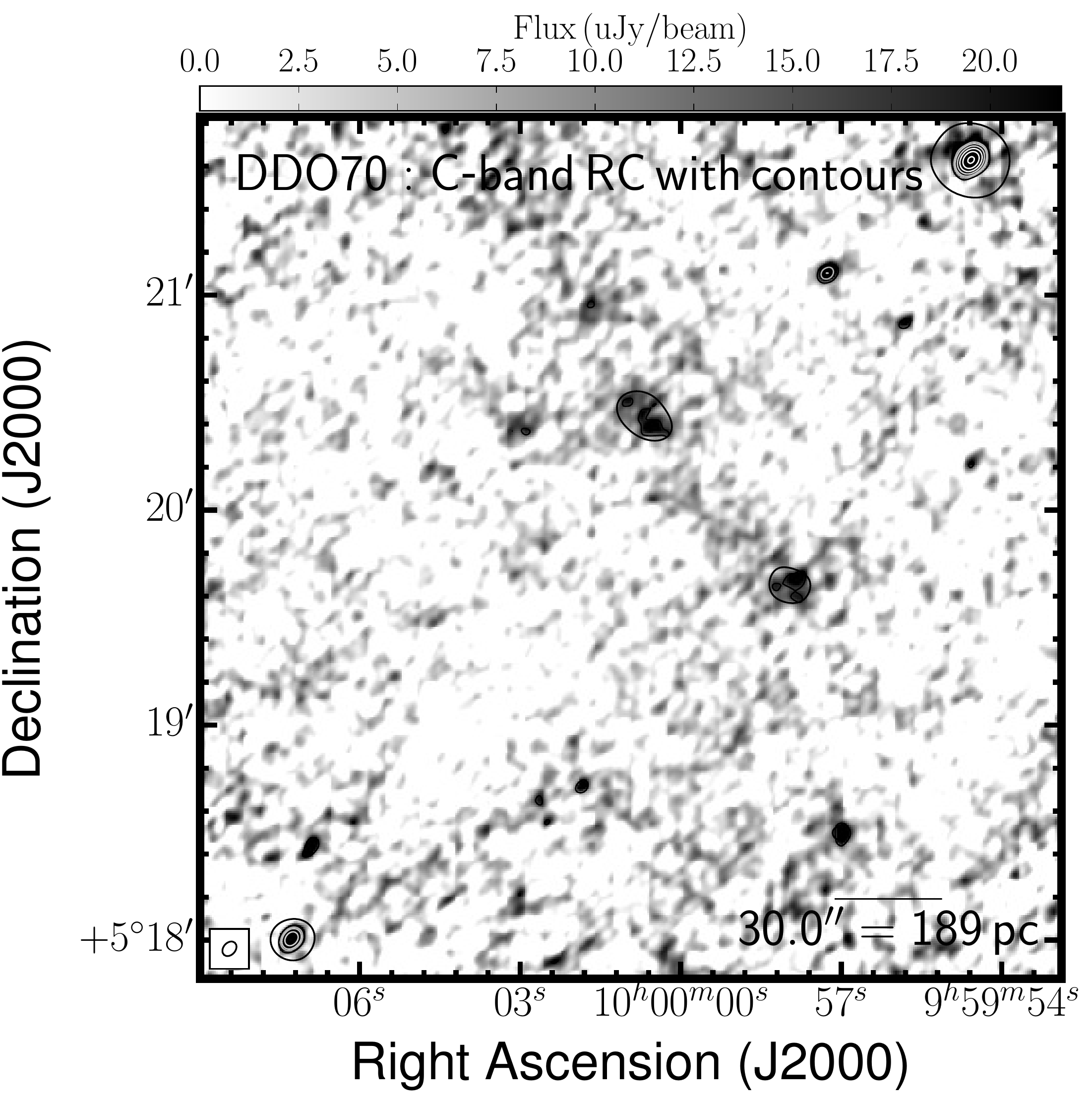} & \ 
    \includegraphics[width=0.31\linewidth,clip]{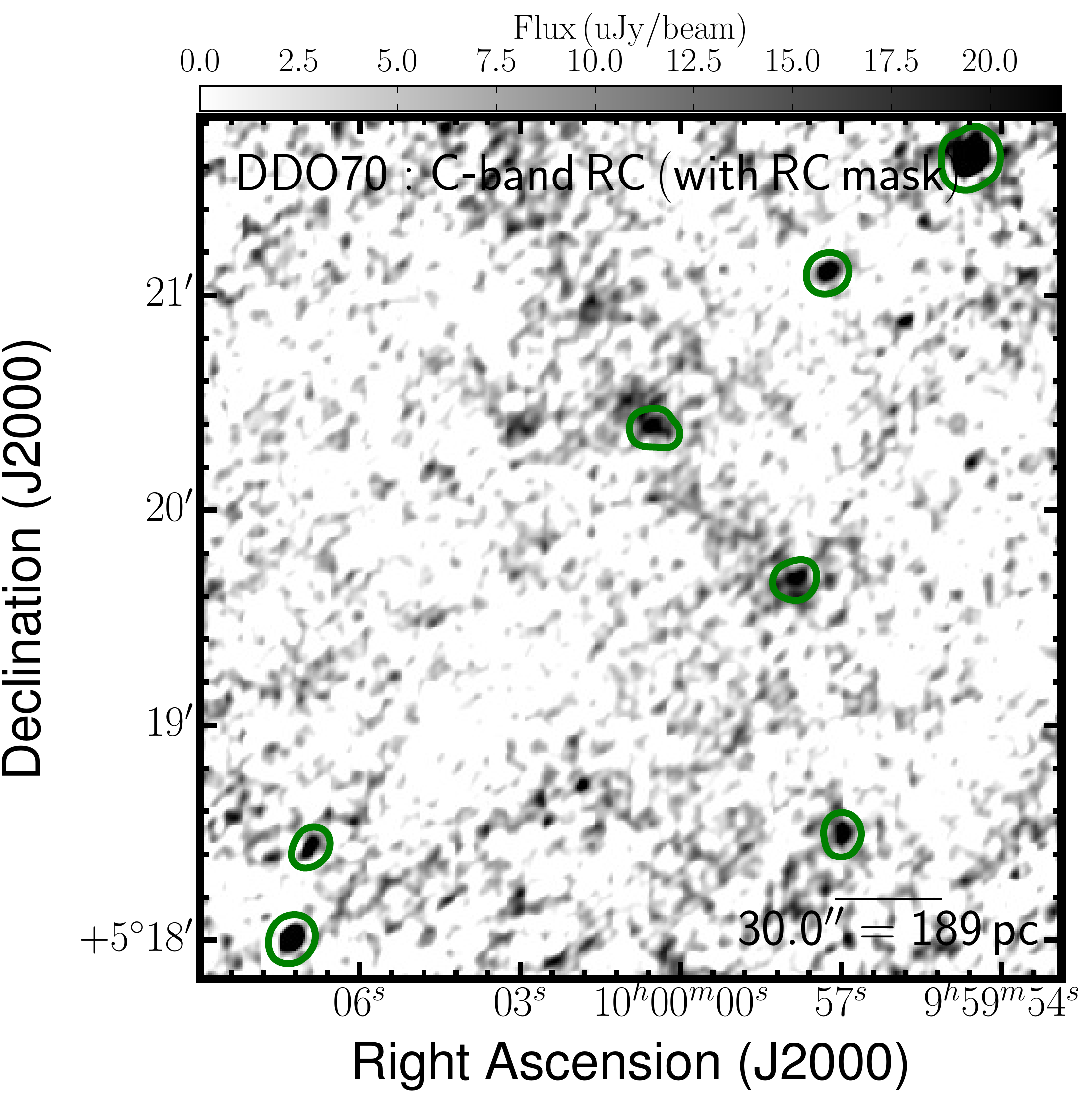} \\
    \includegraphics[width=0.31\linewidth,clip]{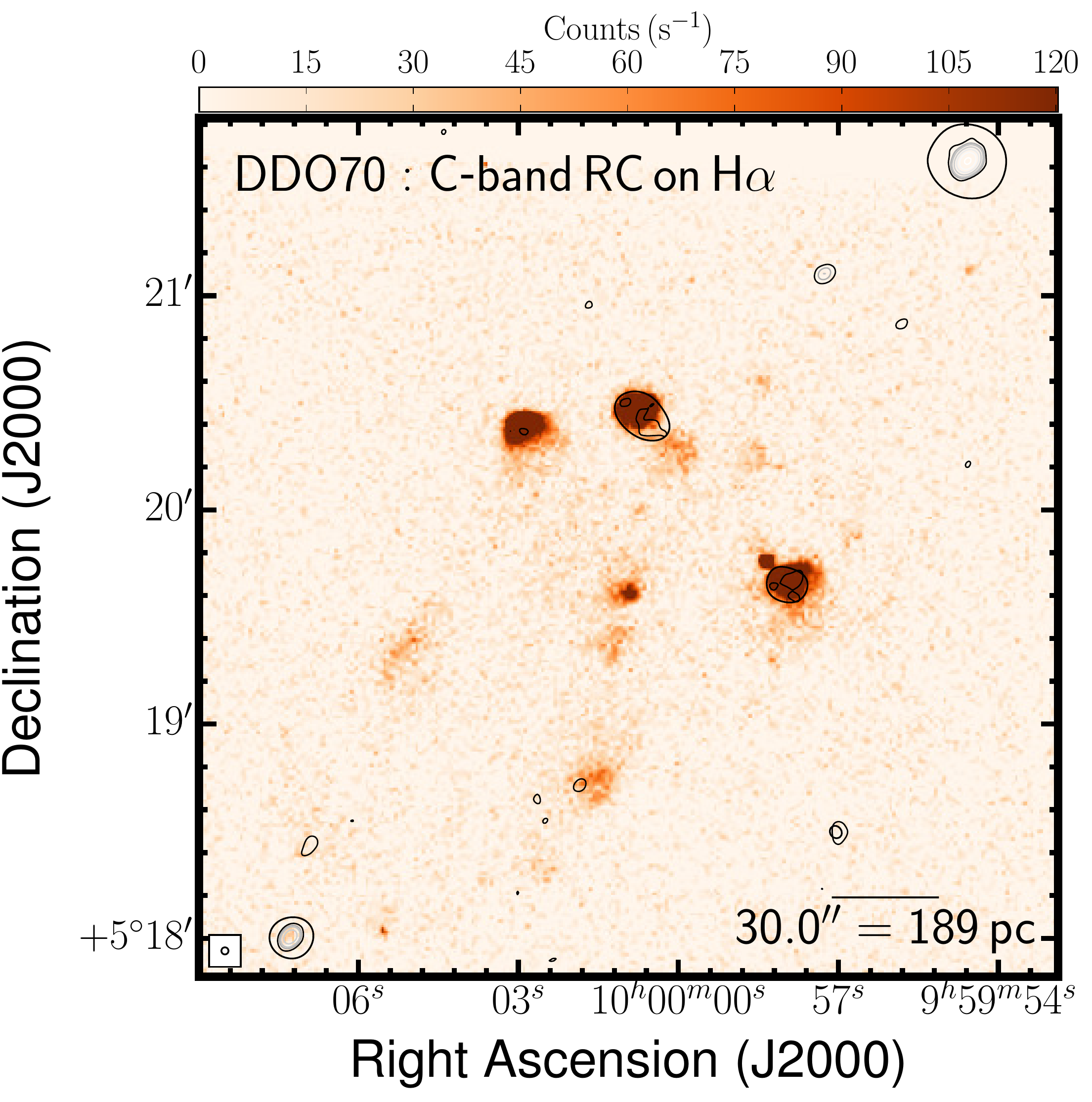} & \ 
    \includegraphics[width=0.31\linewidth,clip]{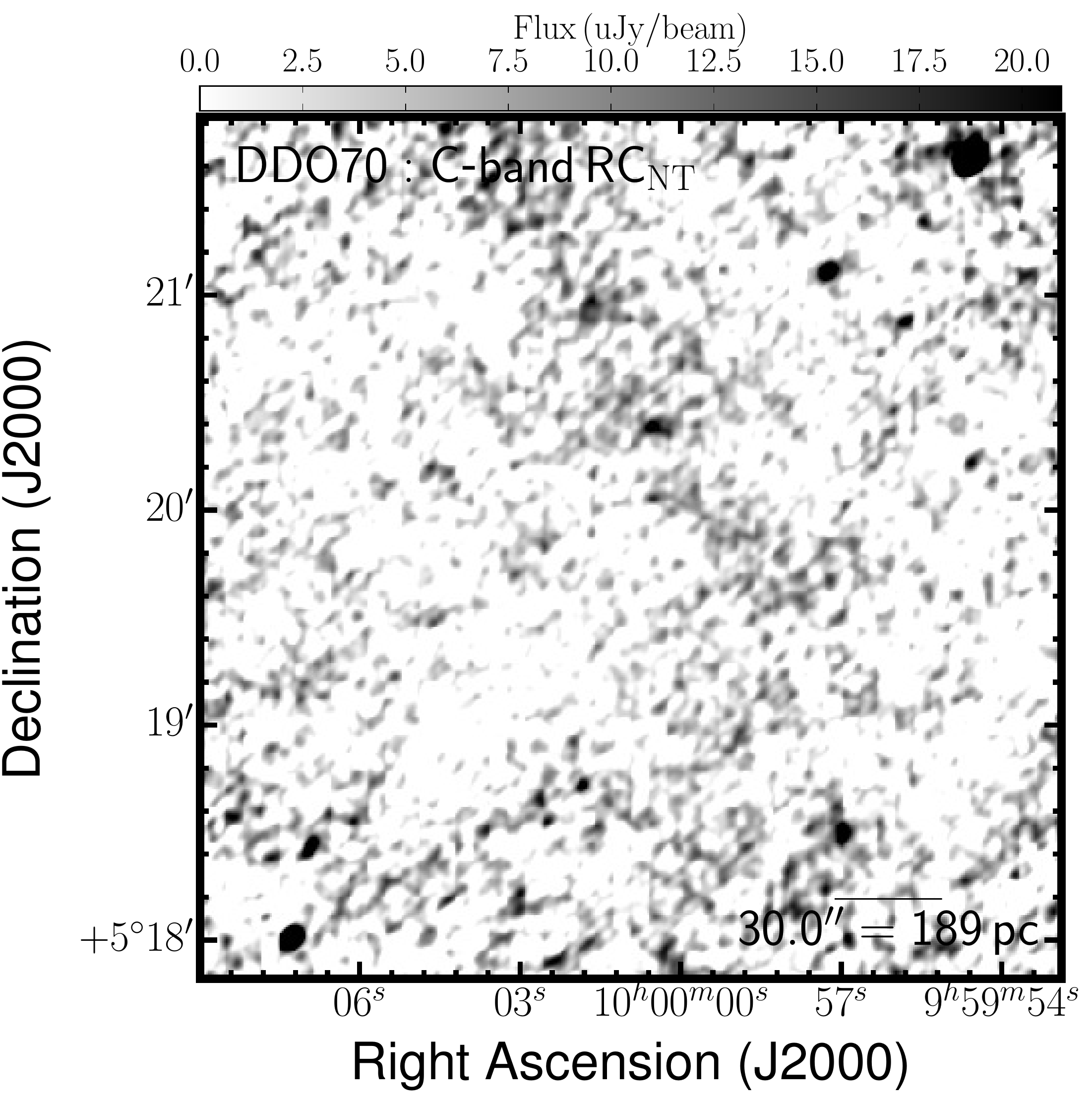} & \ 
    \includegraphics[width=0.31\linewidth,clip]{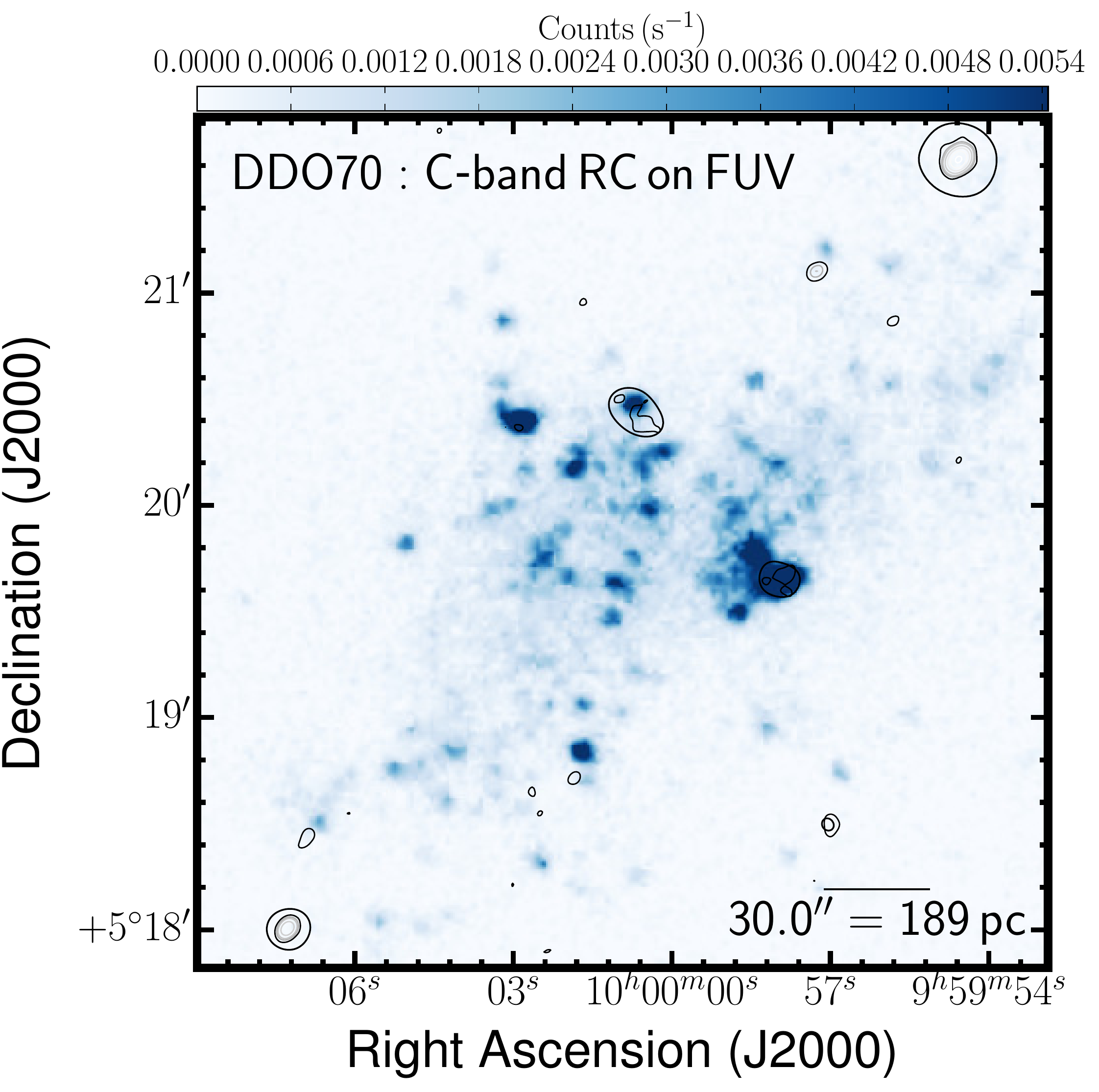} \\
    \includegraphics[width=0.31\linewidth,clip]{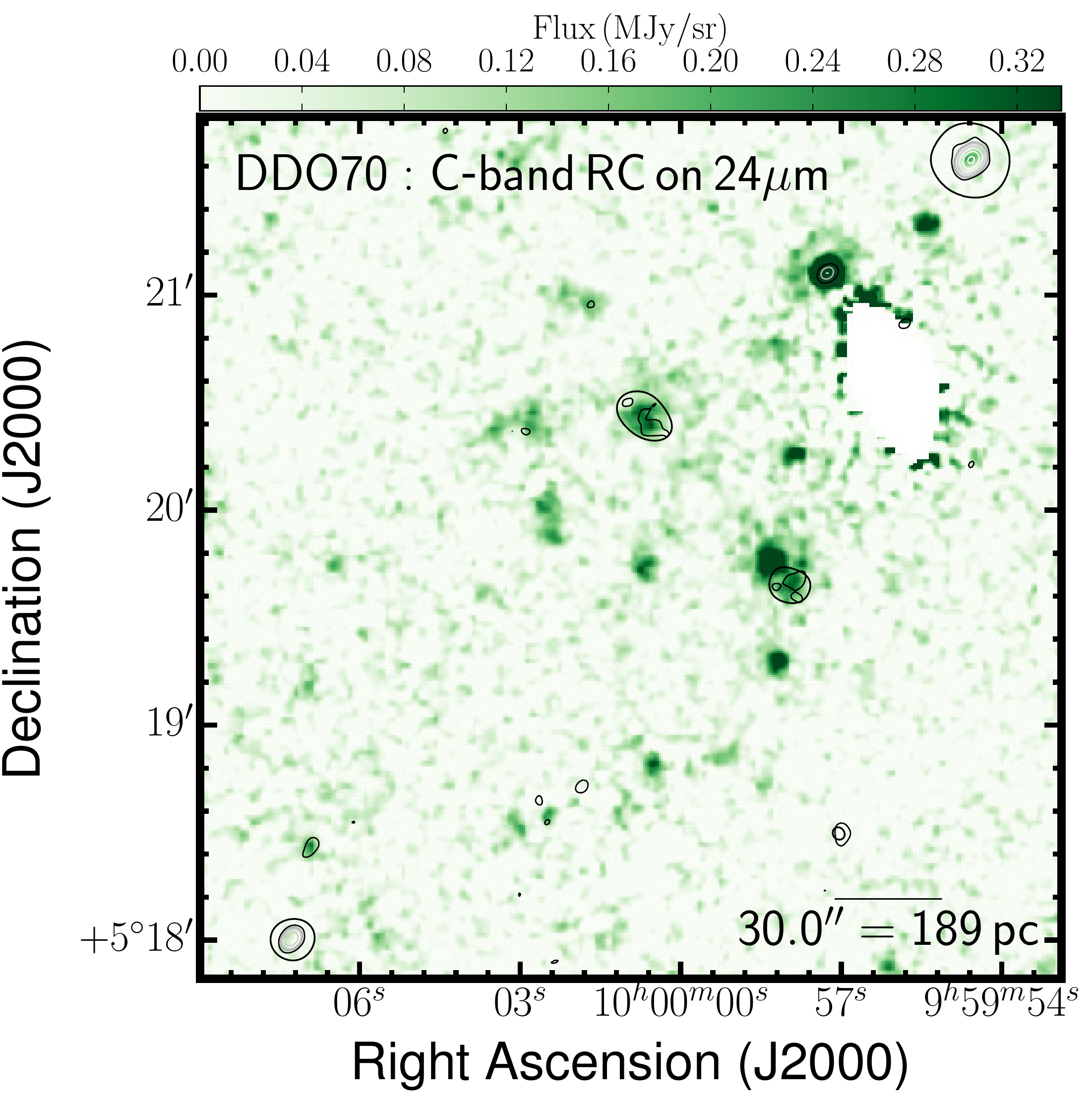} & \ 
    \includegraphics[width=0.31\linewidth,clip]{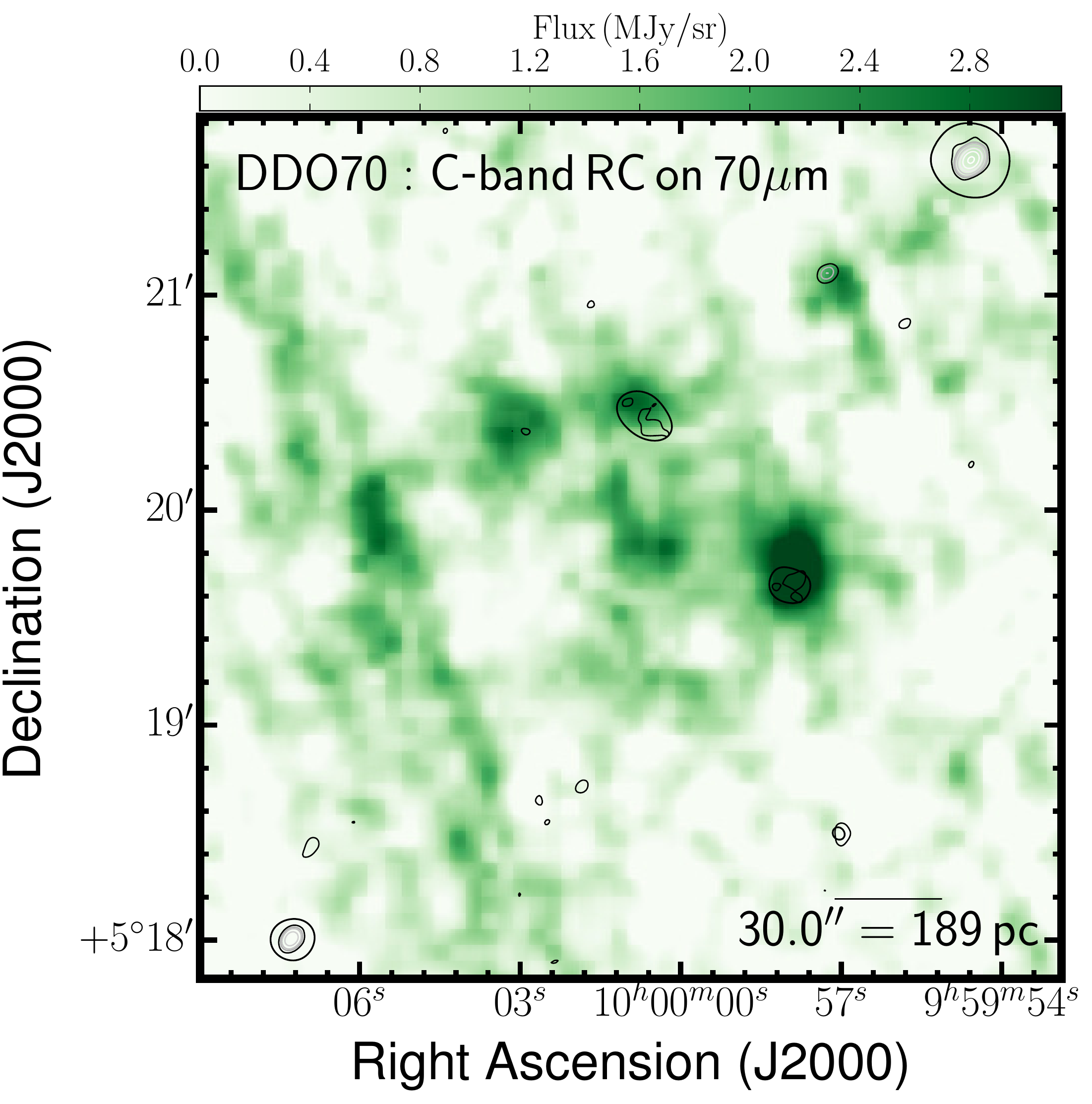} & \ 
    \includegraphics[width=0.31\linewidth,clip]{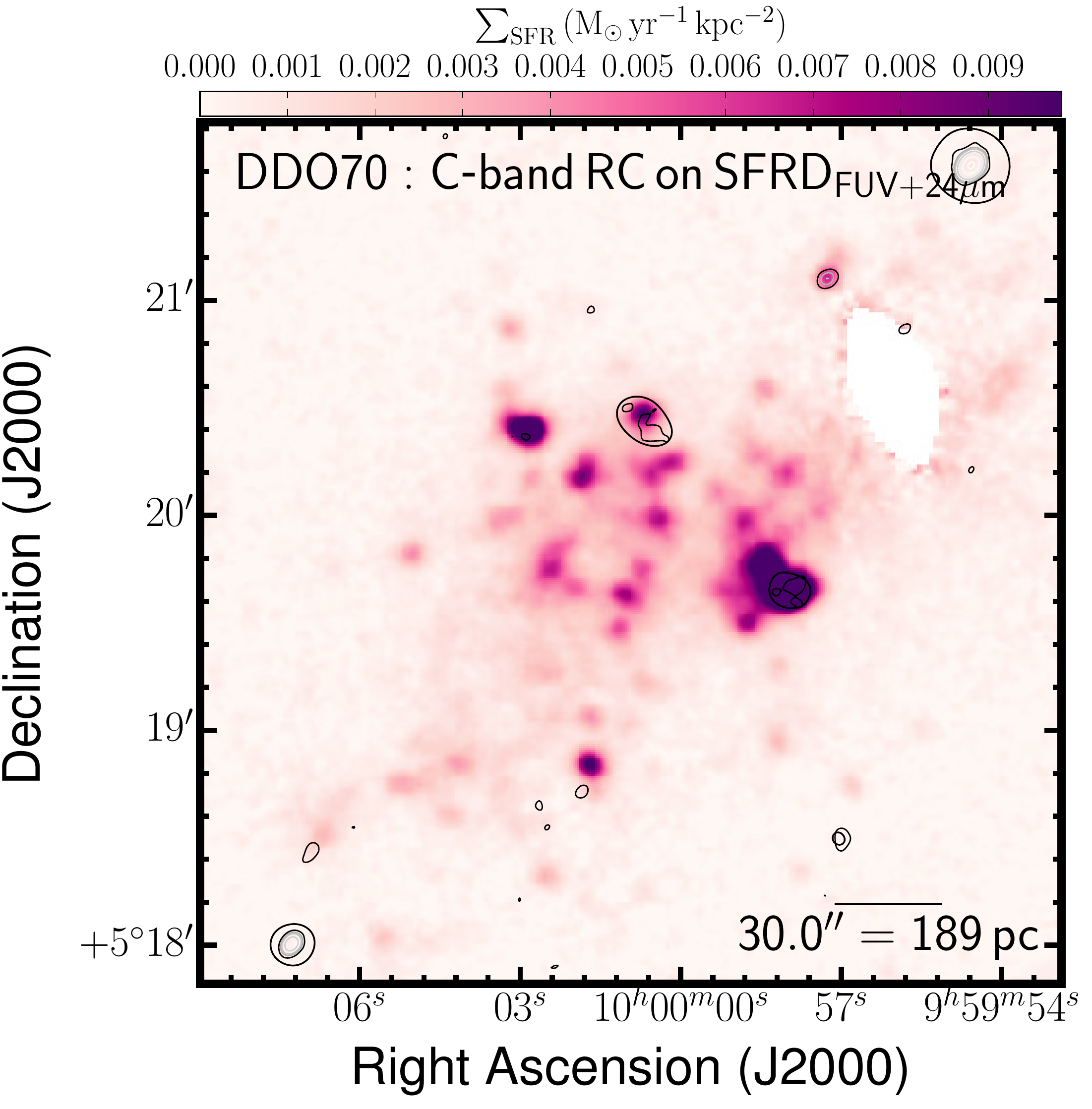} \\
  \end{tabular}
\caption[DDO\,70 images: RC, IR, optical, and FUV]{Multi-wavelength coverage of DDO 70 displaying a $4.0^\prime \times 4.0^\prime$ area. We show total RC flux density at the native resolution (top-left) and again with contours (top-centre). The RC contours are superposed on ancillary LITTLE THINGS images where possible: \halpha\ (middle-left); \RCNT\ obtained by subtracting the expected \RCT\ based on the \halpha-\RCT\ scaling factor of \cite{Deeg1997} from the total RC; {\em GALEX} FUV (middle-right); {\em Spitzer} 24\micron\ (bottom-left); {\em Spitzer} 70\micron\ (bottom-centre); FUV$+24{\rm \mu m}$--inferred SFRD from \citealp{Leroy2012} (bottom-right). We also show the RC that was isolated by the RC--based masking technique (top-right).}
  \label{figure:ddo70Cc_maps}
\end{figure}

\begin{figure}
  \begin{tabular}{ccc}
    \includegraphics[width=0.31\linewidth,clip]{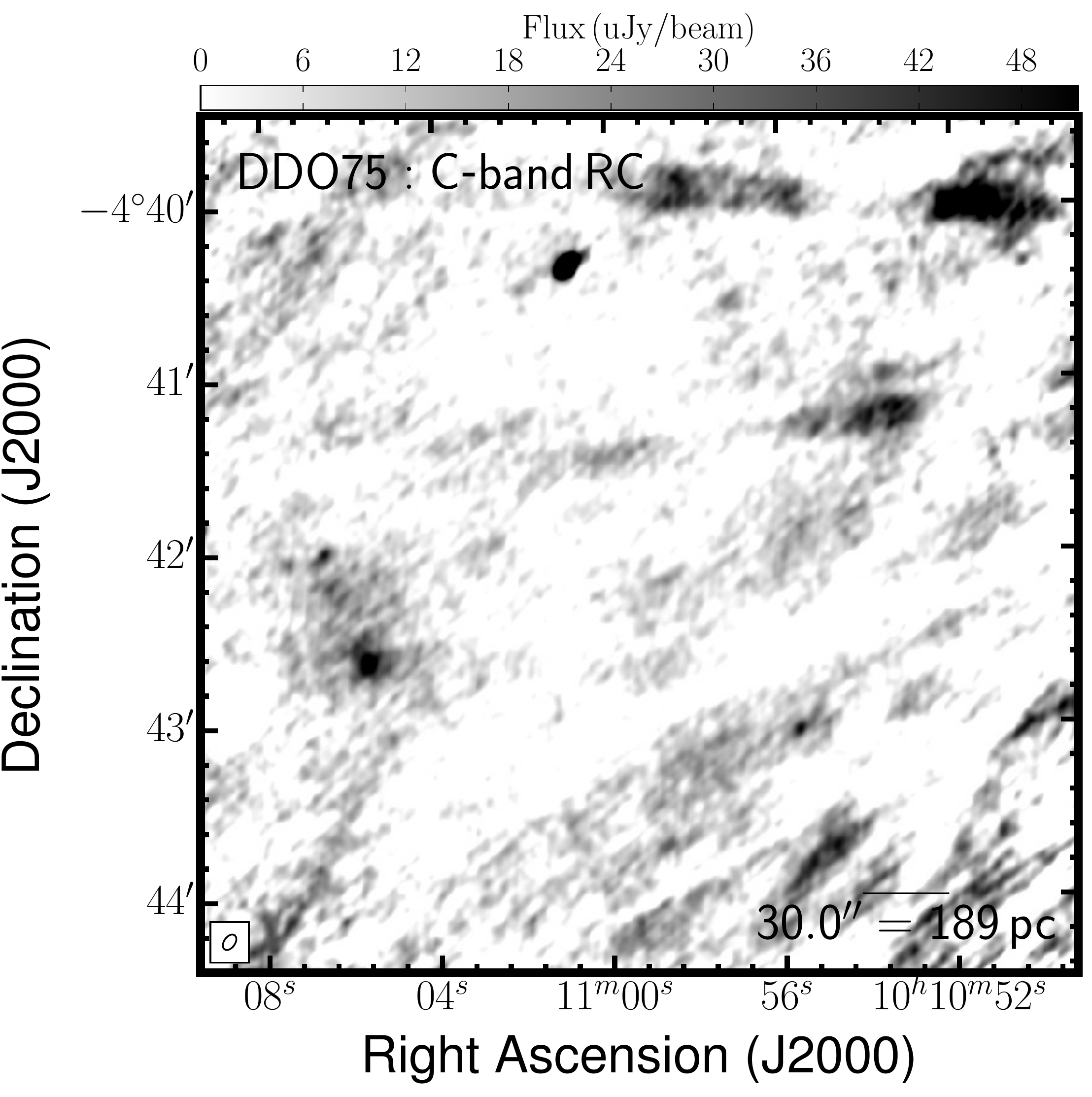} & \ 
    \includegraphics[width=0.31\linewidth,clip]{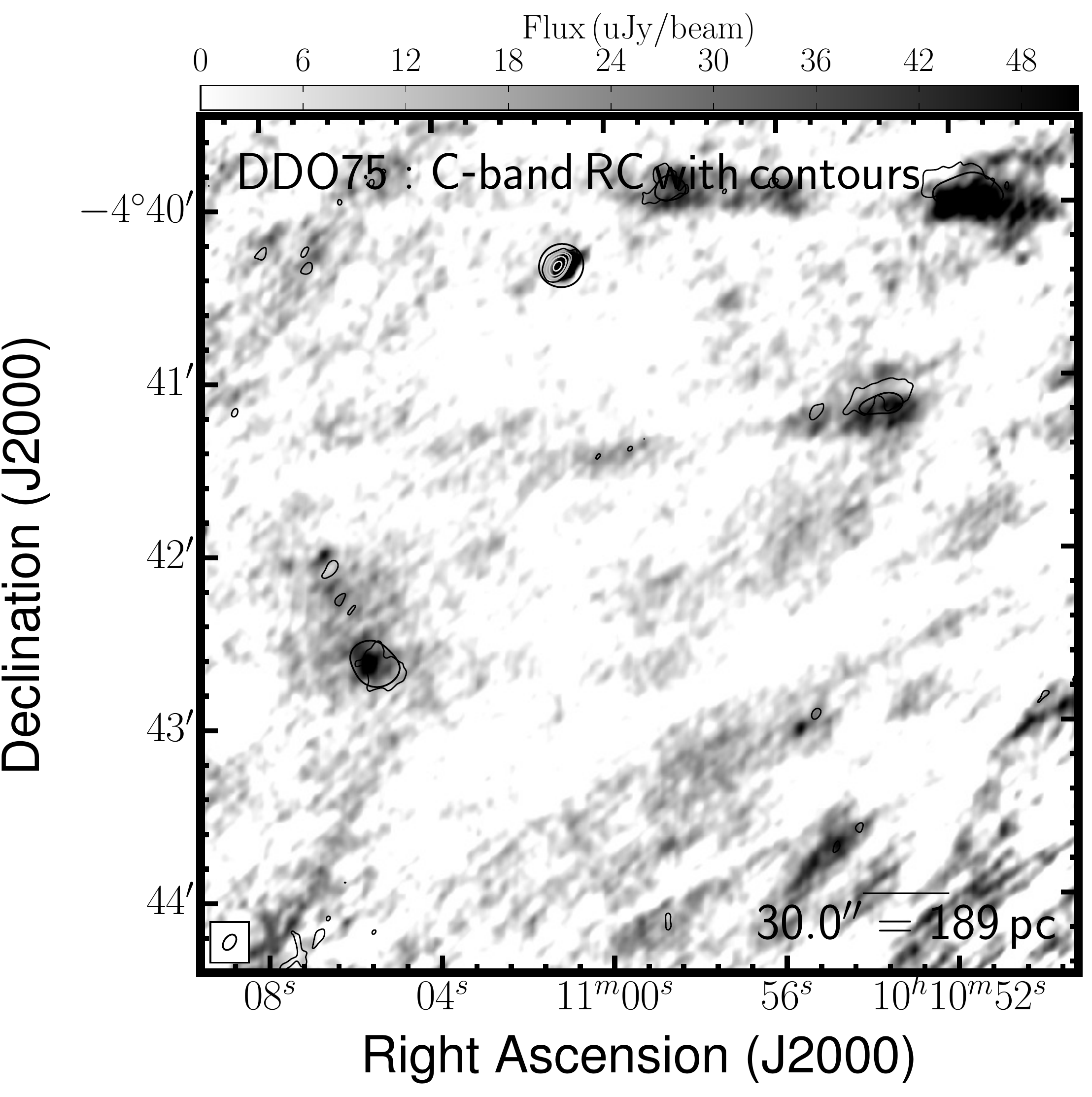} & \ 
    \includegraphics[width=0.31\linewidth,clip]{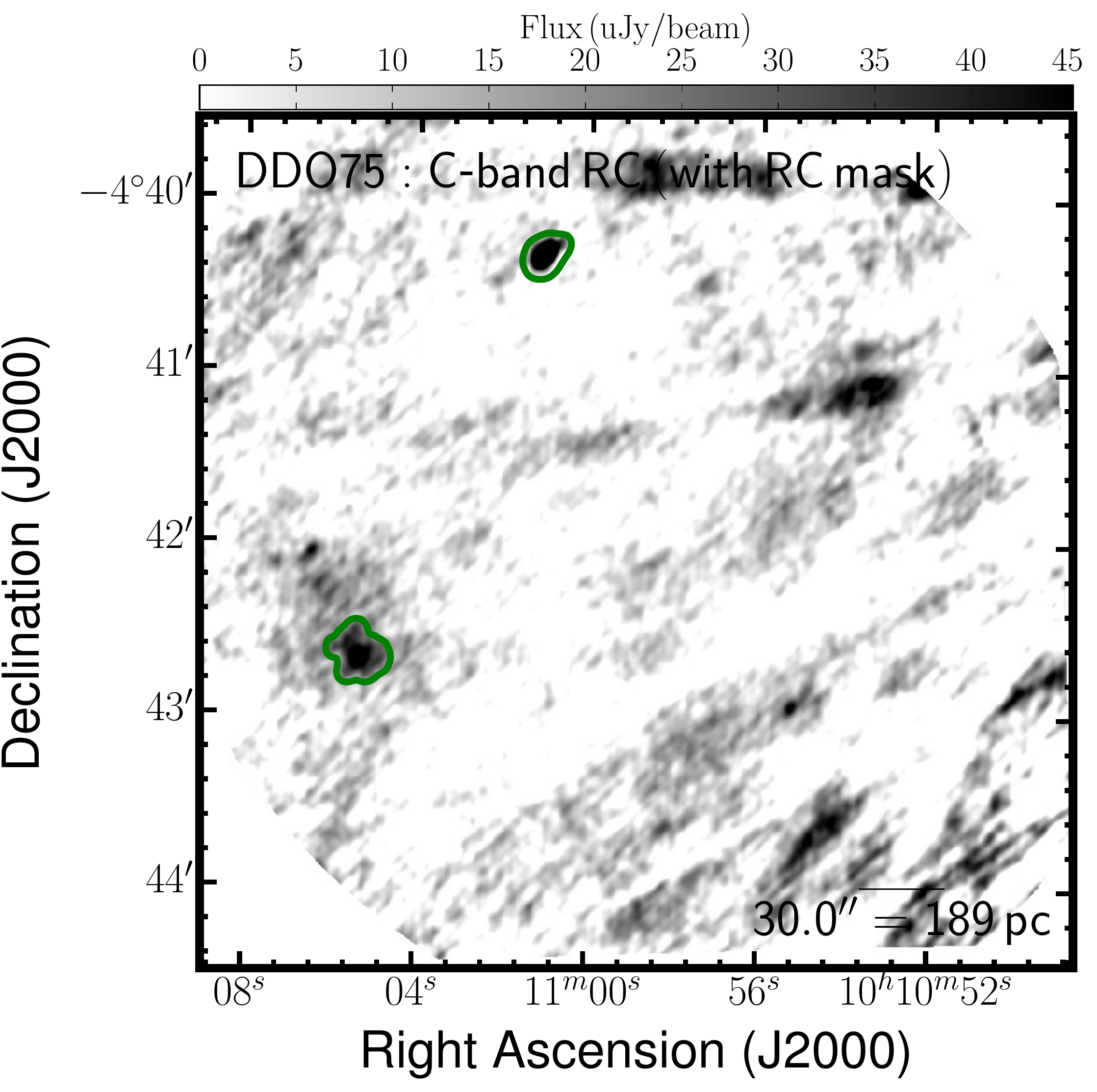} \\
    \includegraphics[width=0.31\linewidth,clip]{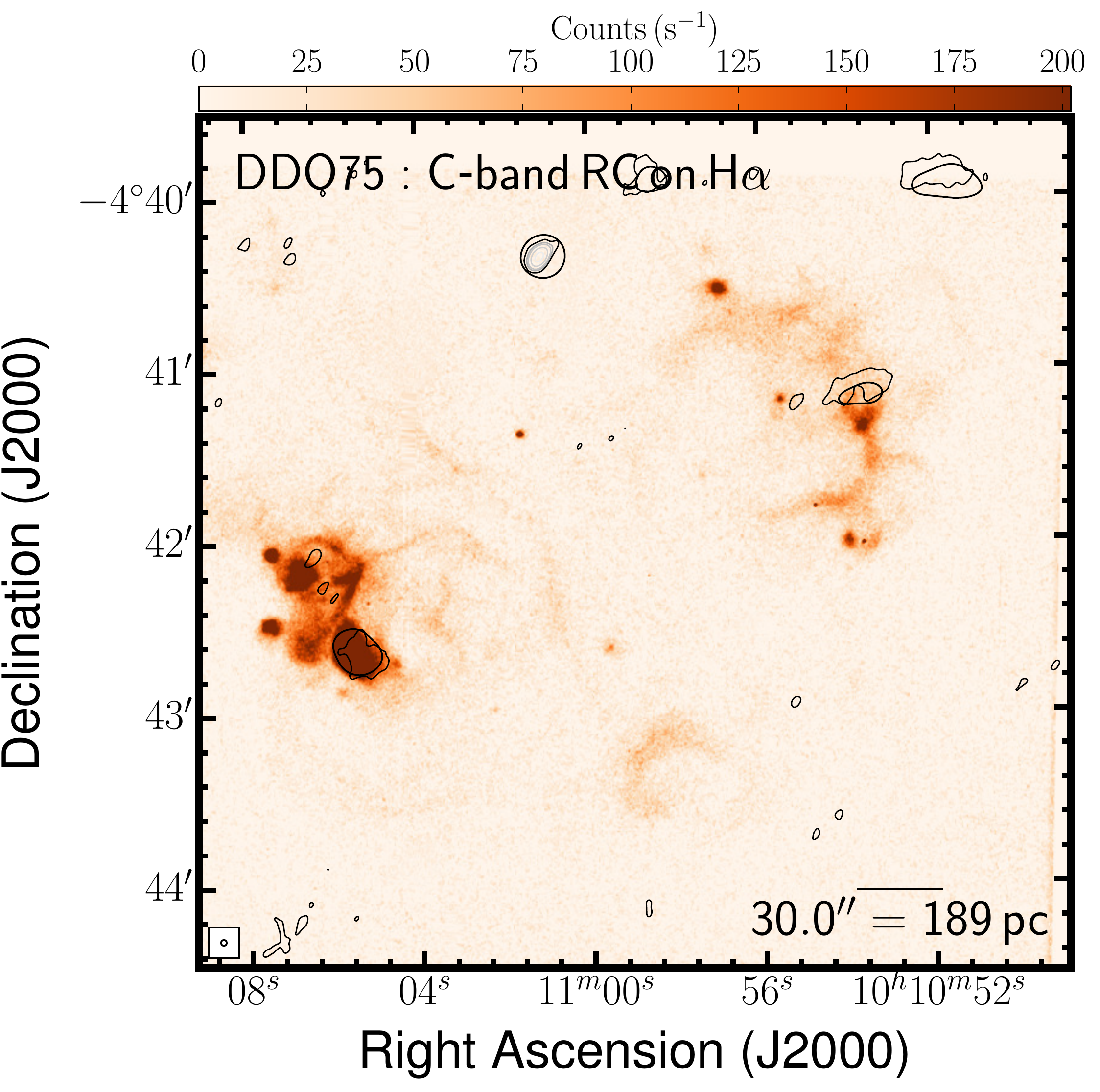} & \ 
    \includegraphics[width=0.31\linewidth,clip]{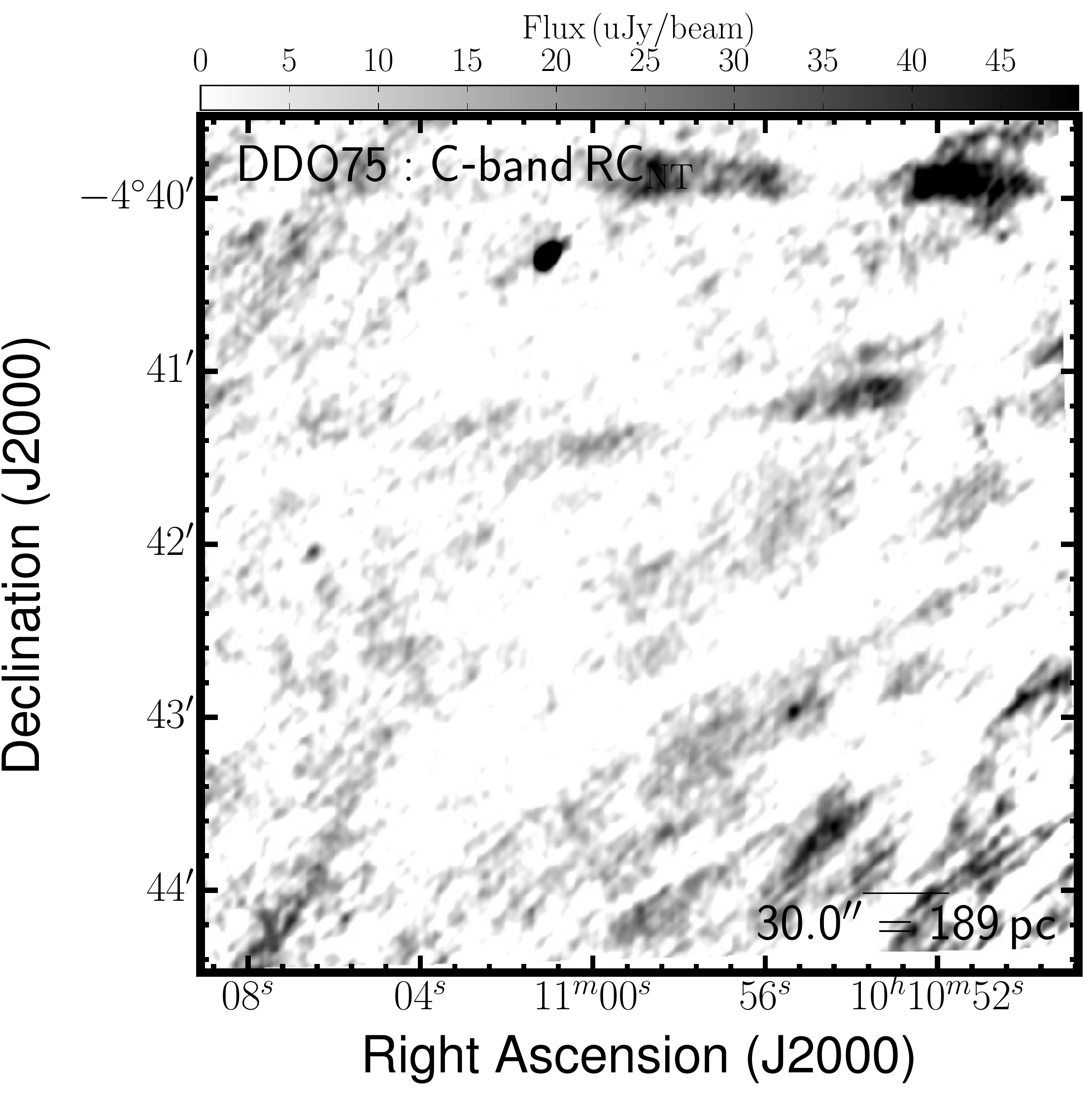} & \ 
    \includegraphics[width=0.31\linewidth,clip]{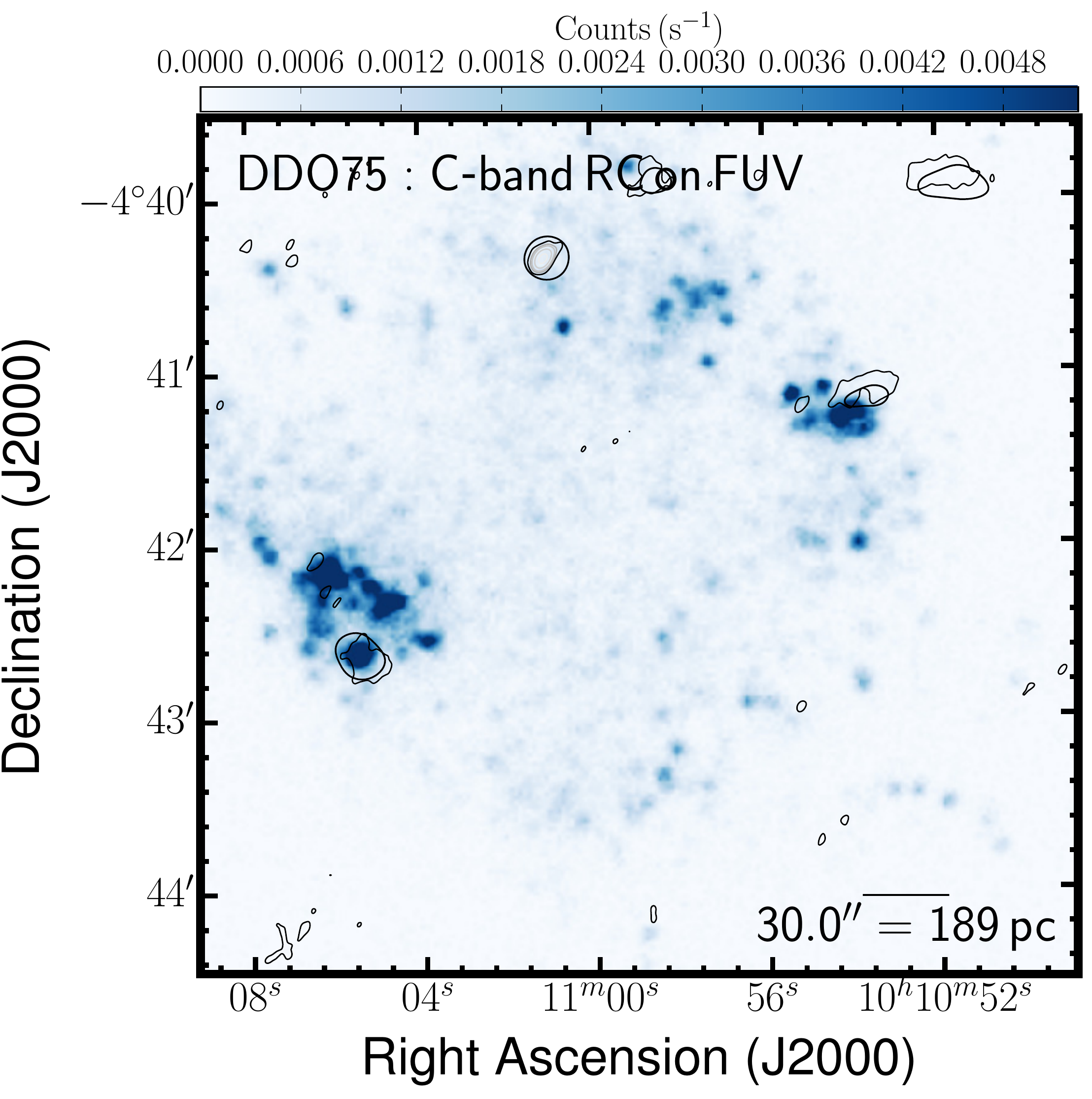} \\
    \includegraphics[width=0.31\linewidth,clip]{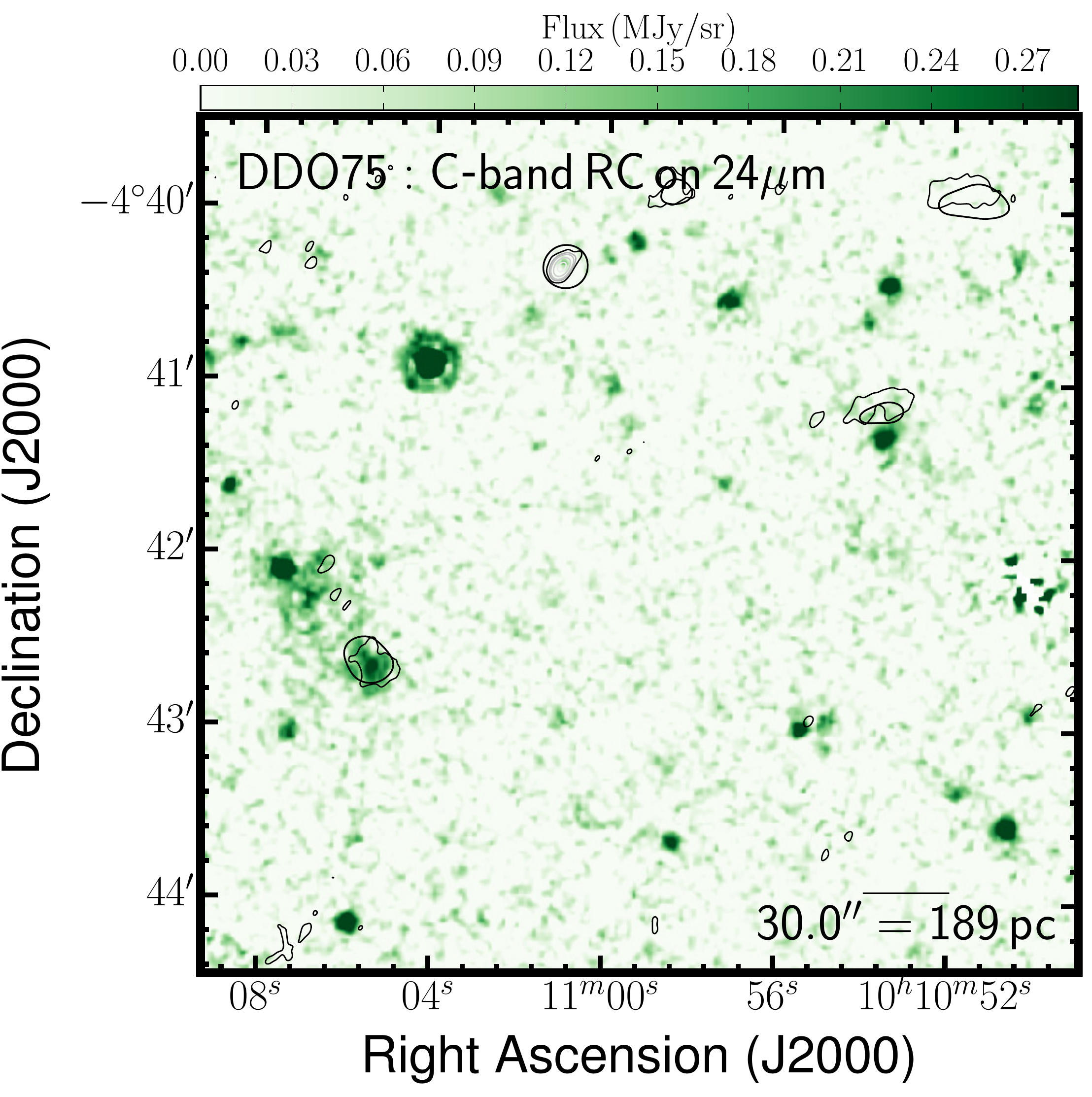} & \ 
    \includegraphics[width=0.31\linewidth,clip]{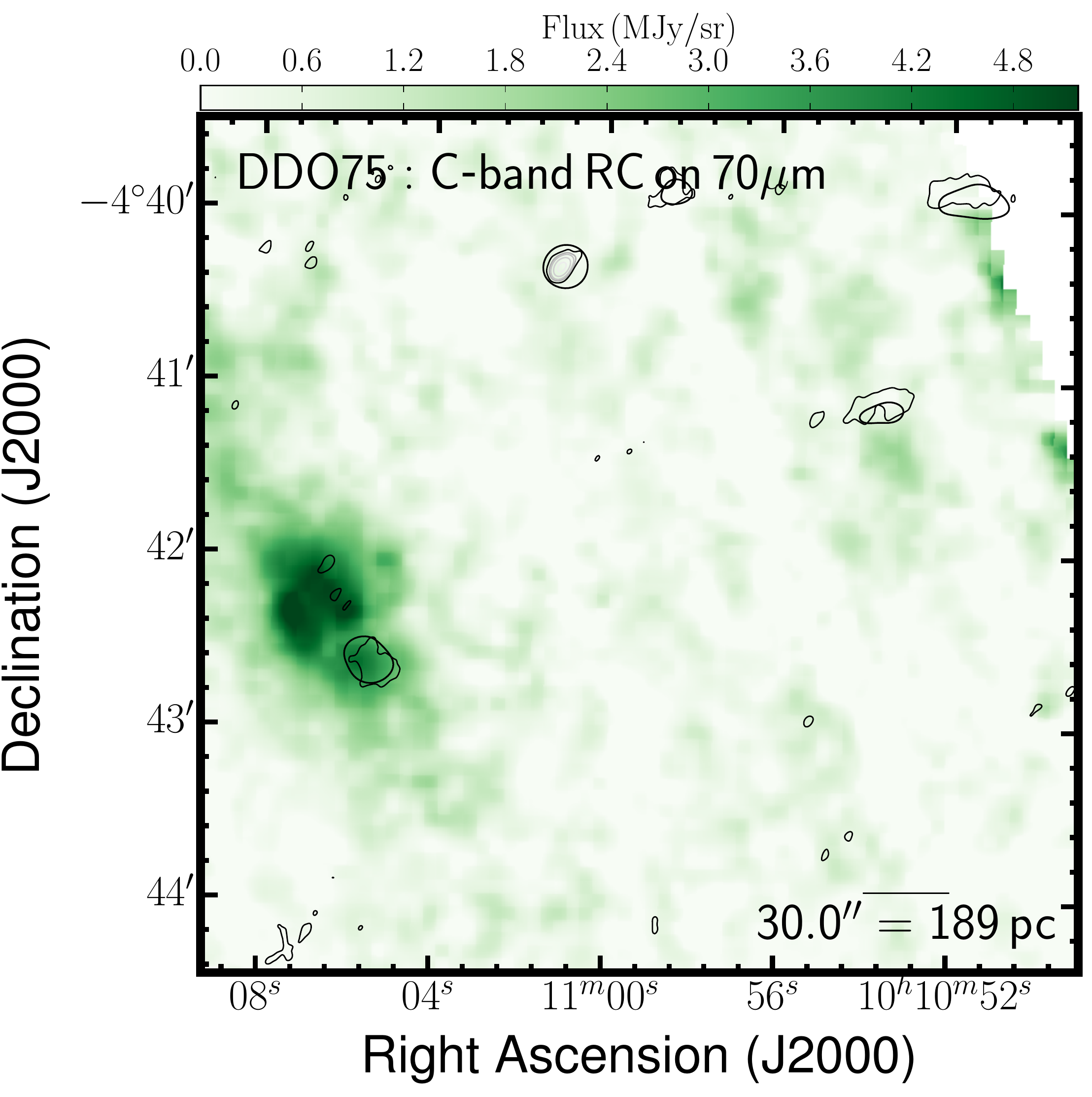} & \ 
    \includegraphics[width=0.31\linewidth,clip]{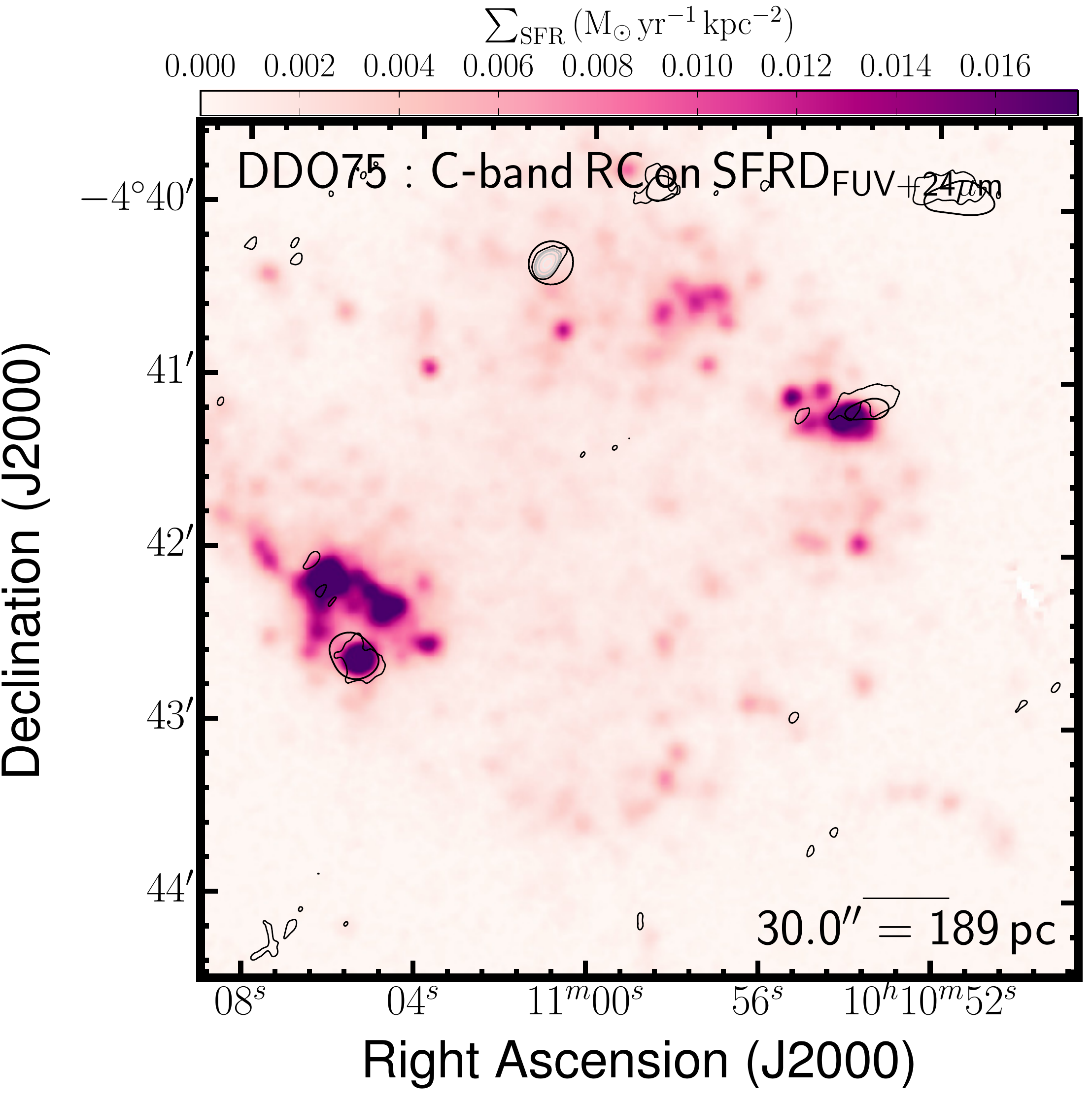} \\
  \end{tabular}
\caption[DDO\,75 images: RC, IR, optical, and FUV]{Multi-wavelength coverage of DDO 75 displaying a $6.3^\prime \times 6.3^\prime$ area. We show total RC flux density at the native resolution (top-left) and again with contours (top-centre). The RC contours are superposed on ancillary LITTLE THINGS images where possible: \halpha\ (middle-left); \RCNT\ obtained by subtracting the expected \RCT\ based on the \halpha-\RCT\ scaling factor of \cite{Deeg1997} from the total RC; {\em GALEX} FUV (middle-right); {\em Spitzer} 24\micron\ (bottom-left); {\em Spitzer} 70\micron\ (bottom-centre); FUV$+24{\rm \mu m}$--inferred SFRD from \citealp{Leroy2012} (bottom-right). We also show the RC that was isolated by the RC--based masking technique (top-right).}
  \label{figure:ddo75Cc_maps}
\end{figure}

\begin{figure}
  \begin{tabular}{ccc}
    \includegraphics[width=0.31\linewidth,clip]{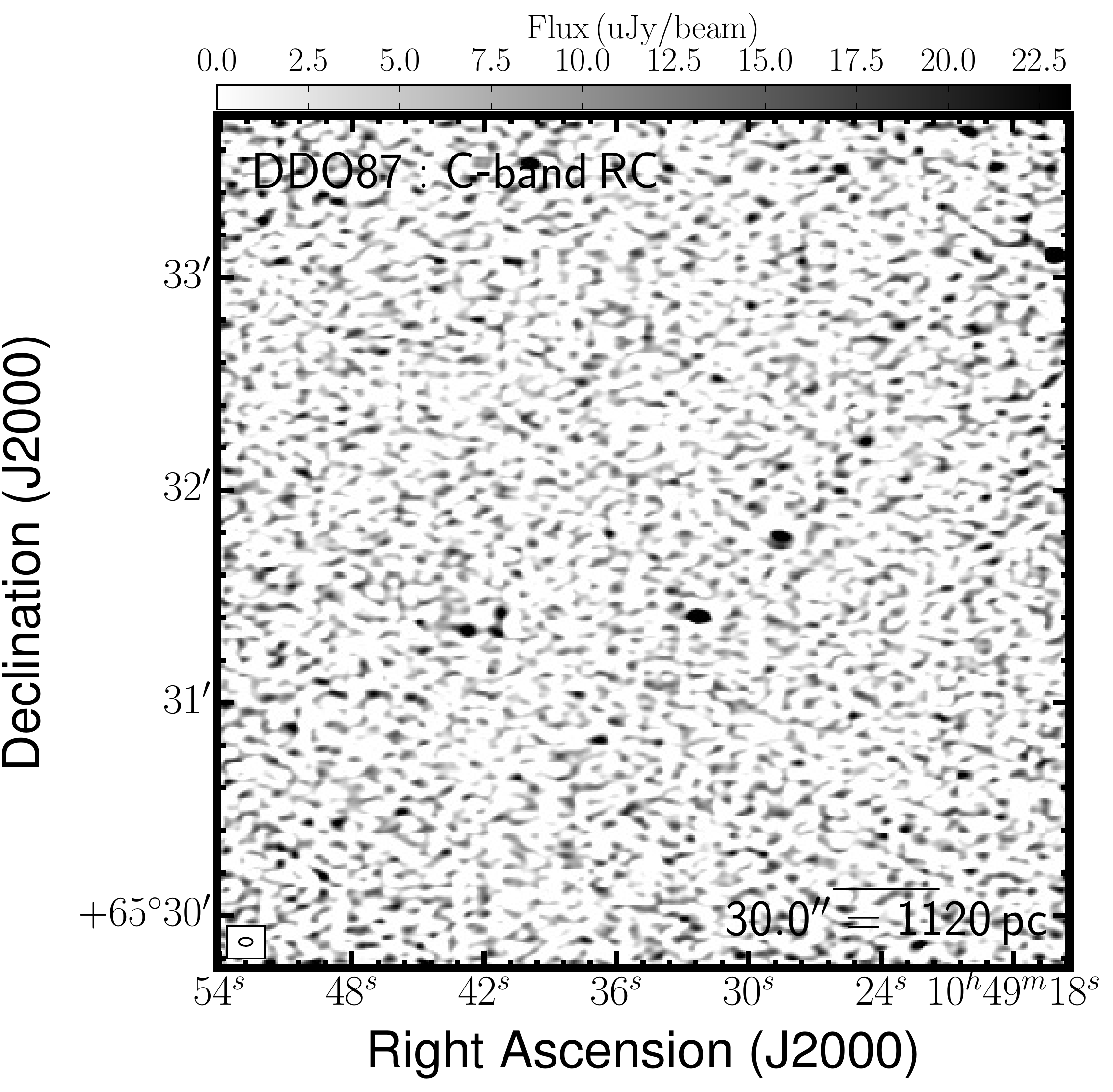} & \ 
    \includegraphics[width=0.31\linewidth,clip]{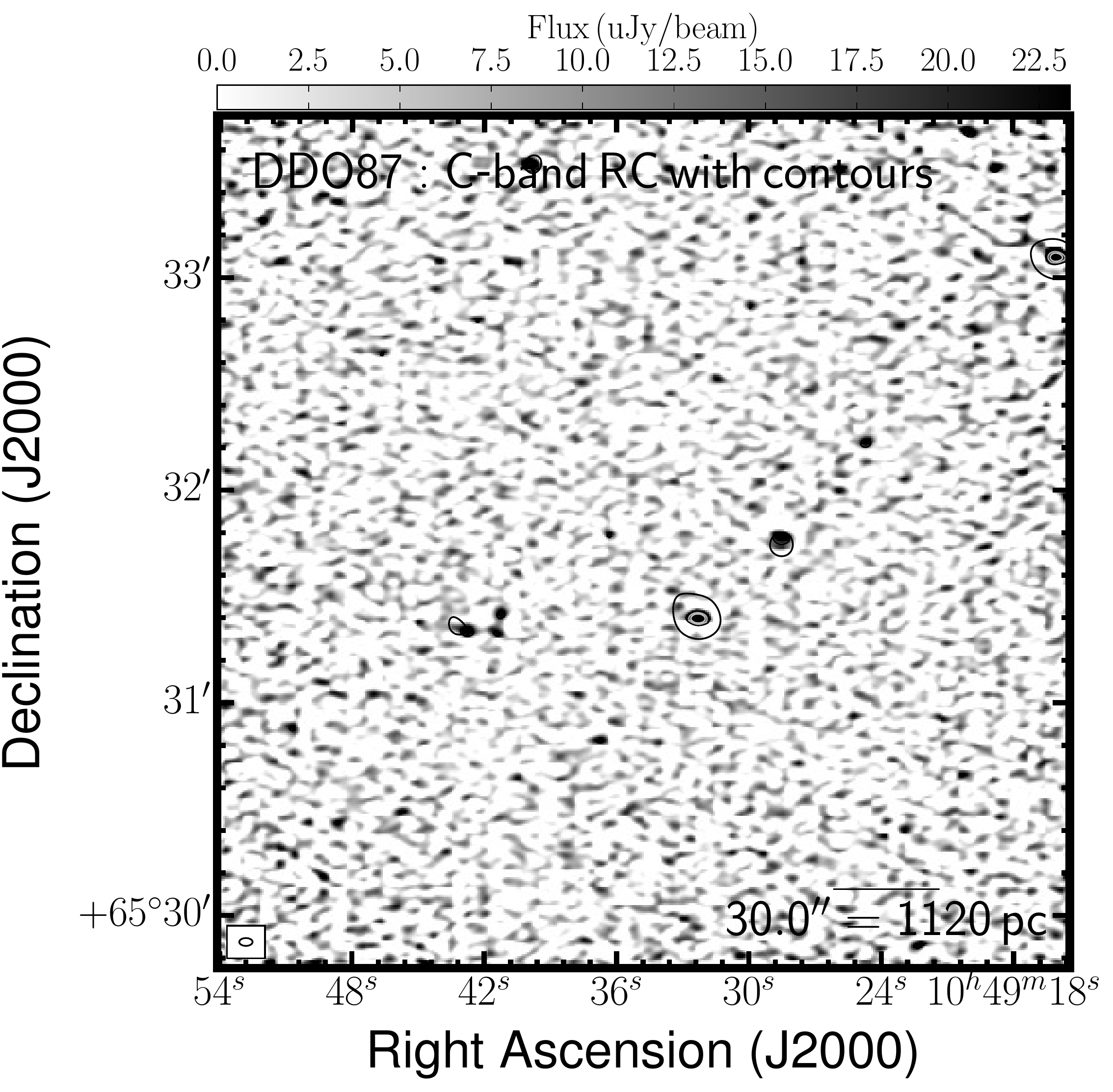} & \ 
    \includegraphics[width=0.31\linewidth,clip]{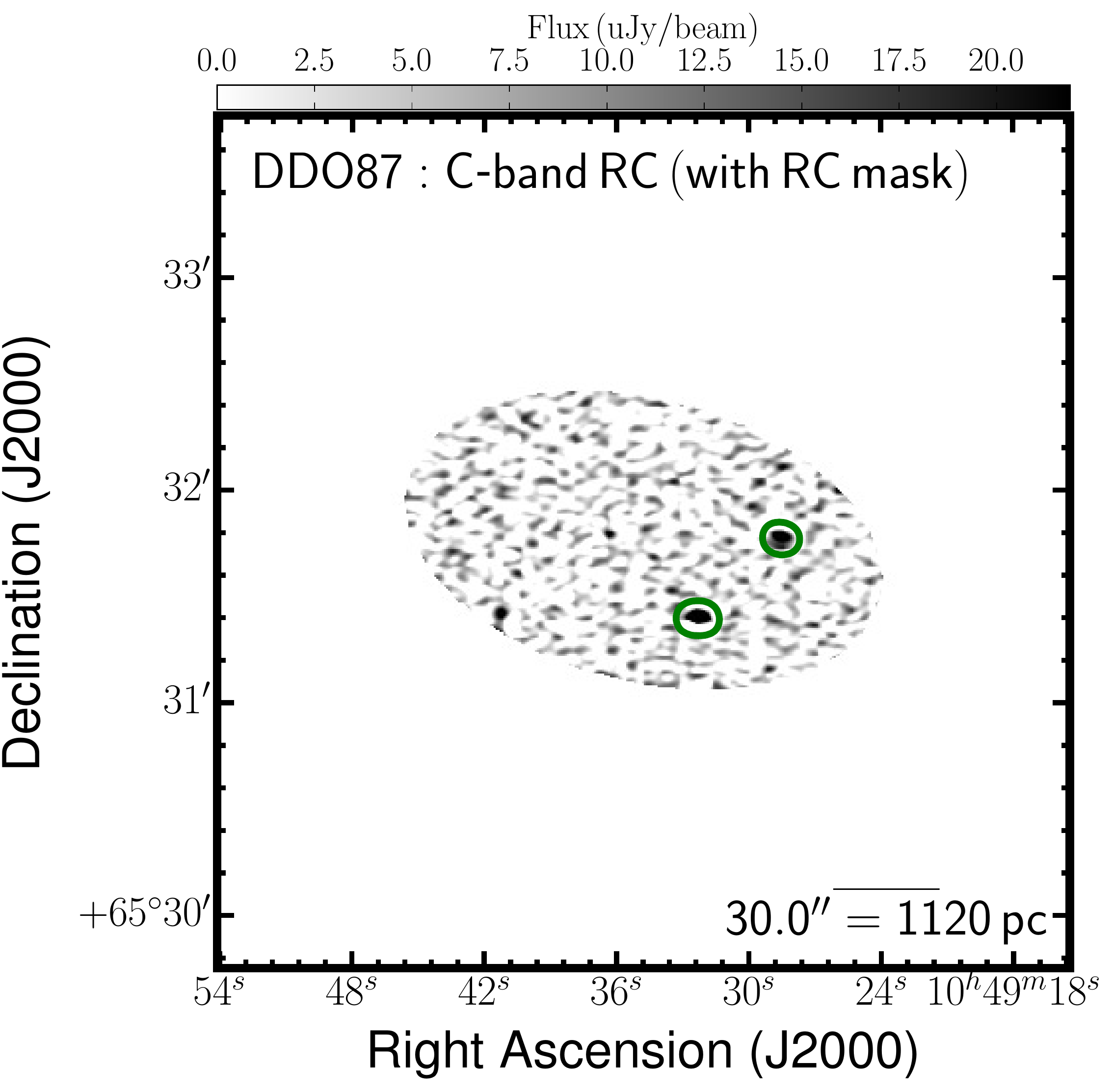} \\
    \includegraphics[width=0.31\linewidth,clip]{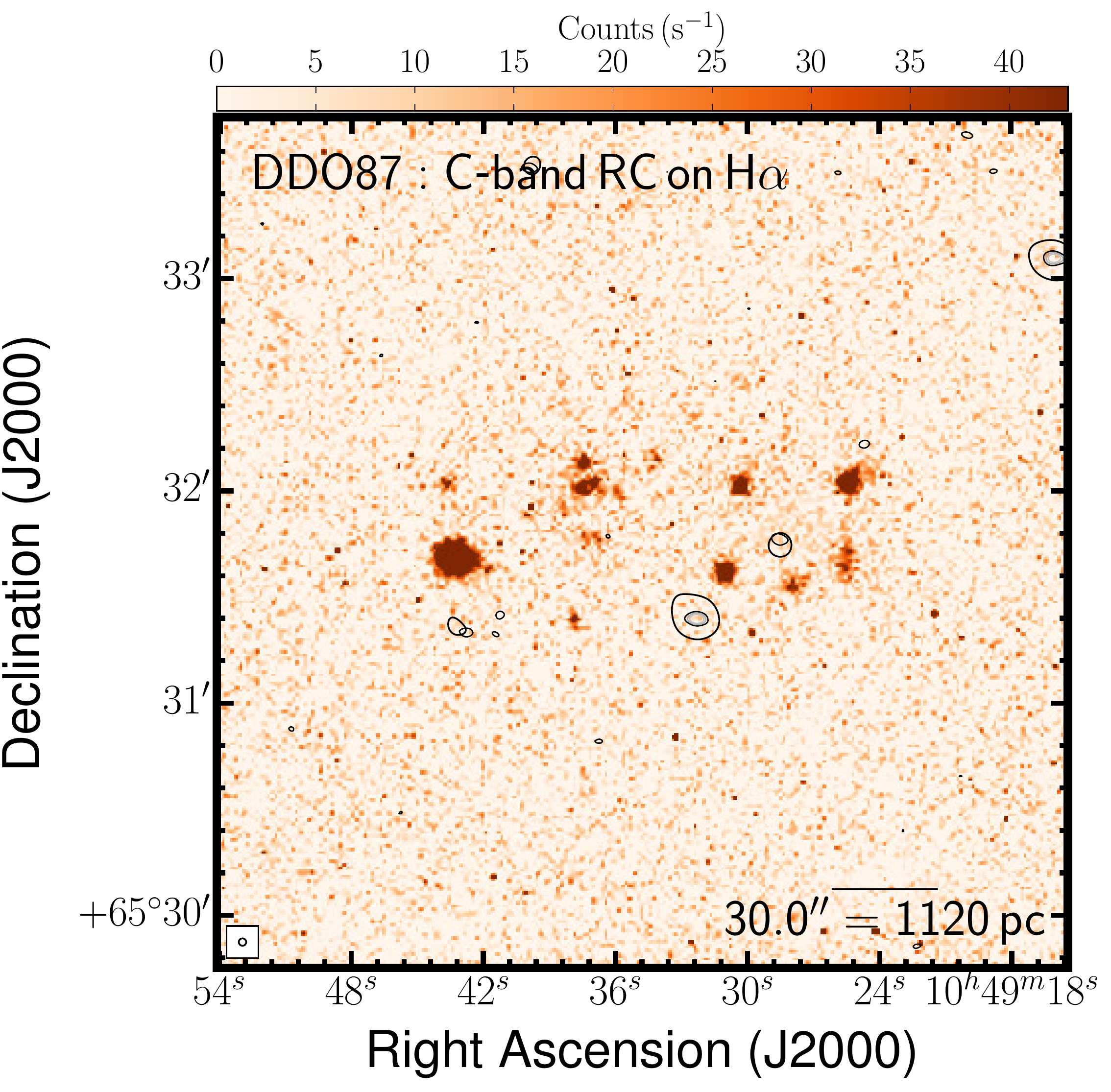} & \ 
    \includegraphics[width=0.31\linewidth,clip]{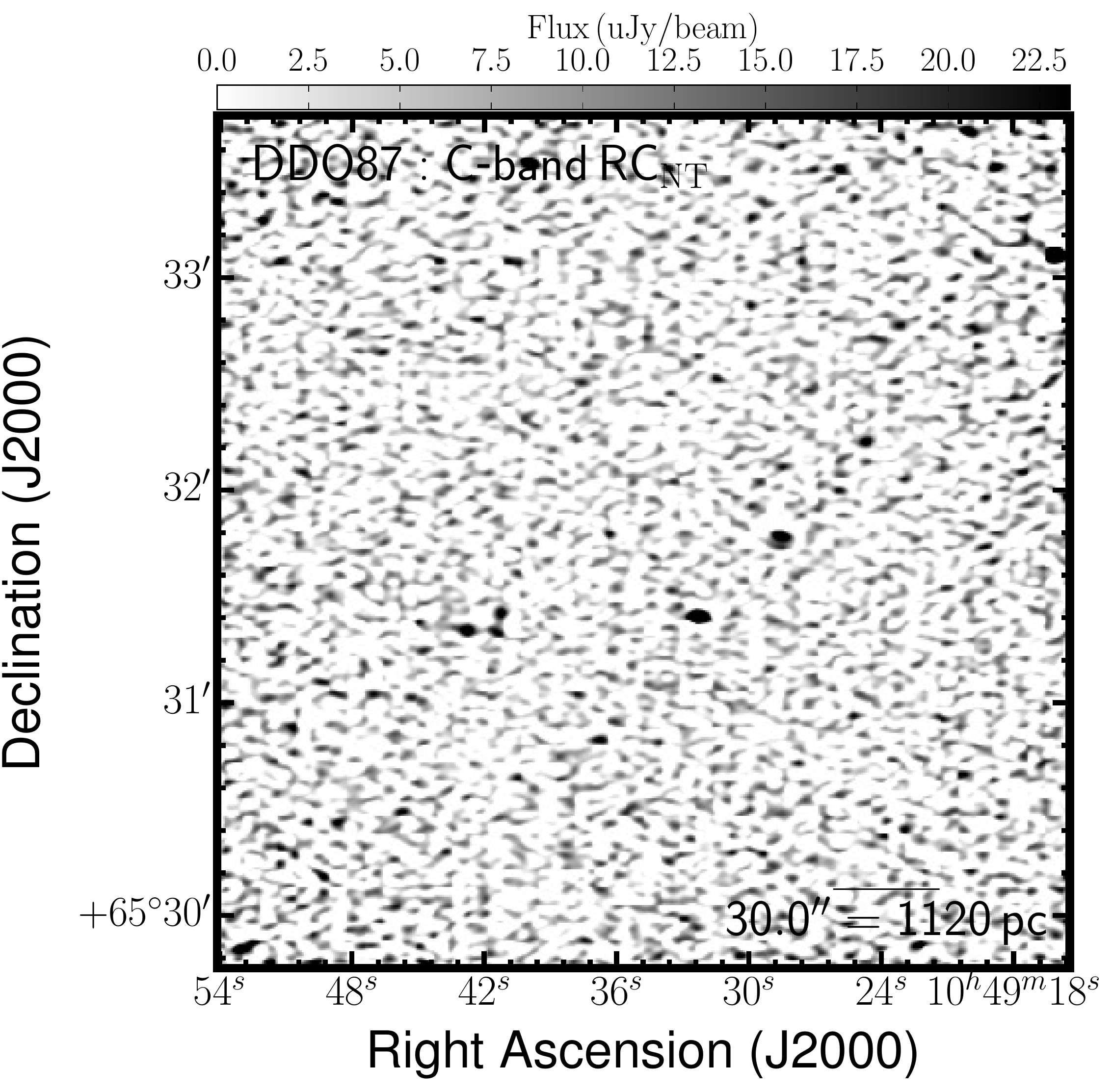} & \ 
    \includegraphics[width=0.31\linewidth,clip]{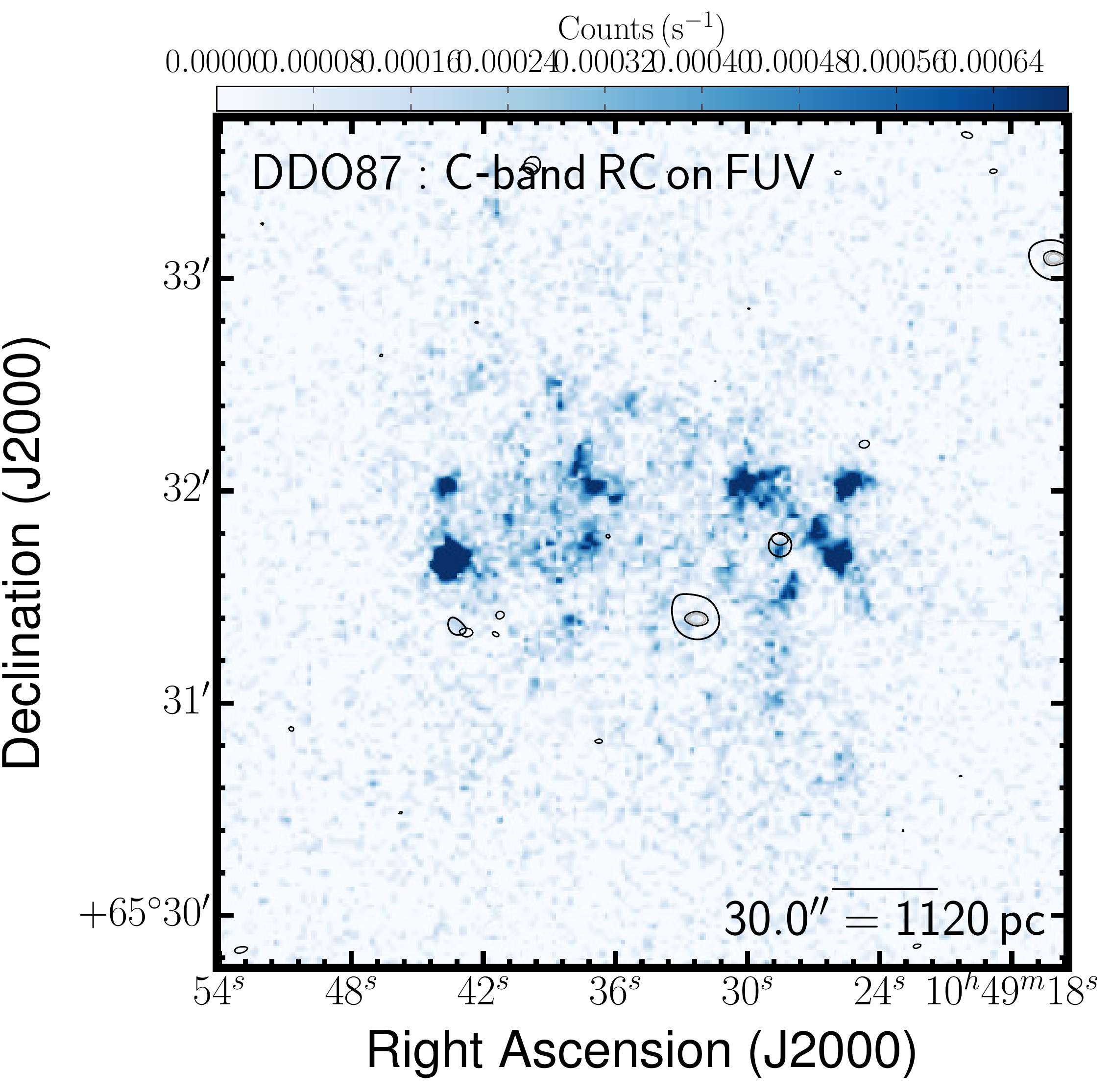} \\
    \includegraphics[width=0.31\linewidth,clip]{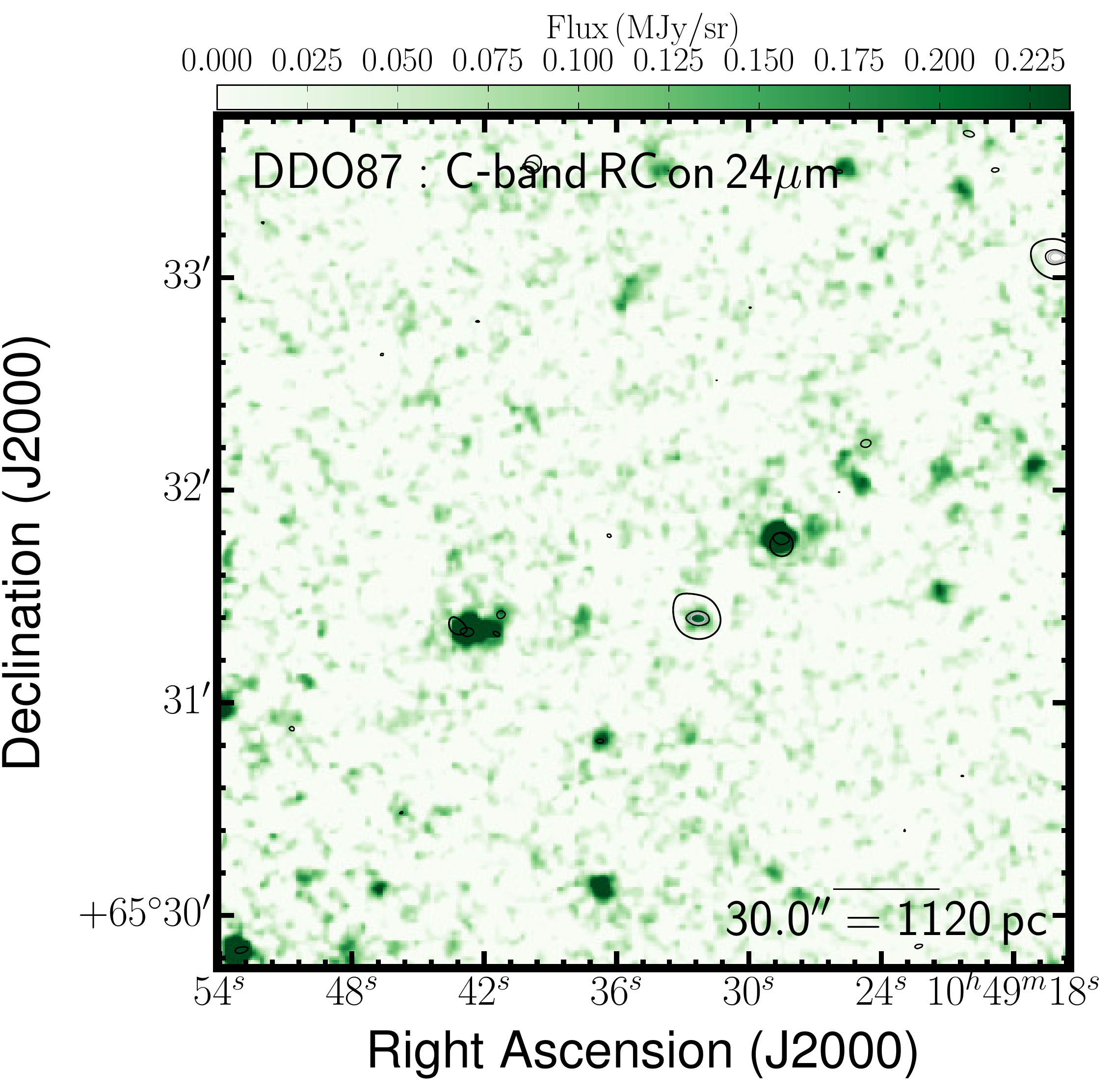} & \ 
    \includegraphics[width=0.31\linewidth,clip]{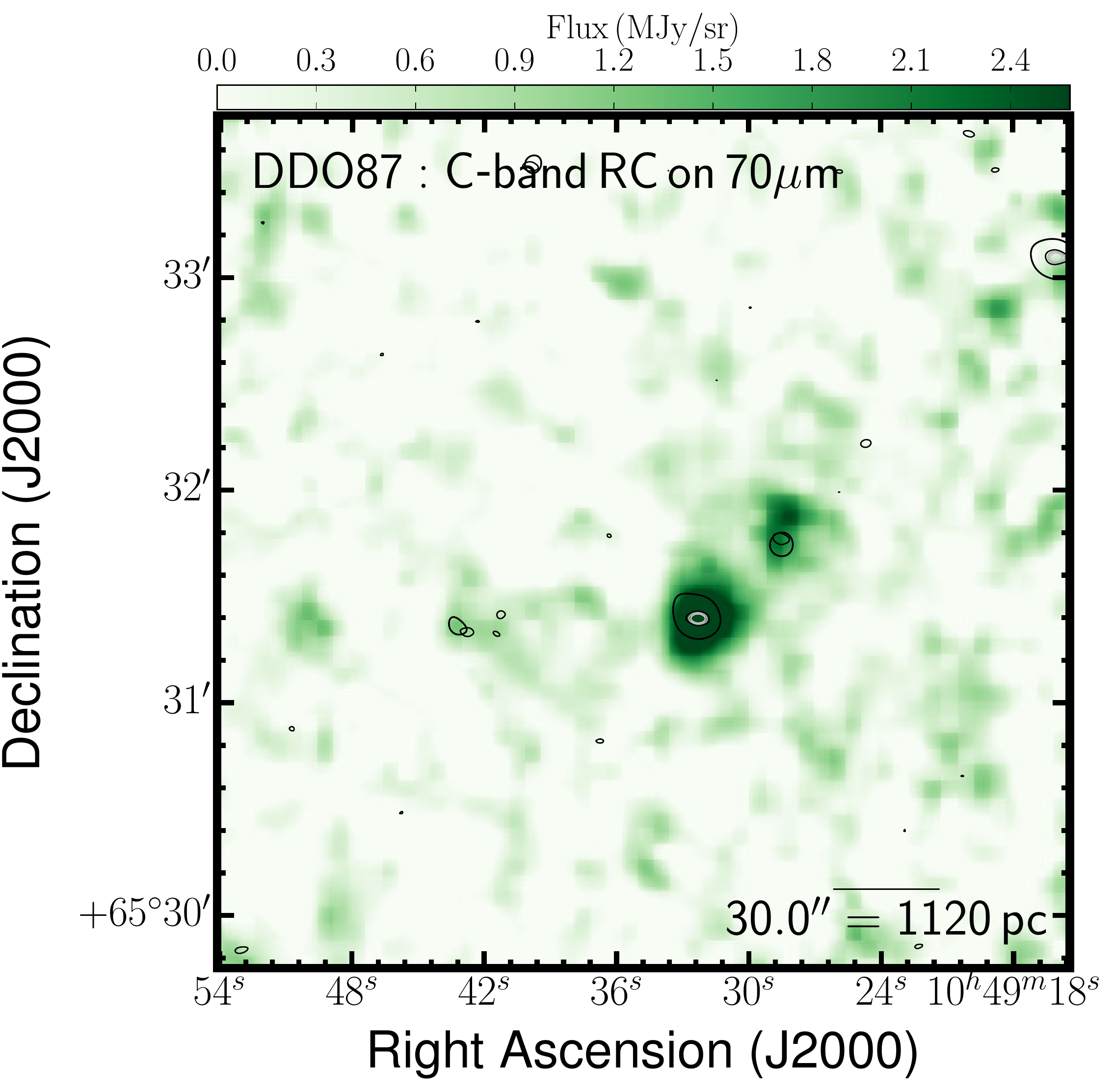} & \ 
    \includegraphics[width=0.31\linewidth,clip]{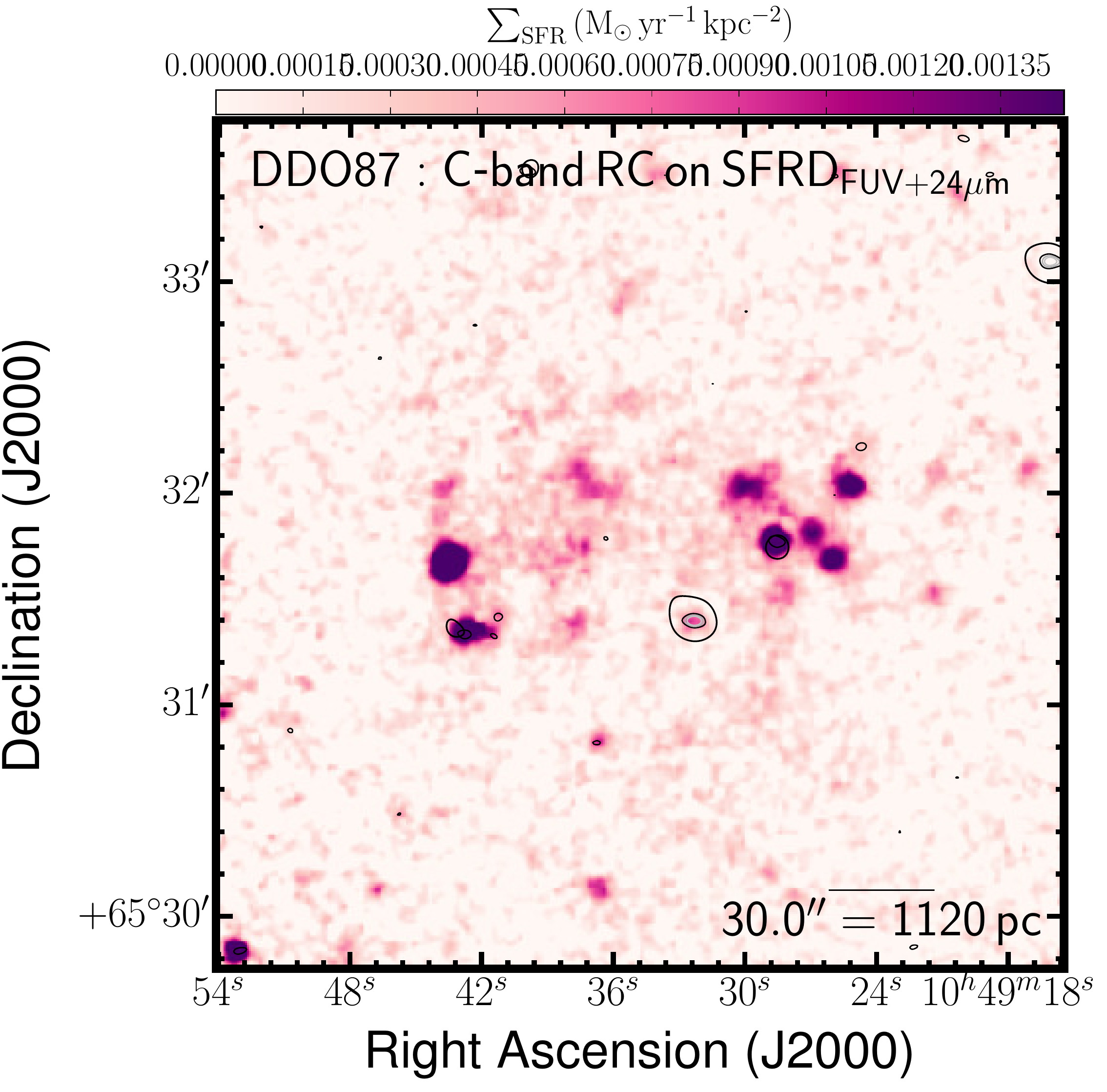} \\
  \end{tabular}
\caption[DDO\,87 images: RC, IR, optical, and FUV]{Multi-wavelength coverage of DDO 87 displaying a $4.0^\prime \times 4.0^\prime$ area. We show total RC flux density at the native resolution (top-left) and again with contours (top-centre). The RC contours are superposed on ancillary LITTLE THINGS images where possible: \halpha\ (middle-left); \RCNT\ obtained by subtracting the expected \RCT\ based on the \halpha-\RCT\ scaling factor of \cite{Deeg1997} from the total RC; {\em GALEX} FUV (middle-right); {\em Spitzer} 24\micron\ (bottom-left); {\em Spitzer} 70\micron\ (bottom-centre); FUV$+24{\rm \mu m}$--inferred SFRD from \citealp{Leroy2012} (bottom-right). We also show the RC that was isolated by the RC--based masking technique (top-right).}
  \label{figure:ddo87Cc_maps}
\end{figure}

\begin{figure}
  \begin{tabular}{ccc}
    \includegraphics[width=0.31\linewidth,clip]{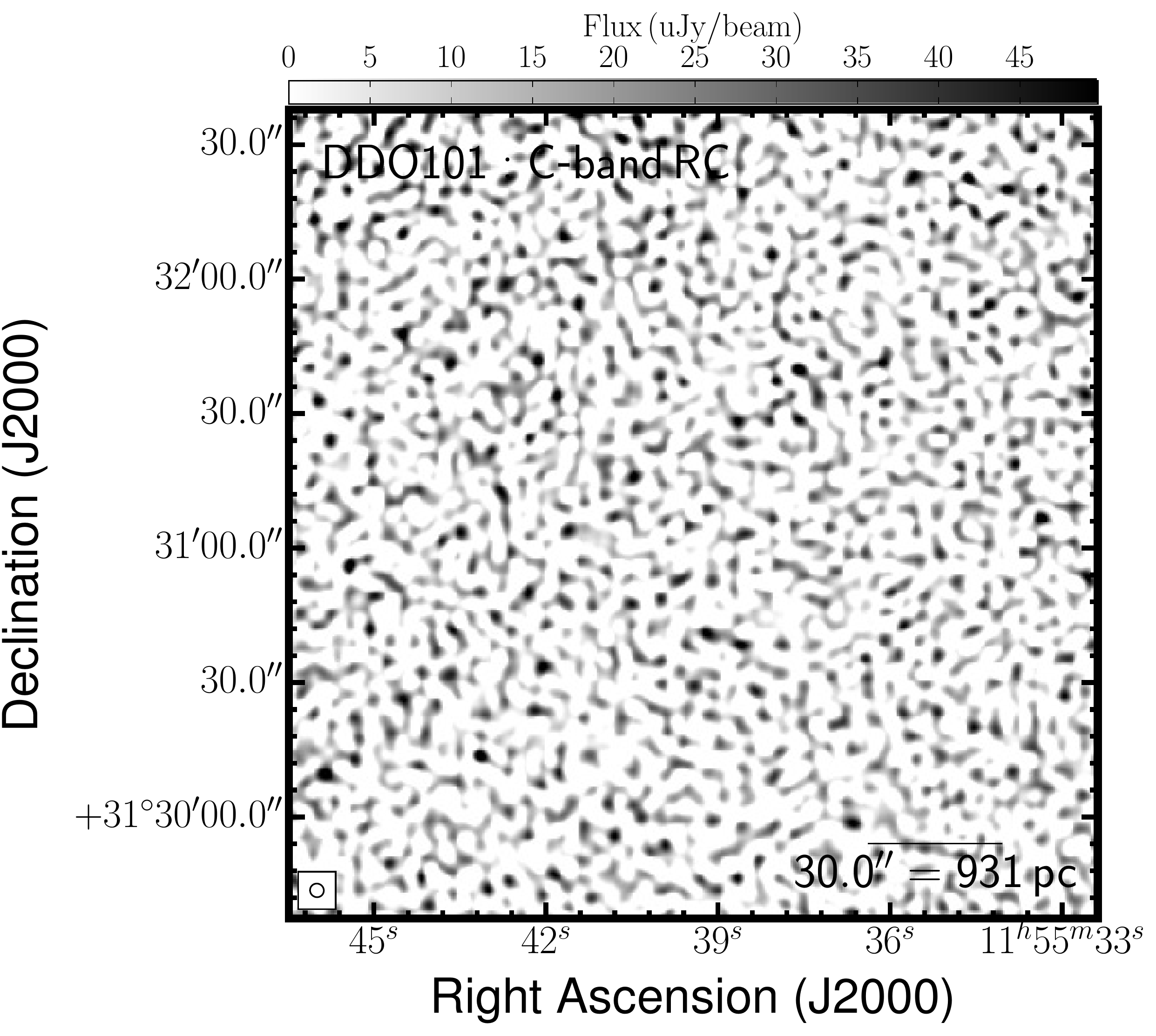} & \ 
    \includegraphics[width=0.31\linewidth,clip]{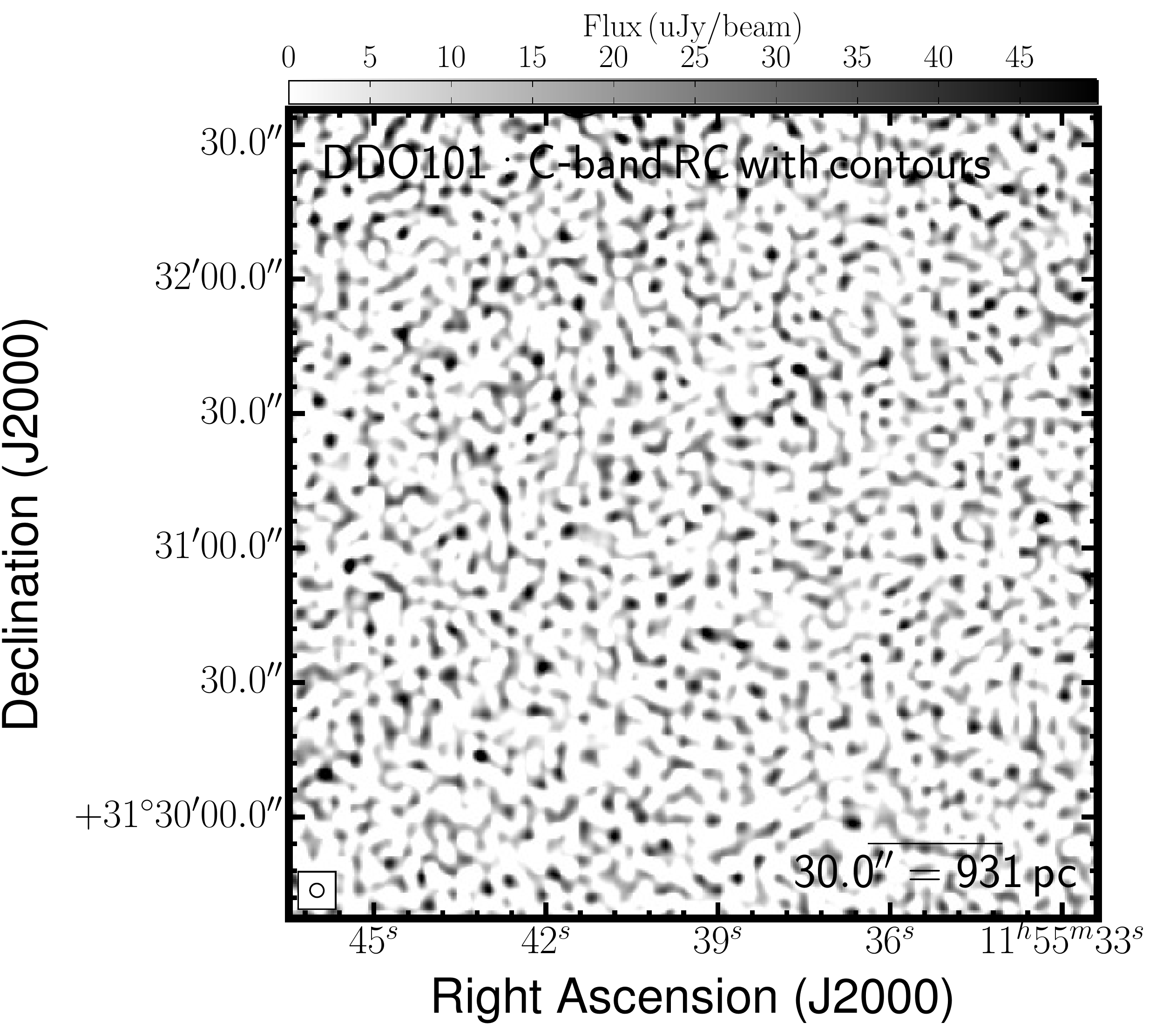} & \ 
    \includegraphics[width=0.31\linewidth,clip]{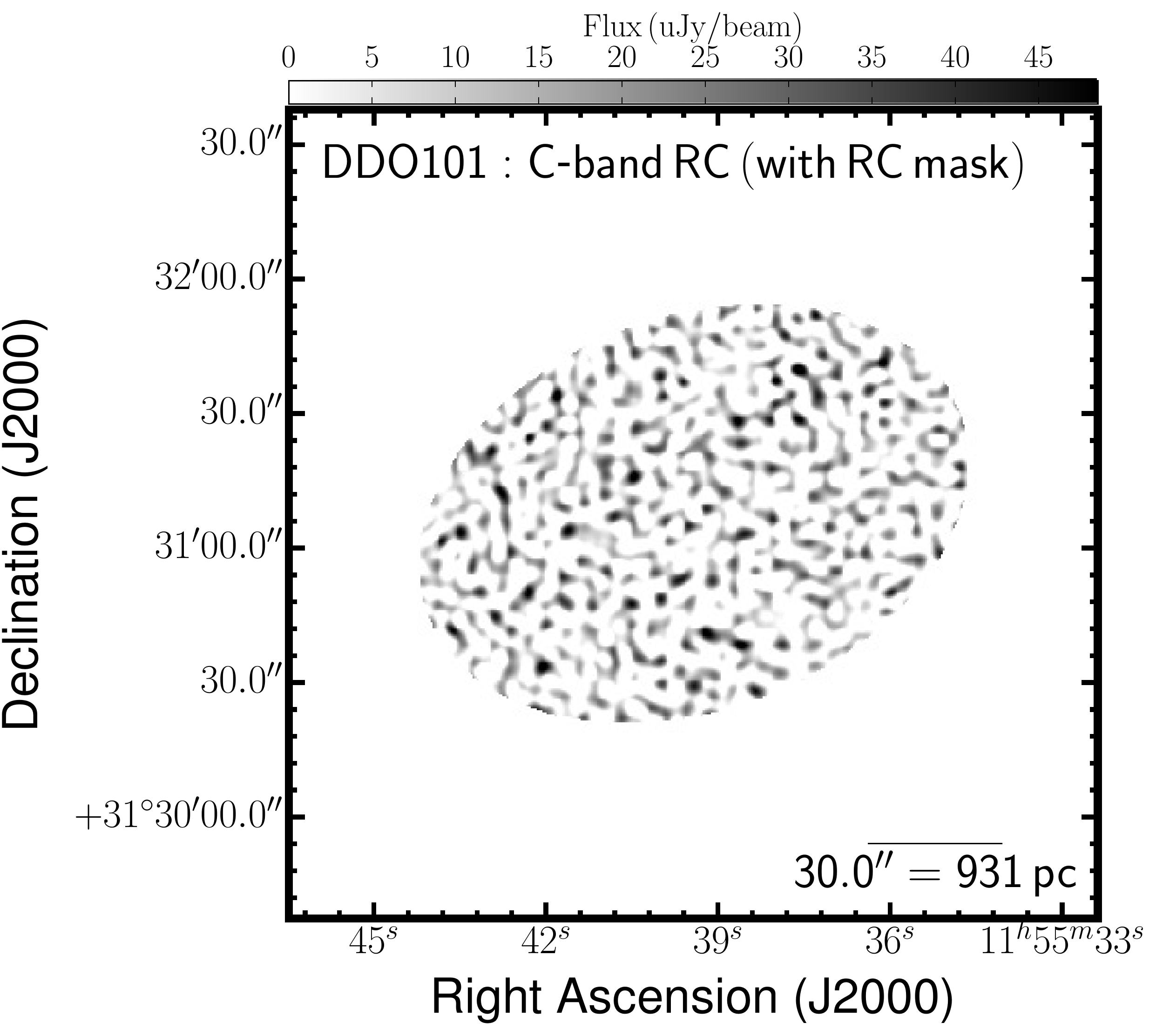} \\
    \includegraphics[width=0.31\linewidth,clip]{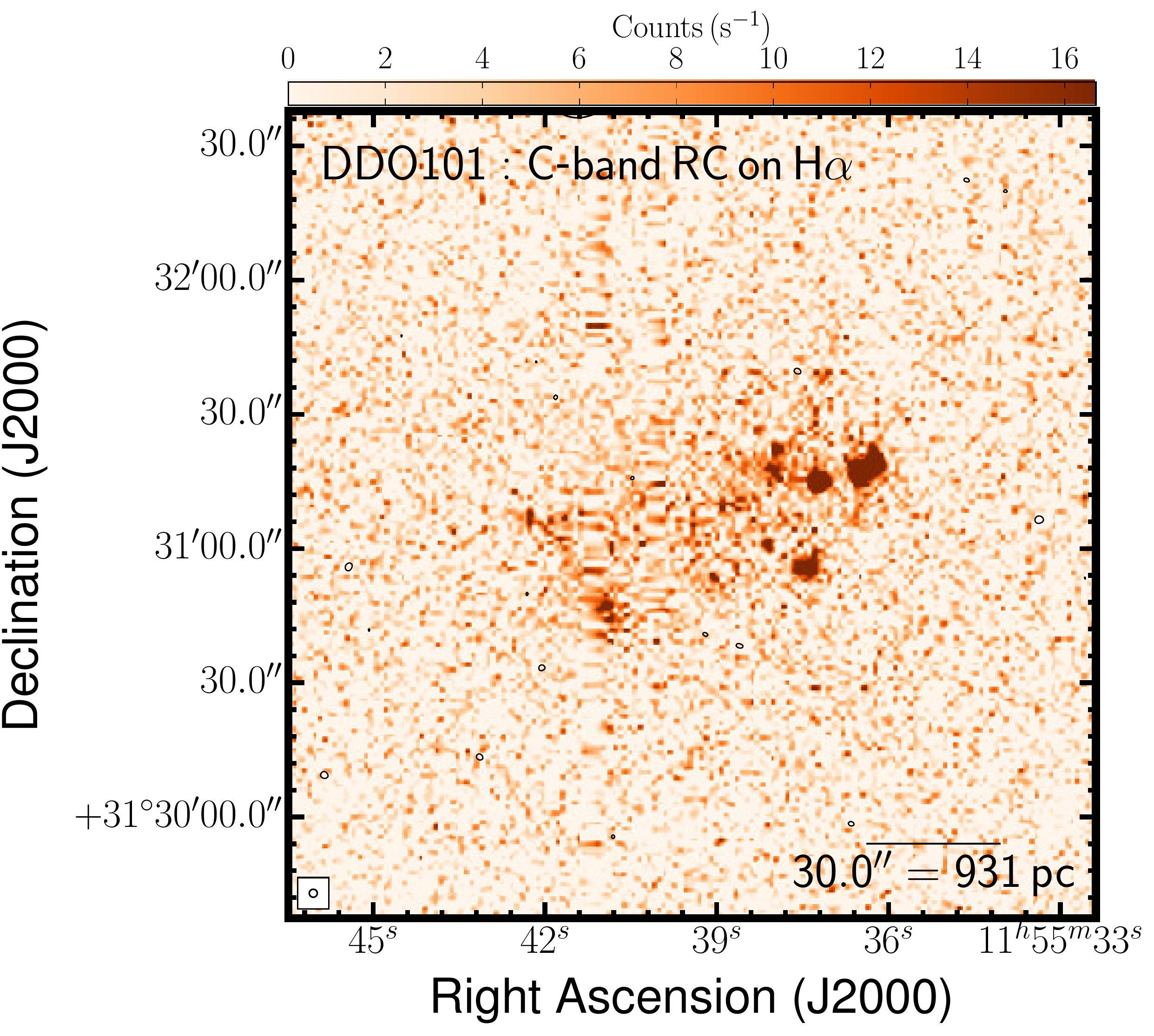} & \ 
    \includegraphics[width=0.31\linewidth,clip]{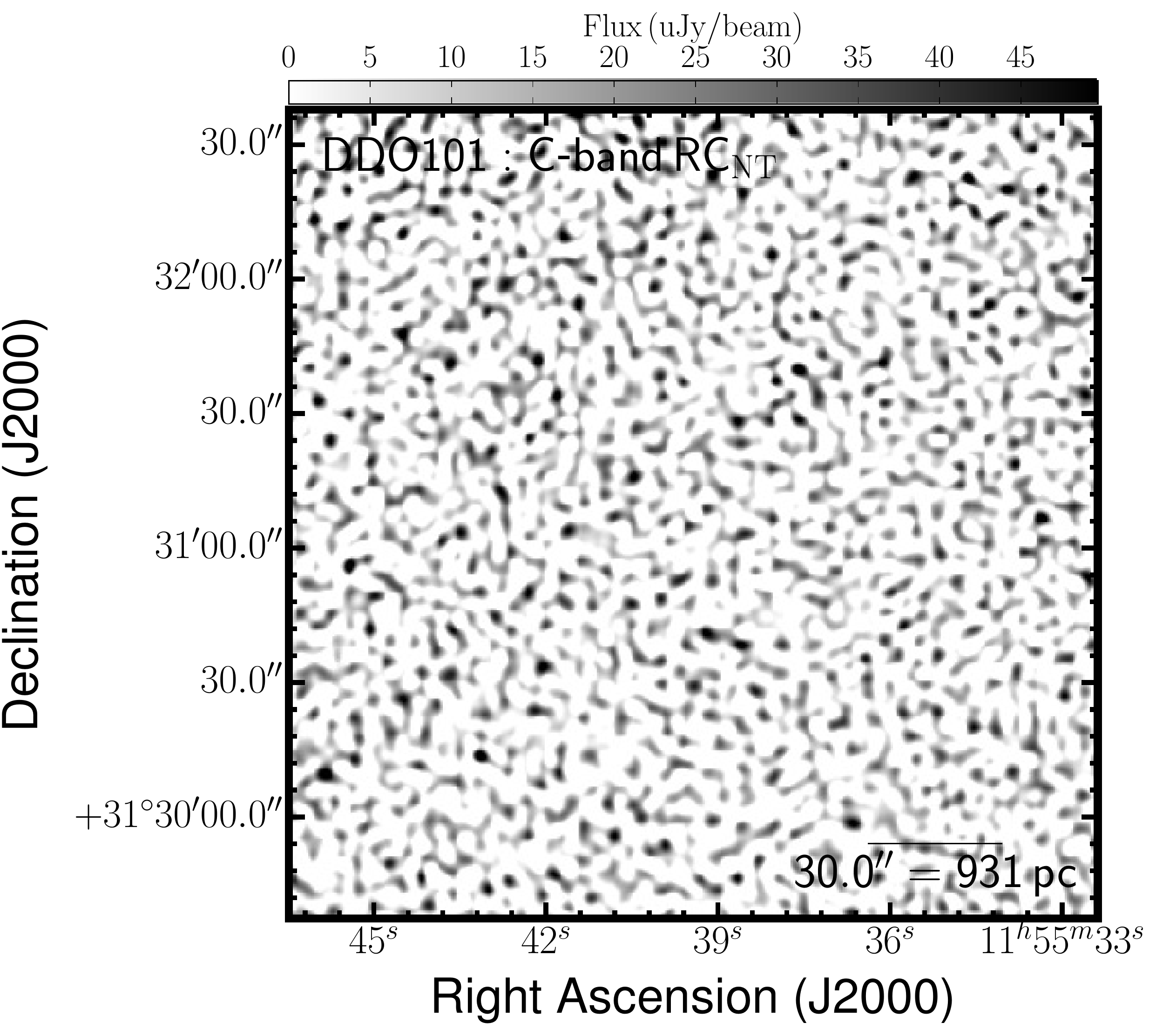} & \ 
    \includegraphics[width=0.31\linewidth,clip]{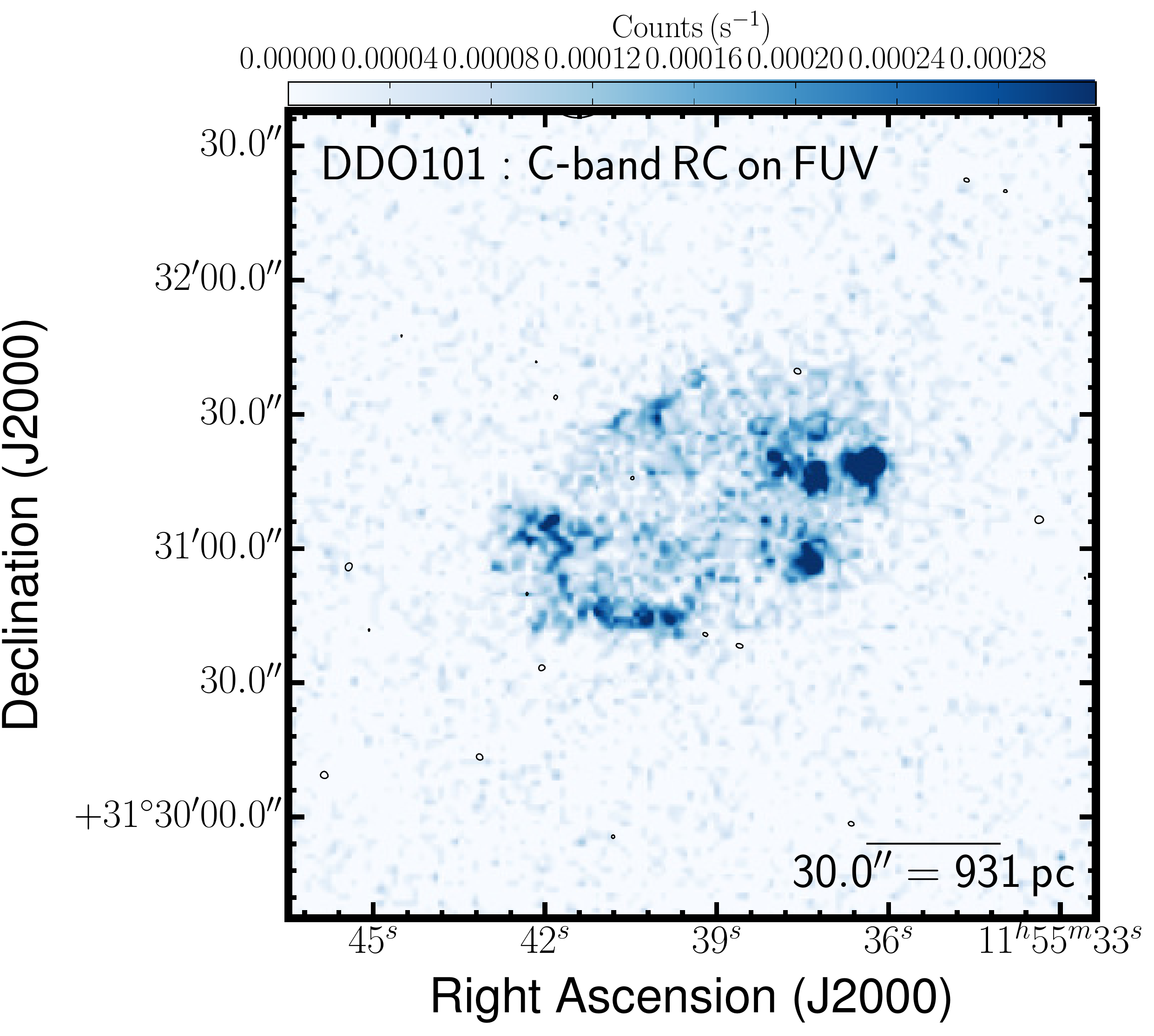} \\
    \includegraphics[width=0.31\linewidth,clip]{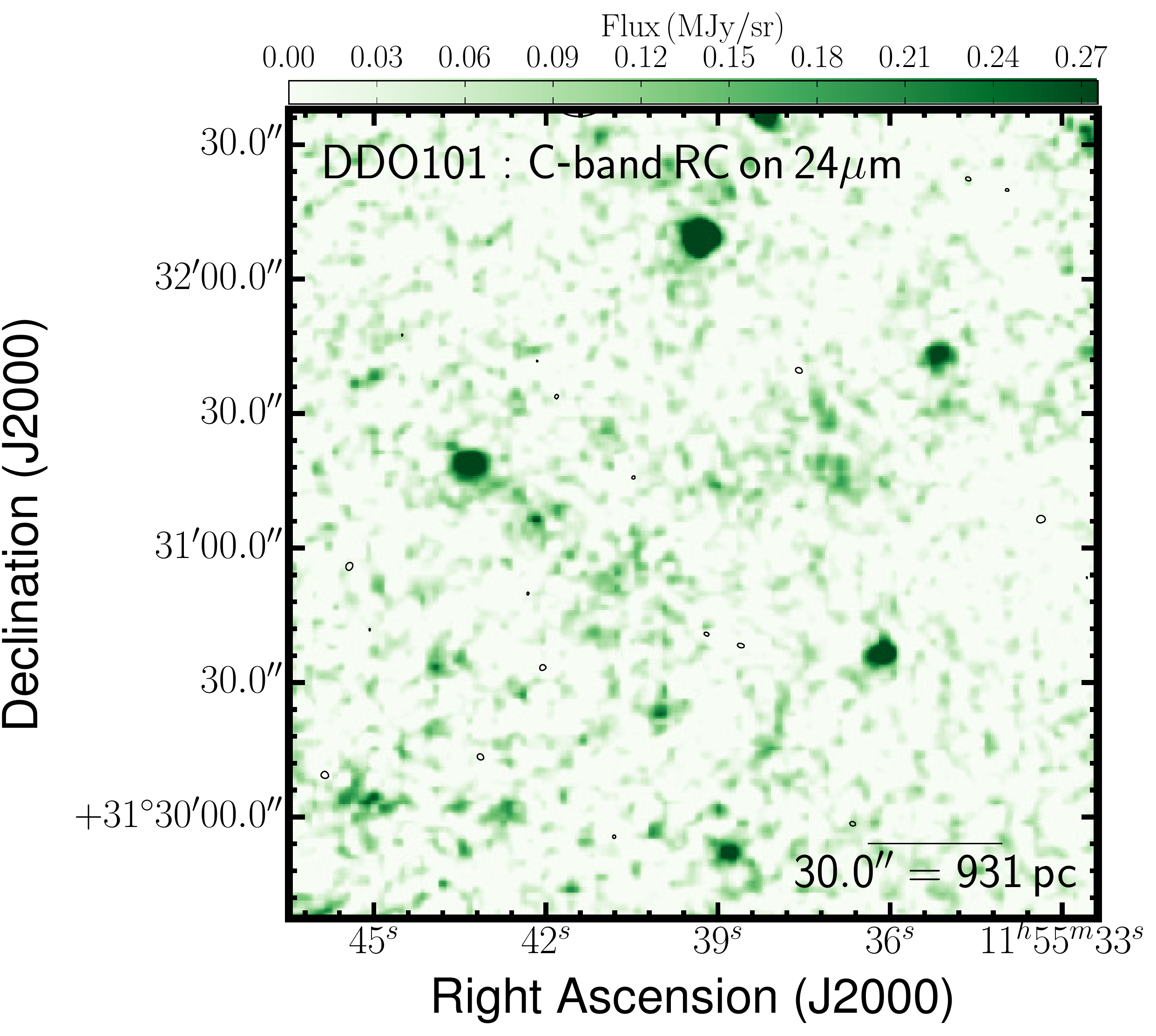} & \ 
    \includegraphics[width=0.31\linewidth,clip]{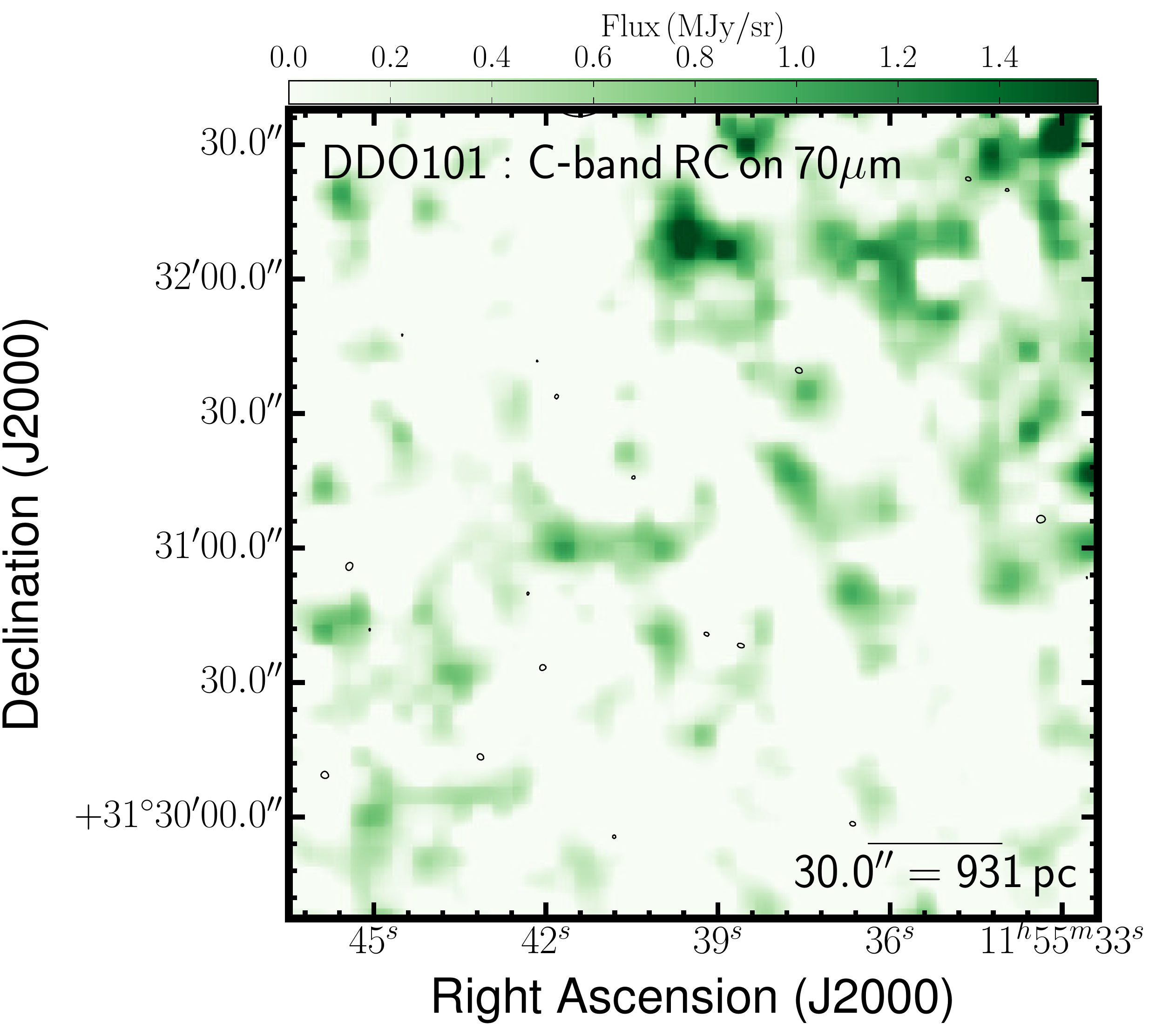} & \ 
    \includegraphics[width=0.31\linewidth,clip]{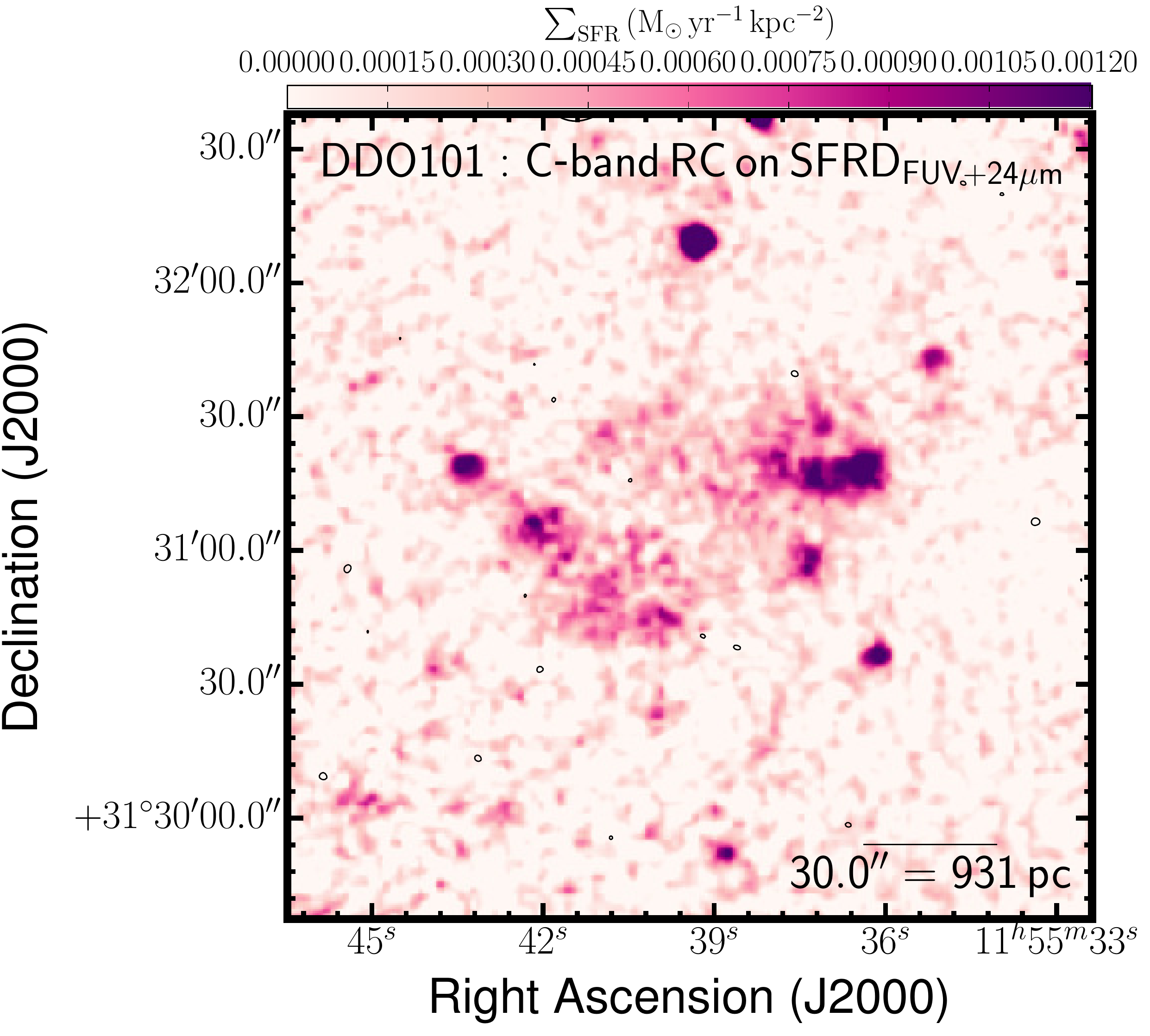} \\
  \end{tabular}
\caption[DDO\,101 images: RC, IR, optical, and FUV]{Multi-wavelength coverage of DDO 101 displaying a $3.0^\prime \times 3.0^\prime$ area. We show total RC flux density at the native resolution (top-left) and again with contours (top-centre). The RC contours are superposed on ancillary LITTLE THINGS images where possible: \halpha\ (middle-left); \RCNT\ obtained by subtracting the expected \RCT\ based on the \halpha-\RCT\ scaling factor of \cite{Deeg1997} from the total RC; {\em GALEX} FUV (middle-right); {\em Spitzer} 24\micron\ (bottom-left); {\em Spitzer} 70\micron\ (bottom-centre); FUV$+24{\rm \mu m}$--inferred SFRD from \citealp{Leroy2012} (bottom-right). We also show the RC that was isolated by the RC--based masking technique (top-right).}
  \label{figure:ddo101Cc_maps}
\end{figure}

\begin{figure}
  \begin{tabular}{ccc}
    \includegraphics[width=0.31\linewidth,clip]{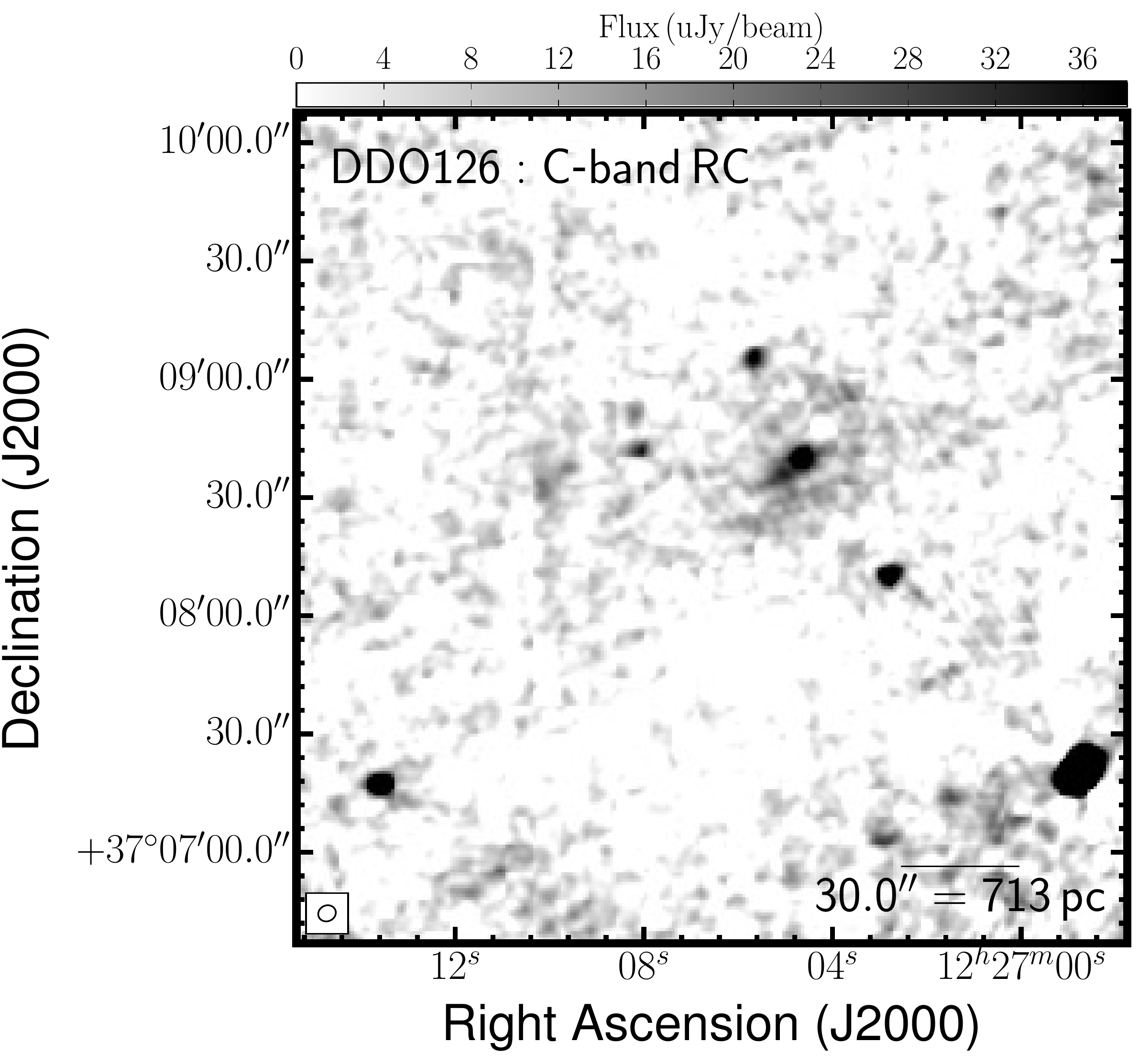} & \ 
    \includegraphics[width=0.31\linewidth,clip]{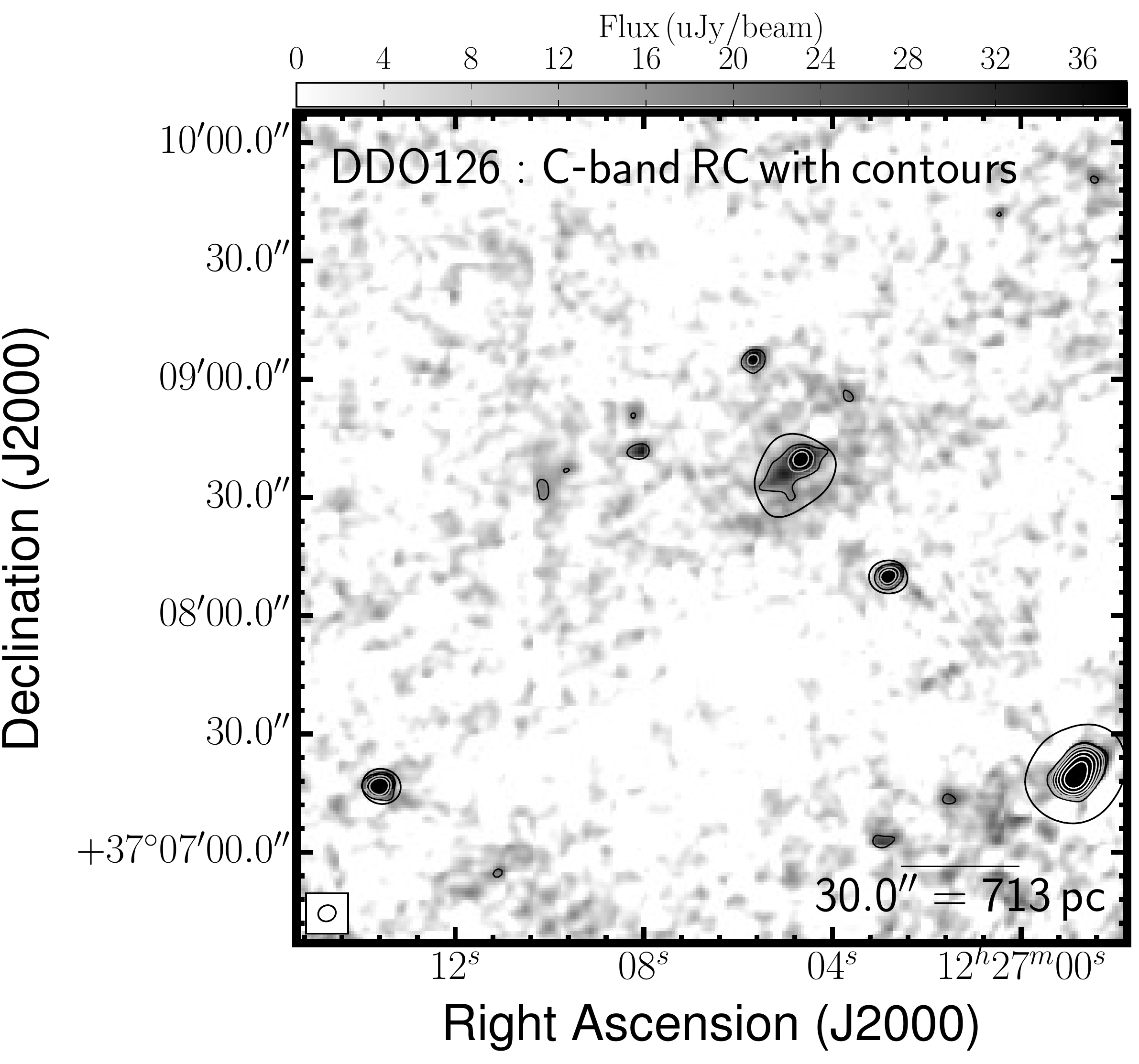} & \ 
    \includegraphics[width=0.31\linewidth,clip]{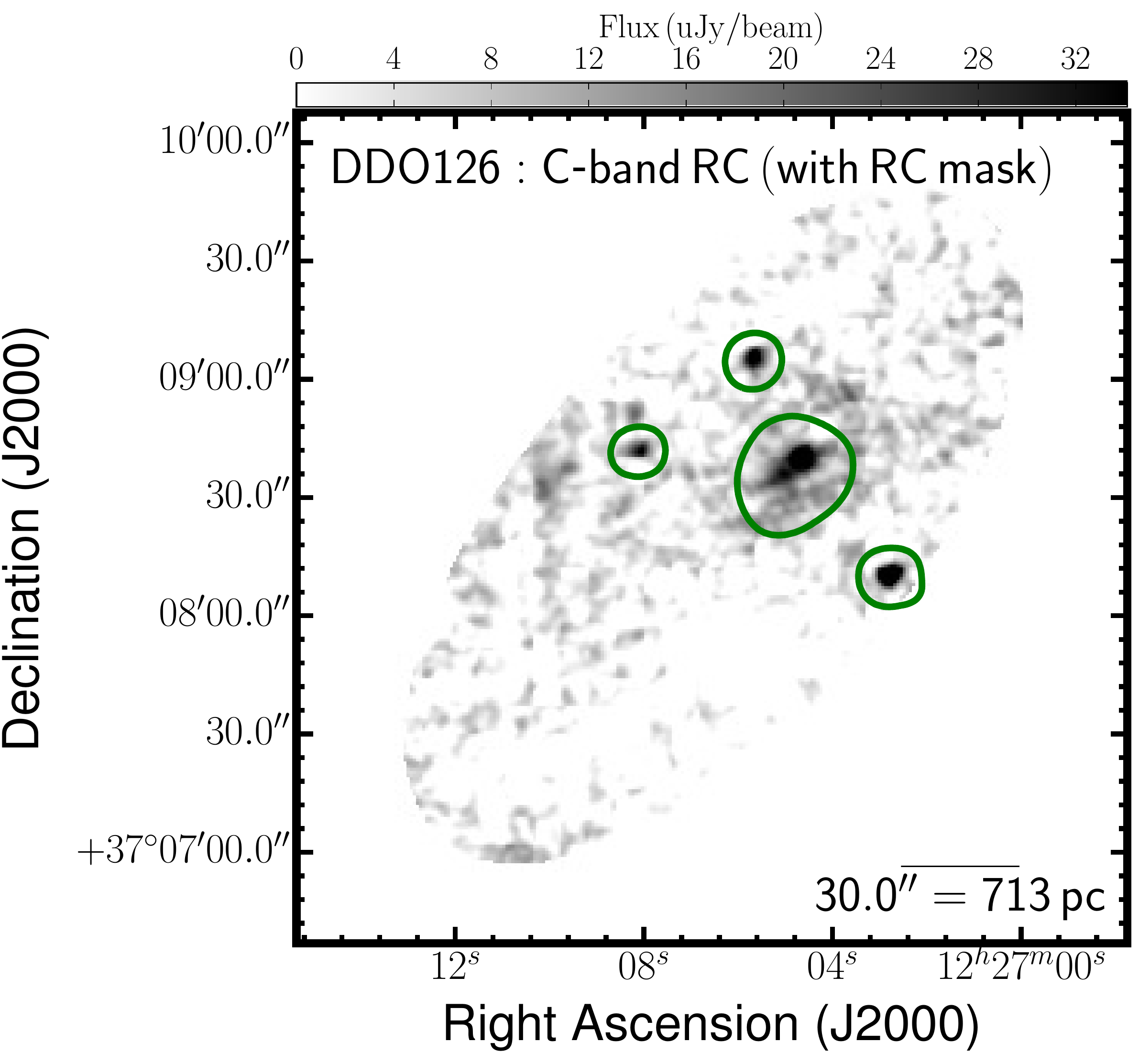} \\
    \includegraphics[width=0.31\linewidth,clip]{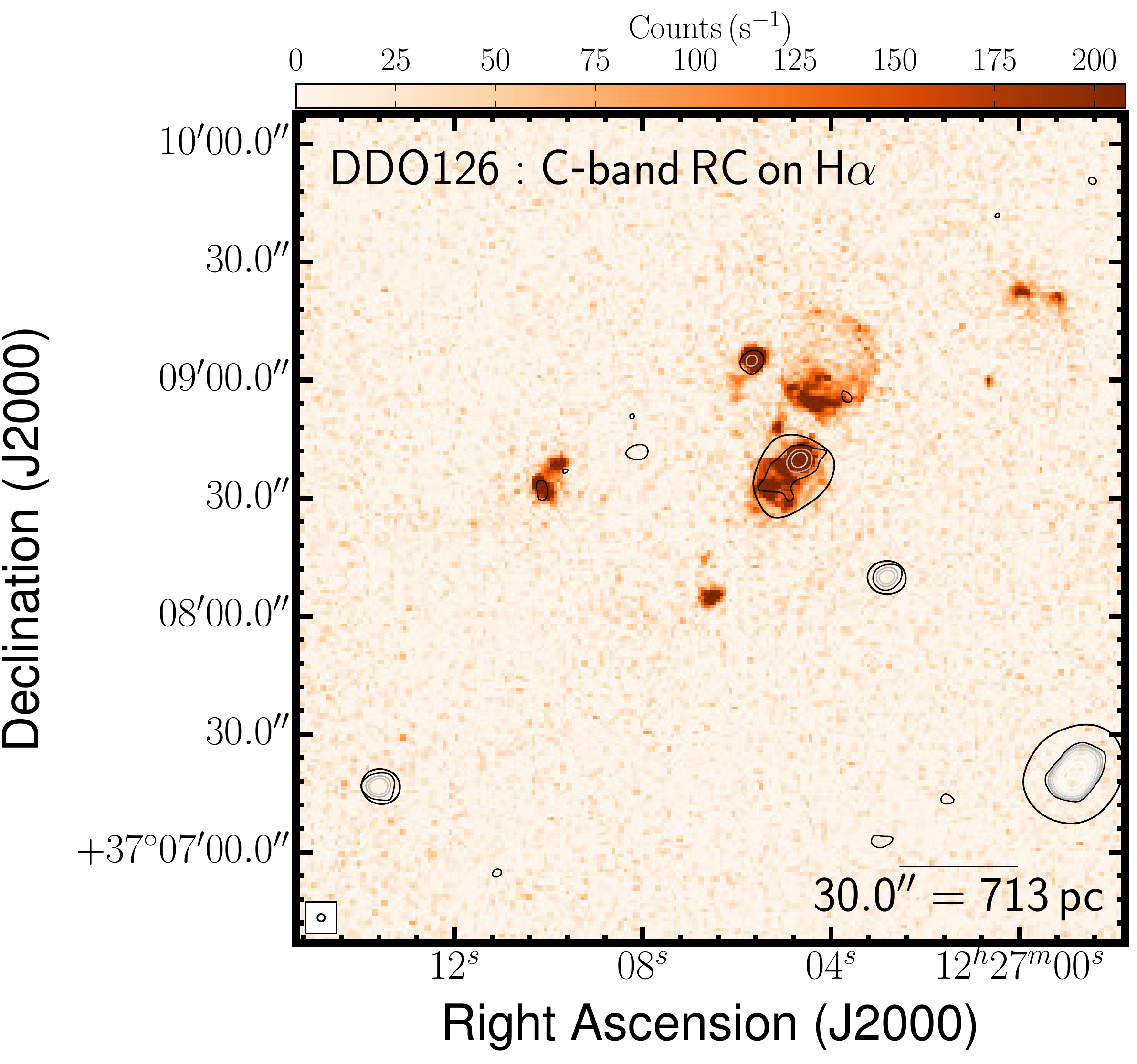} & \ 
    \includegraphics[width=0.31\linewidth,clip]{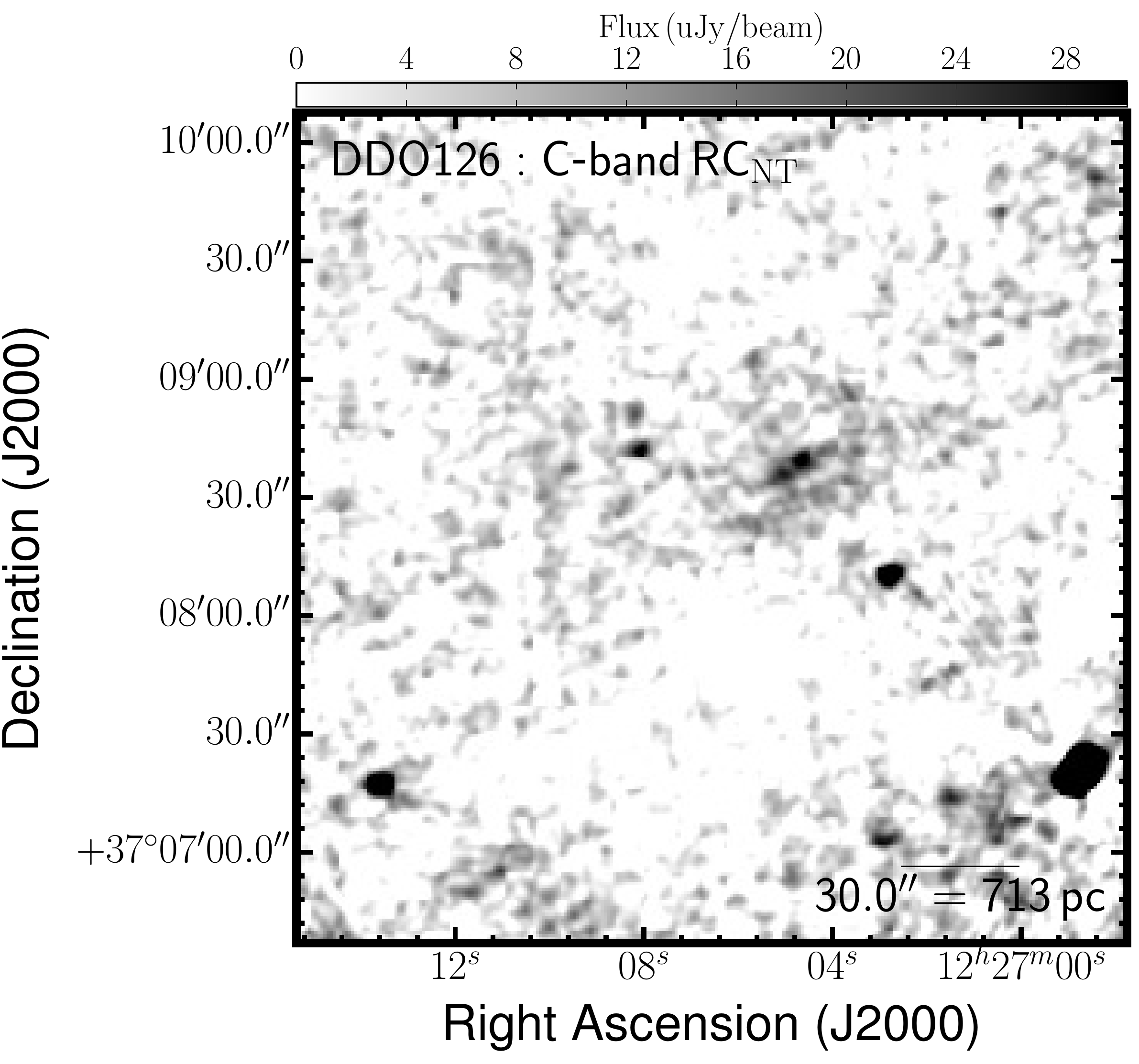} & \ 
    \includegraphics[width=0.31\linewidth,clip]{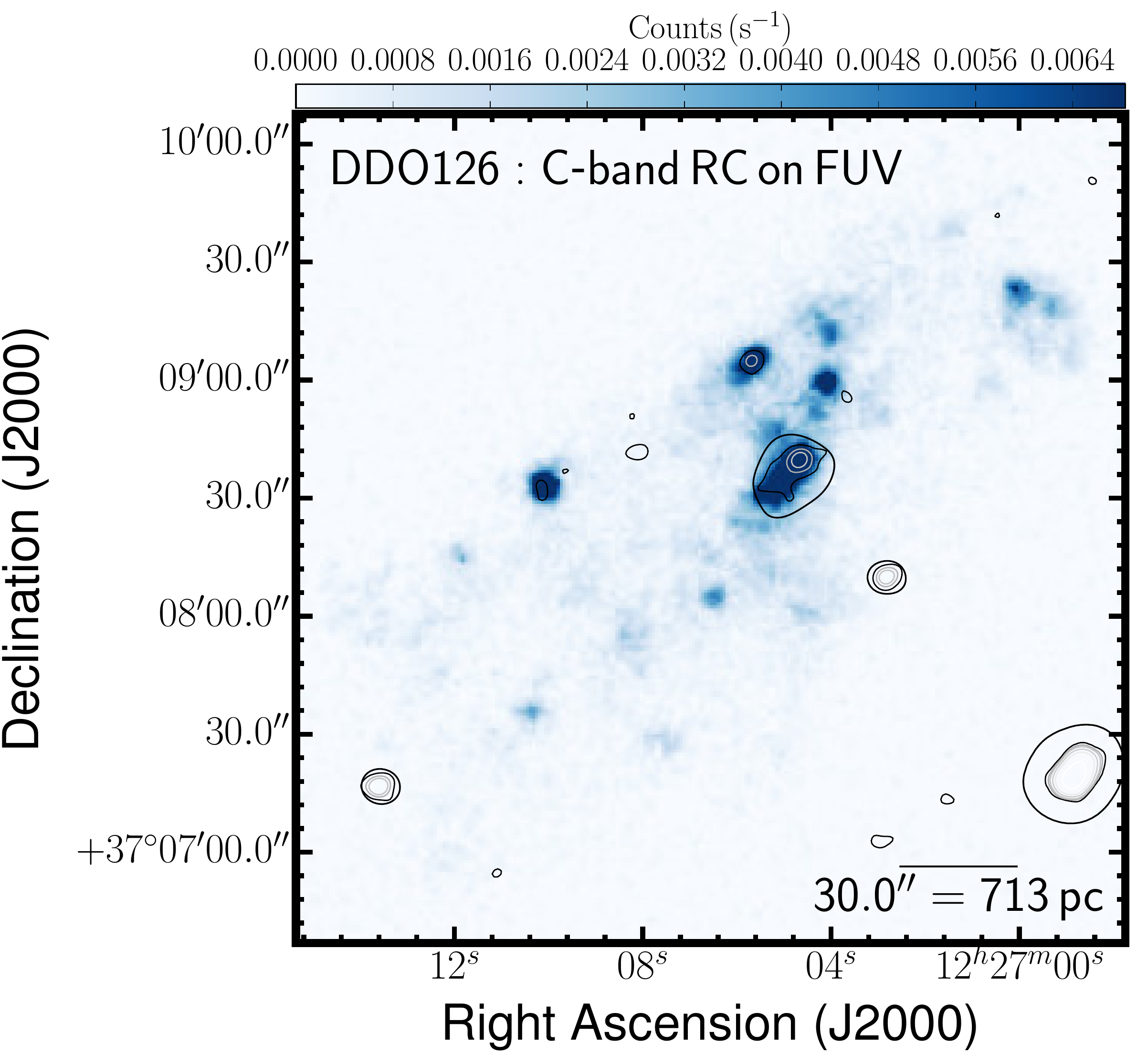} \\
    \includegraphics[width=0.31\linewidth,clip]{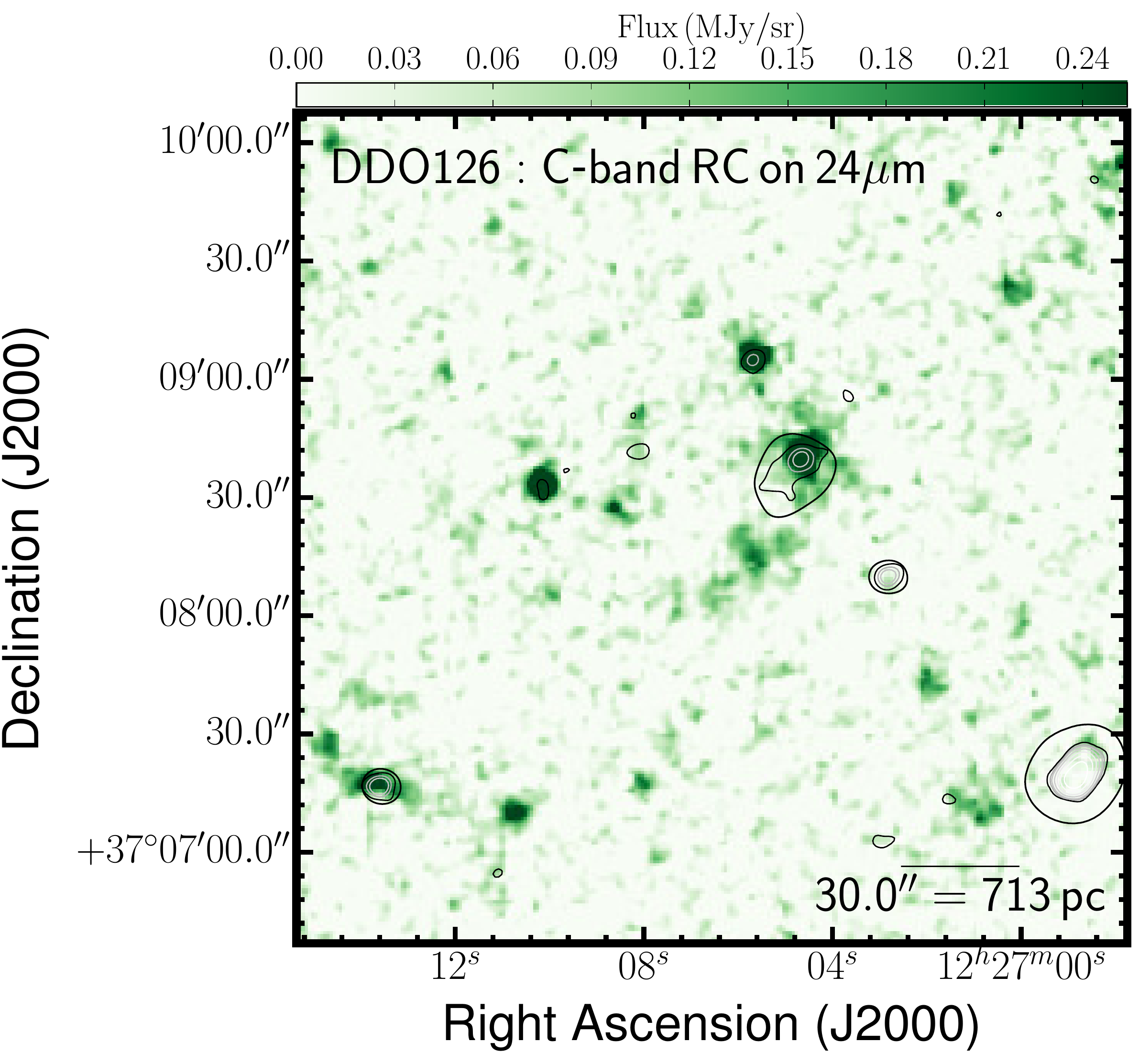} & \ 
    \includegraphics[width=0.31\linewidth,clip]{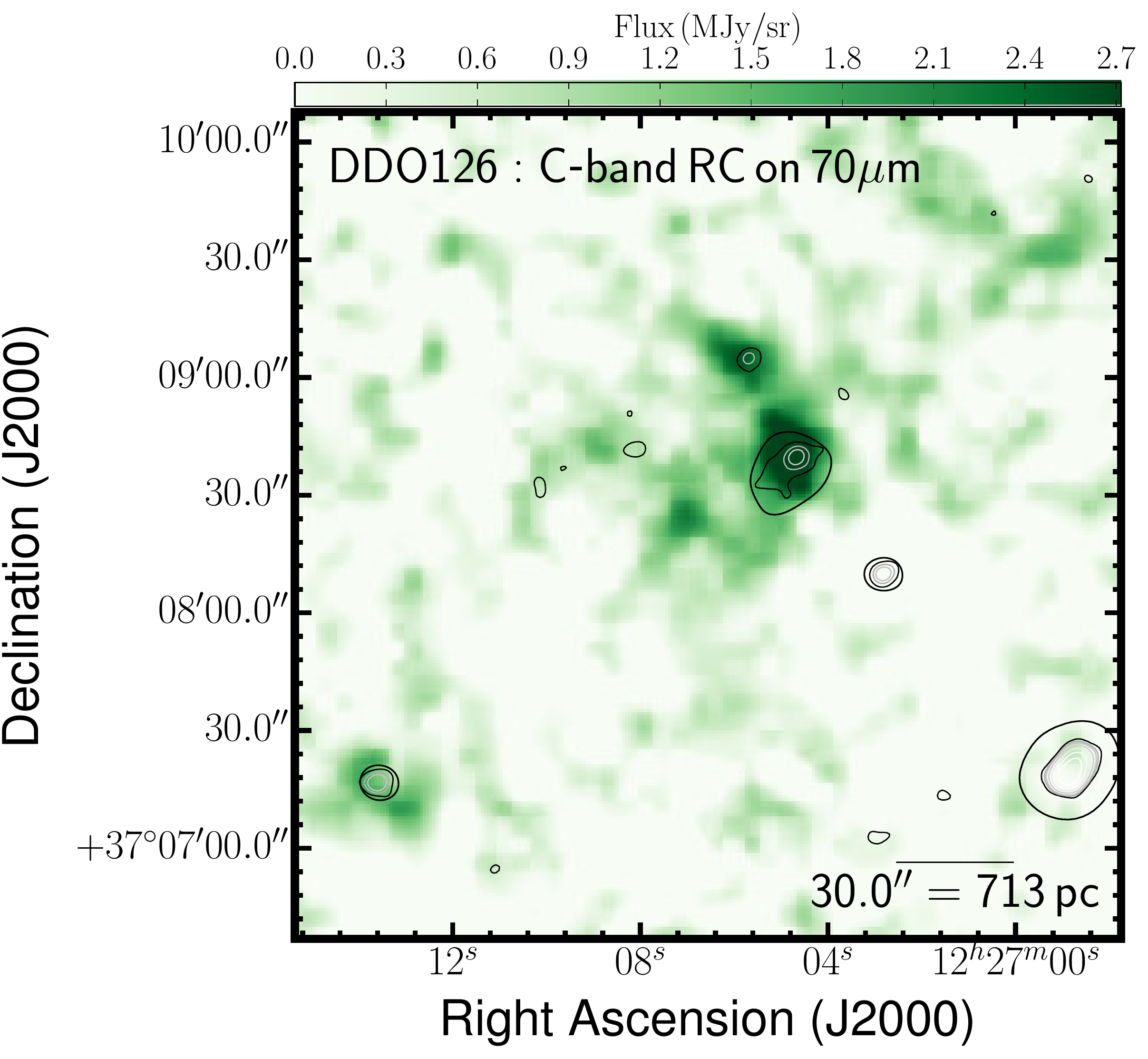} & \ 
    \includegraphics[width=0.31\linewidth,clip]{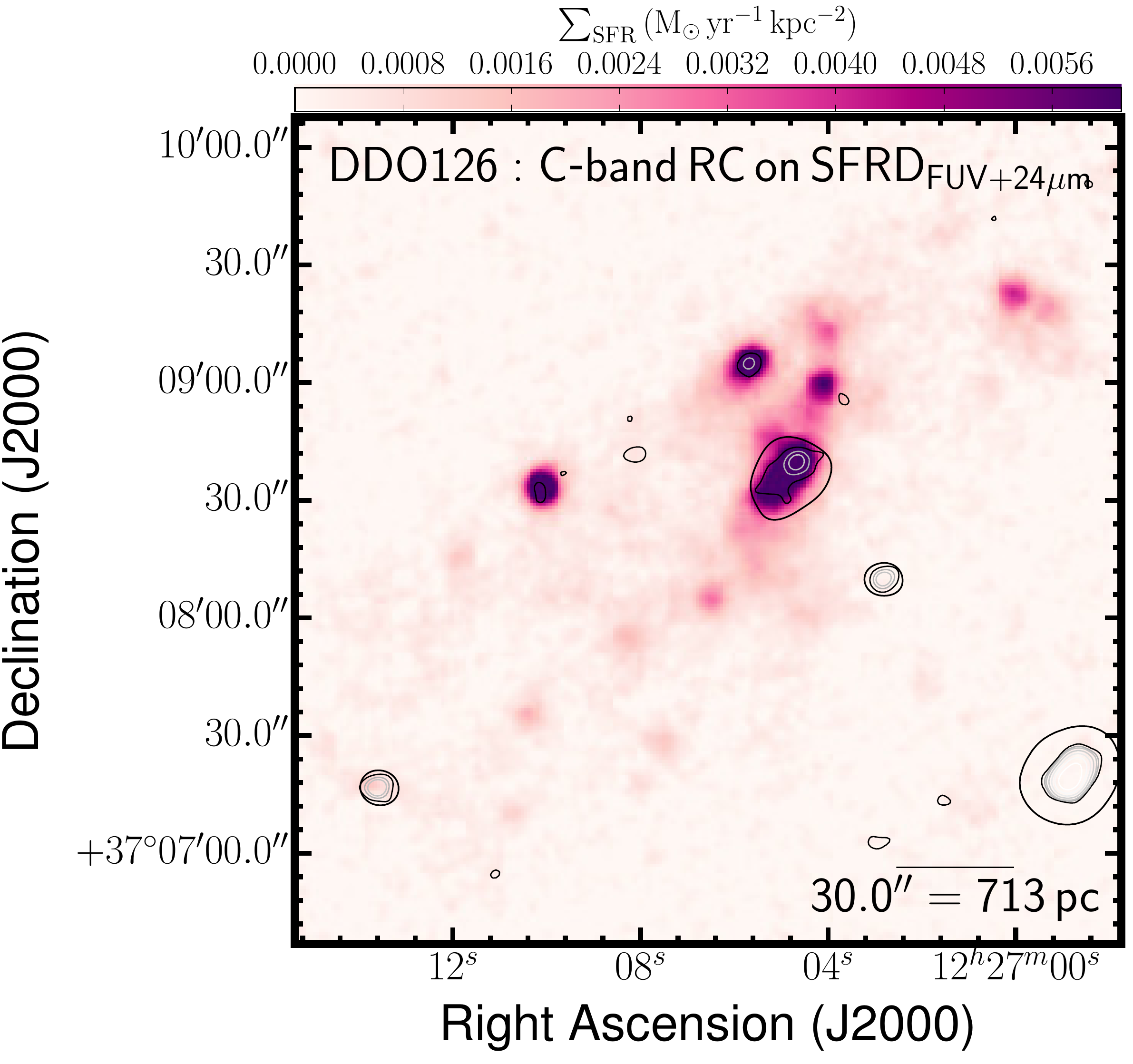} \\
  \end{tabular}
\caption[DDO\,126 images: RC, IR, optical, and FUV]{Multi-wavelength coverage of DDO 126 displaying a $3.5^\prime \times 3.5^\prime$ area. We show total RC flux density at the native resolution (top-left) and again with contours (top-centre). The RC contours are superposed on ancillary LITTLE THINGS images where possible: \halpha\ (middle-left); \RCNT\ obtained by subtracting the expected \RCT\ based on the \halpha-\RCT\ scaling factor of \cite{Deeg1997} from the total RC; {\em GALEX} FUV (middle-right); {\em Spitzer} 24\micron\ (bottom-left); {\em Spitzer} 70\micron\ (bottom-centre); FUV$+24{\rm \mu m}$--inferred SFRD from \citealp{Leroy2012} (bottom-right). We also show the RC that was isolated by the RC--based masking technique (top-right).}
  \label{figure:ddo126Cc_maps}
\end{figure}

\begin{figure}
  \begin{tabular}{ccc}
    \includegraphics[width=0.31\linewidth,clip]{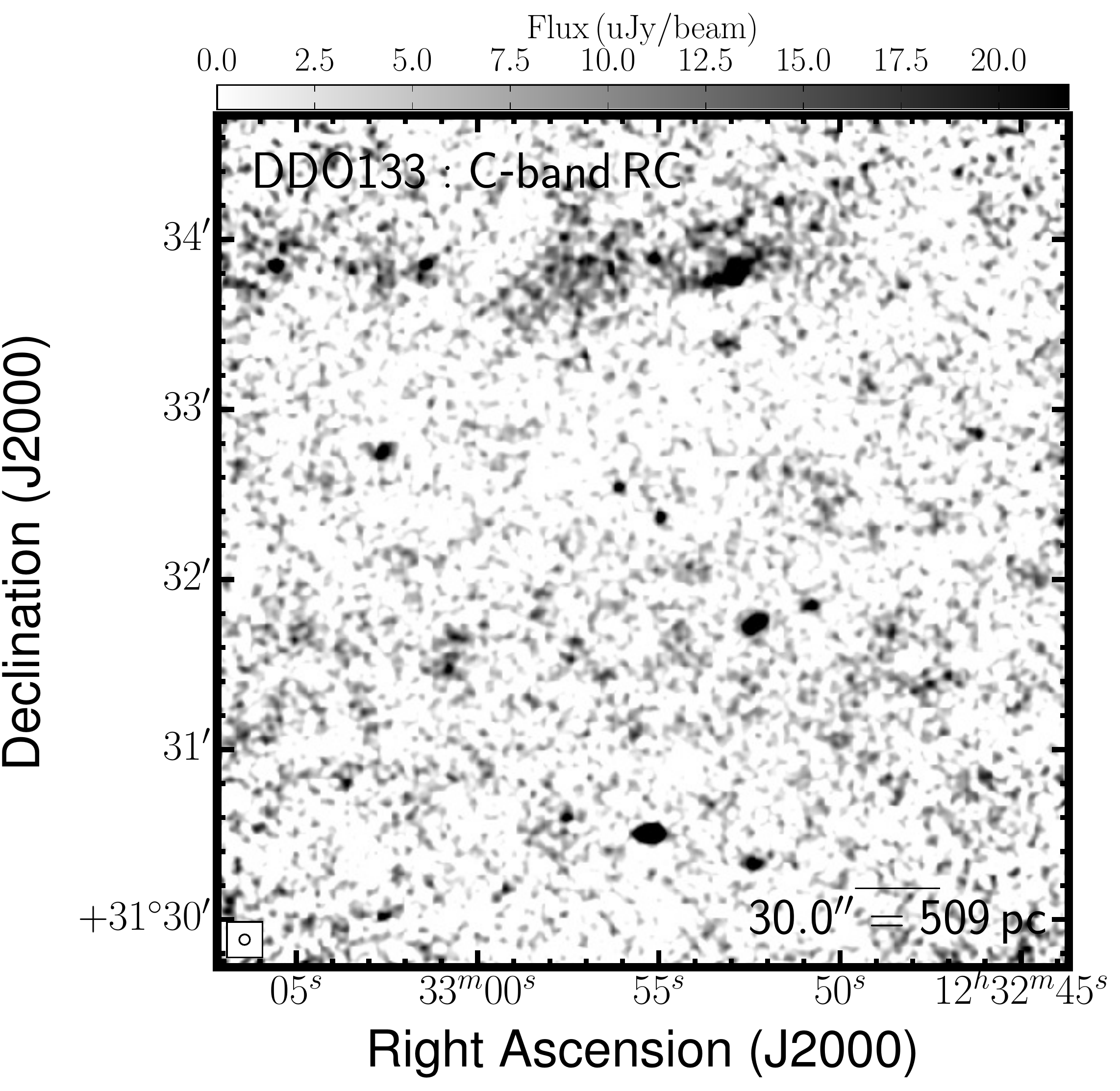} & \ 
    \includegraphics[width=0.31\linewidth,clip]{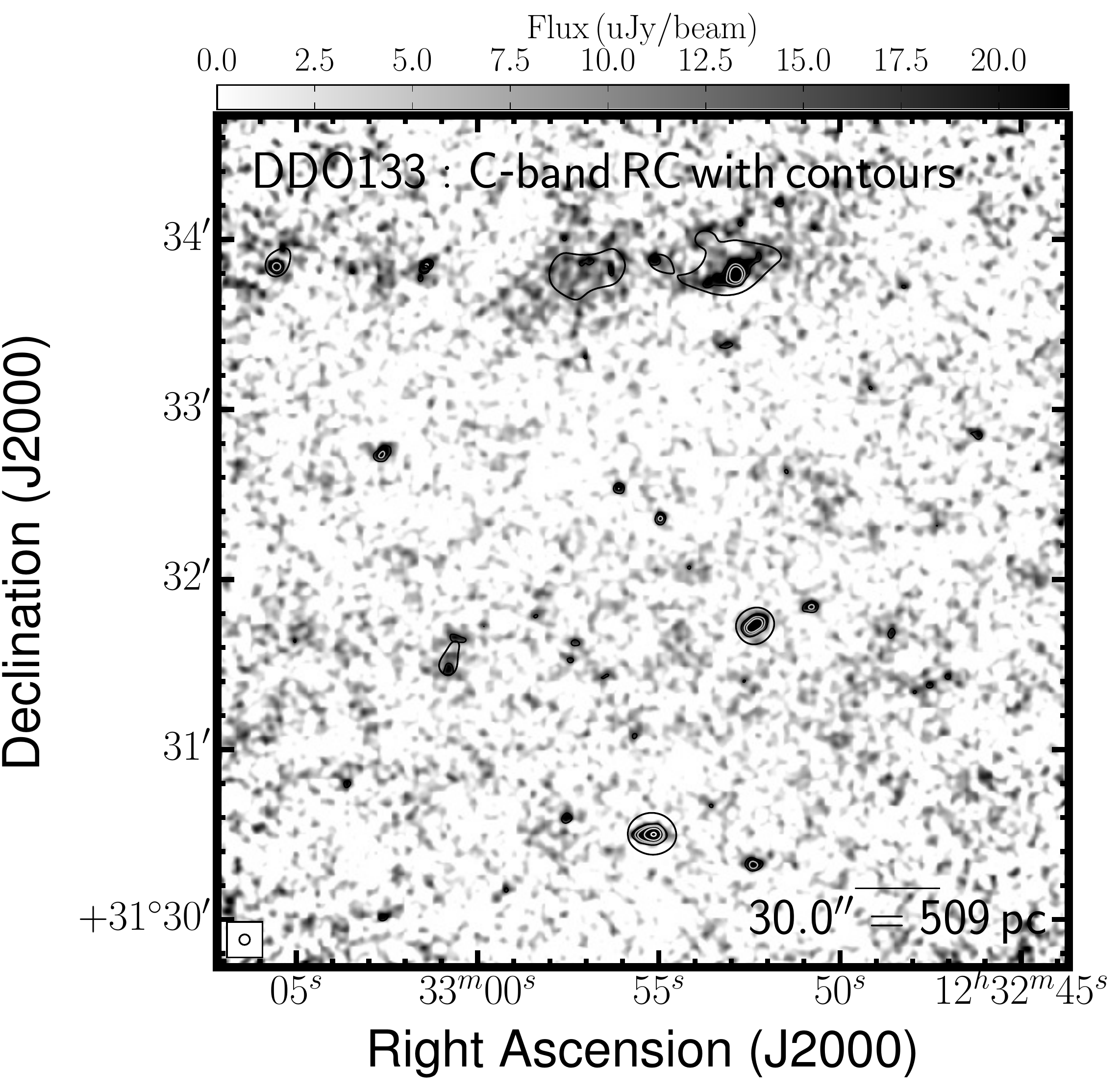} & \ 
    \includegraphics[width=0.31\linewidth,clip]{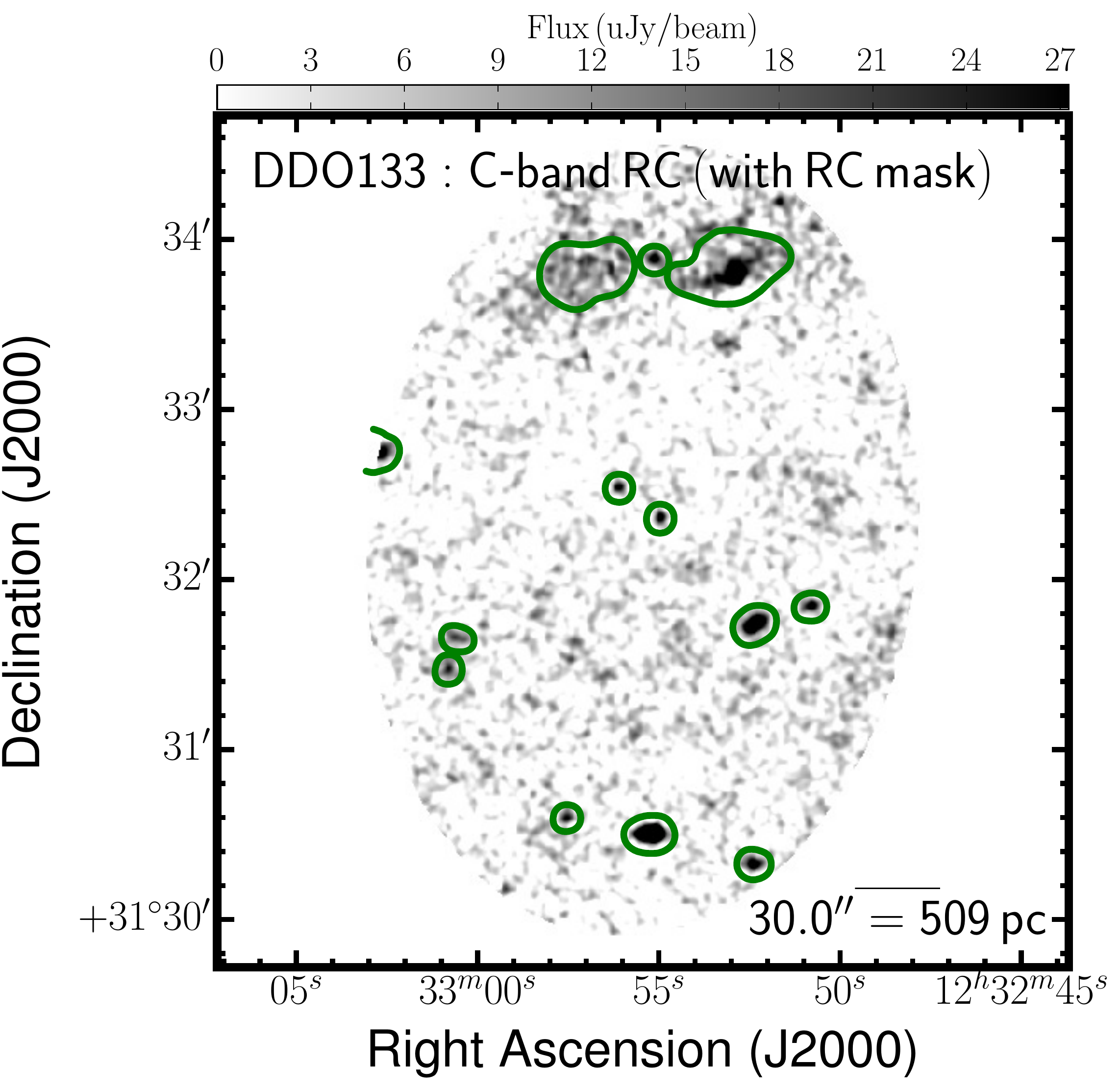} \\
    \includegraphics[width=0.31\linewidth,clip]{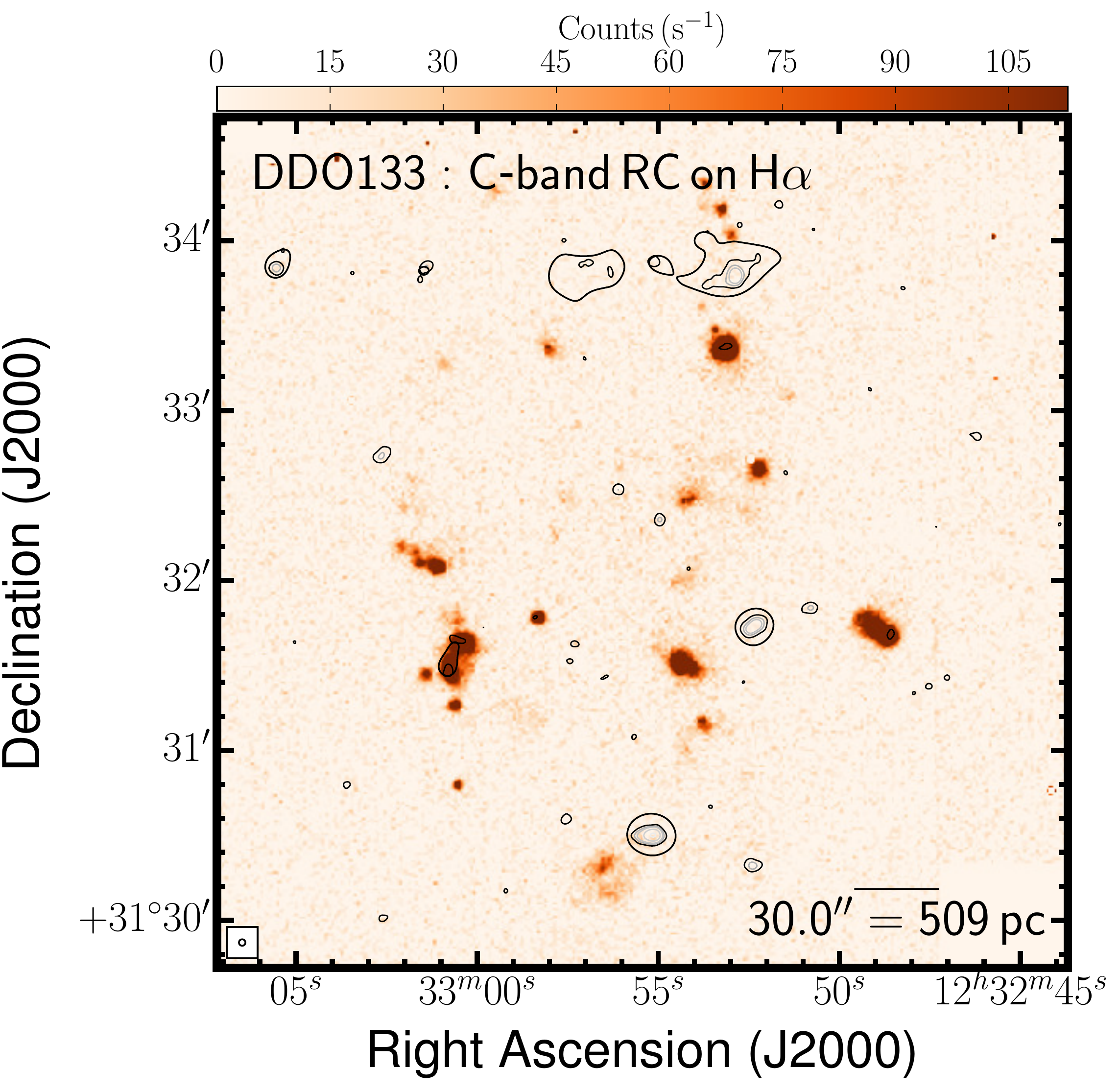} & \ 
    \includegraphics[width=0.31\linewidth,clip]{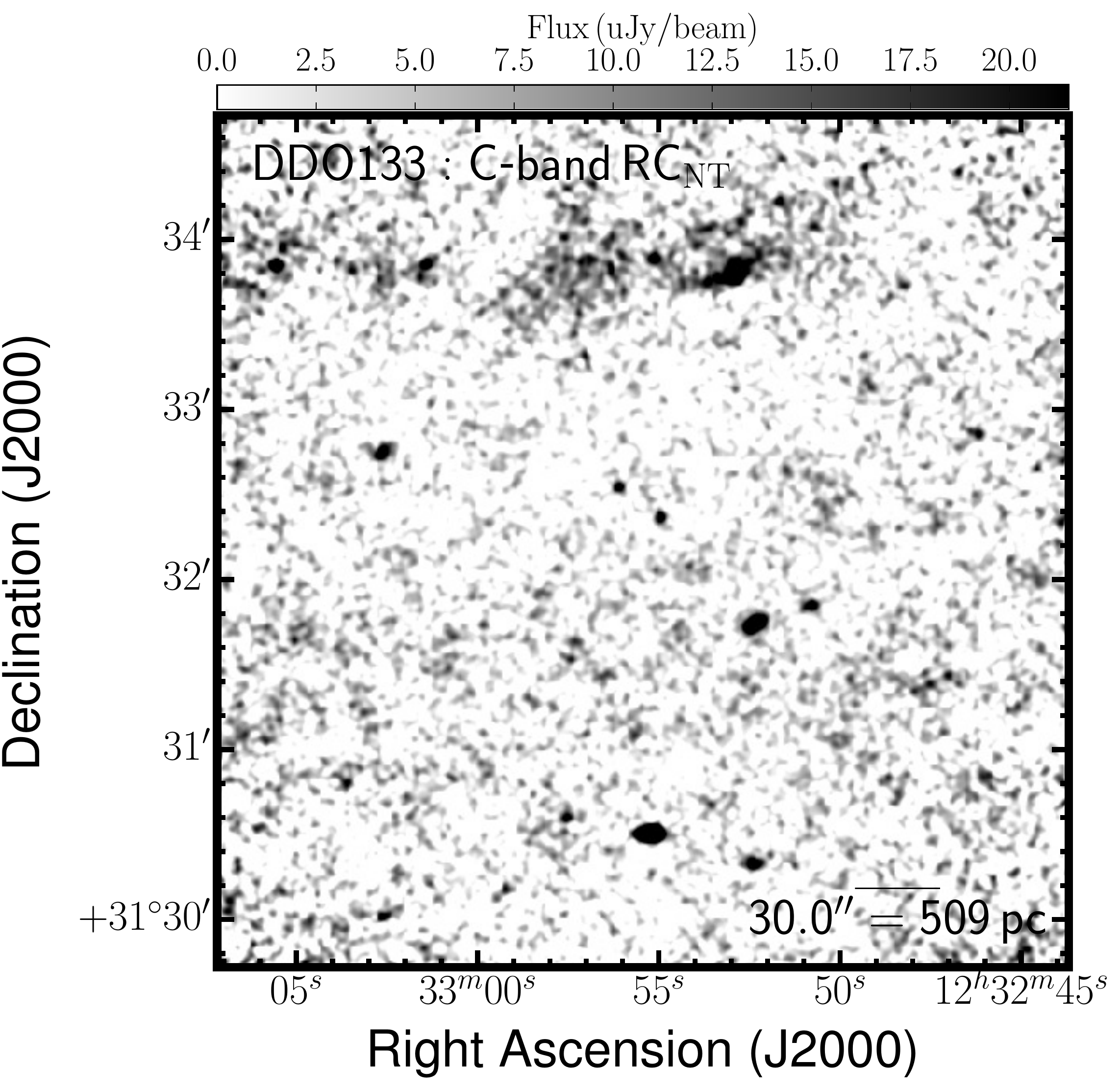} & \ 
    \includegraphics[width=0.31\linewidth,clip]{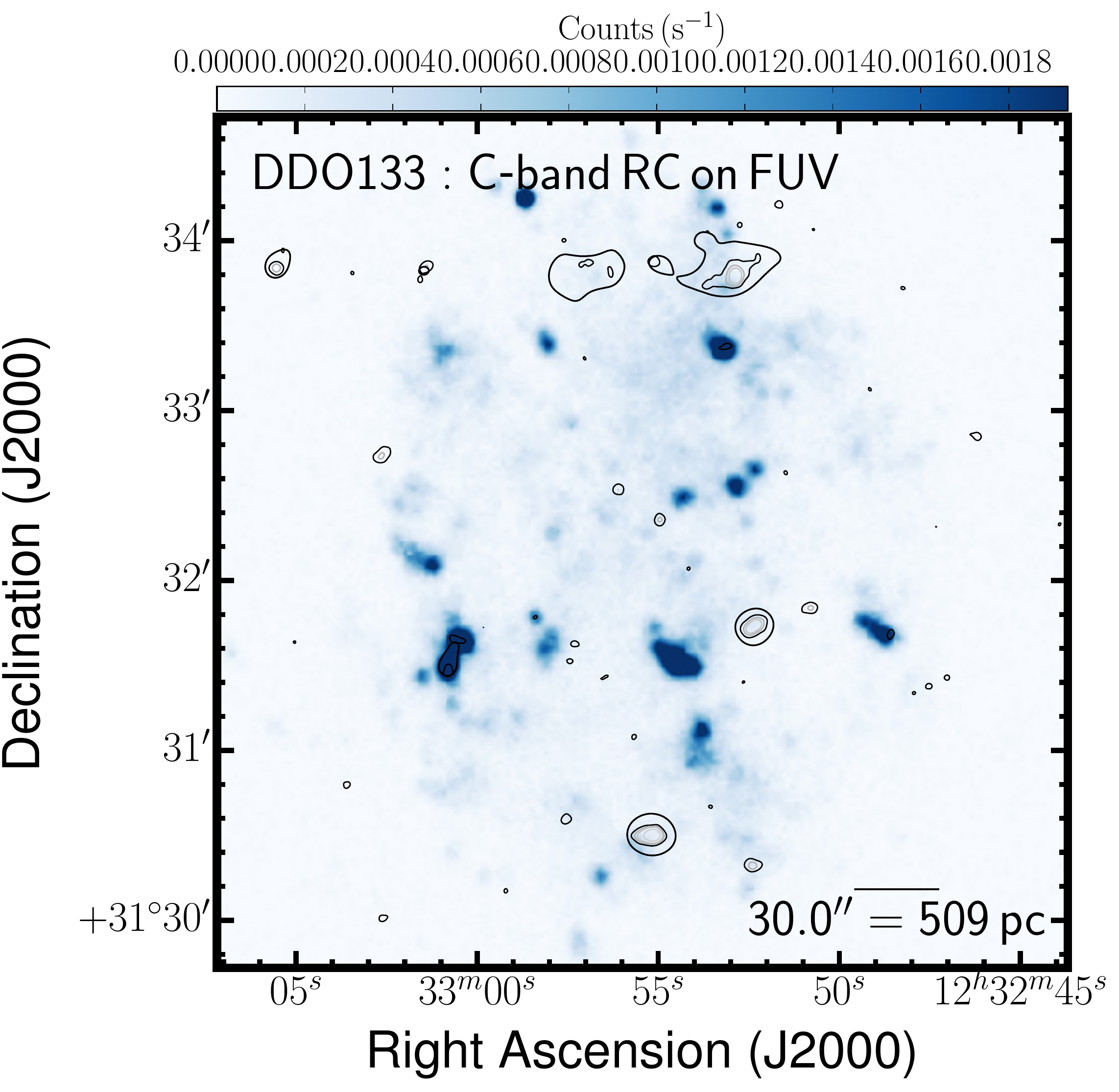} \\
    \includegraphics[width=0.31\linewidth,clip]{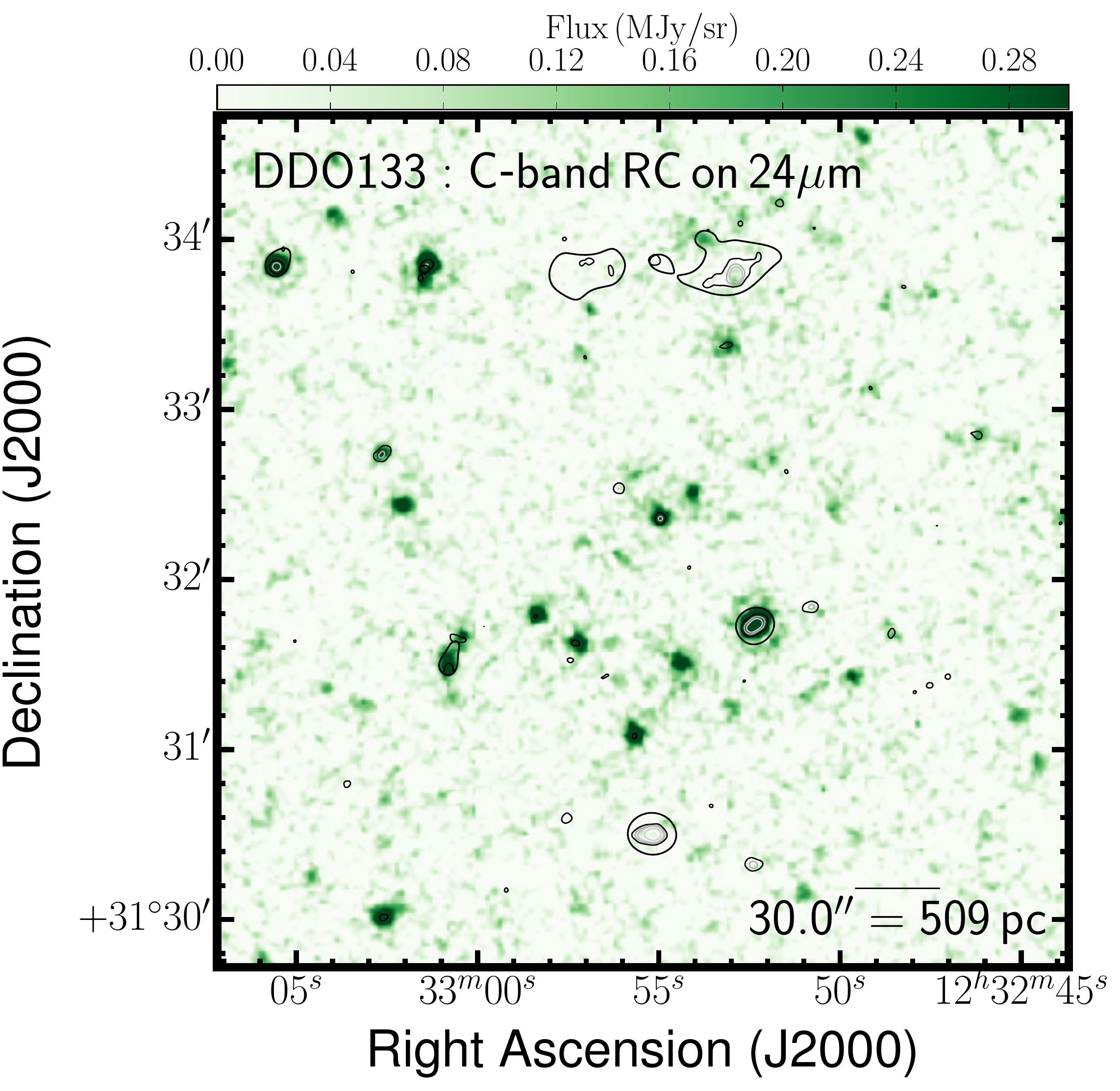} & \ 
    \includegraphics[width=0.31\linewidth,clip]{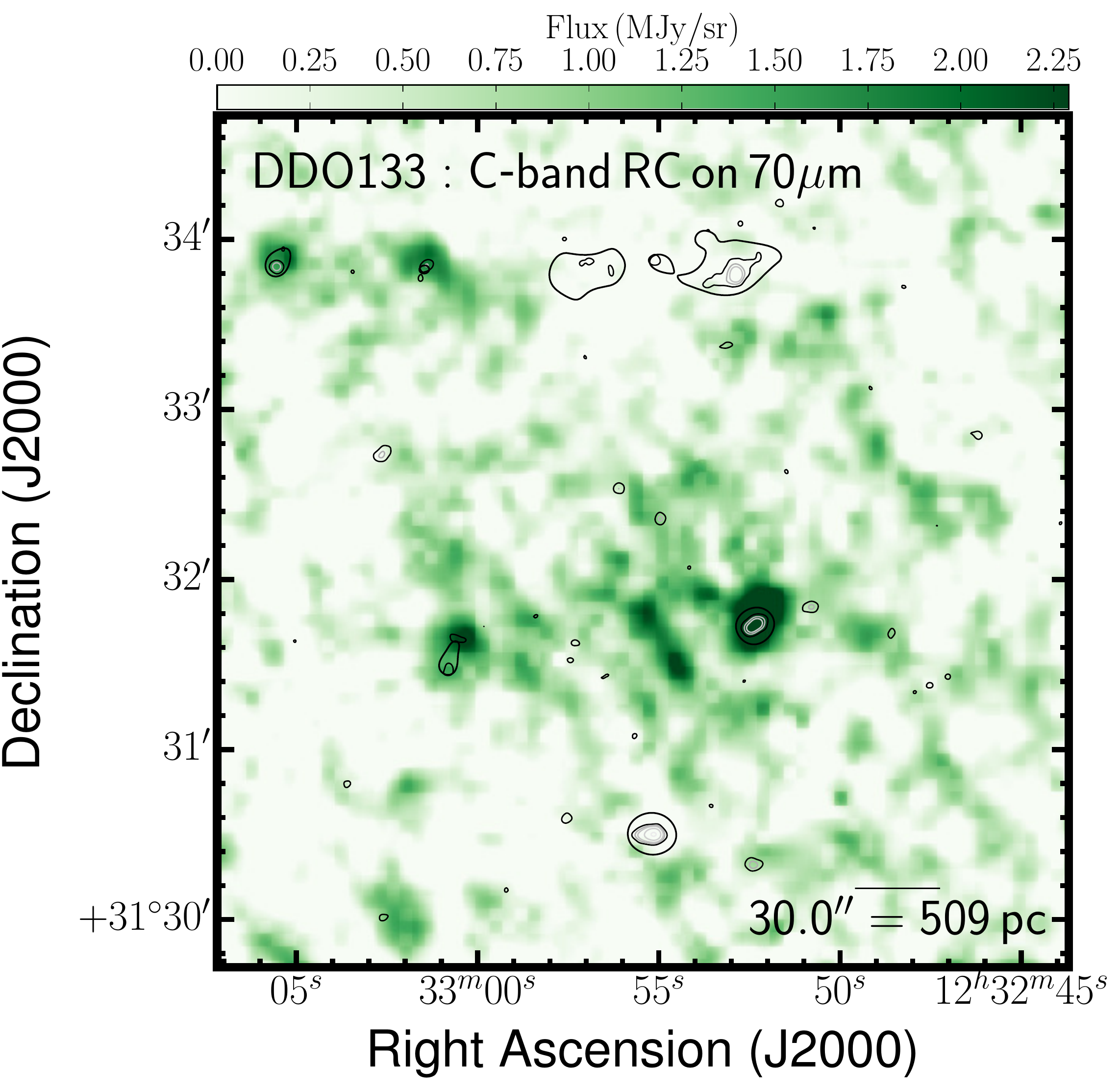} & \ 
    \includegraphics[width=0.31\linewidth,clip]{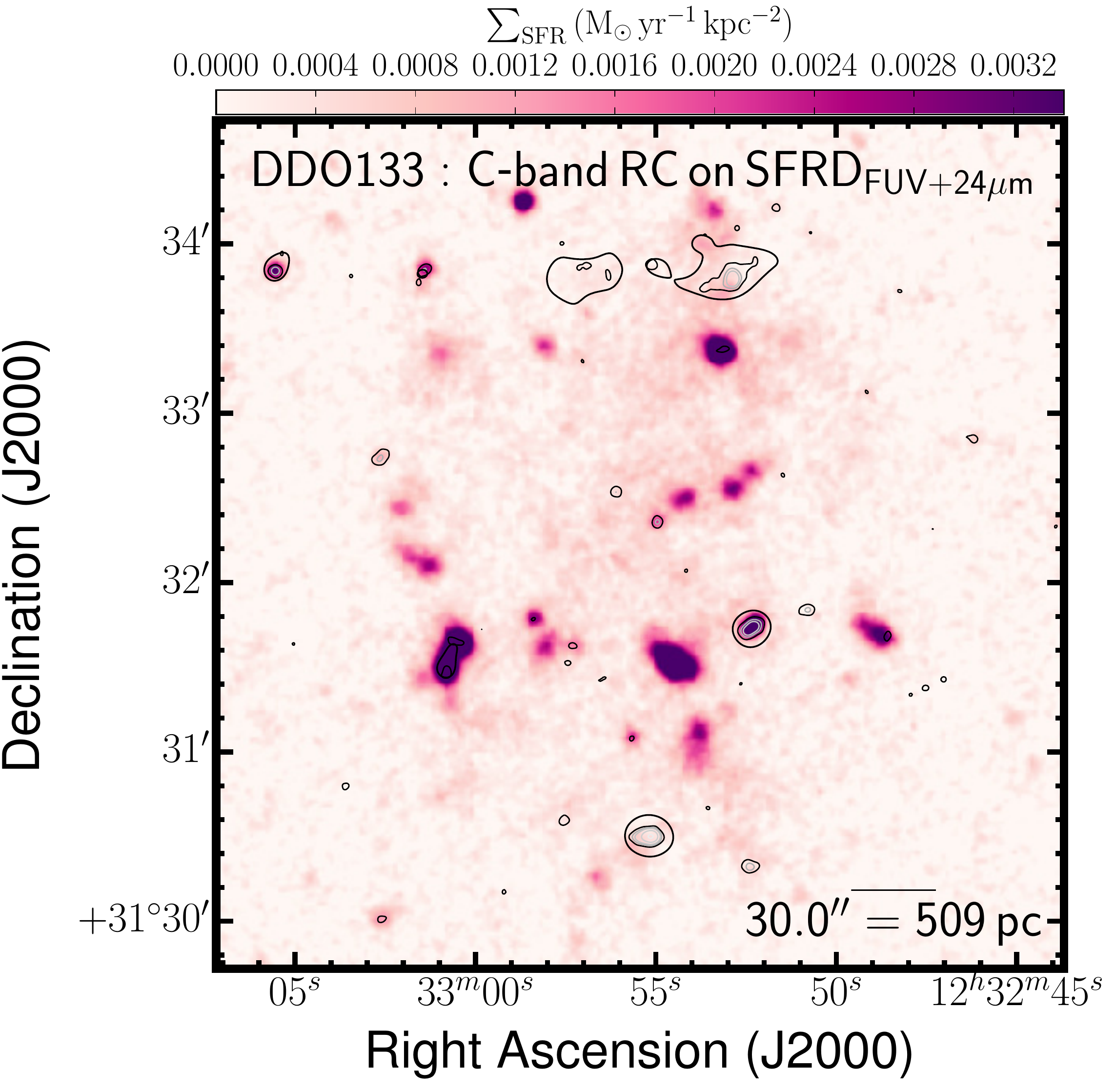} \\
  \end{tabular}
\caption[DDO\,133 images: RC, IR, optical, and FUV]{Multi-wavelength coverage of DDO 133 displaying a $5.0^\prime \times 5.0^\prime$ area. We show total RC flux density at the native resolution (top-left) and again with contours (top-centre). The RC contours are superposed on ancillary LITTLE THINGS images where possible: \halpha\ (middle-left); \RCNT\ obtained by subtracting the expected \RCT\ based on the \halpha-\RCT\ scaling factor of \cite{Deeg1997} from the total RC; {\em GALEX} FUV (middle-right); {\em Spitzer} 24\micron\ (bottom-left); {\em Spitzer} 70\micron\ (bottom-centre); FUV$+24{\rm \mu m}$--inferred SFRD from \citealp{Leroy2012} (bottom-right). We also show the RC that was isolated by the RC--based masking technique (top-right).}
  \label{figure:ddo133Cc_maps}
\end{figure}

\begin{figure}
  \begin{tabular}{ccc}
    \includegraphics[width=0.31\linewidth,clip]{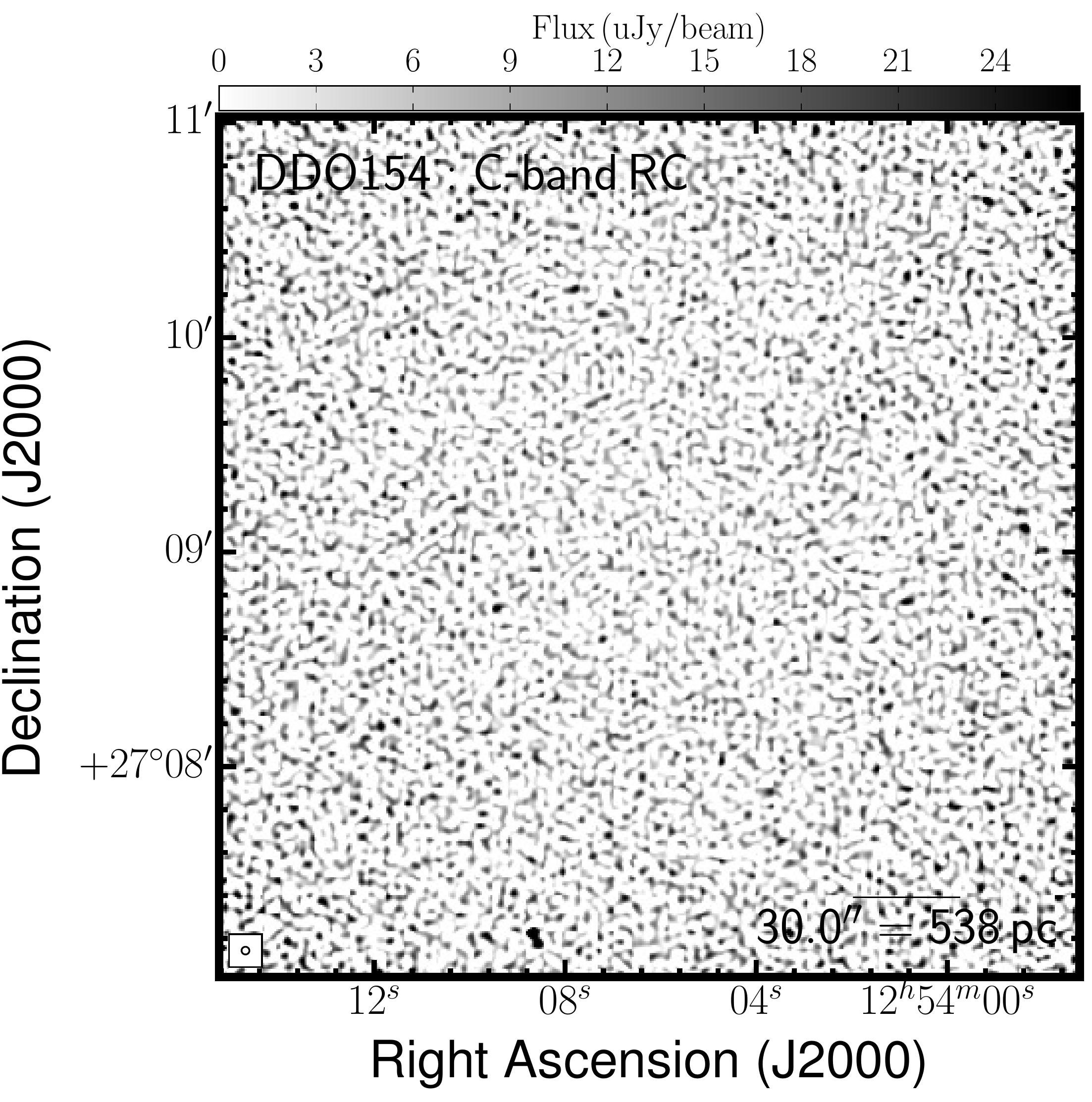} & \ 
    \includegraphics[width=0.31\linewidth,clip]{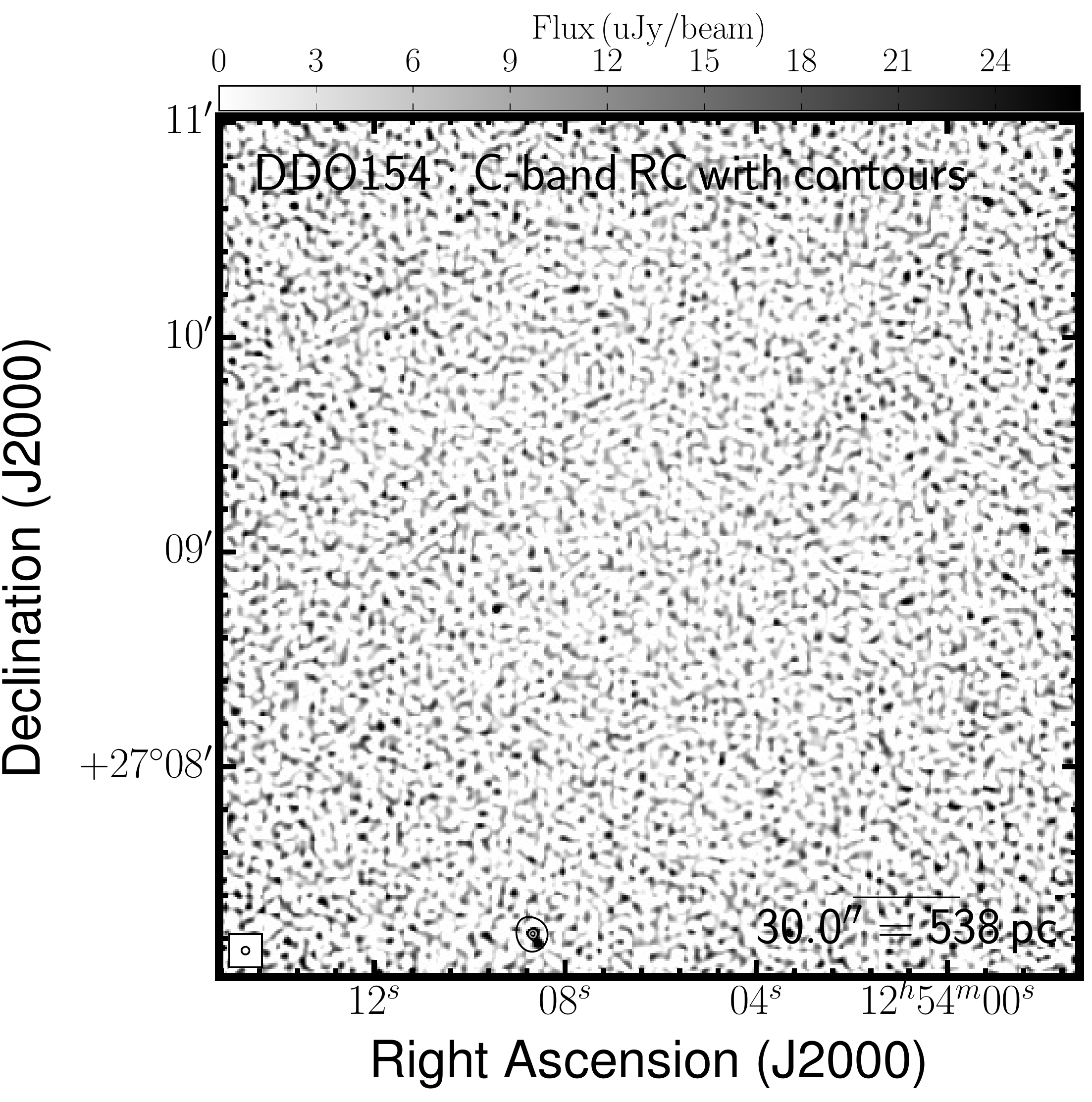} & \ 
    \includegraphics[width=0.31\linewidth,clip]{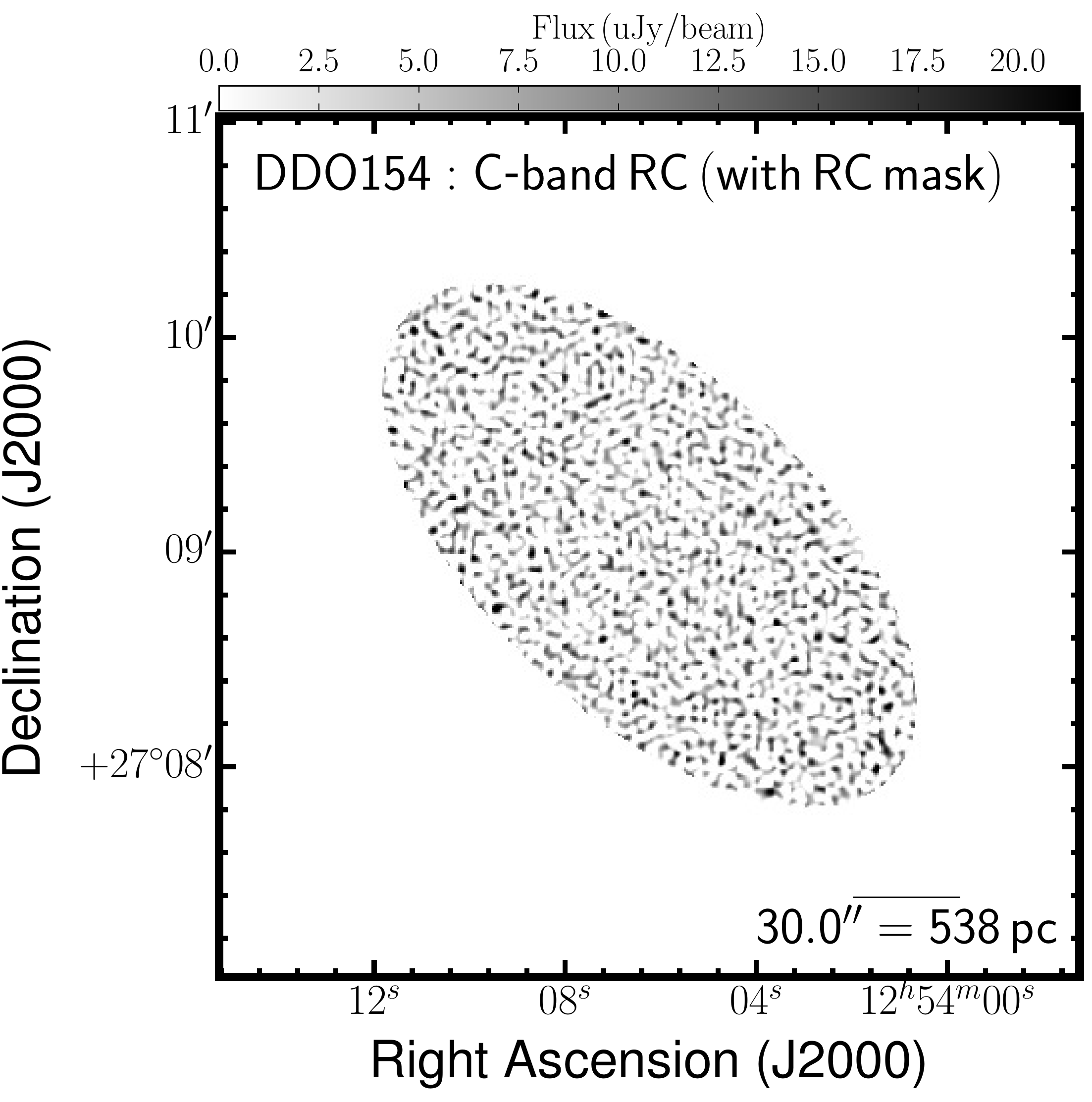} \\
    \includegraphics[width=0.31\linewidth,clip]{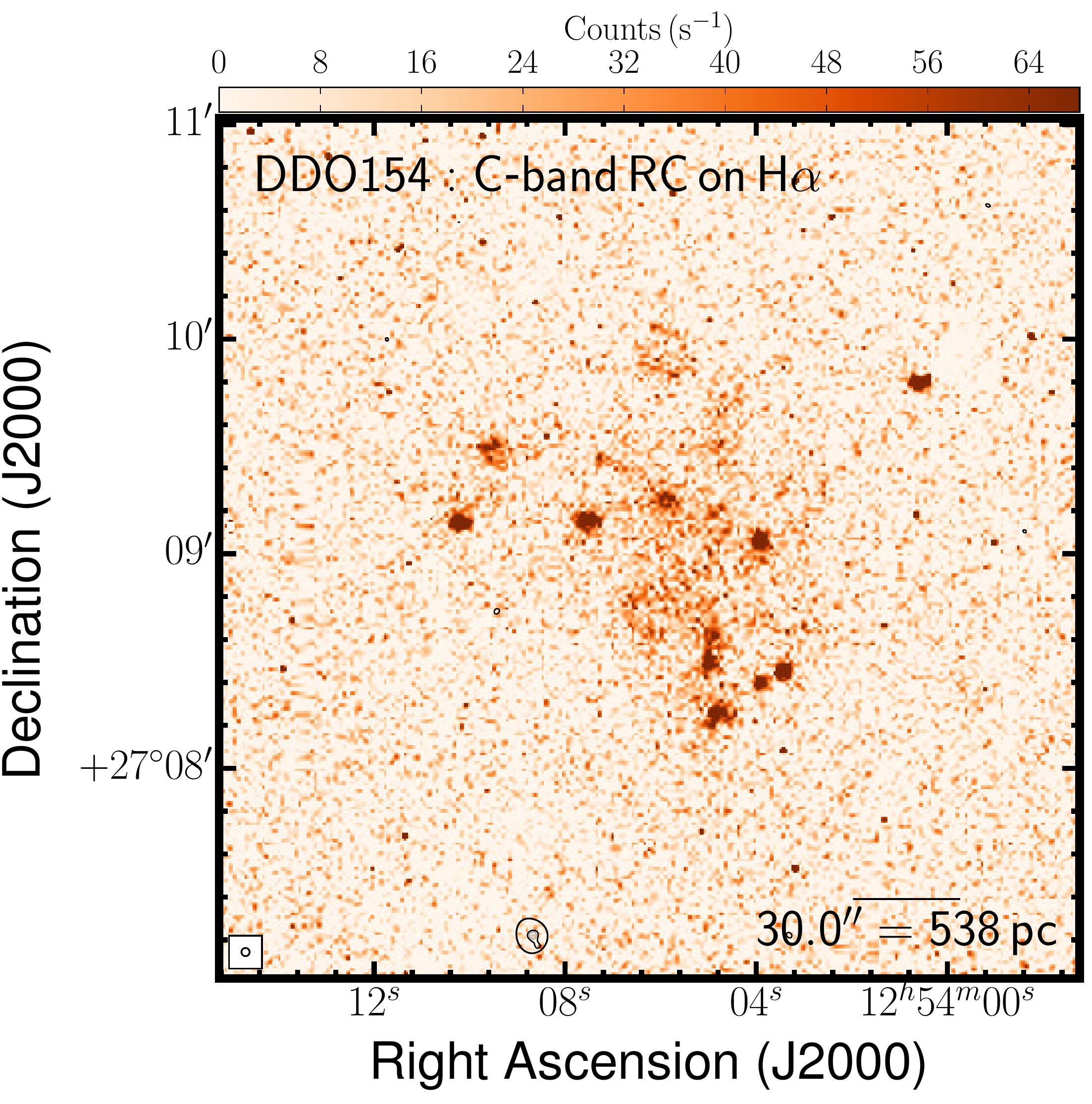} & \ 
    \includegraphics[width=0.31\linewidth,clip]{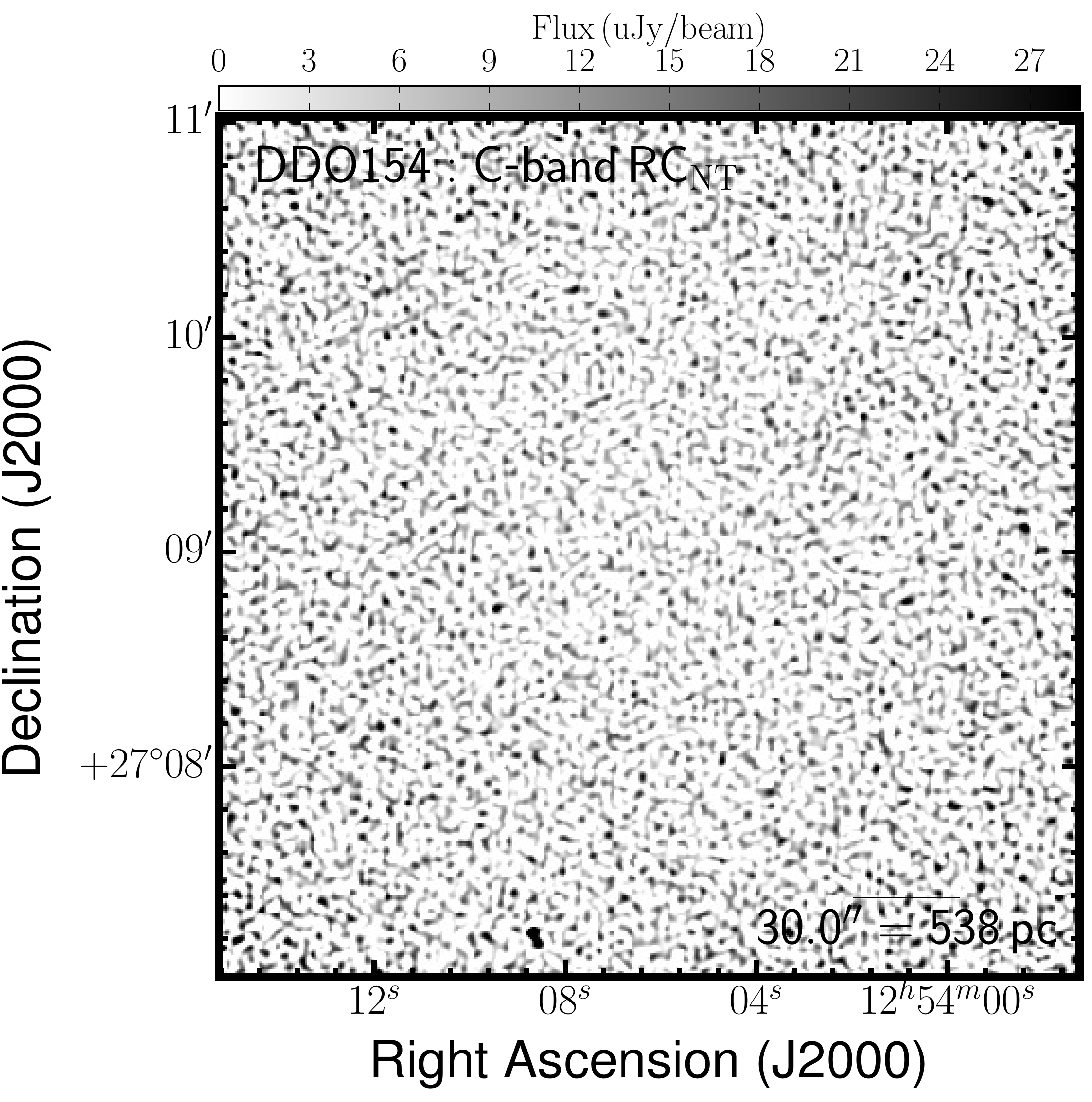} & \ 
    \includegraphics[width=0.31\linewidth,clip]{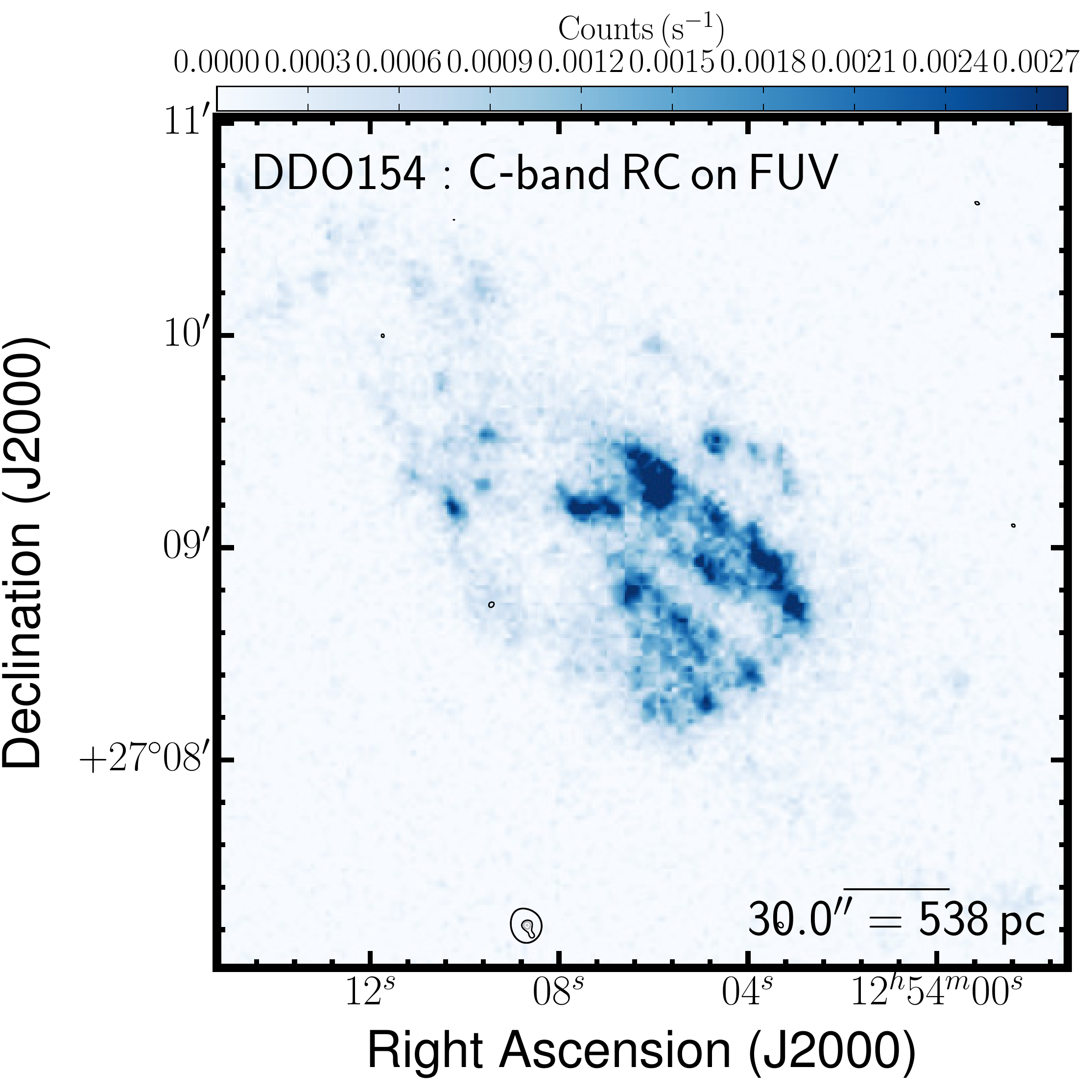} \\
    \includegraphics[width=0.31\linewidth,clip]{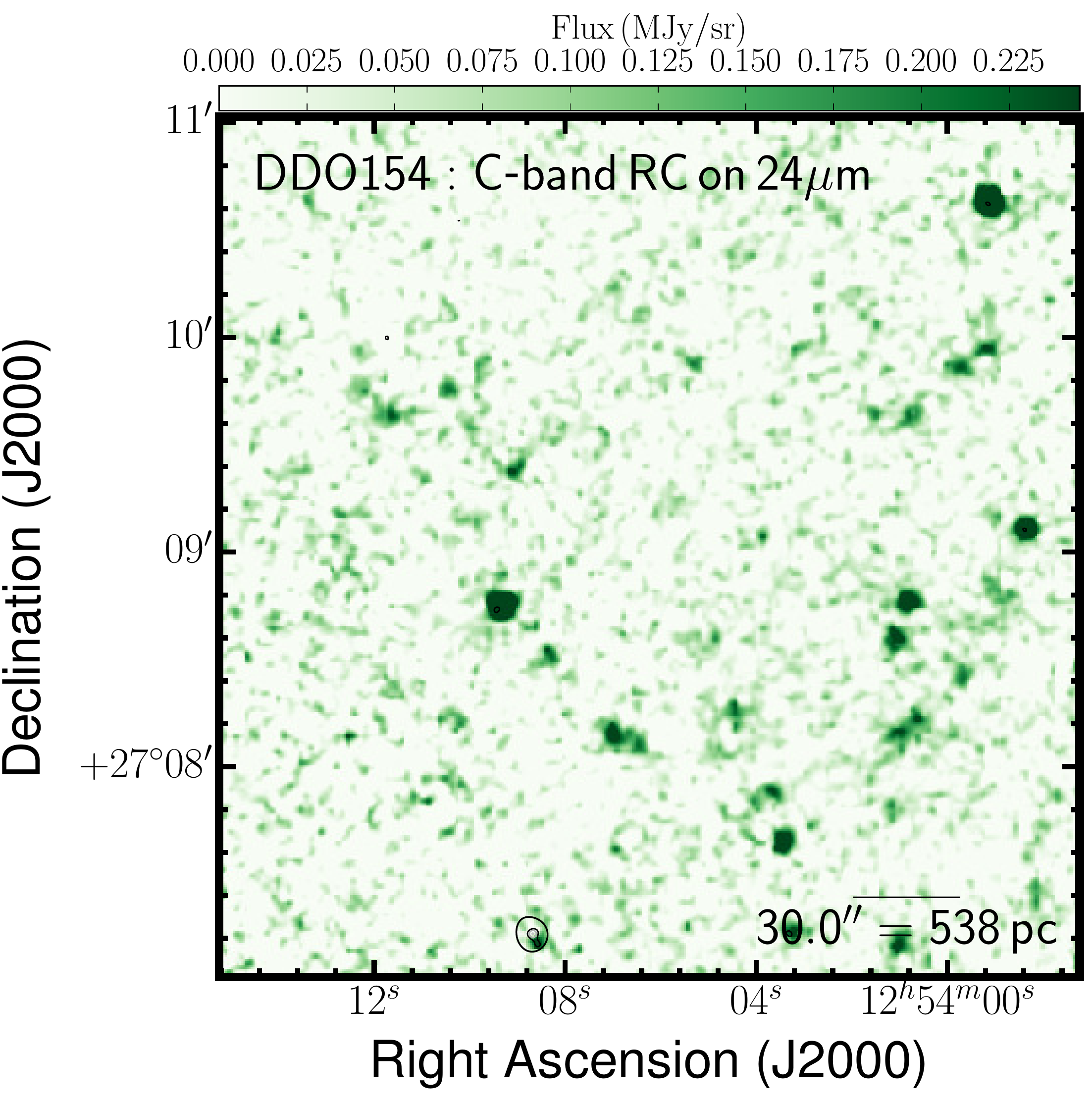} & \ 
    \includegraphics[width=0.31\linewidth,clip]{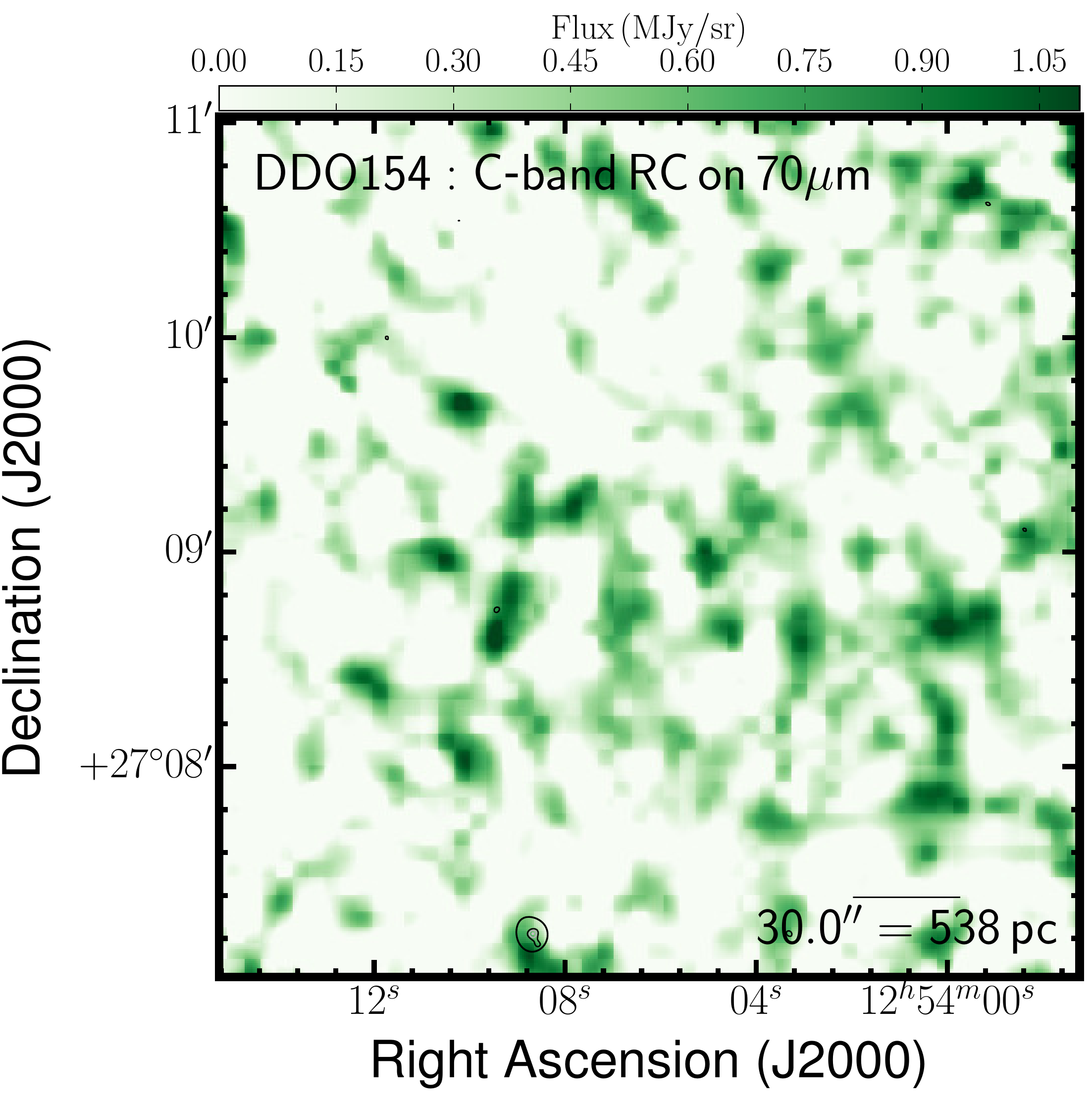} & \ 
    \includegraphics[width=0.31\linewidth,clip]{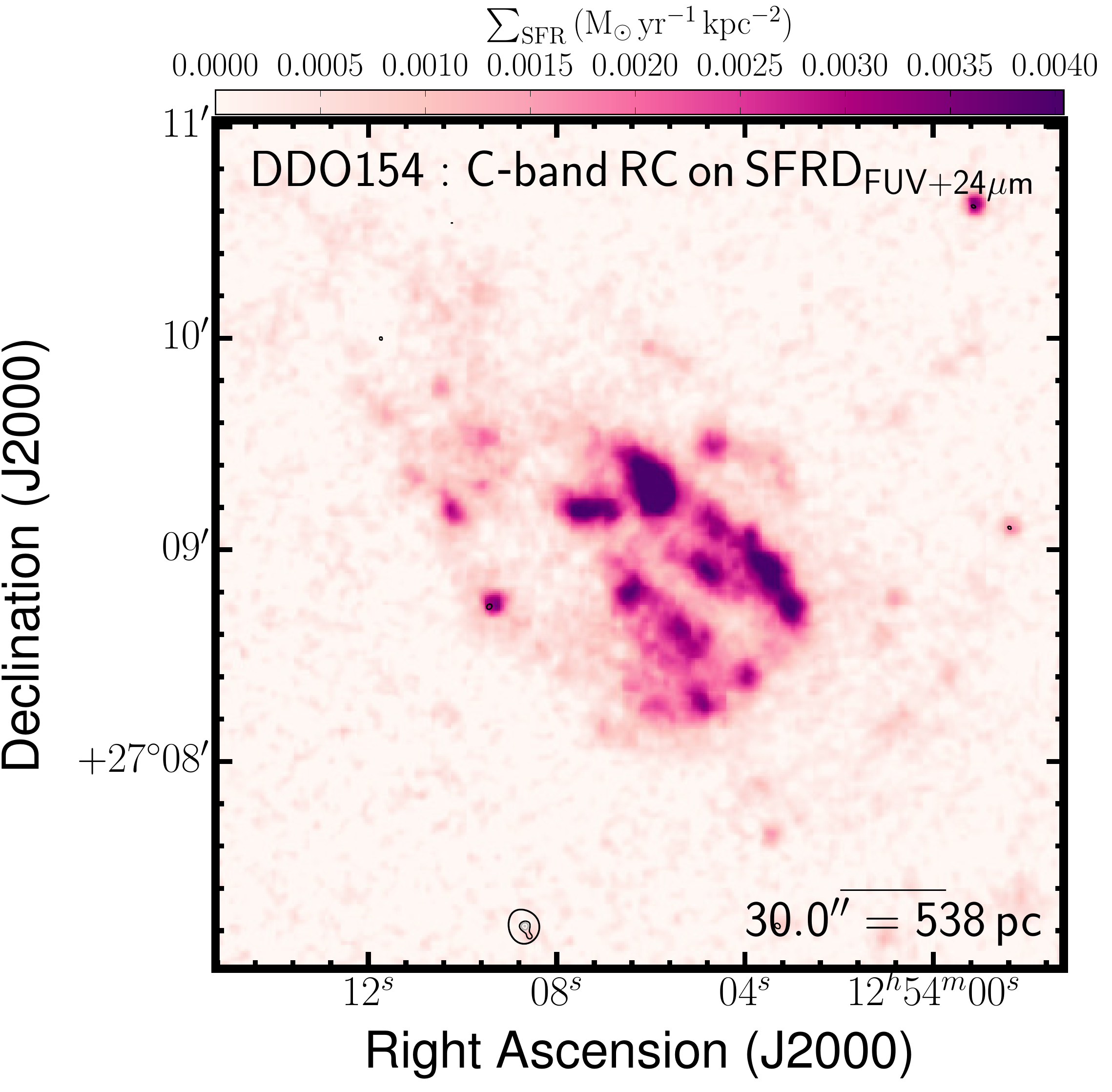} \\
  \end{tabular}
\caption[DDO\,154 images: RC, IR, optical, and FUV]{Multi-wavelength coverage of DDO 154 displaying a $4.0^\prime \times 4.0^\prime$ area. We show total RC flux density at the native resolution (top-left) and again with contours (top-centre). The RC contours are superposed on ancillary LITTLE THINGS images where possible: \halpha\ (middle-left); \RCNT\ obtained by subtracting the expected \RCT\ based on the \halpha-\RCT\ scaling factor of \cite{Deeg1997} from the total RC; {\em GALEX} FUV (middle-right); {\em Spitzer} 24\micron\ (bottom-left); {\em Spitzer} 70\micron\ (bottom-centre); FUV$+24{\rm \mu m}$--inferred SFRD from \citealp{Leroy2012} (bottom-right). We also show the RC that was isolated by the RC--based masking technique (top-right).}
  \label{figure:ddo154Cc_maps}
\end{figure}

\begin{figure}
  \begin{tabular}{ccc}
    \includegraphics[width=0.31\linewidth,clip]{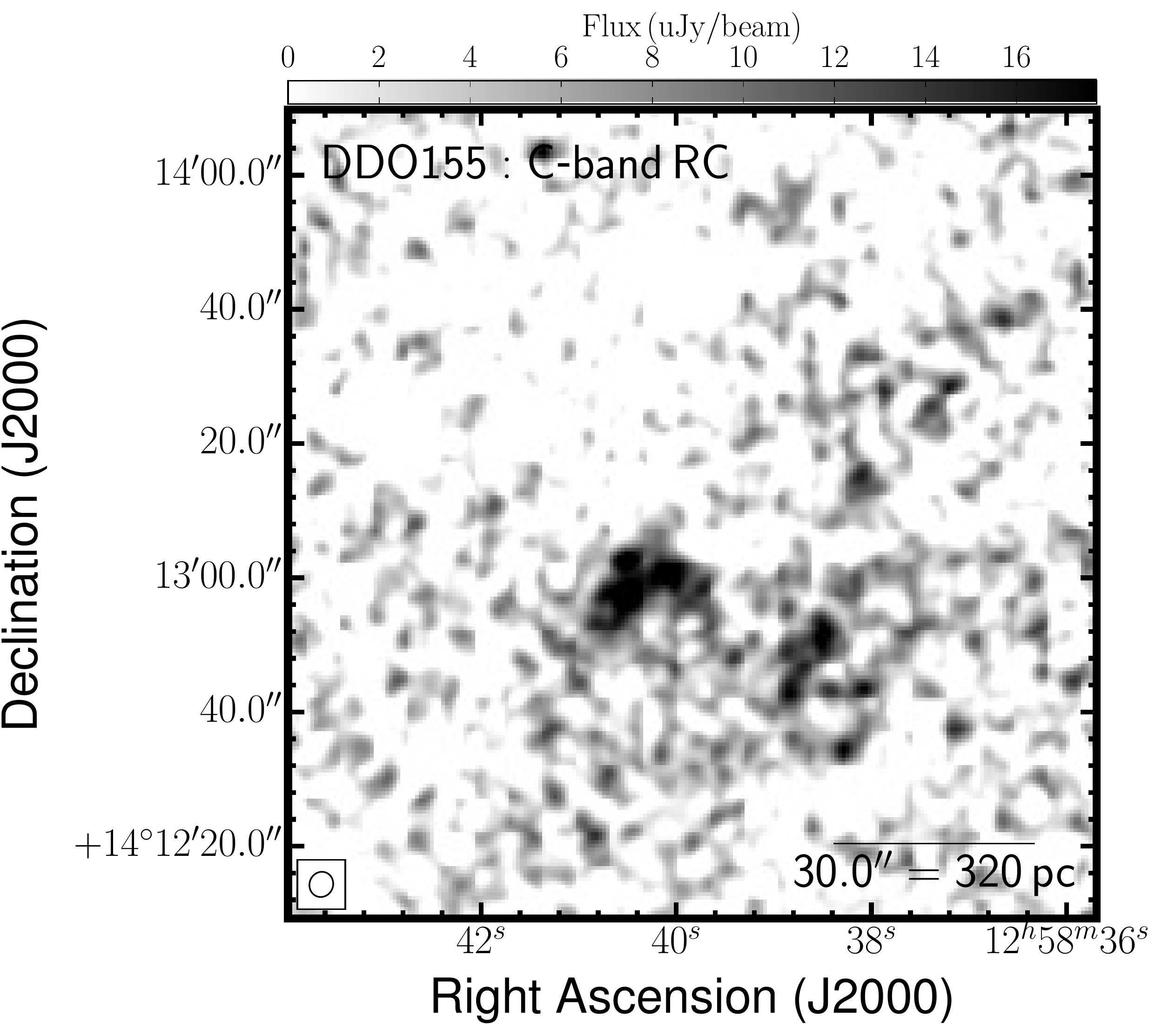} & \ 
    \includegraphics[width=0.31\linewidth,clip]{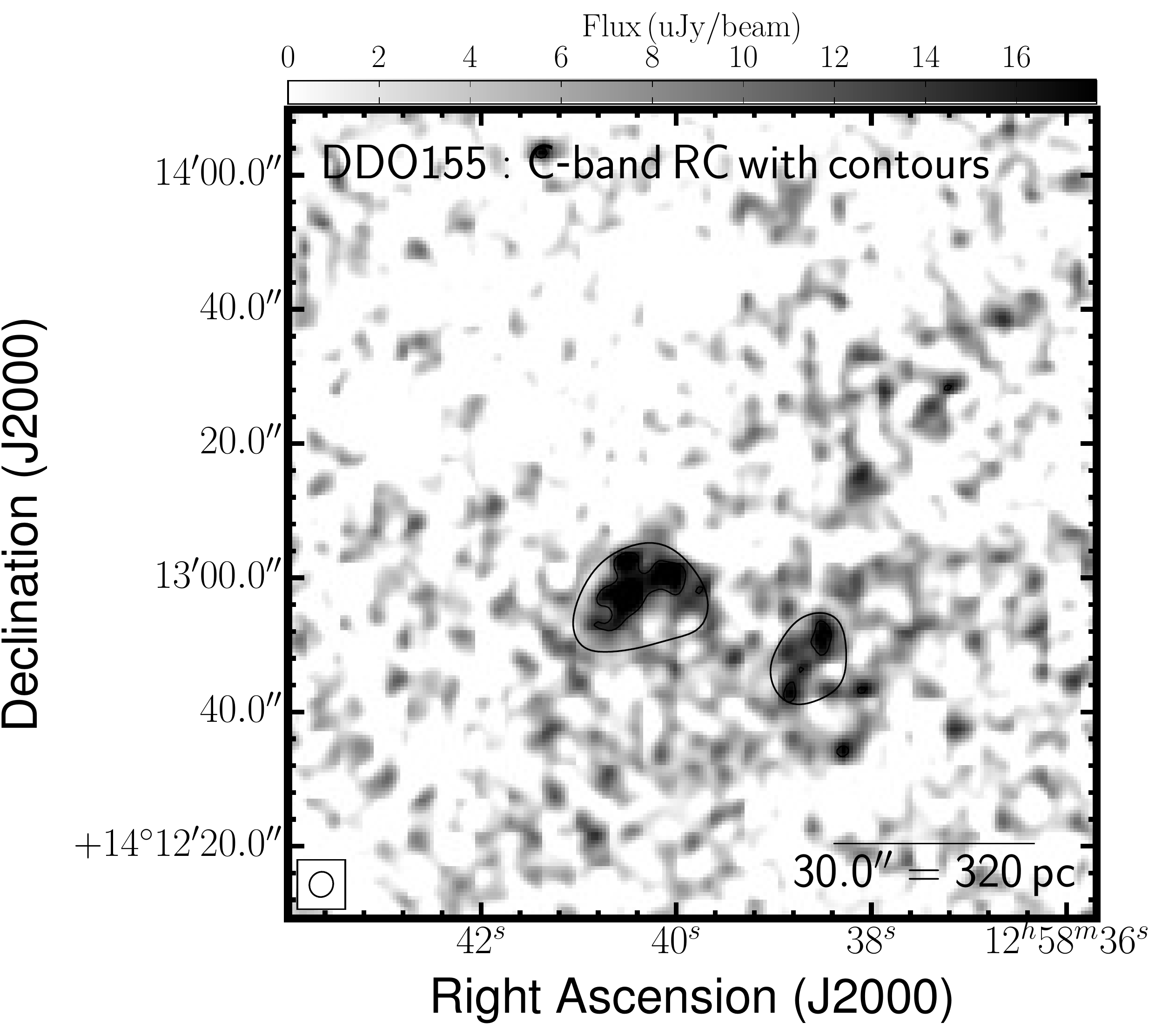} & \ 
    \includegraphics[width=0.31\linewidth,clip]{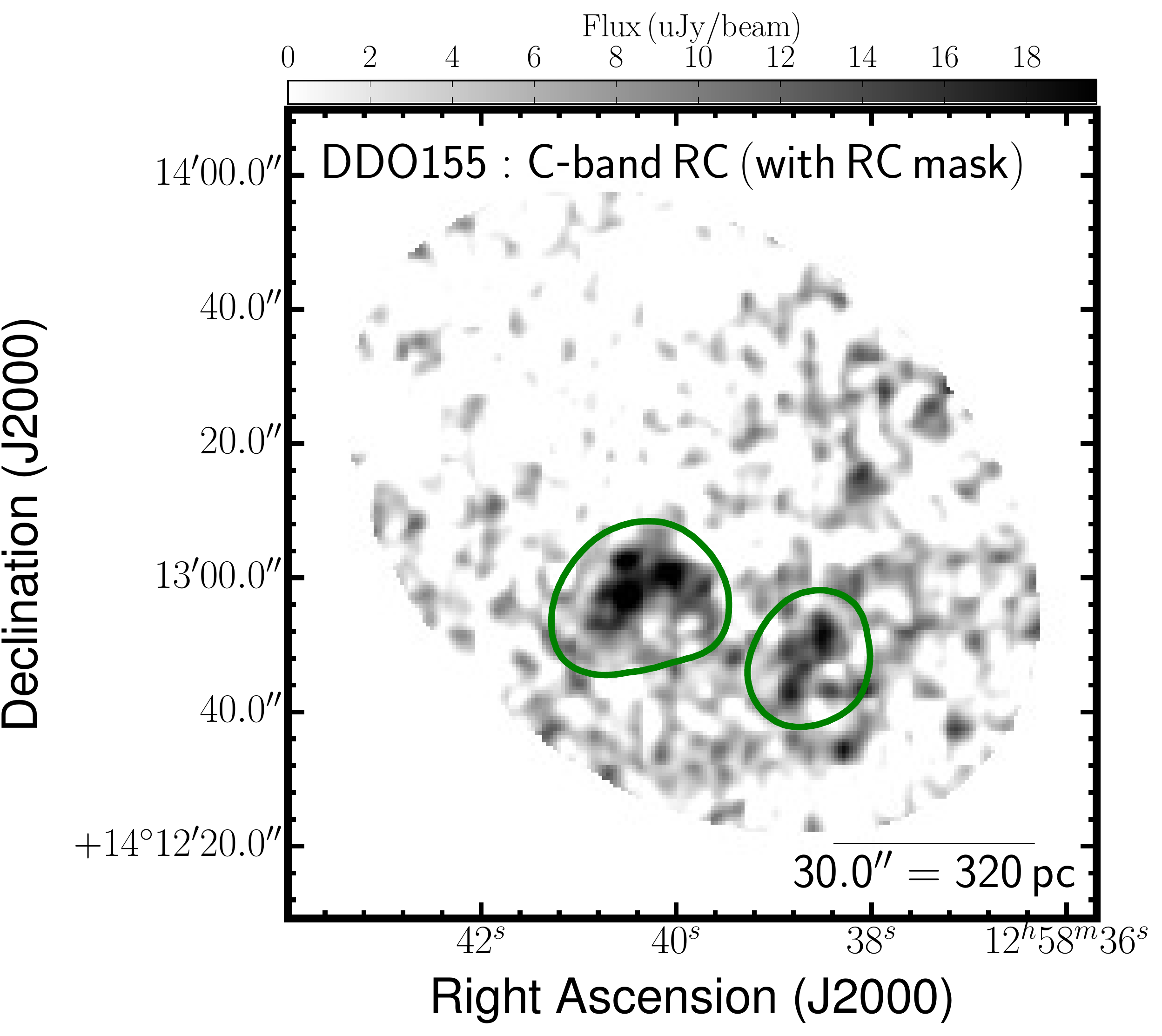} \\
    \includegraphics[width=0.31\linewidth,clip]{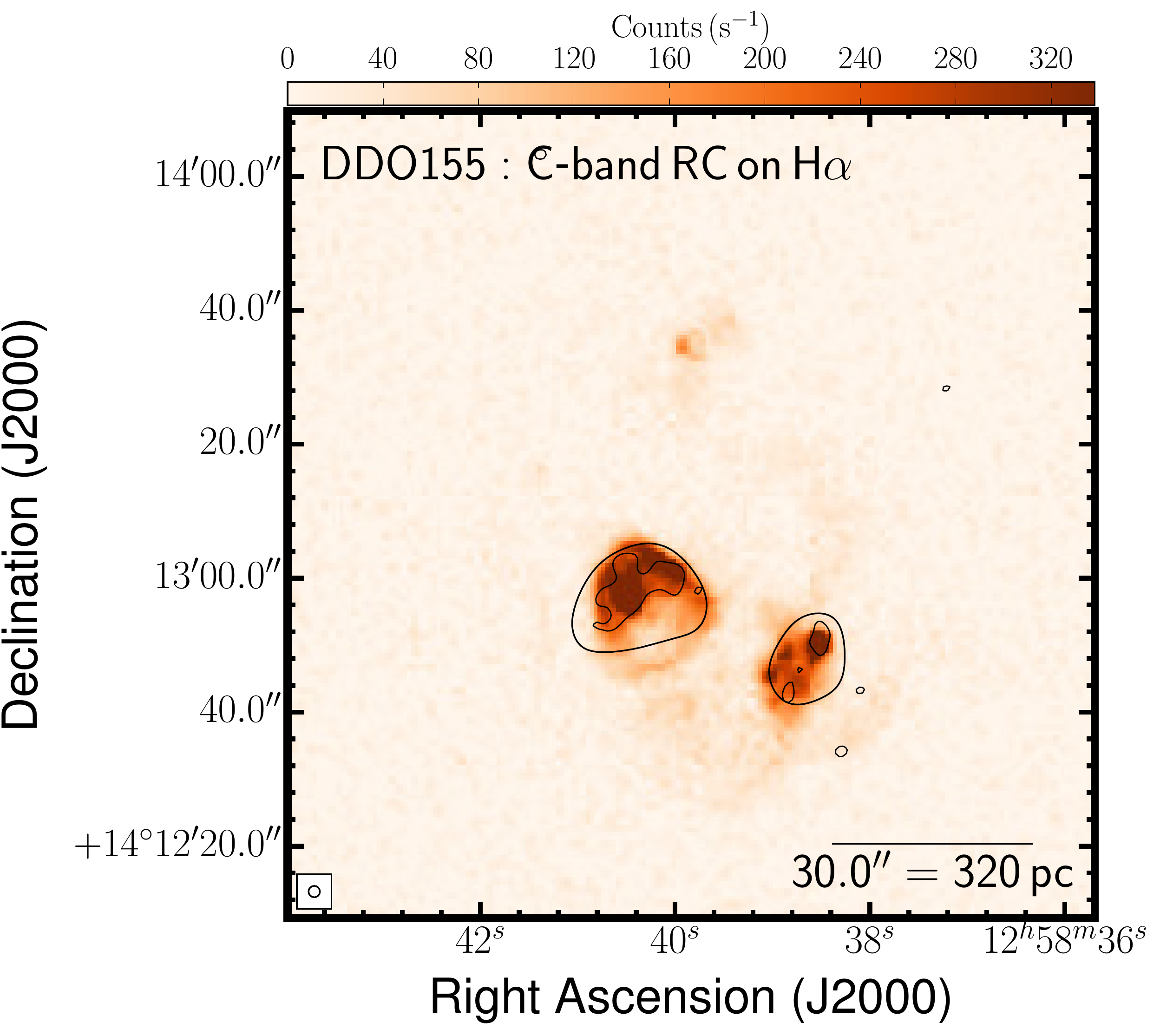} & \ 
    \includegraphics[width=0.31\linewidth,clip]{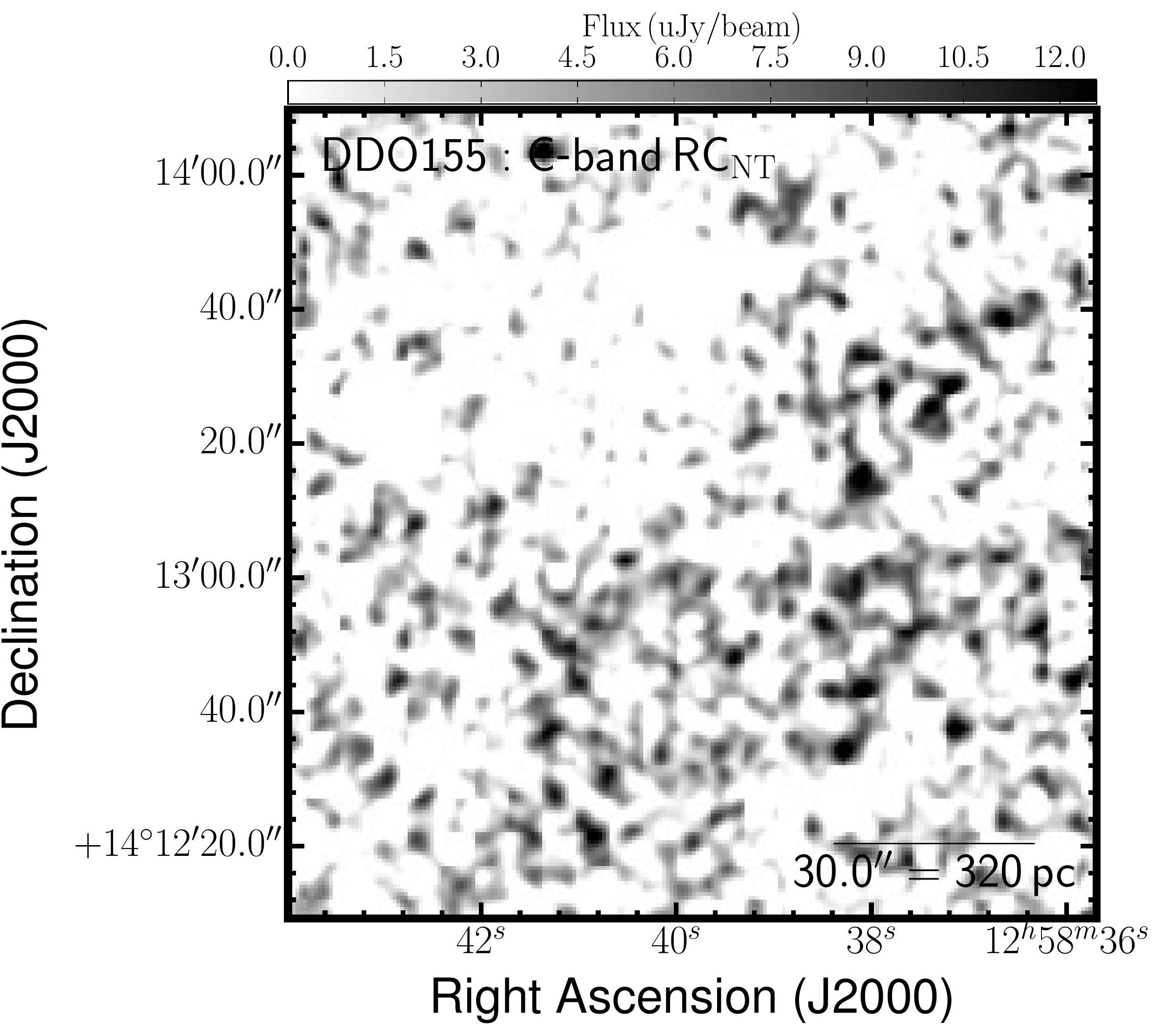} & \ 
    \includegraphics[width=0.31\linewidth,clip]{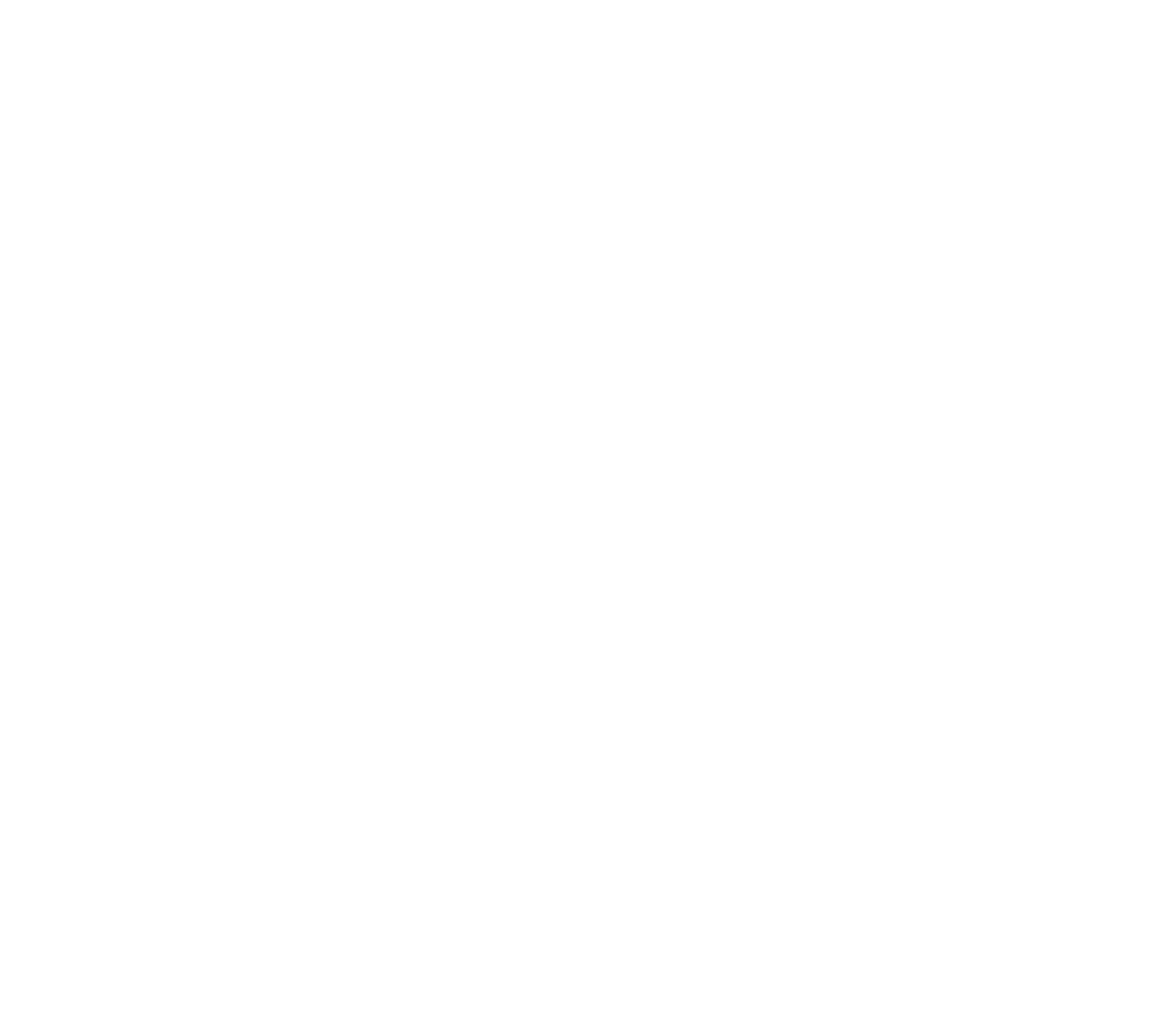} \\
    \includegraphics[width=0.31\linewidth,clip]{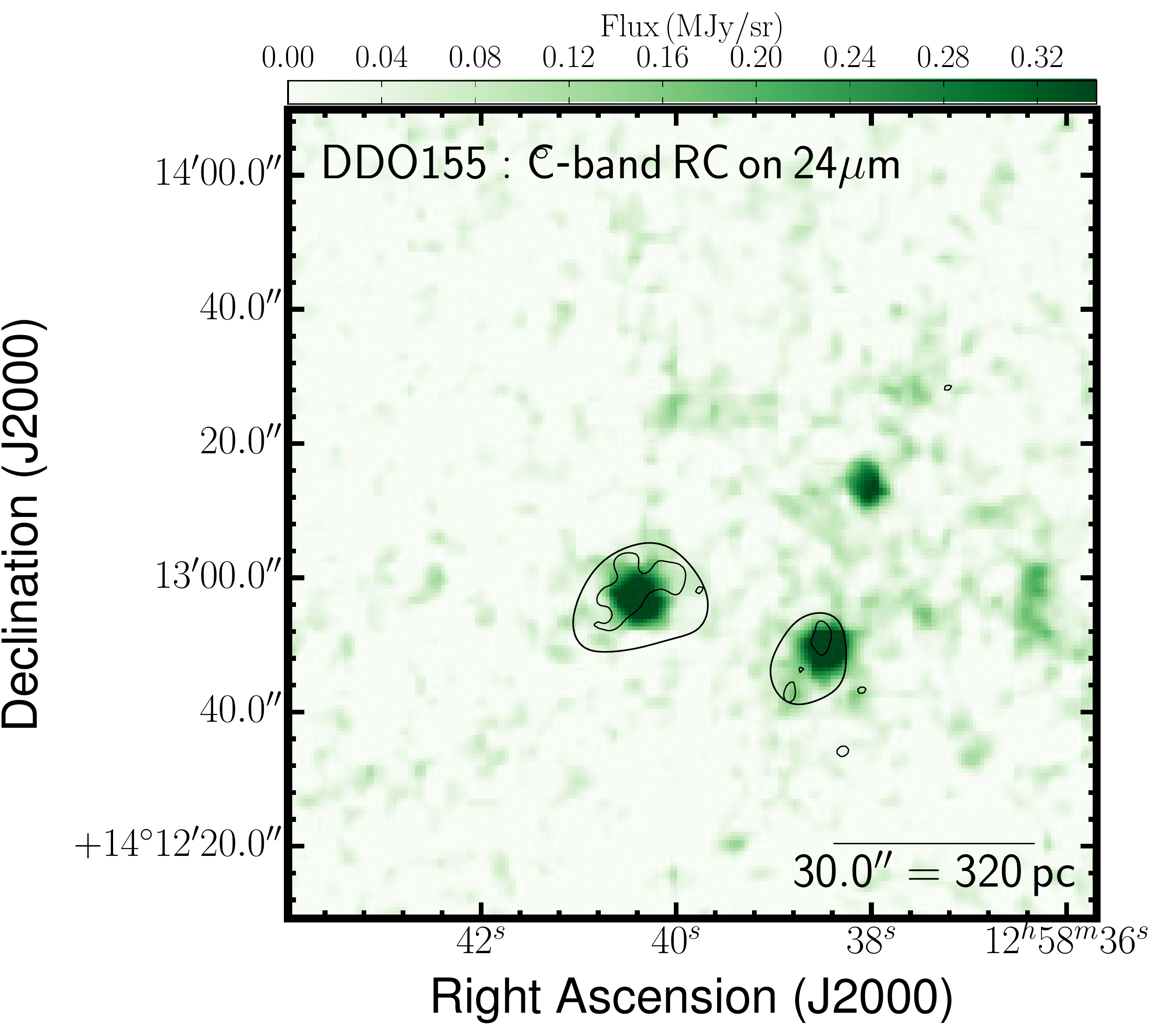} & \ 
    \includegraphics[width=0.31\linewidth,clip]{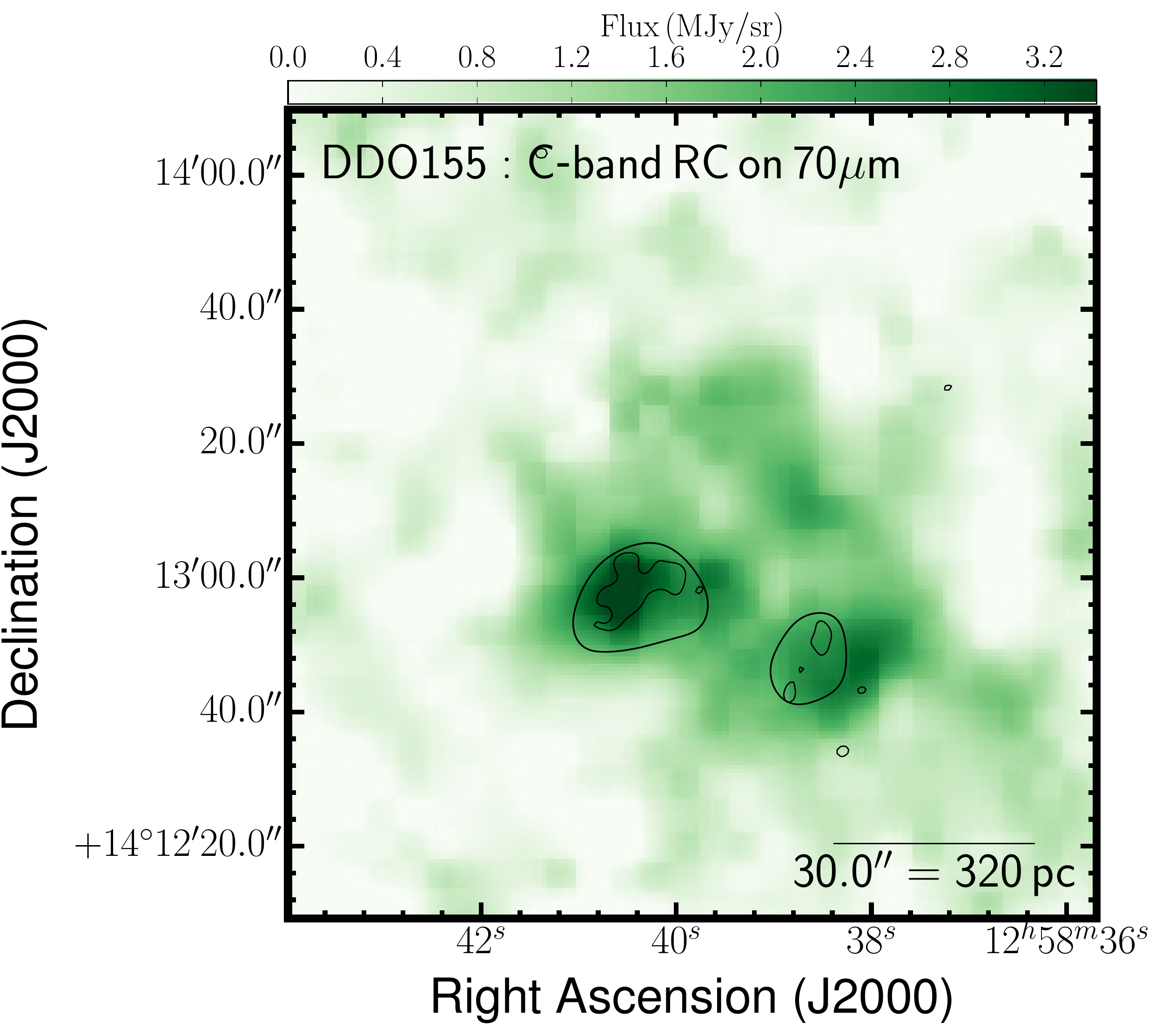} & \ 
    \includegraphics[width=0.31\linewidth,clip]{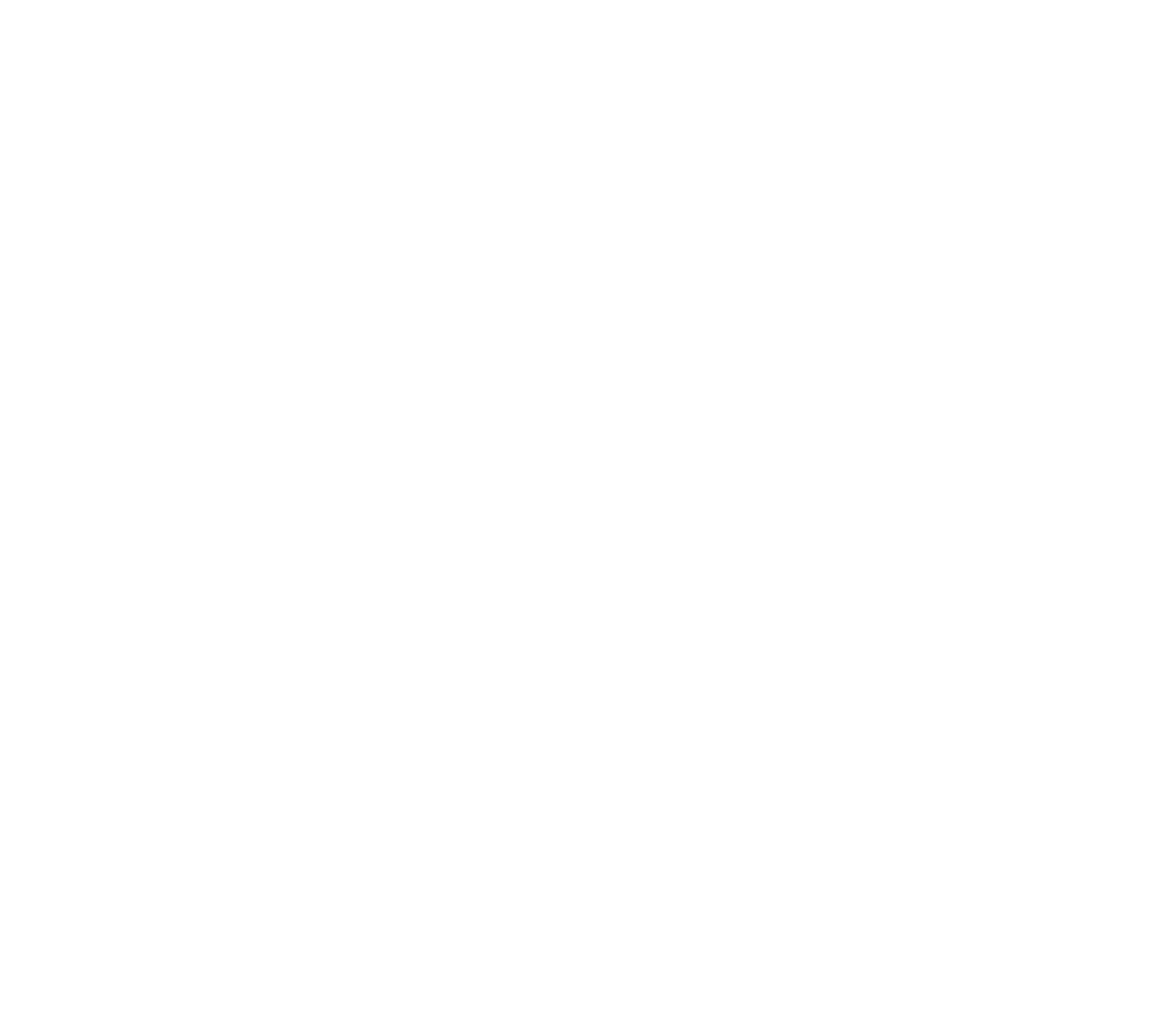} \\
  \end{tabular}
\caption[DDO\,155 images: RC, IR, optical, and FUV]{Multi-wavelength coverage of DDO 155 displaying a $2.0^\prime \times 2.0^\prime$ area. We show total RC flux density at the native resolution (top-left) and again with contours (top-centre). The RC contours are superposed on ancillary LITTLE THINGS images where possible: \halpha\ (middle-left); \RCNT\ obtained by subtracting the expected \RCT\ based on the \halpha-\RCT\ scaling factor of \cite{Deeg1997} from the total RC; {\em GALEX} FUV (middle-right); {\em Spitzer} 24\micron\ (bottom-left); {\em Spitzer} 70\micron\ (bottom-centre); FUV$+24{\rm \mu m}$--inferred SFRD from \citealp{Leroy2012} (bottom-right). We also show the RC that was isolated by the RC--based masking technique (top-right).}
  \label{figure:ddo155Cc_maps}
\end{figure}

\begin{figure}
  \begin{tabular}{ccc}
    \includegraphics[width=0.31\linewidth,clip]{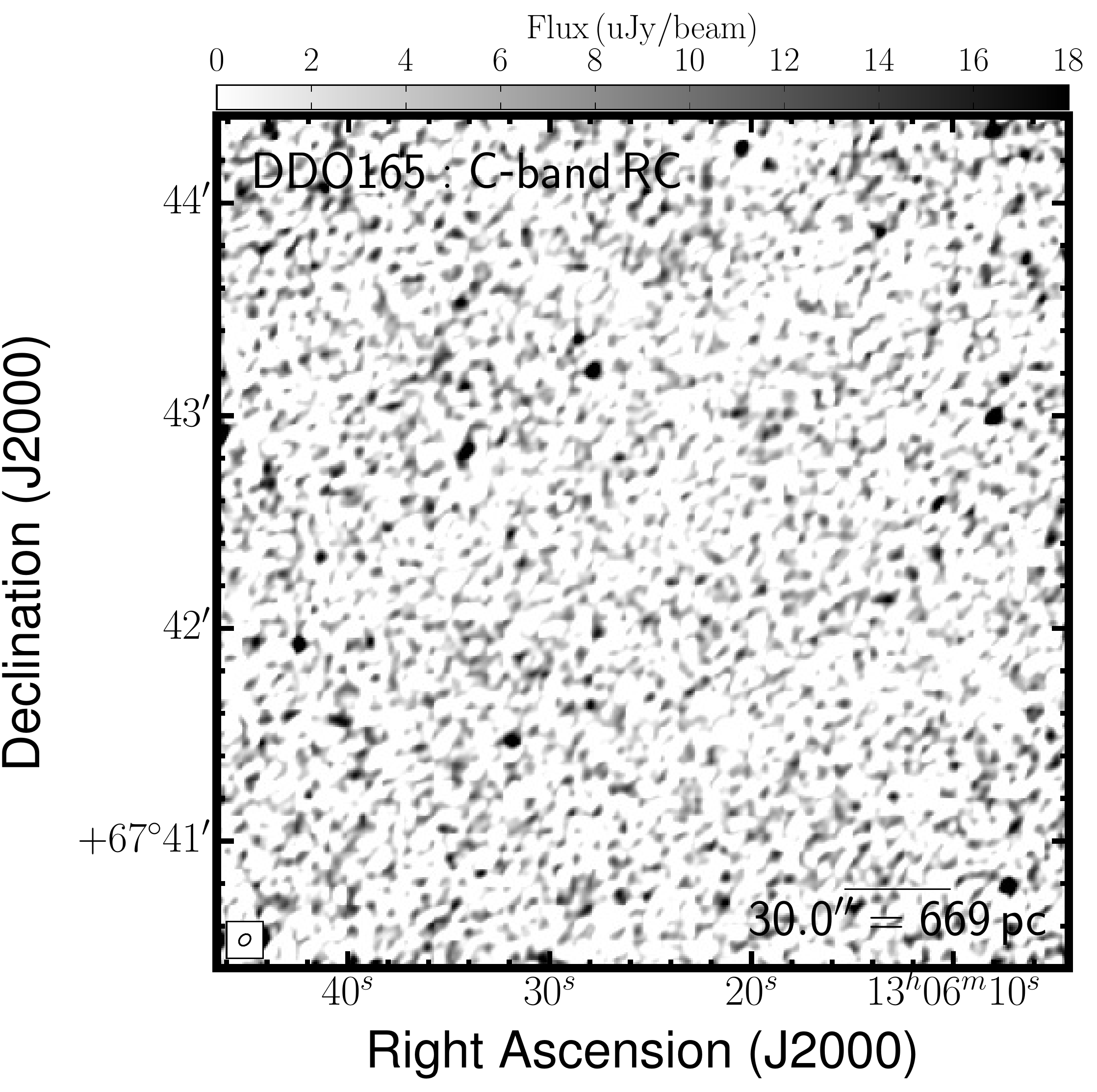} & \ 
    \includegraphics[width=0.31\linewidth,clip]{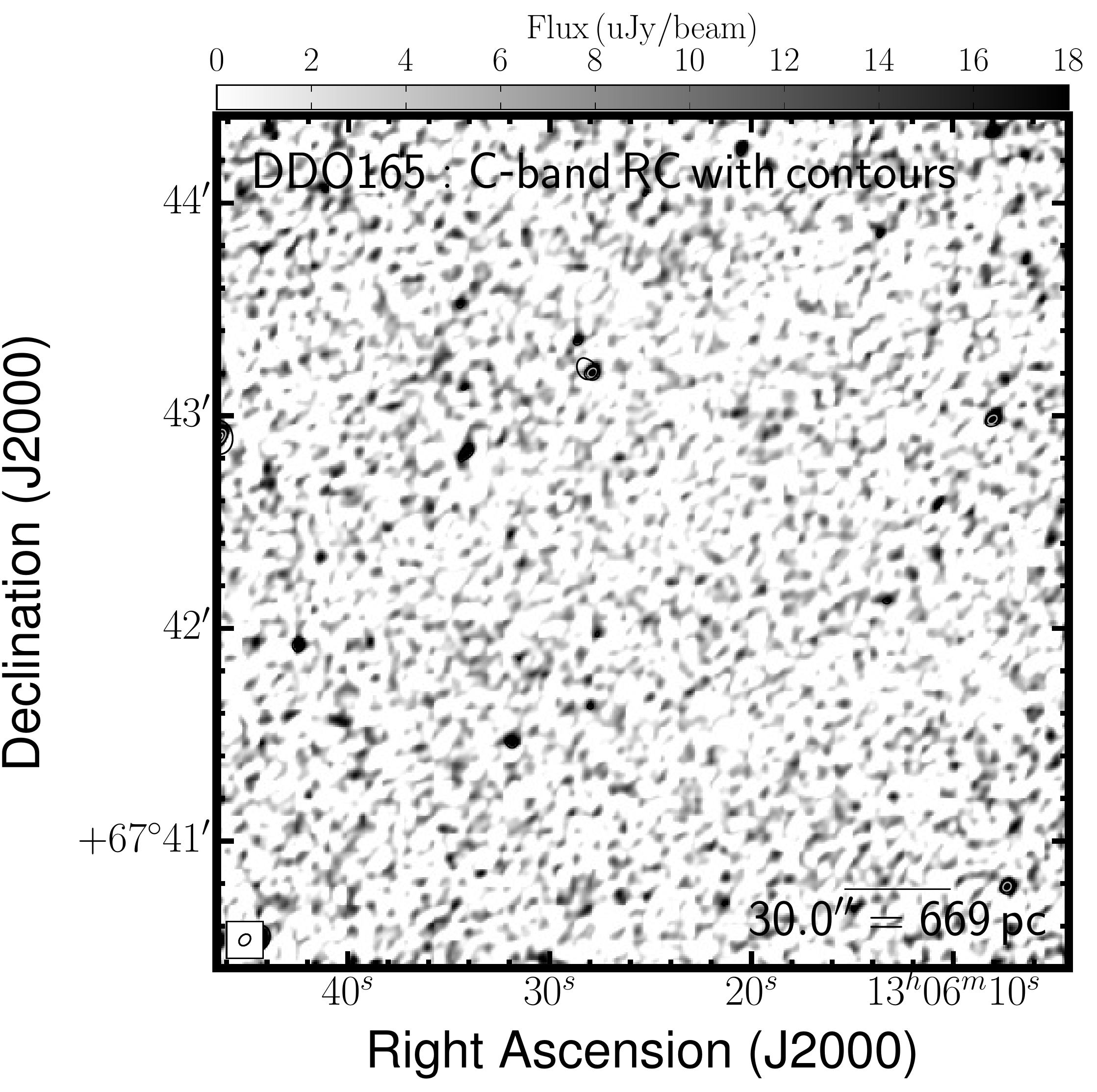} & \ 
    \includegraphics[width=0.31\linewidth,clip]{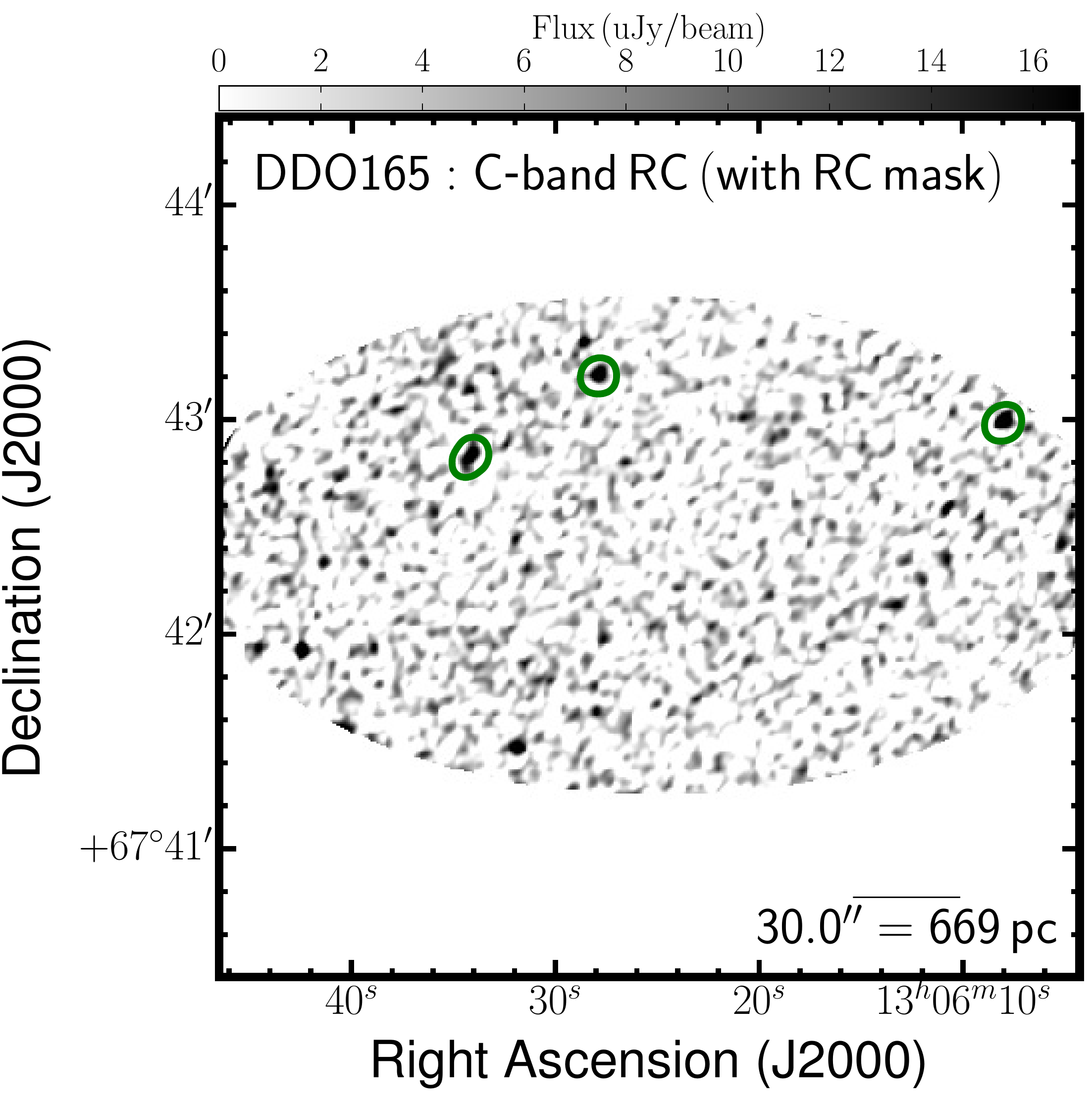} \\
    \includegraphics[width=0.31\linewidth,clip]{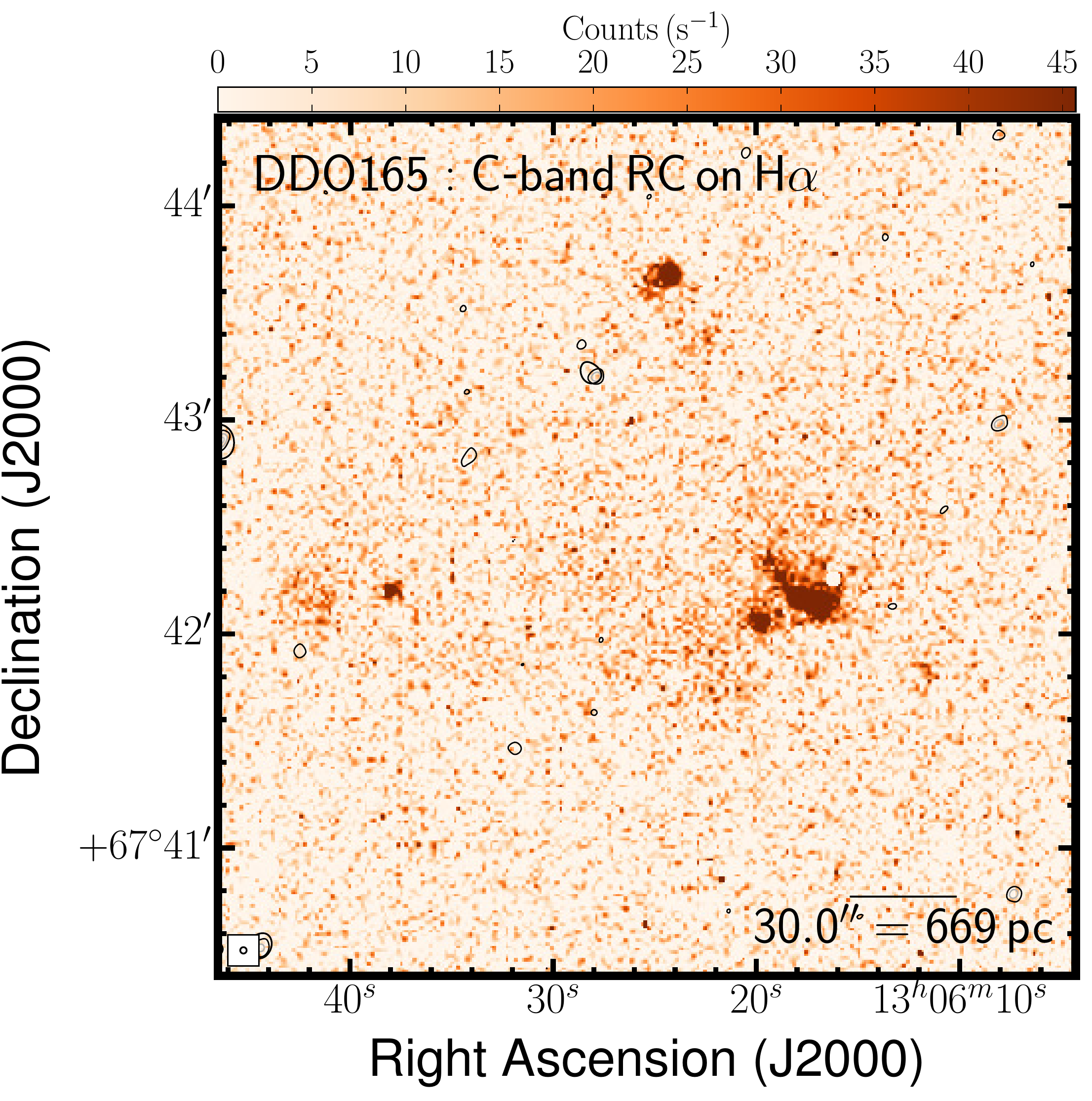} & \ 
    \includegraphics[width=0.31\linewidth,clip]{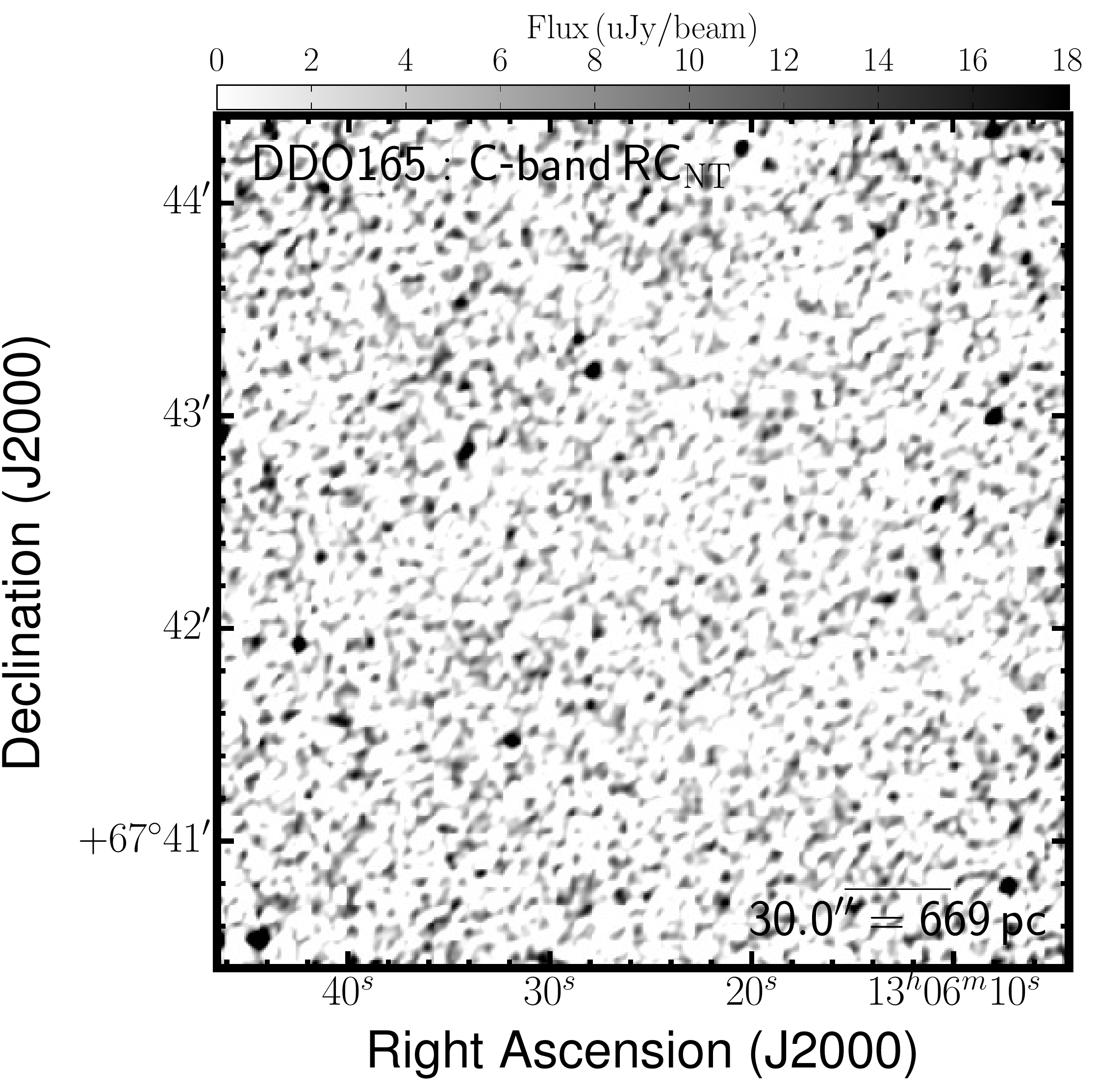} & \ 
    \includegraphics[width=0.31\linewidth,clip]{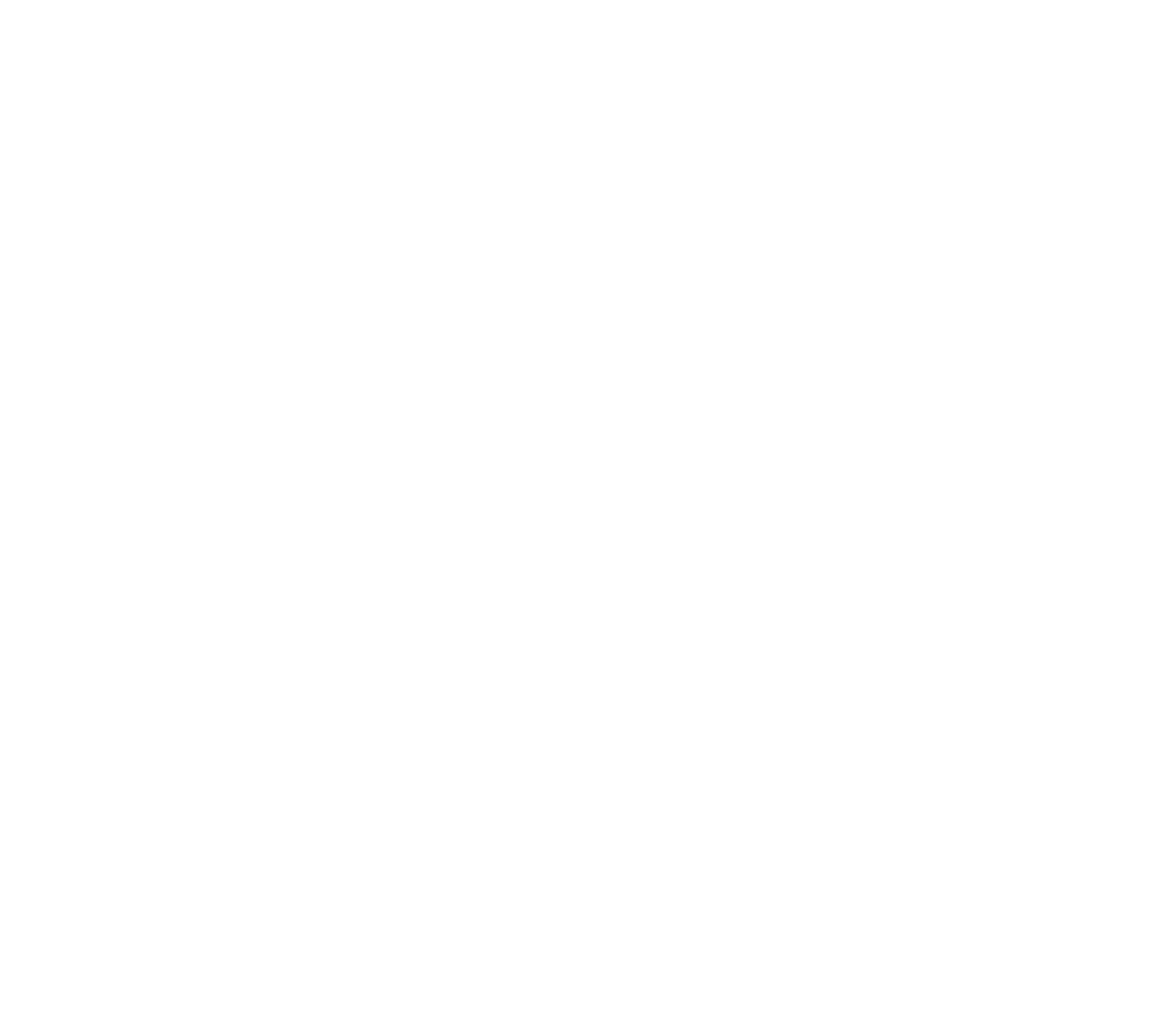} \\
    \includegraphics[width=0.31\linewidth,clip]{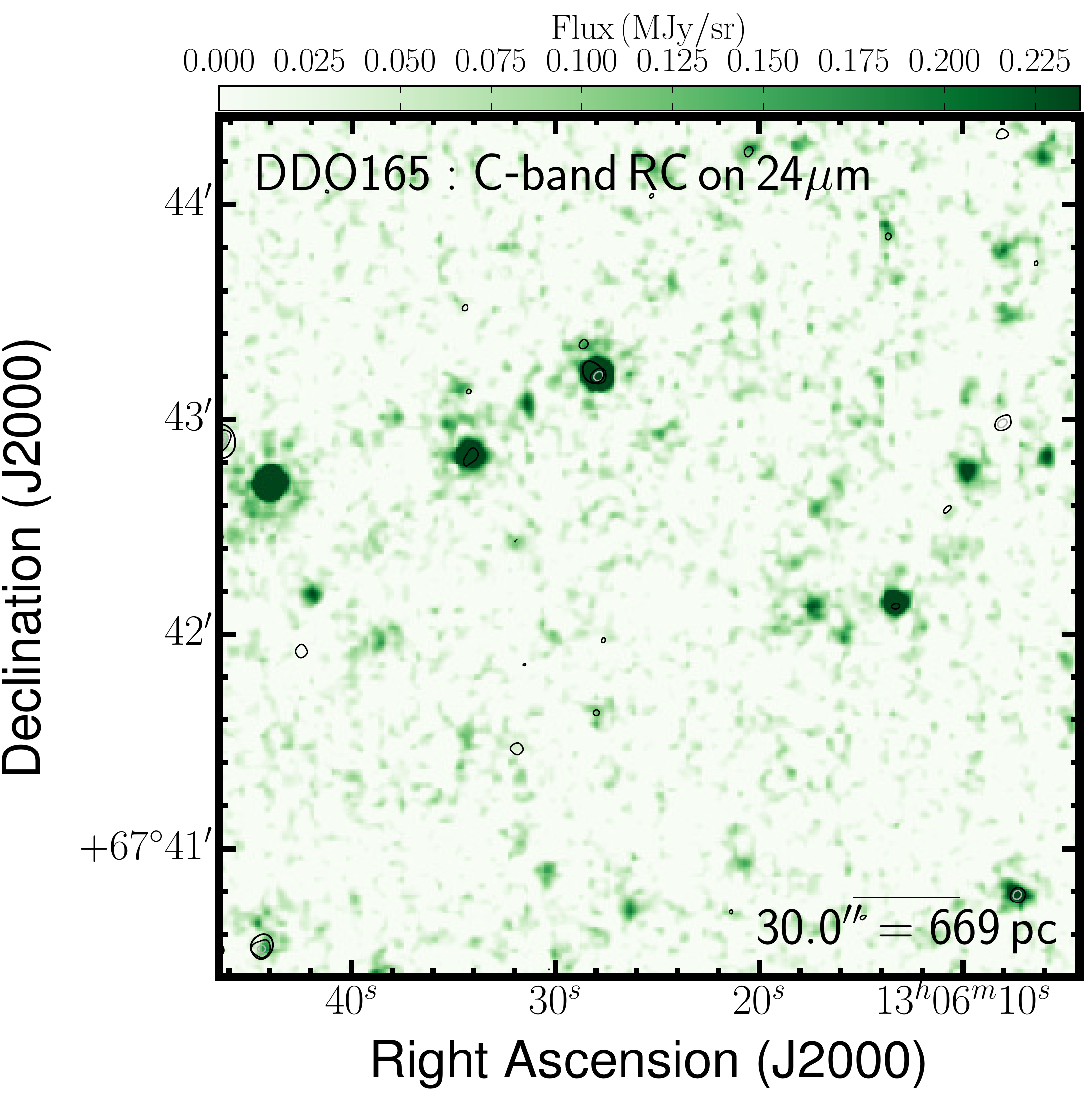} & \ 
    \includegraphics[width=0.31\linewidth,clip]{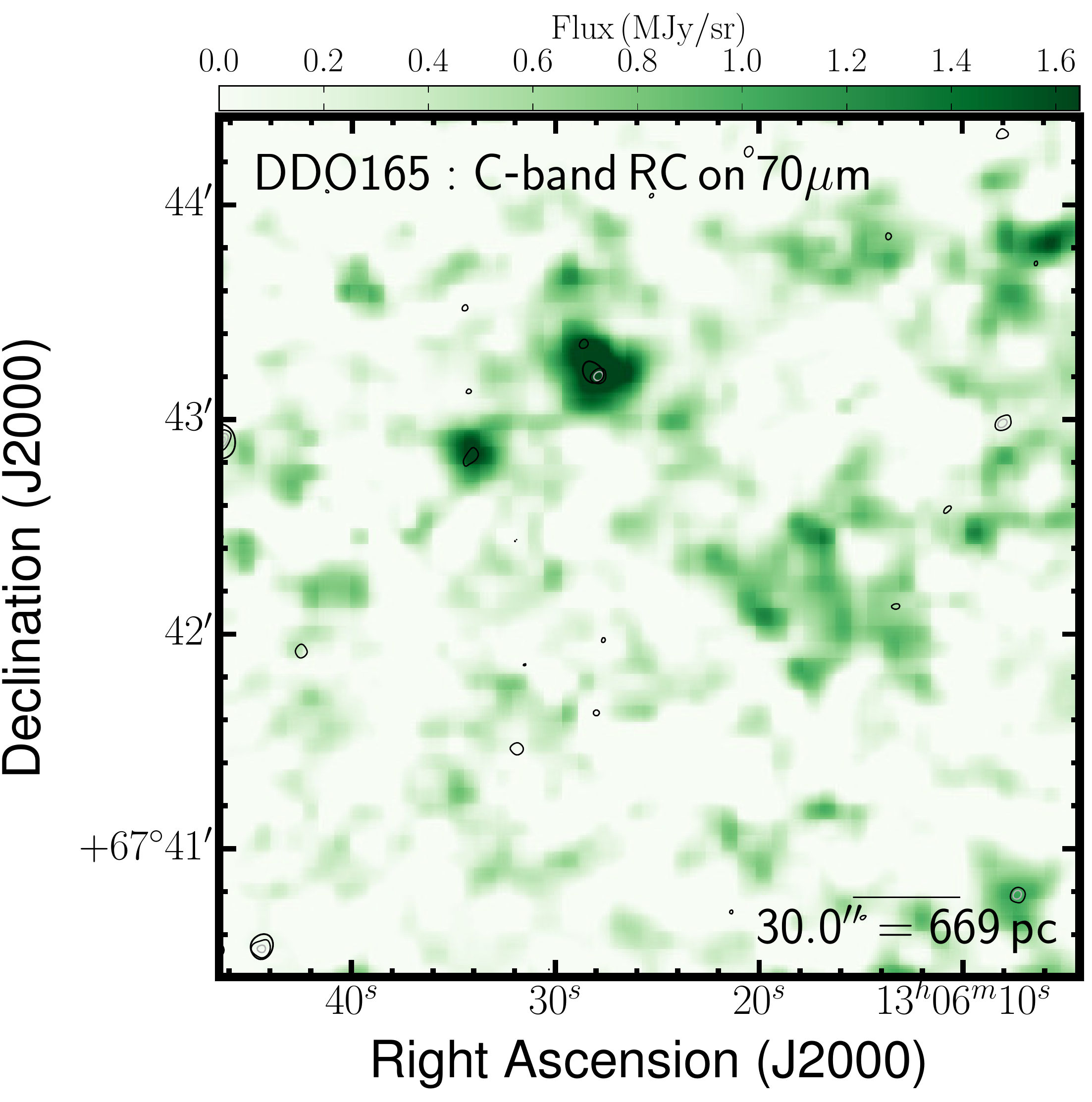} & \ 
    \includegraphics[width=0.31\linewidth,clip]{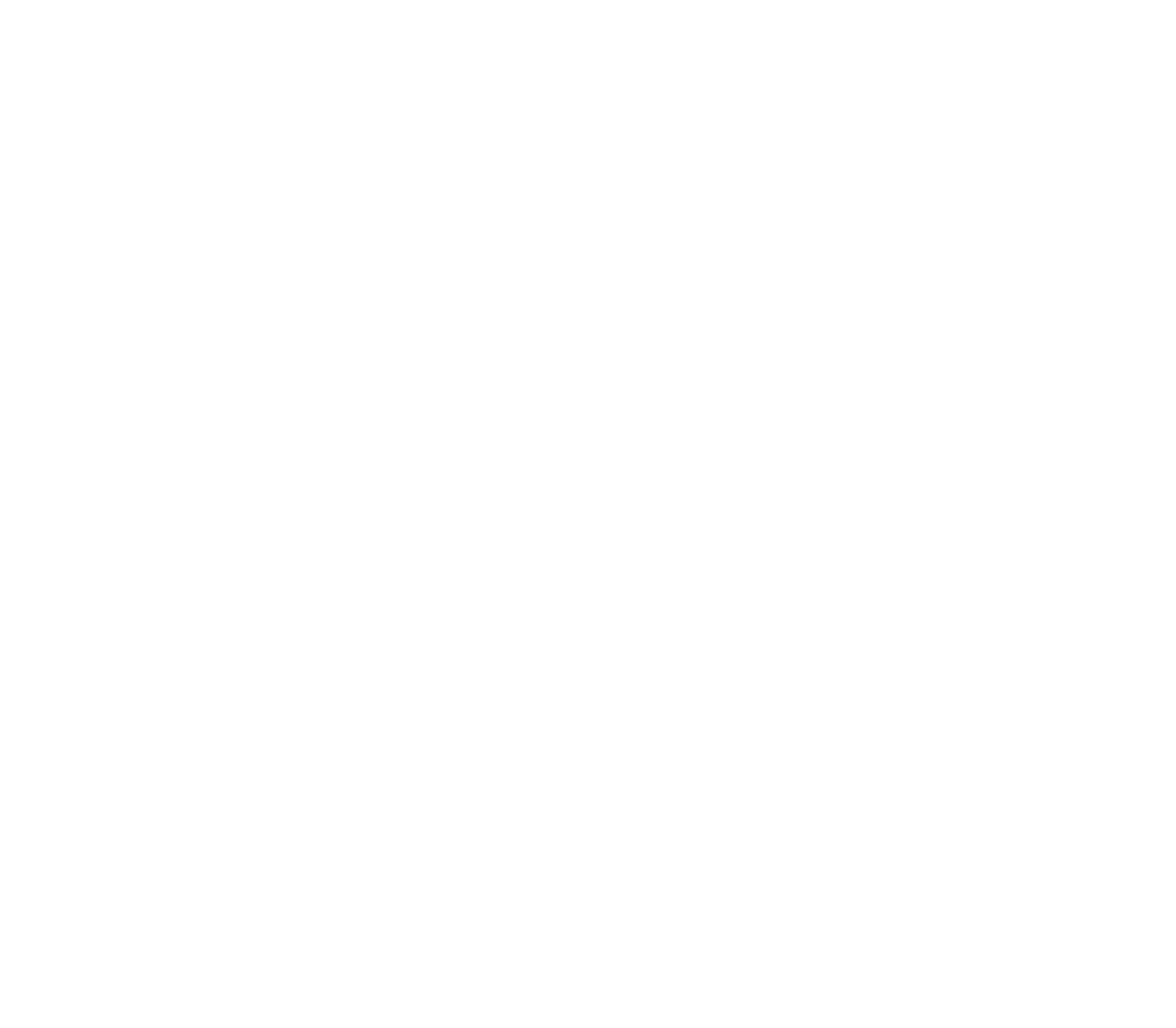} \\
  \end{tabular}
\caption[DDO\,165 images: RC, IR, optical, and FUV]{Multi-wavelength coverage of DDO 165 displaying a $4.0^\prime \times 4.0^\prime$ area. We show total RC flux density at the native resolution (top-left) and again with contours (top-centre). The RC contours are superposed on ancillary LITTLE THINGS images where possible: \halpha\ (middle-left); \RCNT\ obtained by subtracting the expected \RCT\ based on the \halpha-\RCT\ scaling factor of \cite{Deeg1997} from the total RC; {\em GALEX} FUV (middle-right); {\em Spitzer} 24\micron\ (bottom-left); {\em Spitzer} 70\micron\ (bottom-centre); FUV$+24{\rm \mu m}$--inferred SFRD from \citealp{Leroy2012} (bottom-right). We also show the RC that was isolated by the RC--based masking technique (top-right).}
  \label{figure:ddo165Cc_maps}
\end{figure}

\begin{figure}
  \begin{tabular}{ccc}
    \includegraphics[width=0.31\linewidth,clip]{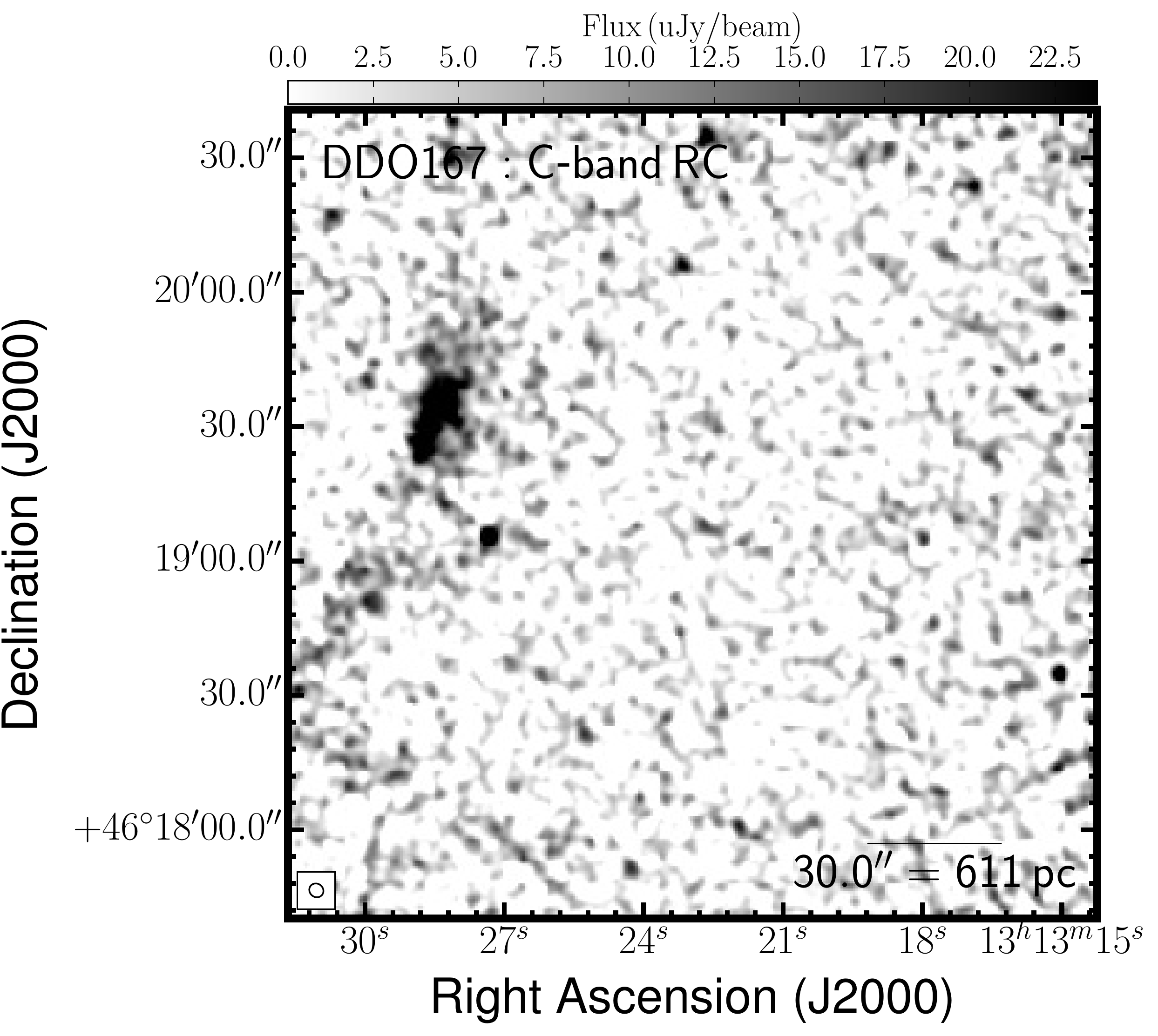} & \ 
    \includegraphics[width=0.31\linewidth,clip]{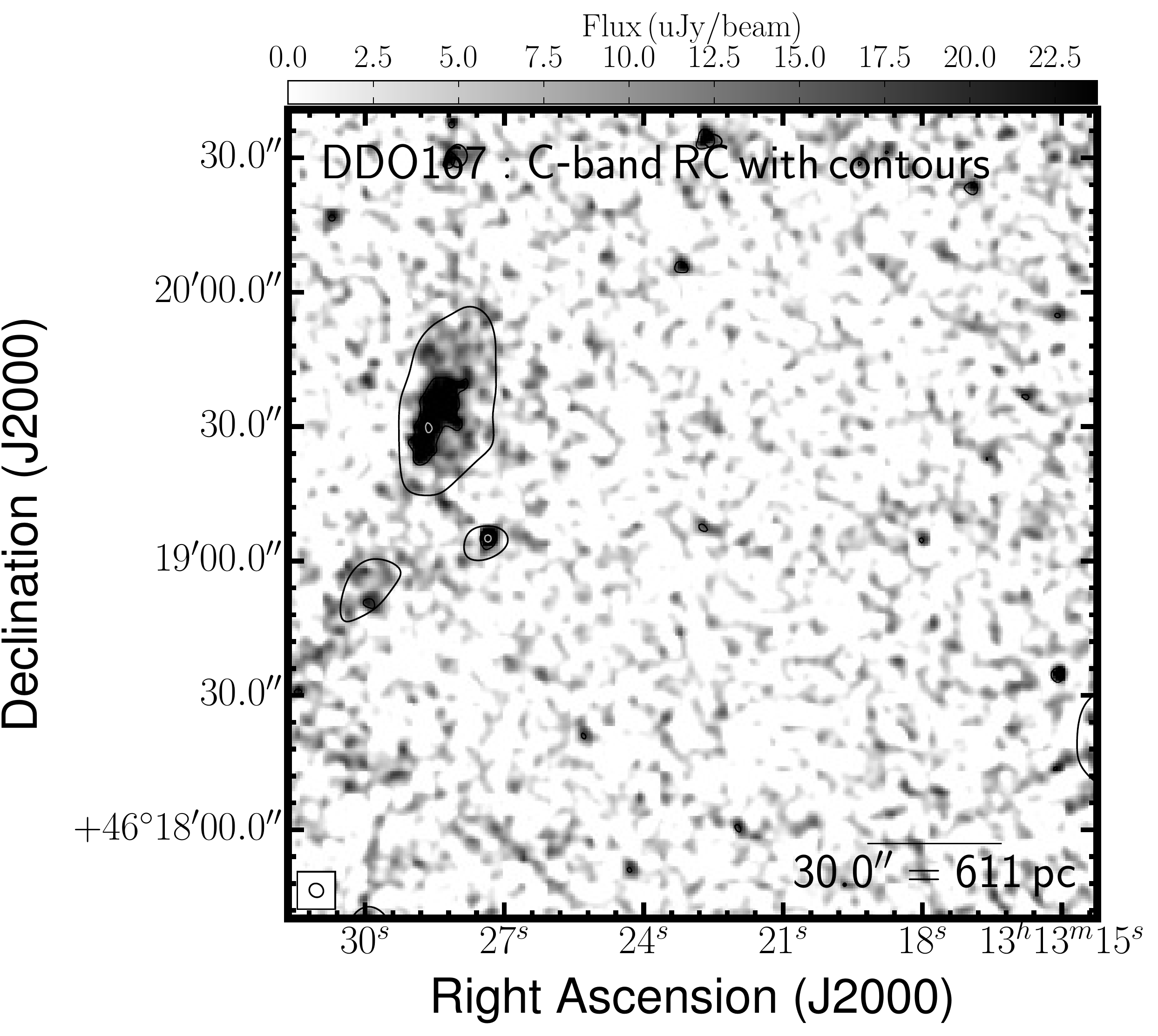} & \ 
    \includegraphics[width=0.31\linewidth,clip]{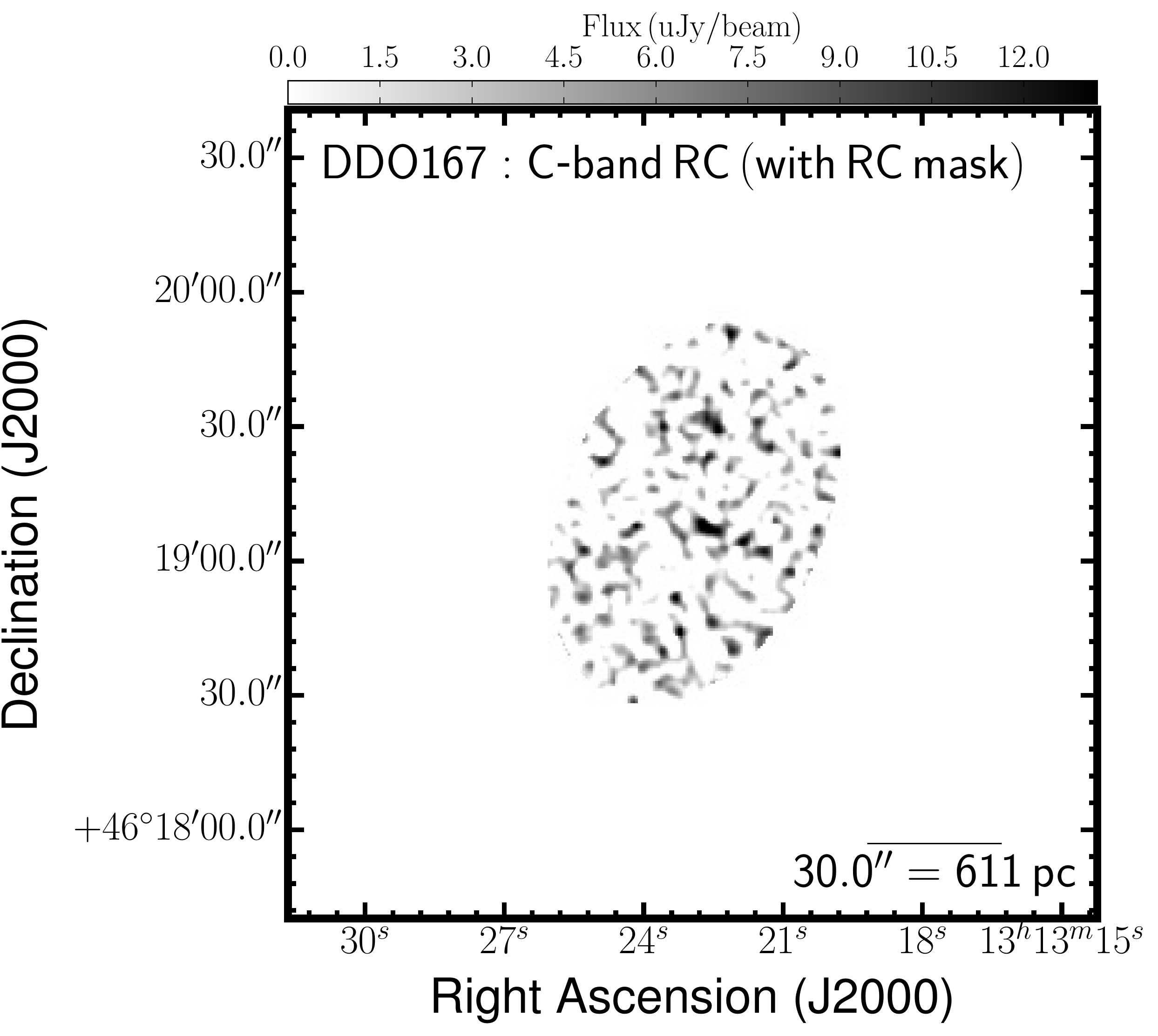} \\
    \includegraphics[width=0.31\linewidth,clip]{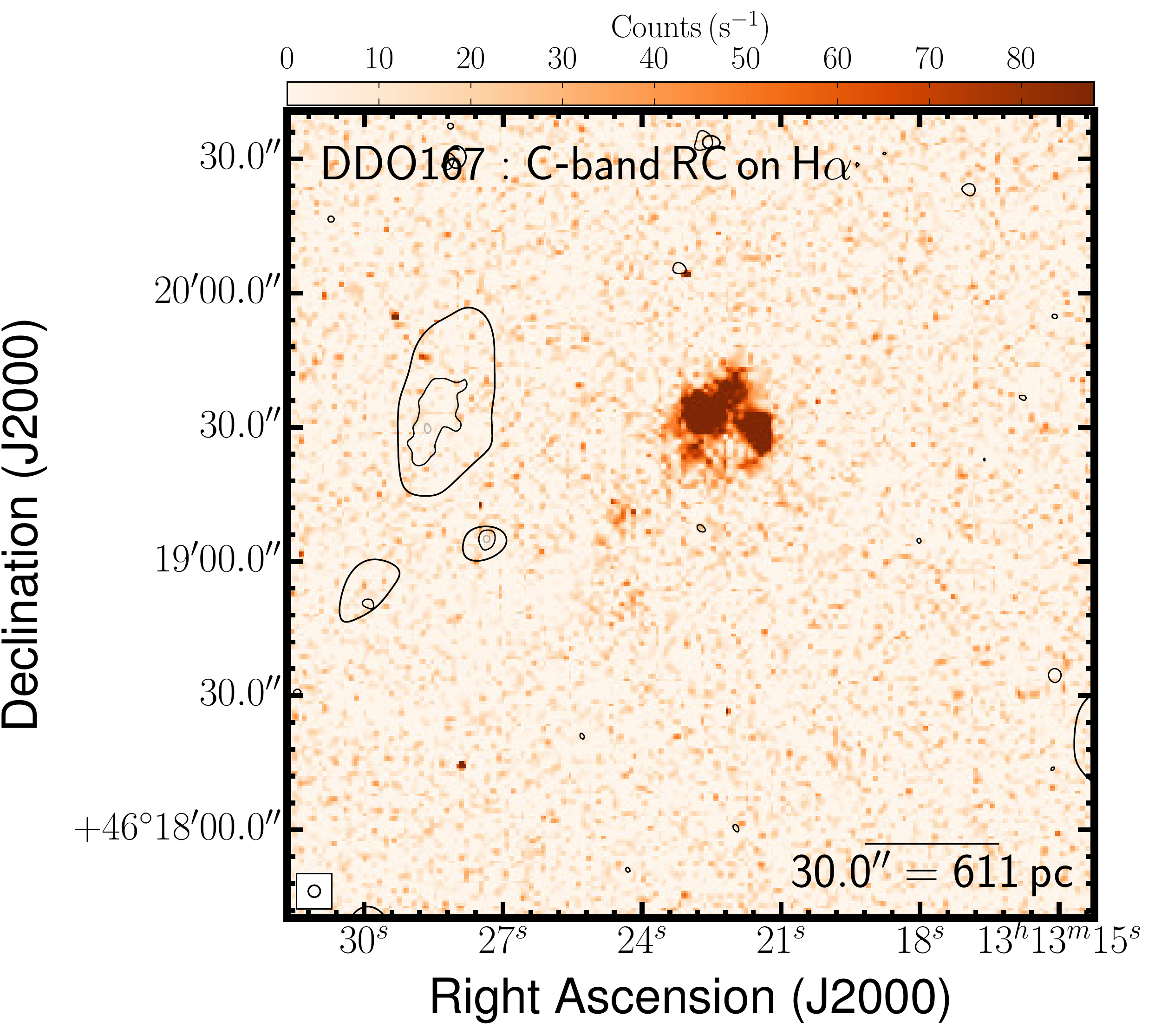} & \ 
    \includegraphics[width=0.31\linewidth,clip]{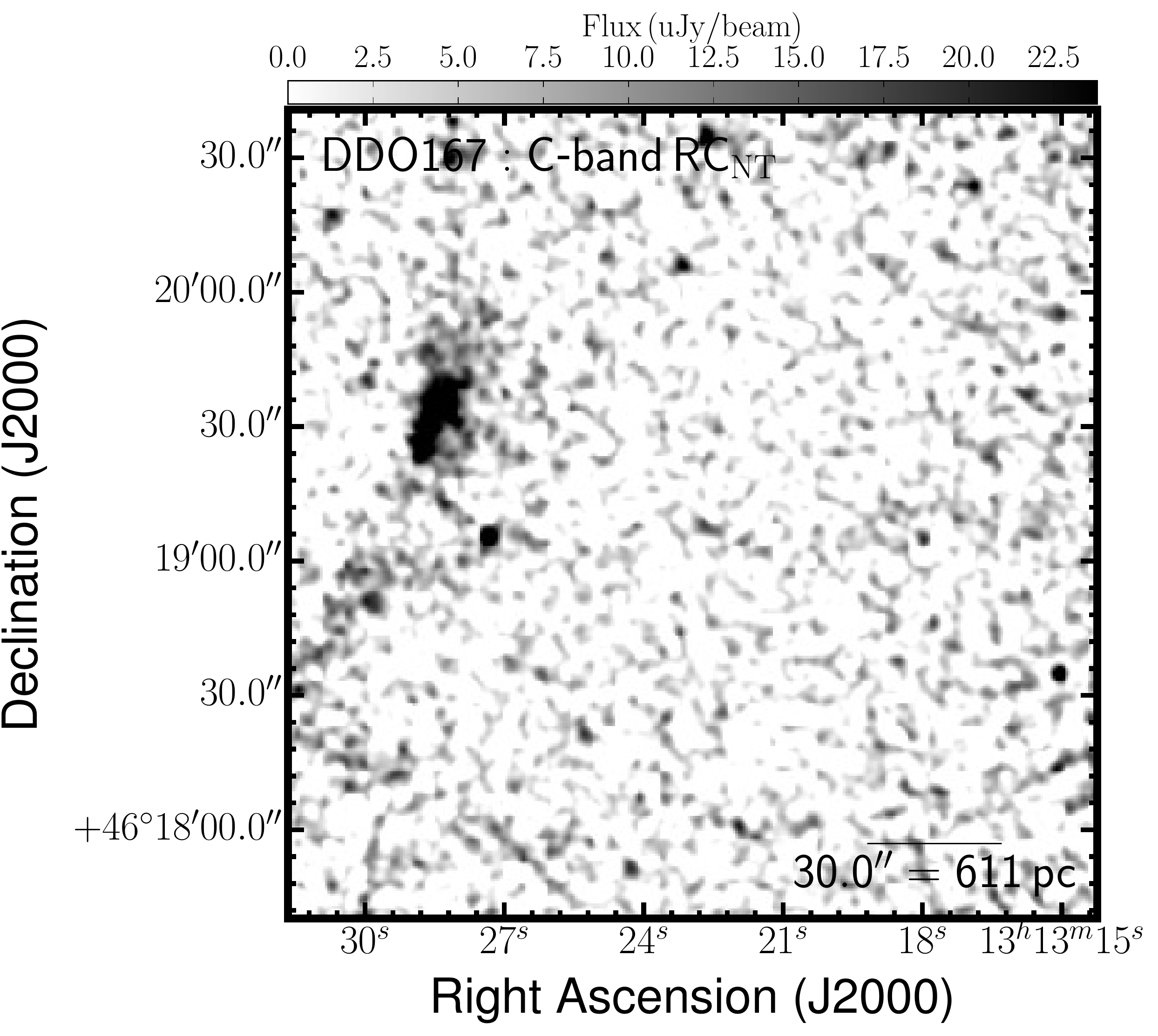} & \ 
    \includegraphics[width=0.31\linewidth,clip]{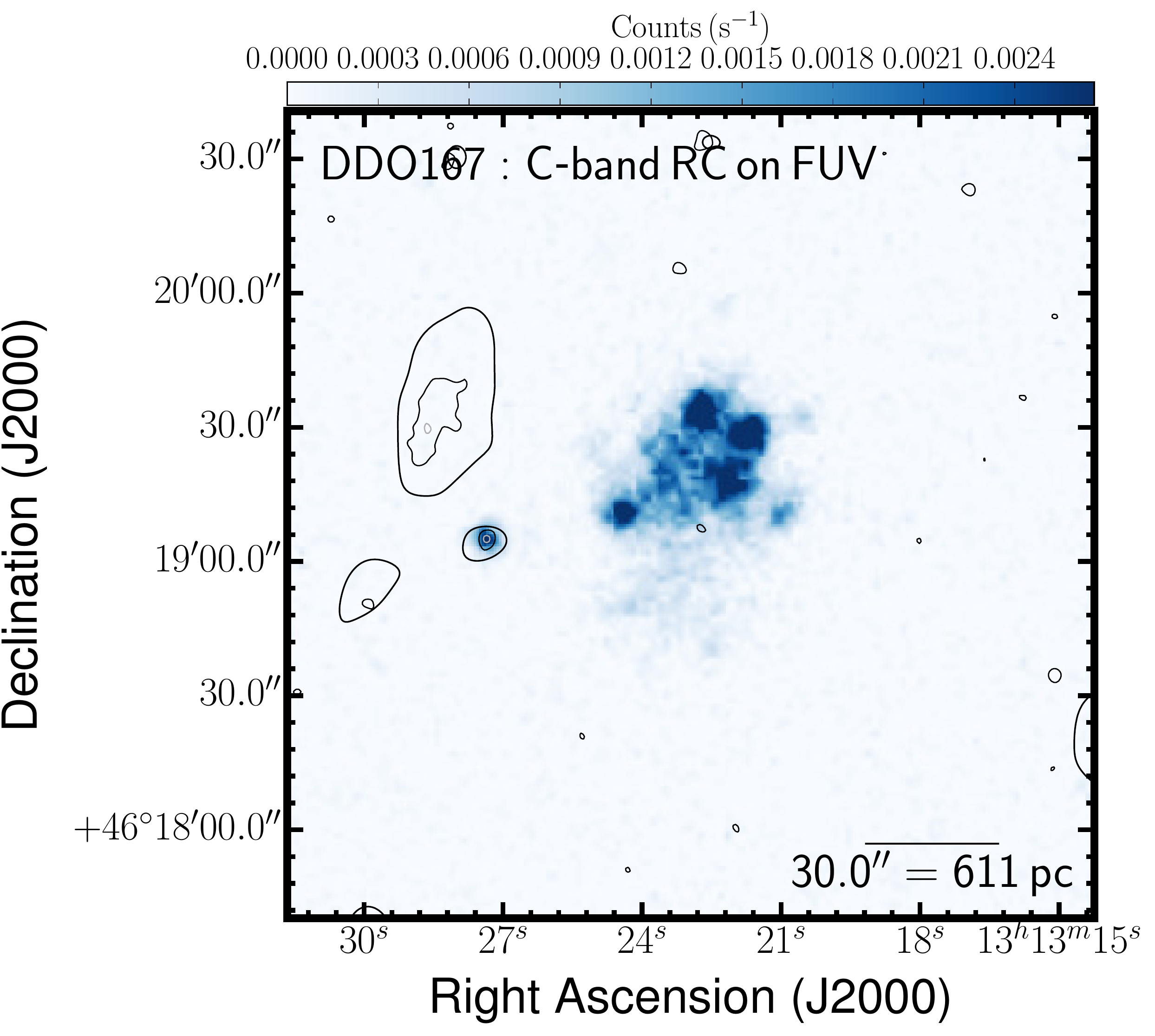} \\
    \includegraphics[width=0.31\linewidth,clip]{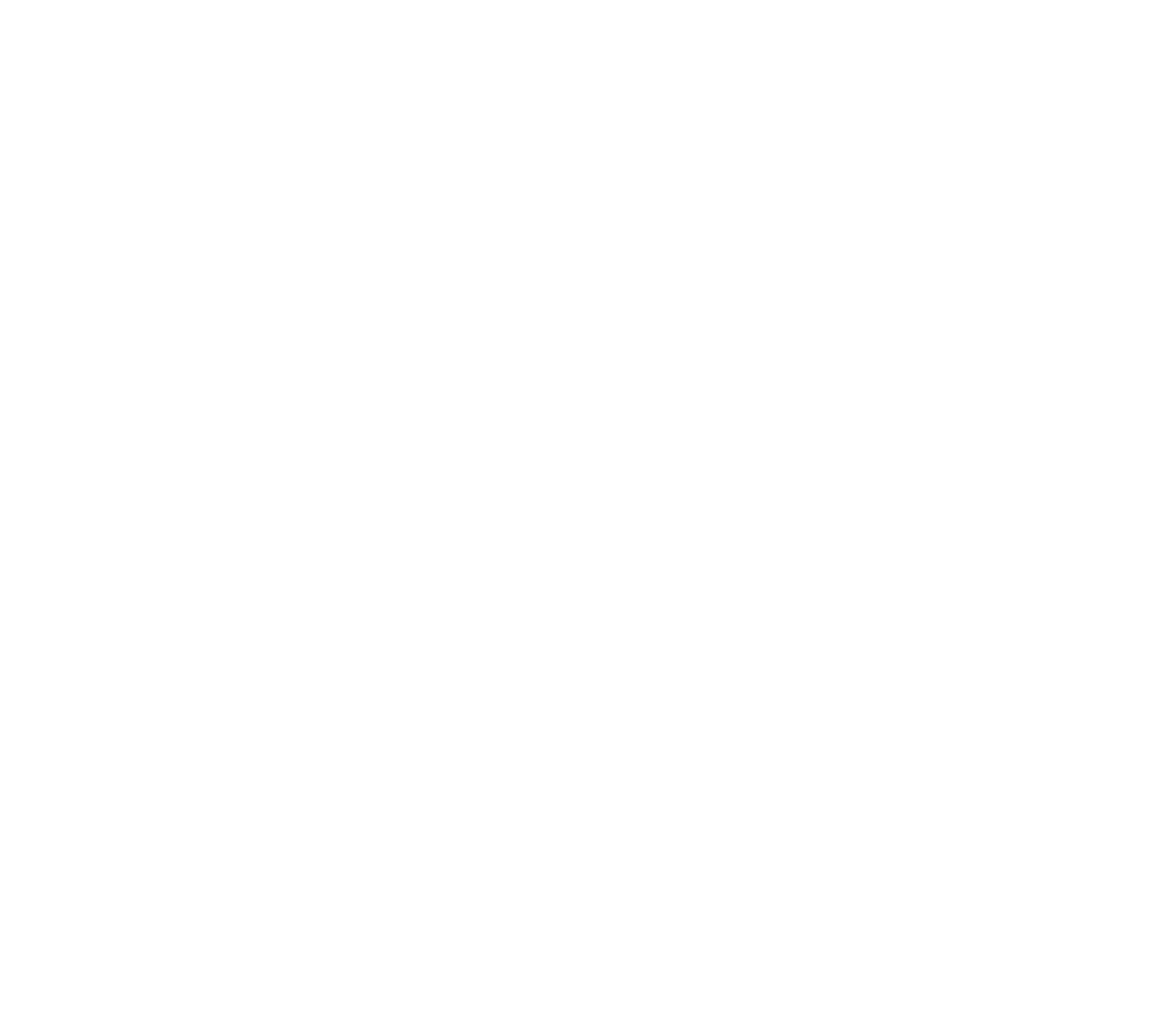} & \ 
    \includegraphics[width=0.31\linewidth,clip]{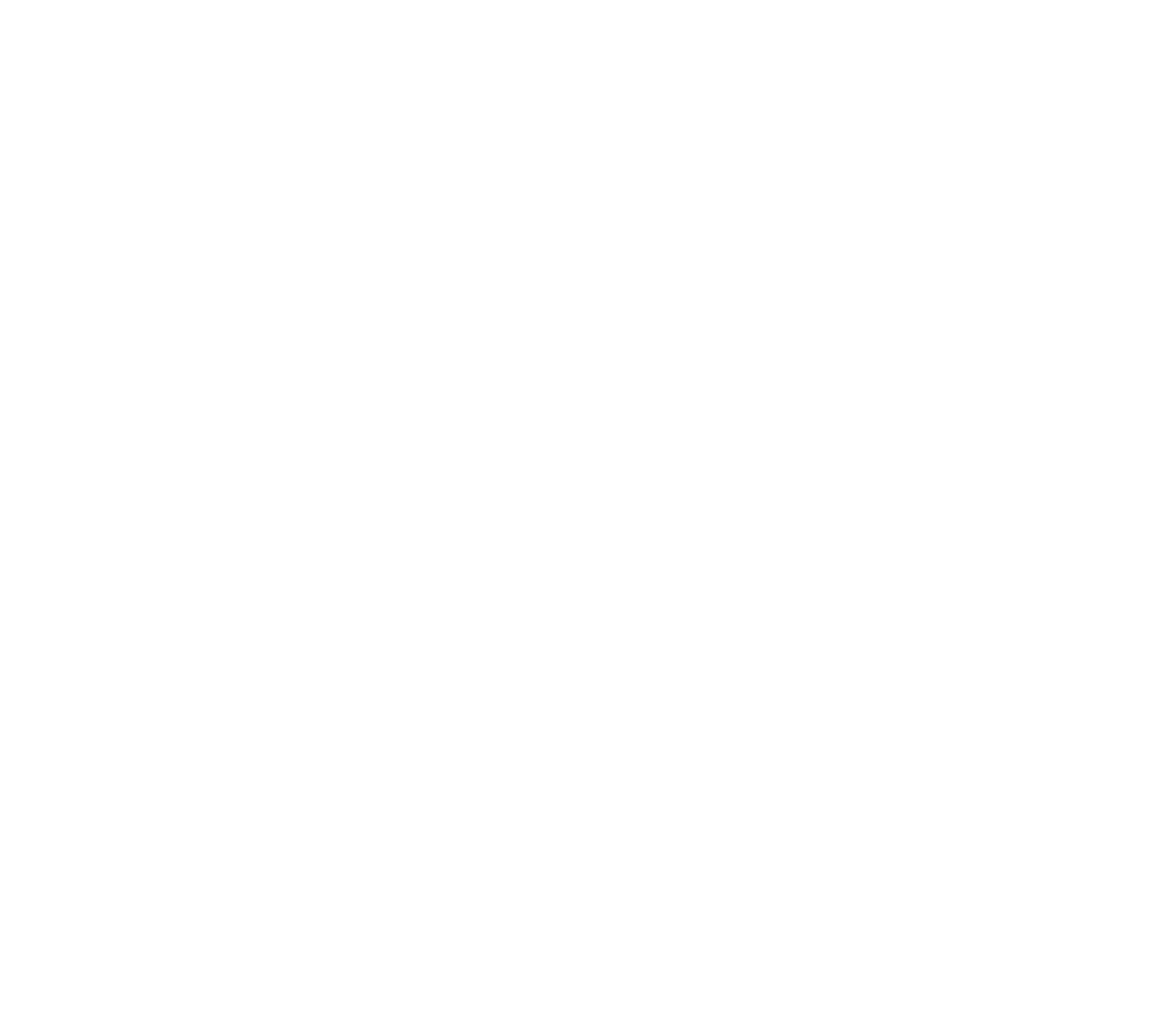} & \ 
    \includegraphics[width=0.31\linewidth,clip]{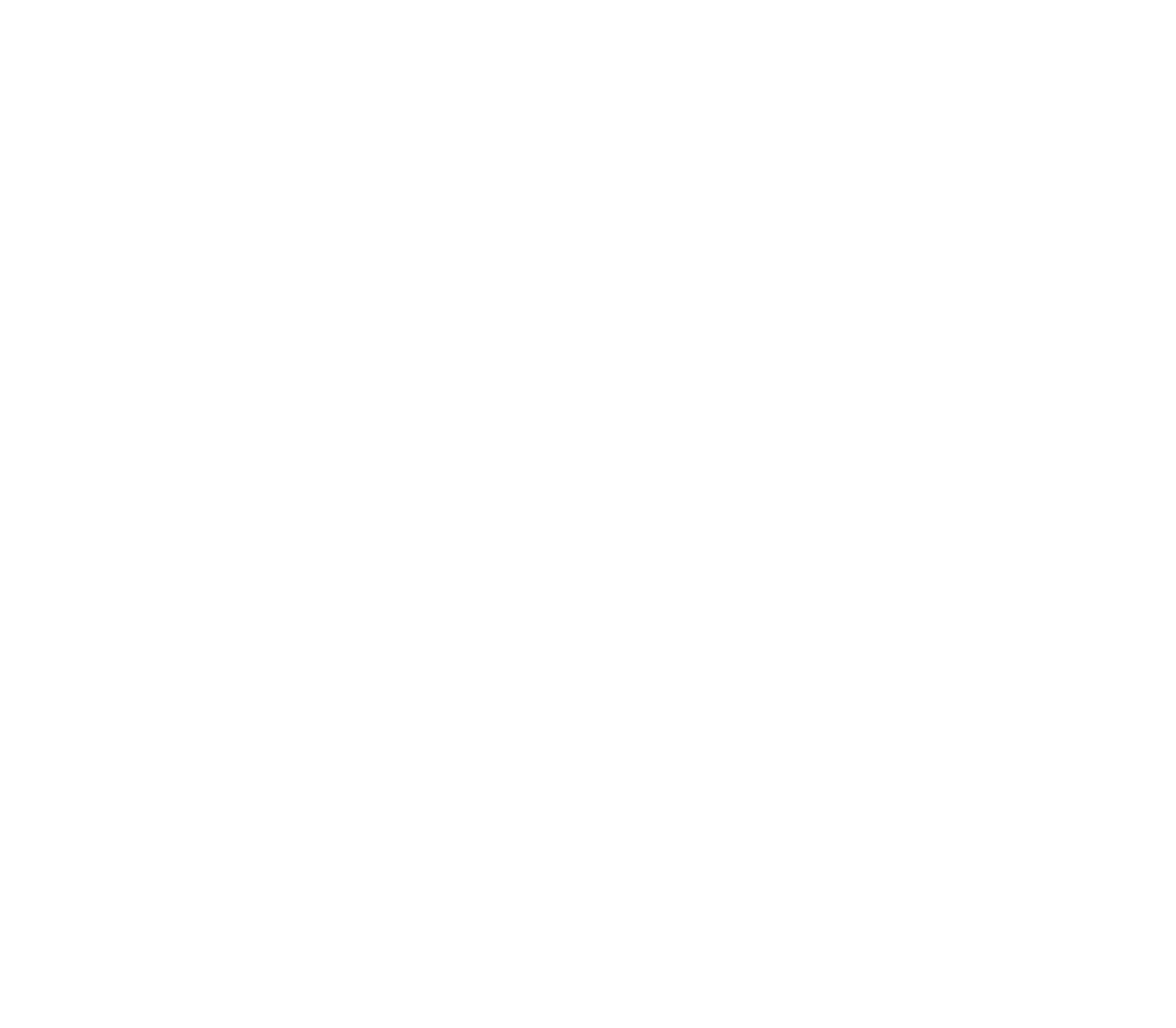} \\
  \end{tabular}
\caption[DDO\,167 images: RC, IR, optical, and FUV]{Multi-wavelength coverage of DDO 167 displaying a $3.0^\prime \times 3.0^\prime$ area. We show total RC flux density at the native resolution (top-left) and again with contours (top-centre). The RC contours are superposed on ancillary LITTLE THINGS images where possible: \halpha\ (middle-left); \RCNT\ obtained by subtracting the expected \RCT\ based on the \halpha-\RCT\ scaling factor of \cite{Deeg1997} from the total RC; {\em GALEX} FUV (middle-right); {\em Spitzer} 24\micron\ (bottom-left); {\em Spitzer} 70\micron\ (bottom-centre); FUV$+24{\rm \mu m}$--inferred SFRD from \citealp{Leroy2012} (bottom-right). We also show the RC that was isolated by the RC--based masking technique (top-right).}
  \label{figure:ddo167Cc_maps}
\end{figure}

\begin{figure}
  \begin{tabular}{ccc}
    \includegraphics[width=0.31\linewidth,clip]{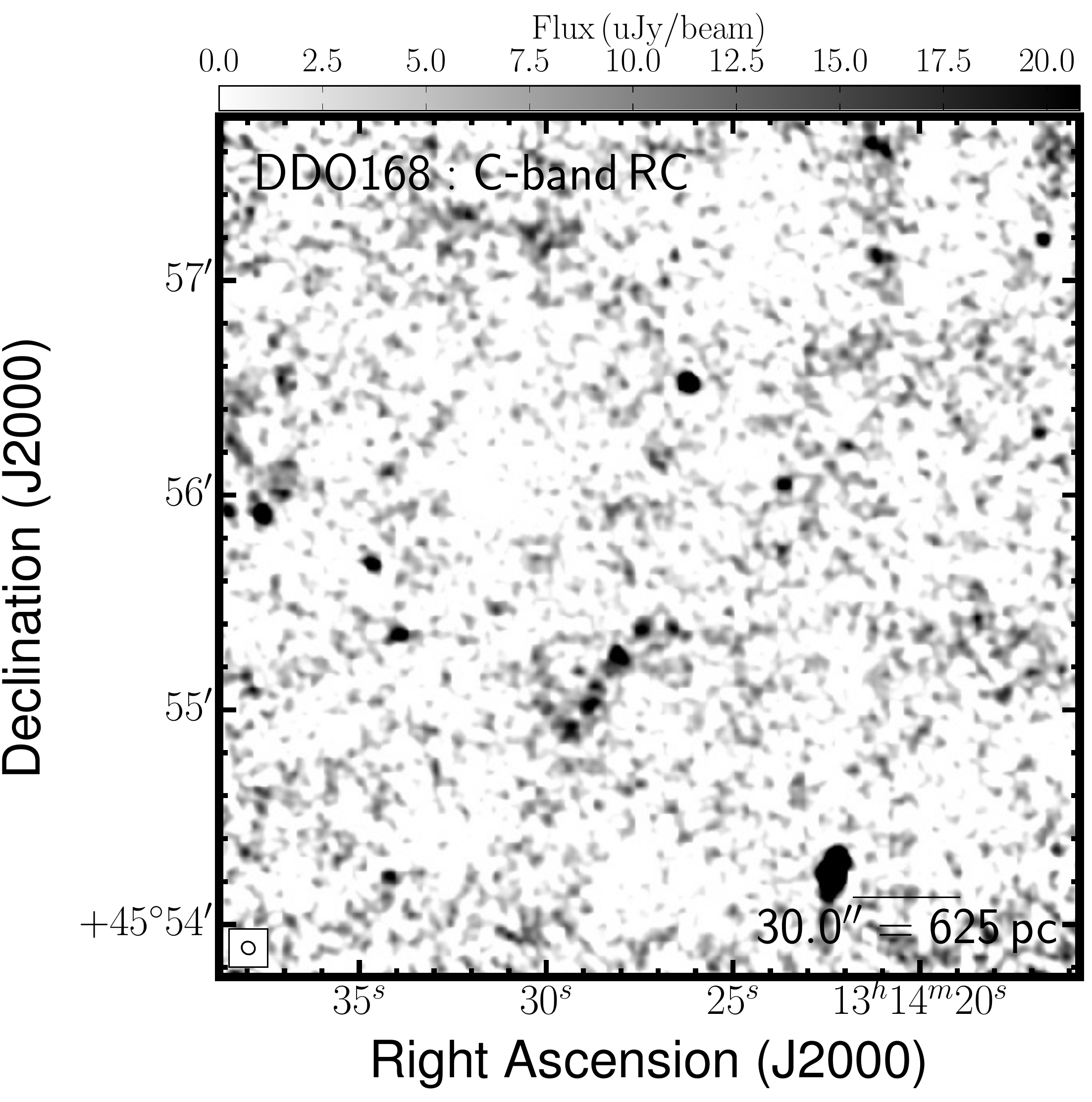} & \ 
    \includegraphics[width=0.31\linewidth,clip]{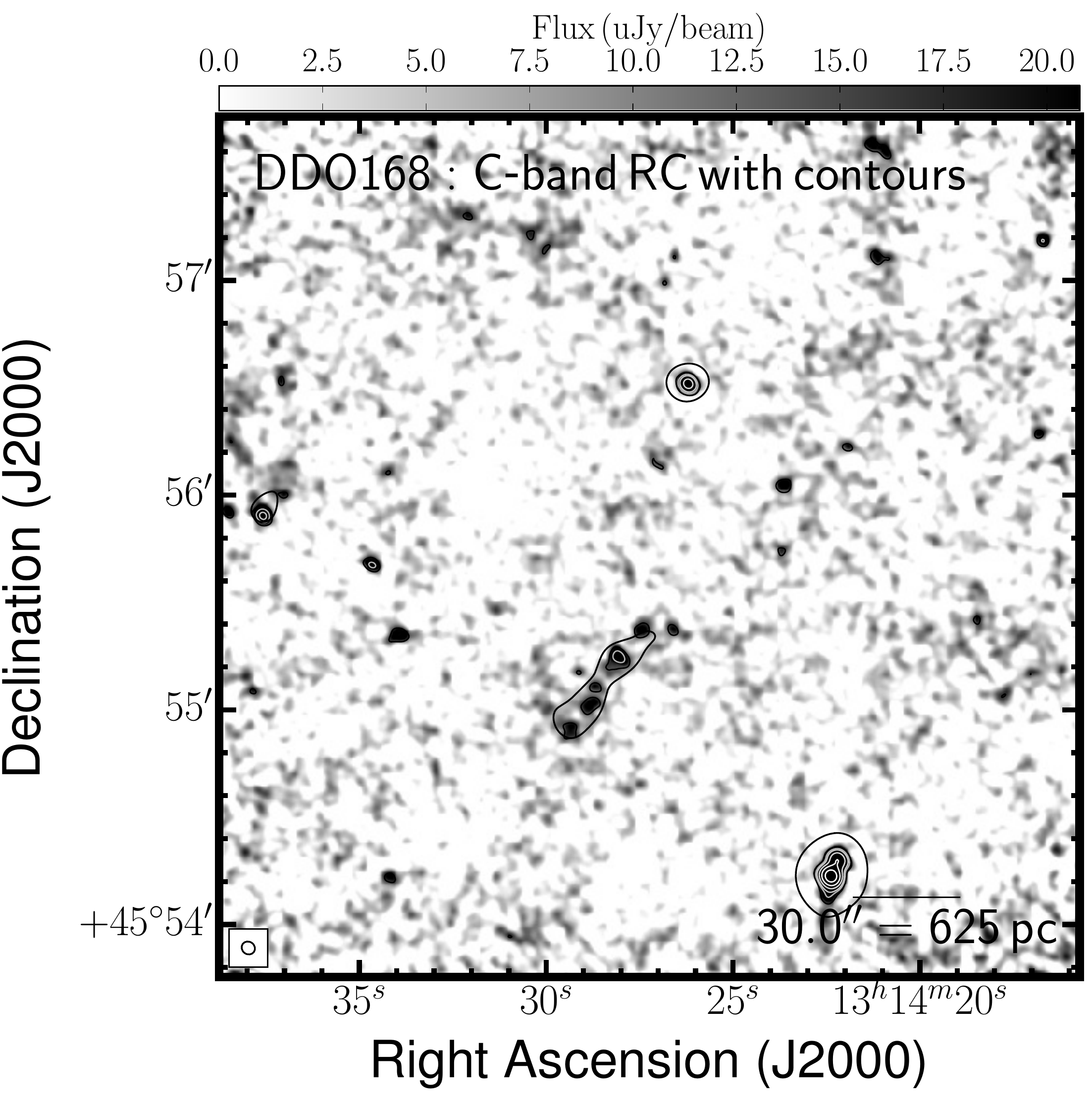} & \ 
    \includegraphics[width=0.31\linewidth,clip]{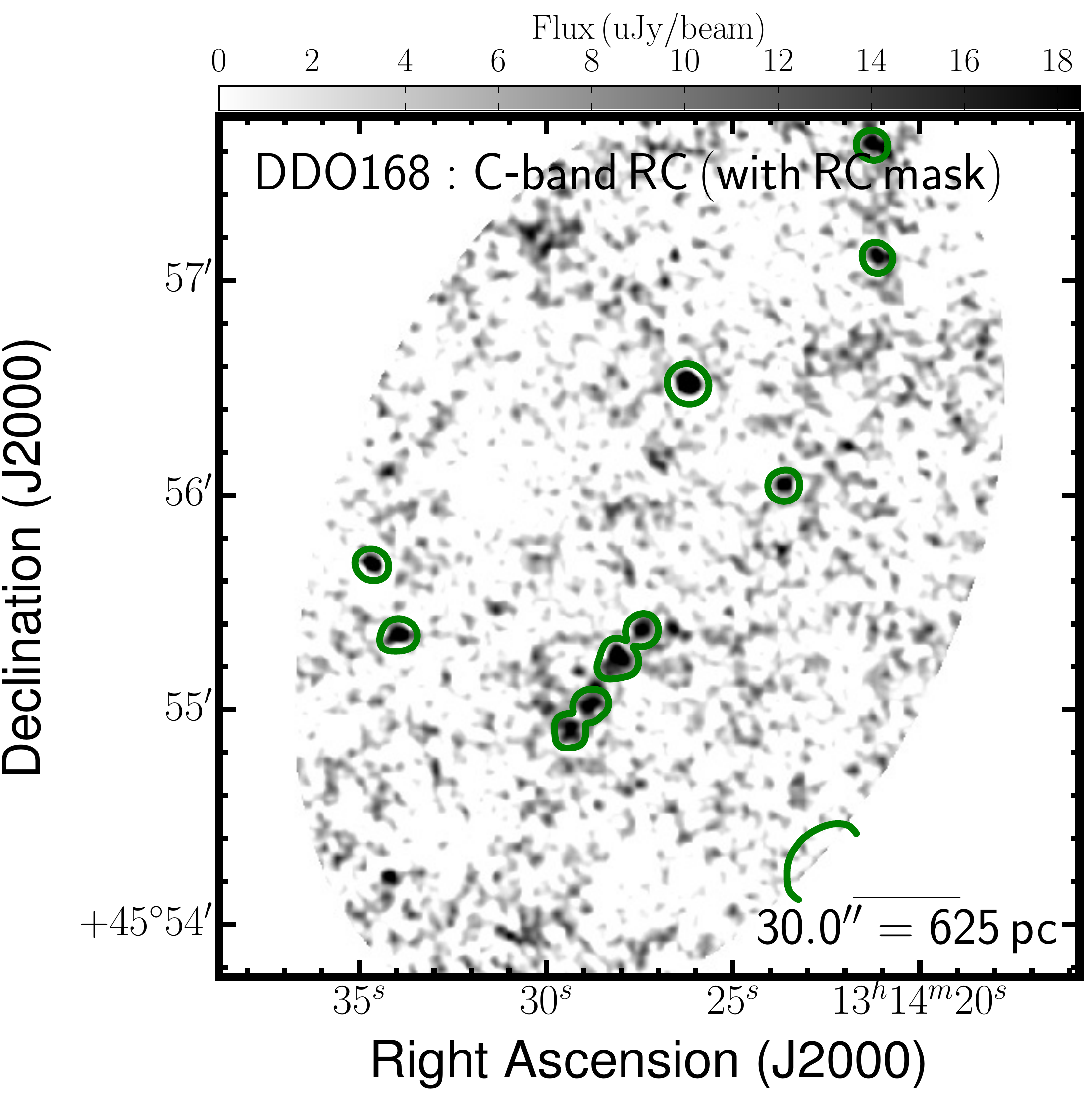} \\
    \includegraphics[width=0.31\linewidth,clip]{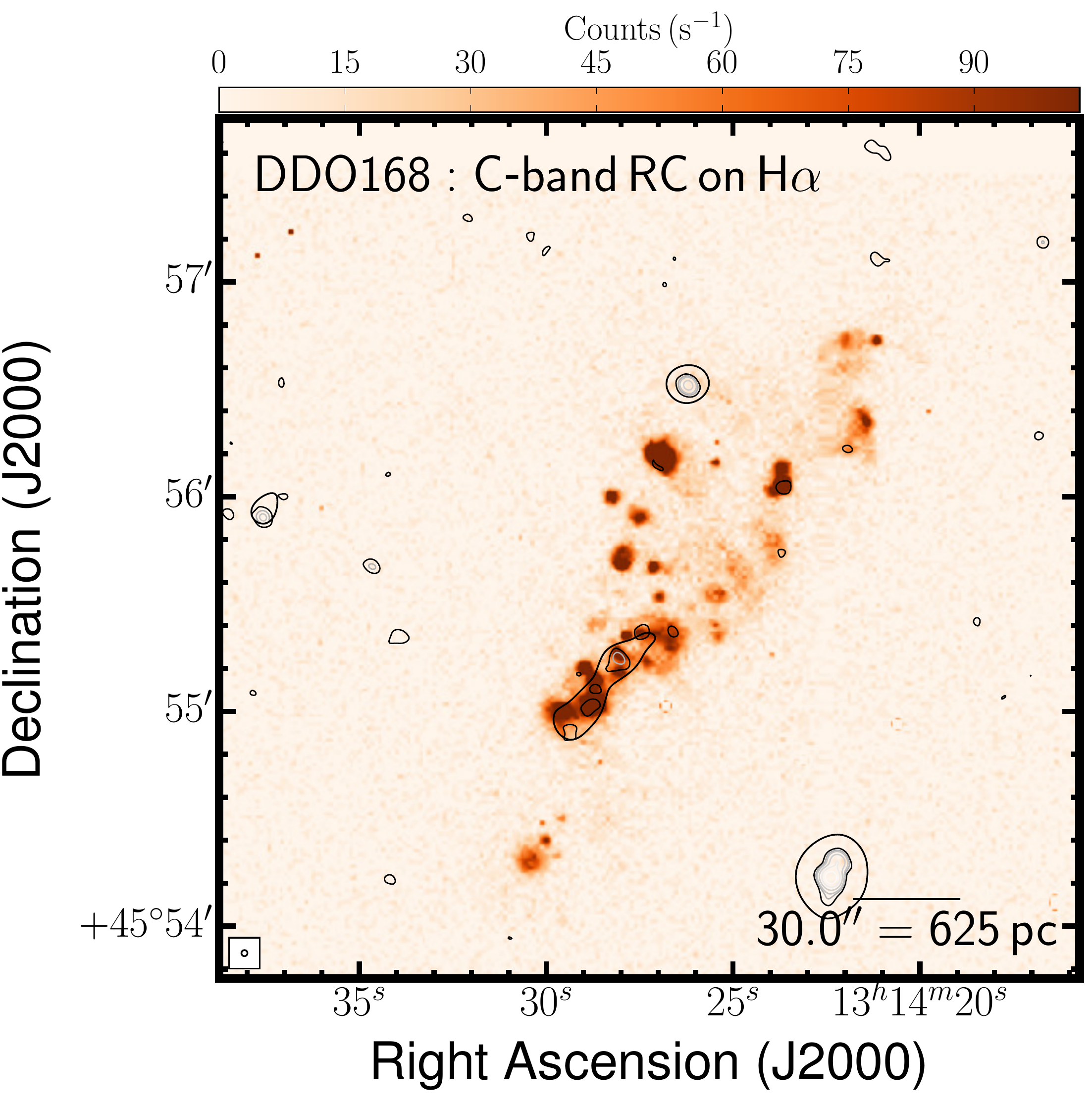} & \ 
    \includegraphics[width=0.31\linewidth,clip]{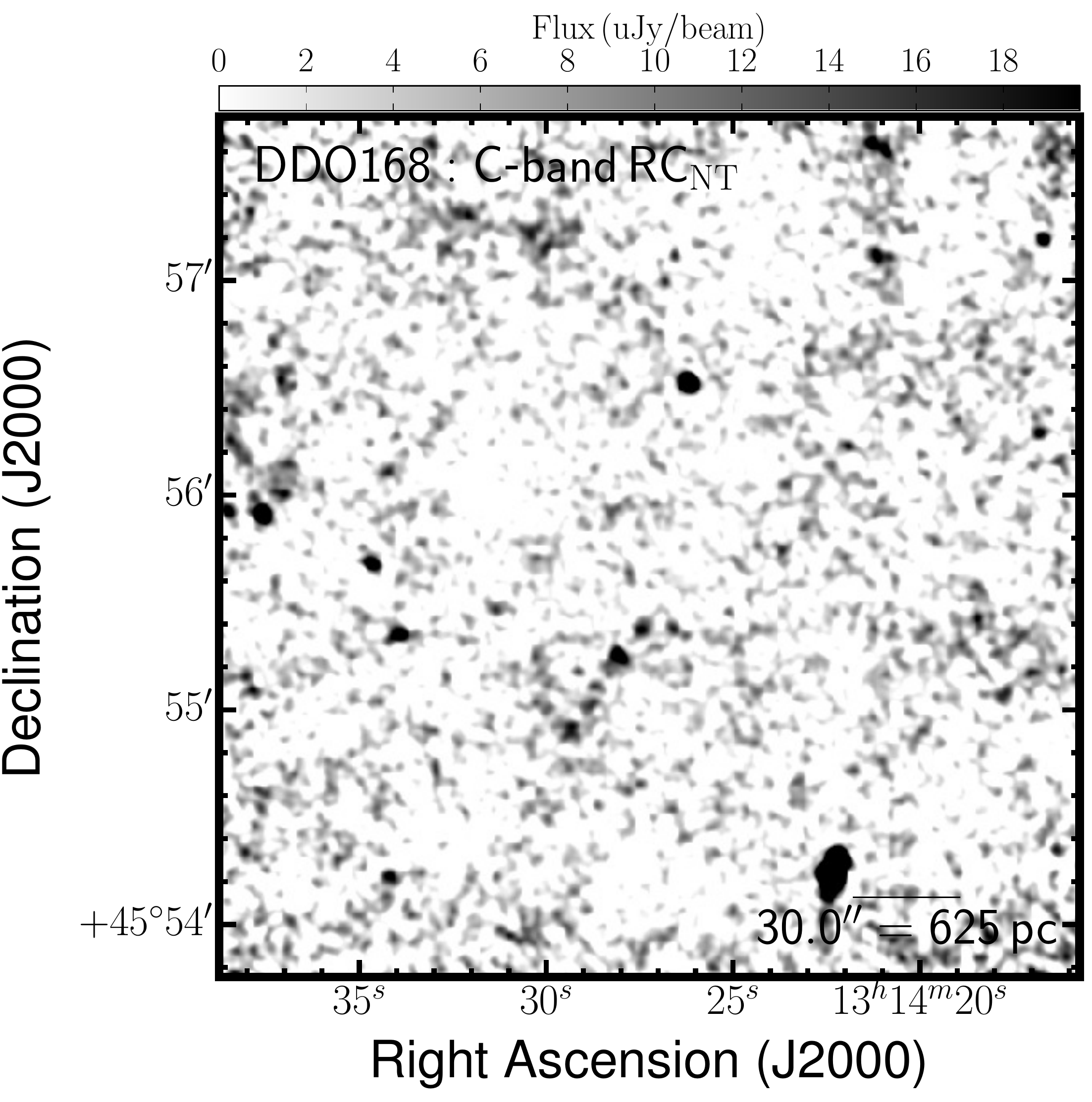} & \ 
    \includegraphics[width=0.31\linewidth,clip]{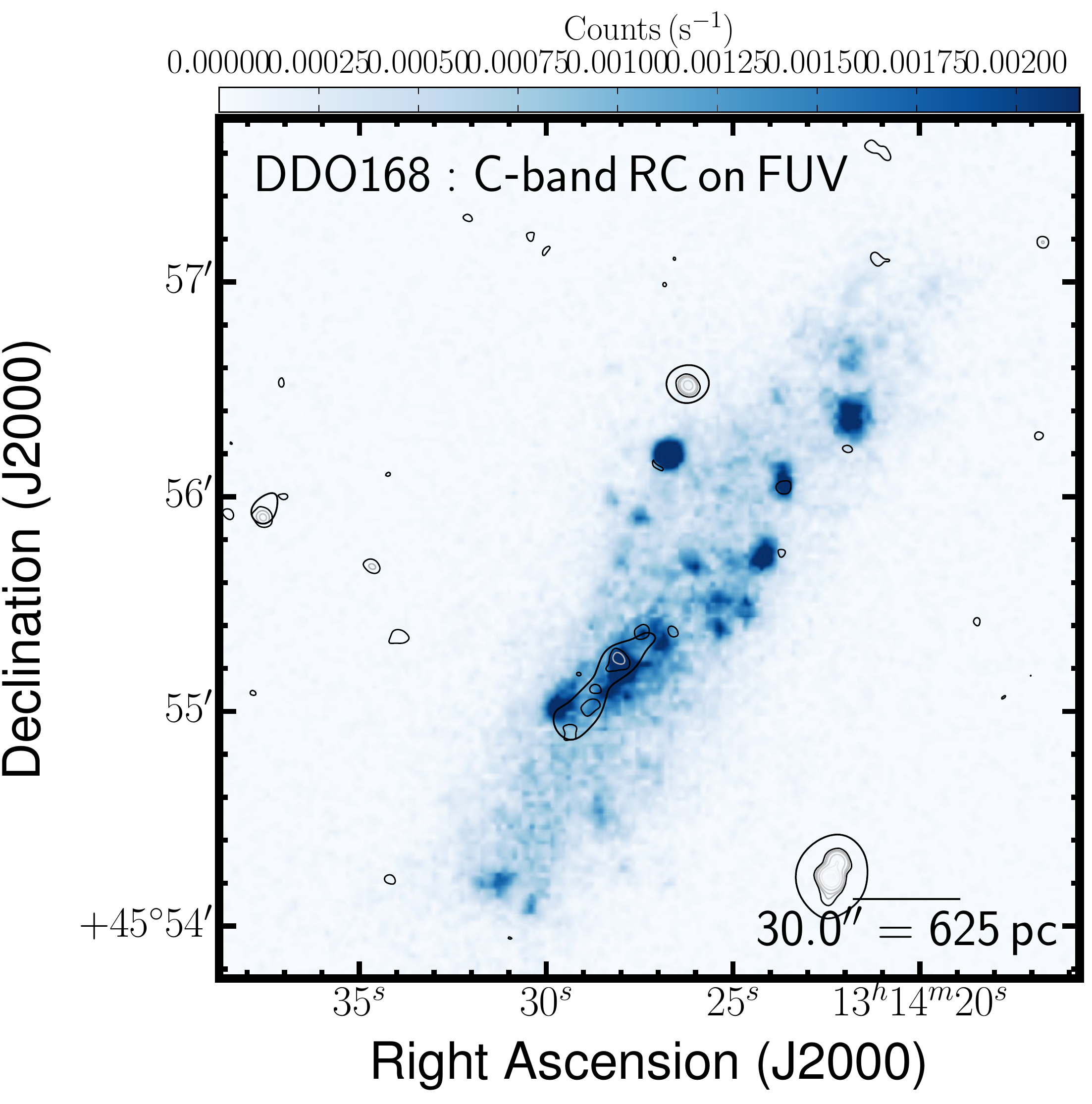} \\
    \includegraphics[width=0.31\linewidth,clip]{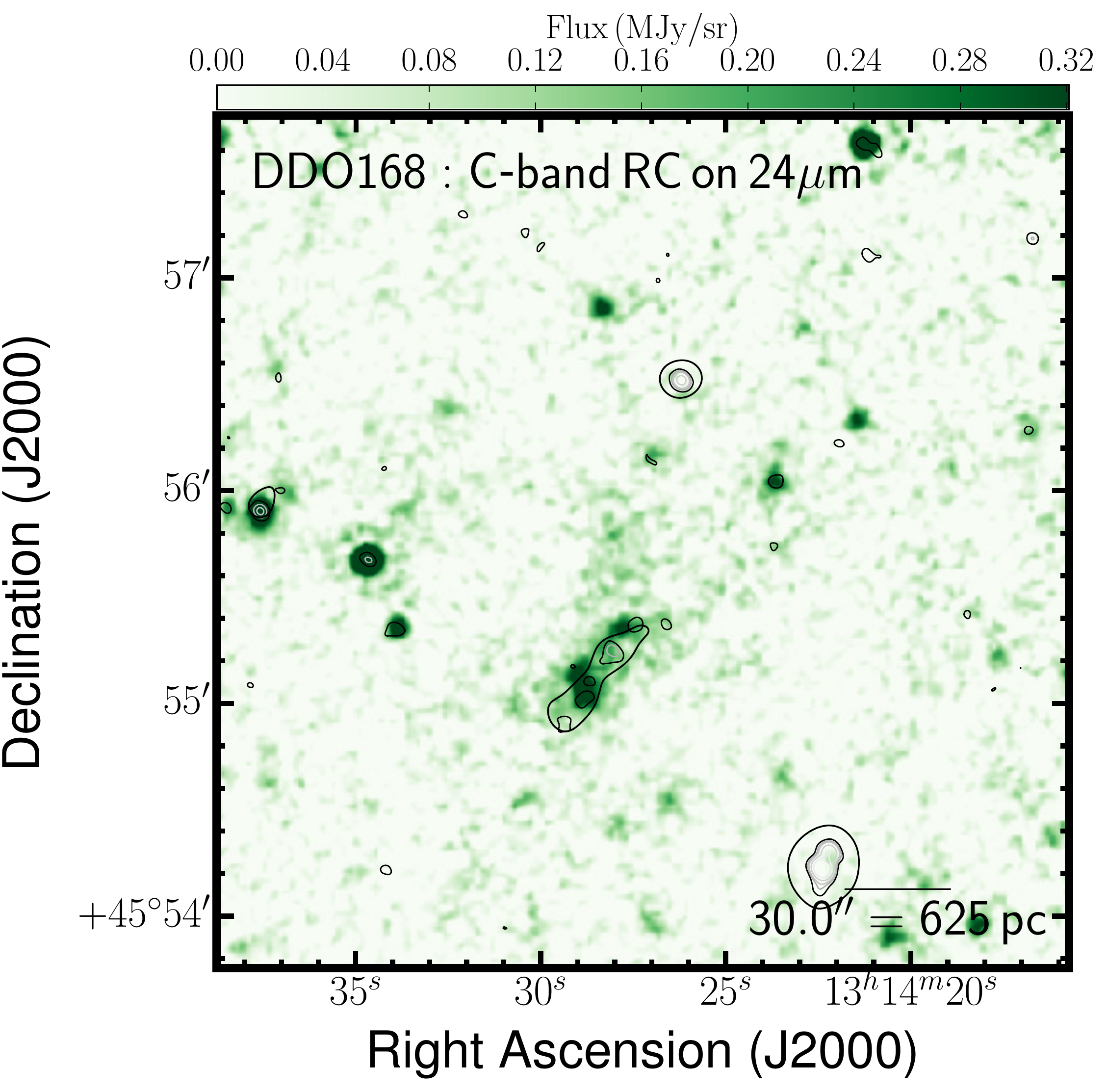} & \ 
    \includegraphics[width=0.31\linewidth,clip]{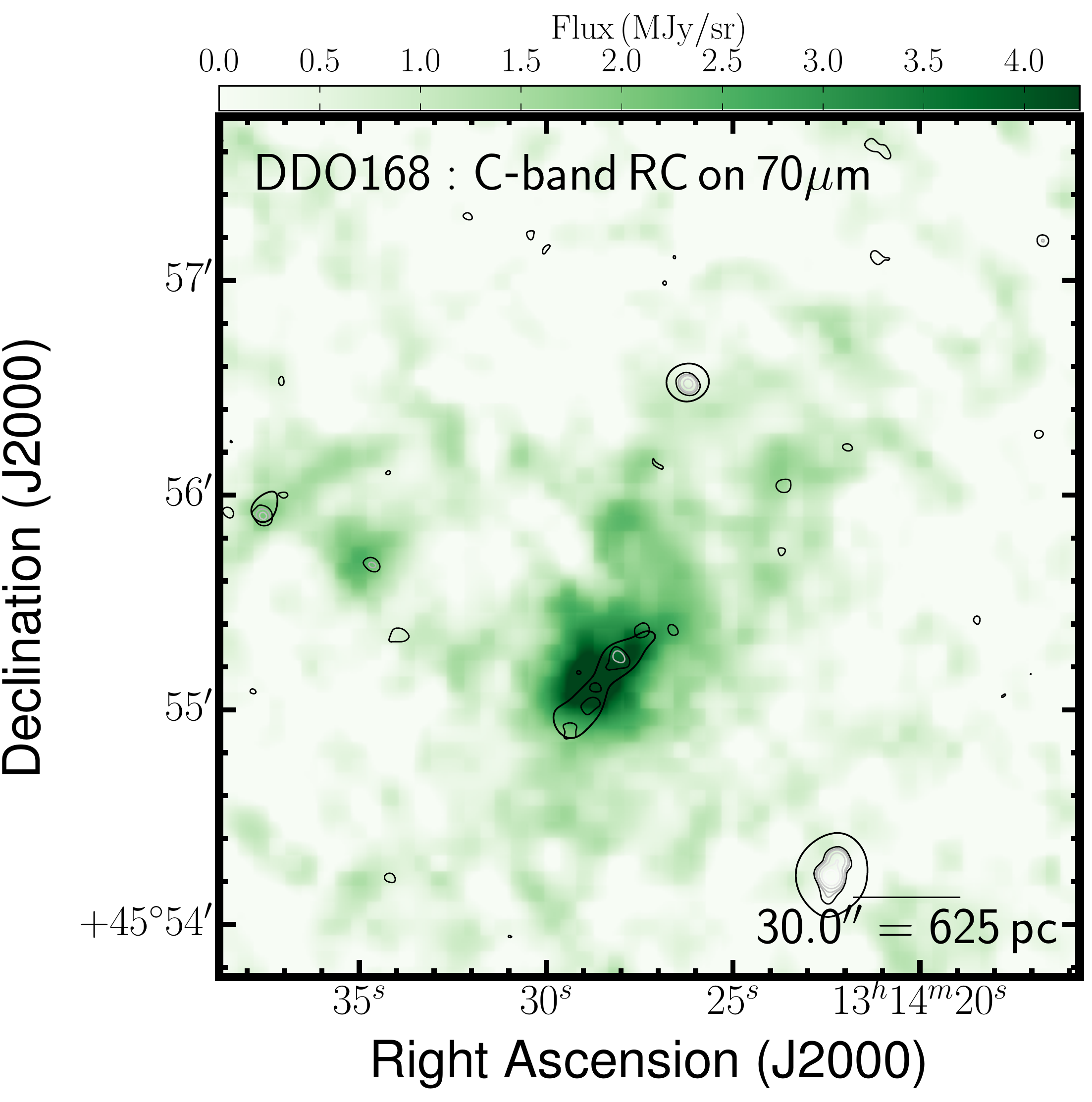} & \ 
    \includegraphics[width=0.31\linewidth,clip]{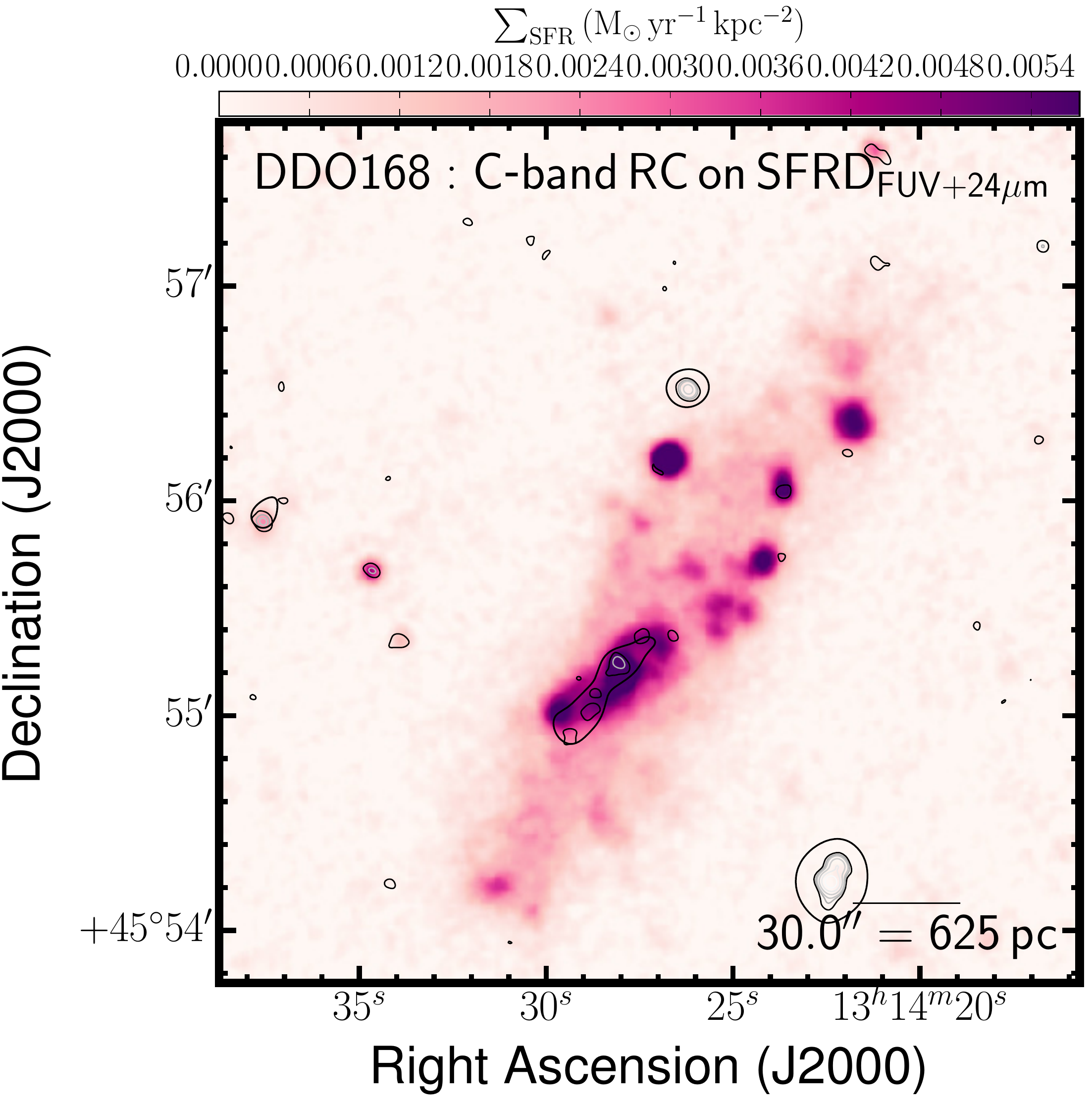} \\
  \end{tabular}
\caption[DDO\,168 images: RC, IR, optical, and FUV]{Multi-wavelength coverage of DDO 168 displaying a $4.0^\prime \times 4.0^\prime$ area. We show total RC flux density at the native resolution (top-left) and again with contours (top-centre). The RC contours are superposed on ancillary LITTLE THINGS images where possible: \halpha\ (middle-left); \RCNT\ obtained by subtracting the expected \RCT\ based on the \halpha-\RCT\ scaling factor of \cite{Deeg1997} from the total RC; {\em GALEX} FUV (middle-right); {\em Spitzer} 24\micron\ (bottom-left); {\em Spitzer} 70\micron\ (bottom-centre); FUV$+24{\rm \mu m}$--inferred SFRD from \citealp{Leroy2012} (bottom-right). We also show the RC that was isolated by the RC--based masking technique (top-right).}
  \label{figure:ddo168Cc_maps}
\end{figure}

\begin{figure}
  \begin{tabular}{ccc}
    \includegraphics[width=0.31\linewidth,clip]{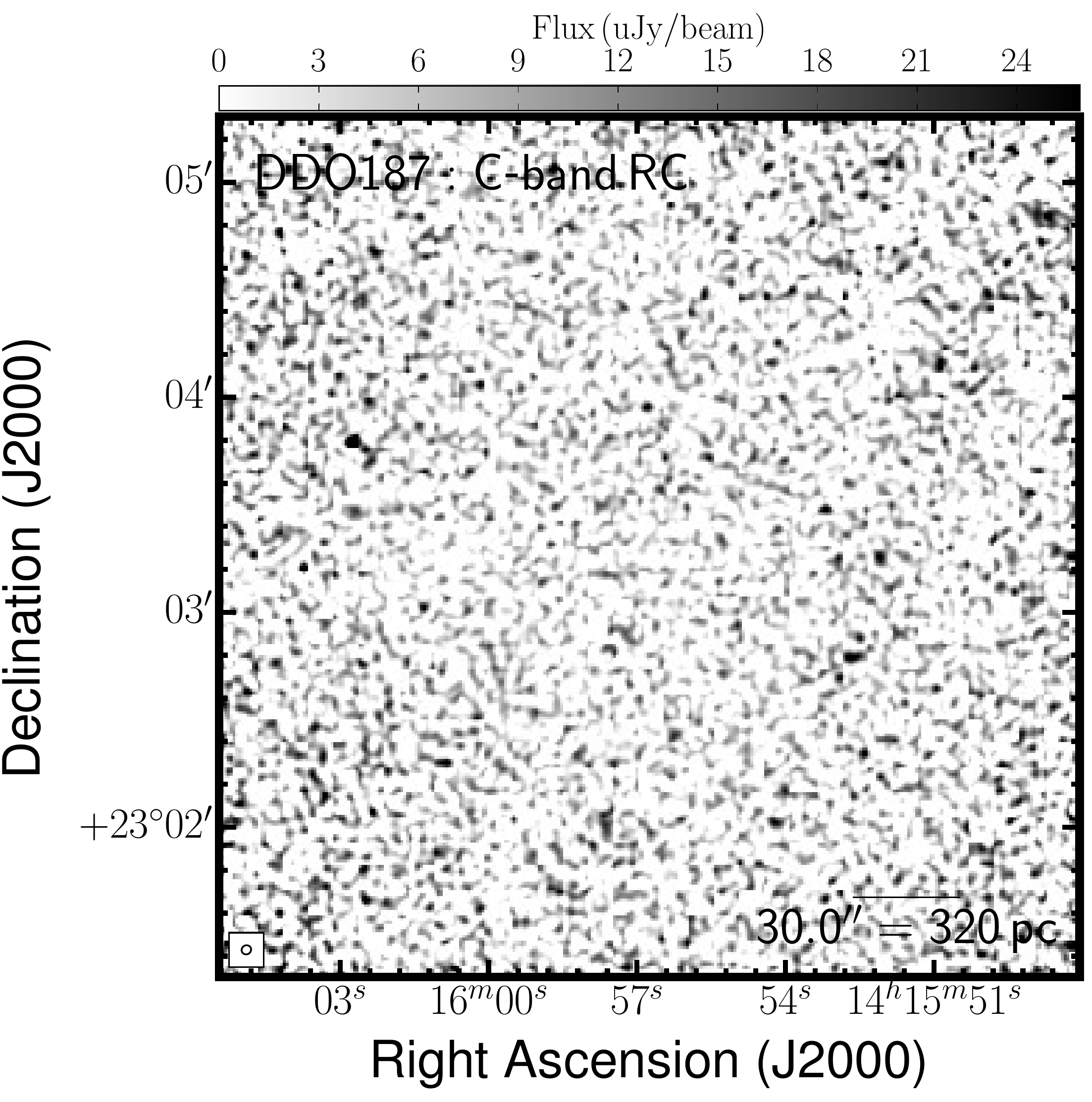} & \ 
    \includegraphics[width=0.31\linewidth,clip]{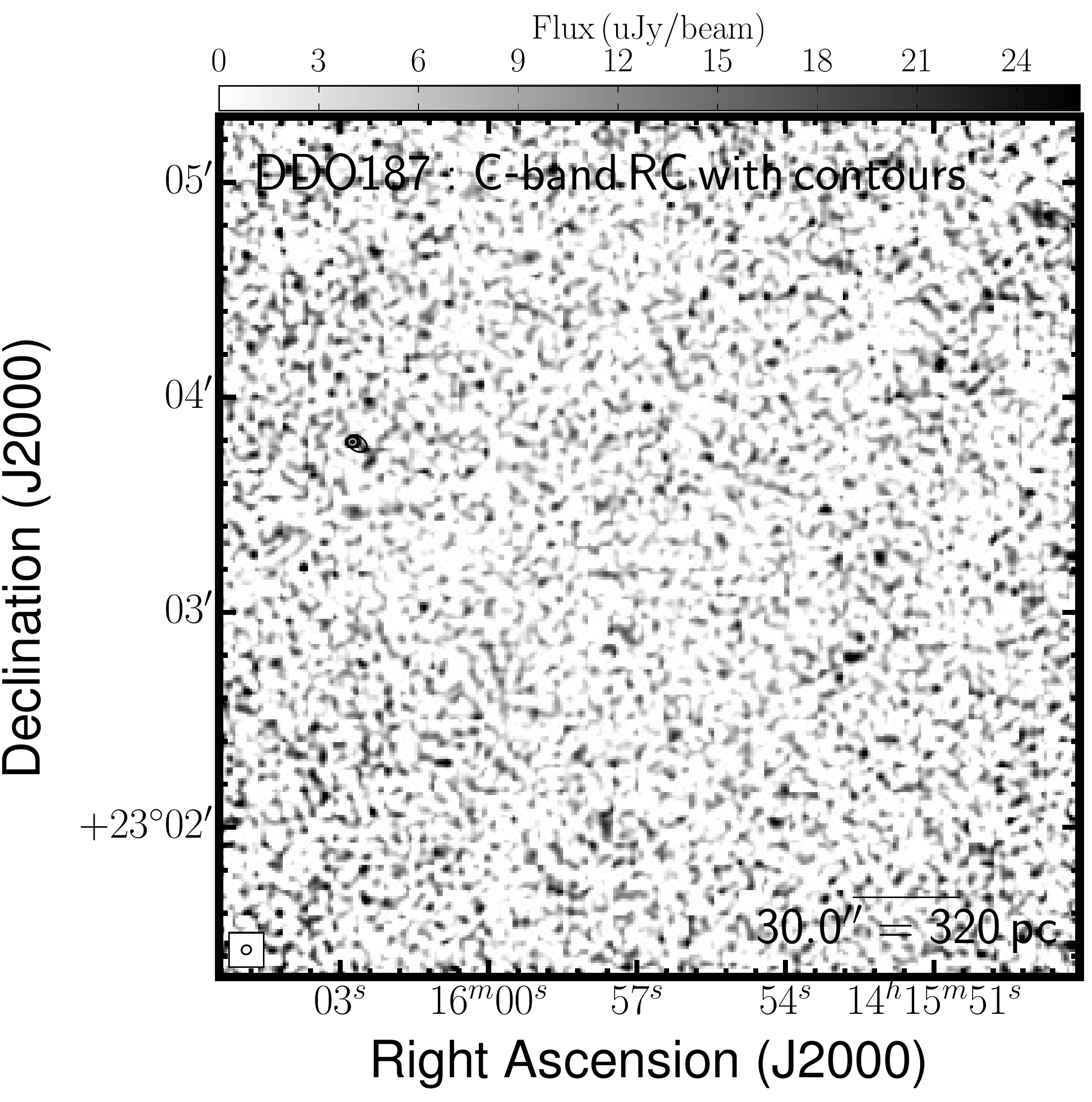} & \ 
    \includegraphics[width=0.31\linewidth,clip]{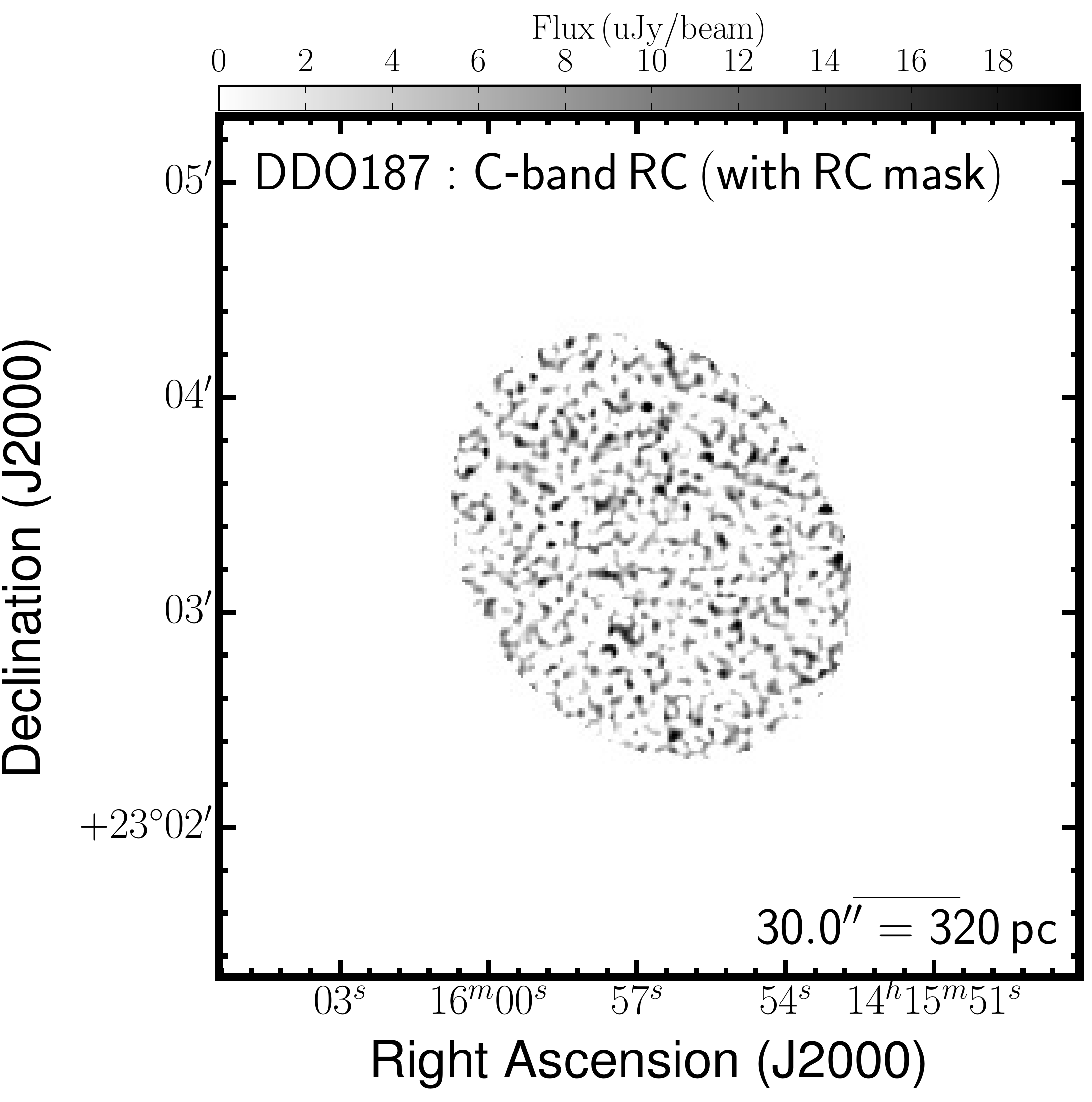} \\
    \includegraphics[width=0.31\linewidth,clip]{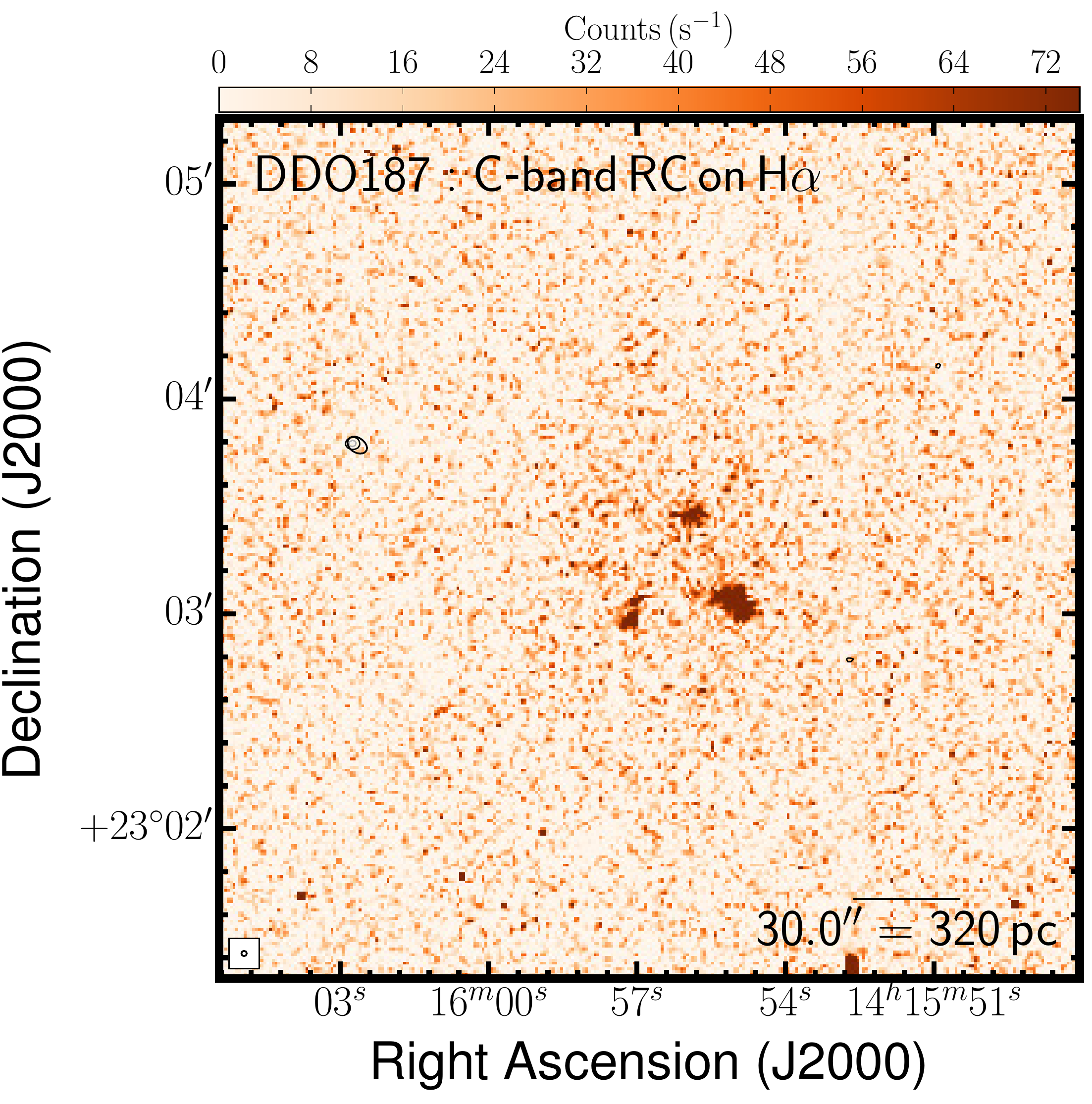} & \ 
    \includegraphics[width=0.31\linewidth,clip]{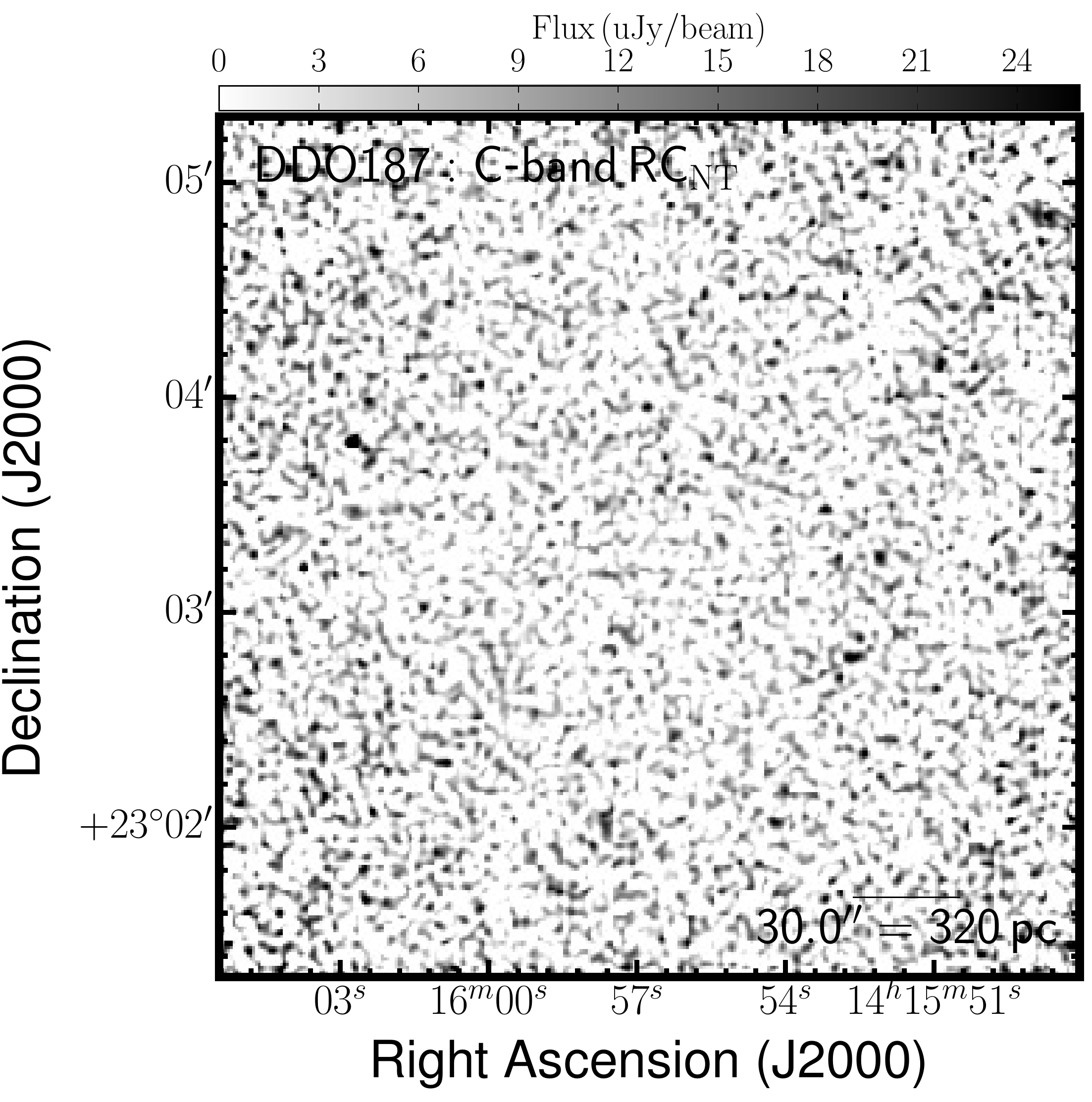} & \ 
    \includegraphics[width=0.31\linewidth,clip]{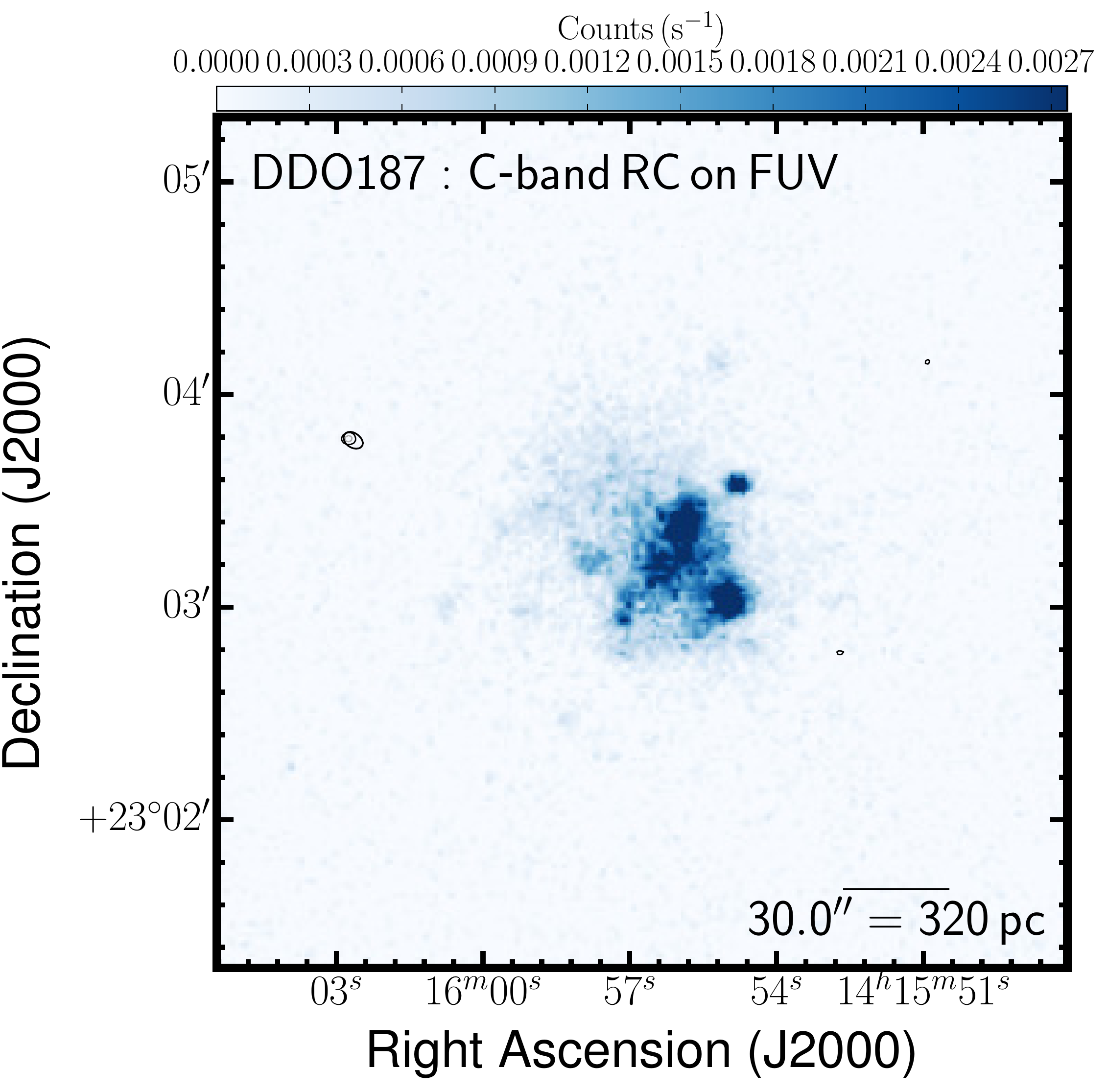} \\
    \includegraphics[width=0.31\linewidth,clip]{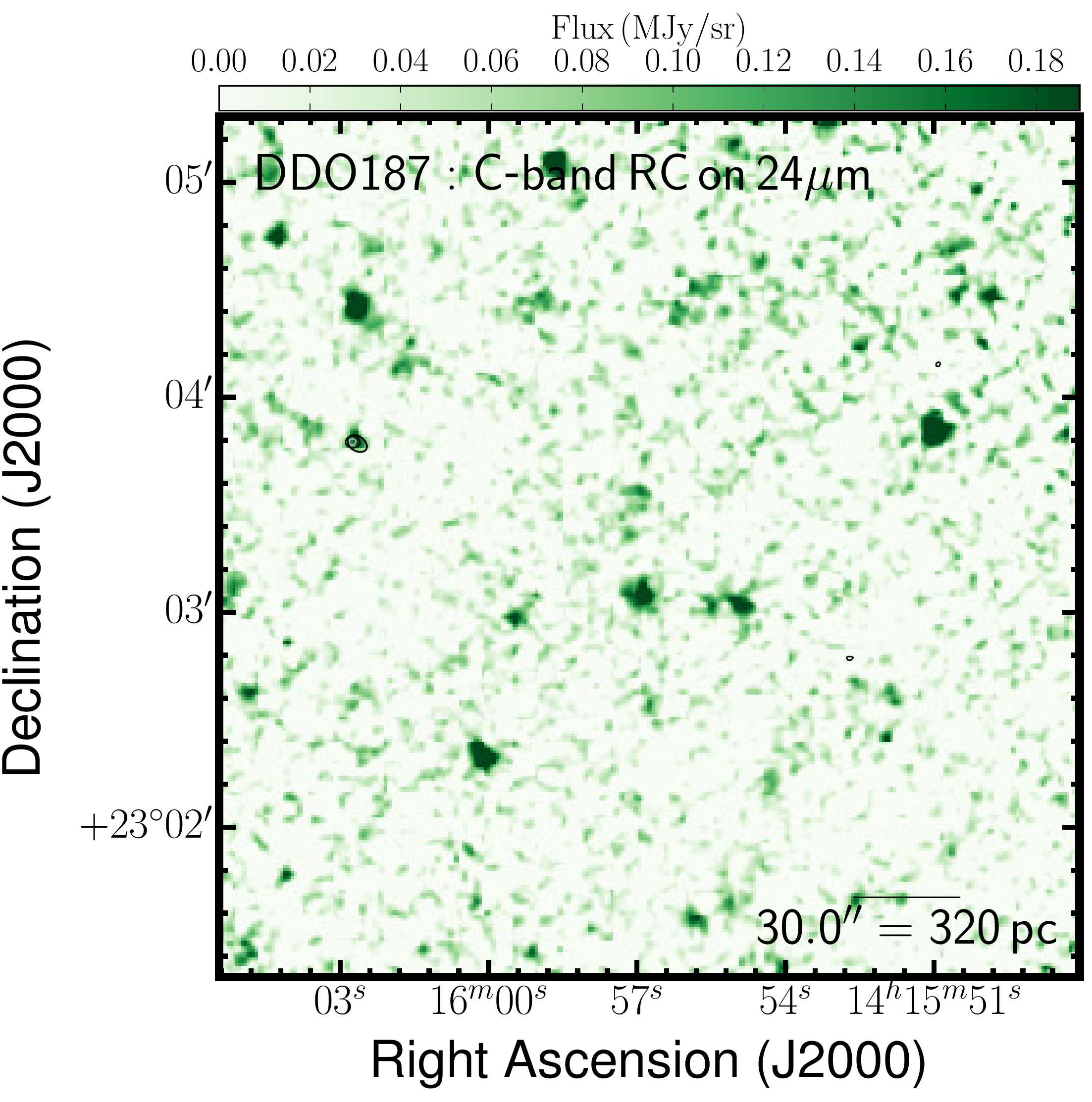} & \ 
    \includegraphics[width=0.31\linewidth,clip]{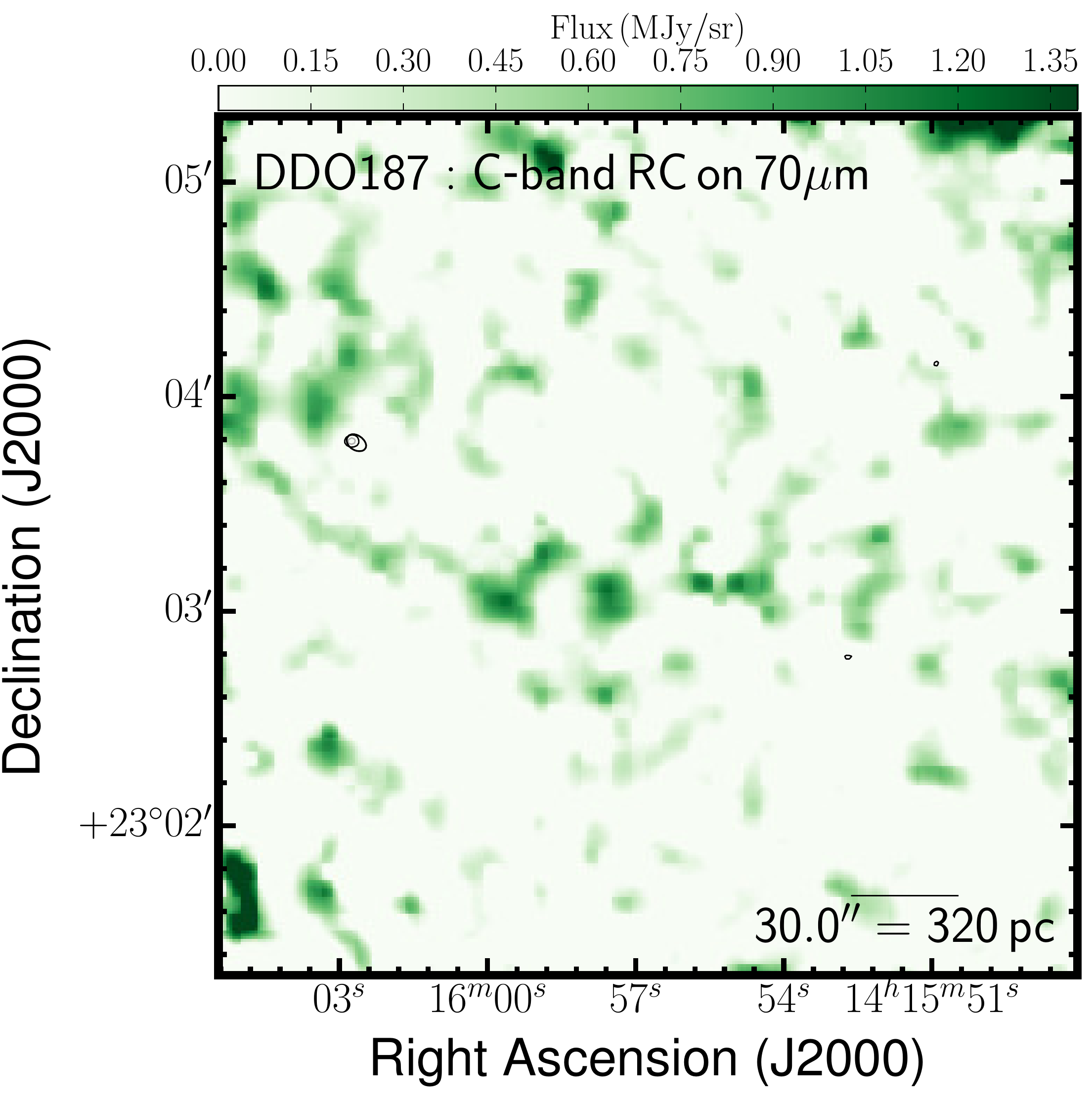} & \ 
    \includegraphics[width=0.31\linewidth,clip]{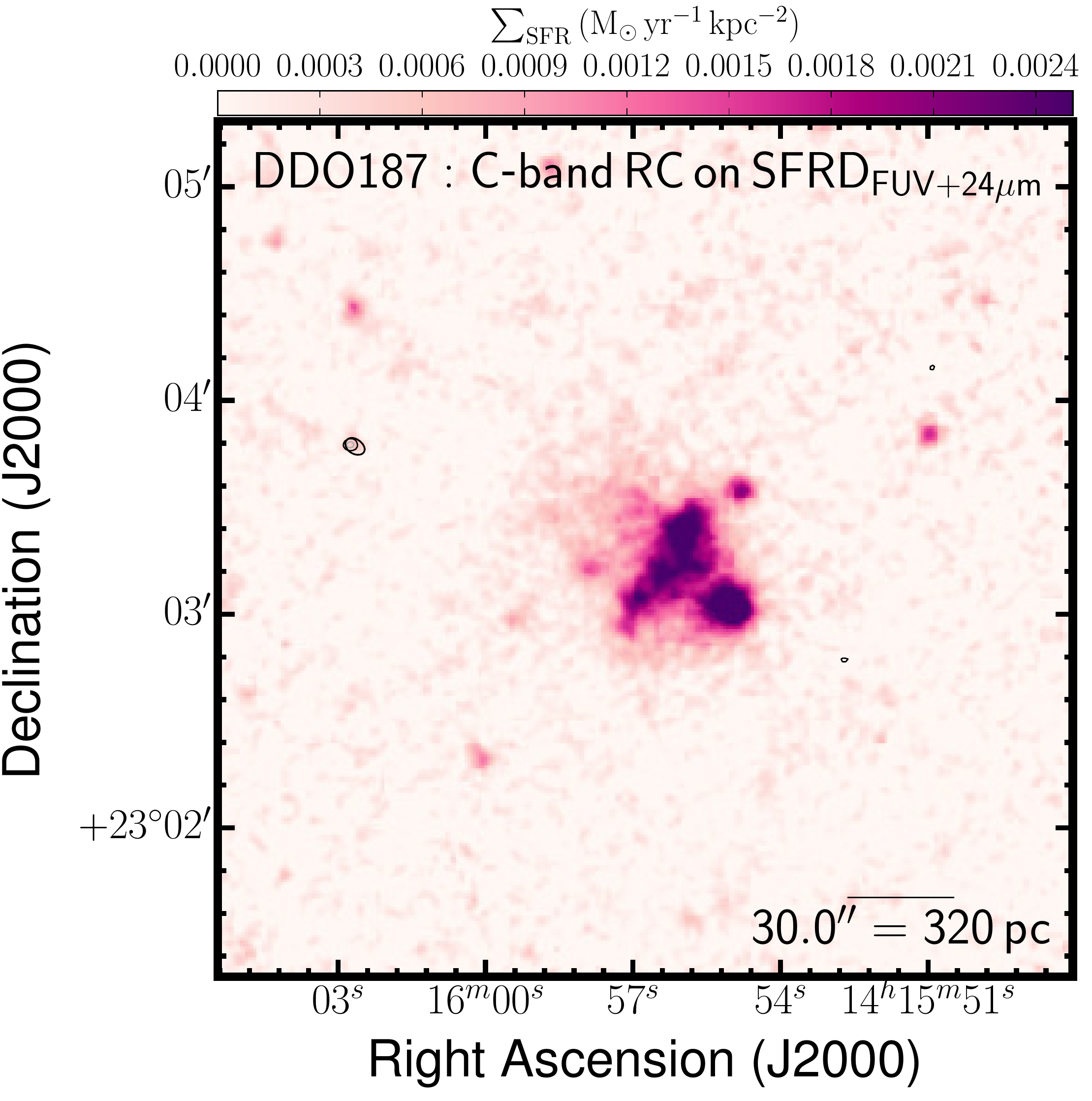} \\
  \end{tabular}
\caption[DDO\,187 images: RC, IR, optical, and FUV]{Multi-wavelength coverage of DDO 187 displaying a $4.0^\prime \times 4.0^\prime$ area. We show total RC flux density at the native resolution (top-left) and again with contours (top-centre). The RC contours are superposed on ancillary LITTLE THINGS images where possible: \halpha\ (middle-left); \RCNT\ obtained by subtracting the expected \RCT\ based on the \halpha-\RCT\ scaling factor of \cite{Deeg1997} from the total RC; {\em GALEX} FUV (middle-right); {\em Spitzer} 24\micron\ (bottom-left); {\em Spitzer} 70\micron\ (bottom-centre); FUV$+24{\rm \mu m}$--inferred SFRD from \citealp{Leroy2012} (bottom-right). We also show the RC that was isolated by the RC--based masking technique (top-right).}
  \label{figure:ddo187Cc_maps}
\end{figure}

\begin{figure}
  \begin{tabular}{ccc}
    \includegraphics[width=0.31\linewidth,clip]{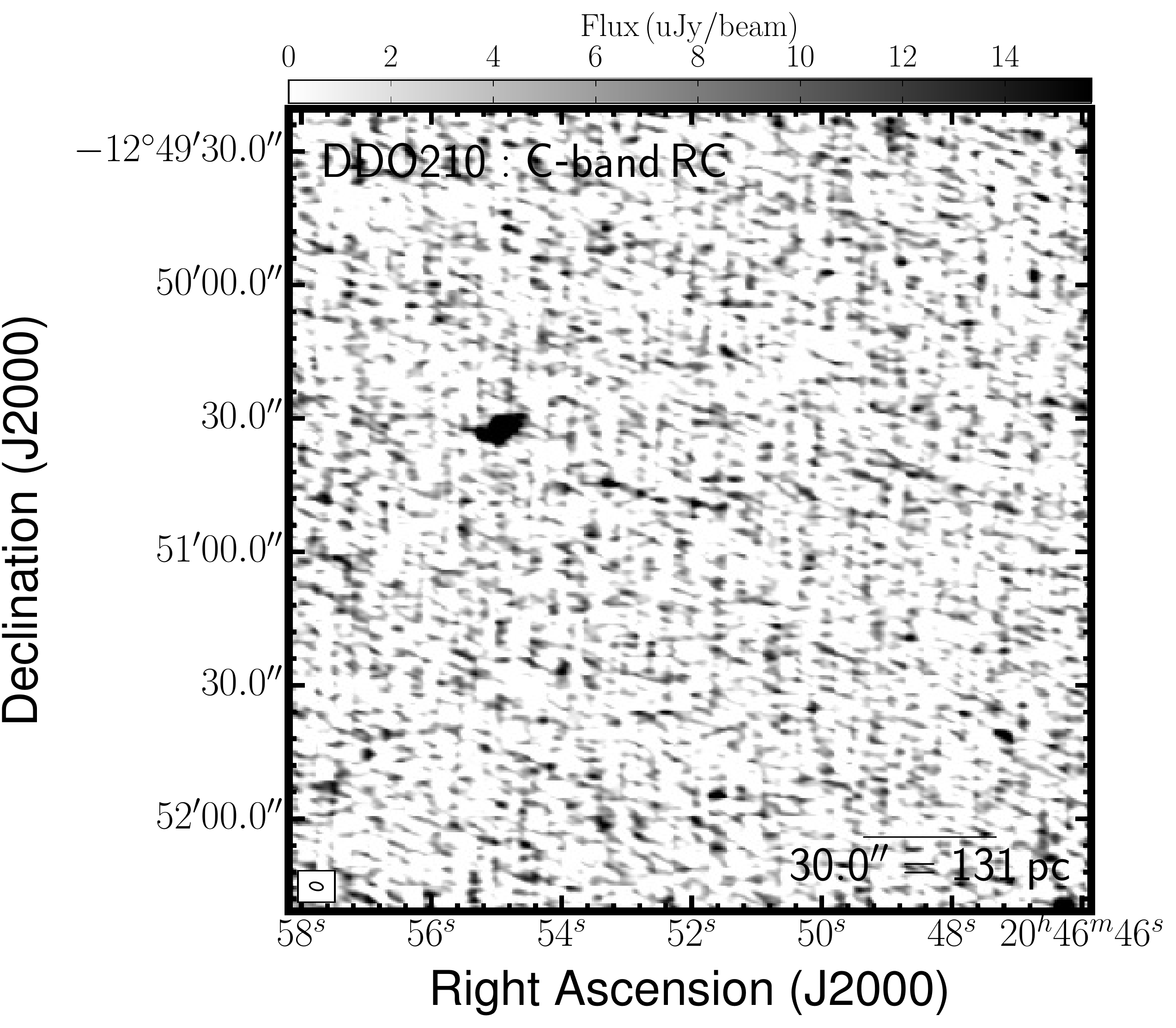} & \ 
    \includegraphics[width=0.31\linewidth,clip]{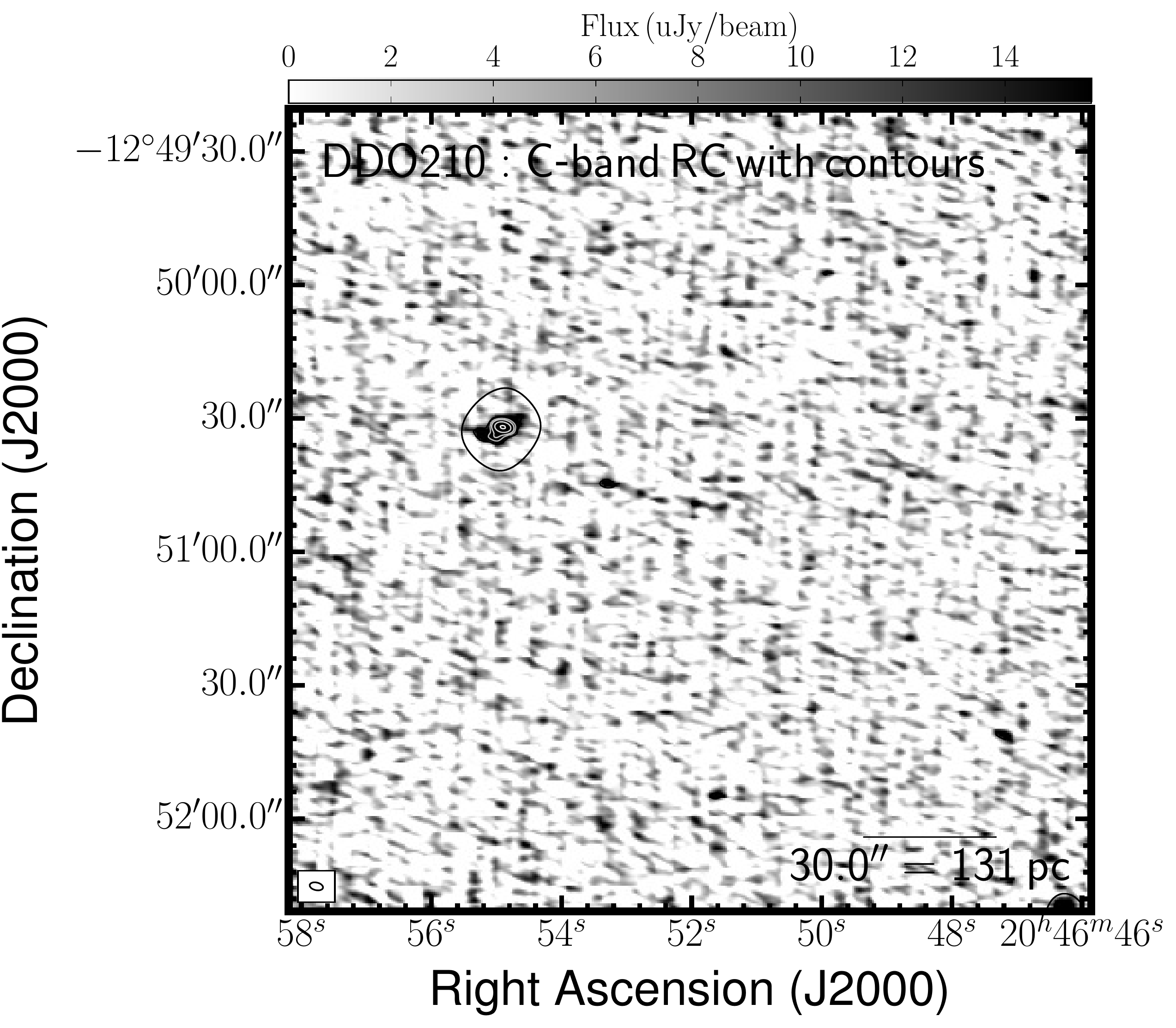} & \ 
    \includegraphics[width=0.31\linewidth,clip]{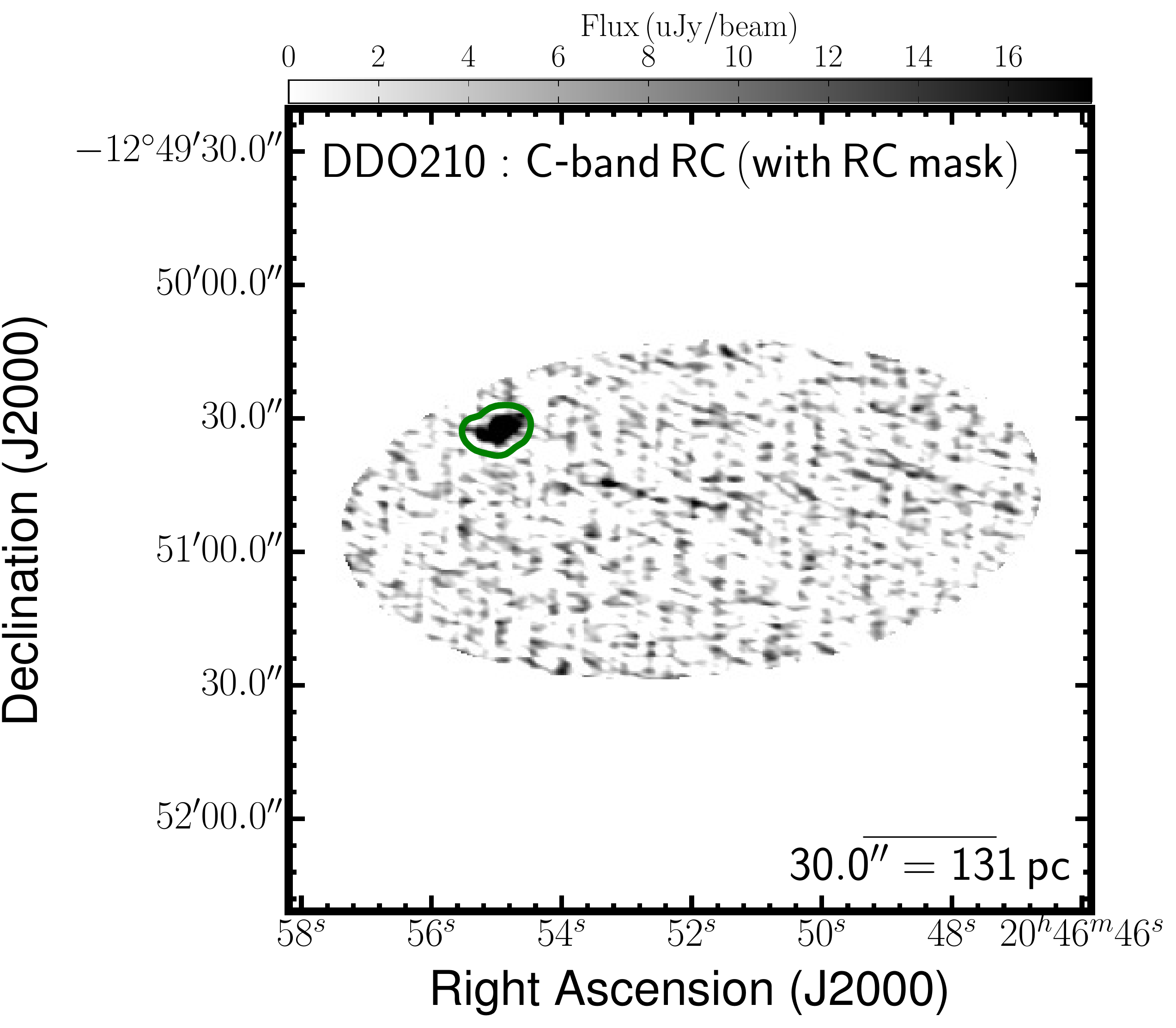} \\
    \includegraphics[width=0.31\linewidth,clip]{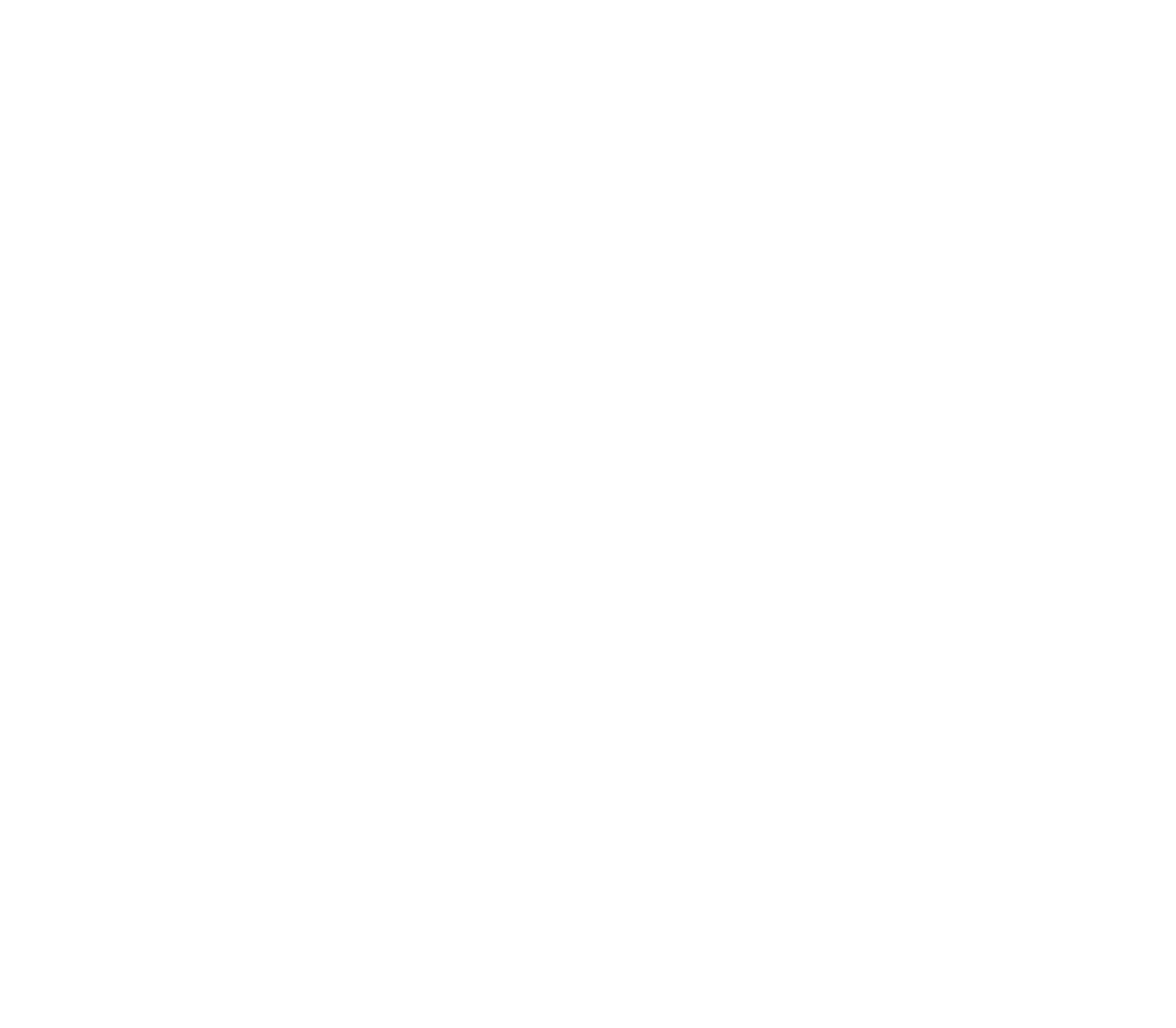} & \ 
    \includegraphics[width=0.31\linewidth,clip]{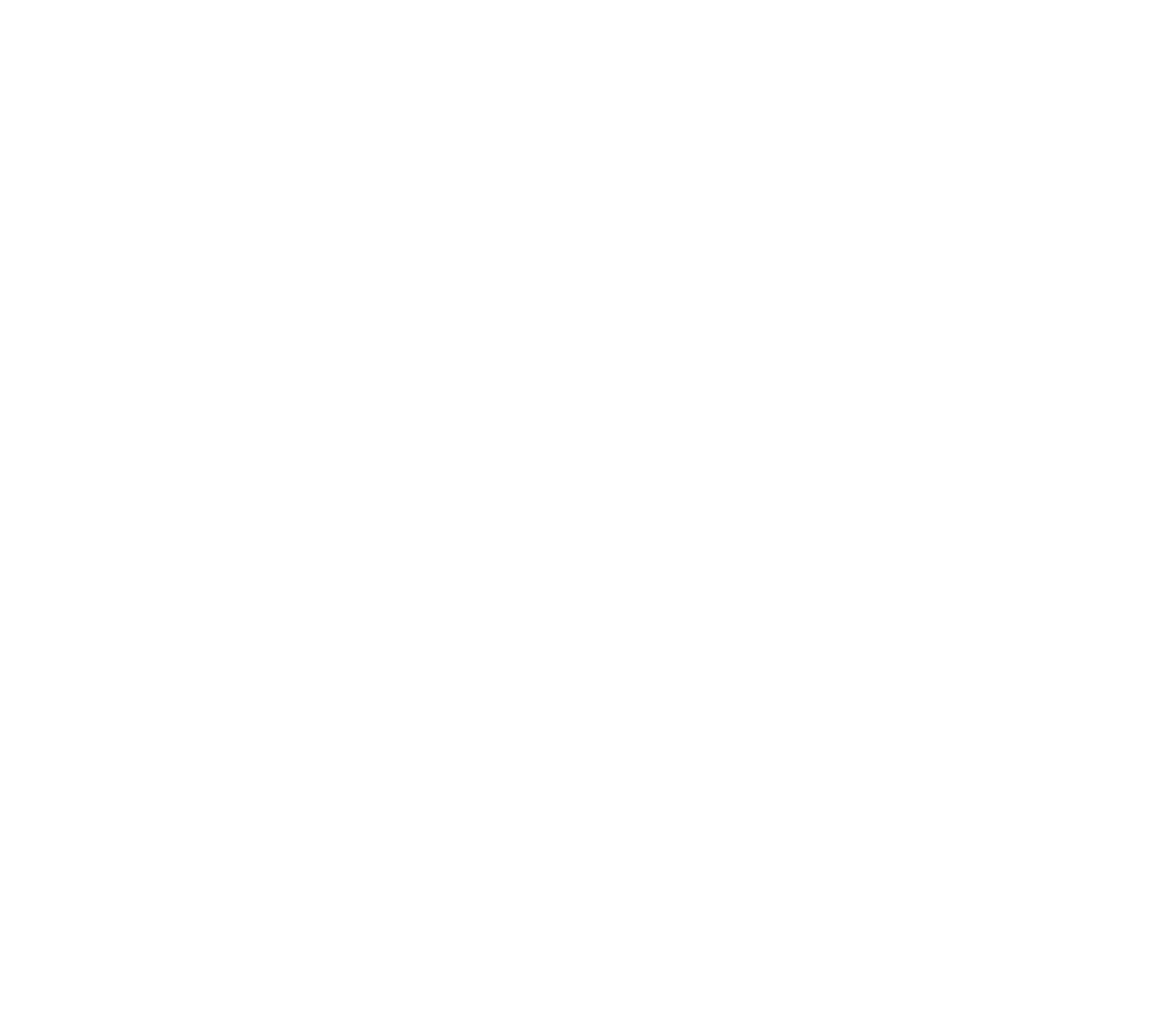} & \ 
    \includegraphics[width=0.31\linewidth,clip]{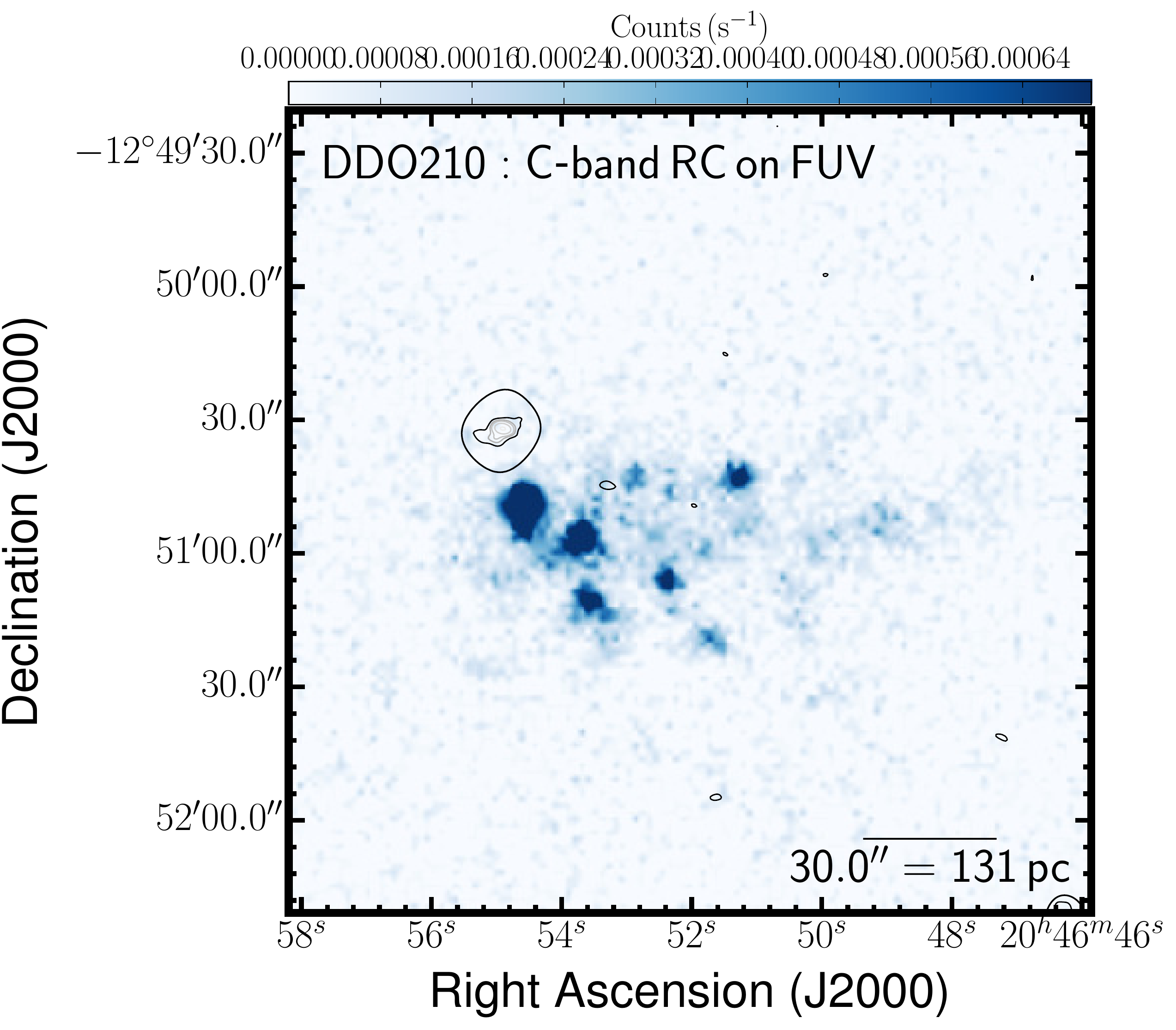} \\
    \includegraphics[width=0.31\linewidth,clip]{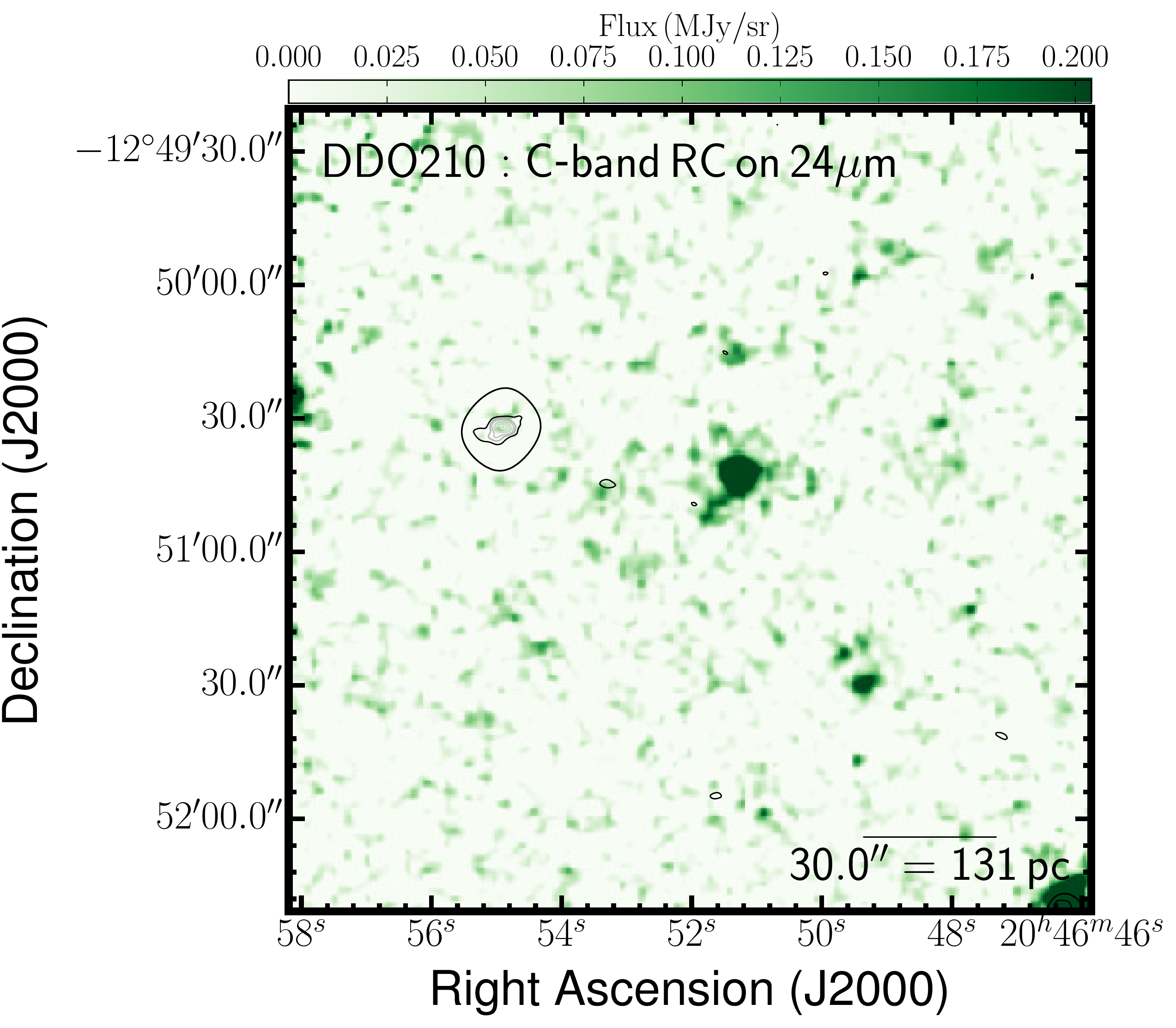} & \ 
    \includegraphics[width=0.31\linewidth,clip]{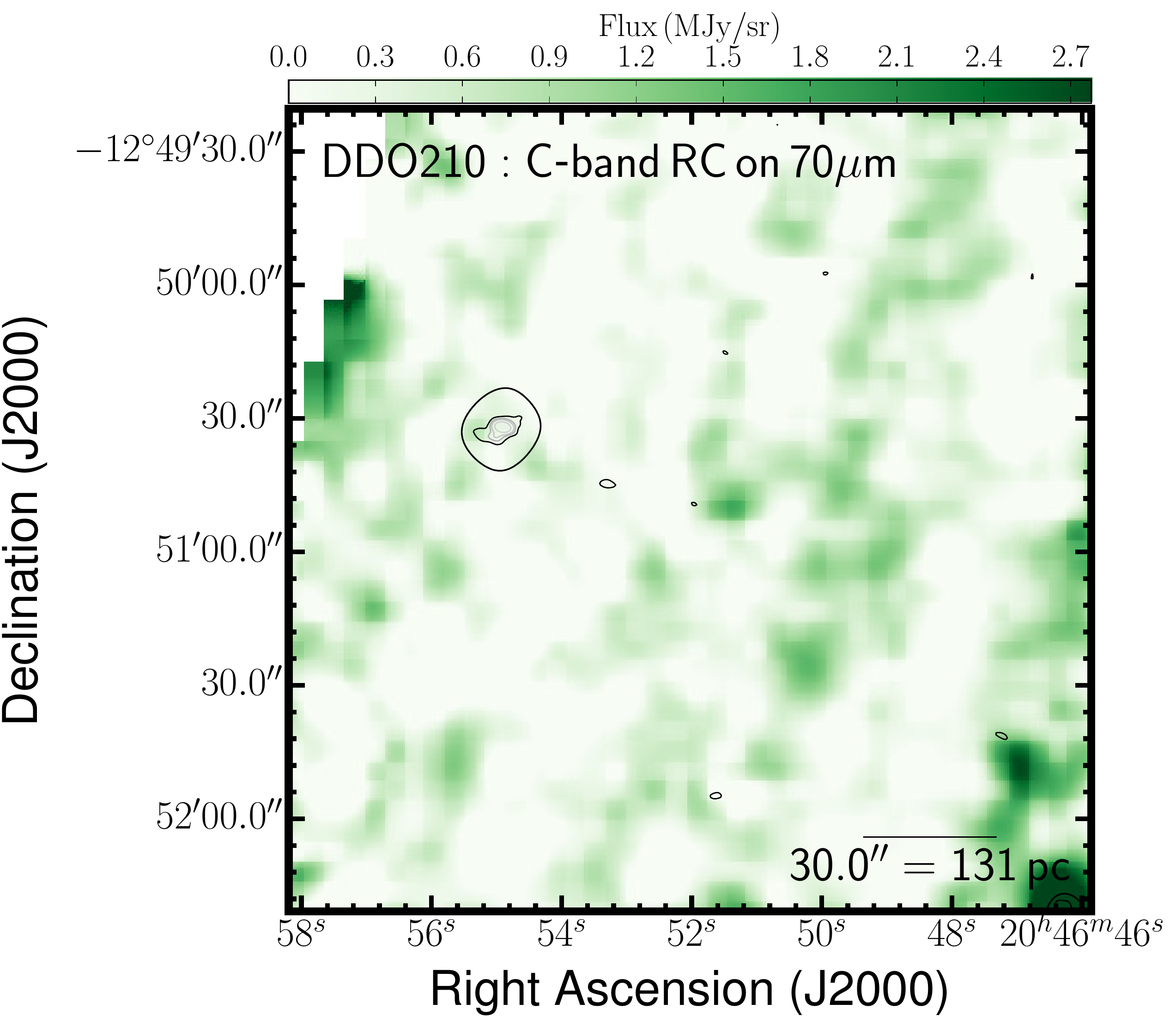} & \ 
    \includegraphics[width=0.31\linewidth,clip]{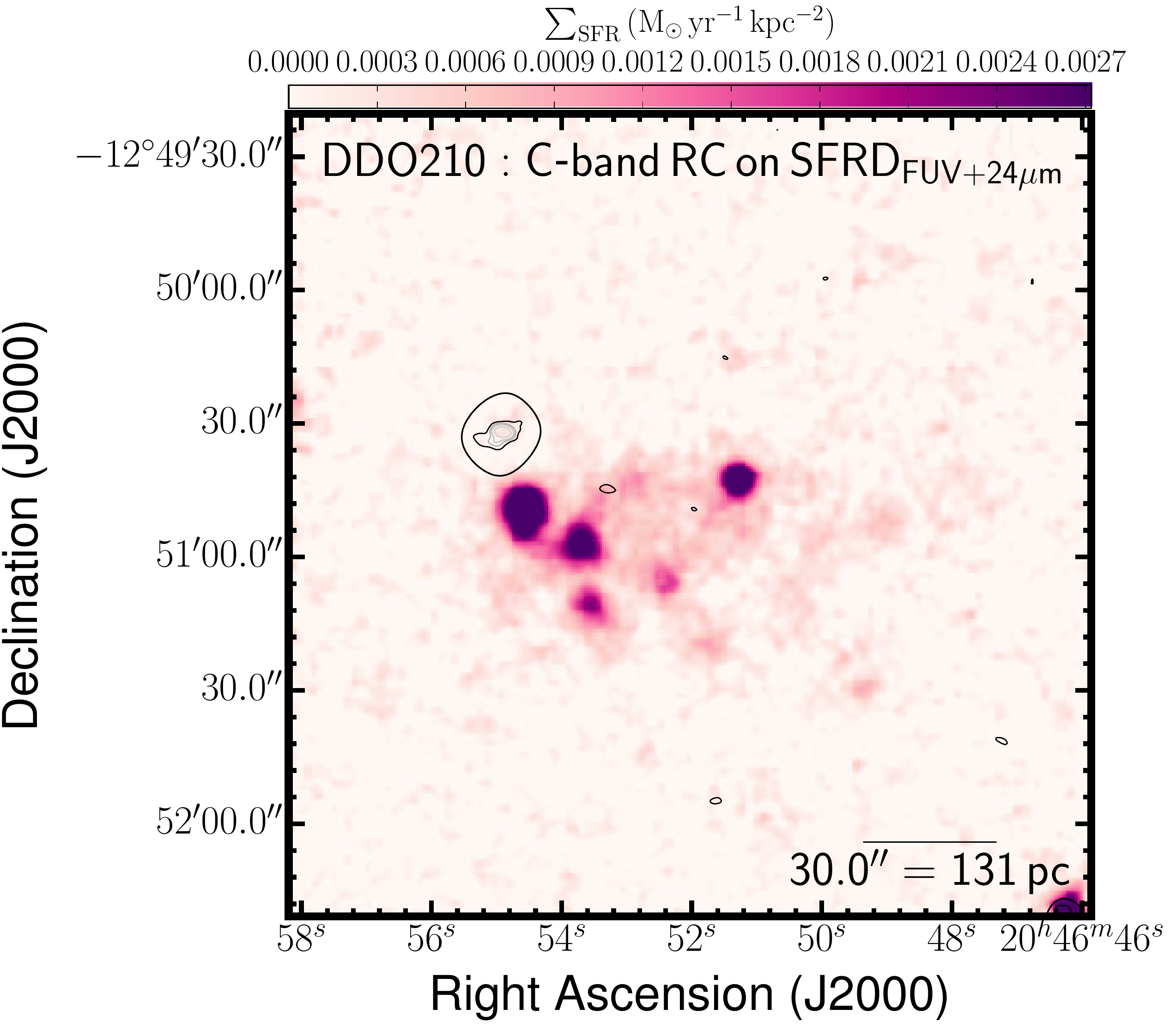} \\
  \end{tabular}
\caption[DDO\,210 images: RC, IR, optical, and FUV]{Multi-wavelength coverage of DDO 210 displaying a $3.0^\prime \times 3.0^\prime$ area. We show total RC flux density at the native resolution (top-left) and again with contours (top-centre). The RC contours are superposed on ancillary LITTLE THINGS images where possible: \halpha\ (middle-left); \RCNT\ obtained by subtracting the expected \RCT\ based on the \halpha-\RCT\ scaling factor of \cite{Deeg1997} from the total RC; {\em GALEX} FUV (middle-right); {\em Spitzer} 24\micron\ (bottom-left); {\em Spitzer} 70\micron\ (bottom-centre); FUV$+24{\rm \mu m}$--inferred SFRD from \citealp{Leroy2012} (bottom-right). We also show the RC that was isolated by the RC--based masking technique (top-right).}
  \label{figure:ddo210Cc_maps}
\end{figure}

\begin{figure}
  \begin{tabular}{ccc}
    \includegraphics[width=0.31\linewidth,clip]{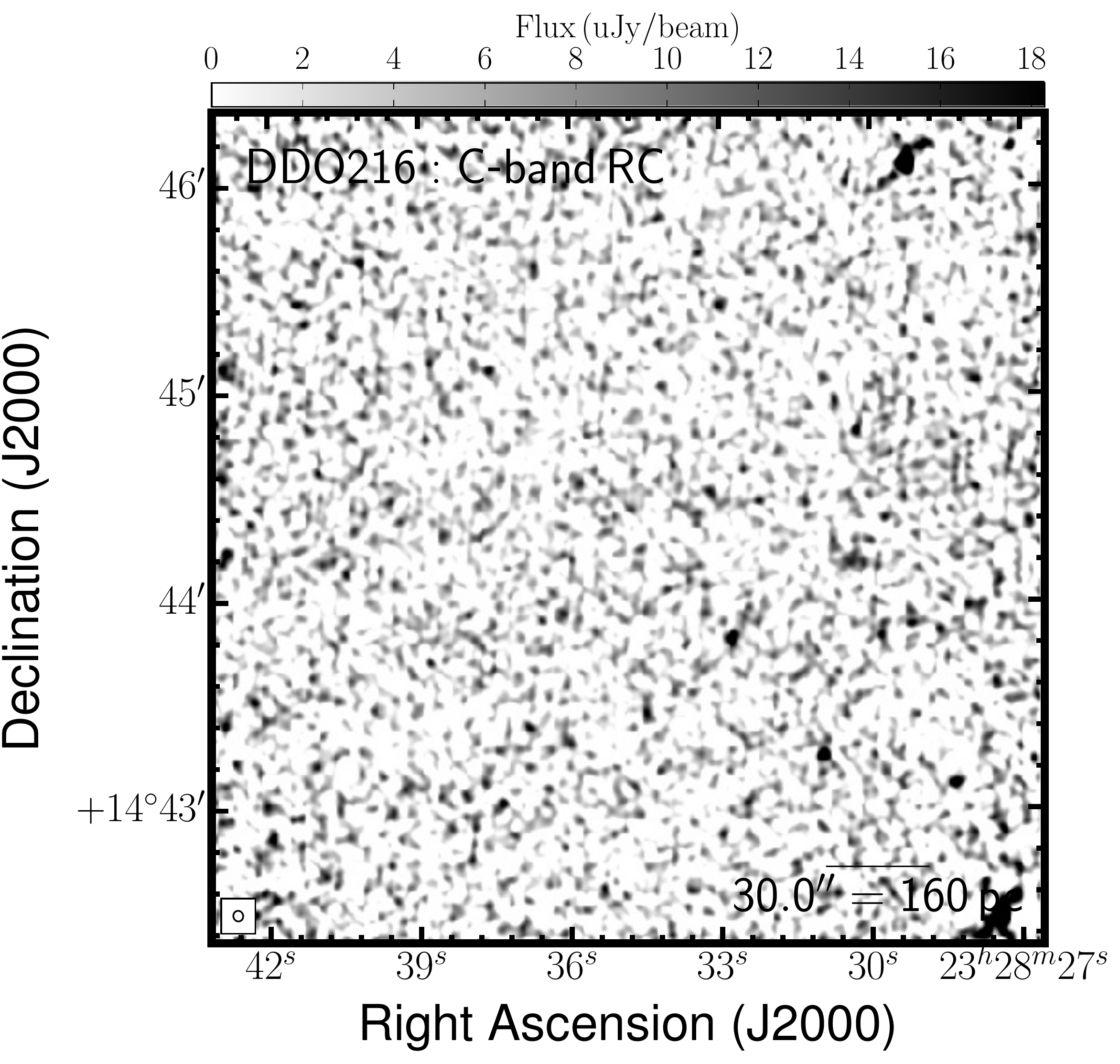} & \ 
    \includegraphics[width=0.31\linewidth,clip]{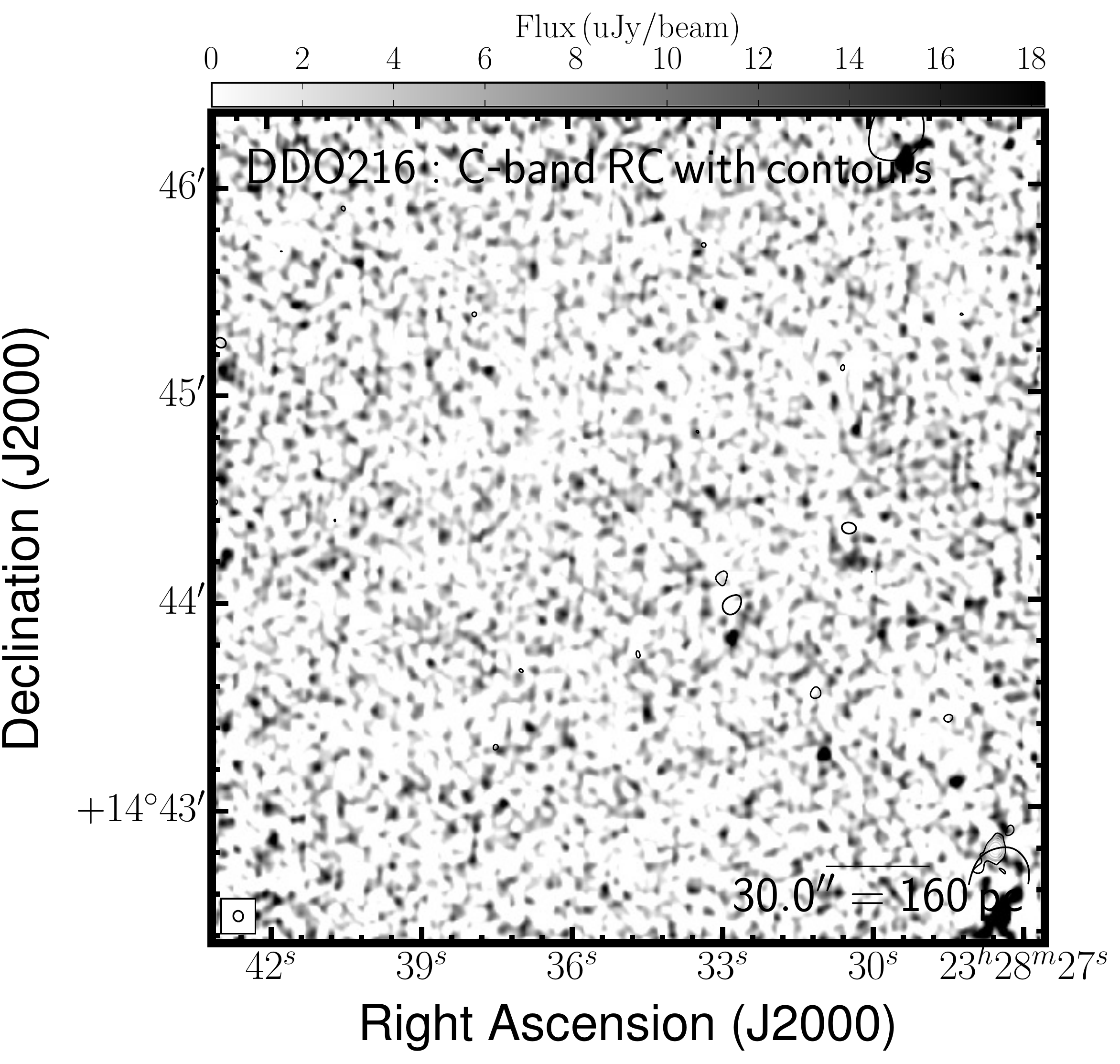} & \ 
    \includegraphics[width=0.31\linewidth,clip]{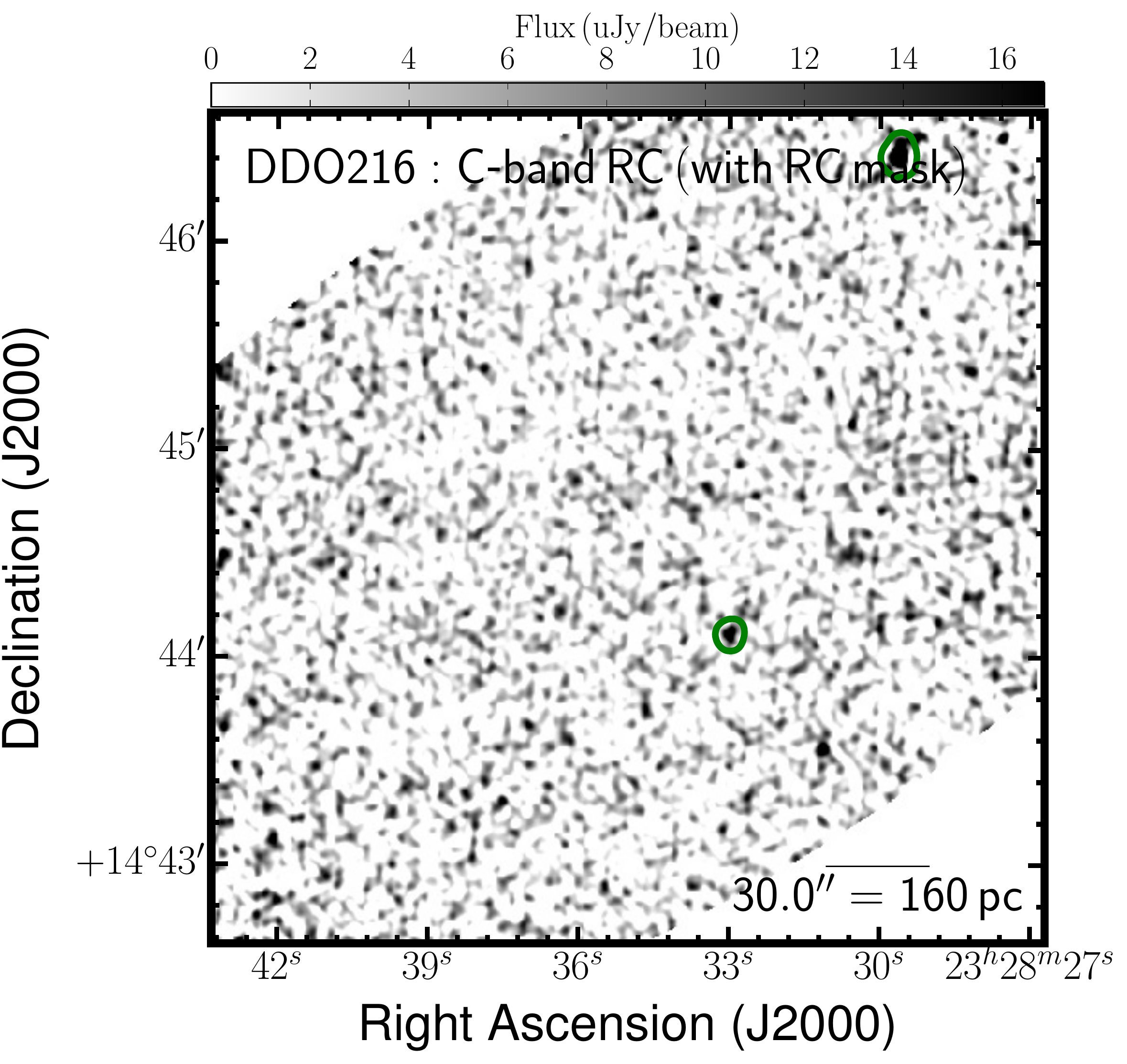} \\
    \includegraphics[width=0.31\linewidth,clip]{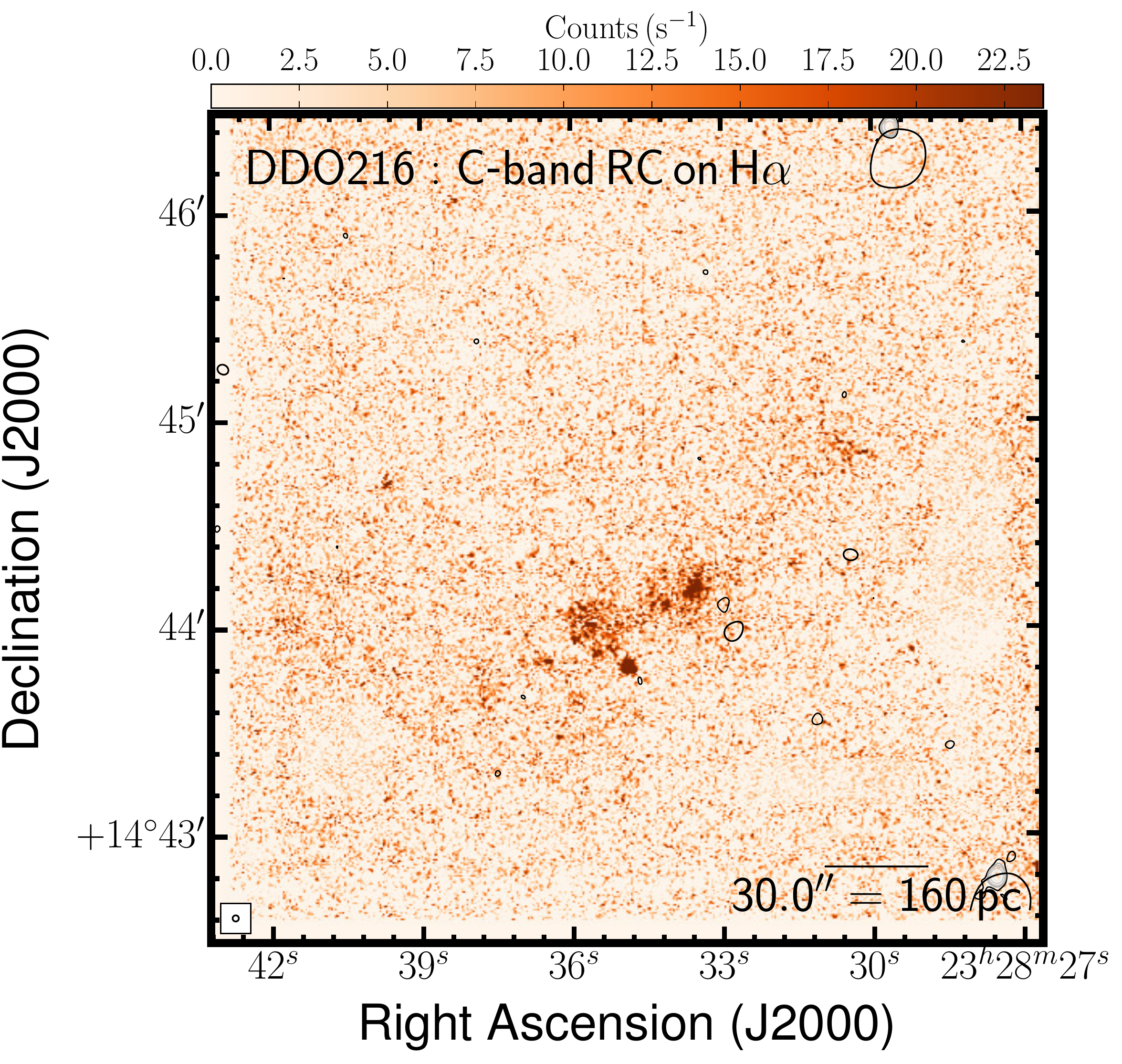} & \ 
    \includegraphics[width=0.31\linewidth,clip]{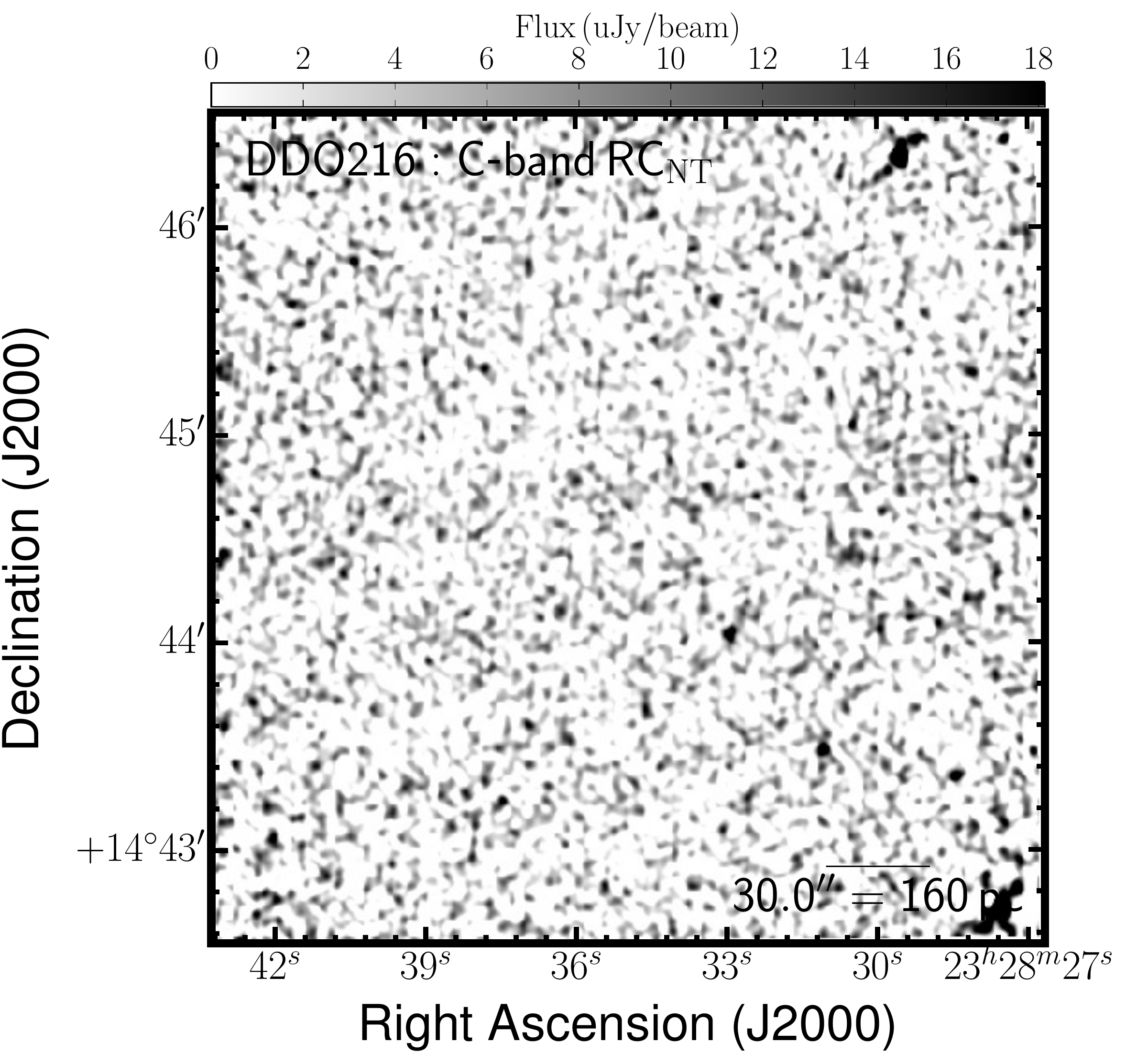} & \ 
    \includegraphics[width=0.31\linewidth,clip]{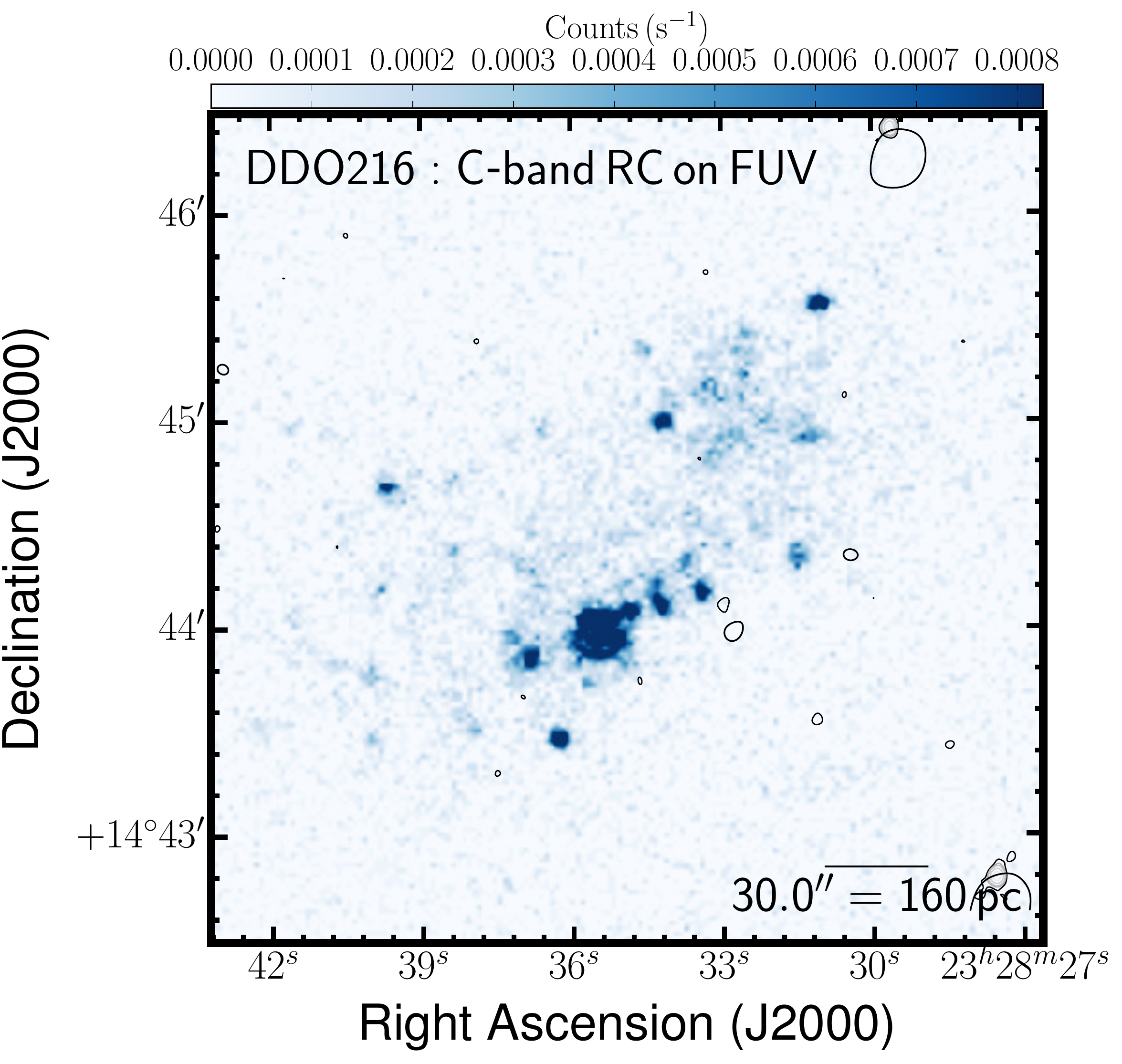} \\
    \includegraphics[width=0.31\linewidth,clip]{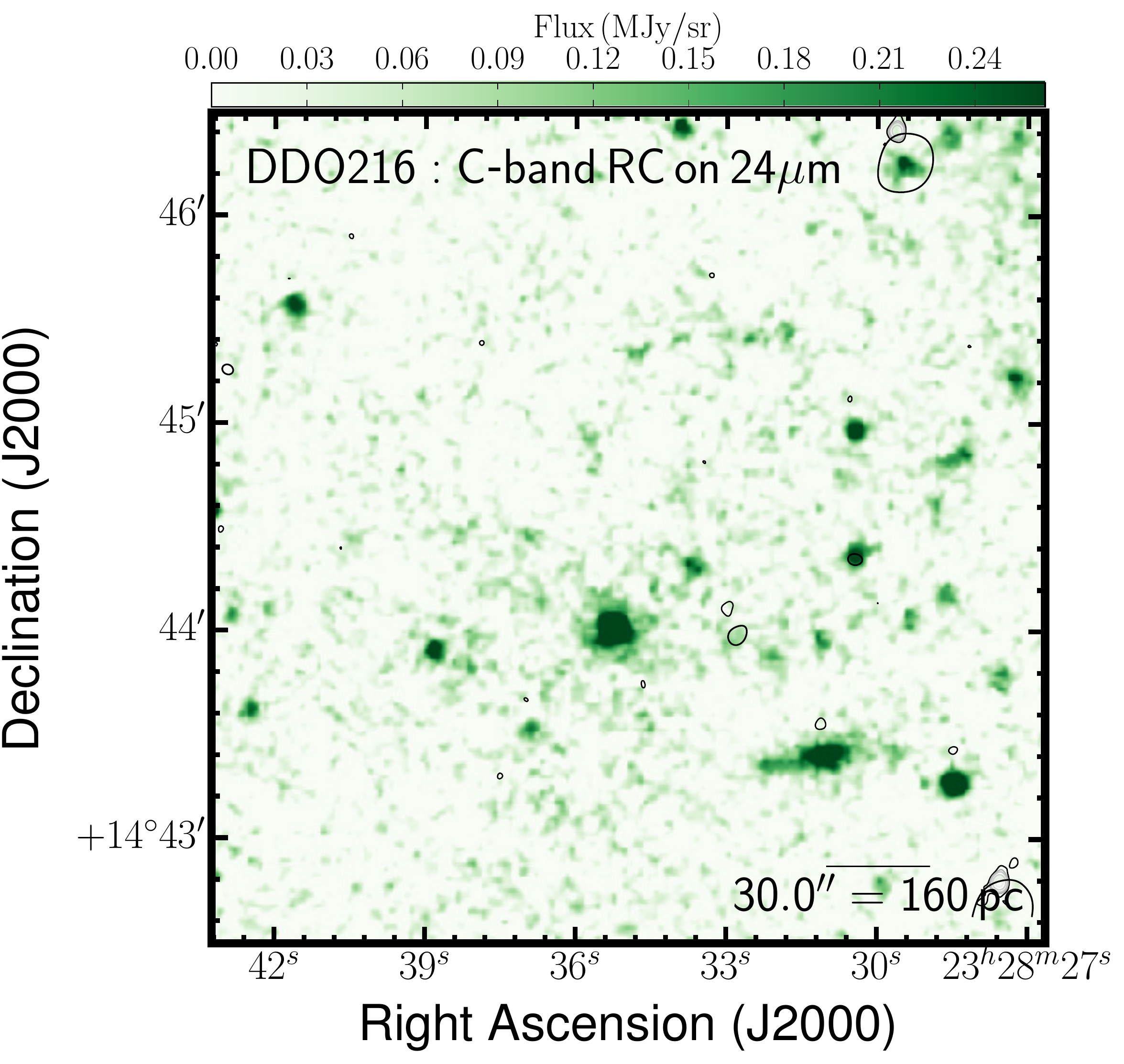} & \ 
    \includegraphics[width=0.31\linewidth,clip]{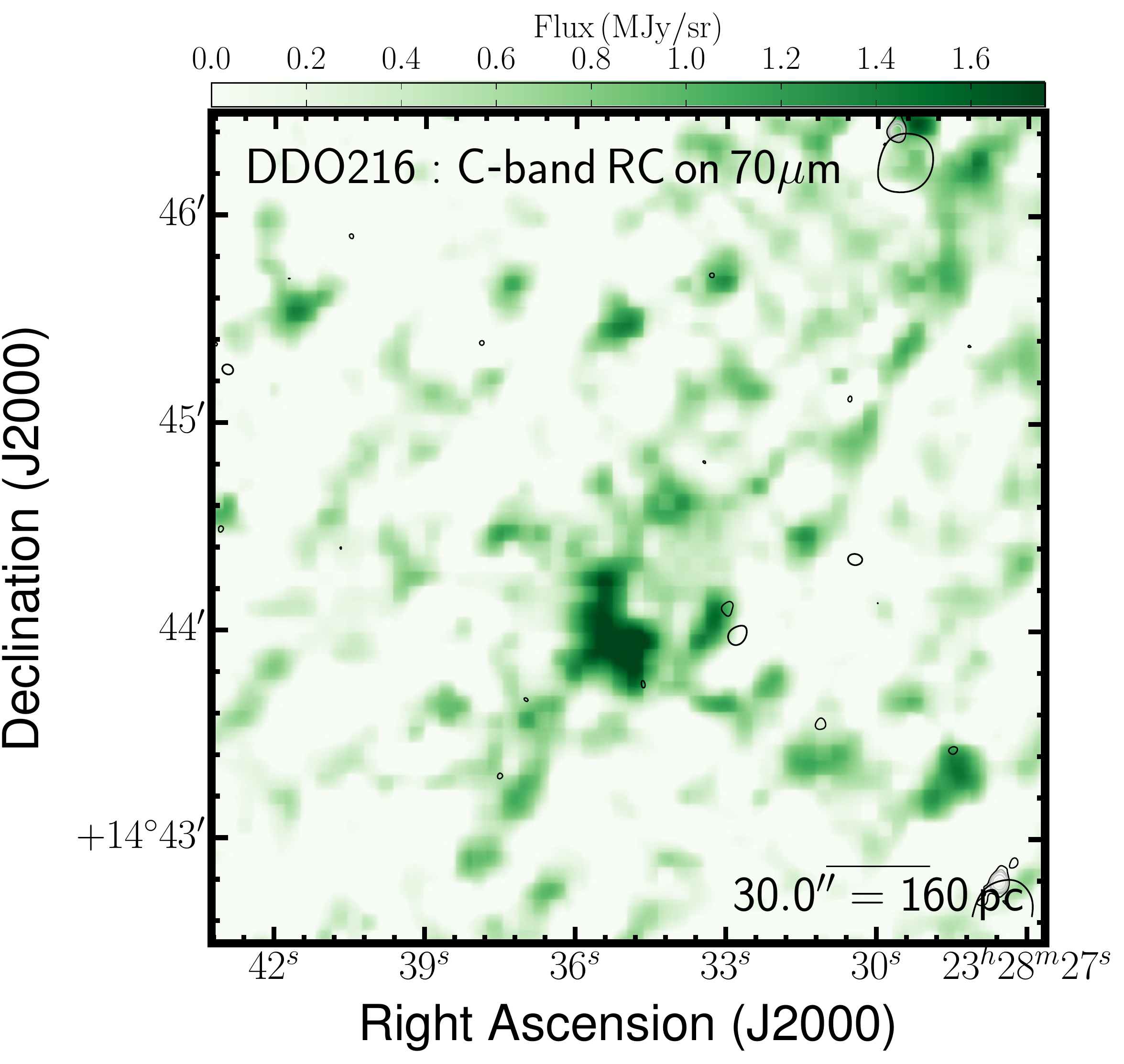} & \ 
    \includegraphics[width=0.31\linewidth,clip]{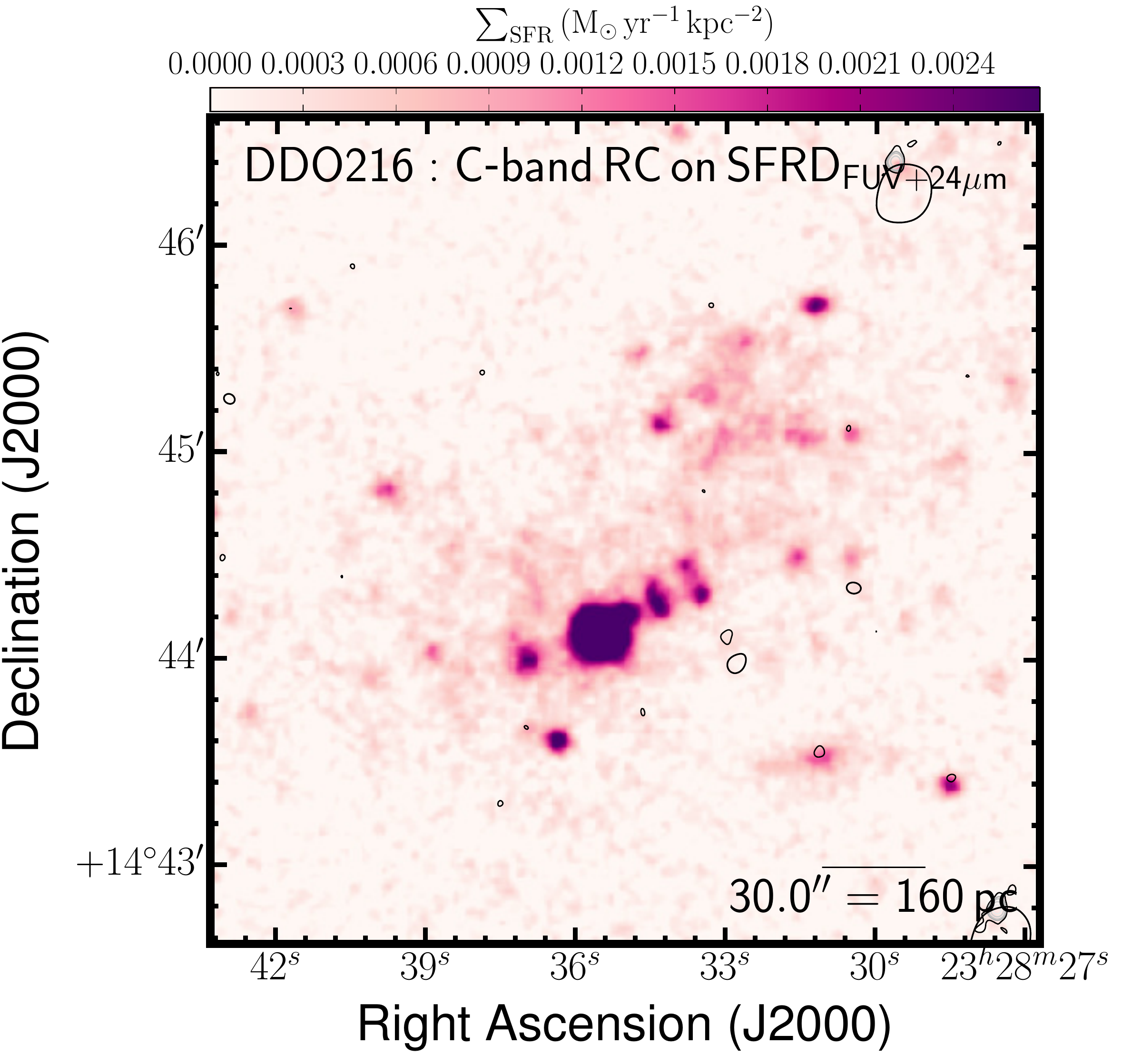} \\
  \end{tabular}
\caption[DDO\,216 images: RC, IR, optical, and FUV]{Multi-wavelength coverage of DDO 216 displaying a $3.7^\prime \times 3.7^\prime$ area. We show total RC flux density at the native resolution (top-left) and again with contours (top-centre). The RC contours are superposed on ancillary LITTLE THINGS images where possible: \halpha\ (middle-left); \RCNT\ obtained by subtracting the expected \RCT\ based on the \halpha-\RCT\ scaling factor of \cite{Deeg1997} from the total RC; {\em GALEX} FUV (middle-right); {\em Spitzer} 24\micron\ (bottom-left); {\em Spitzer} 70\micron\ (bottom-centre); FUV$+24{\rm \mu m}$--inferred SFRD from \citealp{Leroy2012} (bottom-right). We also show the RC that was isolated by the RC--based masking technique (top-right).}
  \label{figure:ddo216Cc_maps}
\end{figure}

\begin{figure}
  \begin{tabular}{ccc}
    \includegraphics[width=0.31\linewidth,clip]{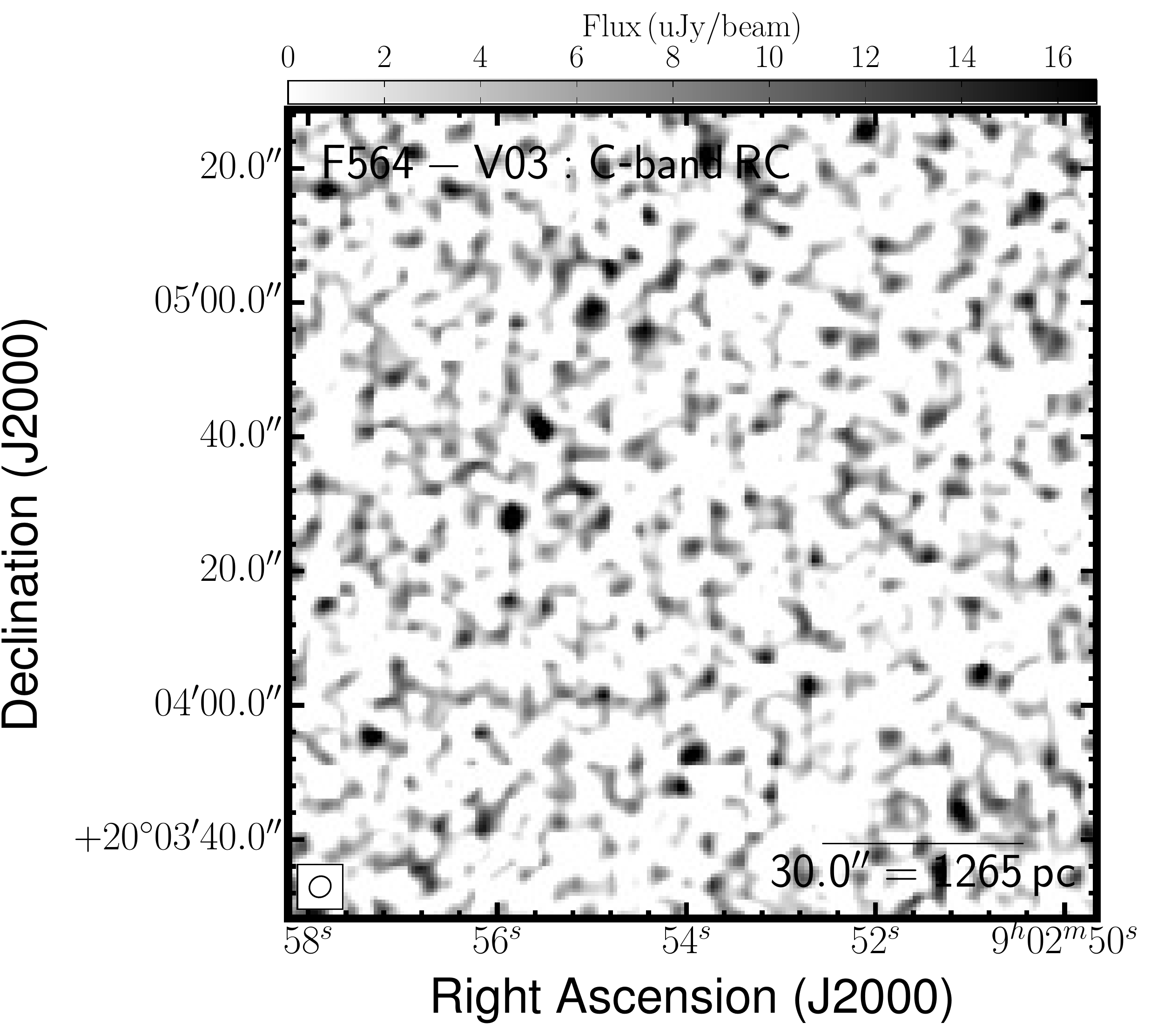} & \ 
    \includegraphics[width=0.31\linewidth,clip]{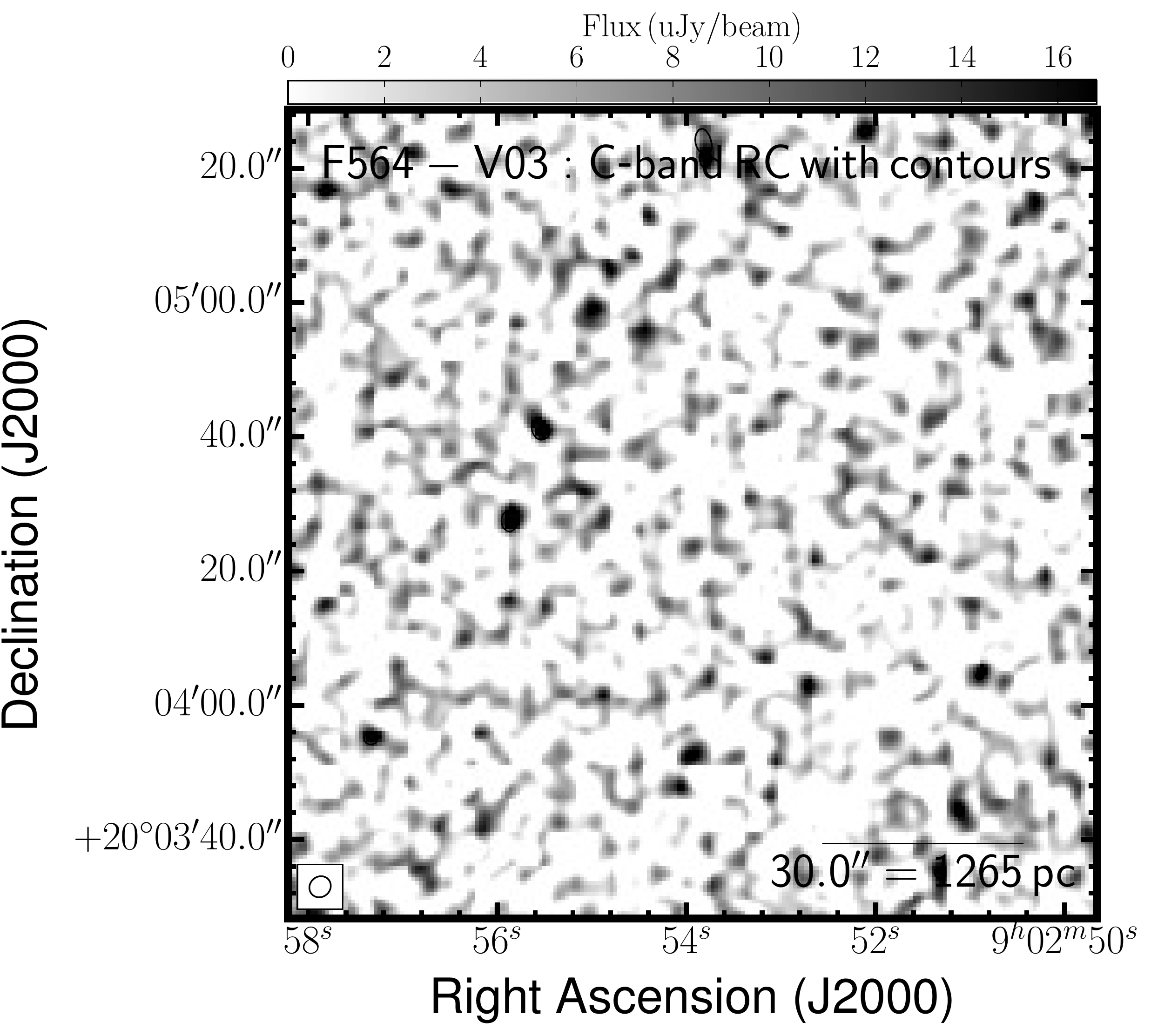} & \ 
    \includegraphics[width=0.31\linewidth,clip]{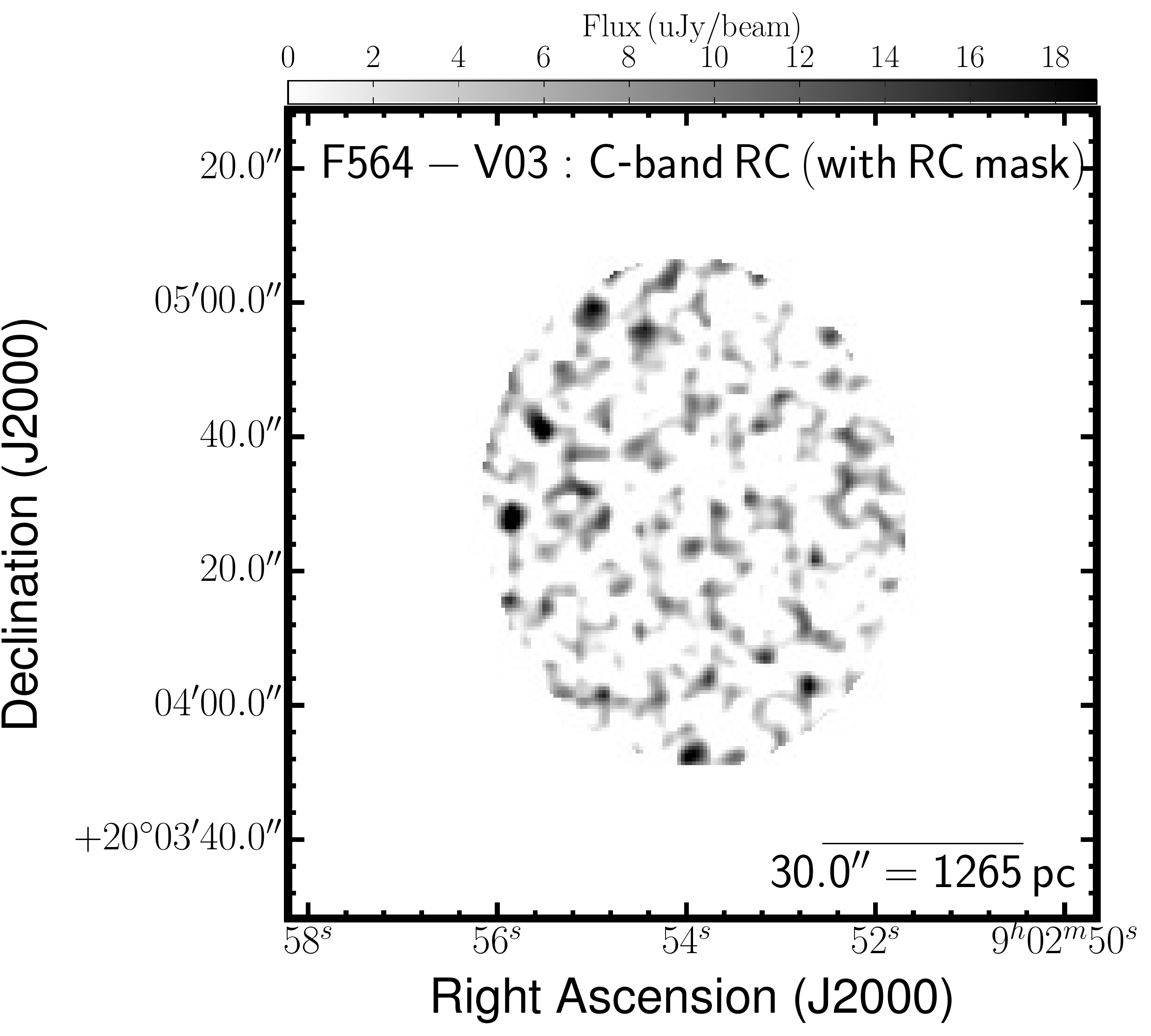} \\
    \includegraphics[width=0.31\linewidth,clip]{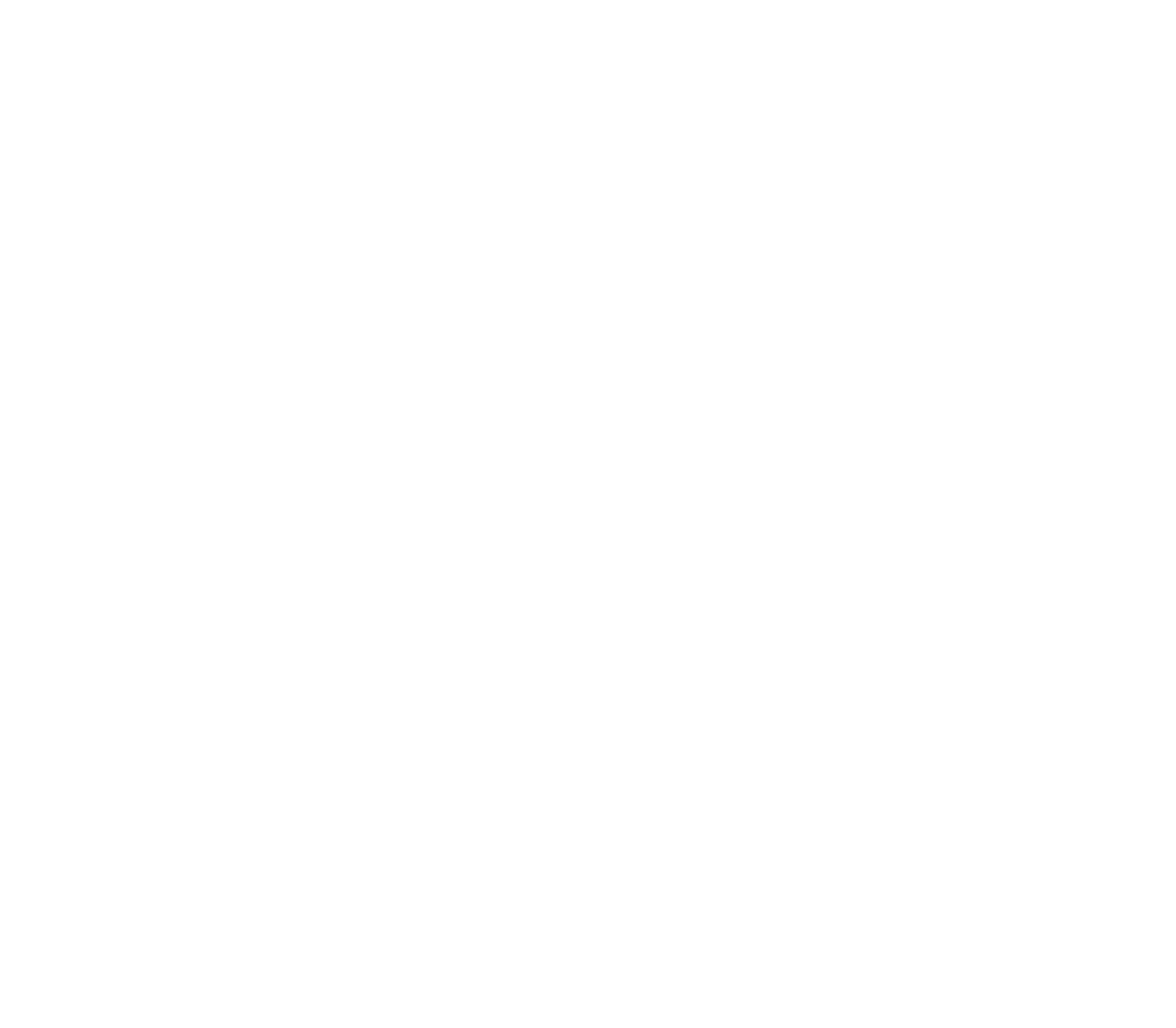} & \ 
    \includegraphics[width=0.31\linewidth,clip]{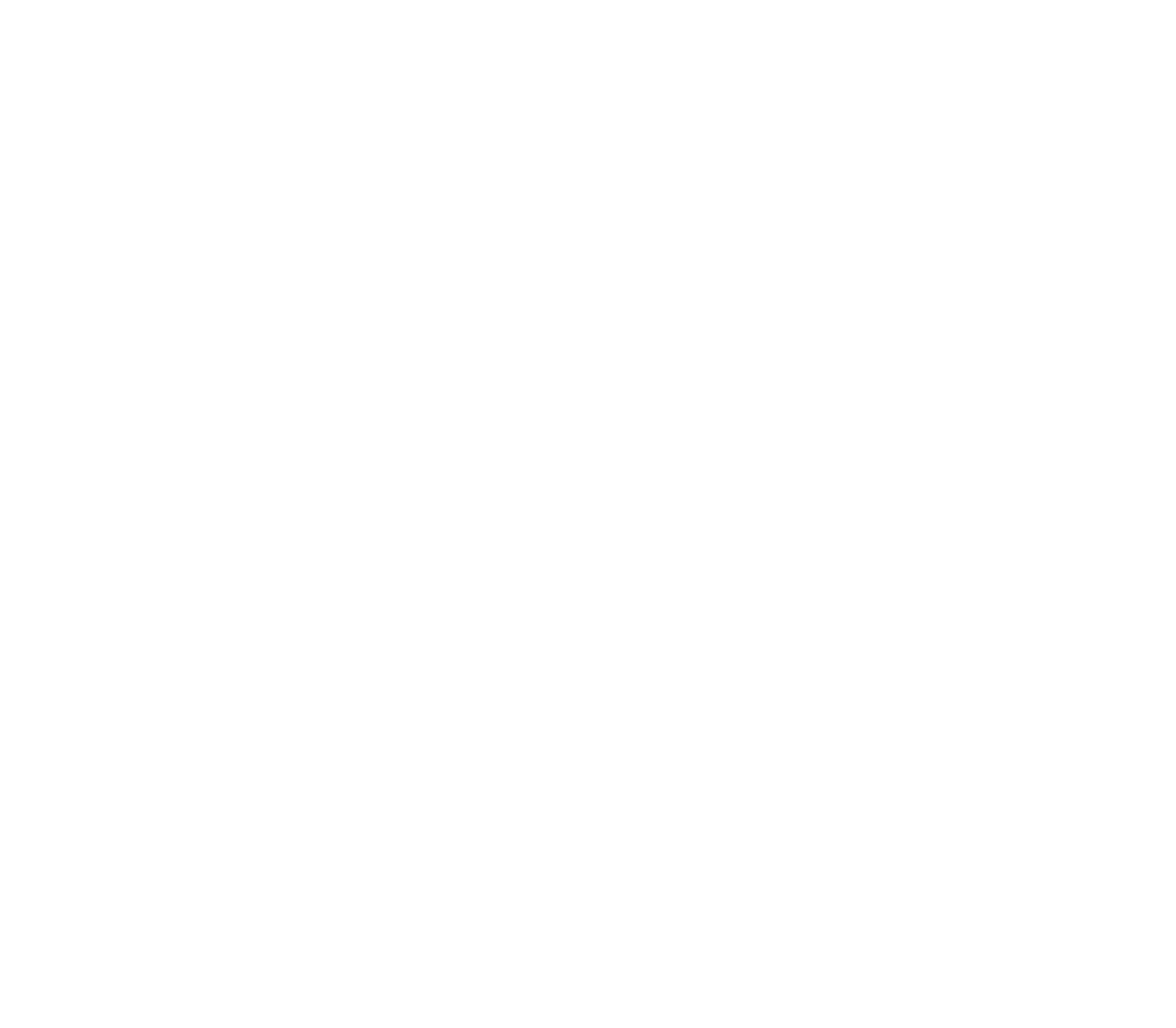} & \ 
    \includegraphics[width=0.31\linewidth,clip]{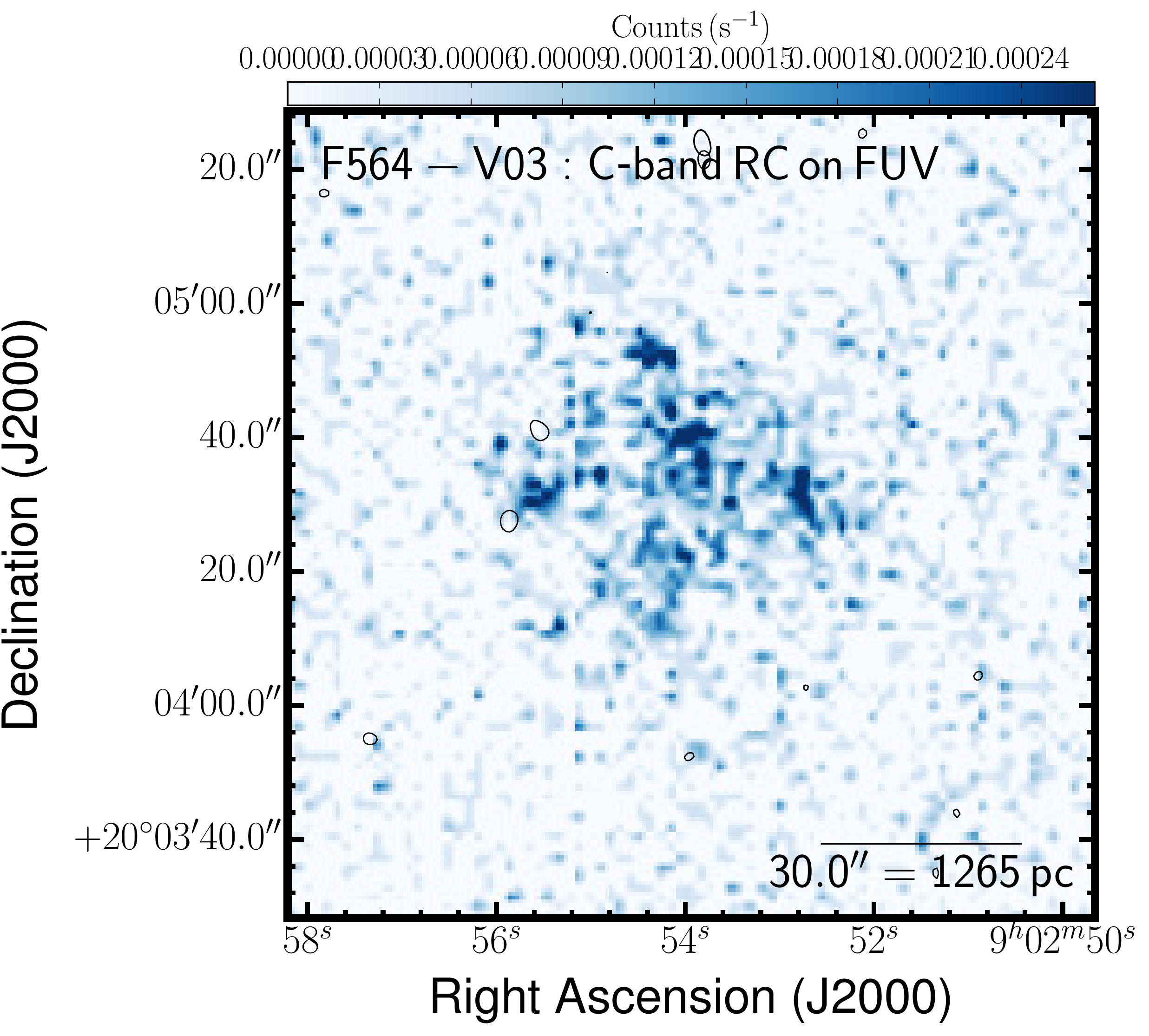} \\
    \includegraphics[width=0.31\linewidth,clip]{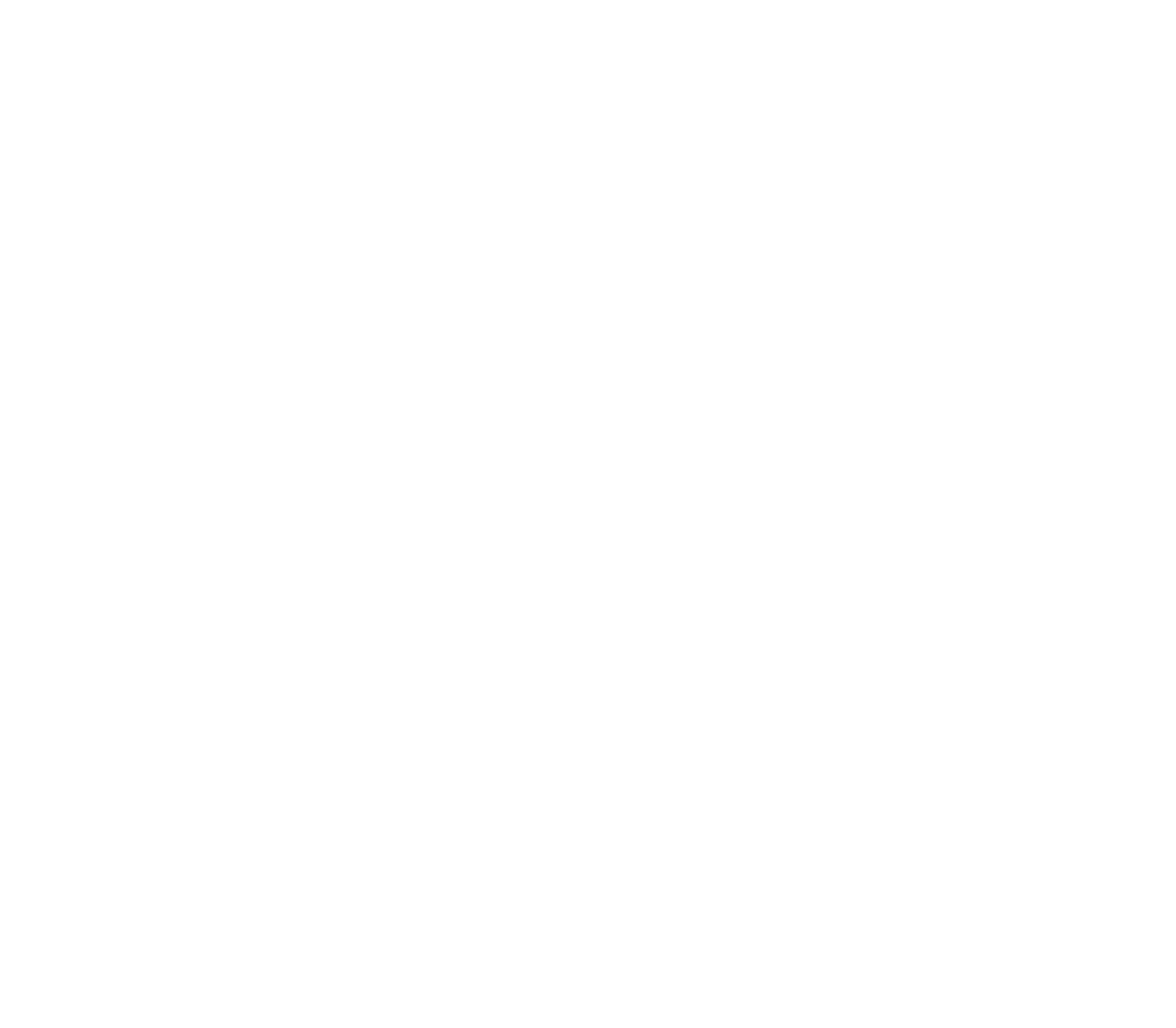} & \ 
    \includegraphics[width=0.31\linewidth,clip]{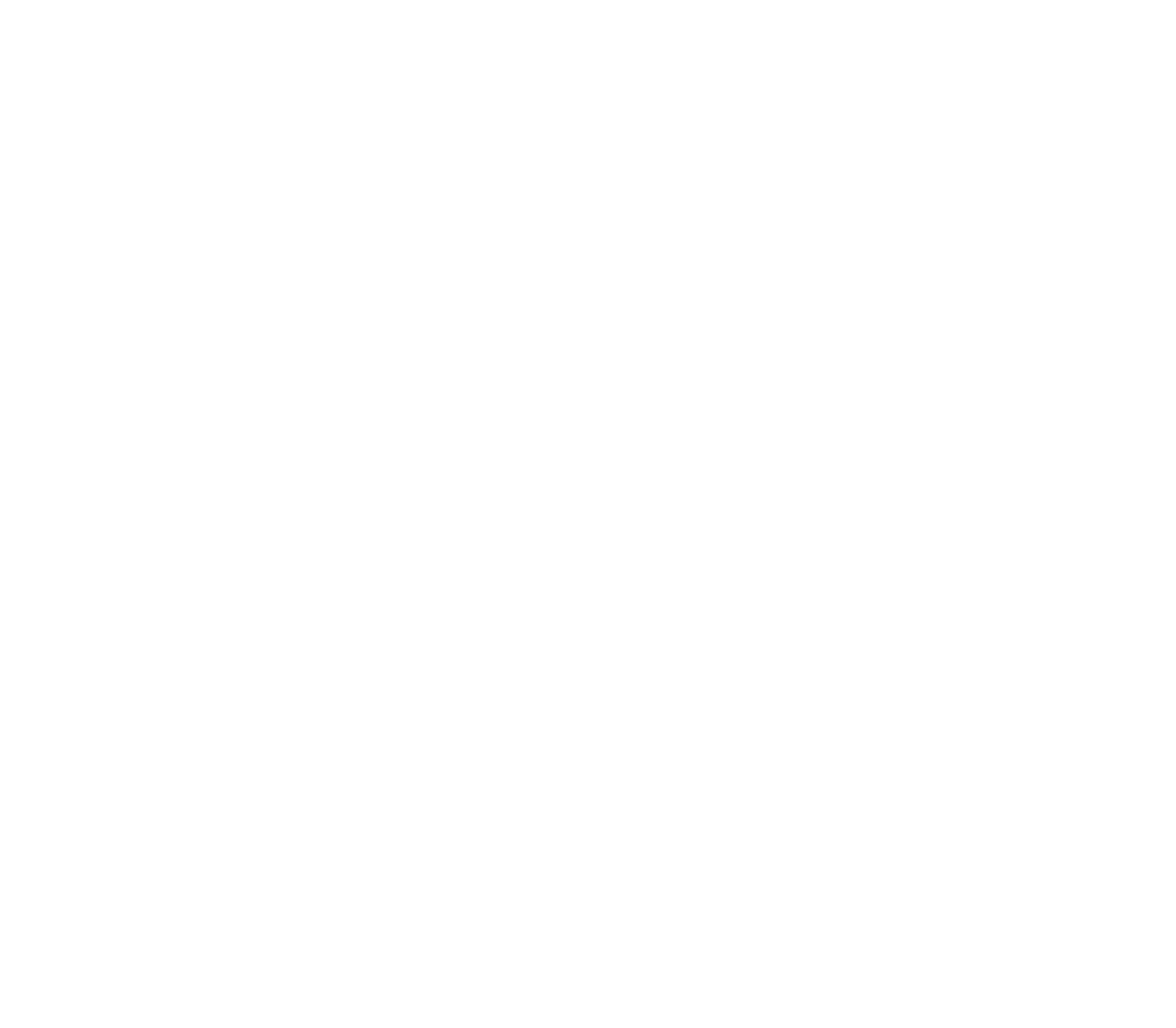} & \ 
    \includegraphics[width=0.31\linewidth,clip]{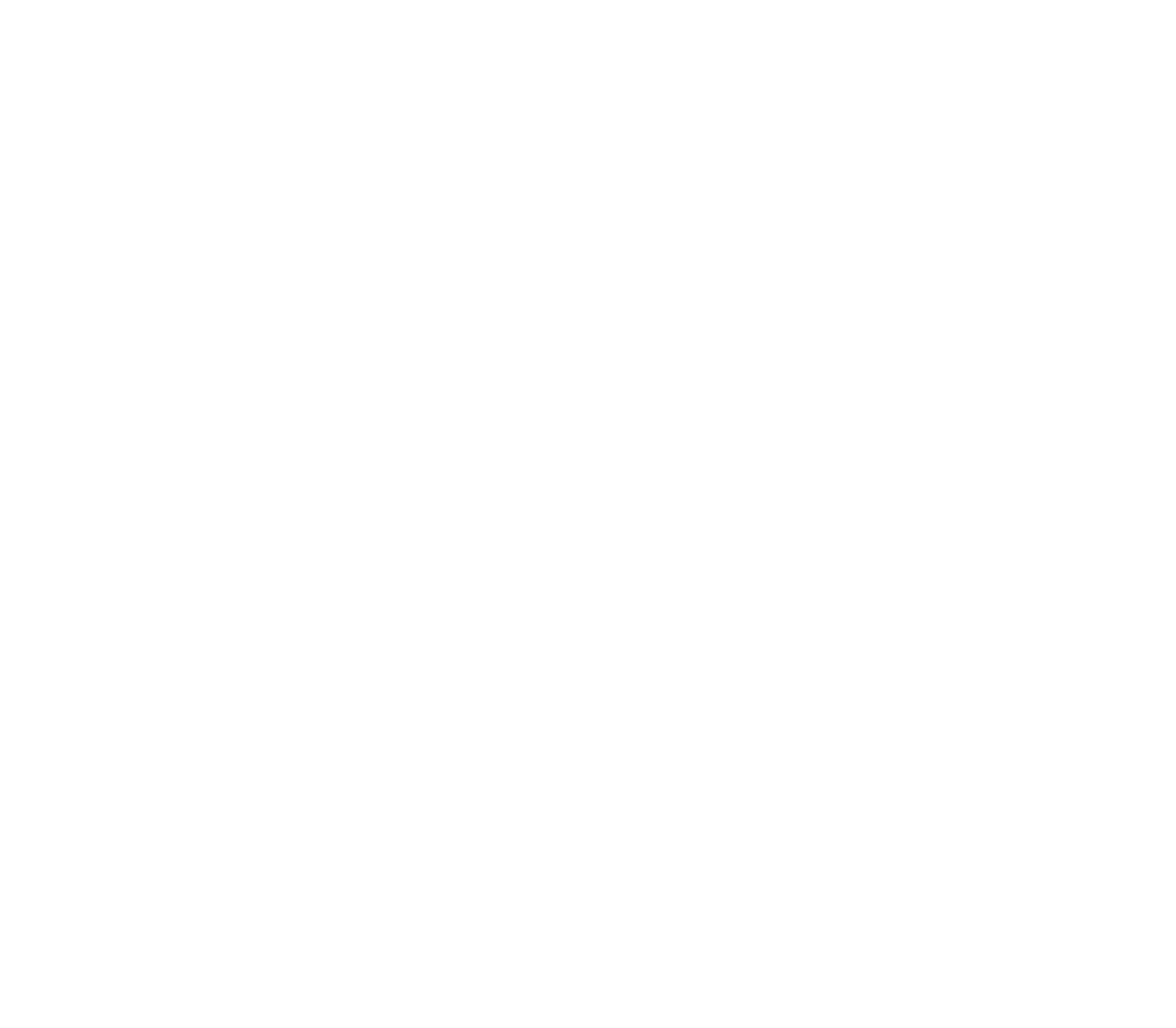} \\
  \end{tabular}
\caption[F564\,V03 images: RC, IR, optical, and FUV]{Multi-wavelength coverage of F564-V03 displaying a $2.0^\prime \times 2.0^\prime$ area. We show total RC flux density at the native resolution (top-left) and again with contours (top-centre). The RC contours are superposed on ancillary LITTLE THINGS images where possible: \halpha\ (middle-left); \RCNT\ obtained by subtracting the expected \RCT\ based on the \halpha-\RCT\ scaling factor of \cite{Deeg1997} from the total RC; {\em GALEX} FUV (middle-right); {\em Spitzer} 24\micron\ (bottom-left); {\em Spitzer} 70\micron\ (bottom-centre); FUV$+24{\rm \mu m}$--inferred SFRD from \citealp{Leroy2012} (bottom-right). We also show the RC that was isolated by the RC--based masking technique (top-right).}
  \label{figure:f564v3Cc_maps}
\end{figure}

\begin{figure}
  \begin{tabular}{ccc}
    \includegraphics[width=0.31\linewidth,clip]{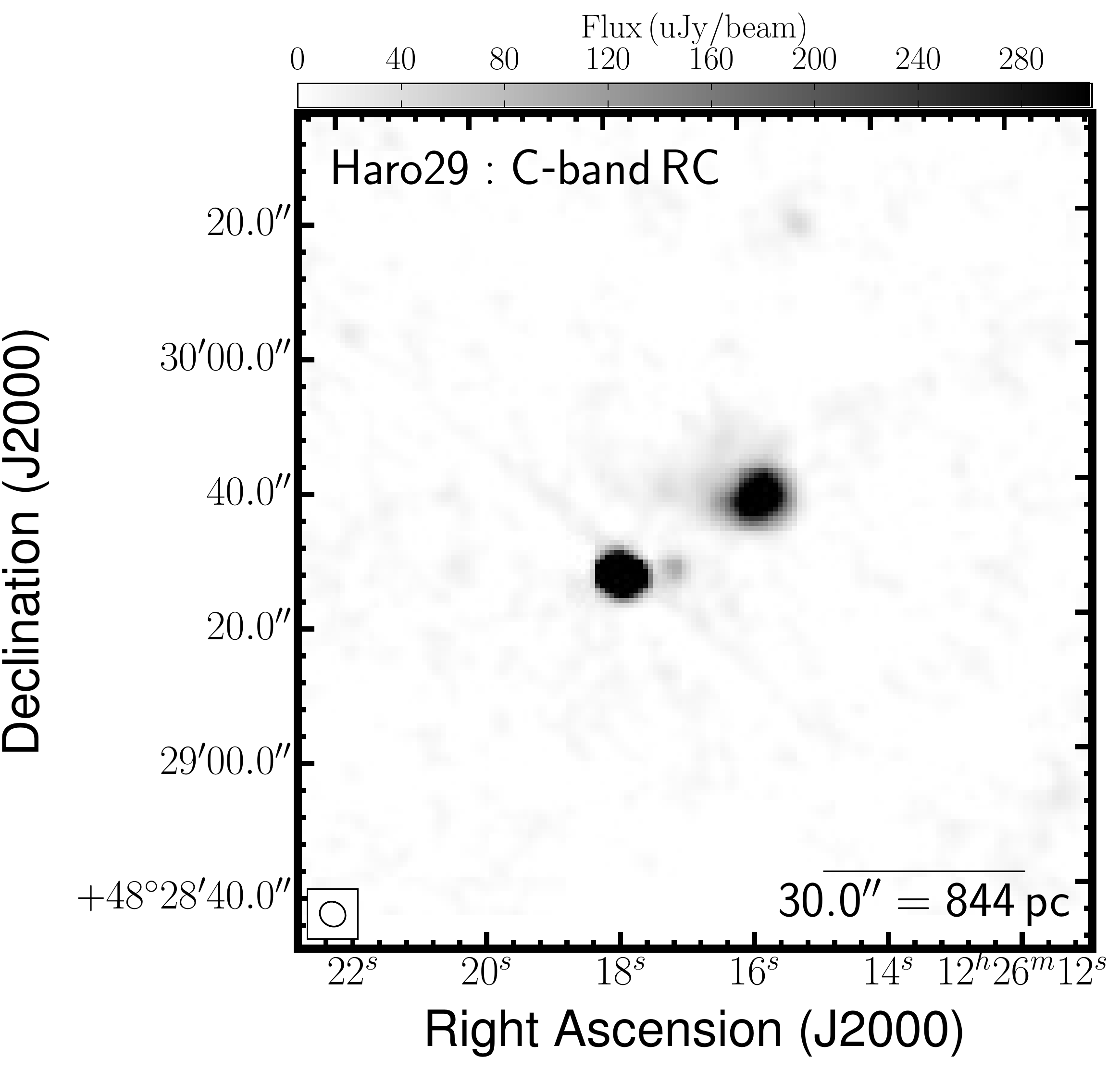} & \ 
    \includegraphics[width=0.31\linewidth,clip]{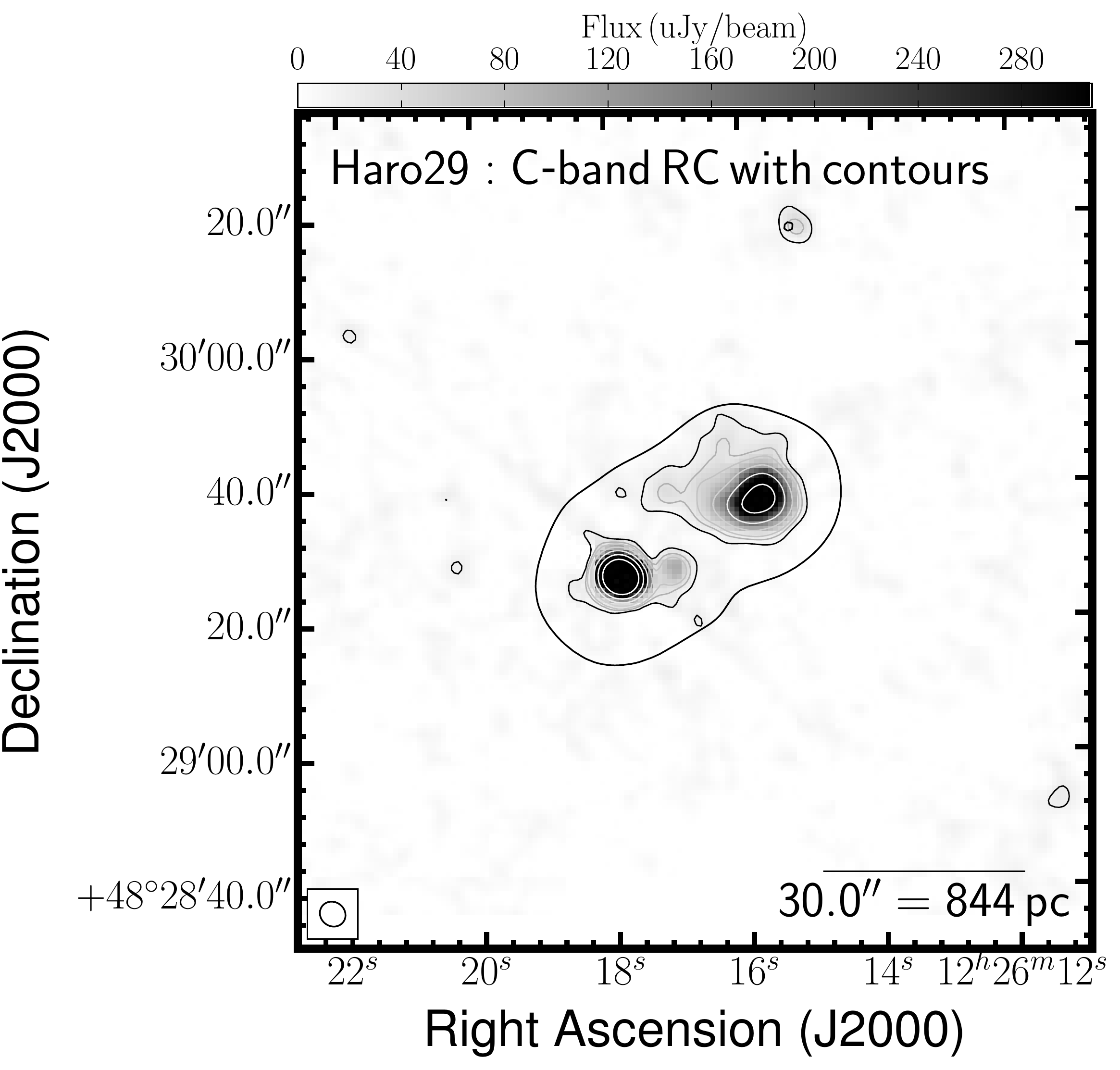} & \ 
    \includegraphics[width=0.31\linewidth,clip]{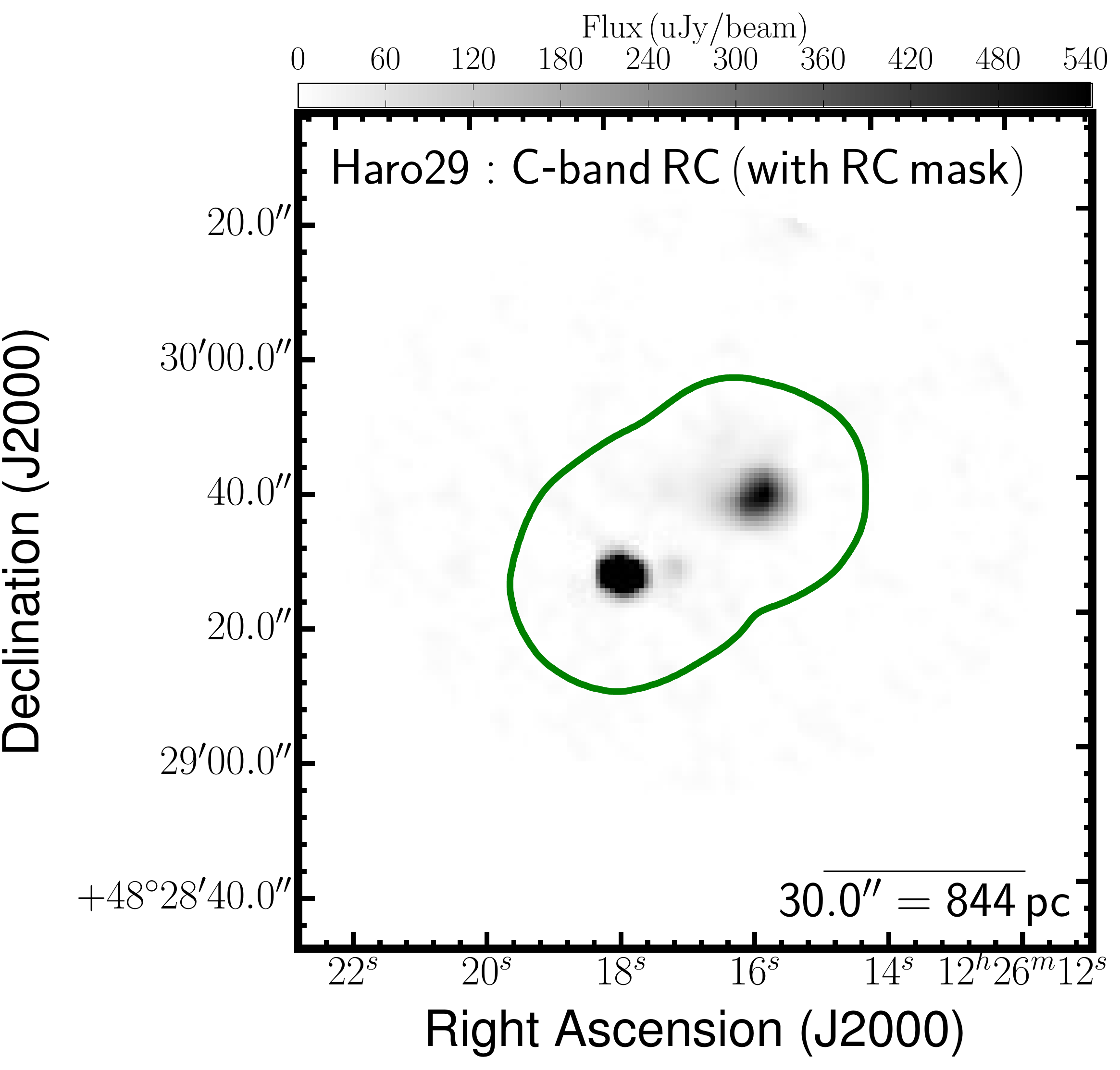} \\
    \includegraphics[width=0.31\linewidth,clip]{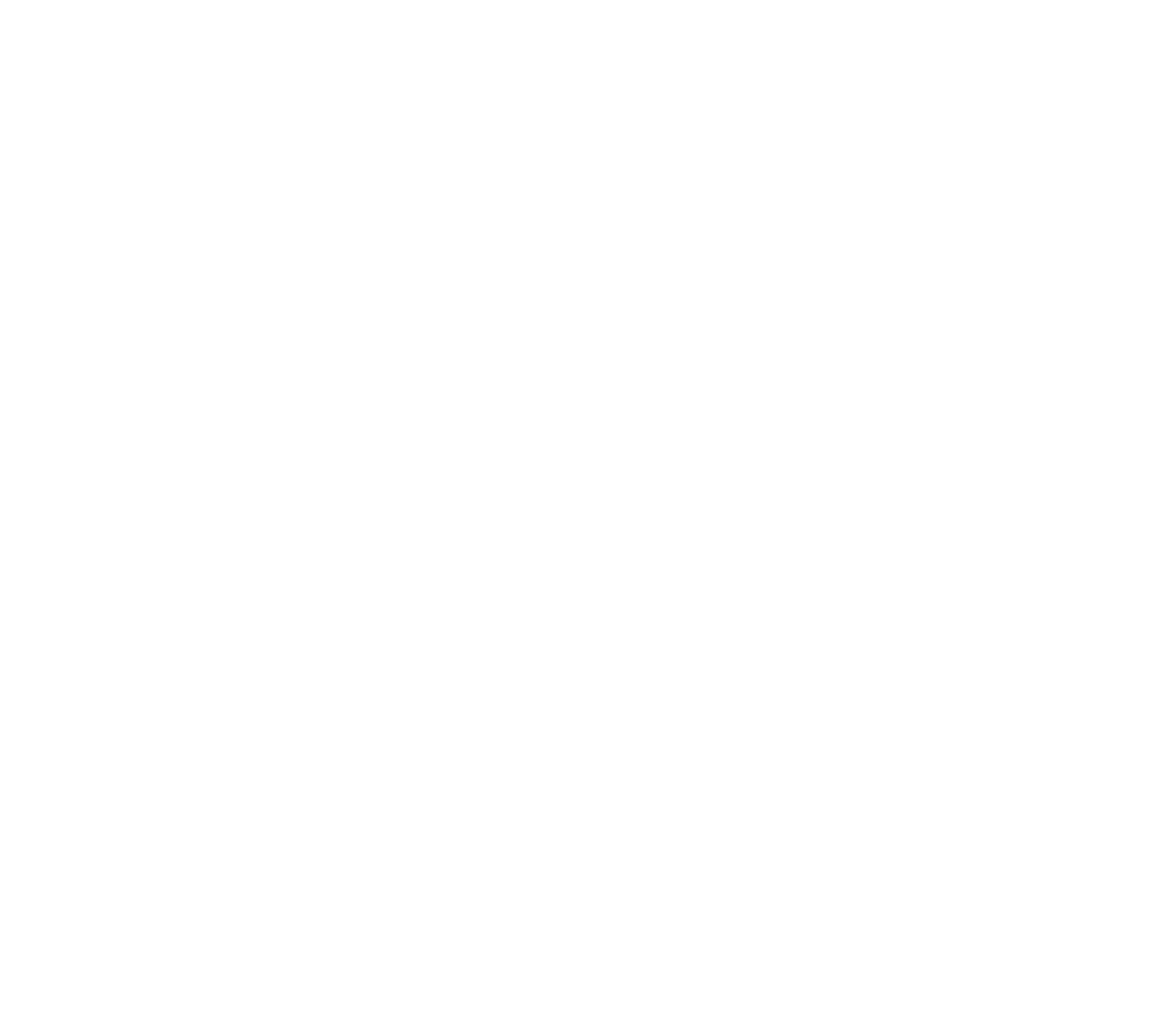} & \
    \includegraphics[width=0.31\linewidth,clip]{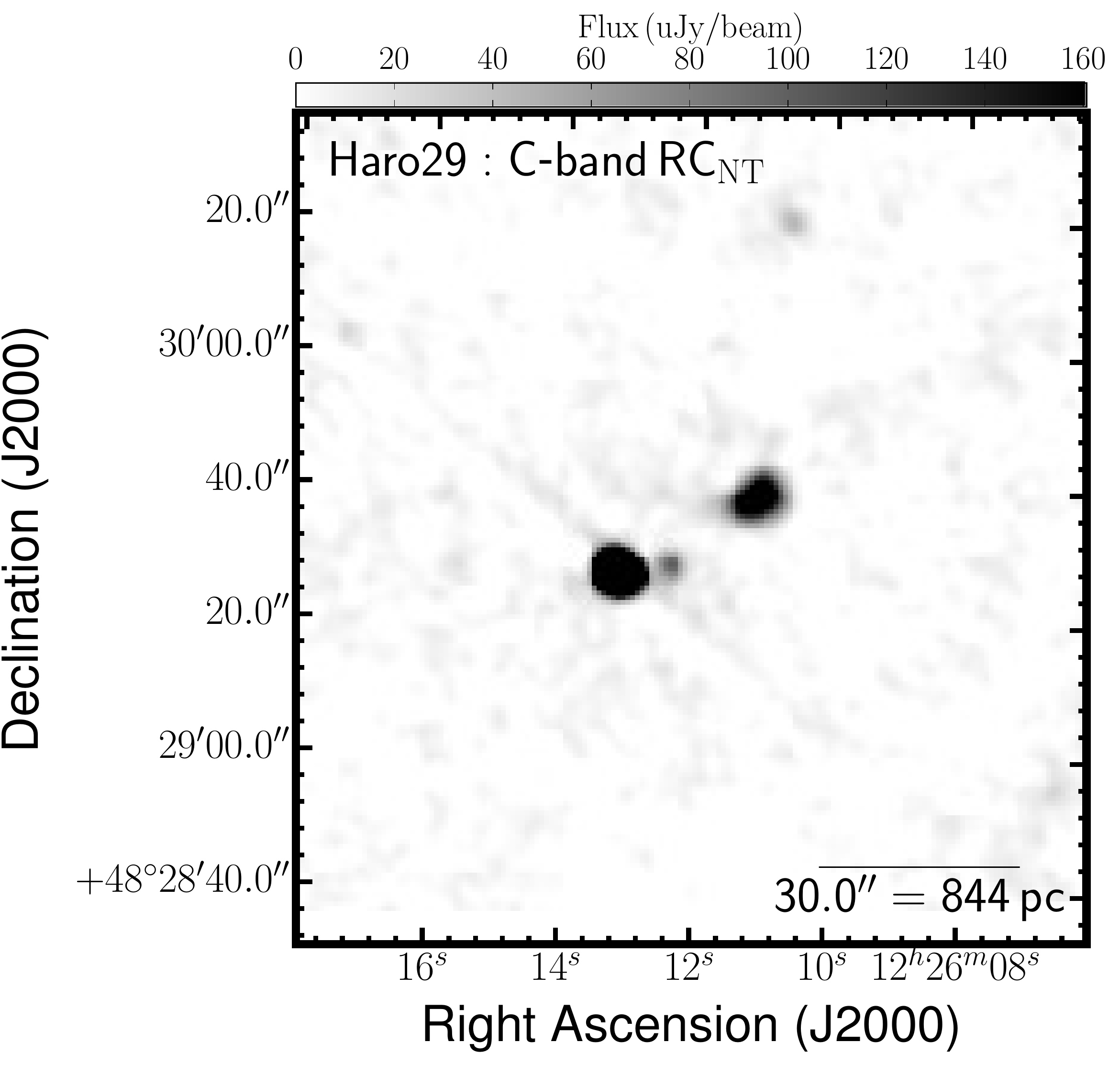} & \ 
    \includegraphics[width=0.31\linewidth,clip]{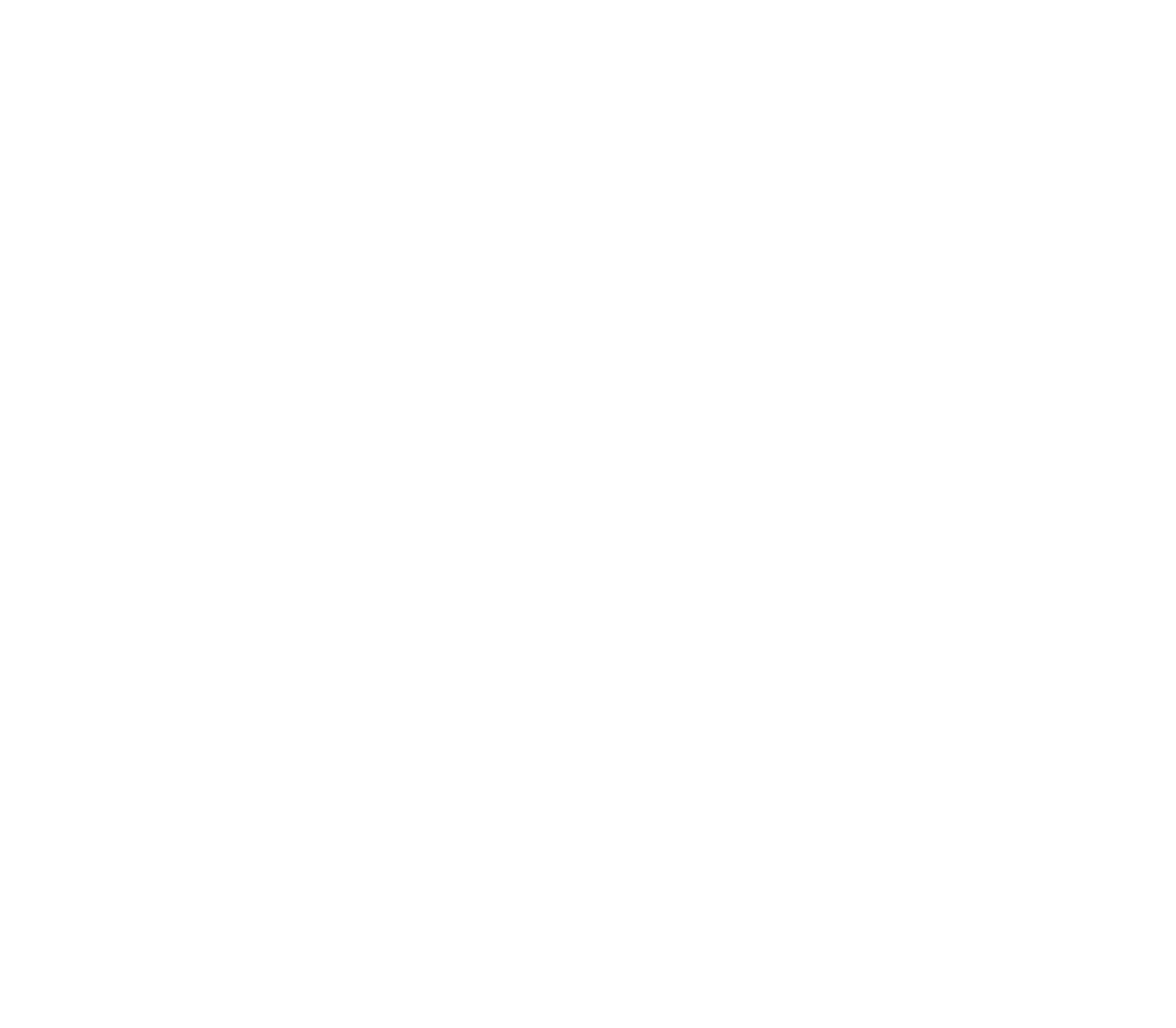} \\ 
    \includegraphics[width=0.31\linewidth,clip]{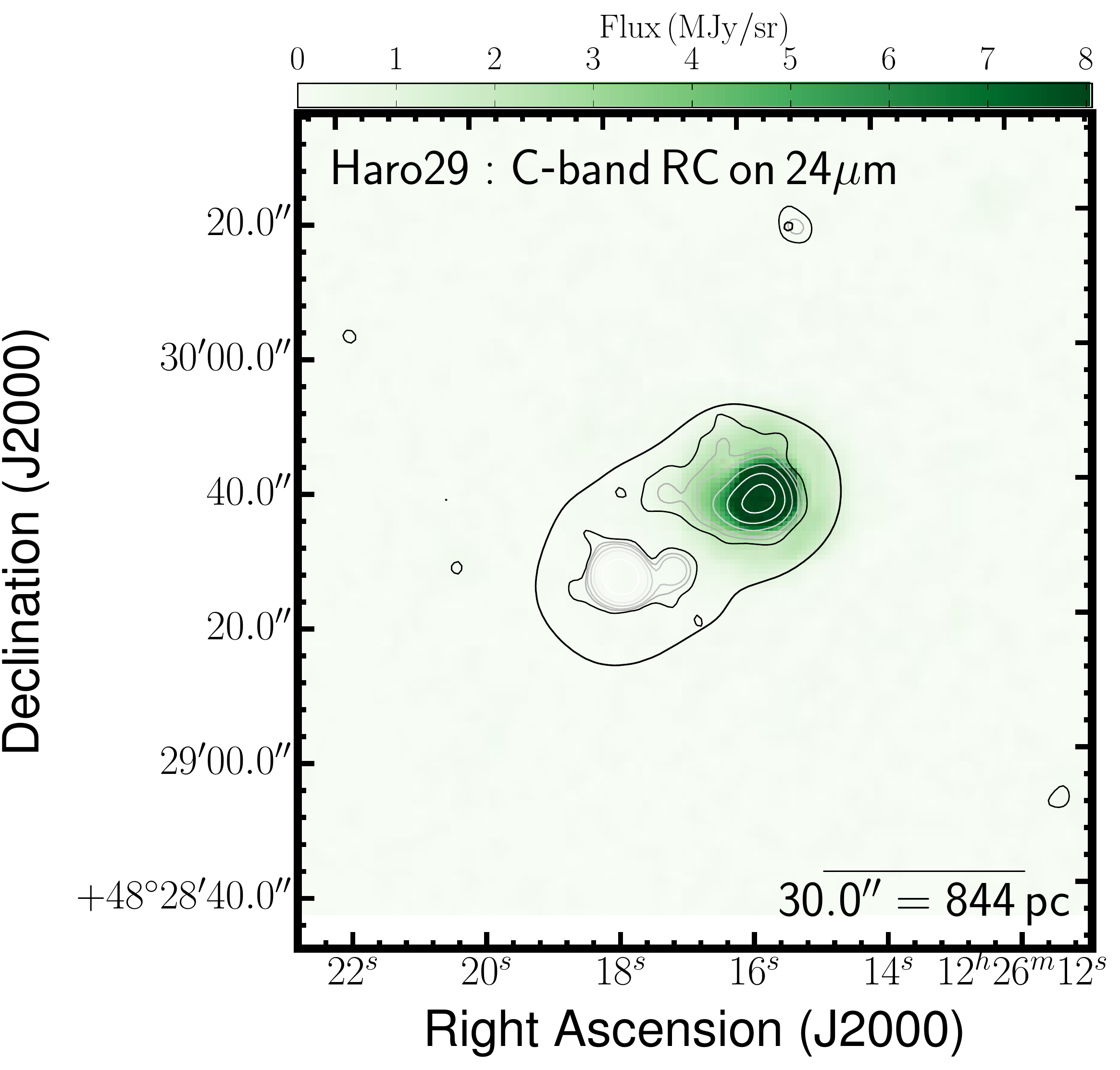} & \ 
    \includegraphics[width=0.31\linewidth,clip]{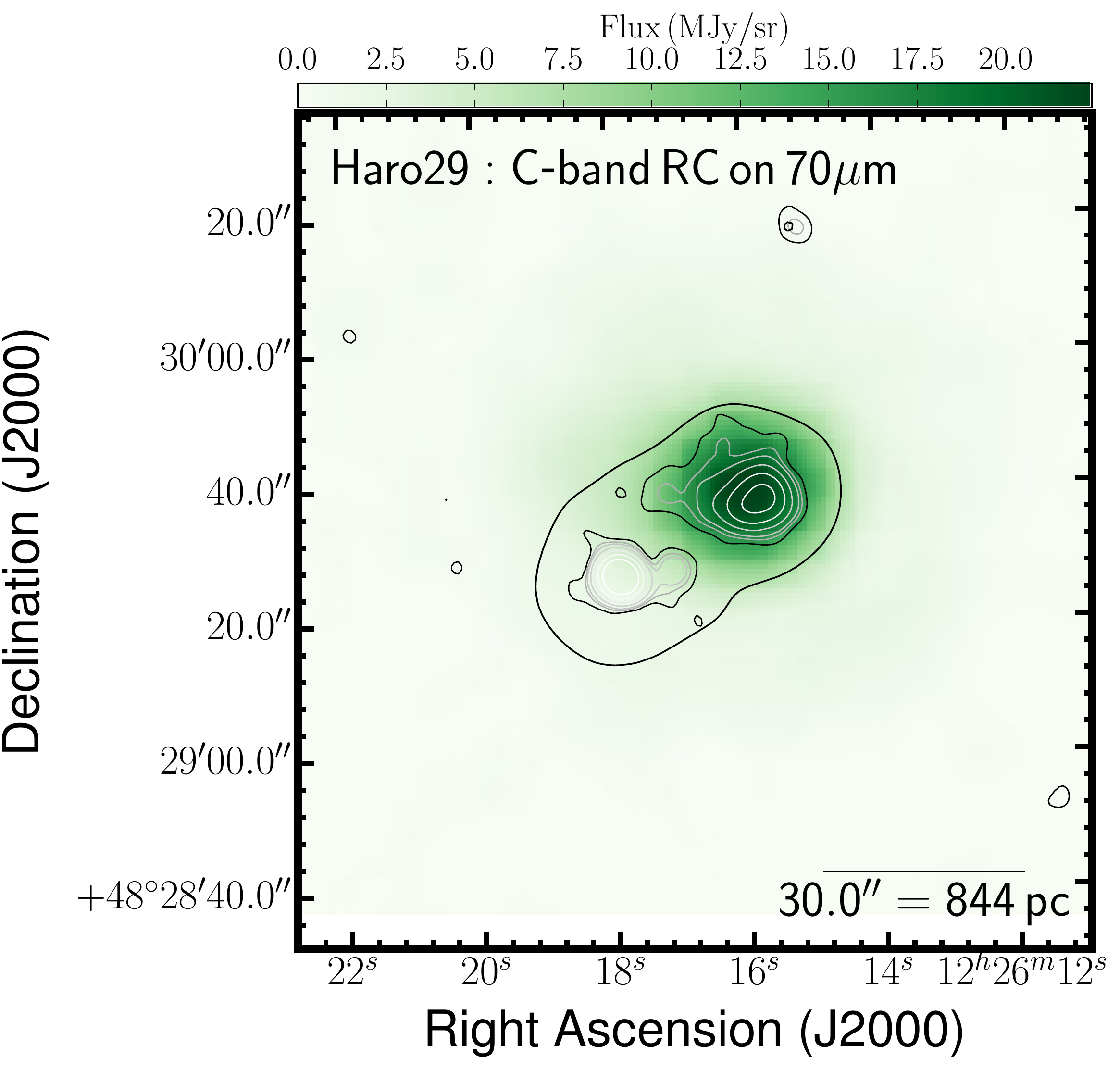} & \ 
    \includegraphics[width=0.31\linewidth,clip]{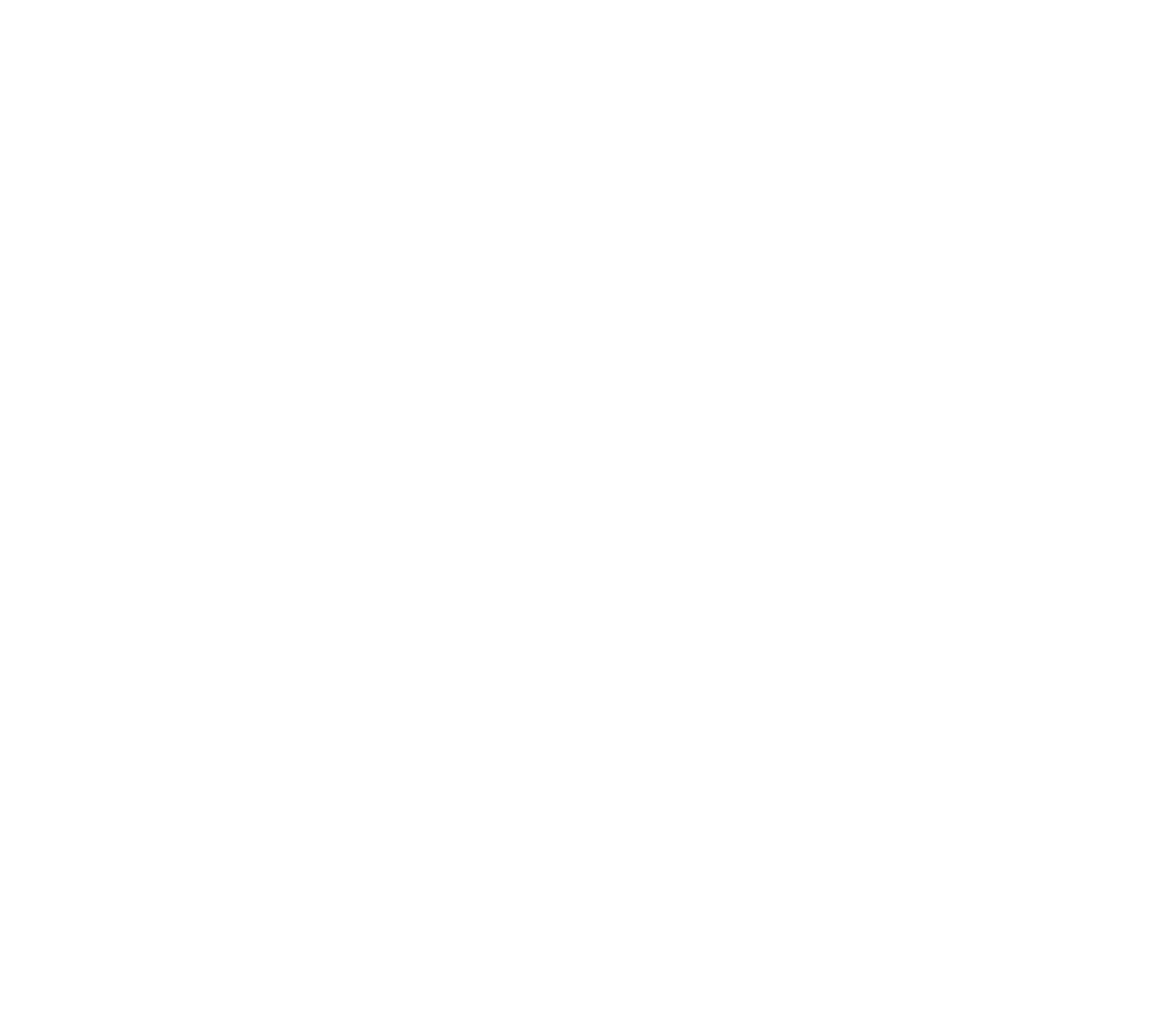} \\ 
  \end{tabular}
\caption[Haro\,29 images: RC, IR, optical, and FUV]{Multi-wavelength coverage of Haro 29 displaying a $2.0^\prime \times 2.0^\prime$ area. We show total RC flux density at the native resolution (top-left) and again with contours (top-centre). The RC contours are superposed on ancillary LITTLE THINGS images where possible: \halpha\ (middle-left); \RCNT\ obtained by subtracting the expected \RCT\ based on the \halpha-\RCT\ scaling factor of \cite{Deeg1997} from the total RC; {\em GALEX} FUV (middle-right); {\em Spitzer} 24\micron\ (bottom-left); {\em Spitzer} 70\micron\ (bottom-centre); FUV$+24{\rm \mu m}$--inferred SFRD from \citealp{Leroy2012} (bottom-right). We also show the RC that was isolated by the RC--based masking technique (top-right).}
  \label{figure:haro29Cc_maps}
\end{figure}

\begin{figure}
  \begin{tabular}{ccc}
    \includegraphics[width=0.31\linewidth,clip]{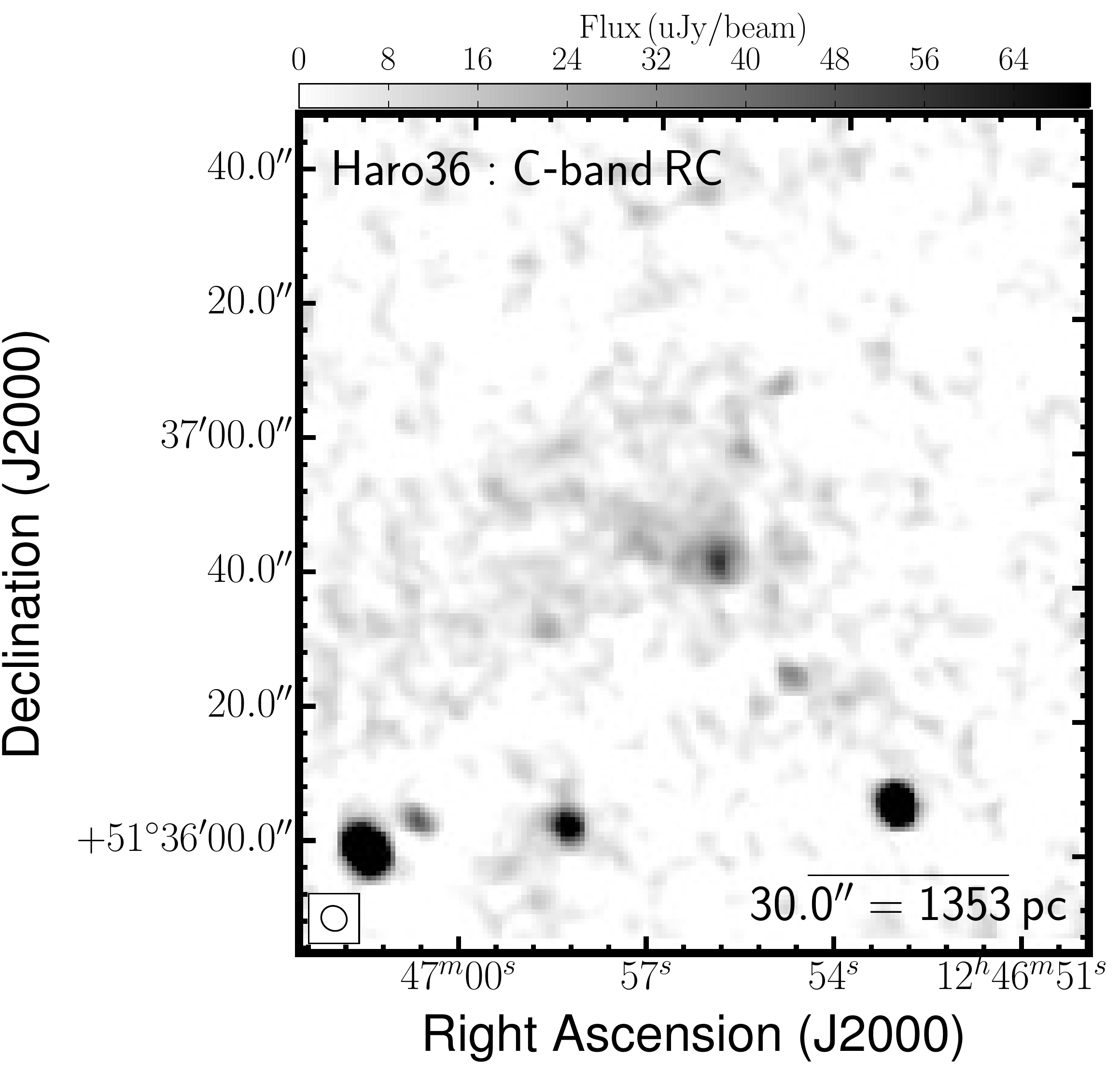} & \ 
    \includegraphics[width=0.31\linewidth,clip]{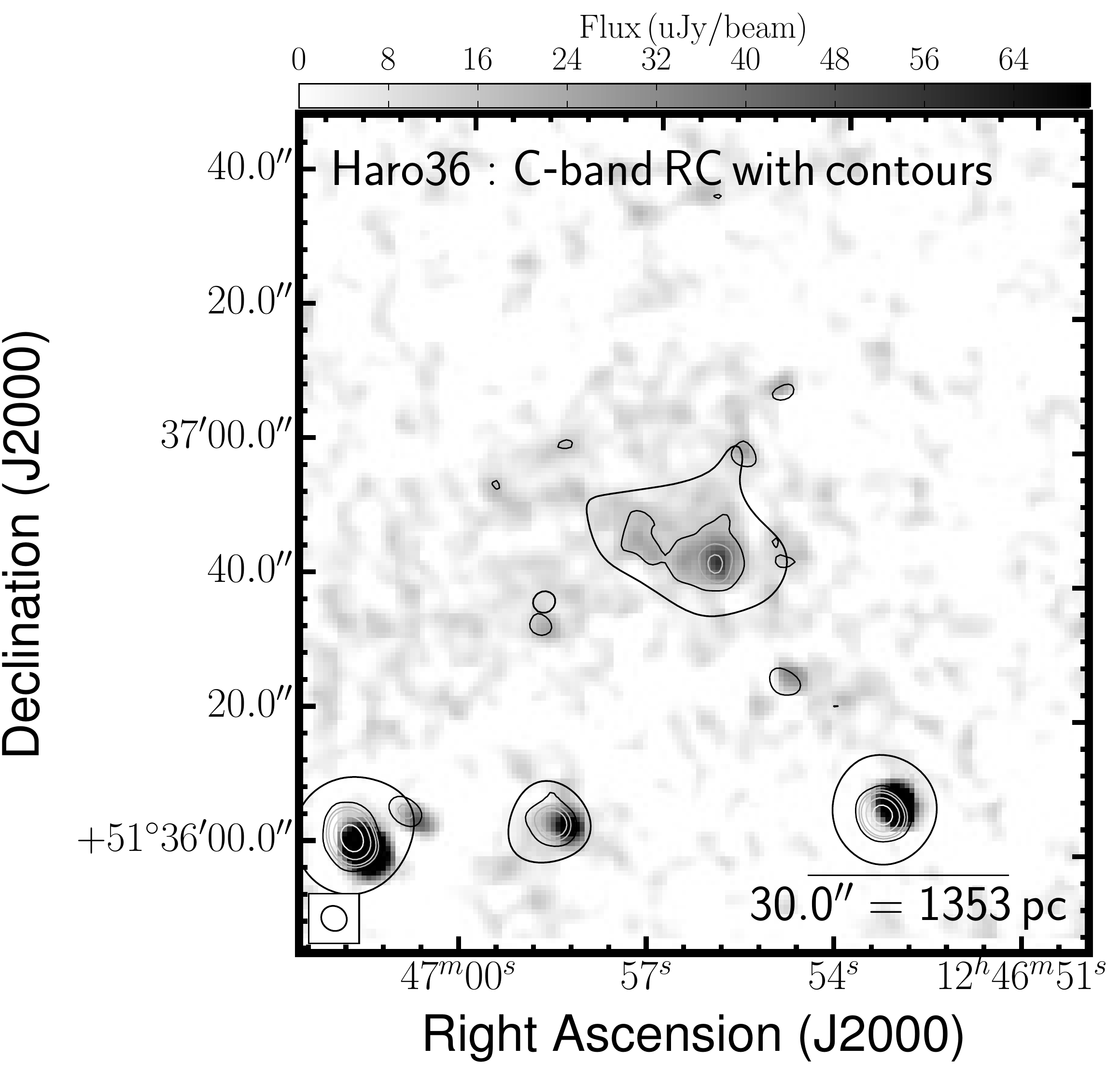} & \ 
    \includegraphics[width=0.31\linewidth,clip]{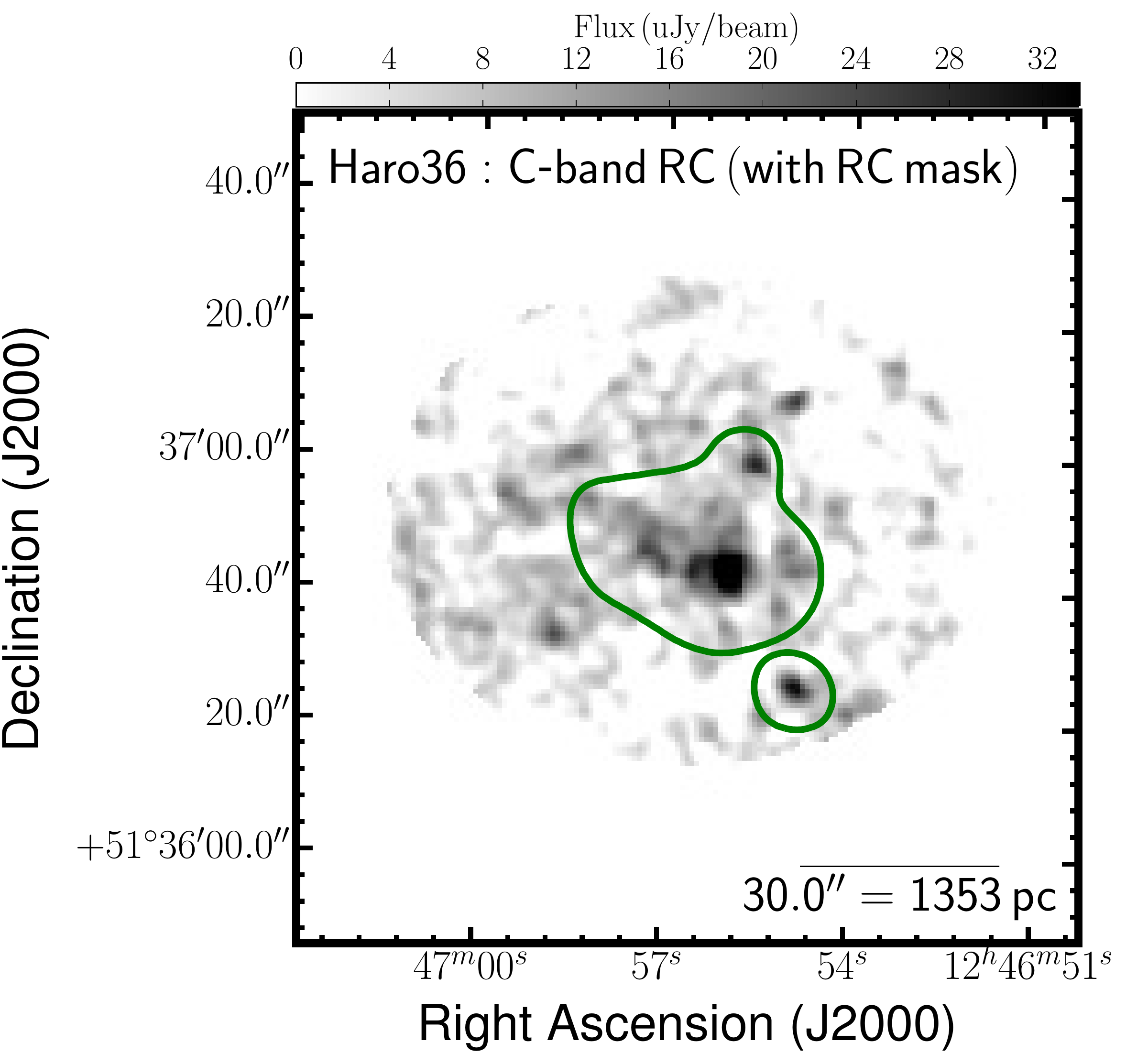} \\
    \includegraphics[width=0.31\linewidth,clip]{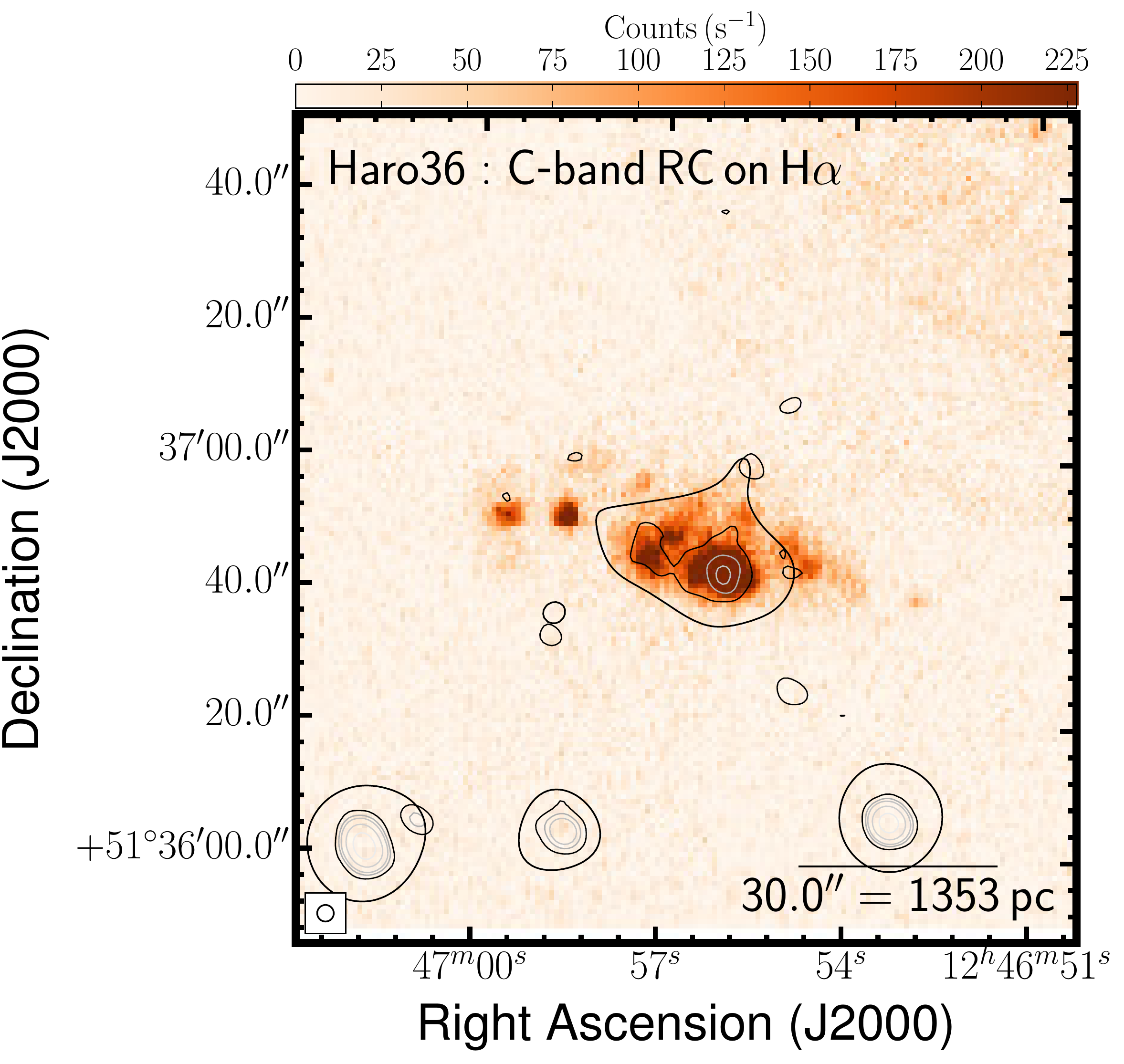} & \ 
    \includegraphics[width=0.31\linewidth,clip]{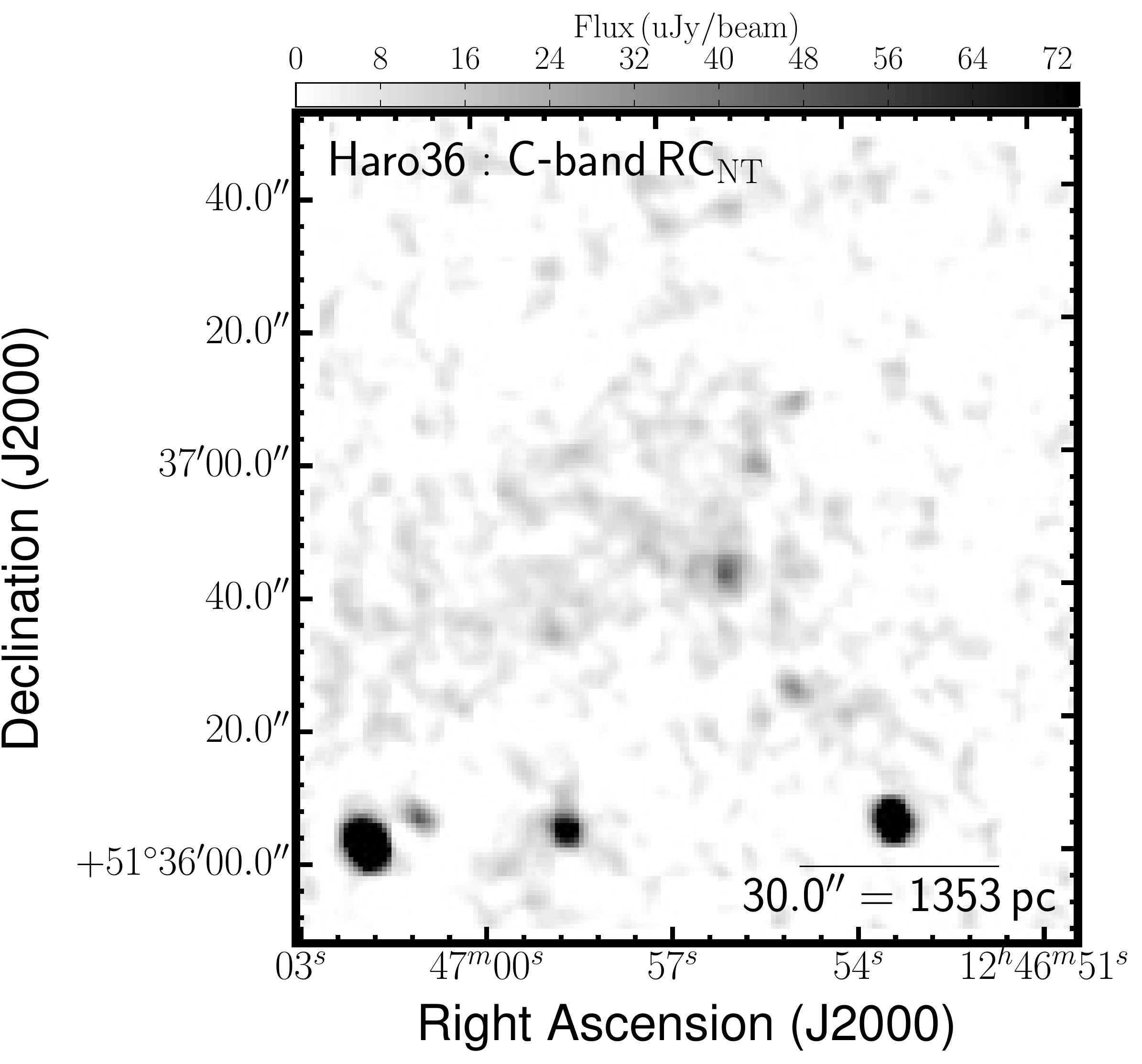} & \ 
    \includegraphics[width=0.31\linewidth,clip]{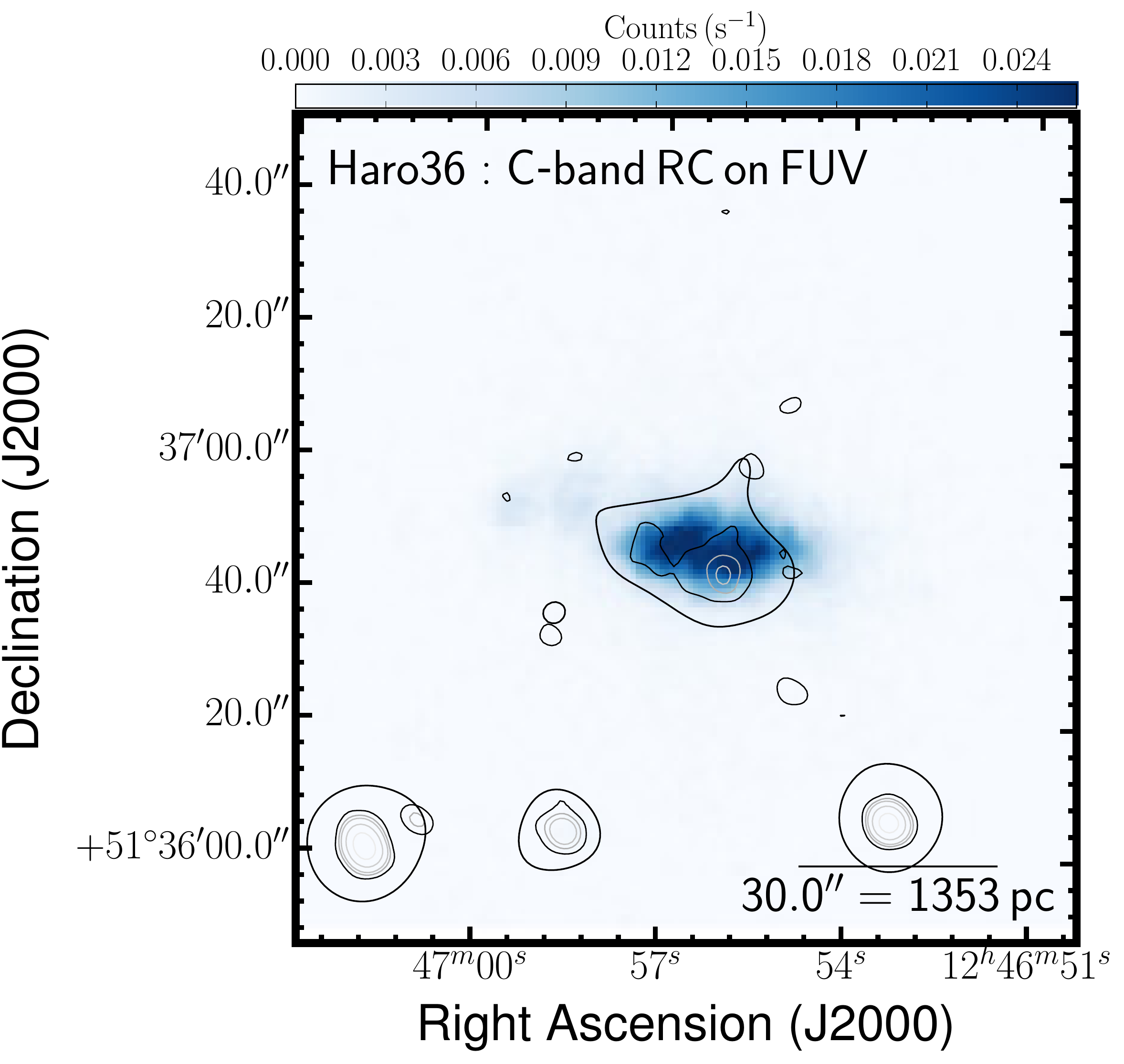} \\
    \includegraphics[width=0.31\linewidth,clip]{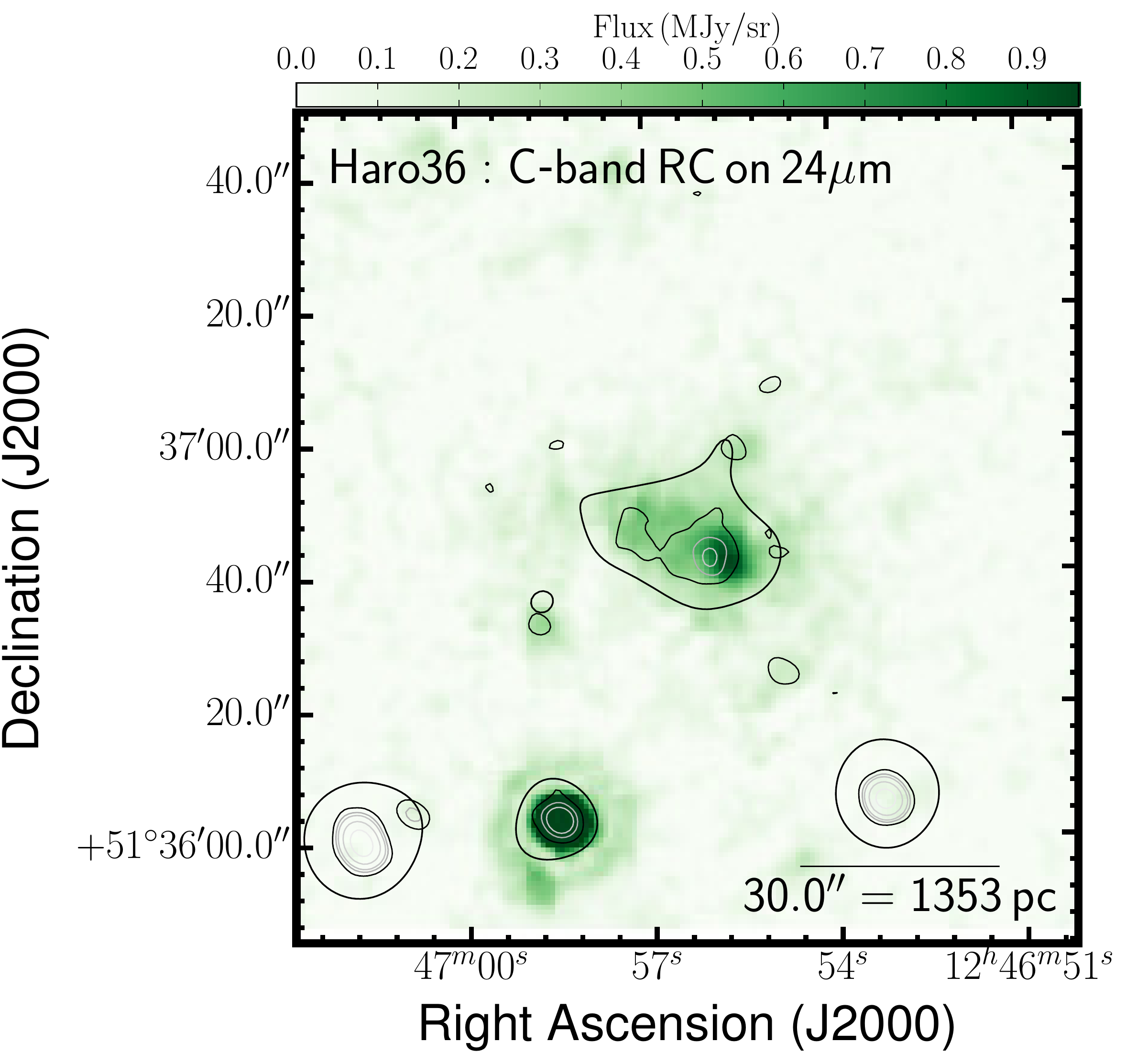} & \ 
    \includegraphics[width=0.31\linewidth,clip]{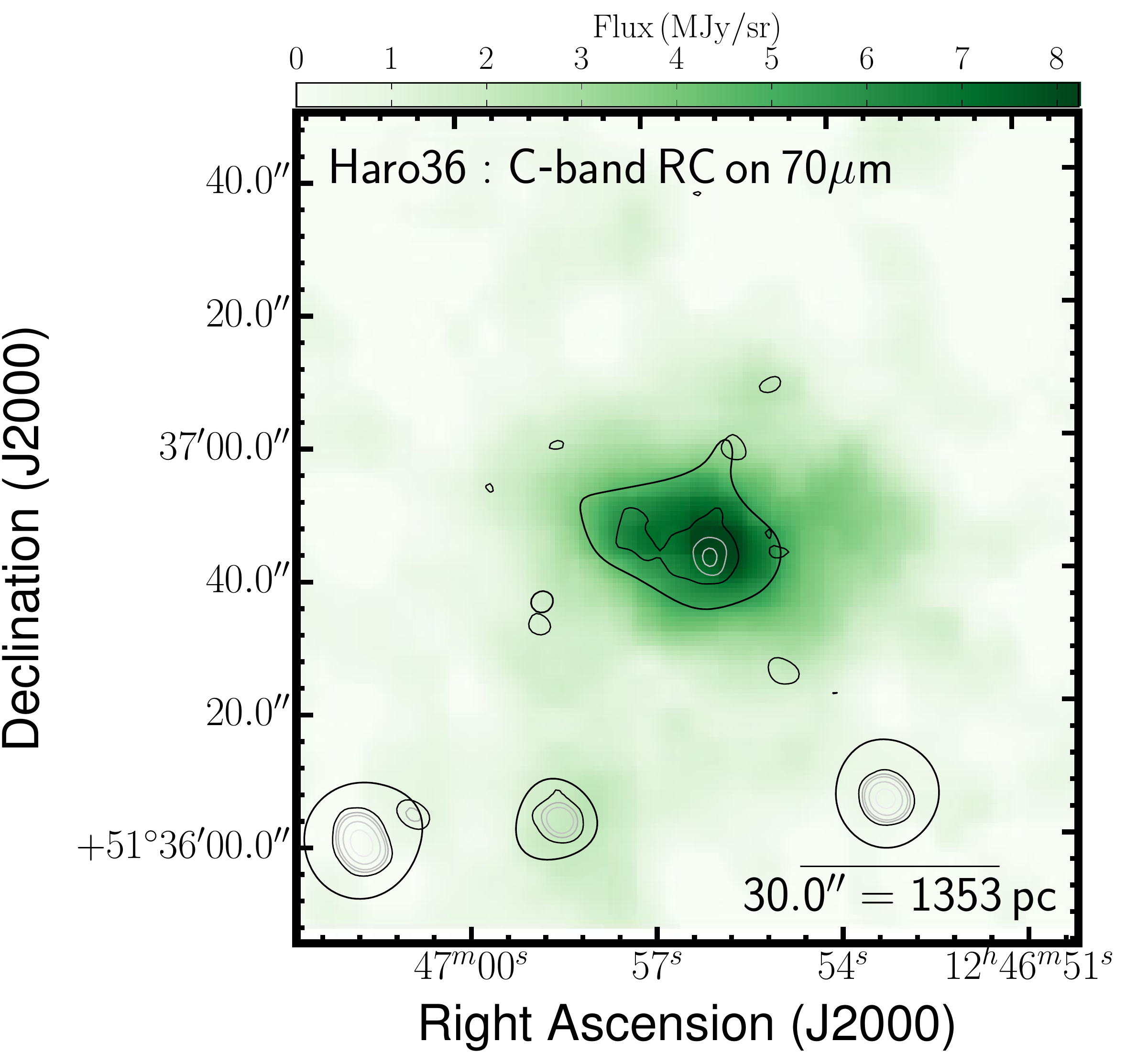} & \ 
    \includegraphics[width=0.31\linewidth,clip]{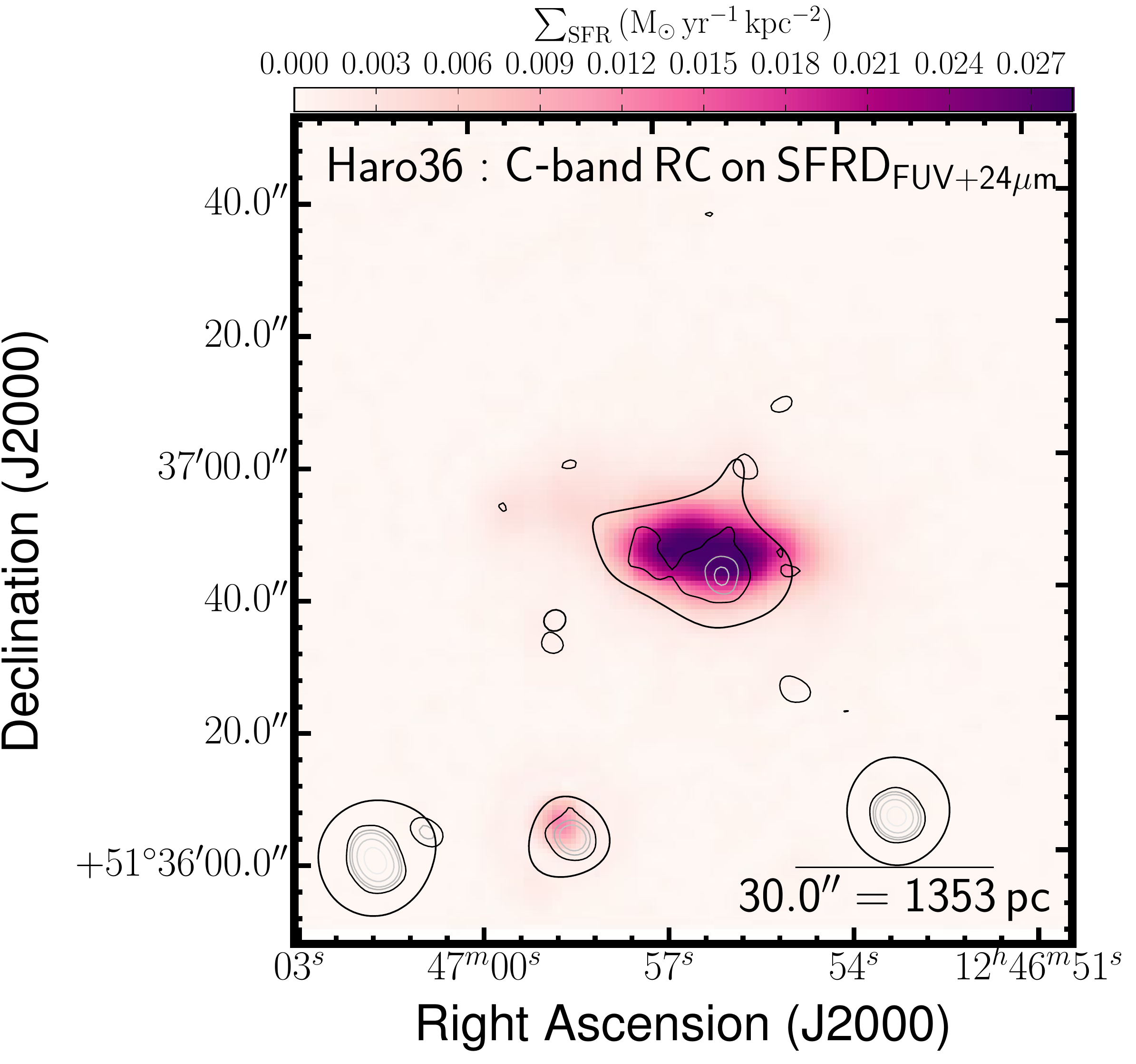} \\
  \end{tabular}
\caption[Haro\,36 images: RC, IR, optical, and FUV]{Multi-wavelength coverage of Haro 36 displaying a $2.0^\prime \times 2.0^\prime$ area. We show total RC flux density at the native resolution (top-left) and again with contours (top-centre). The RC contours are superposed on ancillary LITTLE THINGS images where possible: \halpha\ (middle-left); \RCNT\ obtained by subtracting the expected \RCT\ based on the \halpha-\RCT\ scaling factor of \cite{Deeg1997} from the total RC; {\em GALEX} FUV (middle-right); {\em Spitzer} 24\micron\ (bottom-left); {\em Spitzer} 70\micron\ (bottom-centre); FUV$+24{\rm \mu m}$--inferred SFRD from \citealp{Leroy2012} (bottom-right). We also show the RC that was isolated by the RC--based masking technique (top-right).}
  \label{figure:haro36Cc_maps}
\end{figure}

\begin{figure}
  \begin{tabular}{ccc}
    \includegraphics[width=0.31\linewidth,clip]{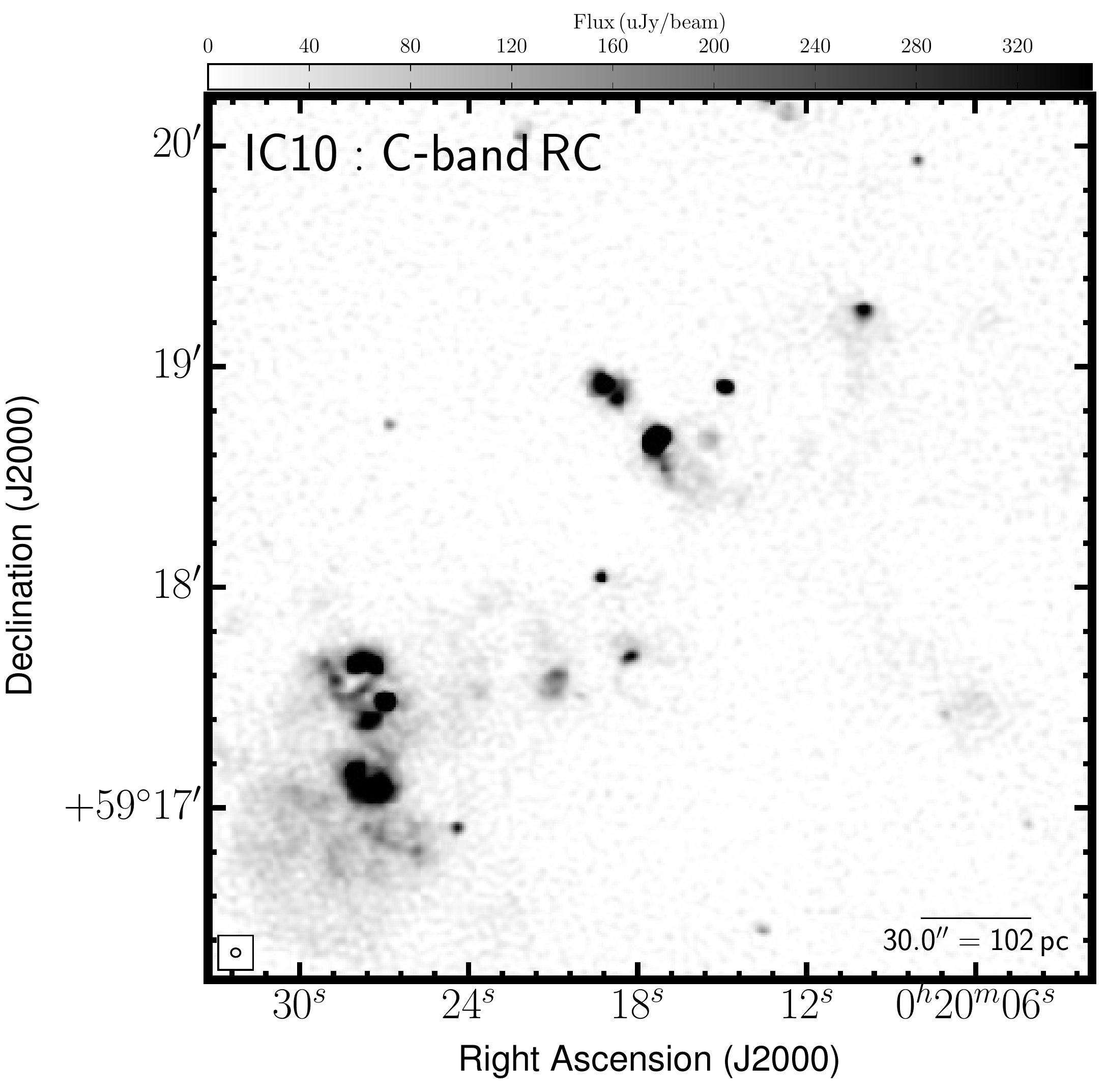} & \ 
    \includegraphics[width=0.31\linewidth,clip]{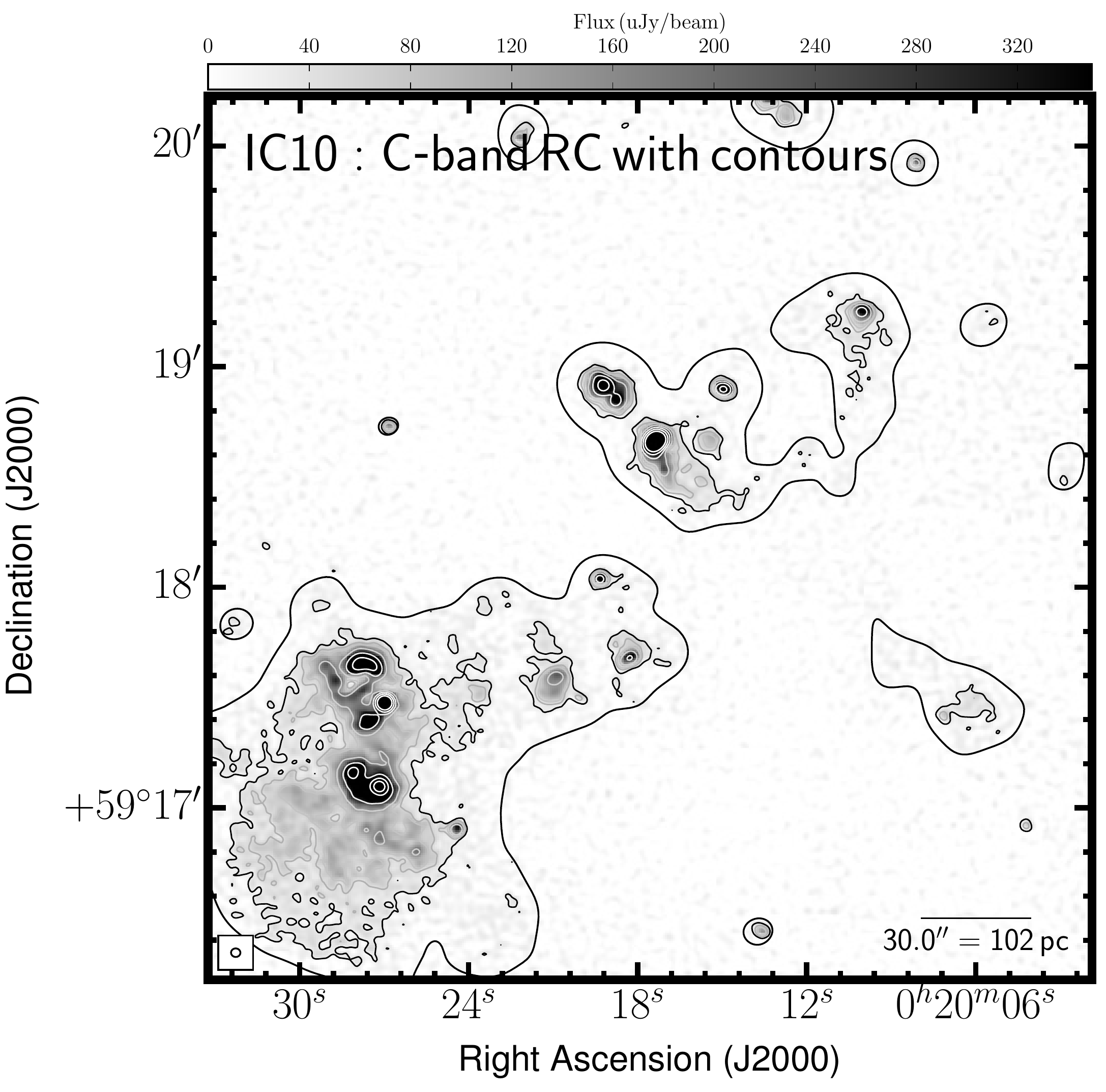} & \ 
    \includegraphics[width=0.31\linewidth,clip]{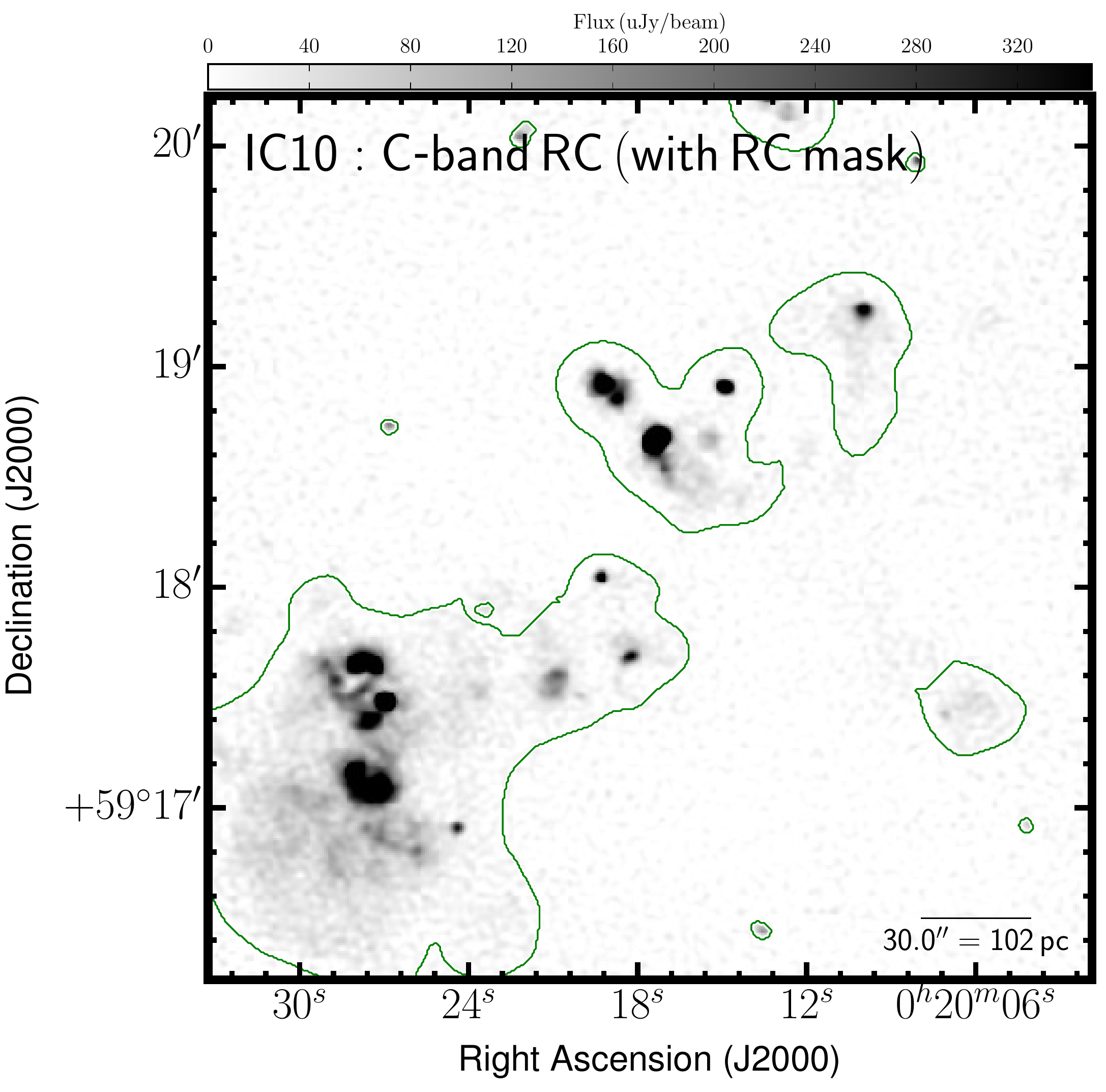} \\
    \includegraphics[width=0.31\linewidth,clip]{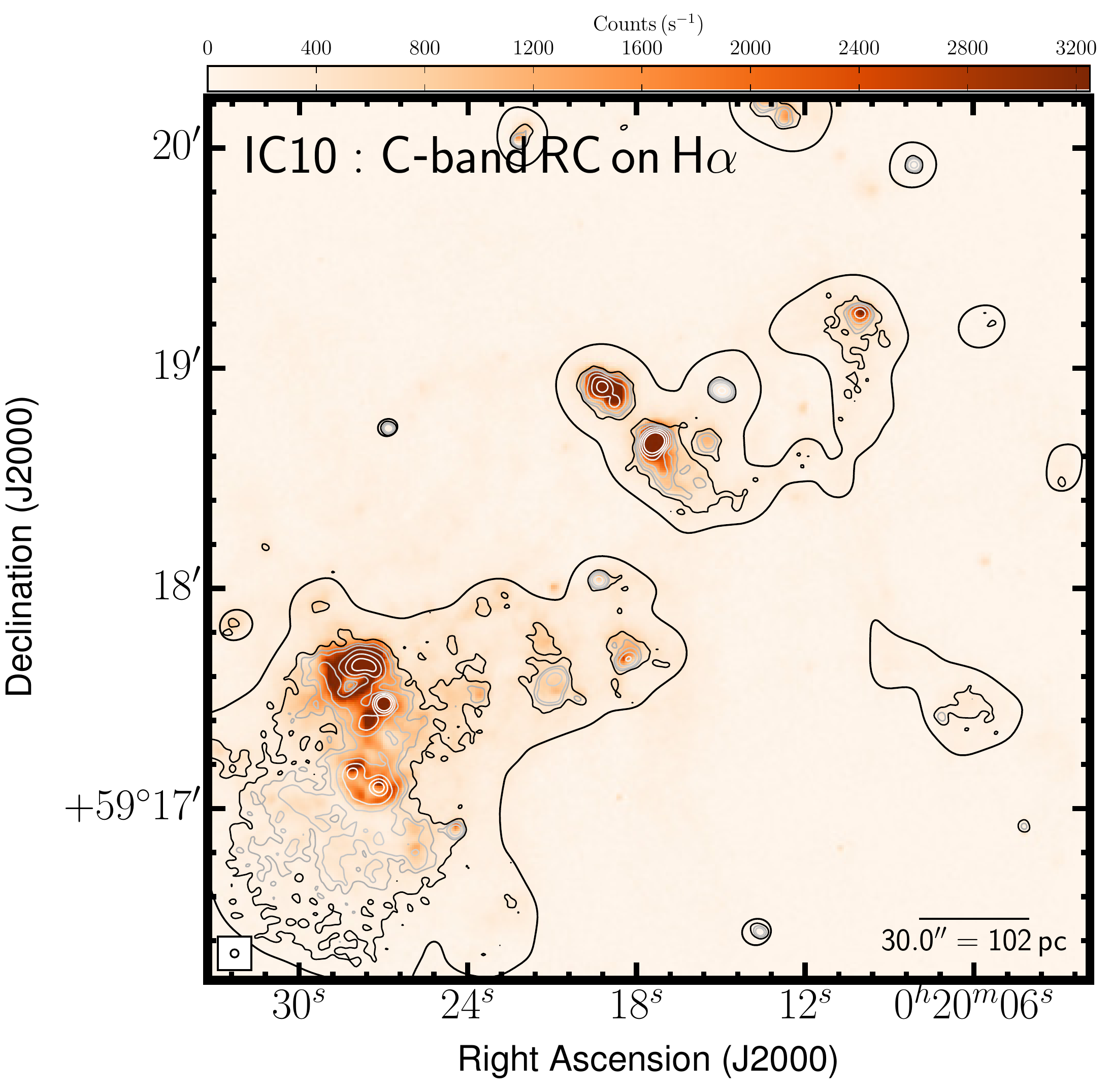} & \ 
    \includegraphics[width=0.31\linewidth,clip]{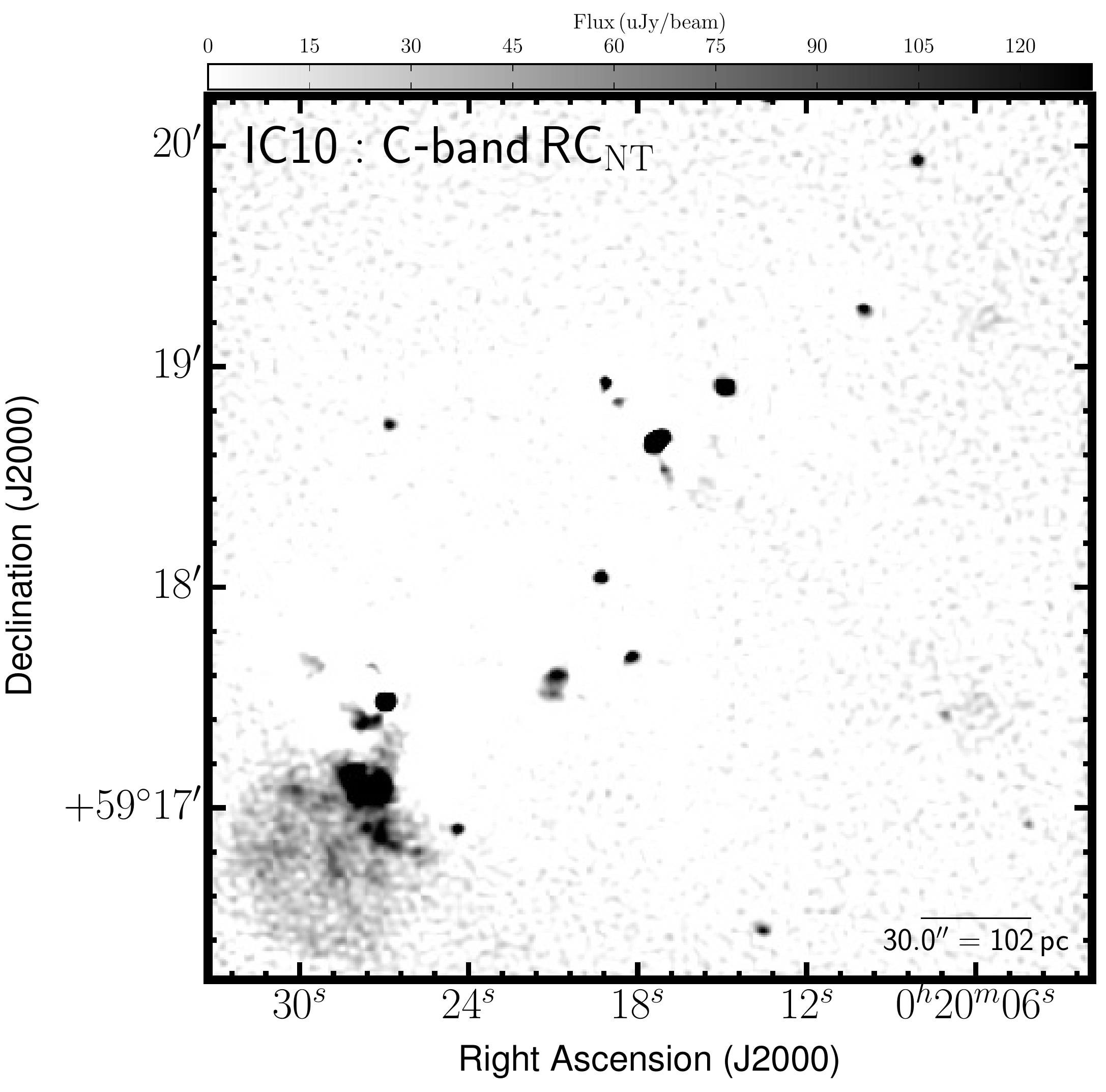} & \ 
    \includegraphics[width=0.31\linewidth,clip]{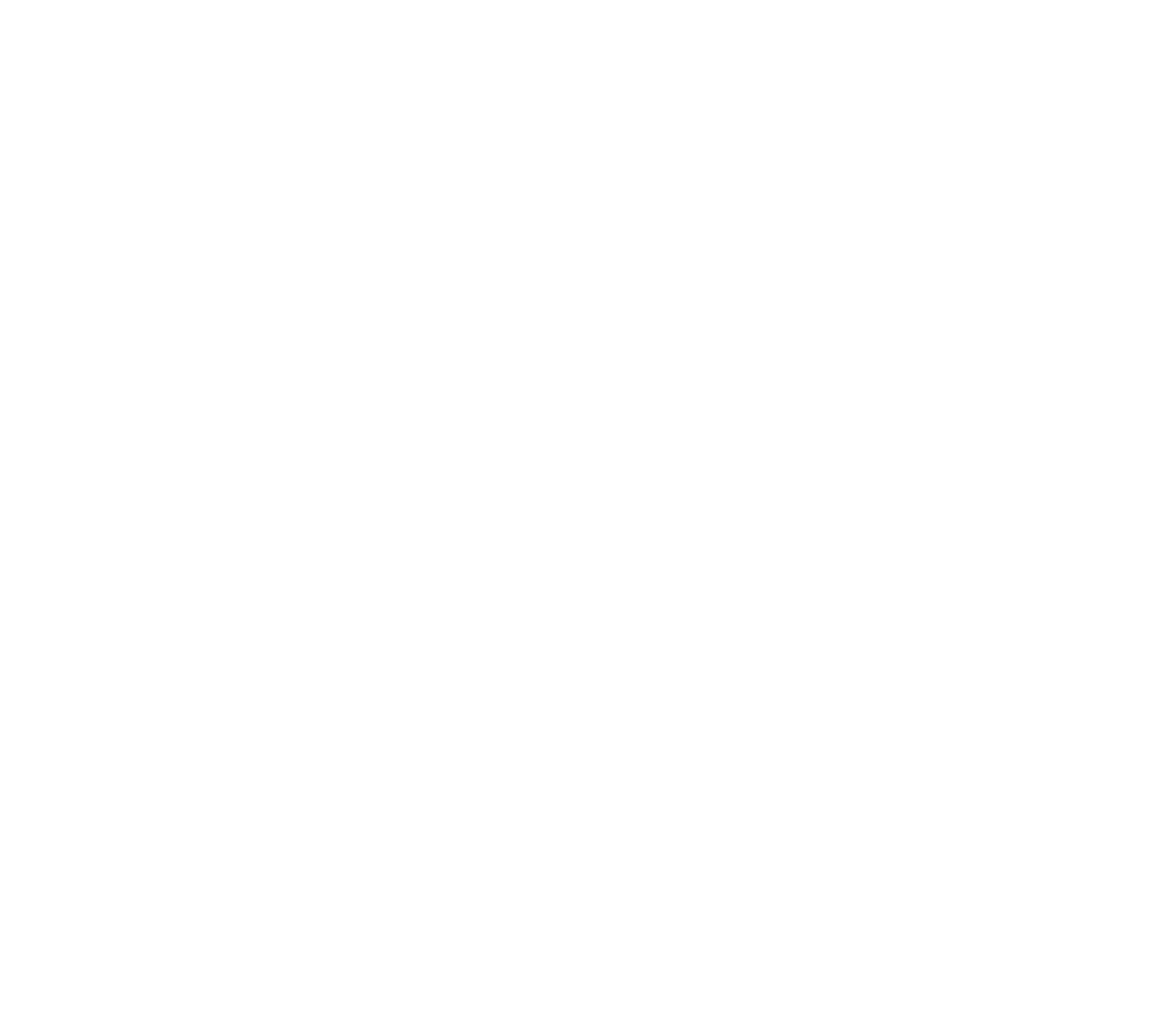} \\
    \includegraphics[width=0.31\linewidth,clip]{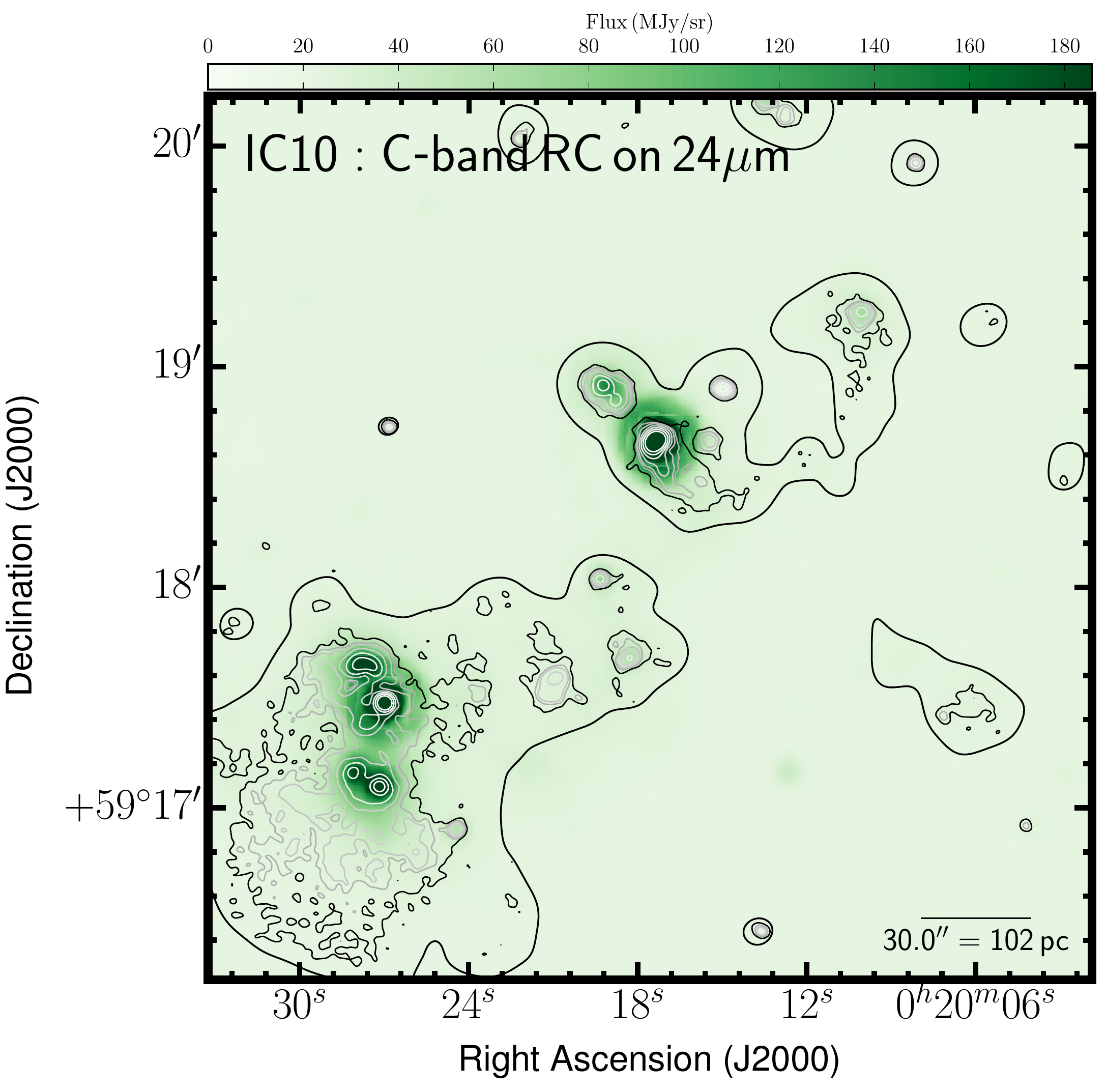} & \ 
    \includegraphics[width=0.31\linewidth,clip]{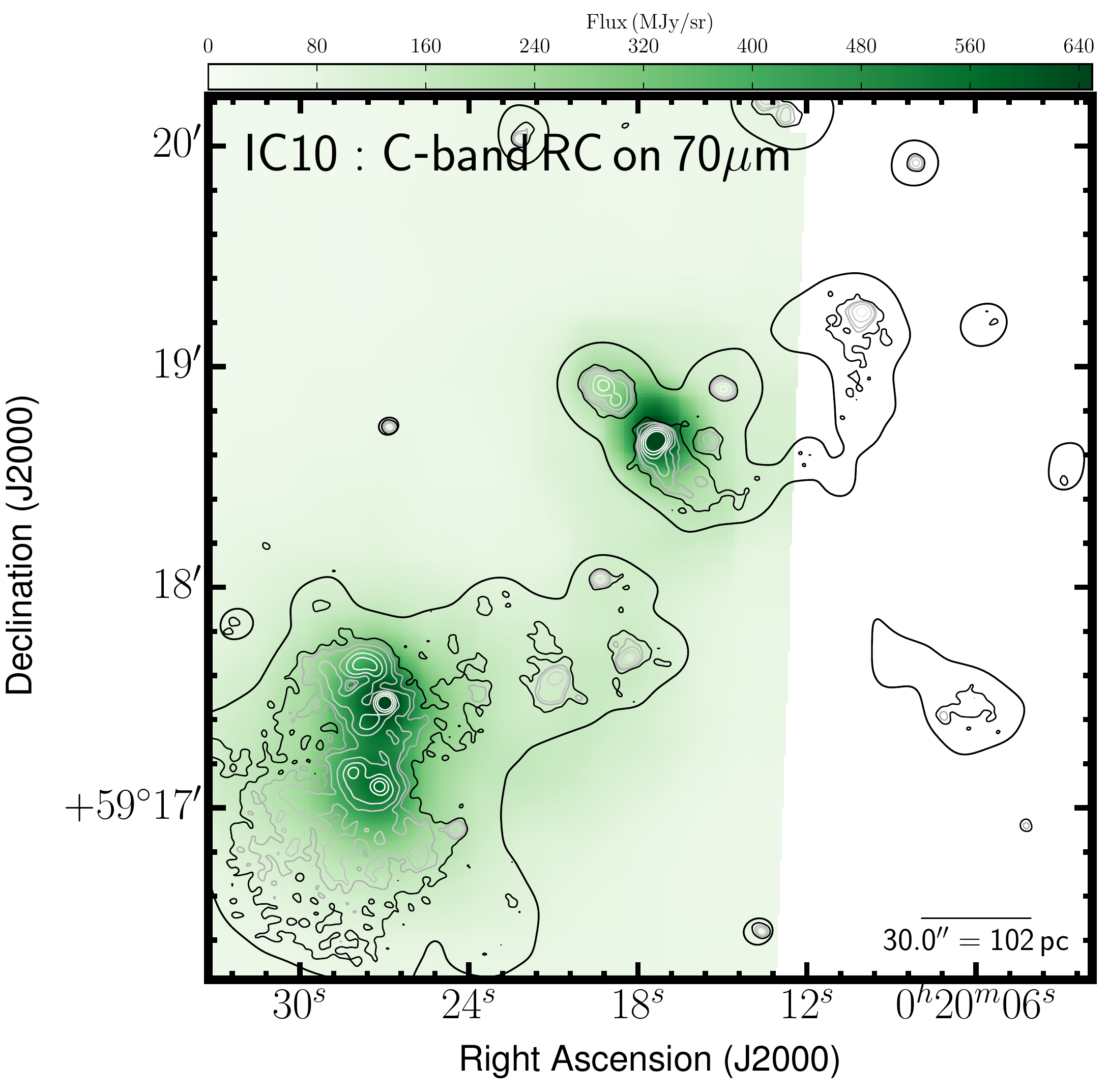} & \ 
    \includegraphics[width=0.31\linewidth,clip]{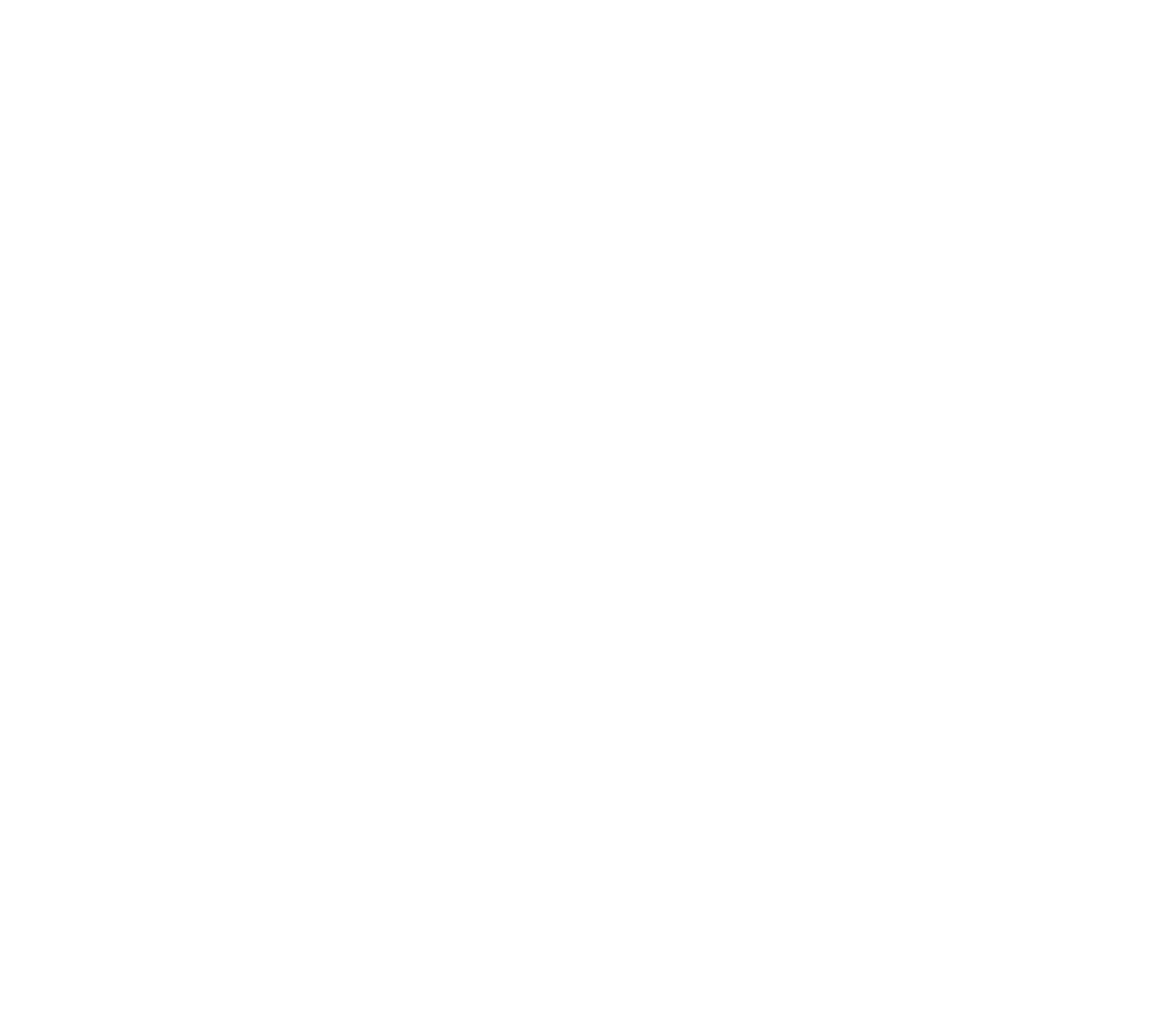} \\
  \end{tabular}
\caption[IC\,10 images: RC, IR, optical, and FUV]{Multi-wavelength coverage of IC 10 displaying a $4.0^\prime \times 4.0^\prime$ area. We show total RC flux density at the native resolution (top-left) and again with contours (top-centre). The RC contours are superposed on ancillary LITTLE THINGS images where possible: \halpha\ (middle-left); \RCNT\ obtained by subtracting the expected \RCT\ based on the \halpha-\RCT\ scaling factor of \cite{Deeg1997} from the total RC; {\em GALEX} FUV (middle-right); {\em Spitzer} 24\micron\ (bottom-left); {\em Spitzer} 70\micron\ (bottom-centre); FUV$+24{\rm \mu m}$--inferred SFRD from \citealp{Leroy2012} (bottom-right). We also show the RC that was isolated by the RC--based masking technique (top-right).}
  \label{figure:ic10Cc_maps}
\end{figure}

\begin{figure}
  \begin{tabular}{ccc}
    \includegraphics[width=0.31\linewidth,clip]{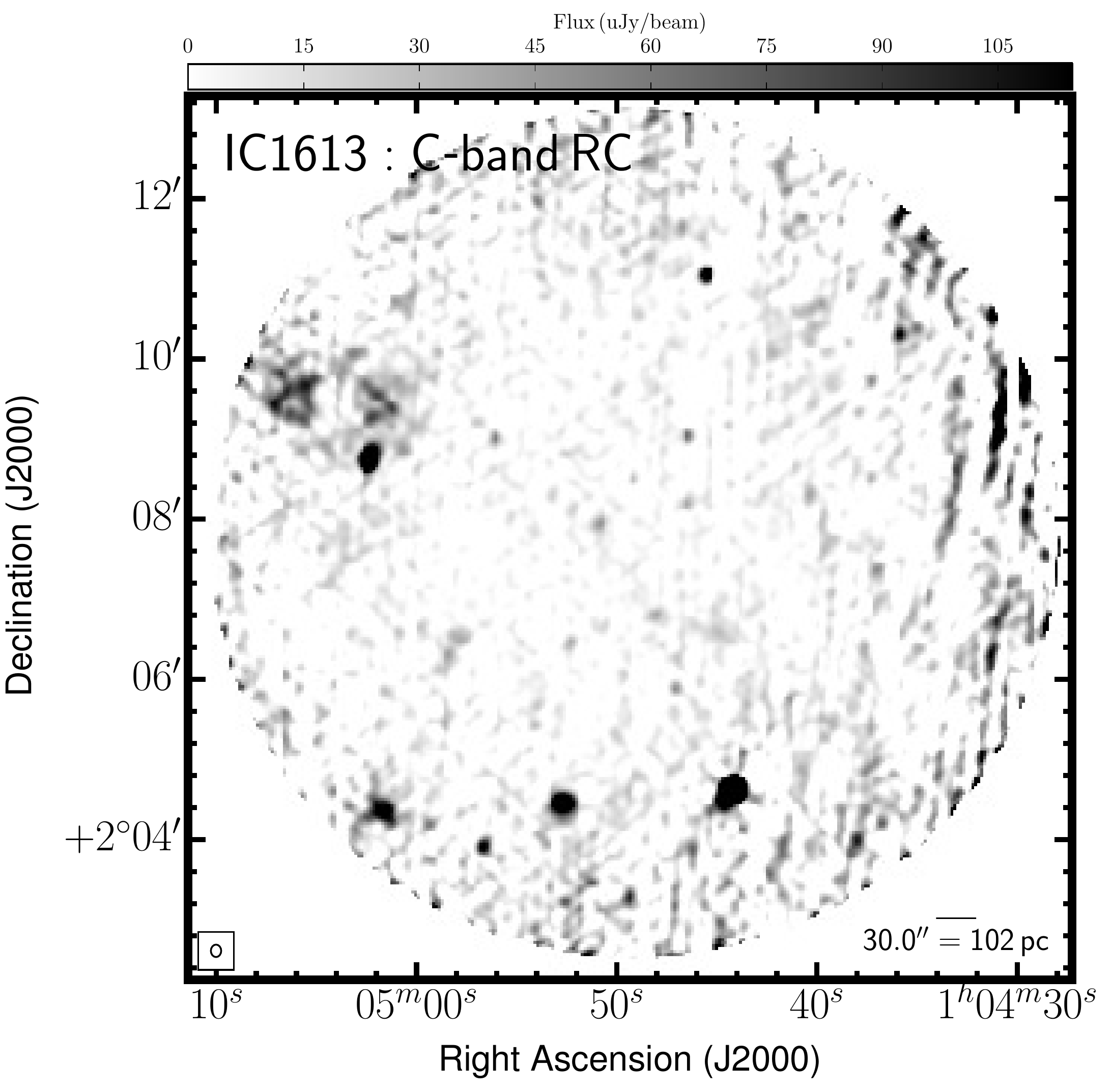} & \ 
    \includegraphics[width=0.31\linewidth,clip]{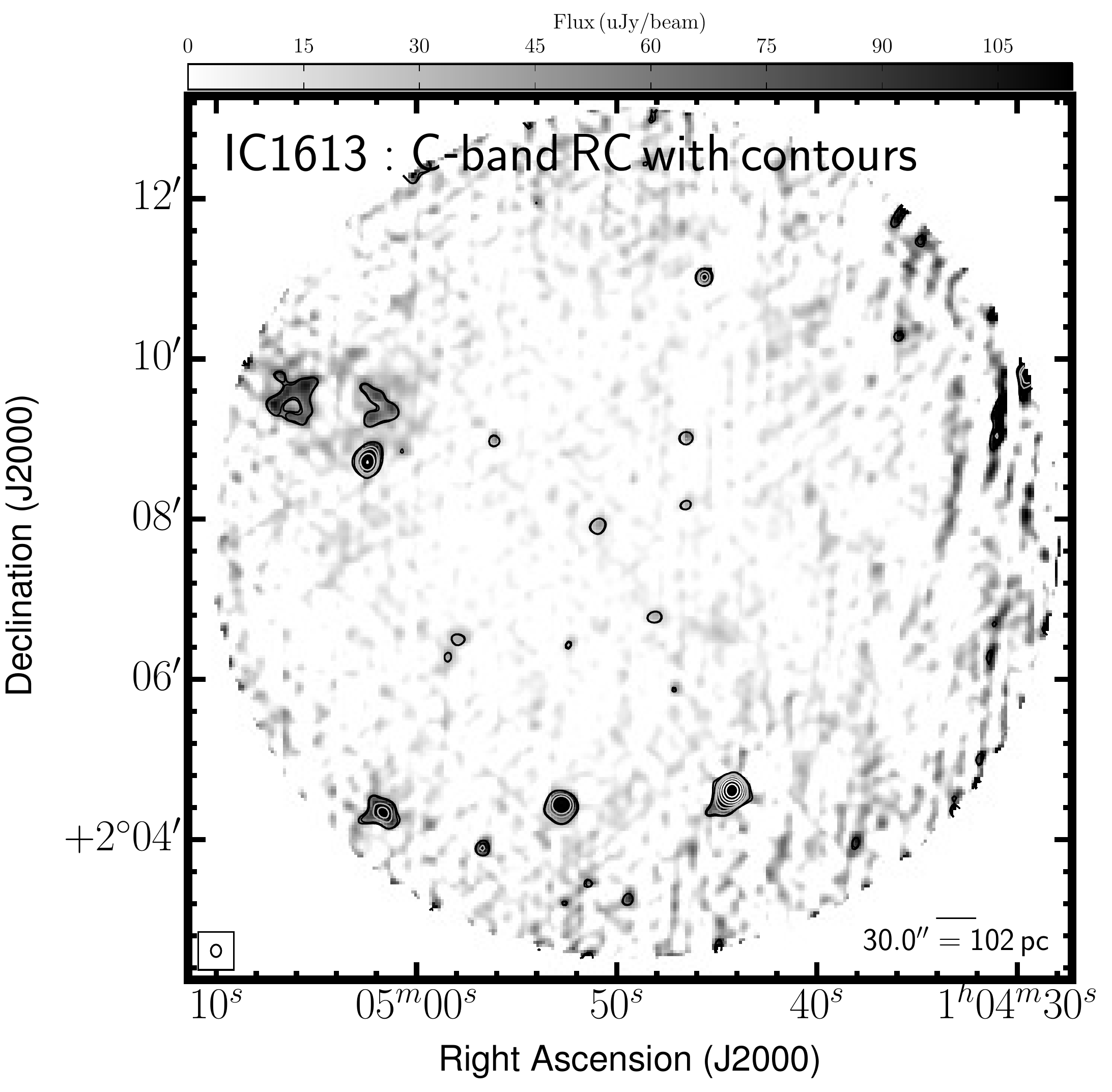} & \ 
    \includegraphics[width=0.31\linewidth,clip]{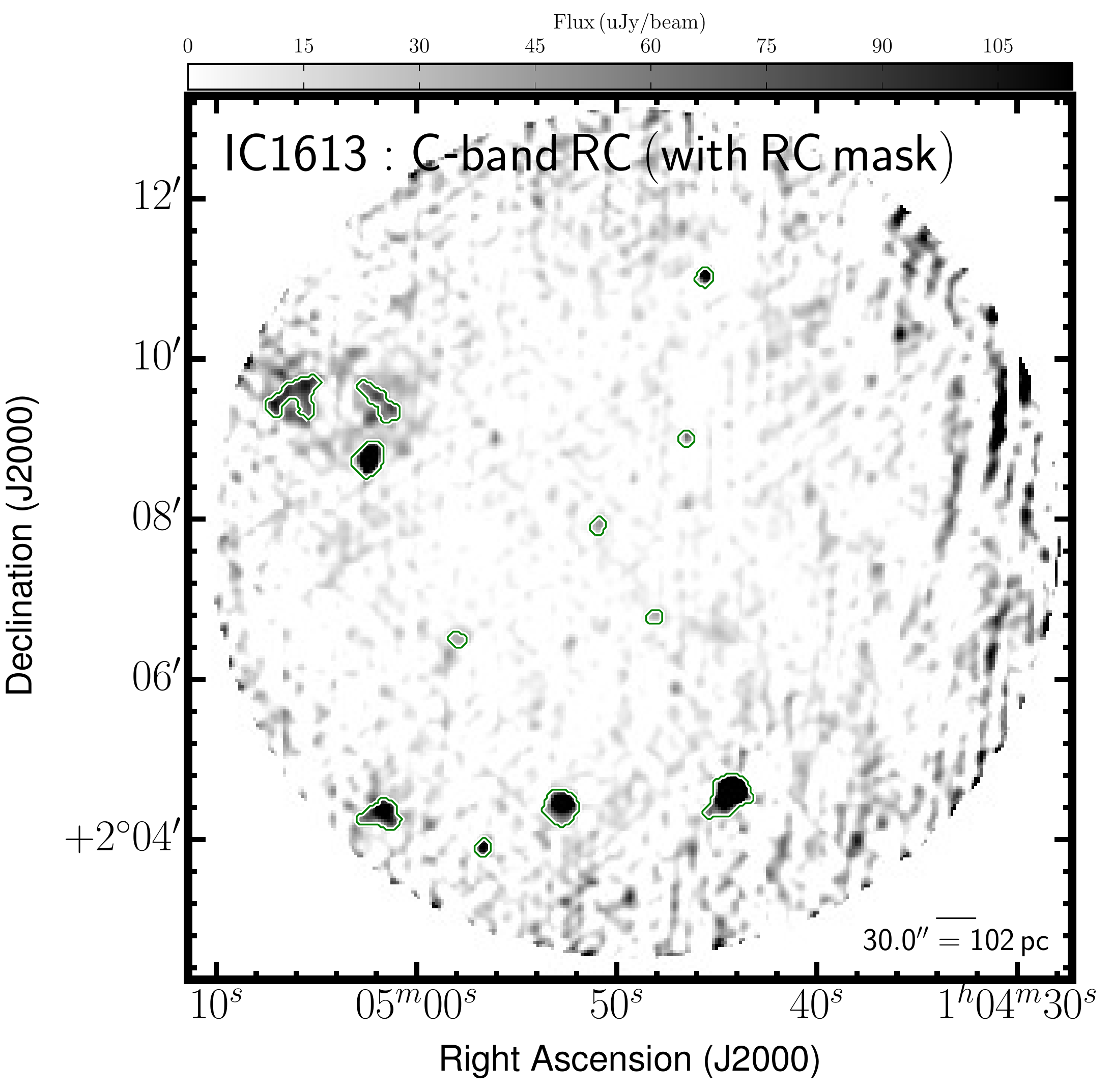} \\
    \includegraphics[width=0.31\linewidth,clip]{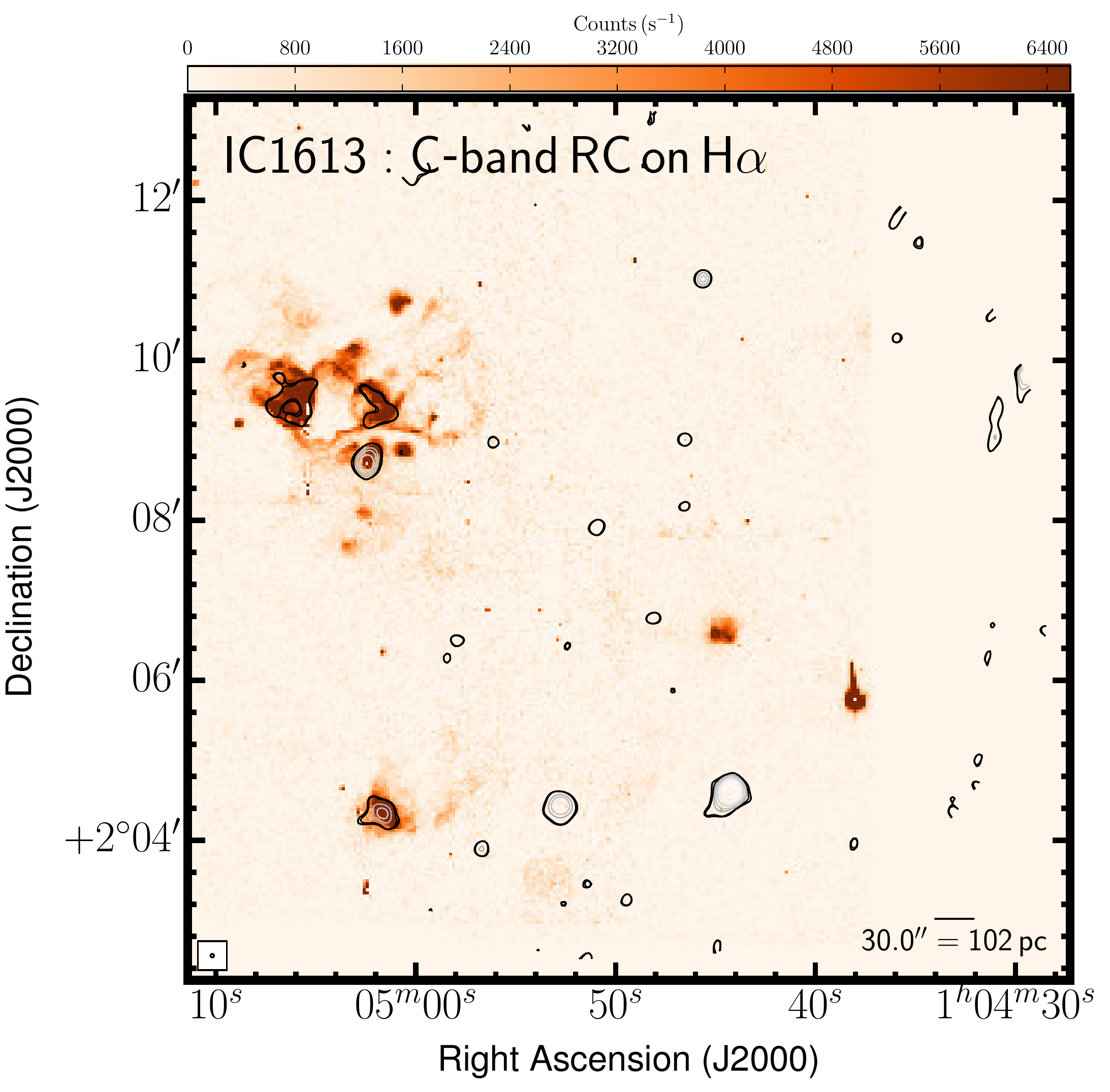} & \ 
    \includegraphics[width=0.31\linewidth,clip]{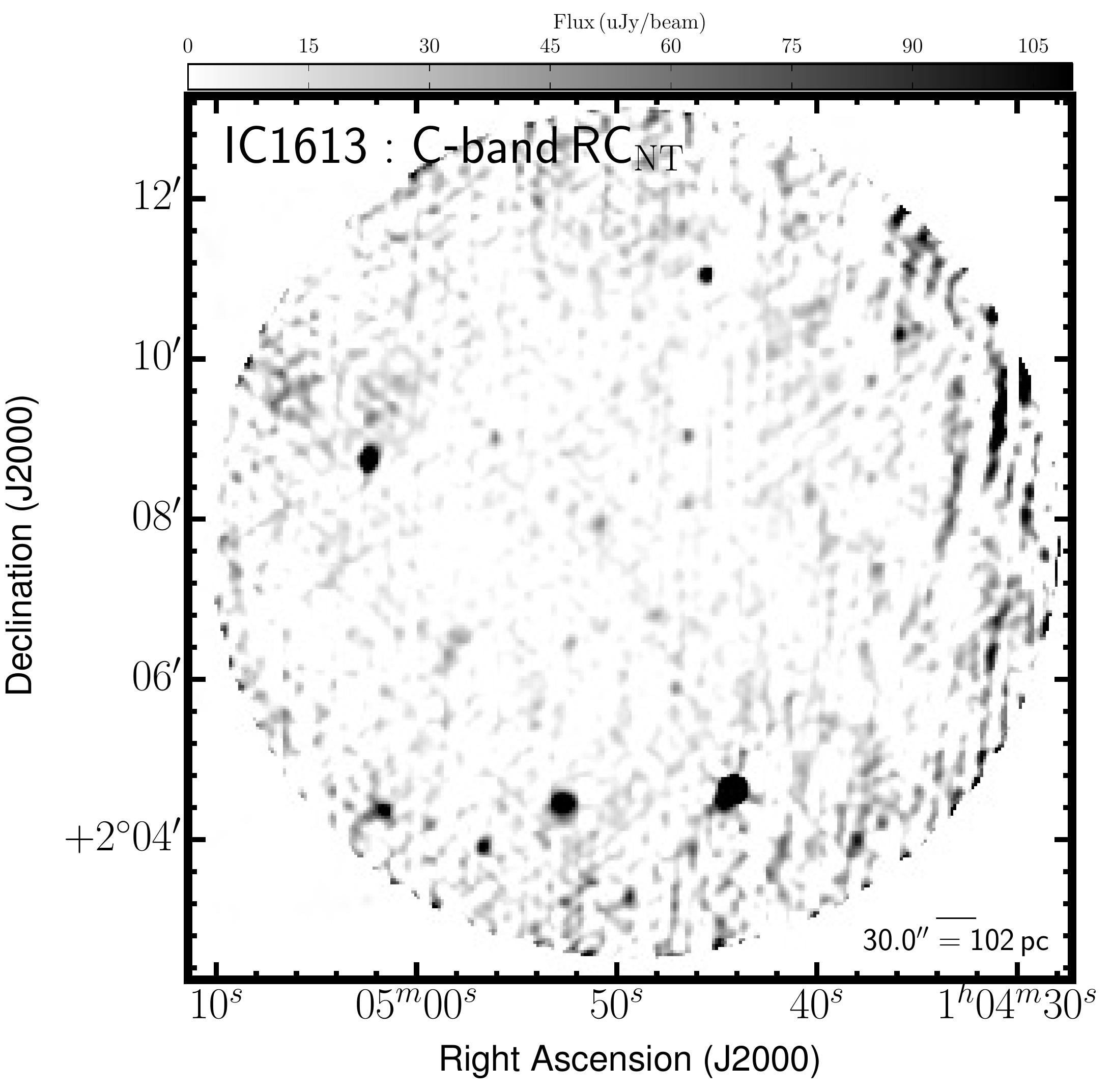} & \ 
    \includegraphics[width=0.31\linewidth,clip]{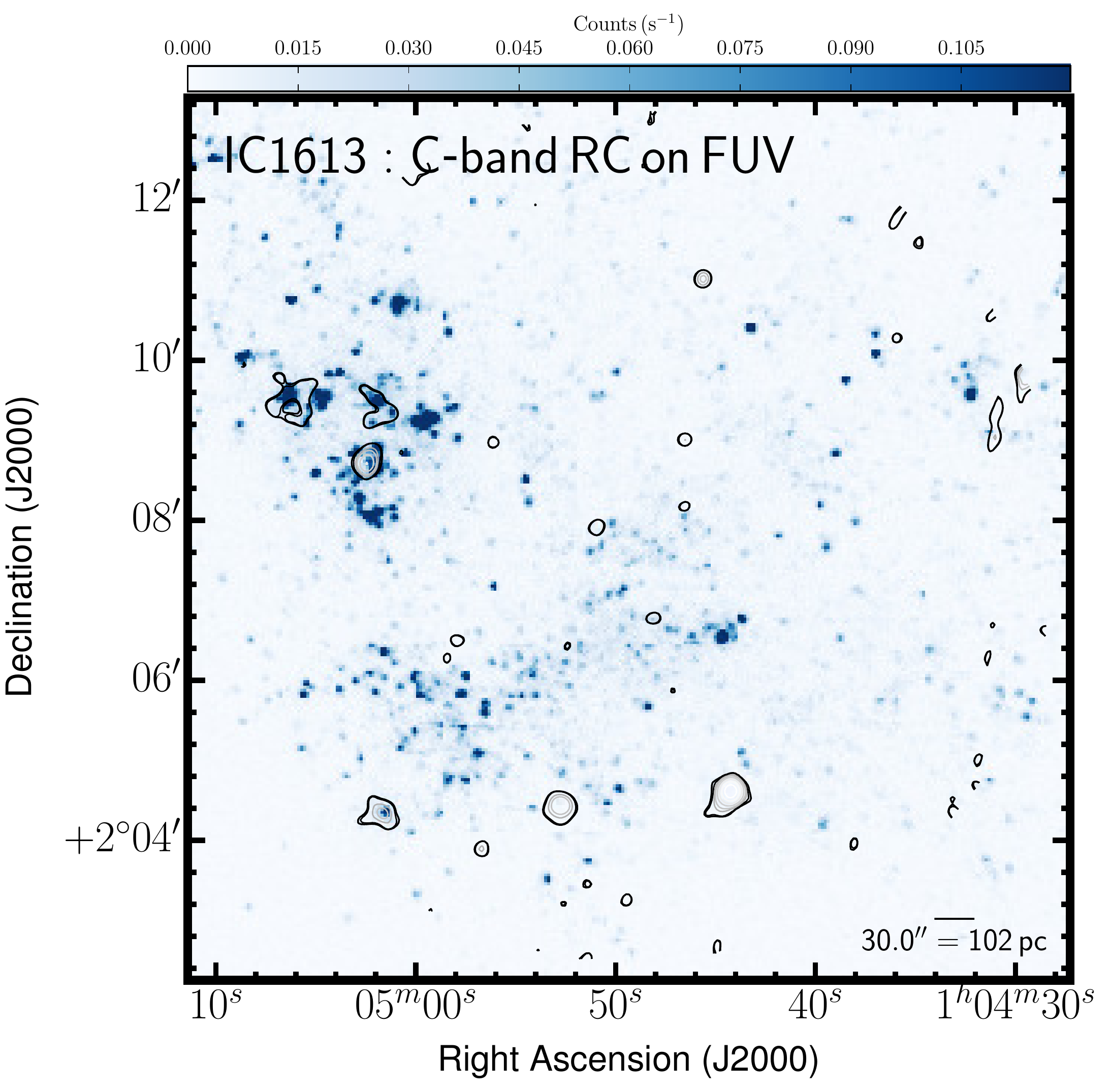} \\
    \includegraphics[width=0.31\linewidth,clip]{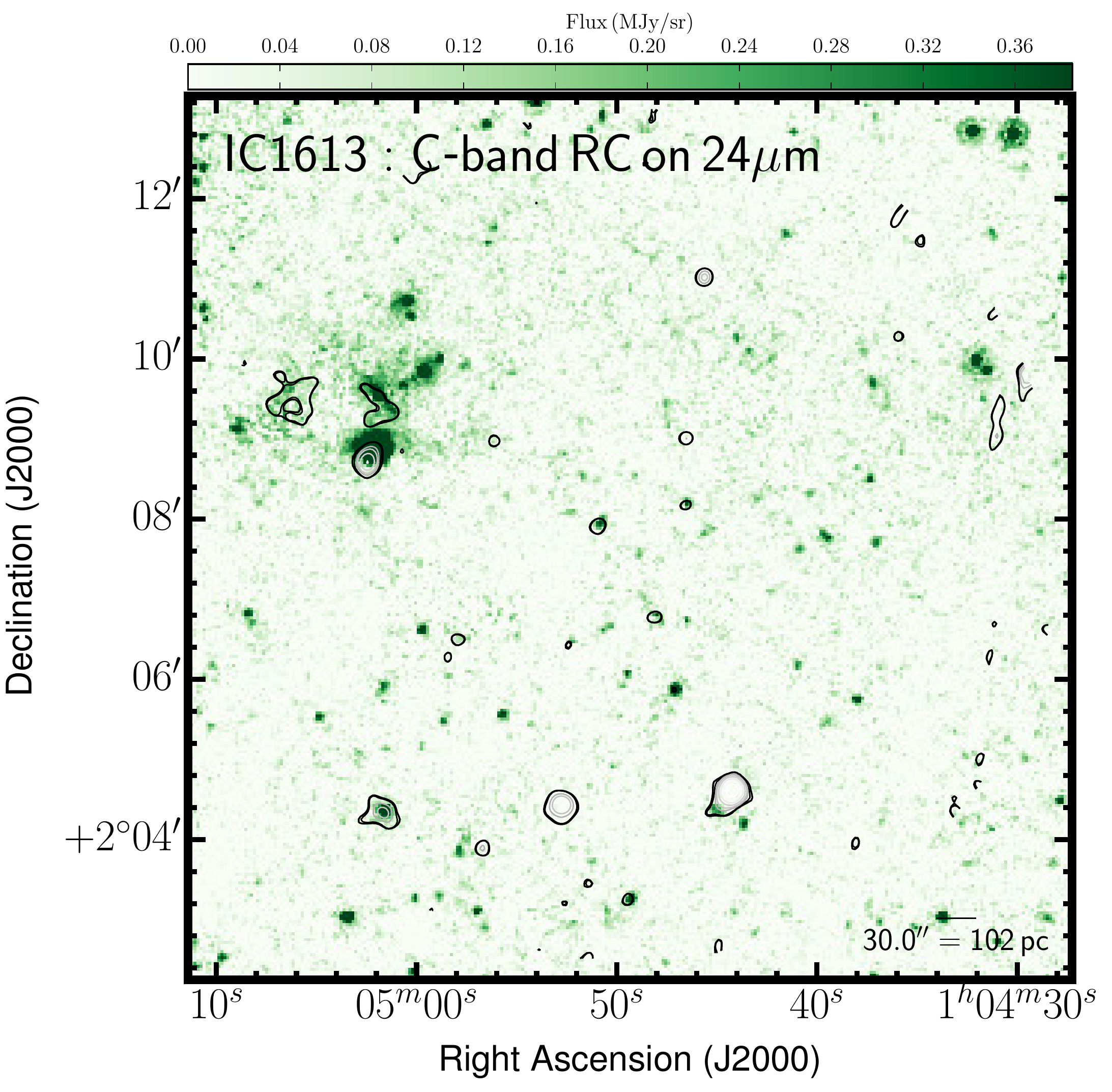} & \ 
    \includegraphics[width=0.31\linewidth,clip]{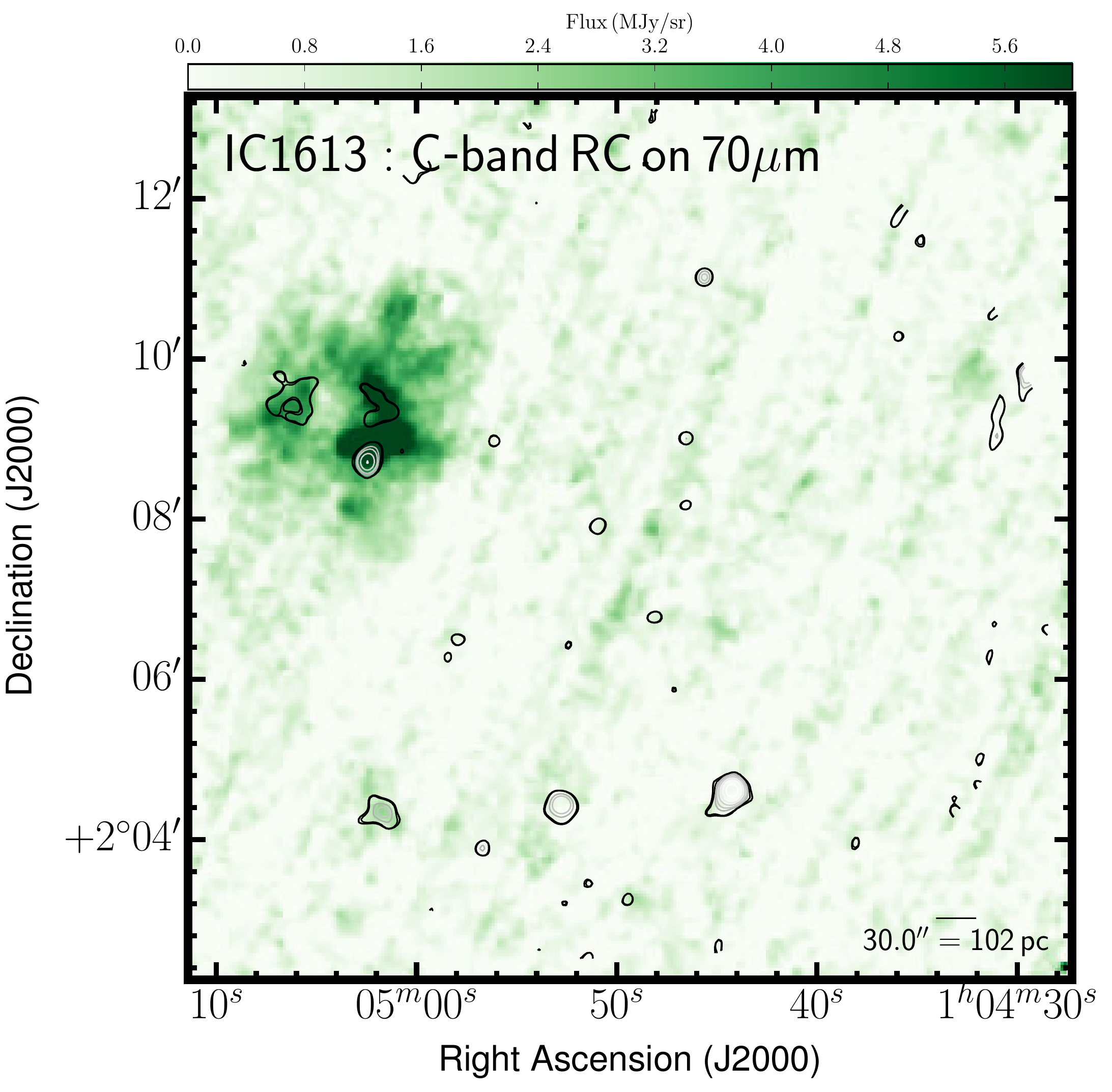} & \ 
    \includegraphics[width=0.31\linewidth,clip]{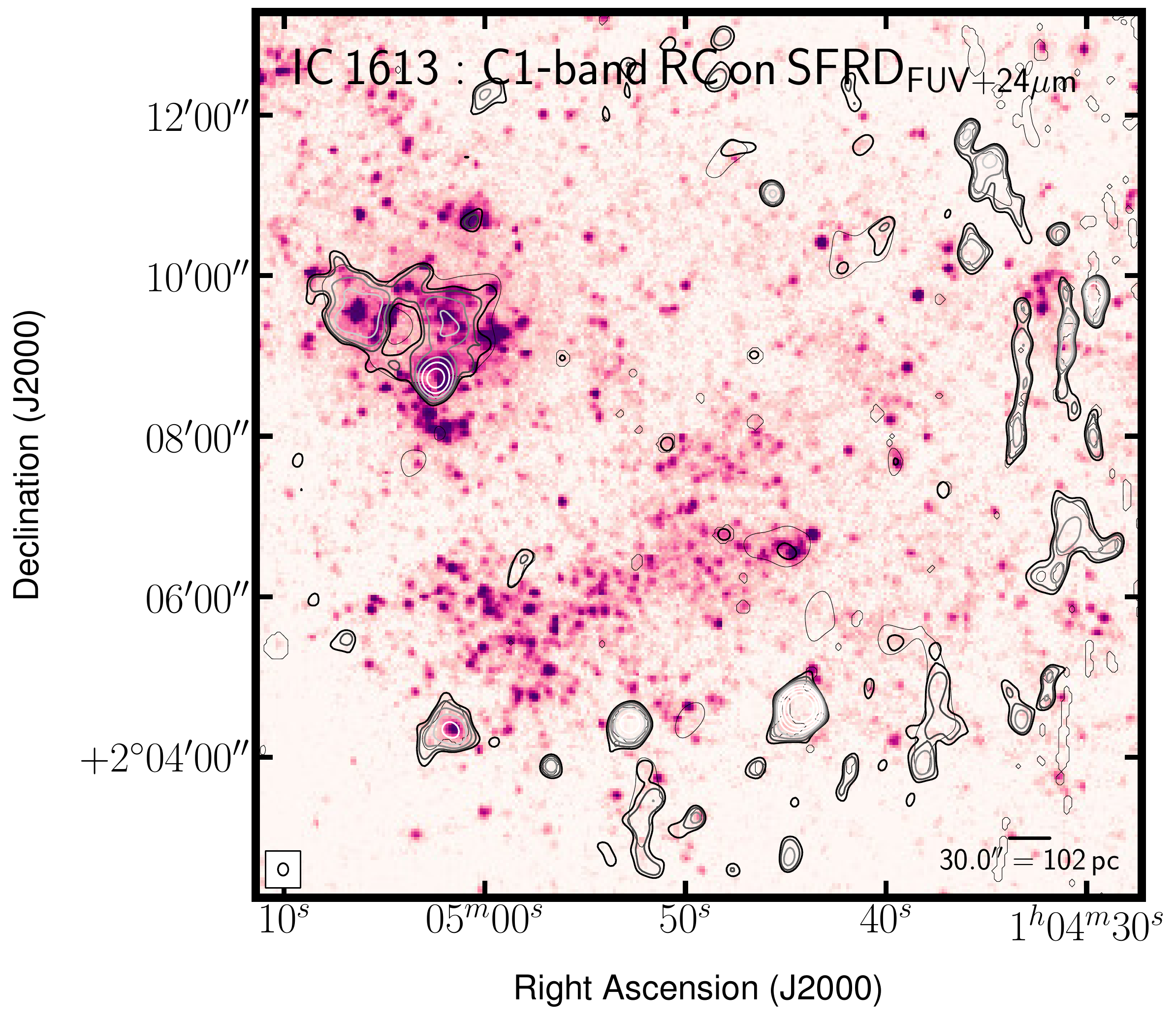} \\
  \end{tabular}
\caption[IC\,1613 images: RC, IR, optical, and FUV]{Multi-wavelength coverage of IC 1613 displaying a $11.1^\prime \times 11.1^\prime$ area. We show total RC flux density at the native resolution (top-left) and again with contours (top-centre). The RC contours are superposed on ancillary LITTLE THINGS images where possible: \halpha\ (middle-left); \RCNT\ obtained by subtracting the expected \RCT\ based on the \halpha-\RCT\ scaling factor of \cite{Deeg1997} from the total RC; {\em GALEX} FUV (middle-right); {\em Spitzer} 24\micron\ (bottom-left); {\em Spitzer} 70\micron\ (bottom-centre); FUV$+24{\rm \mu m}$--inferred SFRD from \citealp{Leroy2012} (bottom-right). We also show the RC that was isolated by the RC--based masking technique (top-right). }
  \label{figure:ic1613C1d1_maps}
\end{figure}

\begin{figure}
  \begin{tabular}{ccc}
    \includegraphics[width=0.31\linewidth,clip]{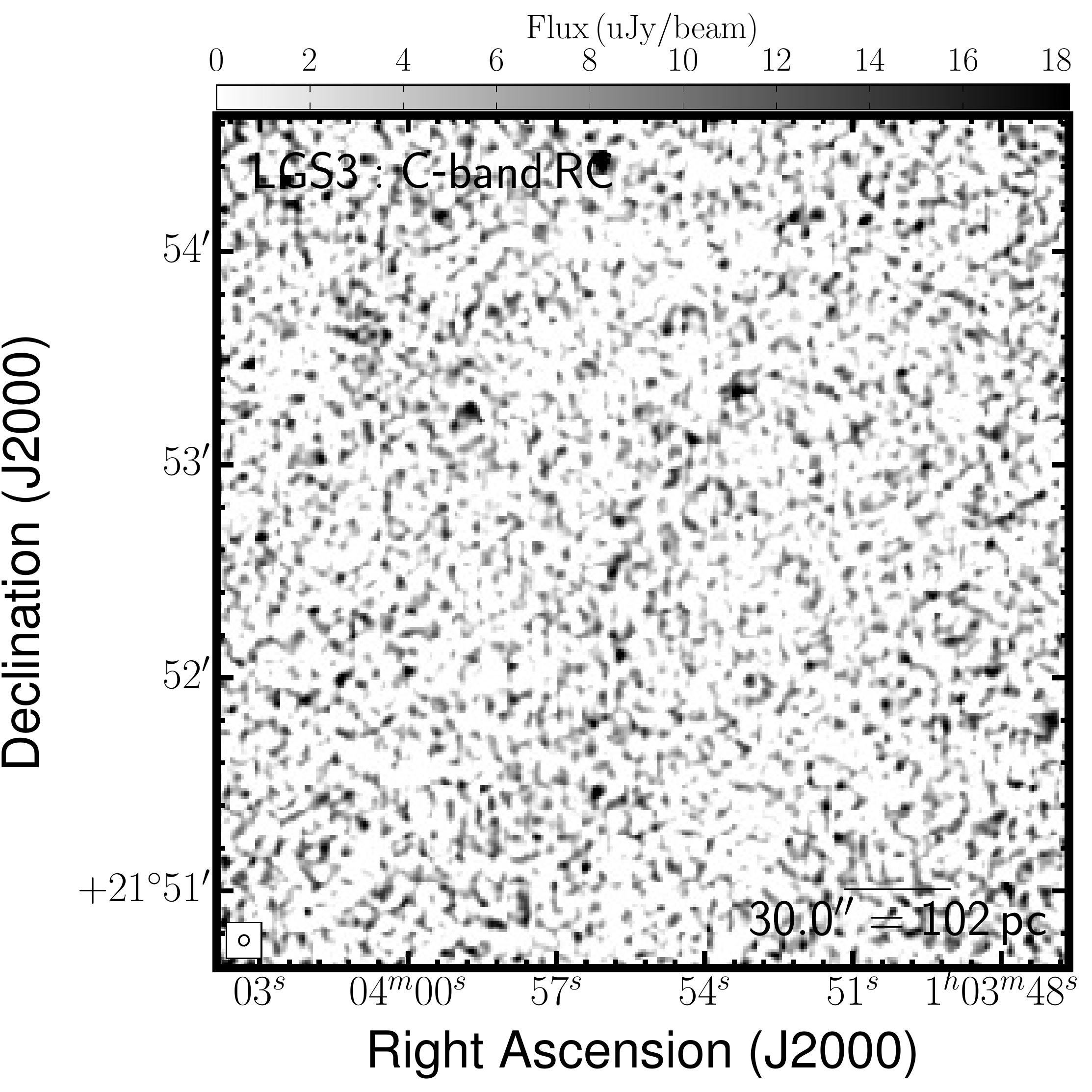} & \ 
    \includegraphics[width=0.31\linewidth,clip]{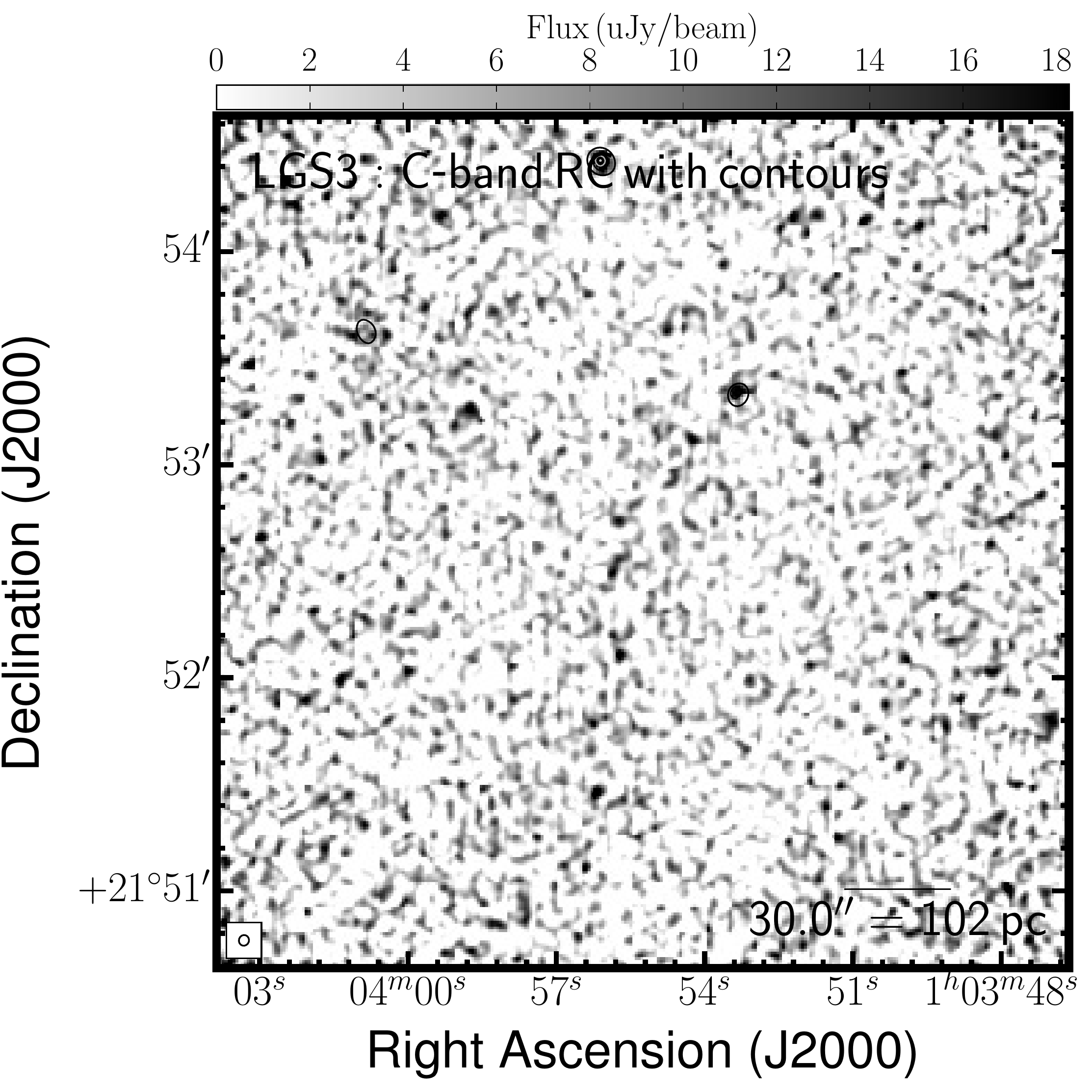} & \ 
    \includegraphics[width=0.31\linewidth,clip]{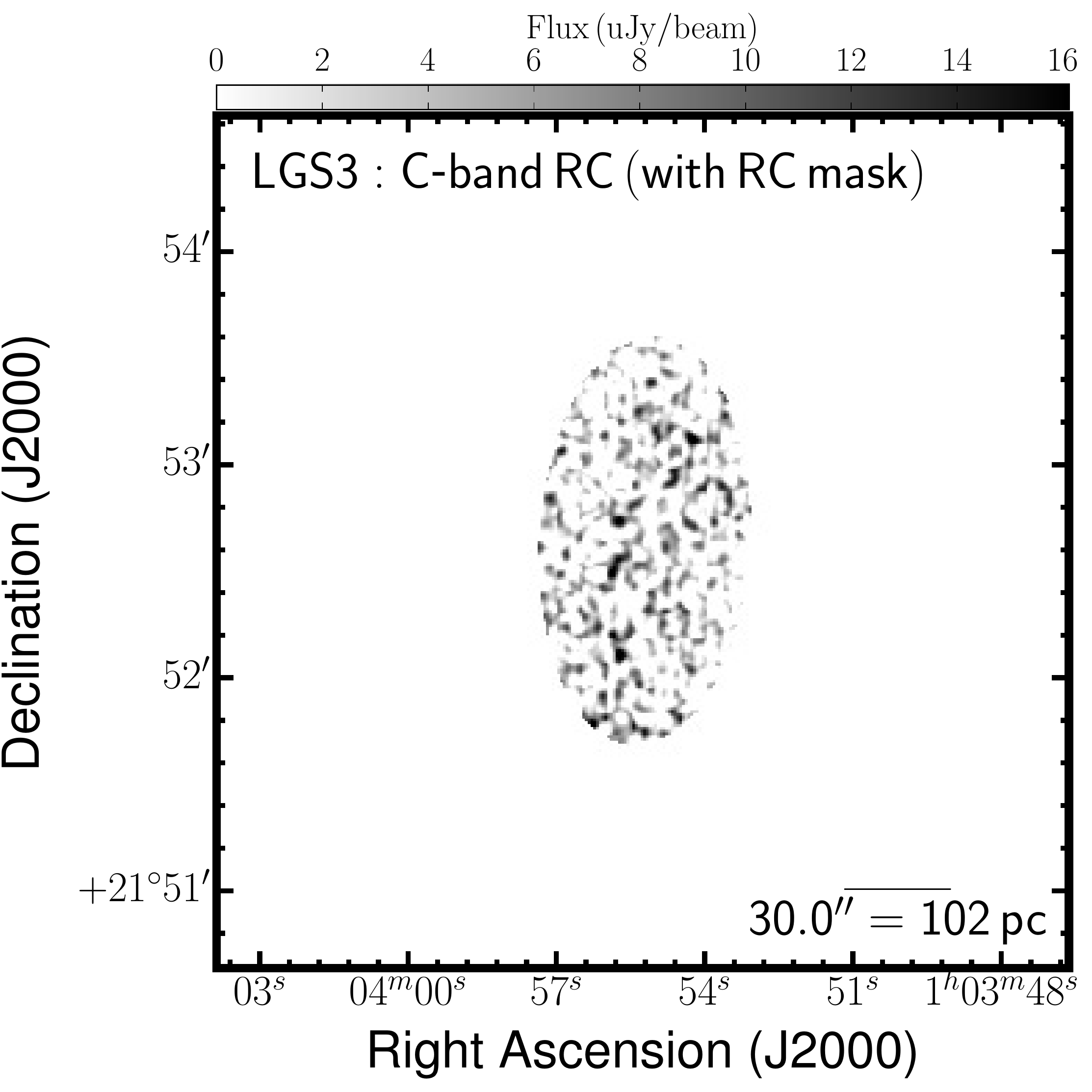} \\
    \includegraphics[width=0.31\linewidth,clip]{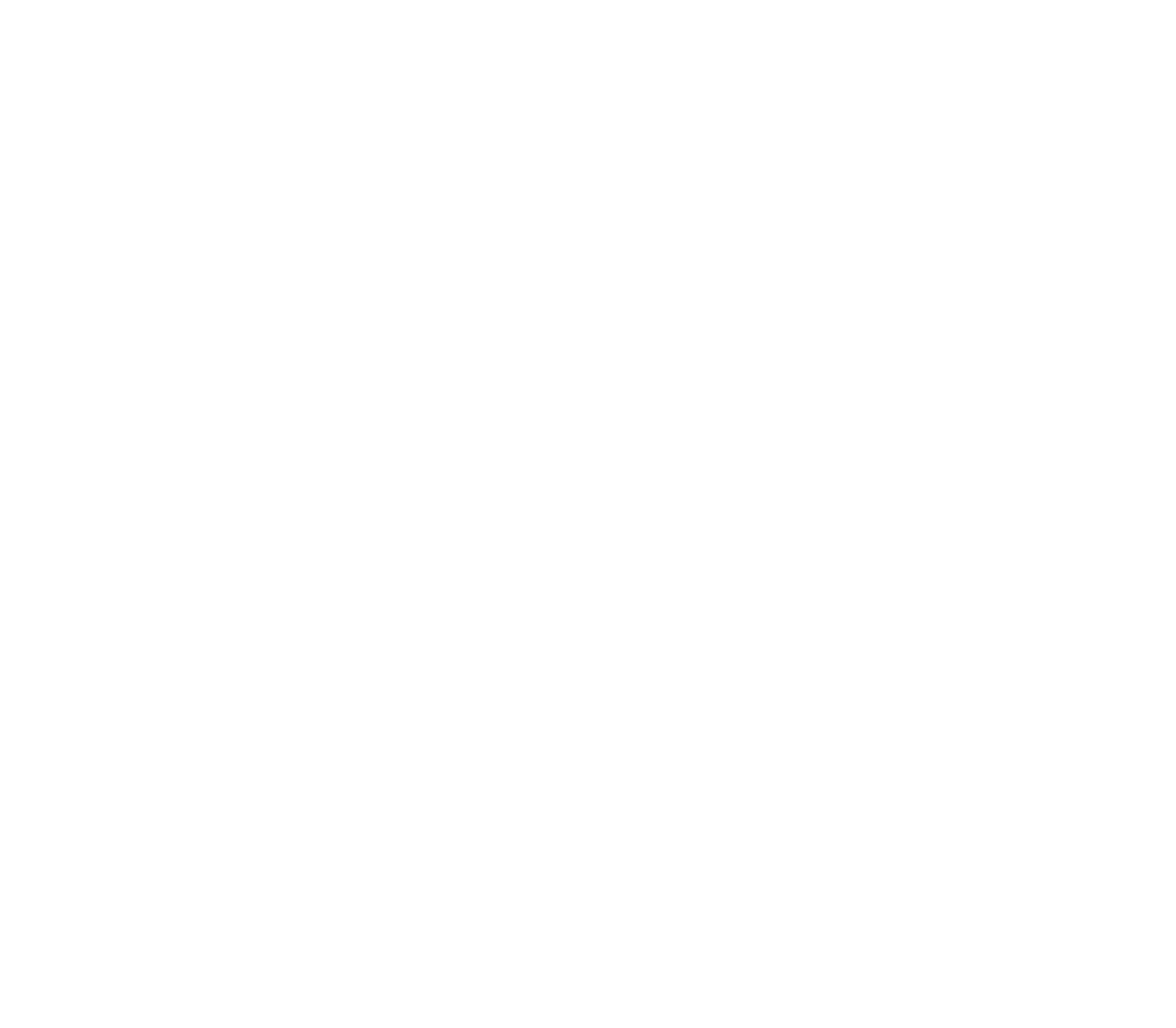} & \ 
    \includegraphics[width=0.31\linewidth,clip]{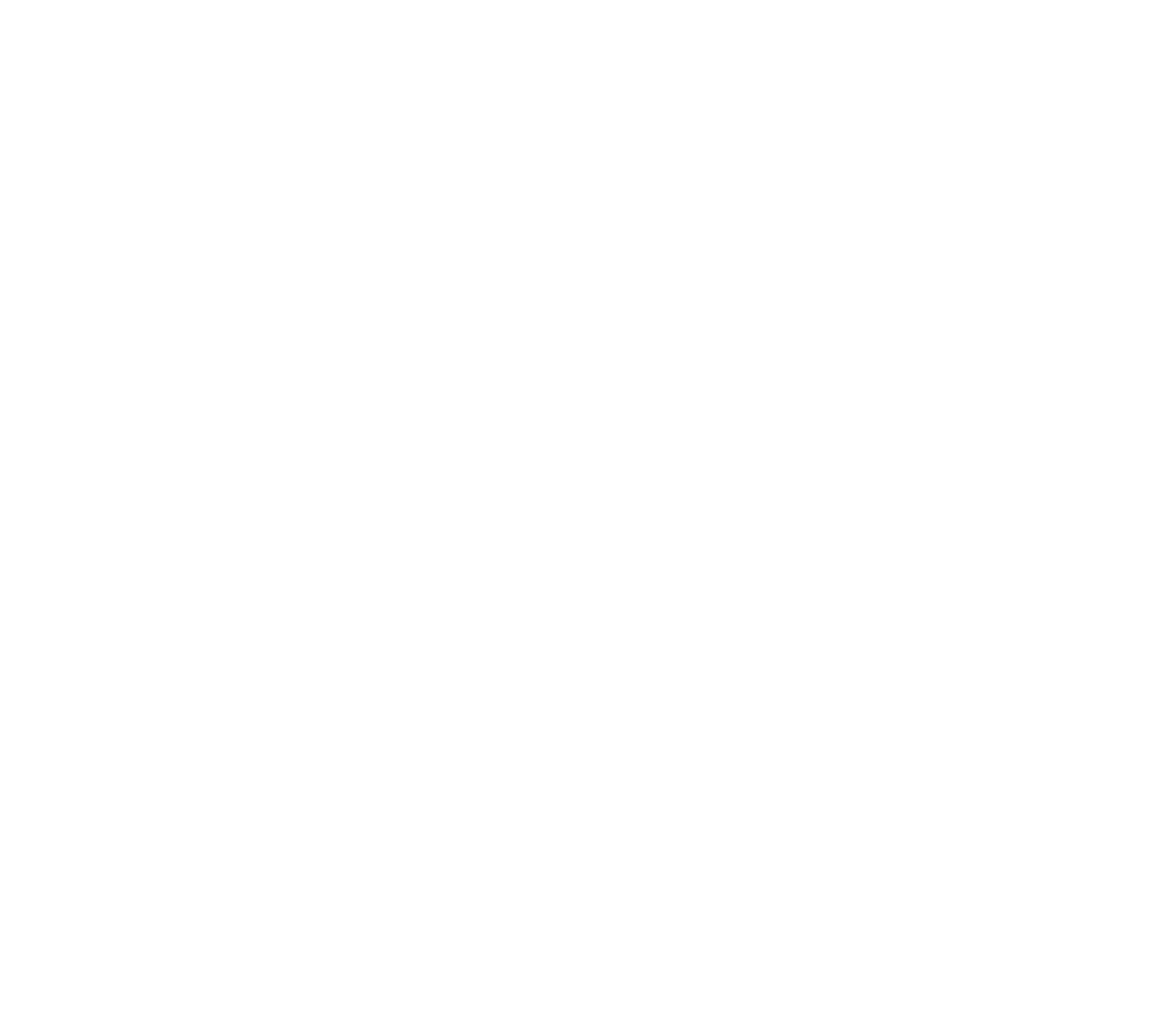} & \ 
    \includegraphics[width=0.31\linewidth,clip]{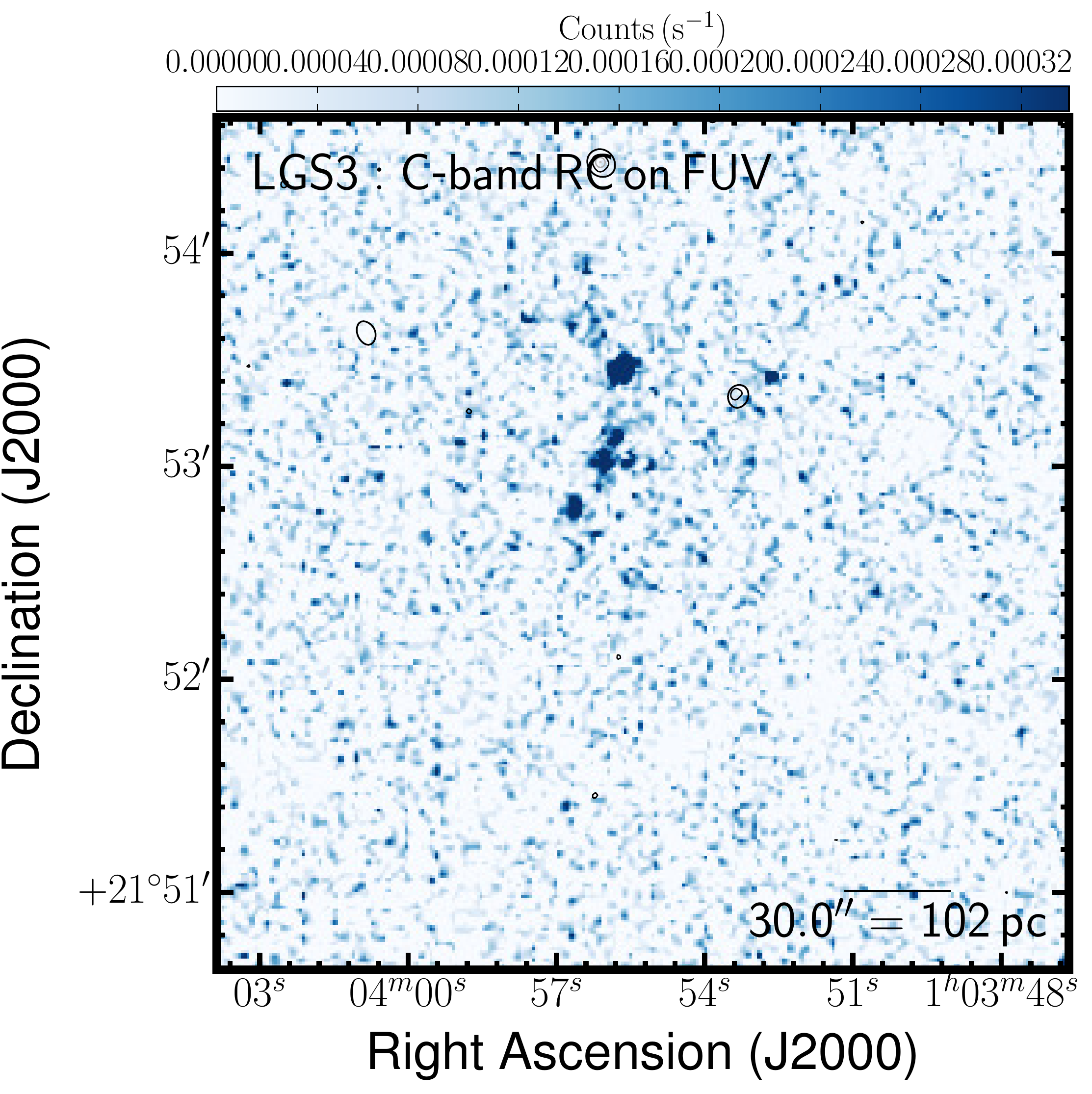} \\
    \includegraphics[width=0.31\linewidth,clip]{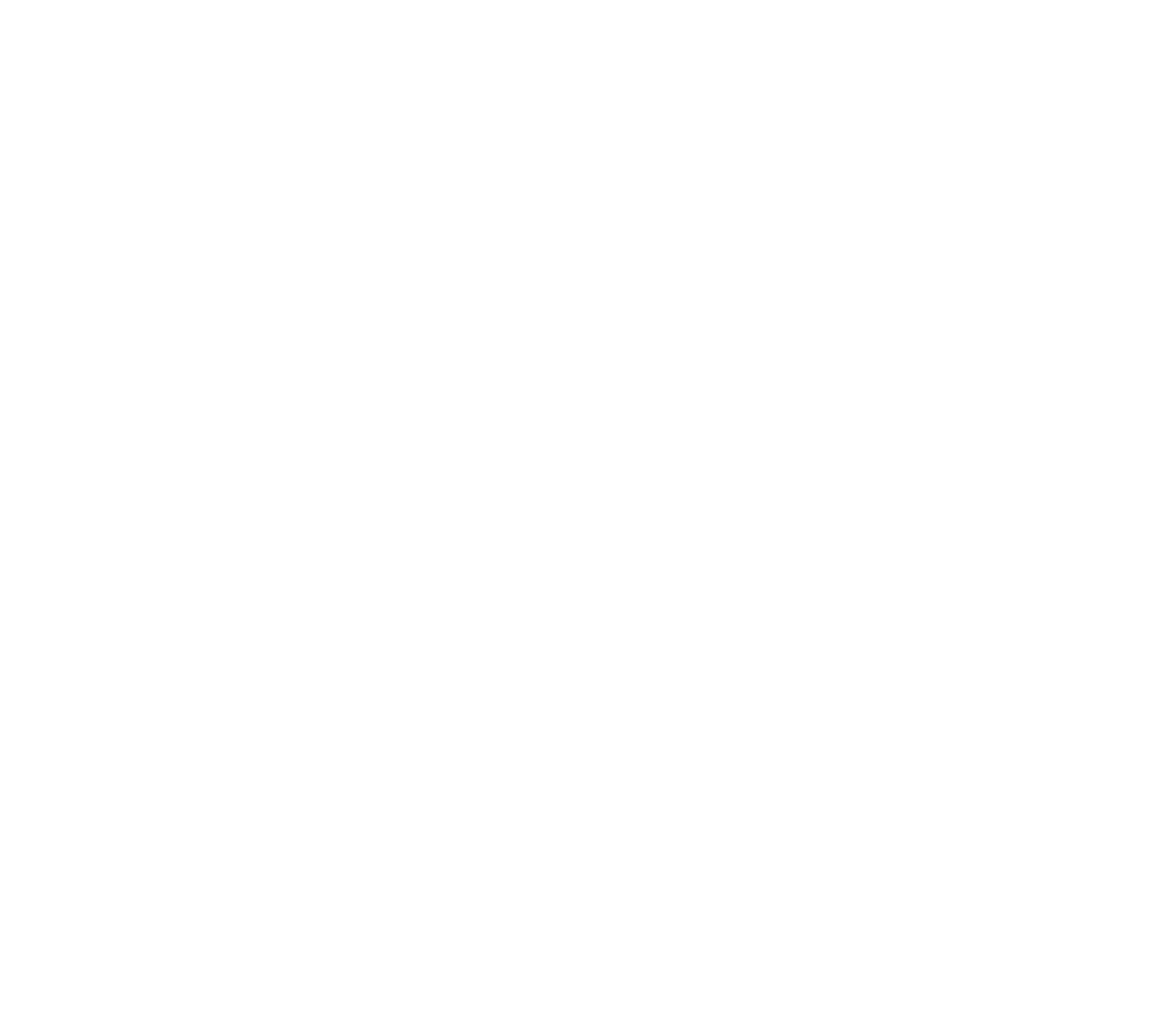} & \ 
    \includegraphics[width=0.31\linewidth,clip]{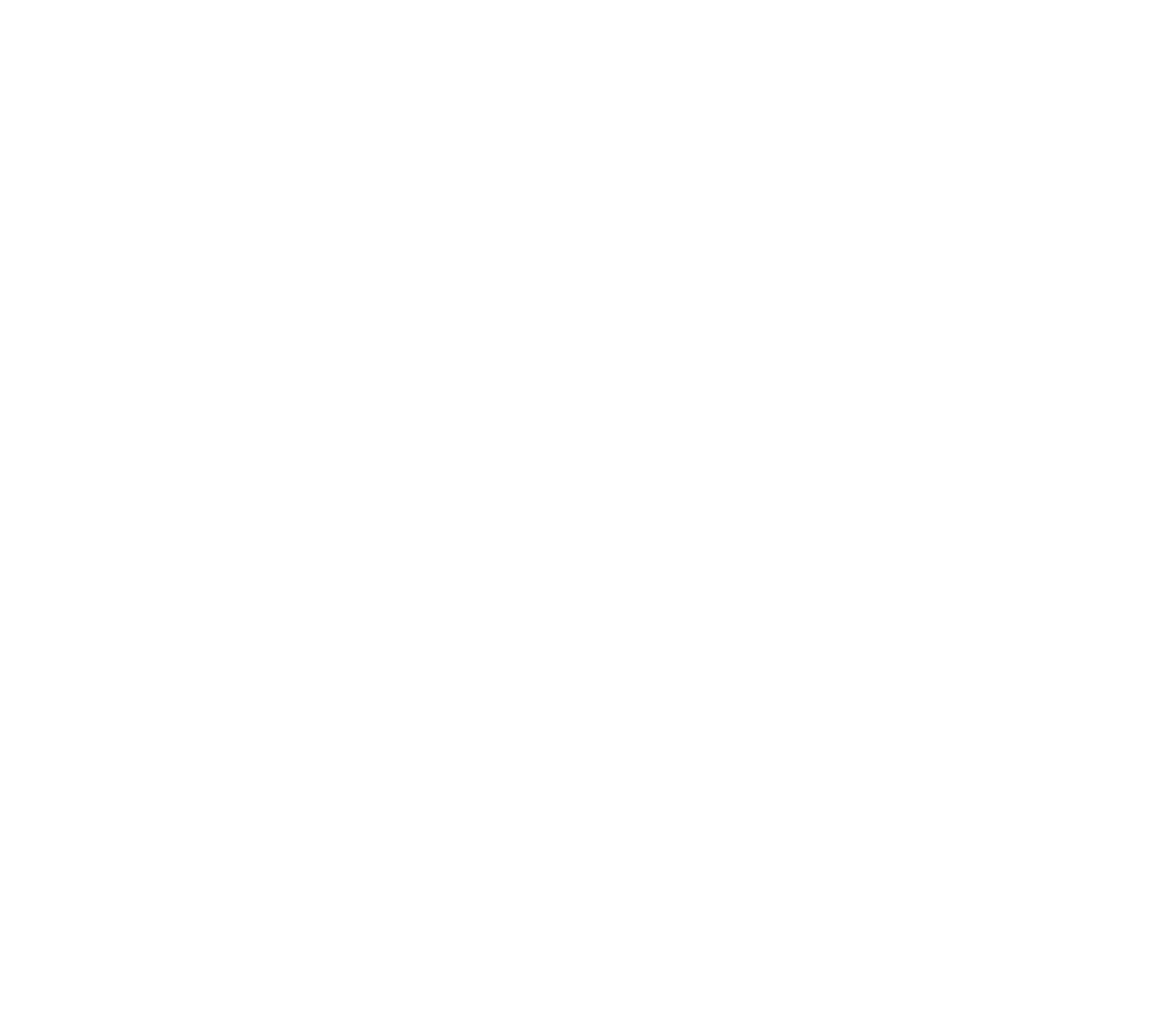} & \ 
    \includegraphics[width=0.31\linewidth,clip]{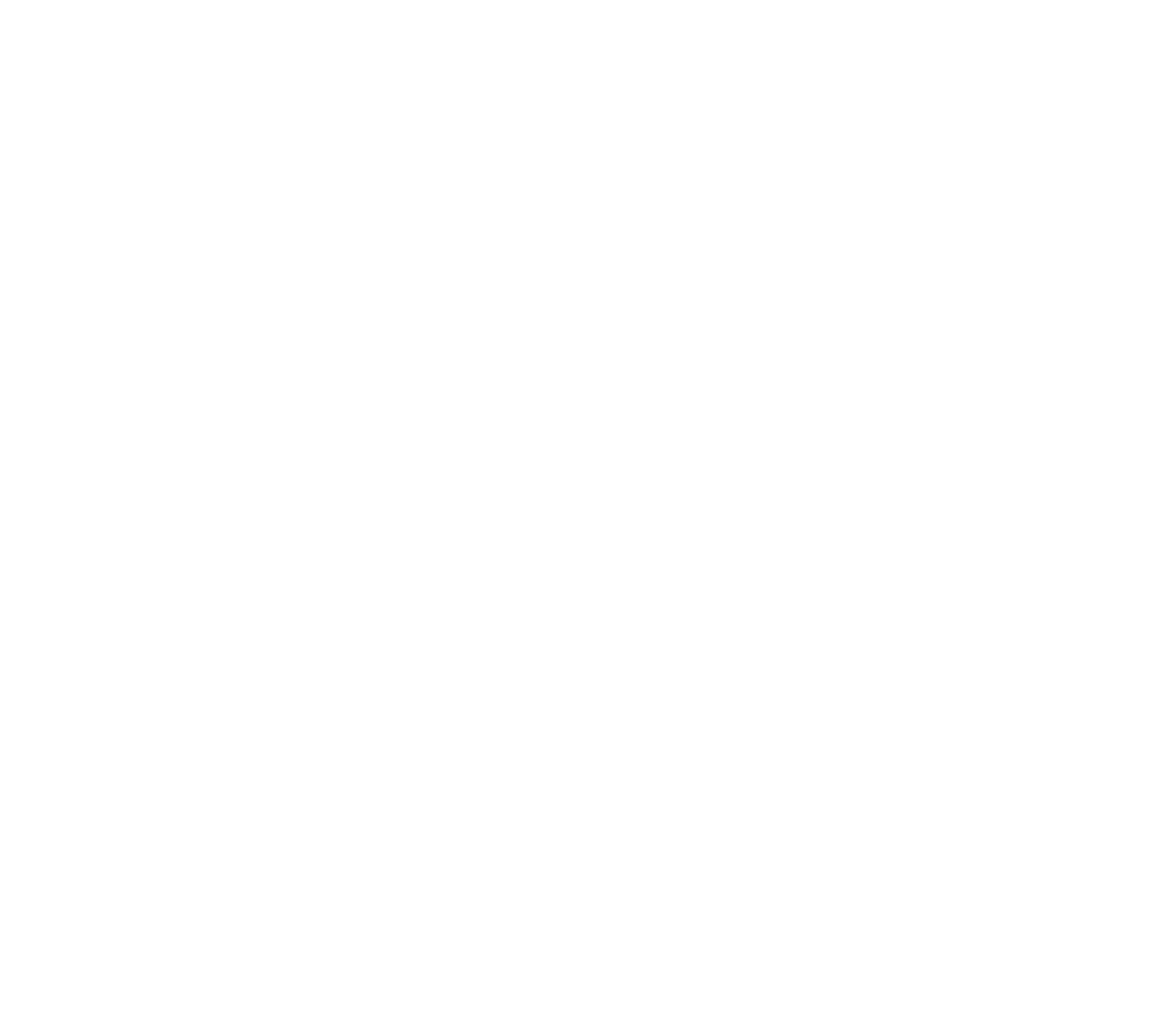} \\
  \end{tabular}
\caption[LGS\,3 images: RC, IR, optical, and FUV]{Multi-wavelength coverage of LGS 3 displaying a $4.0^\prime \times 4.0^\prime$ area. We show total RC flux density at the native resolution (top-left) and again with contours (top-centre). The RC contours are superposed on ancillary LITTLE THINGS images where possible: \halpha\ (middle-left); \RCNT\ obtained by subtracting the expected \RCT\ based on the \halpha-\RCT\ scaling factor of \cite{Deeg1997} from the total RC; {\em GALEX} FUV (middle-right); {\em Spitzer} 24\micron\ (bottom-left); {\em Spitzer} 70\micron\ (bottom-centre); FUV$+24{\rm \mu m}$--inferred SFRD from \citealp{Leroy2012} (bottom-right). We also show the RC that was isolated by the RC--based masking technique (top-right).}
  \label{figure:lgs3Cc_maps}
\end{figure}

\begin{figure}
  \begin{tabular}{ccc}
    \includegraphics[width=0.31\linewidth,clip]{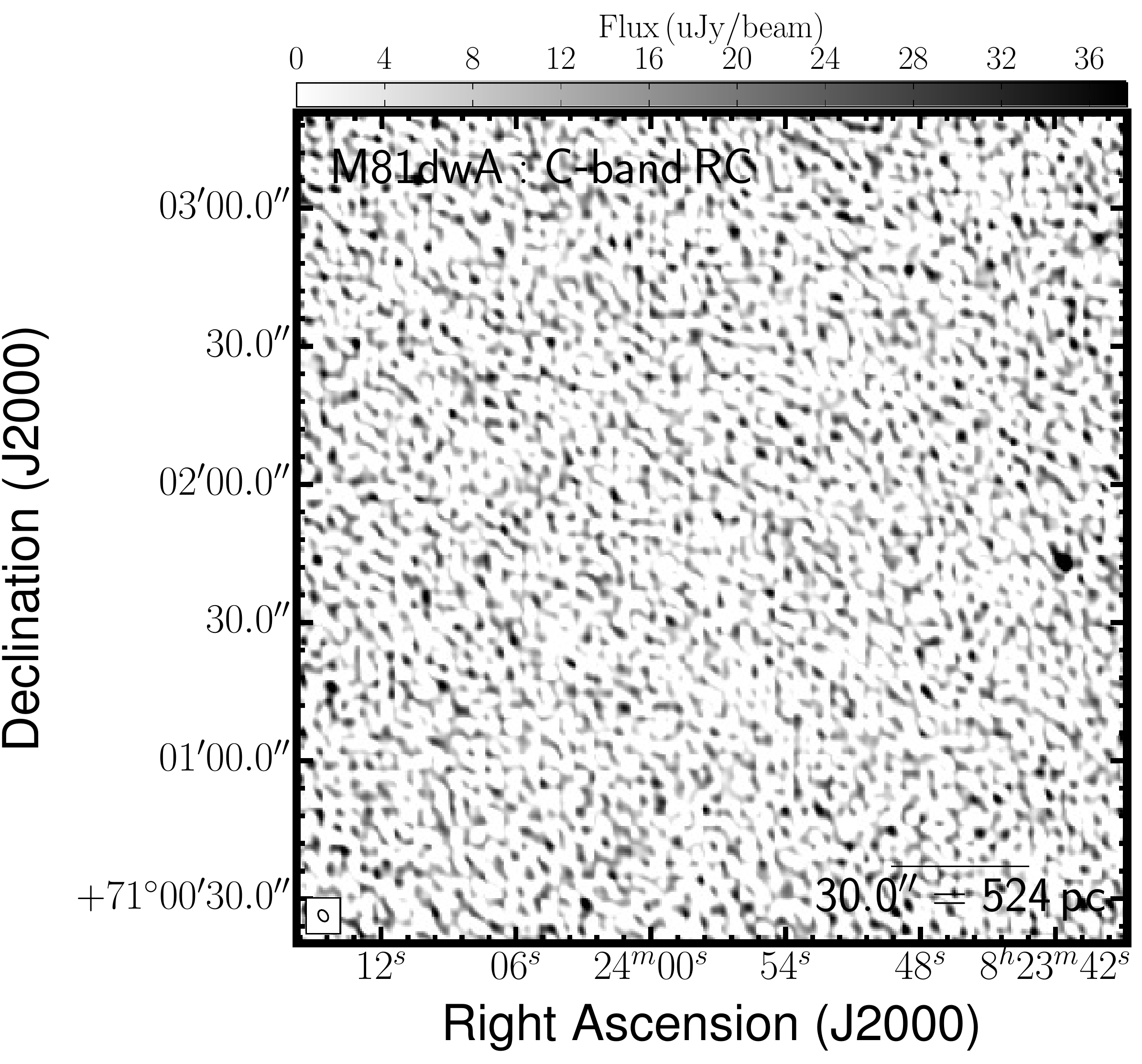} & \ 
    \includegraphics[width=0.31\linewidth,clip]{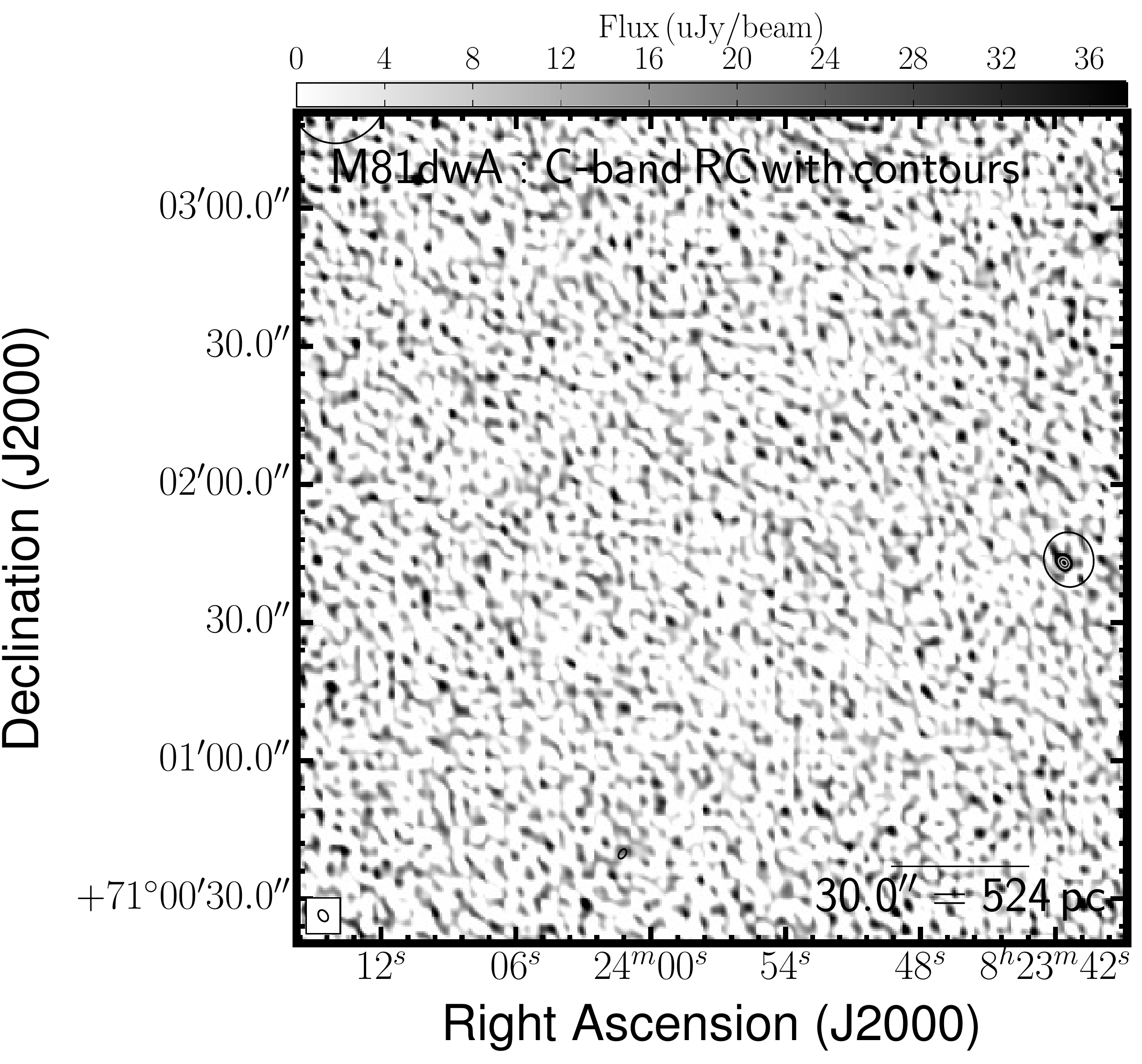} & \ 
    \includegraphics[width=0.31\linewidth,clip]{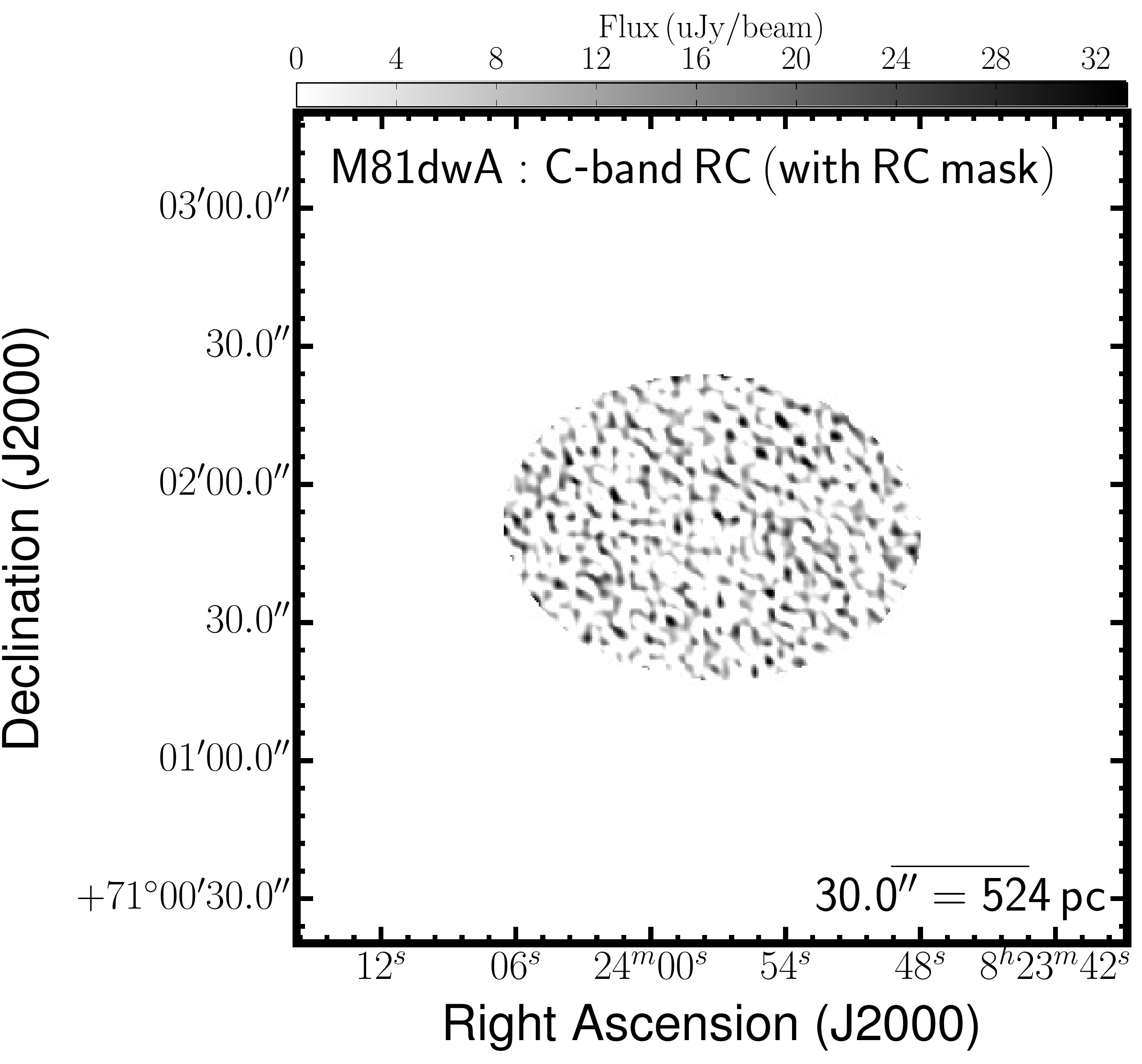} \\
    \includegraphics[width=0.31\linewidth,clip]{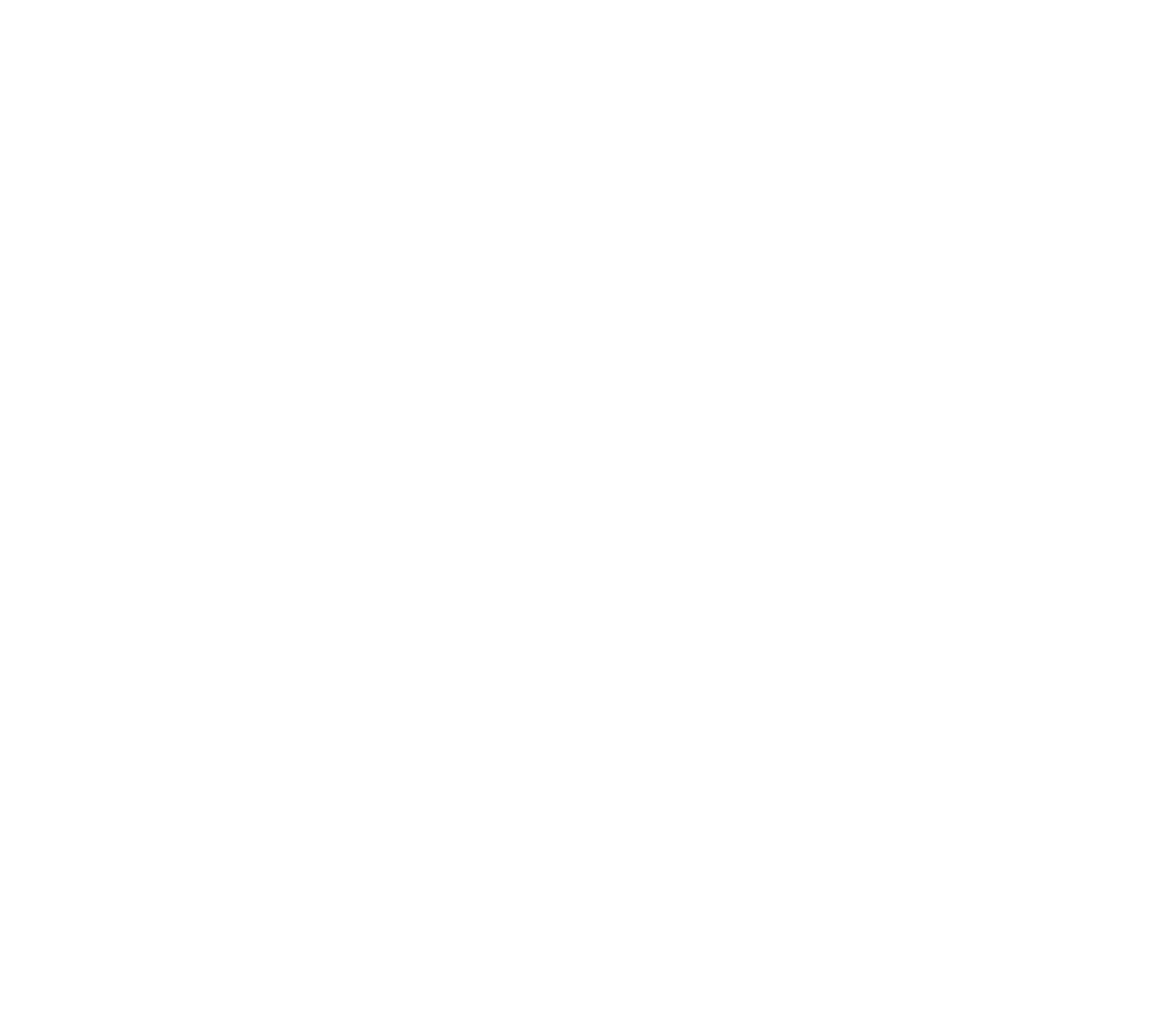} & \ 
    \includegraphics[width=0.31\linewidth,clip]{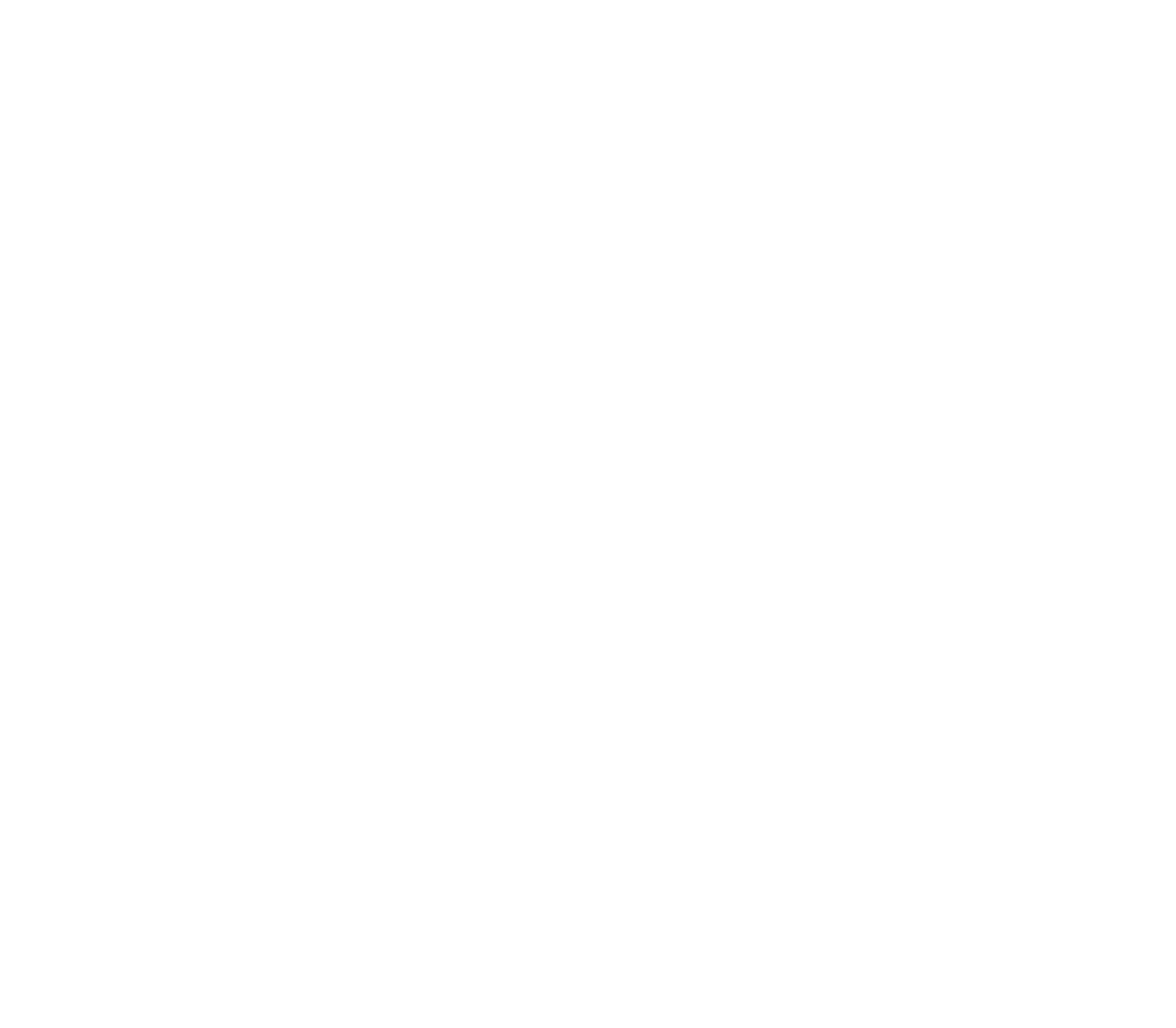} & \ 
    \includegraphics[width=0.31\linewidth,clip]{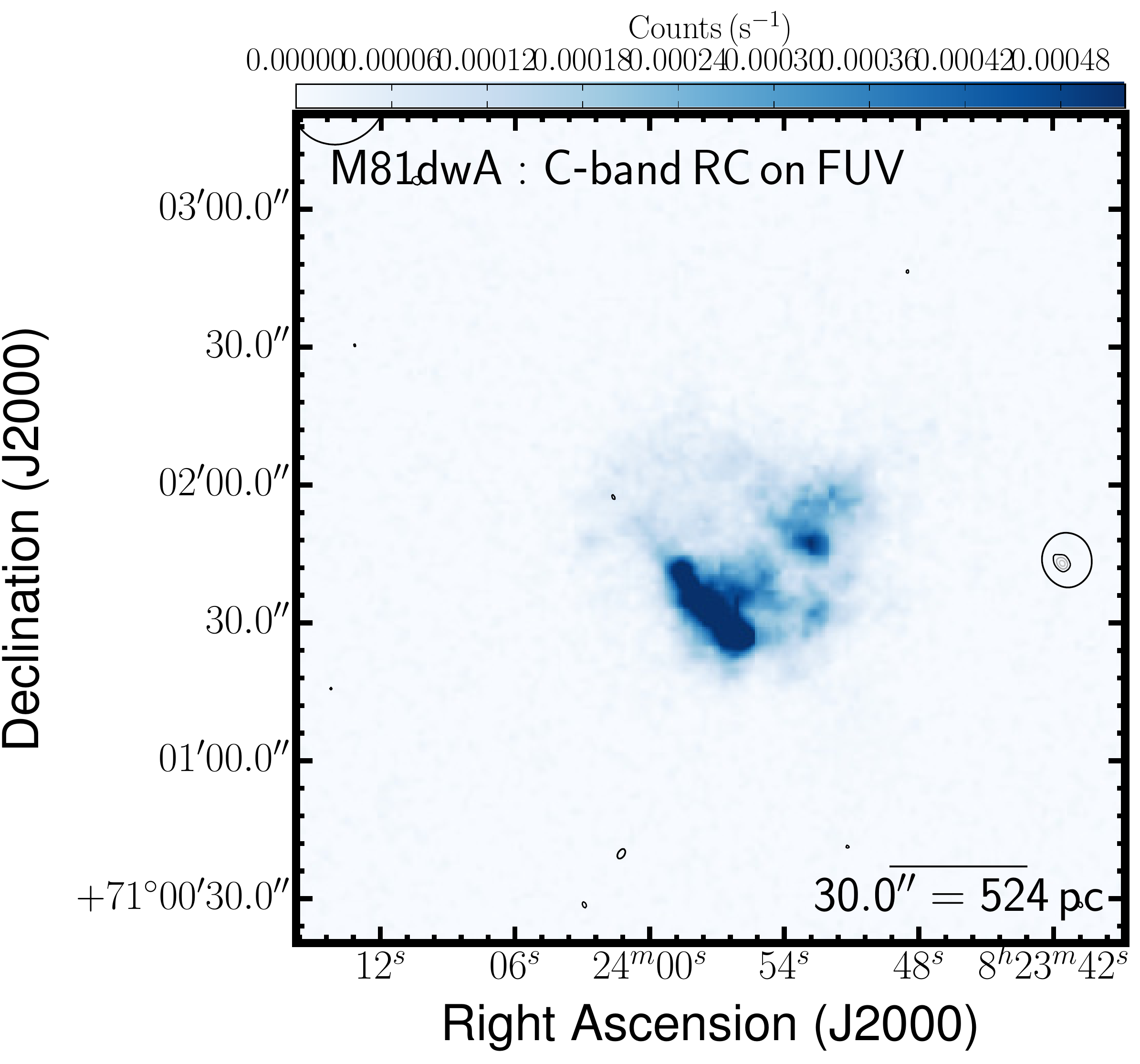} \\
    \includegraphics[width=0.31\linewidth,clip]{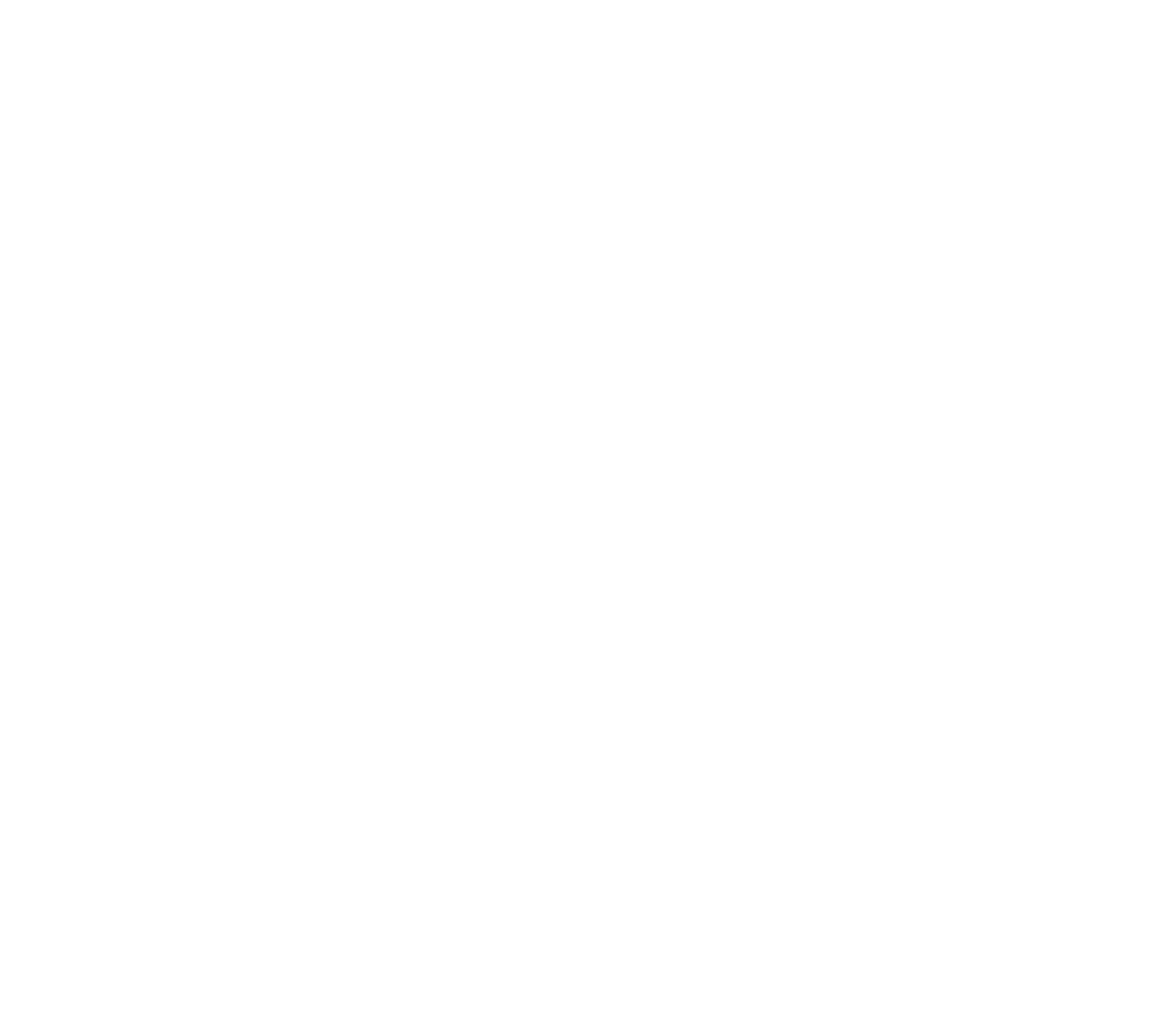} & \ 
    \includegraphics[width=0.31\linewidth,clip]{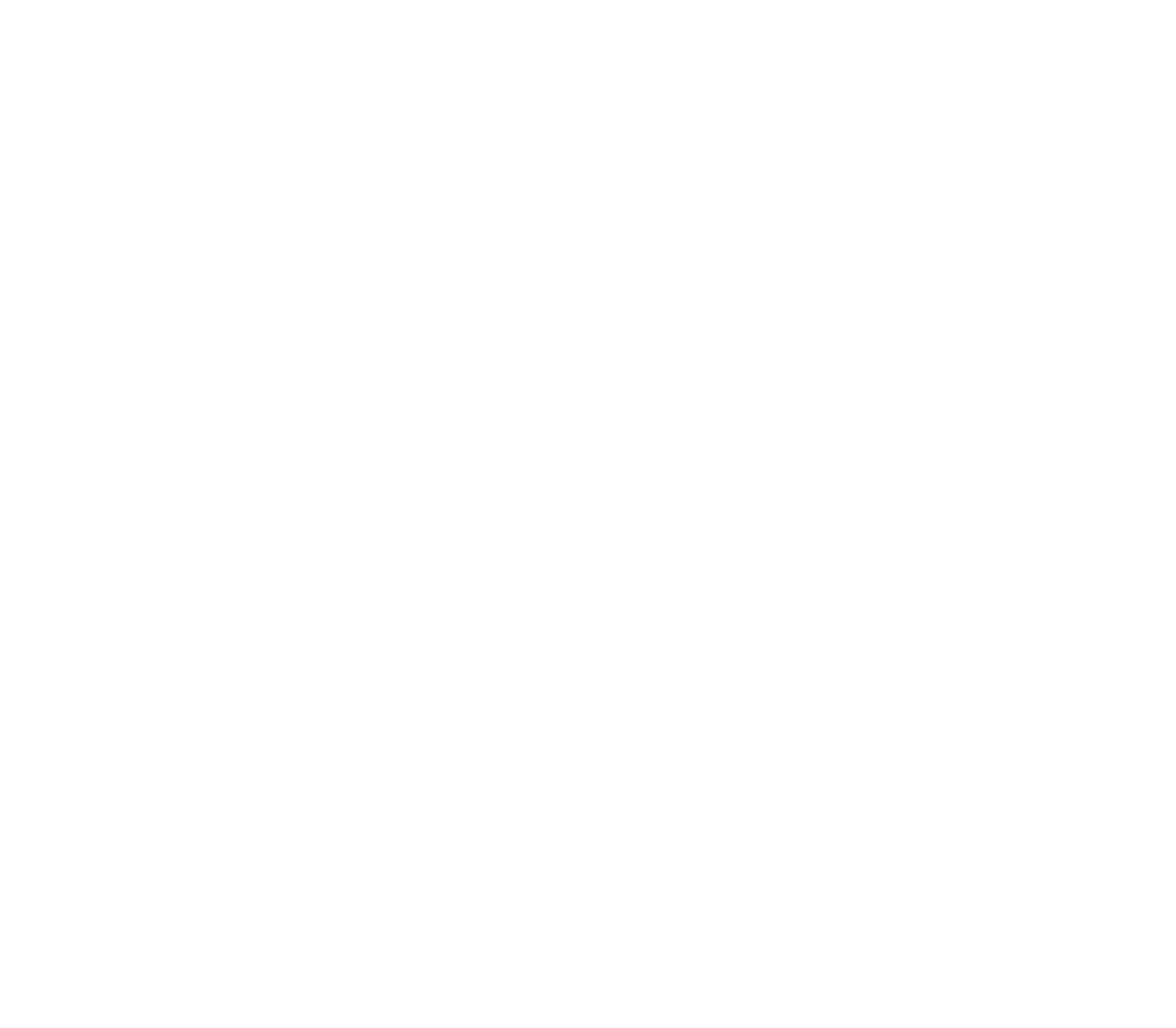} & \ 
    \includegraphics[width=0.31\linewidth,clip]{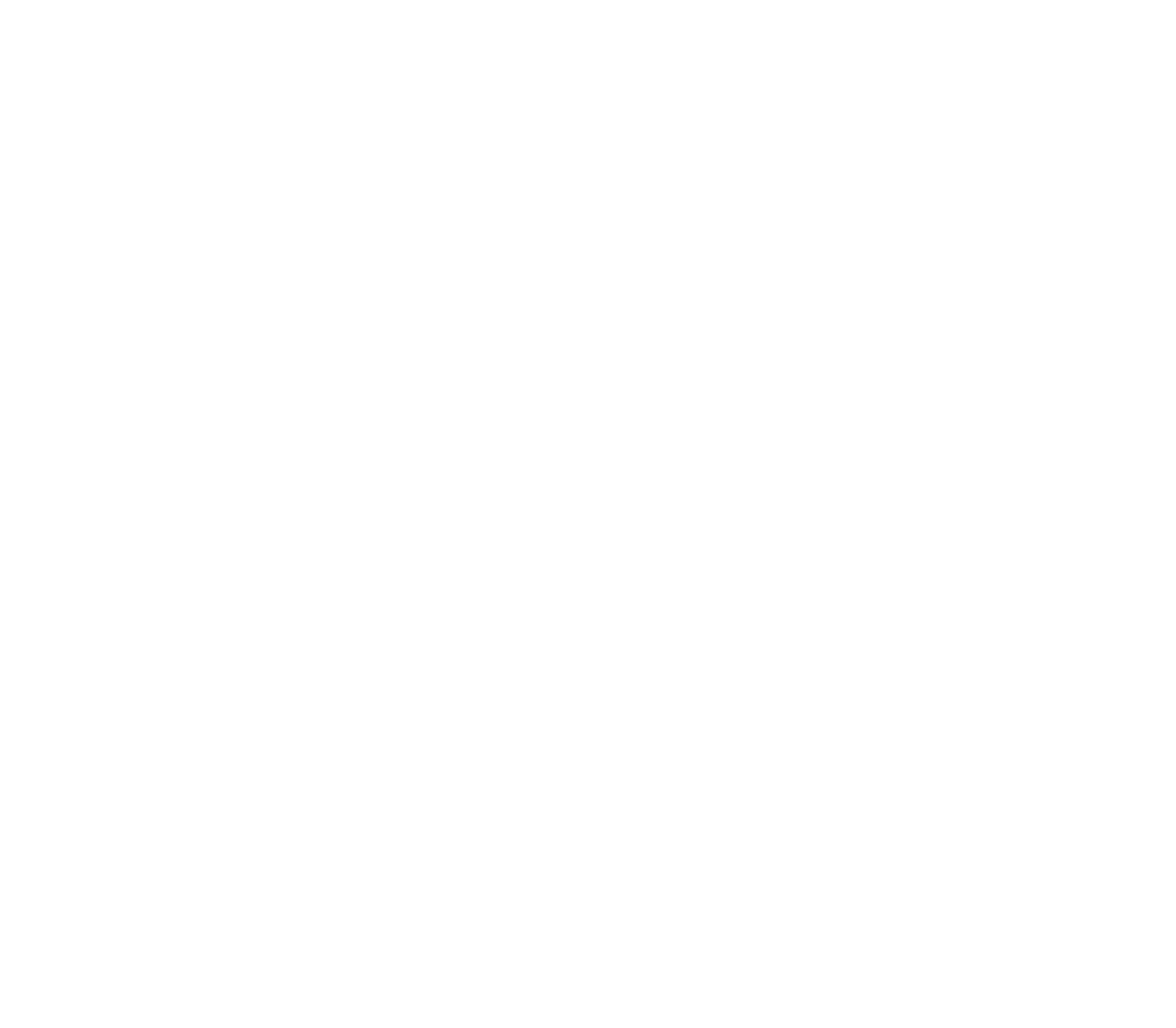} \\
  \end{tabular}
\caption[M81\,DwA images: RC, IR, optical, and FUV]{Multi-wavelength coverage of M81 dwA displaying a $3.0^\prime \times 3.0^\prime$ area. We show total RC flux density at the native resolution (top-left) and again with contours (top-centre). The RC contours are superposed on ancillary LITTLE THINGS images where possible: \halpha\ (middle-left); \RCNT\ obtained by subtracting the expected \RCT\ based on the \halpha-\RCT\ scaling factor of \cite{Deeg1997} from the total RC; {\em GALEX} FUV (middle-right); {\em Spitzer} 24\micron\ (bottom-left); {\em Spitzer} 70\micron\ (bottom-centre); FUV$+24{\rm \mu m}$--inferred SFRD from \citealp{Leroy2012} (bottom-right). We also show the RC that was isolated by the RC--based masking technique (top-right).}
  \label{figure:m81dwaCc_maps}
\end{figure}

\begin{figure}
  \begin{tabular}{ccc}
    \includegraphics[width=0.31\linewidth,clip]{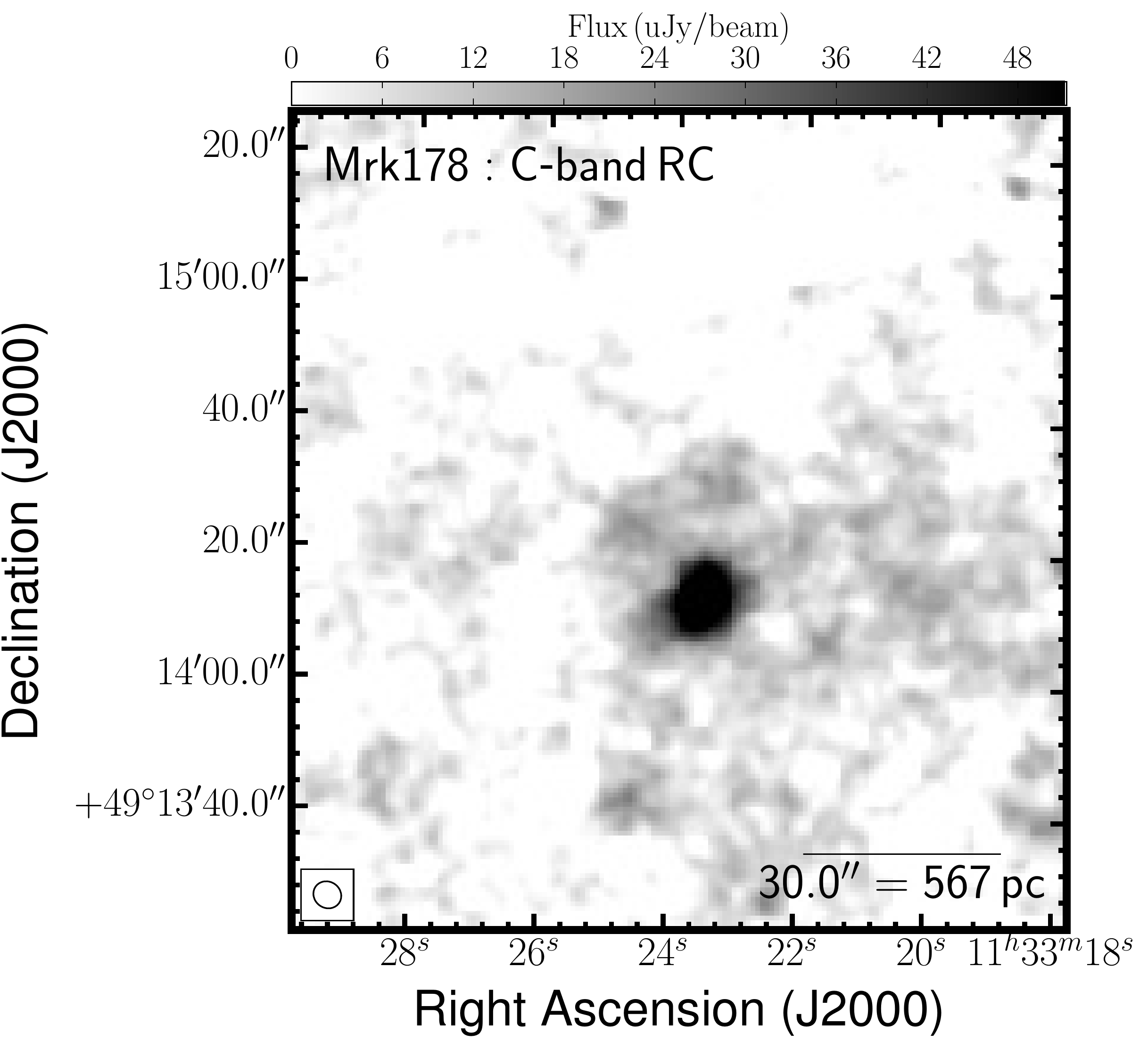} & \ 
    \includegraphics[width=0.31\linewidth,clip]{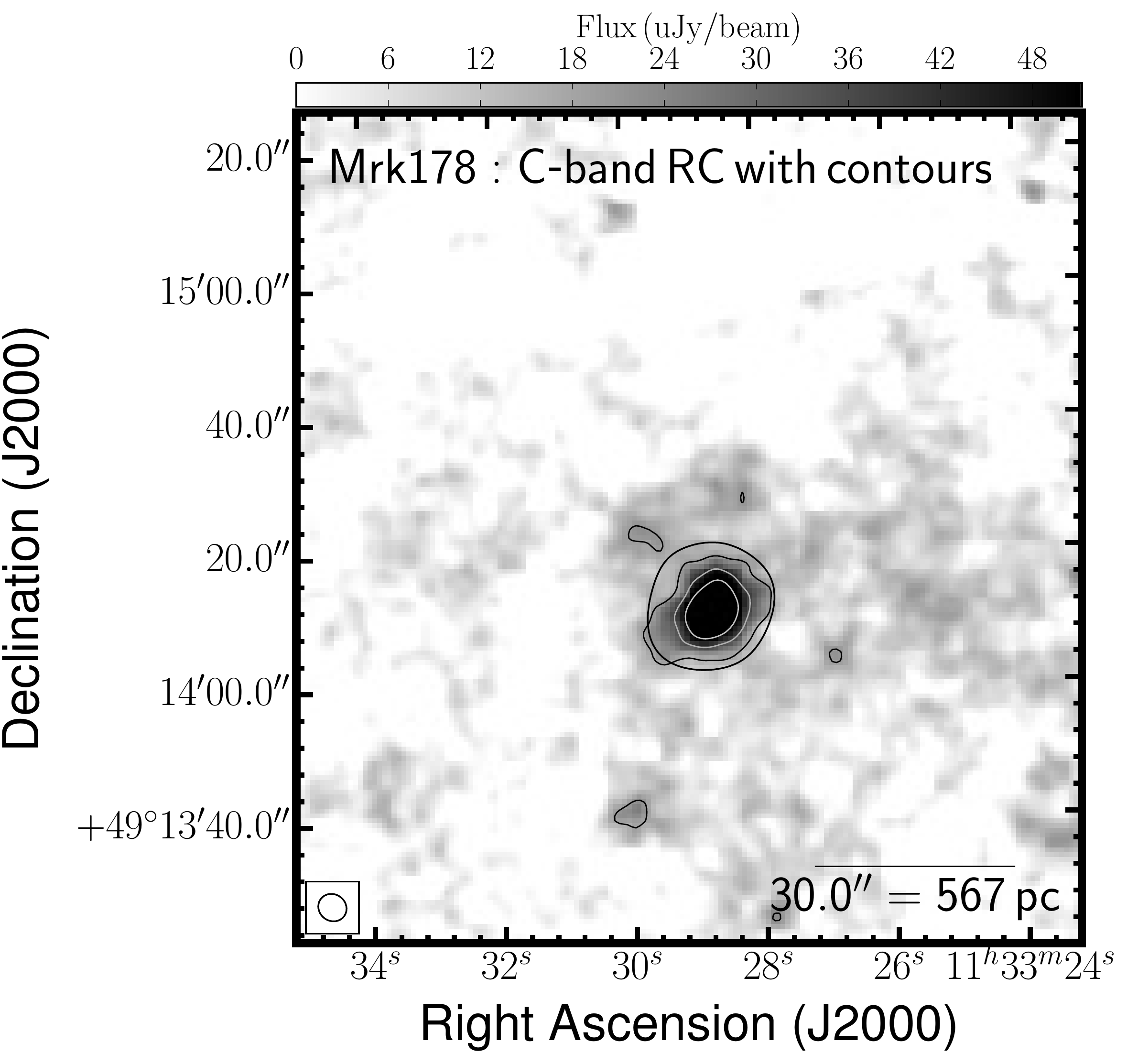} & \ 
    \includegraphics[width=0.31\linewidth,clip]{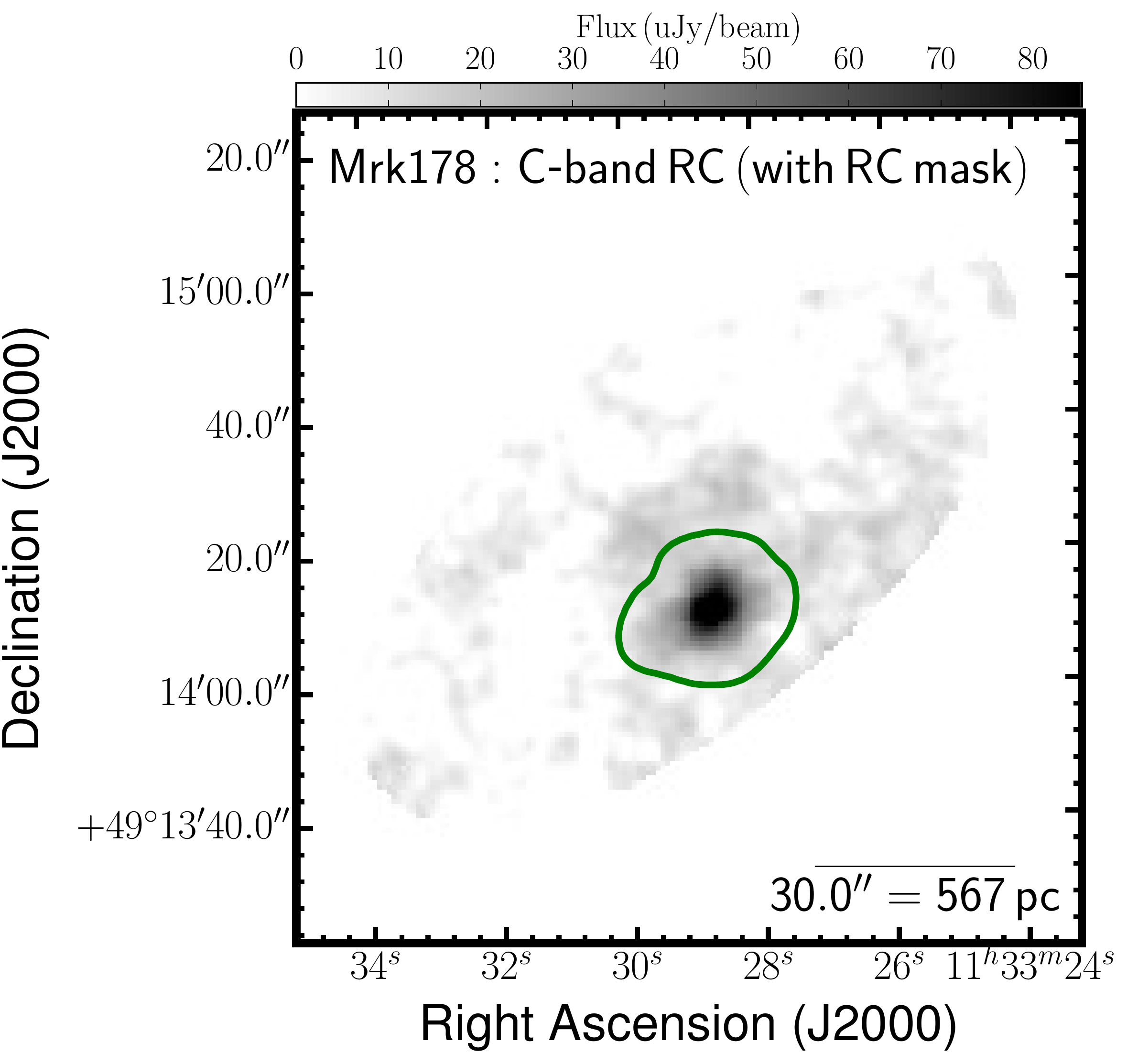} \\
    \includegraphics[width=0.31\linewidth,clip]{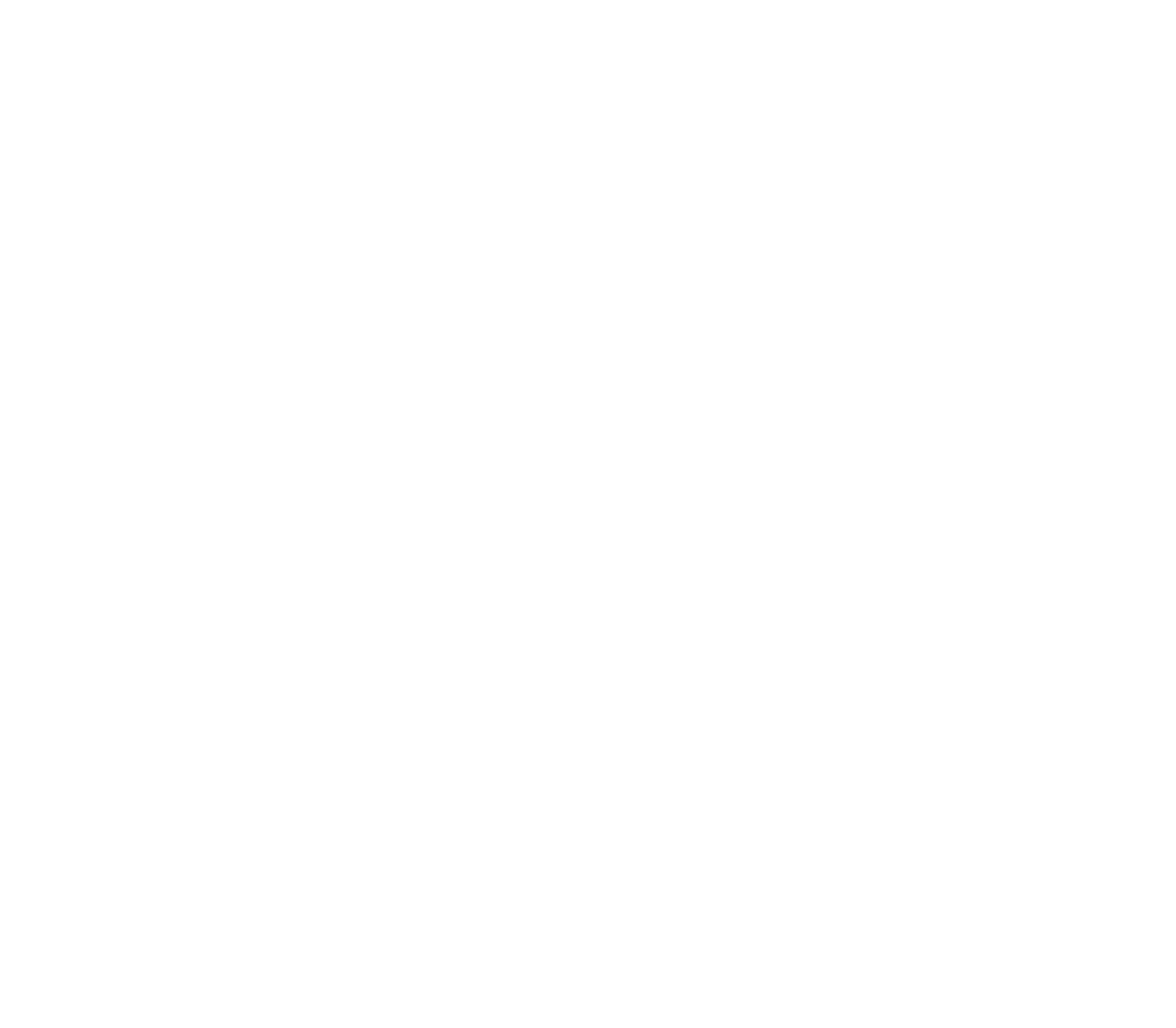} & \ 
    \includegraphics[width=0.31\linewidth,clip]{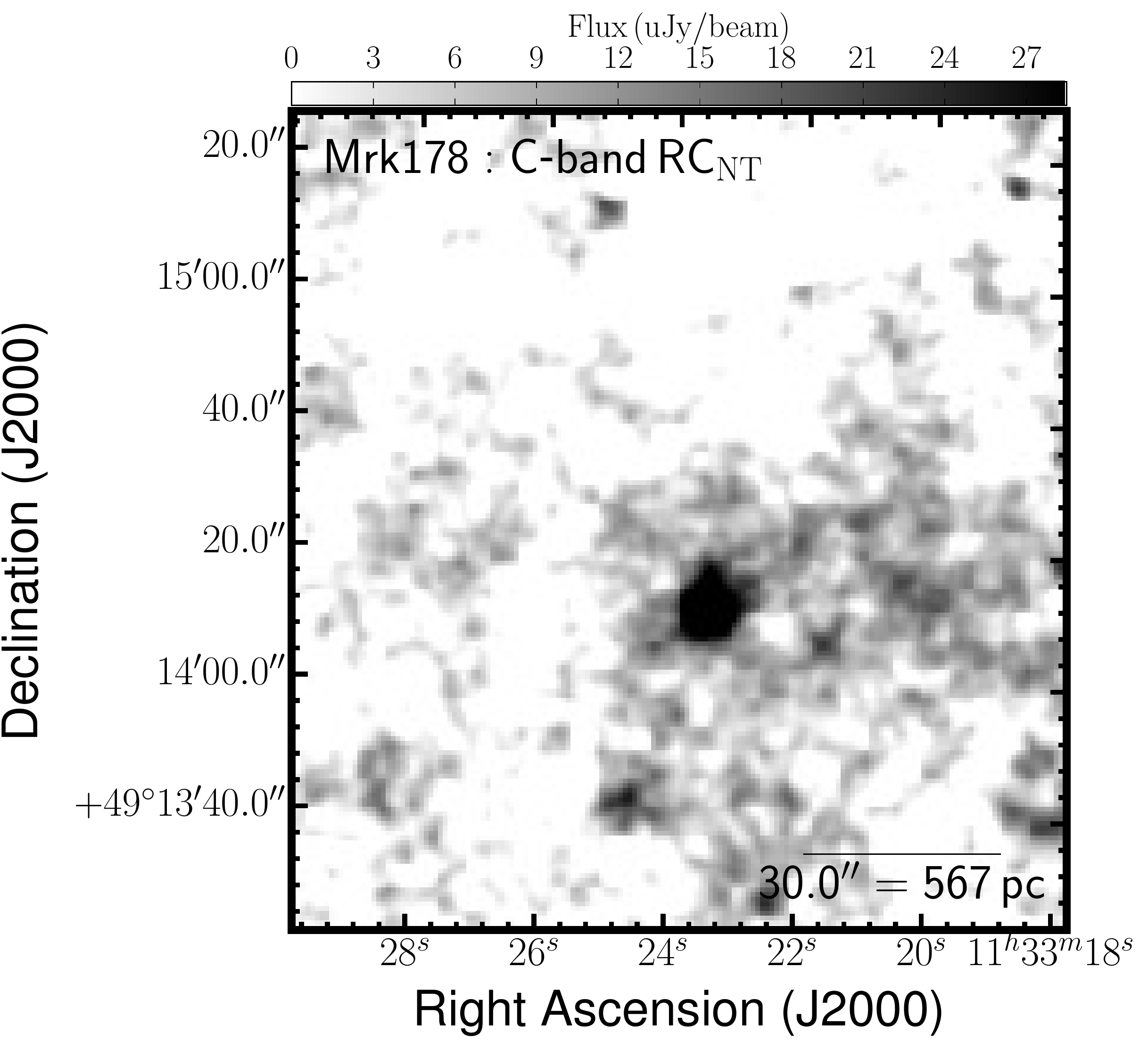} & \ 
    \includegraphics[width=0.31\linewidth,clip]{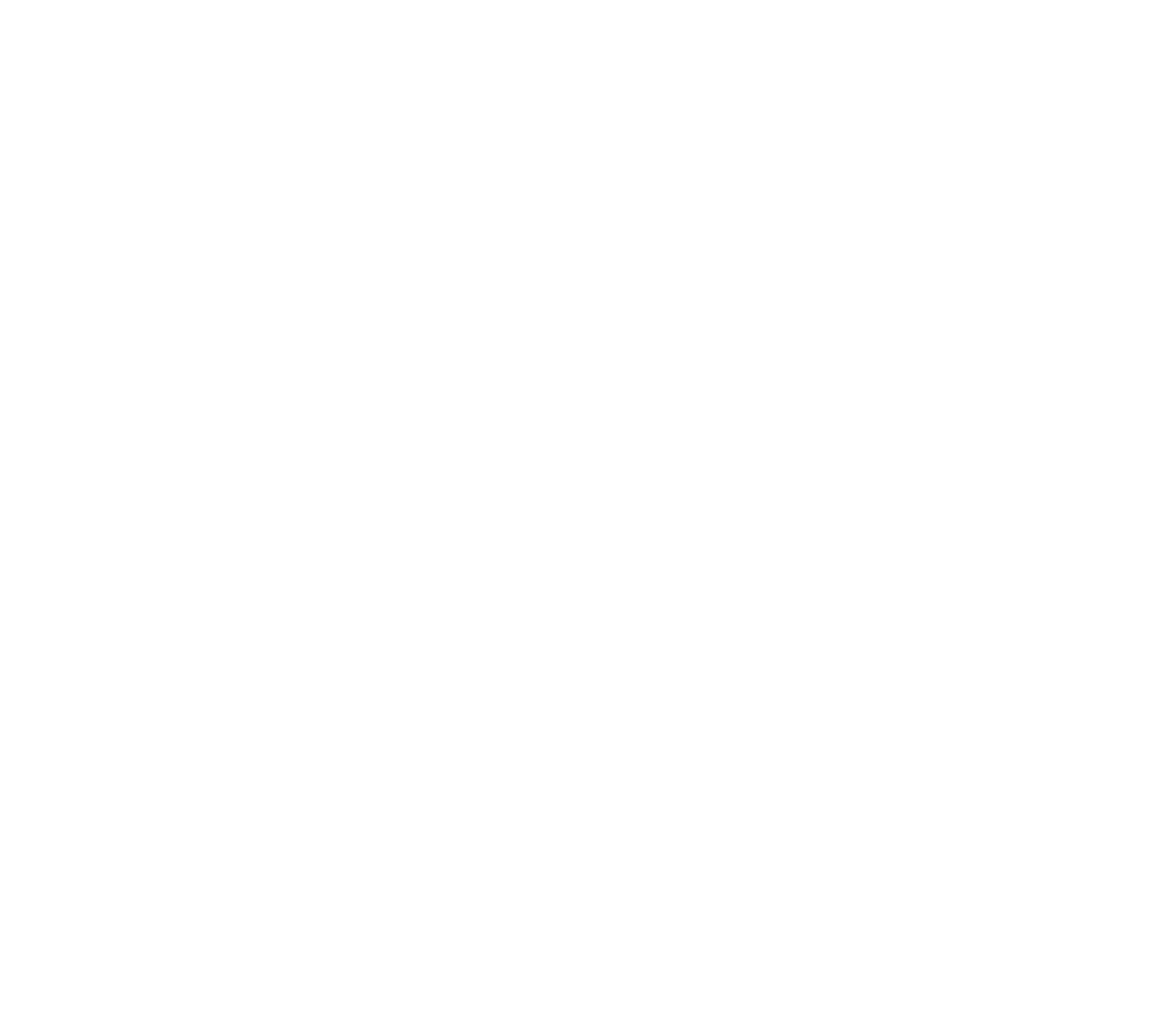} \\
    \includegraphics[width=0.31\linewidth,clip]{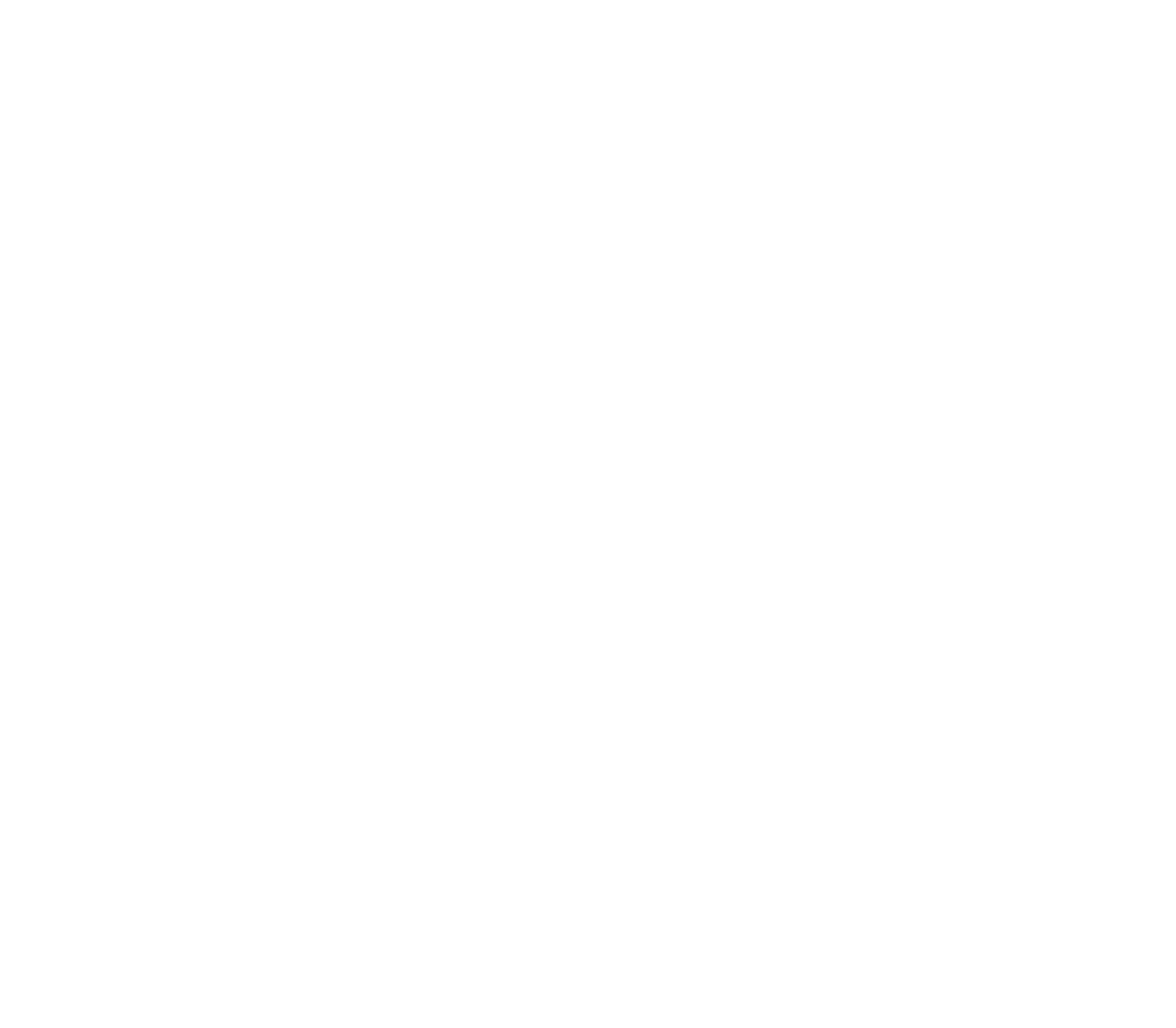} & \ 
    \includegraphics[width=0.31\linewidth,clip]{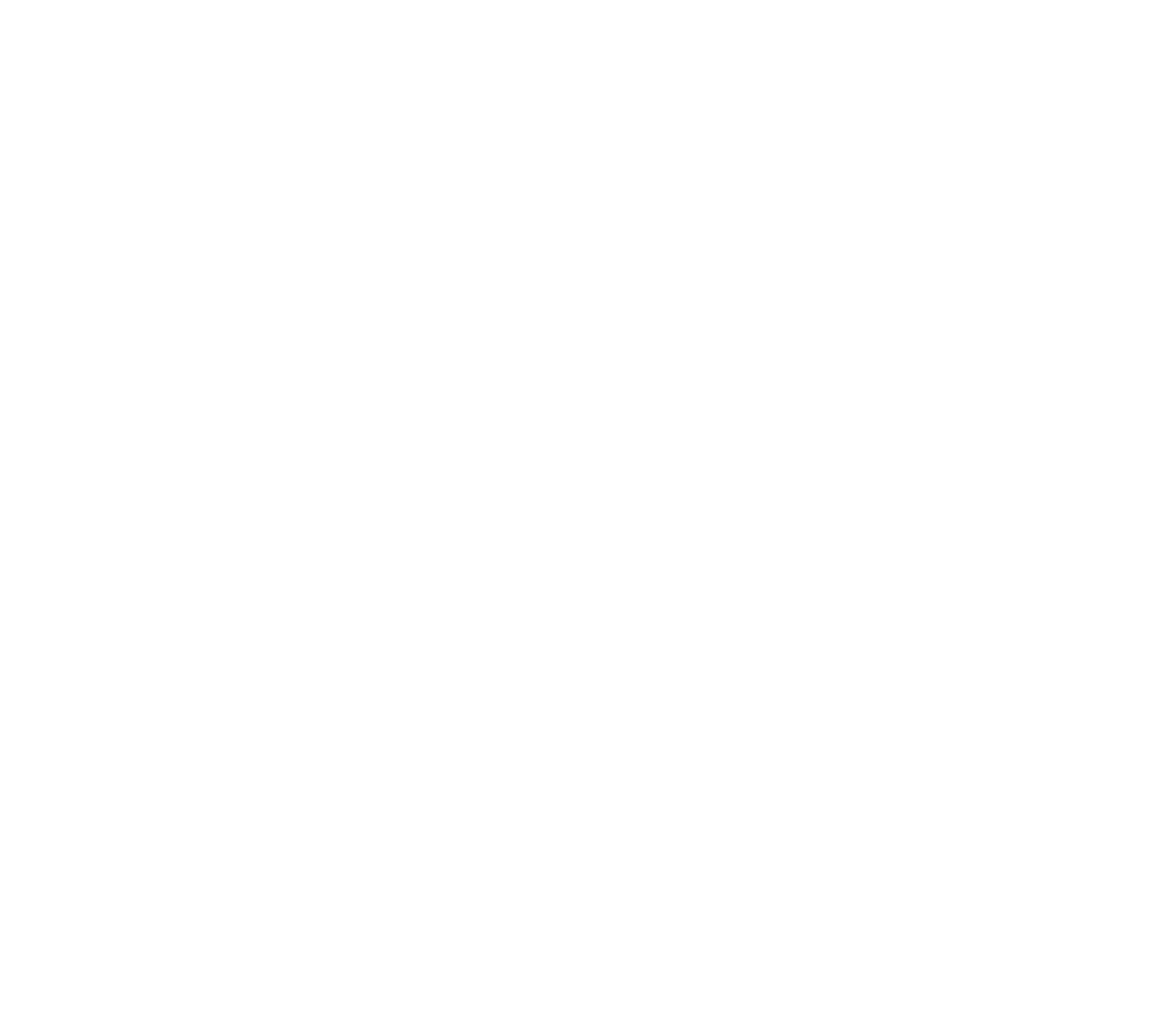} & \ 
    \includegraphics[width=0.31\linewidth,clip]{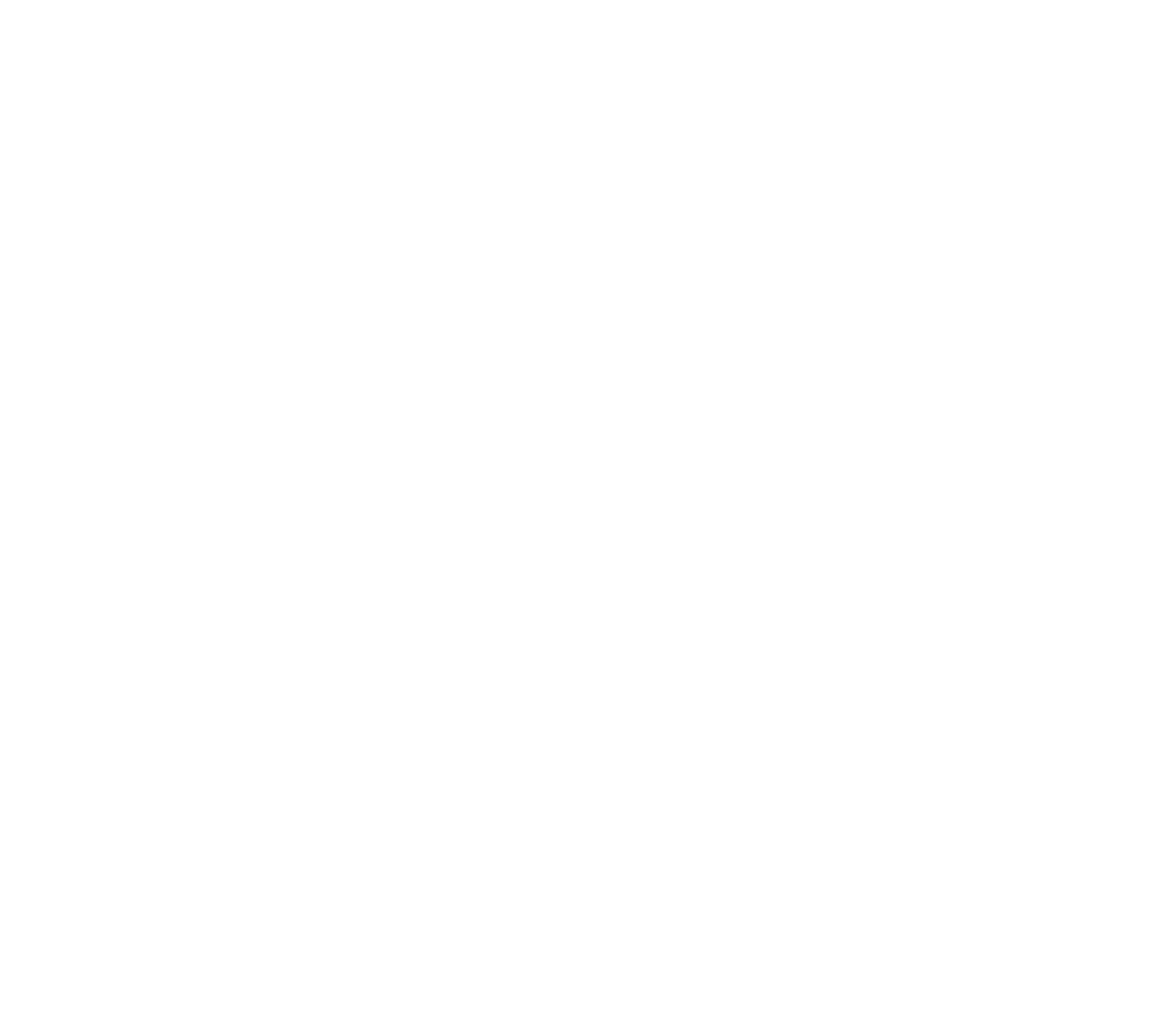} \\
  \end{tabular}
\caption[Mrk\,178 images: RC, IR, optical, and FUV]{Multi-wavelength coverage of Mrk 178 displaying a $2.0^\prime \times 2.0^\prime$ area.We show total RC flux density at the native resolution (top-left) and again with contours (top-centre). The RC contours are superposed on ancillary LITTLE THINGS images where possible: \halpha\ (middle-left); \RCNT\ obtained by subtracting the expected \RCT\ based on the \halpha-\RCT\ scaling factor of \cite{Deeg1997} from the total RC; {\em GALEX} FUV (middle-right); {\em Spitzer} 24\micron\ (bottom-left); {\em Spitzer} 70\micron\ (bottom-centre); FUV$+24{\rm \mu m}$--inferred SFRD from \citealp{Leroy2012} (bottom-right). We also show the RC that was isolated by the RC--based masking technique (top-right).}
  \label{figure:mrk178Cc_maps}
\end{figure}

\begin{figure}
  \begin{tabular}{ccc}
    \includegraphics[width=0.31\linewidth,clip]{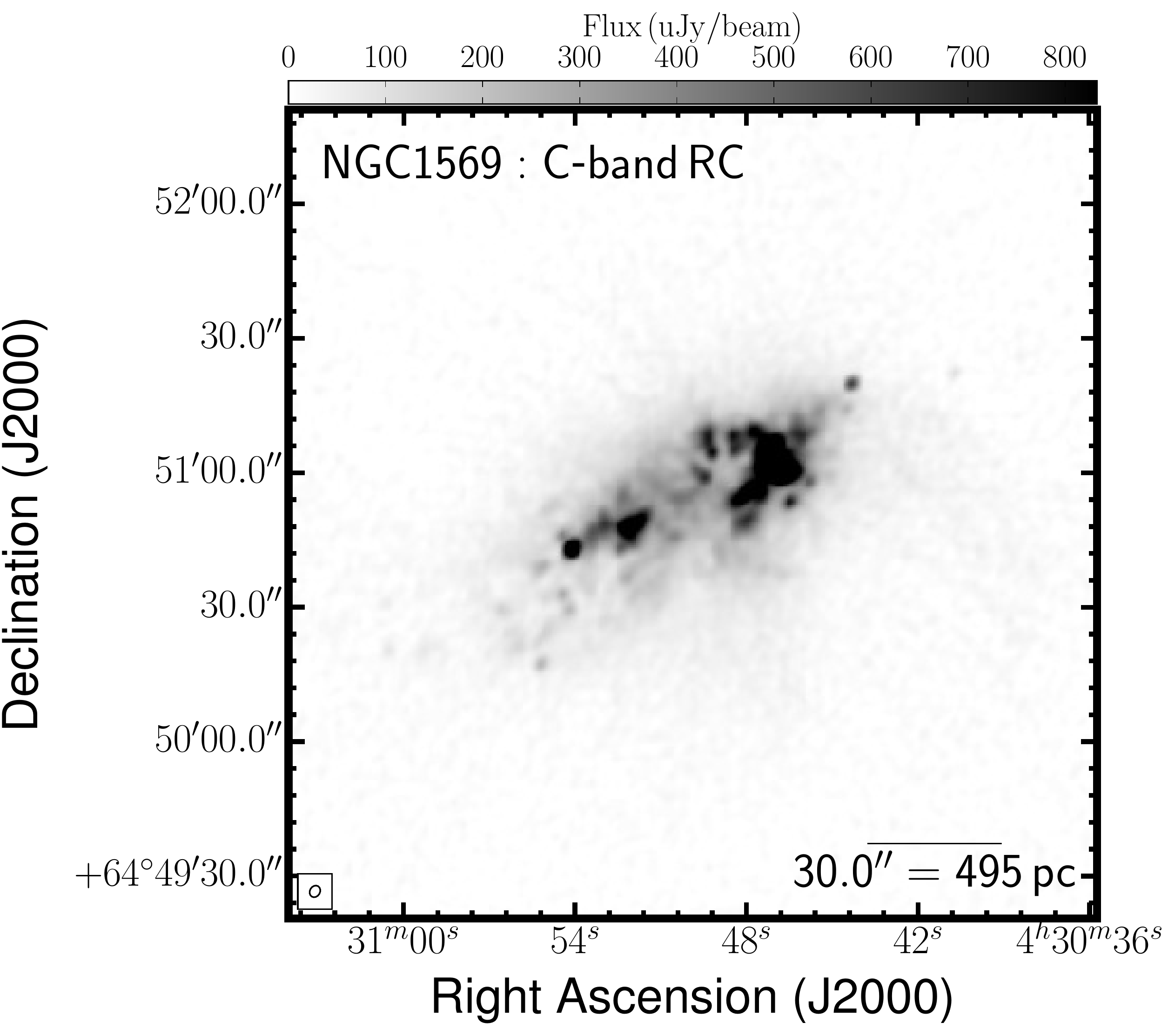} & \ 
    \includegraphics[width=0.31\linewidth,clip]{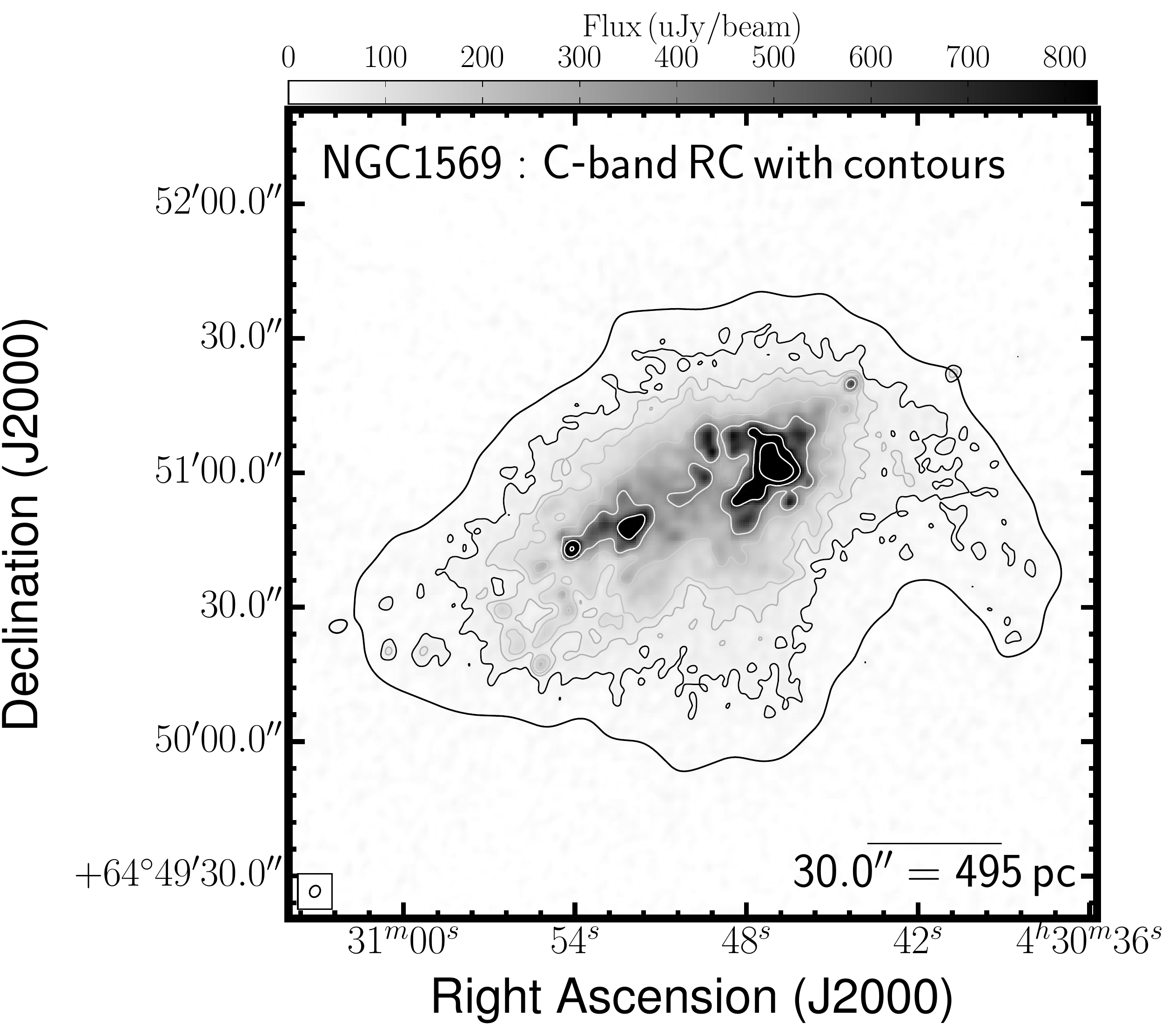} & \ 
    \includegraphics[width=0.31\linewidth,clip]{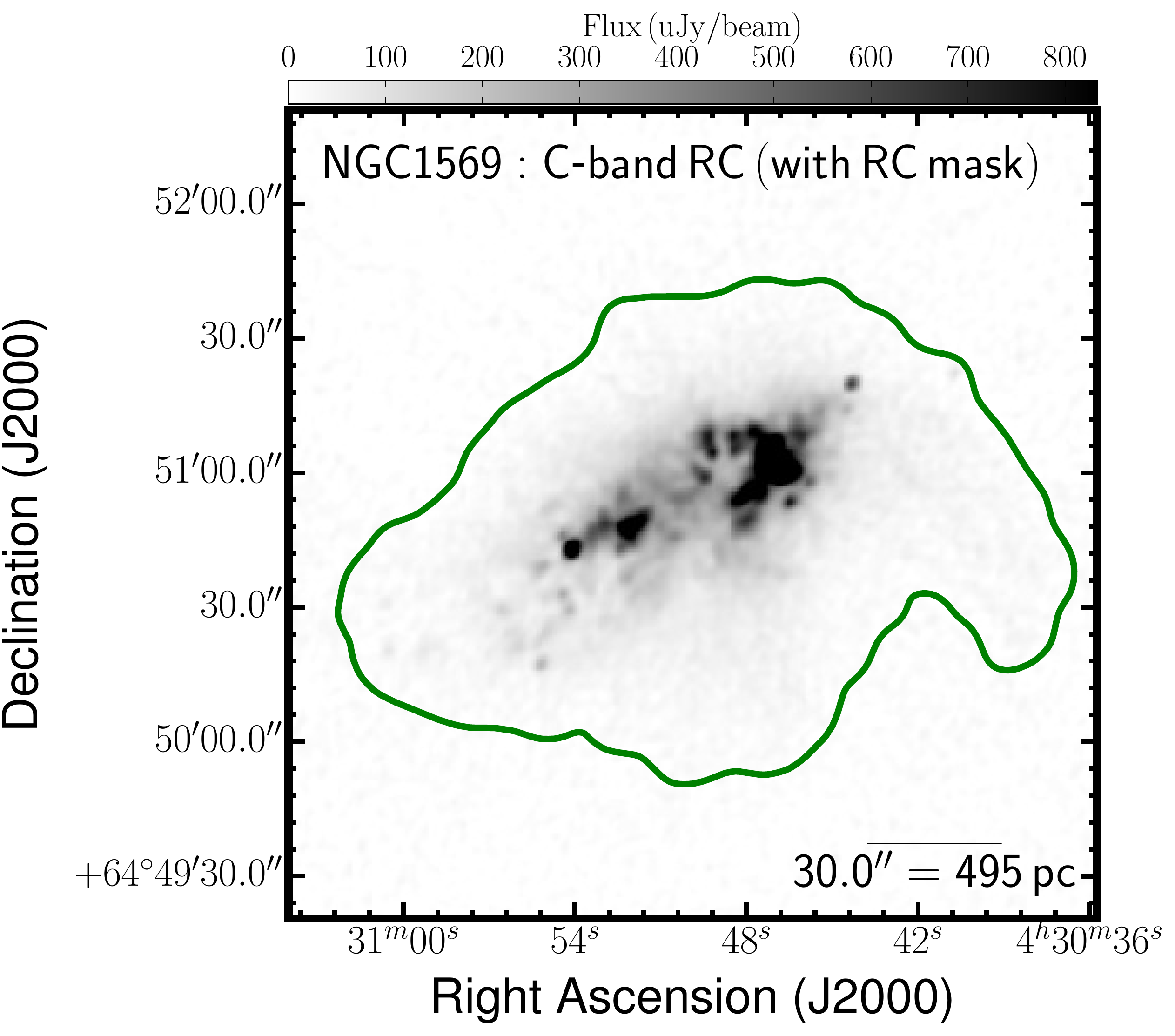} \\
    \includegraphics[width=0.31\linewidth,clip]{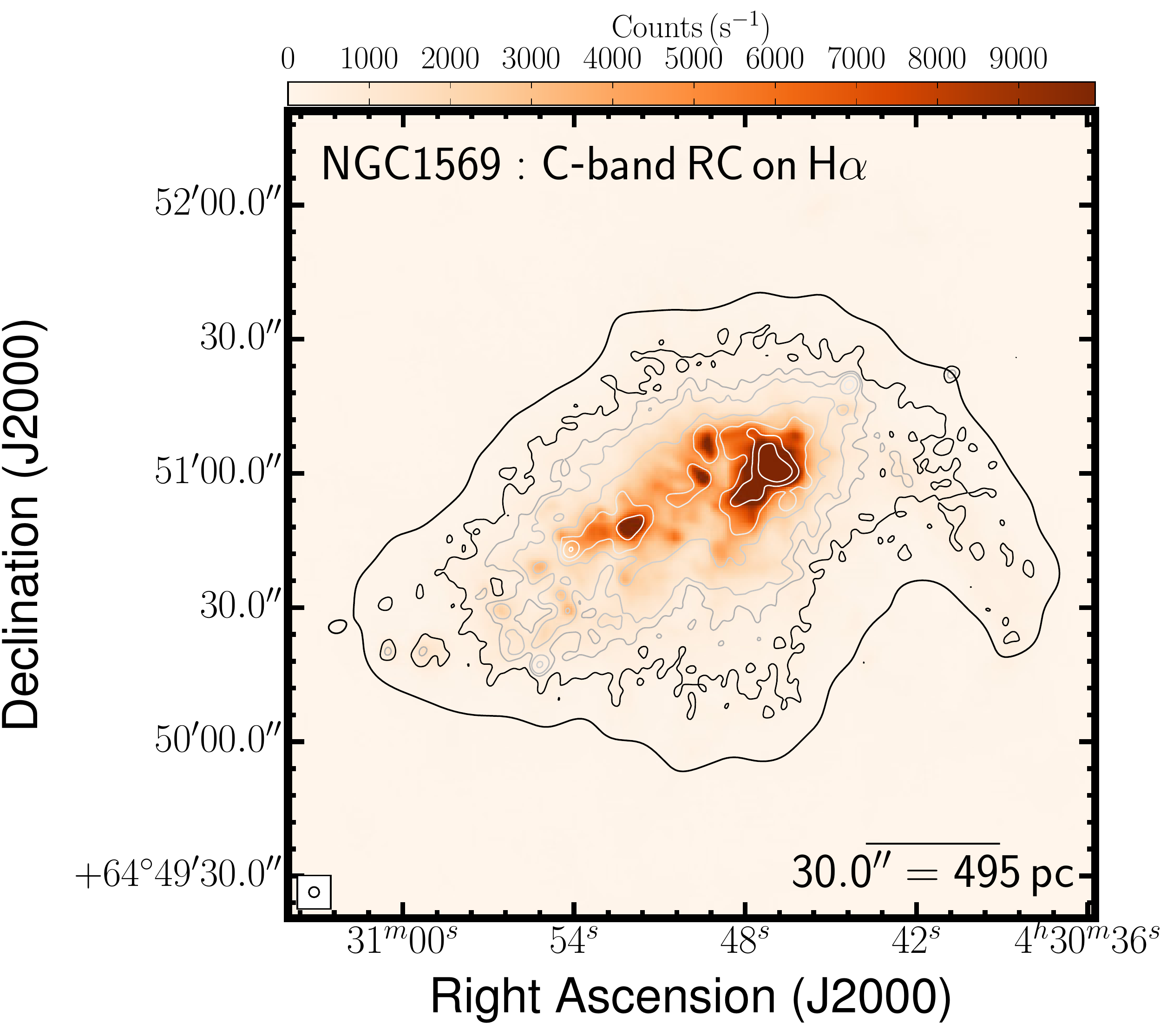} & \ 
    \includegraphics[width=0.31\linewidth,clip]{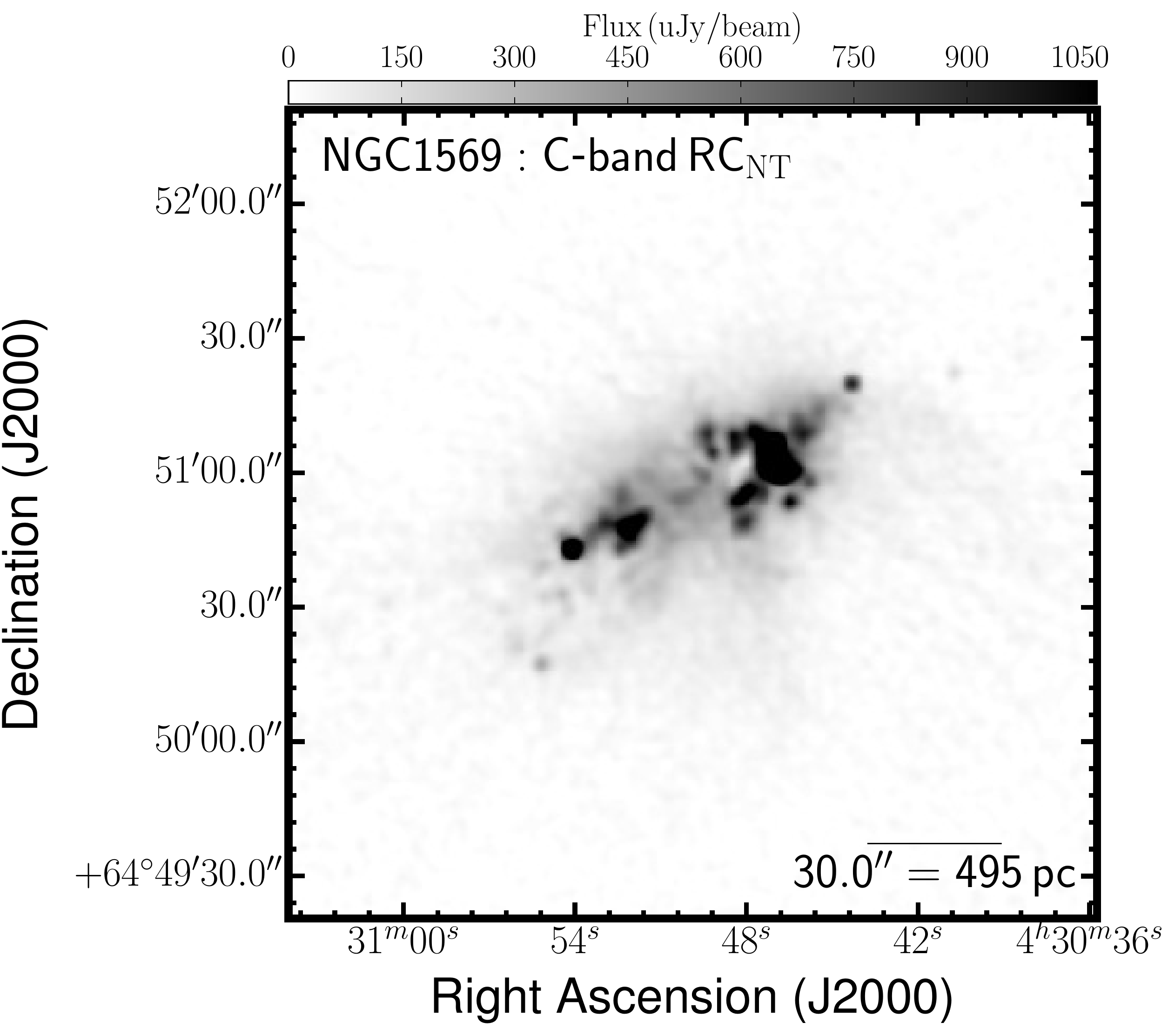} & \ 
    \includegraphics[width=0.31\linewidth,clip]{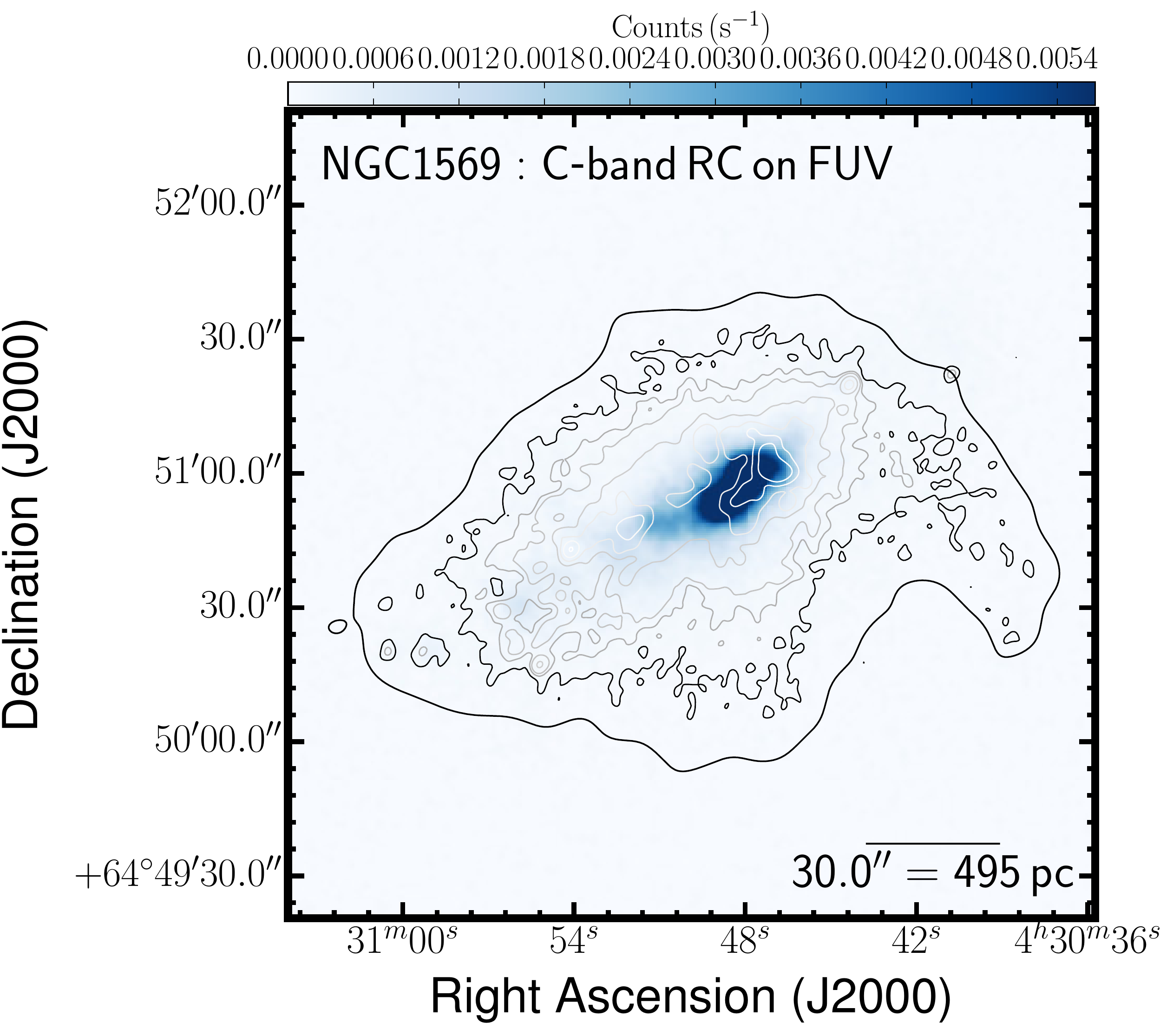} \\
    \includegraphics[width=0.31\linewidth,clip]{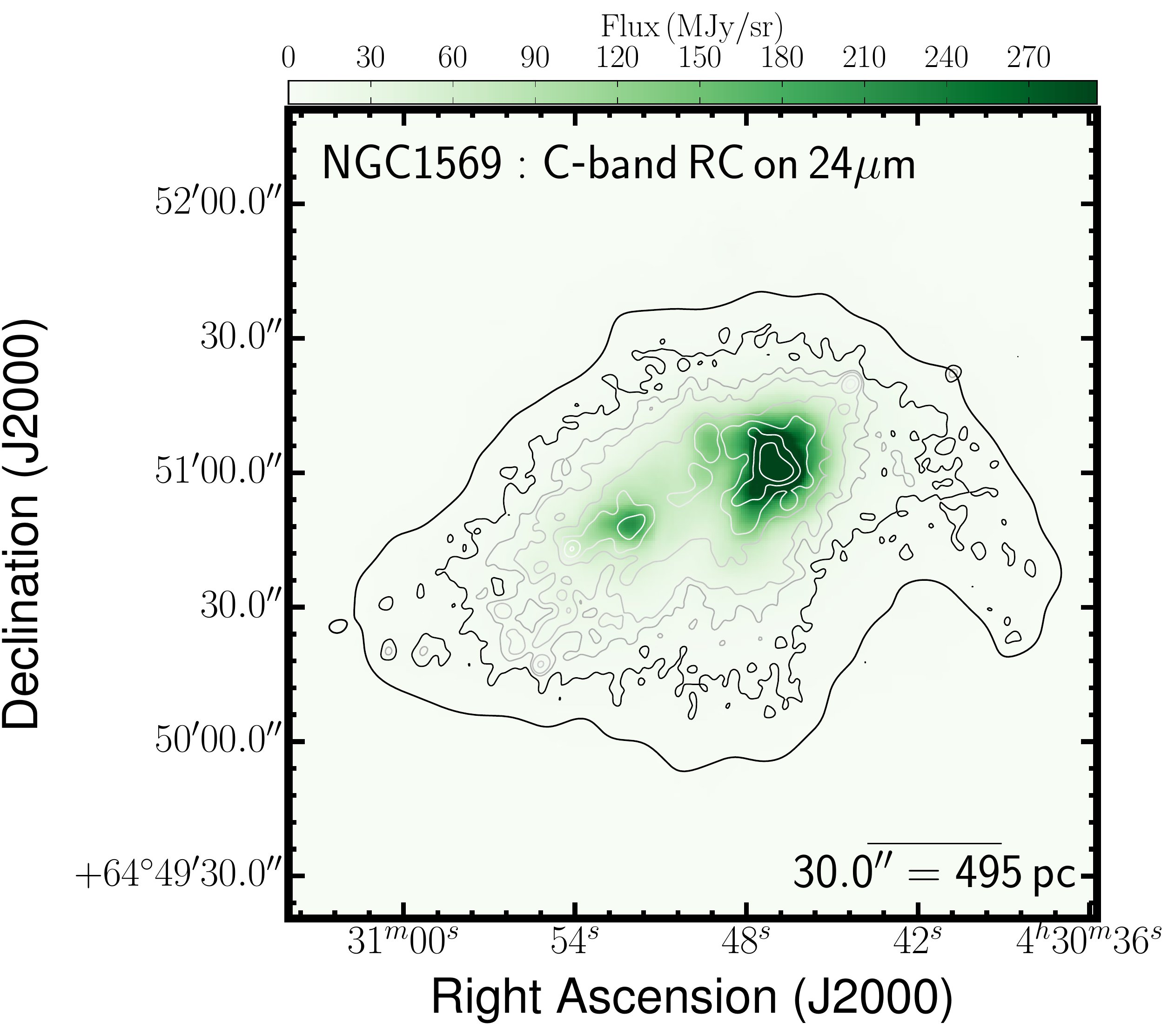} & \ 
    \includegraphics[width=0.31\linewidth,clip]{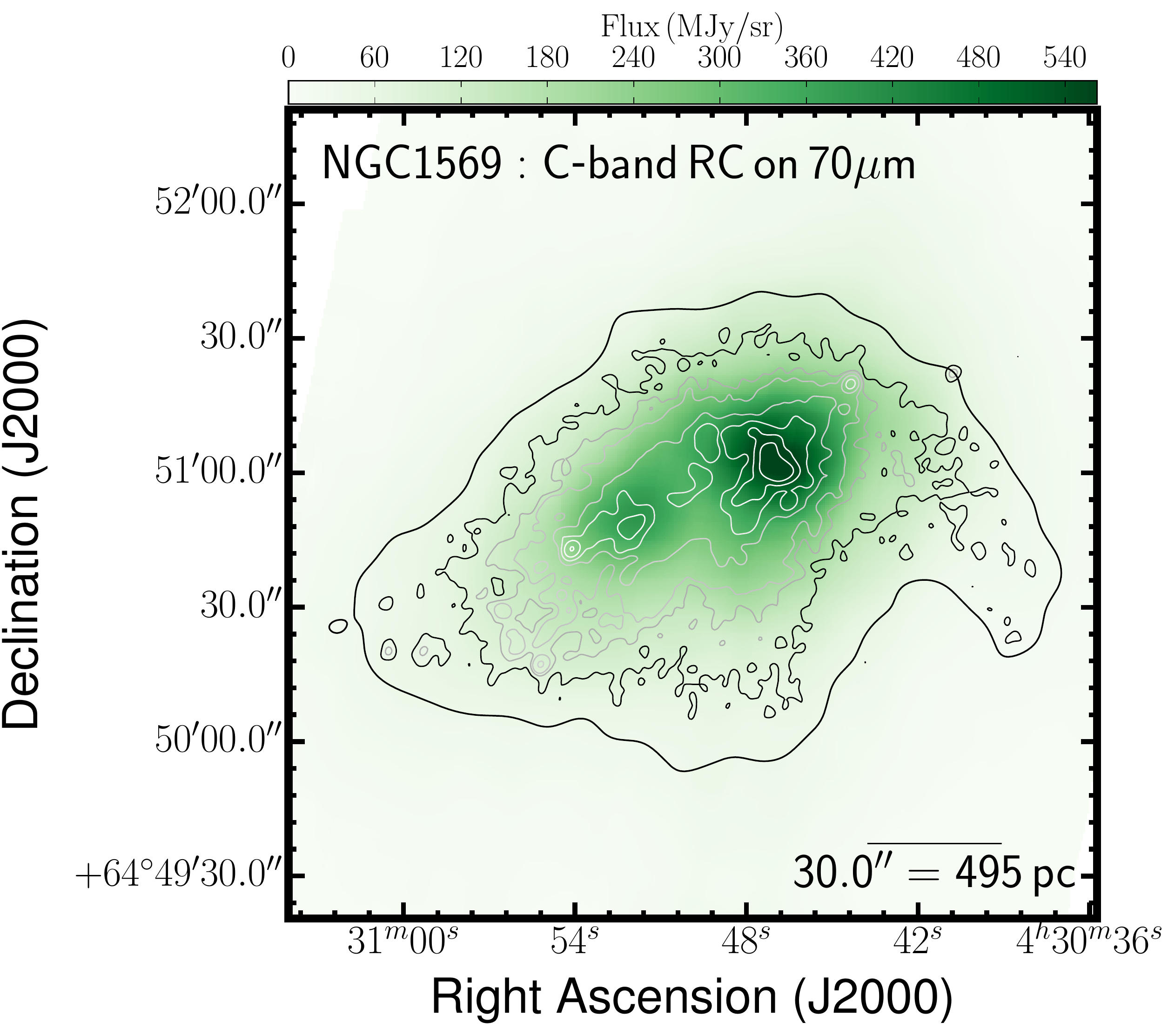} & \ 
    \includegraphics[width=0.31\linewidth,clip]{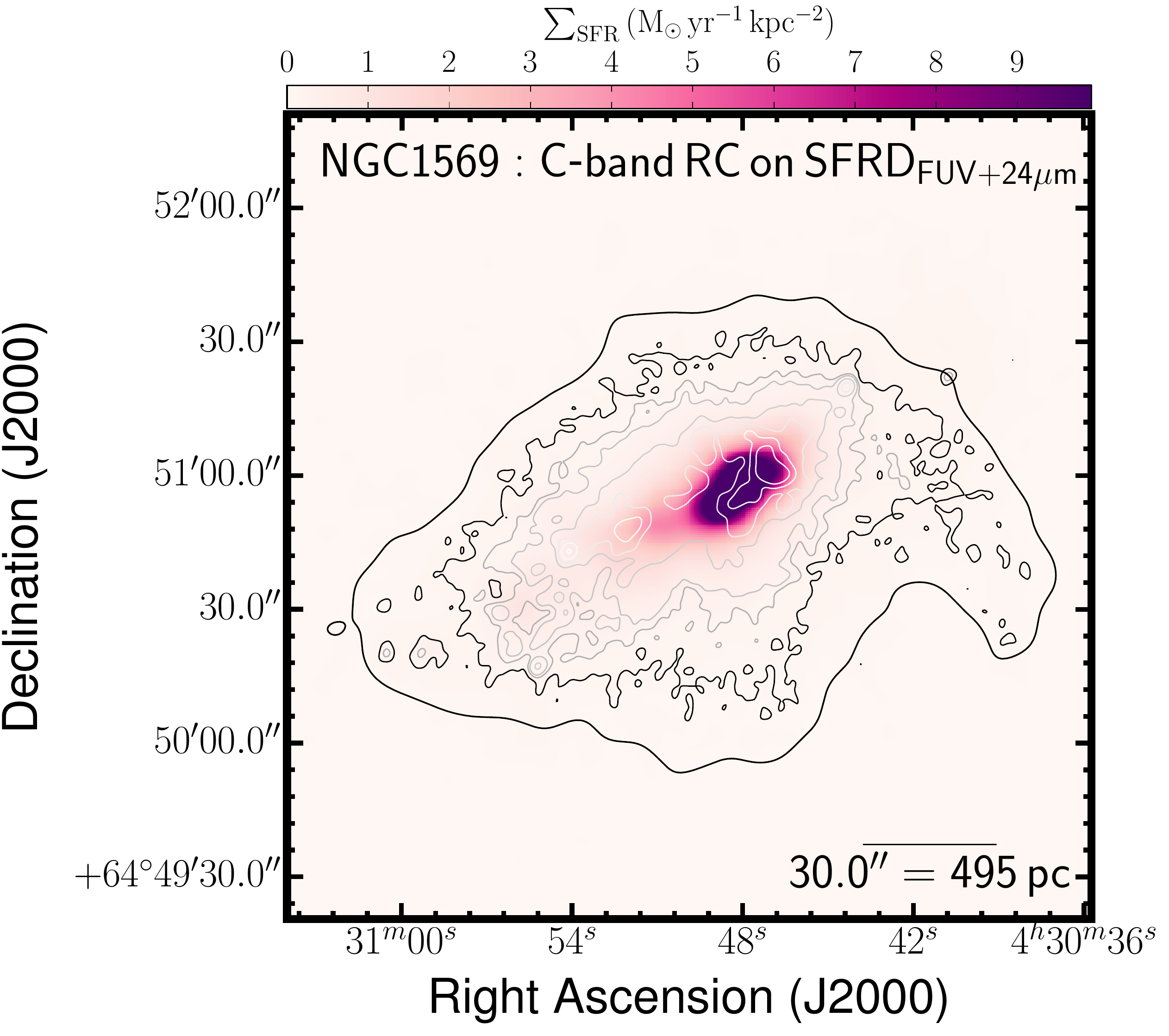} \\
  \end{tabular}
\caption[NGC\,1569 images: RC, IR, optical, and FUV]{Multi-wavelength coverage of NGC 1569 displaying a $3.0^\prime \times 3.0^\prime$ area. We show total RC flux density at the native resolution (top-left) and again with contours (top-centre). The RC contours are superposed on ancillary LITTLE THINGS images where possible: \halpha\ (middle-left); \RCNT\ obtained by subtracting the expected \RCT\ based on the \halpha-\RCT\ scaling factor of \cite{Deeg1997} from the total RC; {\em GALEX} FUV (middle-right); {\em Spitzer} 24\micron\ (bottom-left); {\em Spitzer} 70\micron\ (bottom-centre); FUV$+24{\rm \mu m}$--inferred SFRD from \citealp{Leroy2012} (bottom-right). We also show the RC that was isolated by the RC--based masking technique (top-right).}
  \label{figure:ngc1569Cc_maps}
\end{figure}

\begin{figure}
  \begin{tabular}{ccc}
    \includegraphics[width=0.31\linewidth,clip]{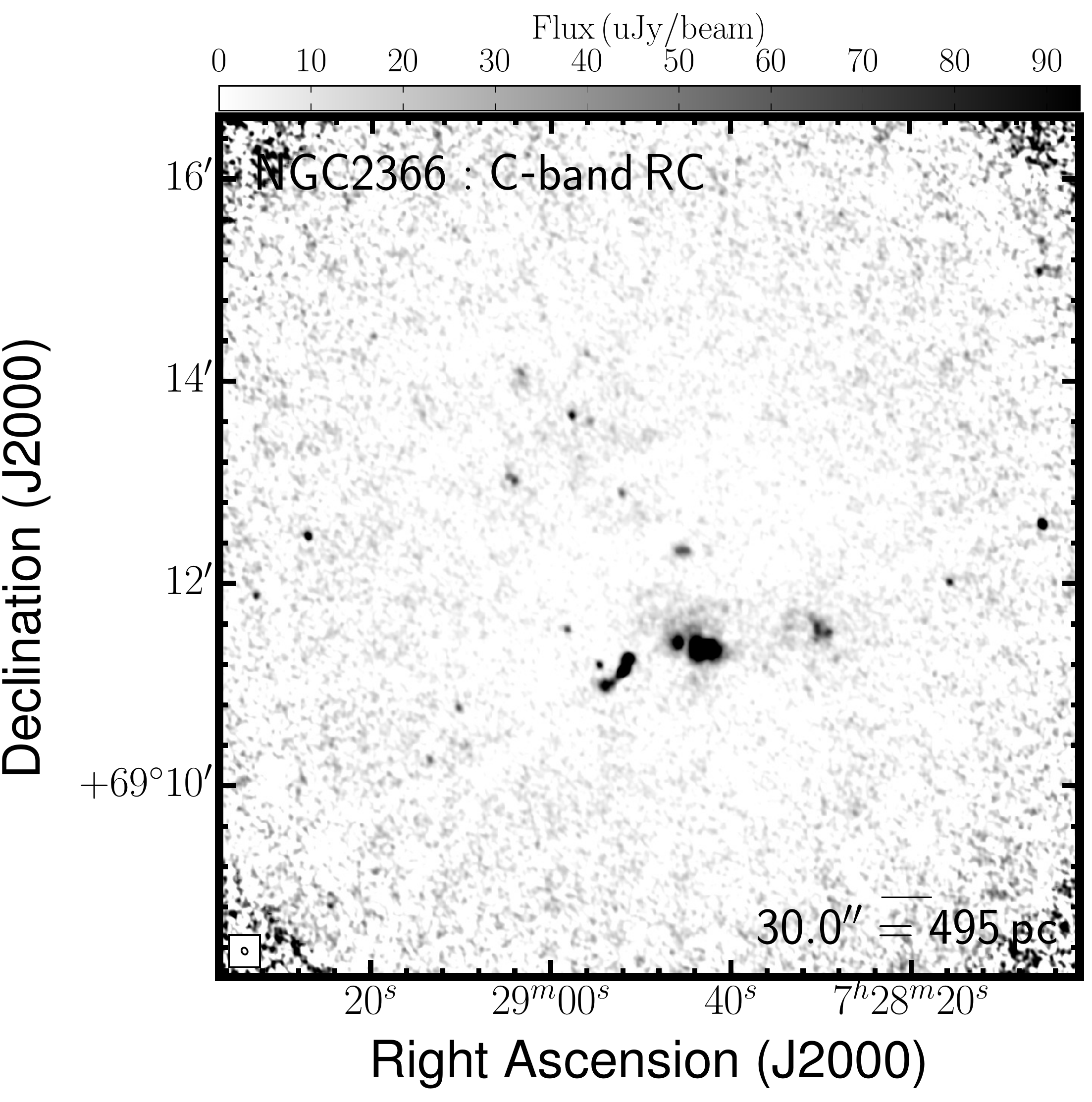} & \ 
    \includegraphics[width=0.31\linewidth,clip]{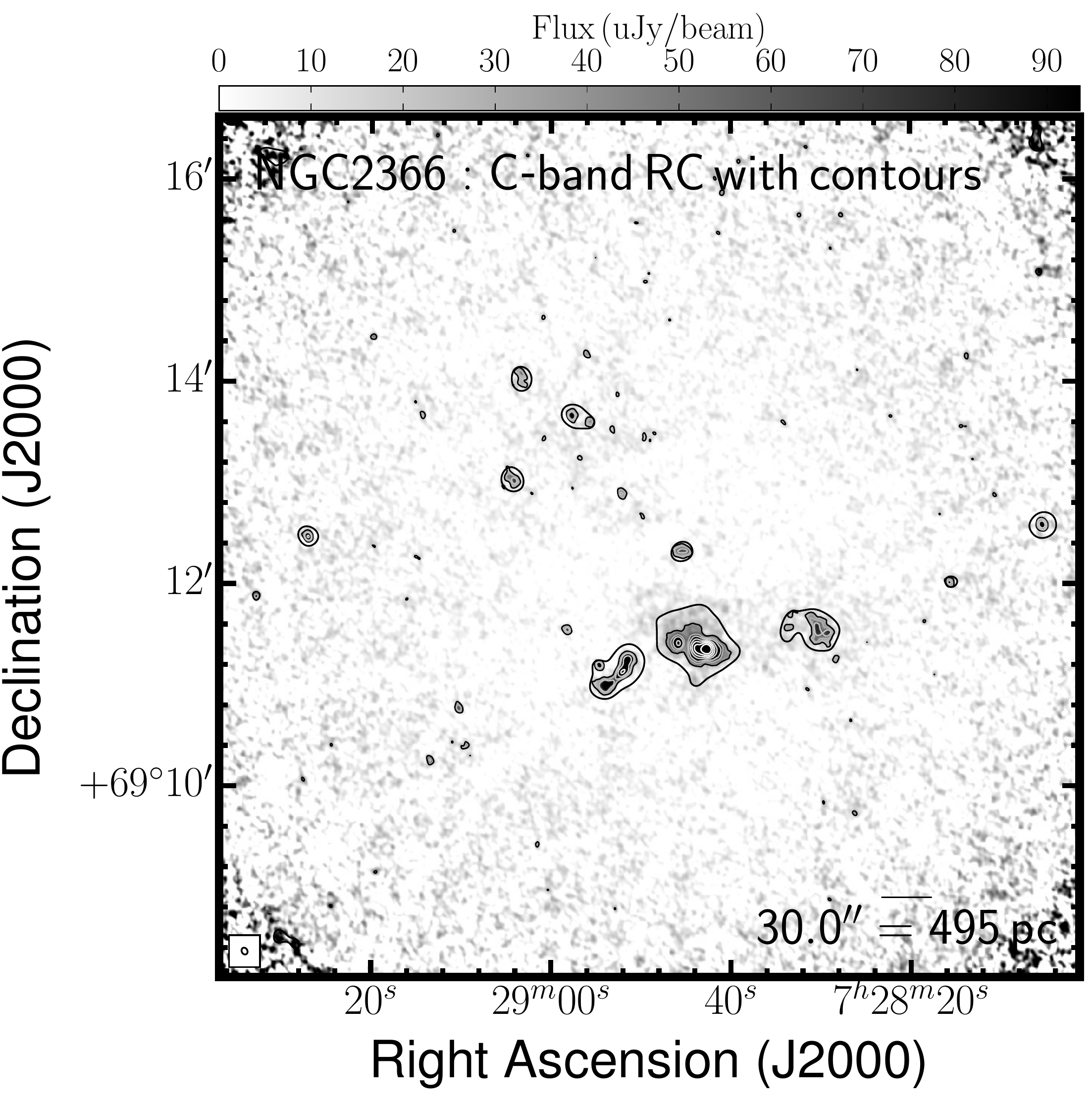} & \ 
    \includegraphics[width=0.31\linewidth,clip]{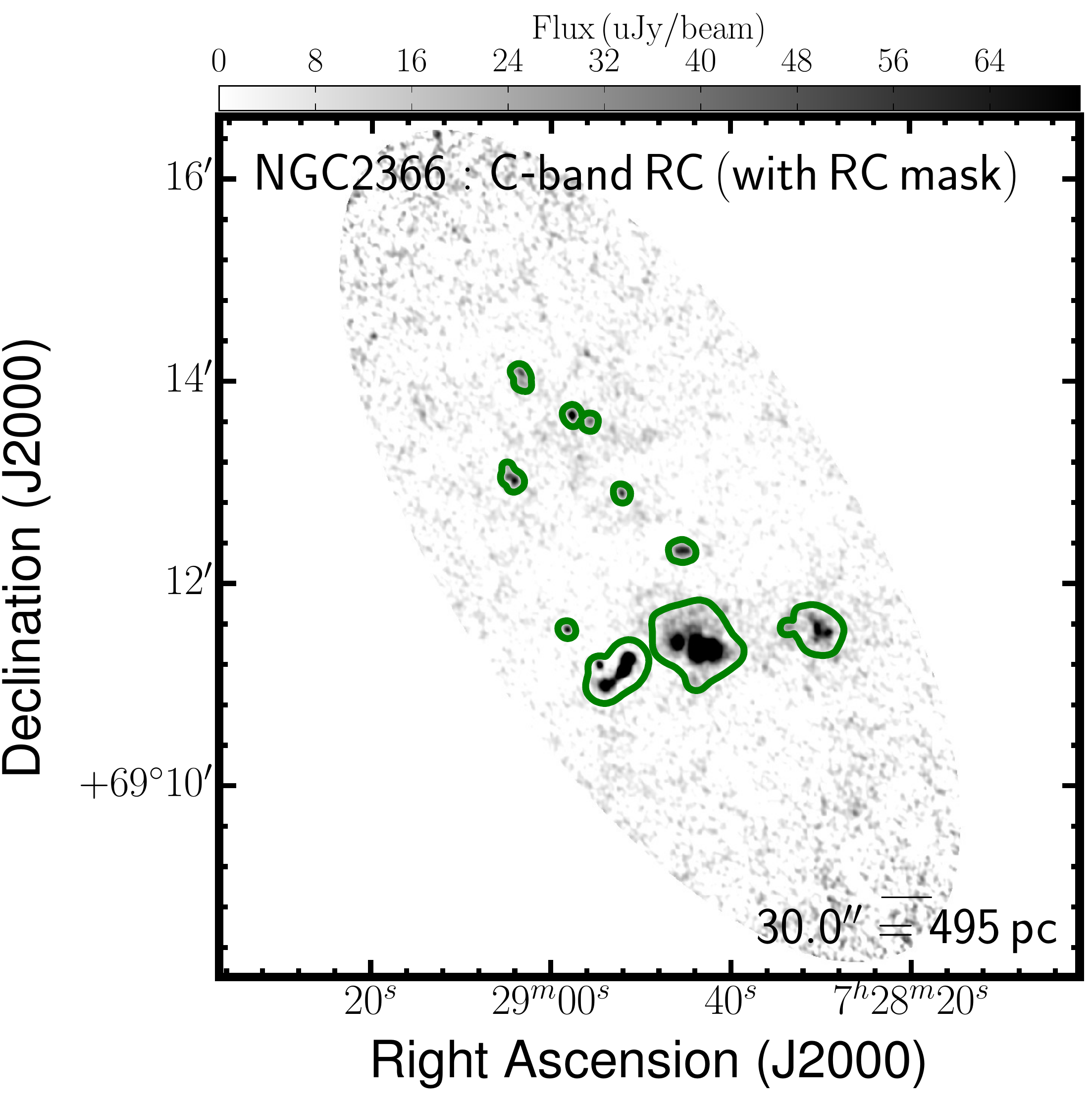} \\
    \includegraphics[width=0.31\linewidth,clip]{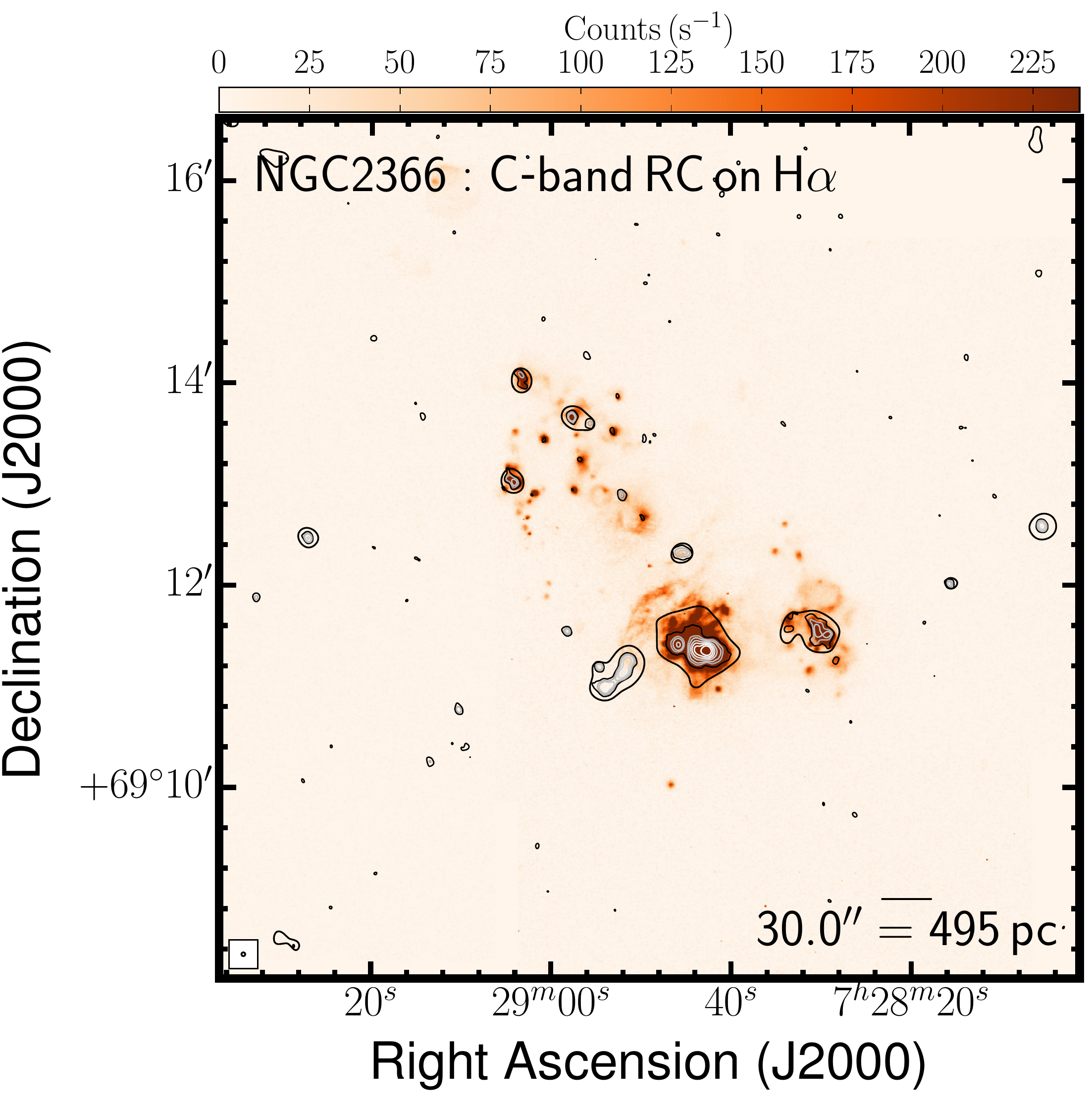} & \ 
    \includegraphics[width=0.31\linewidth,clip]{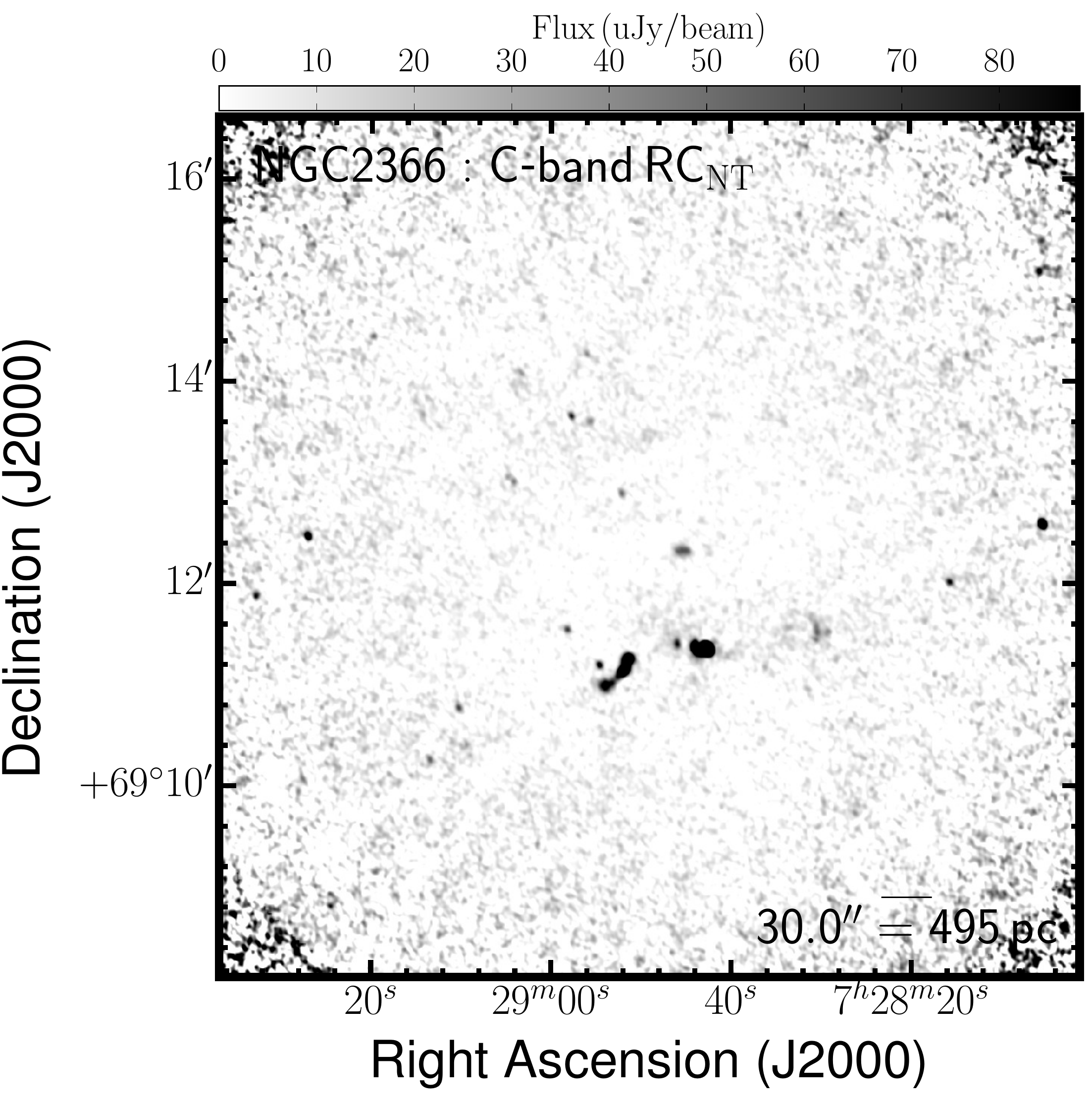} & \ 
    \includegraphics[width=0.31\linewidth,clip]{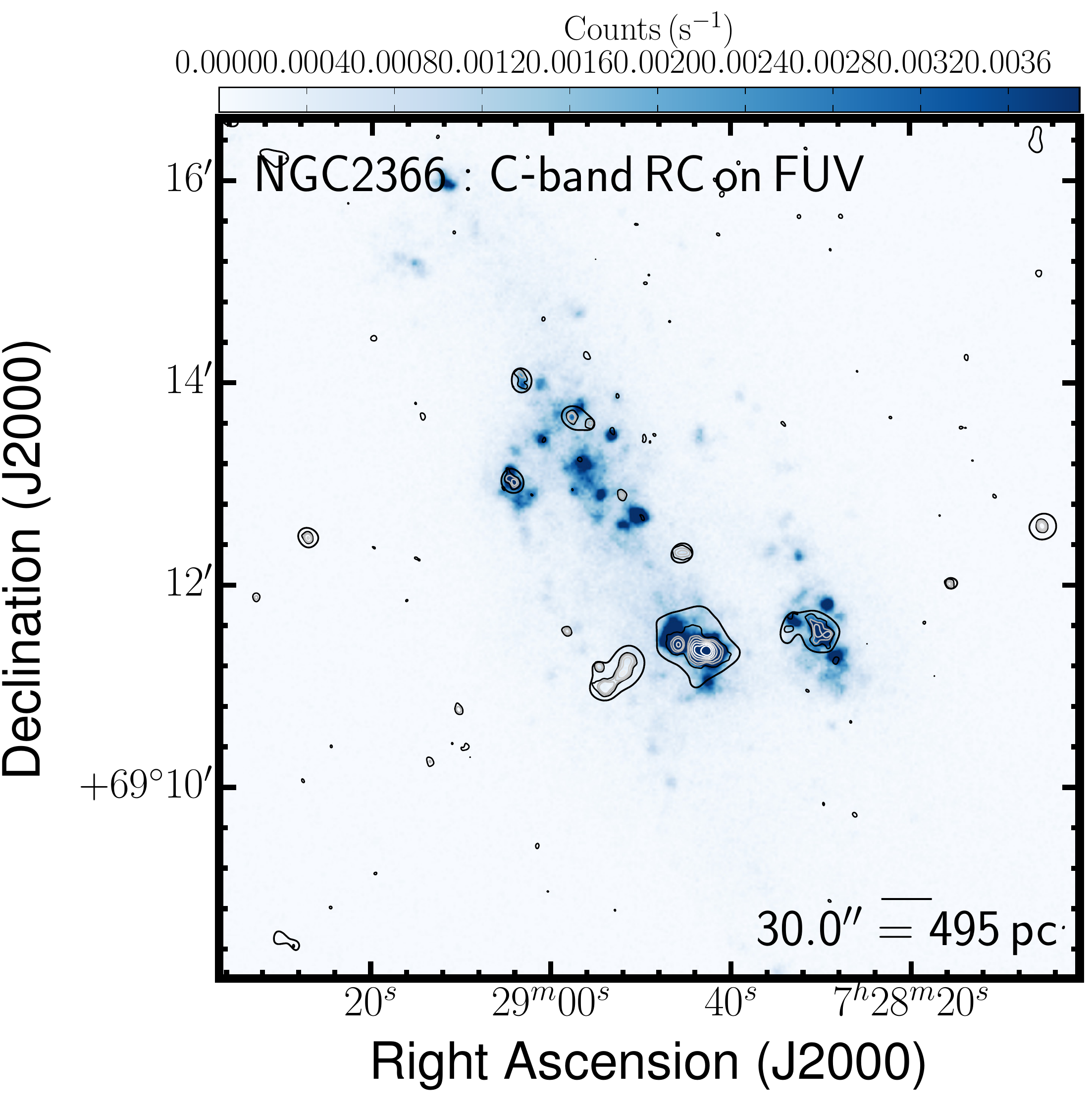} \\
    \includegraphics[width=0.31\linewidth,clip]{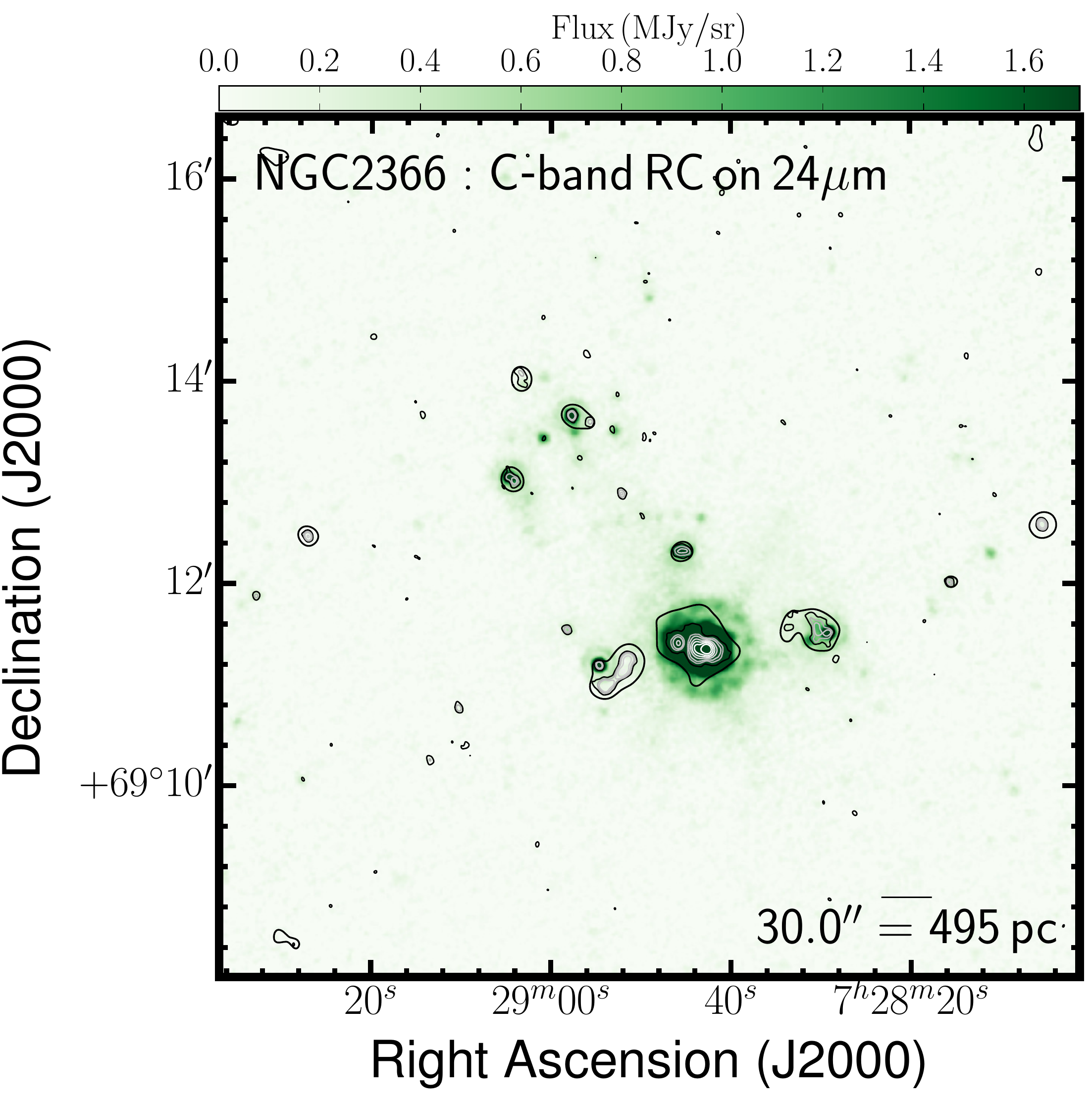} & \ 
    \includegraphics[width=0.31\linewidth,clip]{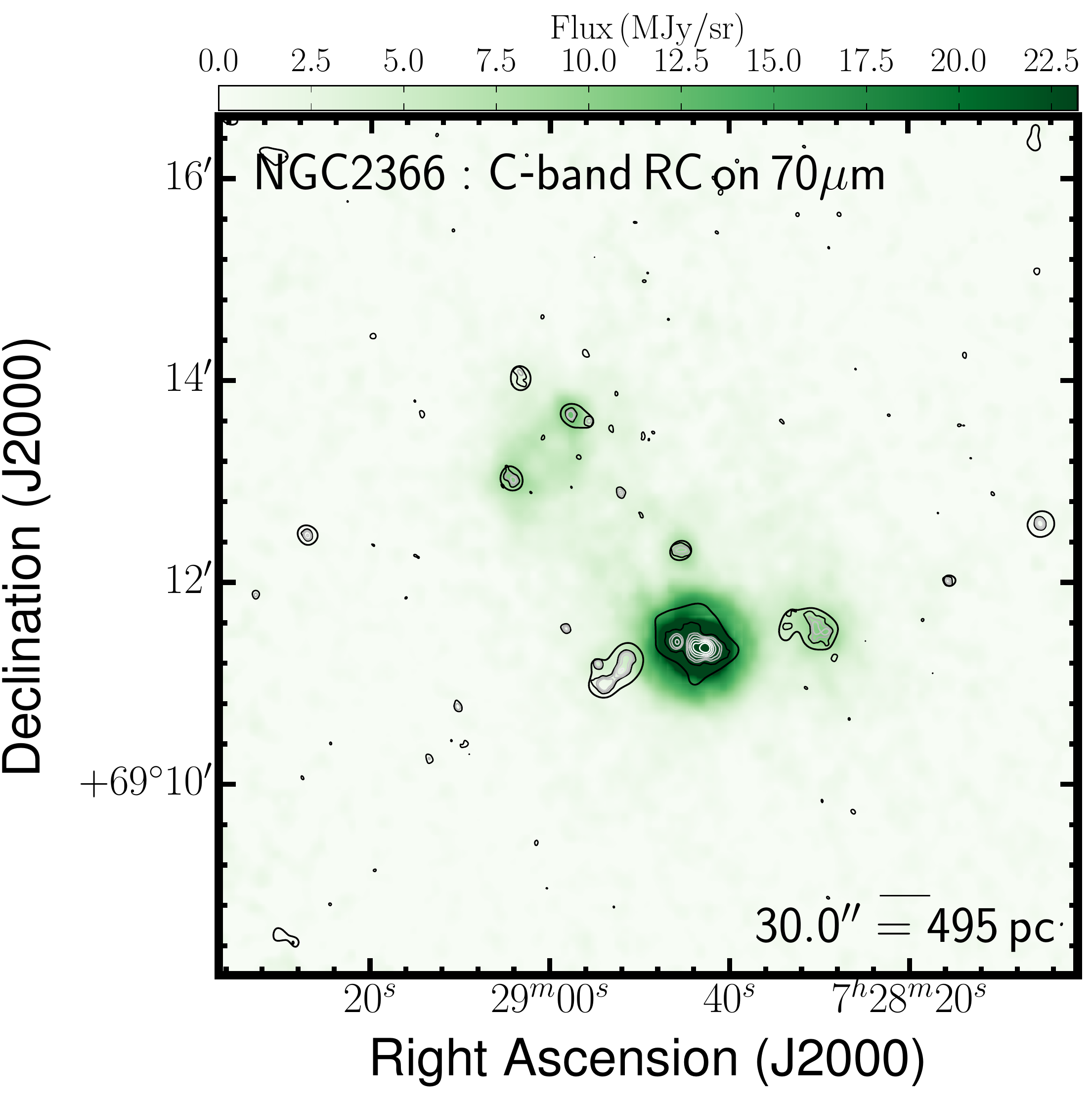} & \ 
    \includegraphics[width=0.31\linewidth,clip]{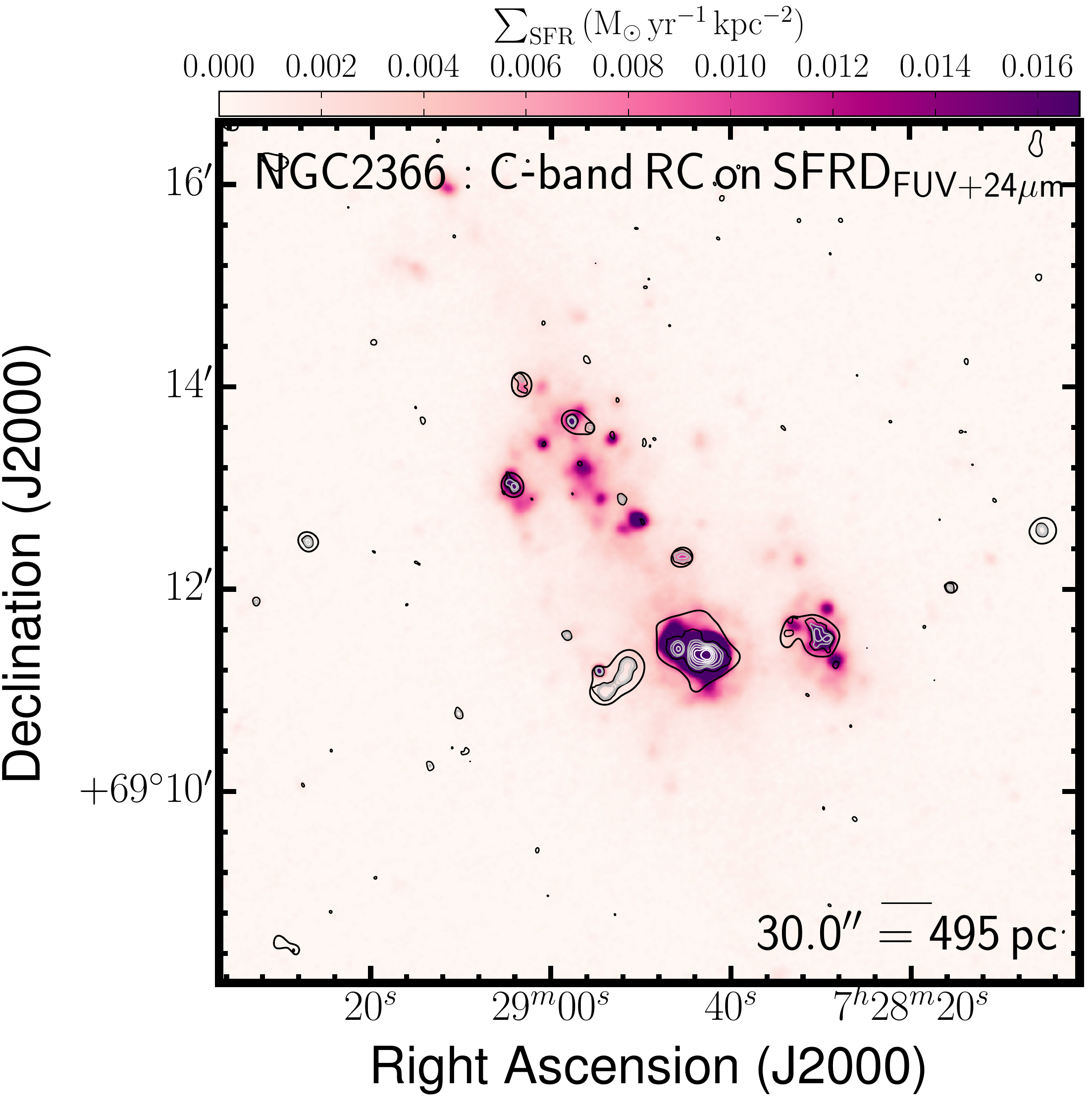} \\
  \end{tabular}
\caption[NGC\,2366 images: RC, IR, optical, and FUV]{Multi-wavelength coverage of NGC 2366 displaying a $8.5^\prime \times 8.5^\prime$ area. We show total RC flux density at the native resolution (top-left) and again with contours (top-centre). The RC contours are superposed on ancillary LITTLE THINGS images where possible: \halpha\ (middle-left); \RCNT\ obtained by subtracting the expected \RCT\ based on the \halpha-\RCT\ scaling factor of \cite{Deeg1997} from the total RC; {\em GALEX} FUV (middle-right); {\em Spitzer} 24\micron\ (bottom-left); {\em Spitzer} 70\micron\ (bottom-centre); FUV$+24{\rm \mu m}$--inferred SFRD from \citealp{Leroy2012} (bottom-right). We also show the RC that was isolated by the RC--based masking technique (top-right).}
  \label{figure:ngc2366Cc_maps}
\end{figure}

\begin{figure}
  \begin{tabular}{ccc}
    \includegraphics[width=0.31\linewidth,clip]{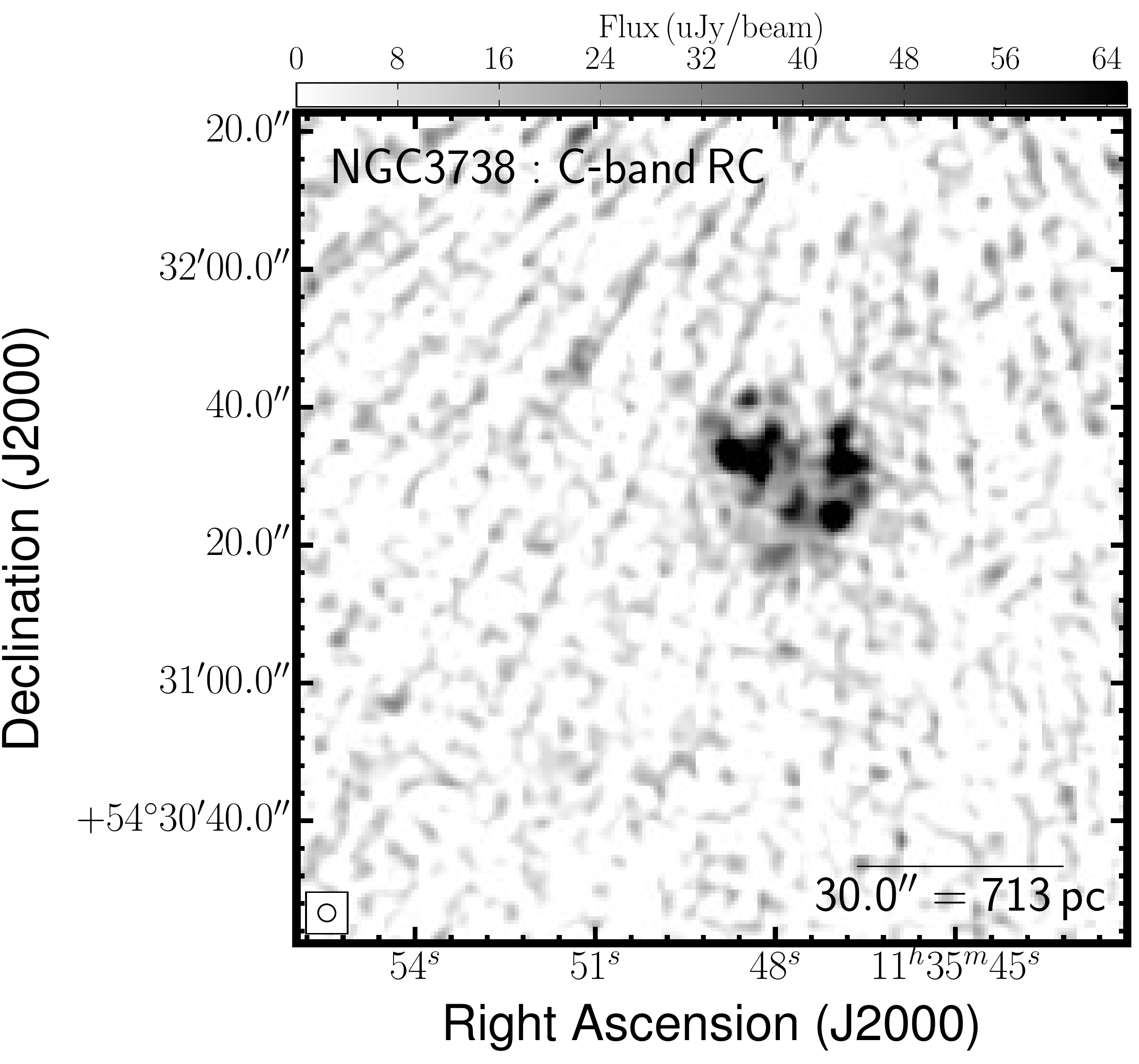} & \ 
    \includegraphics[width=0.31\linewidth,clip]{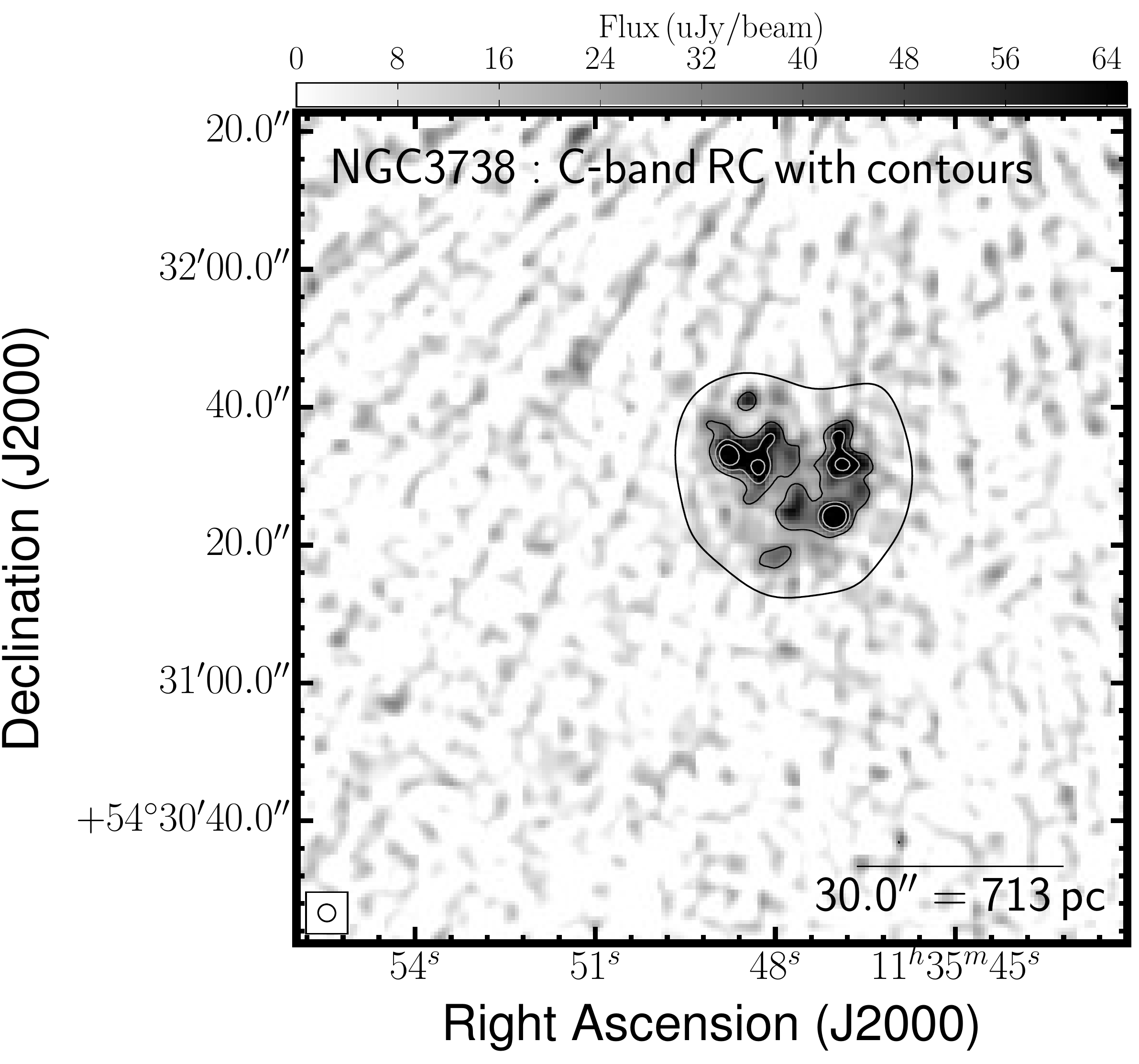} & \ 
    \includegraphics[width=0.31\linewidth,clip]{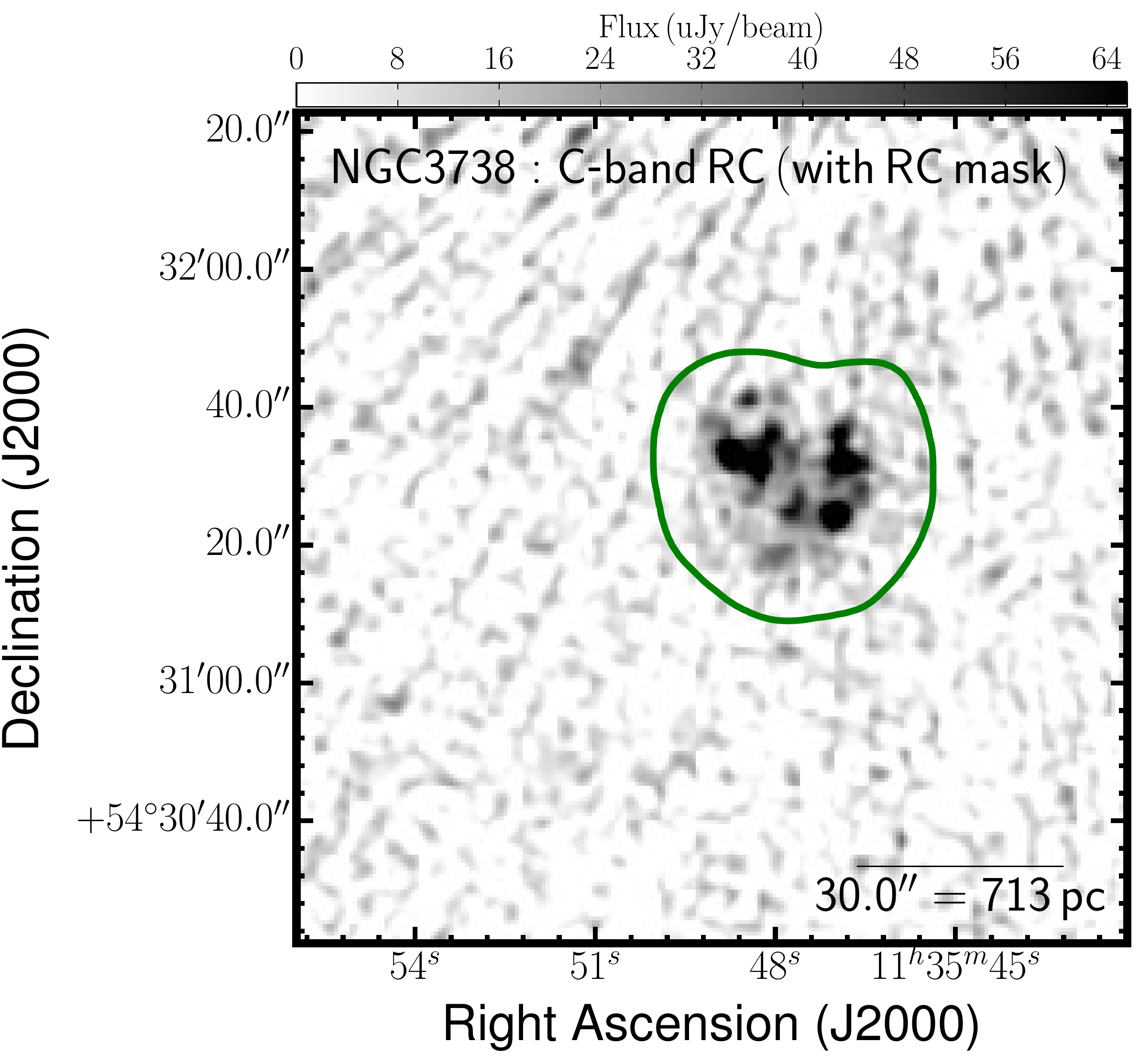} \\
    \includegraphics[width=0.31\linewidth,clip]{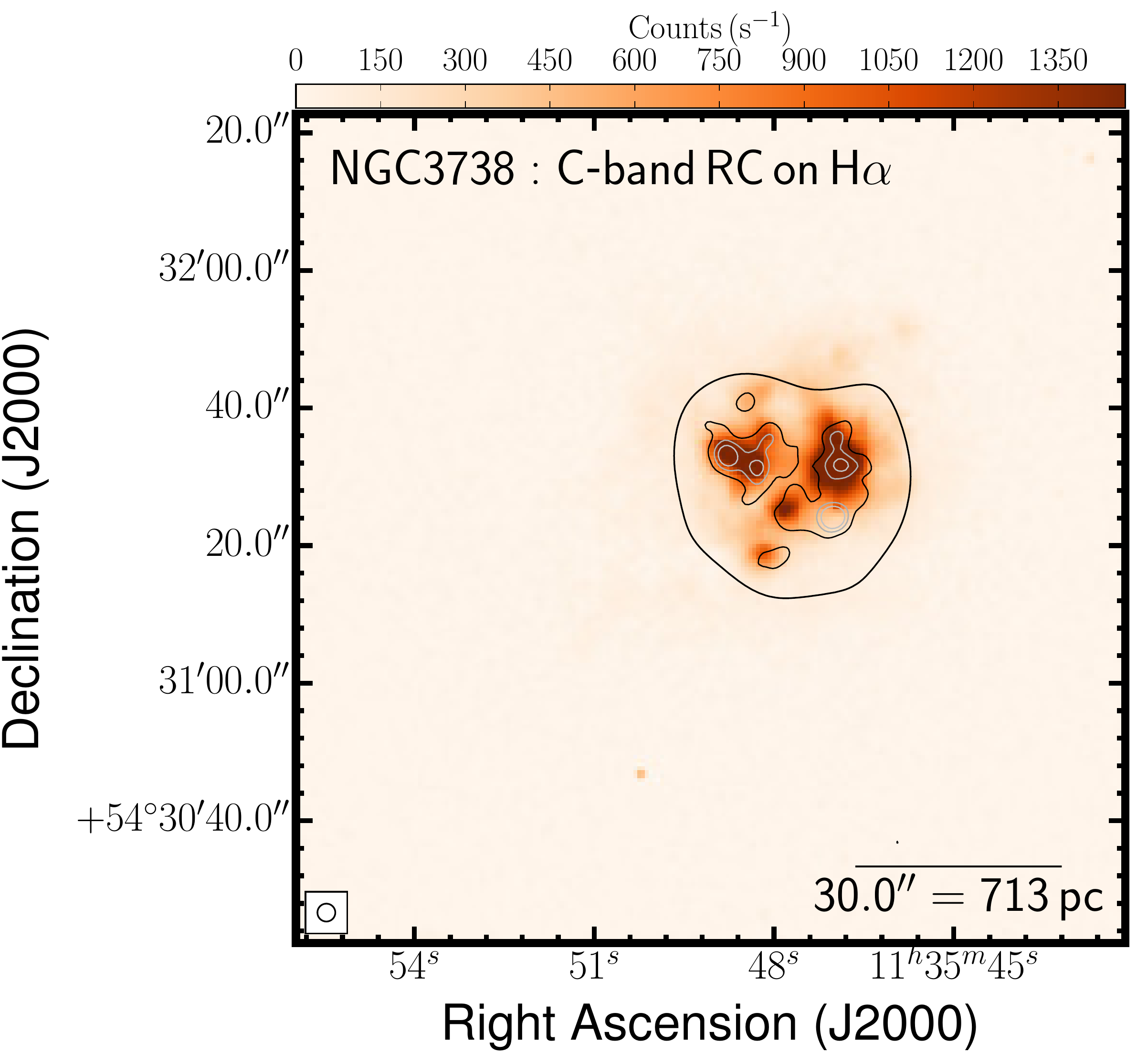} & \ 
    \includegraphics[width=0.31\linewidth,clip]{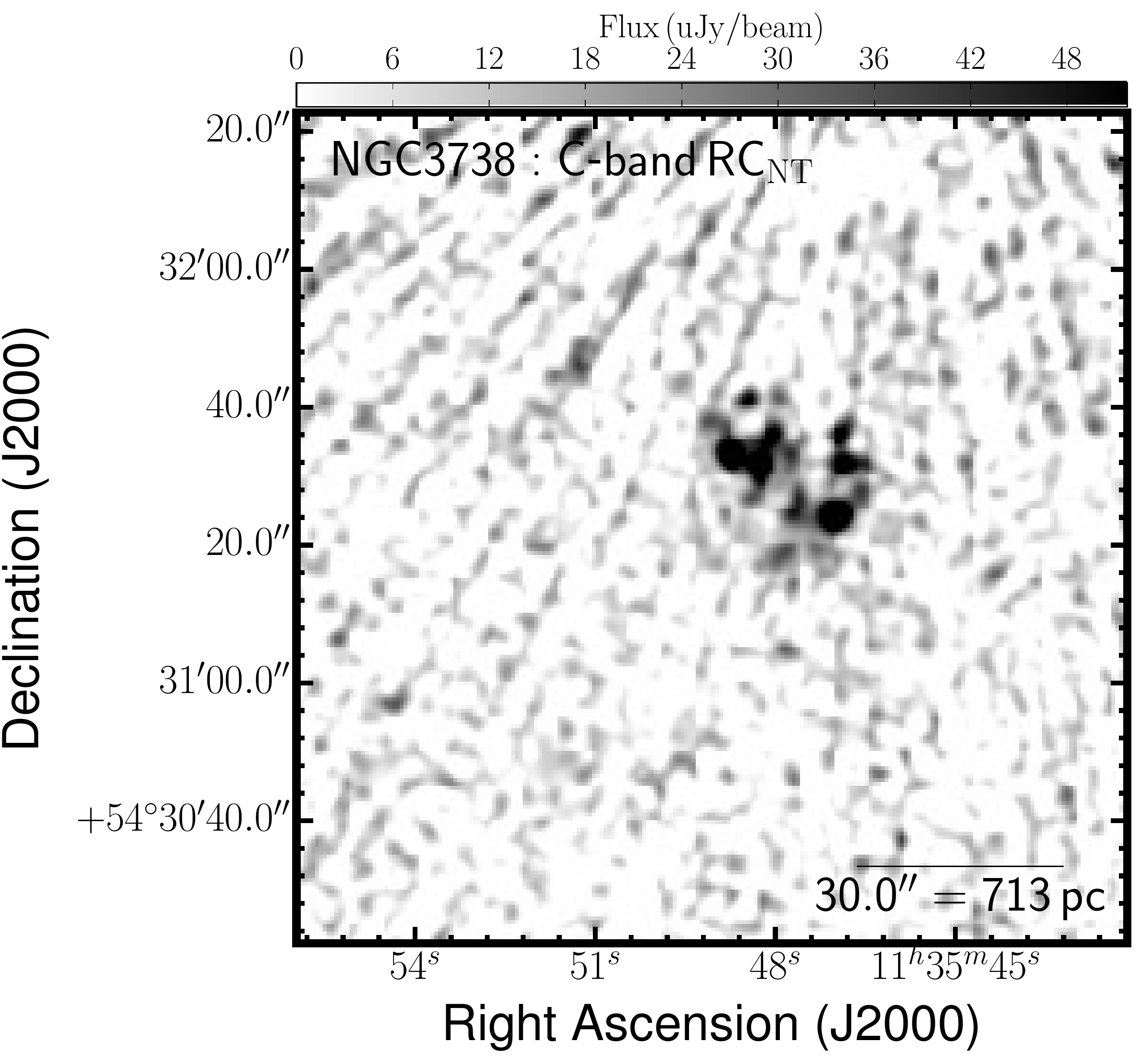} & \ 
    \includegraphics[width=0.31\linewidth,clip]{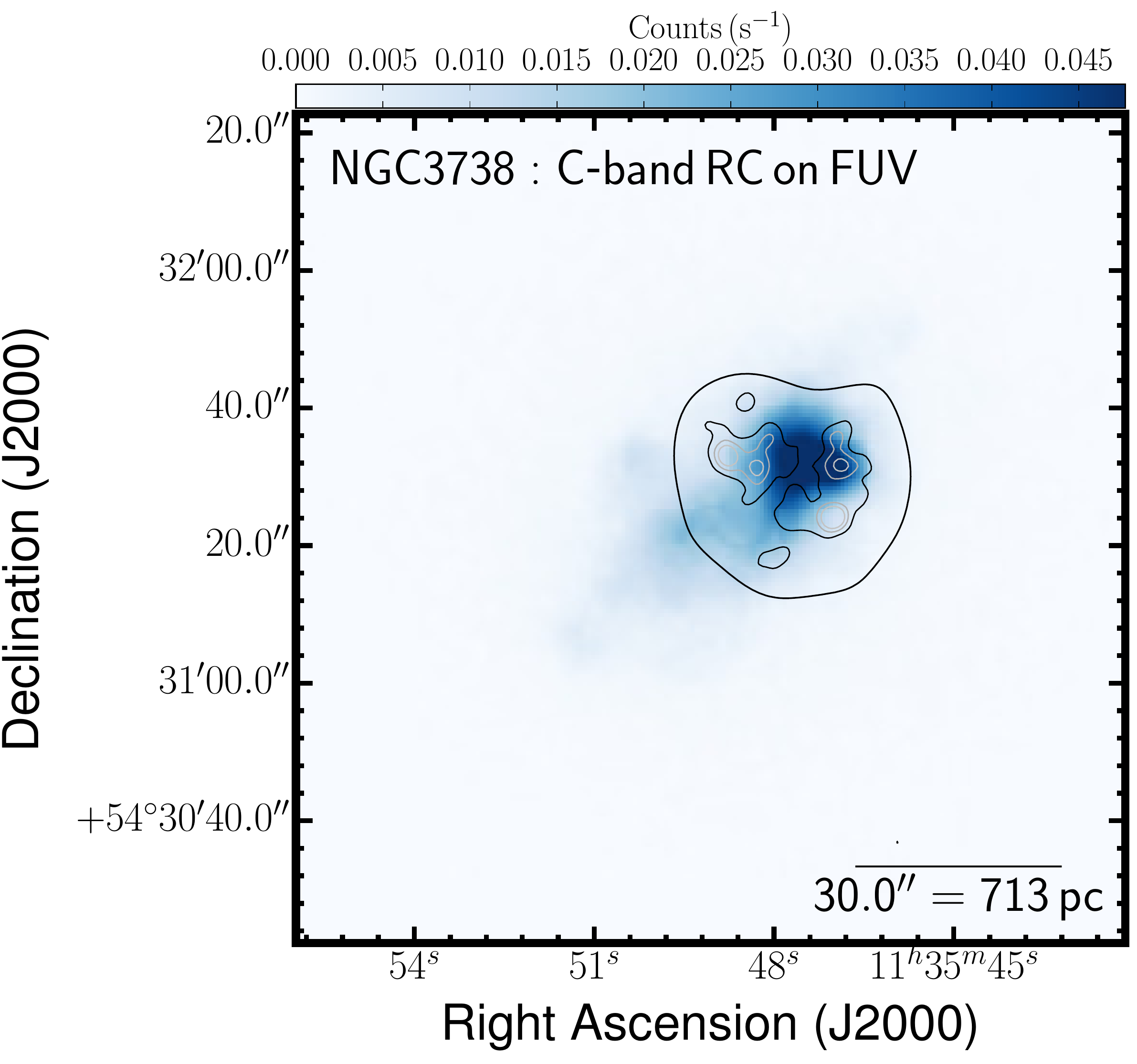} \\
    \includegraphics[width=0.31\linewidth,clip]{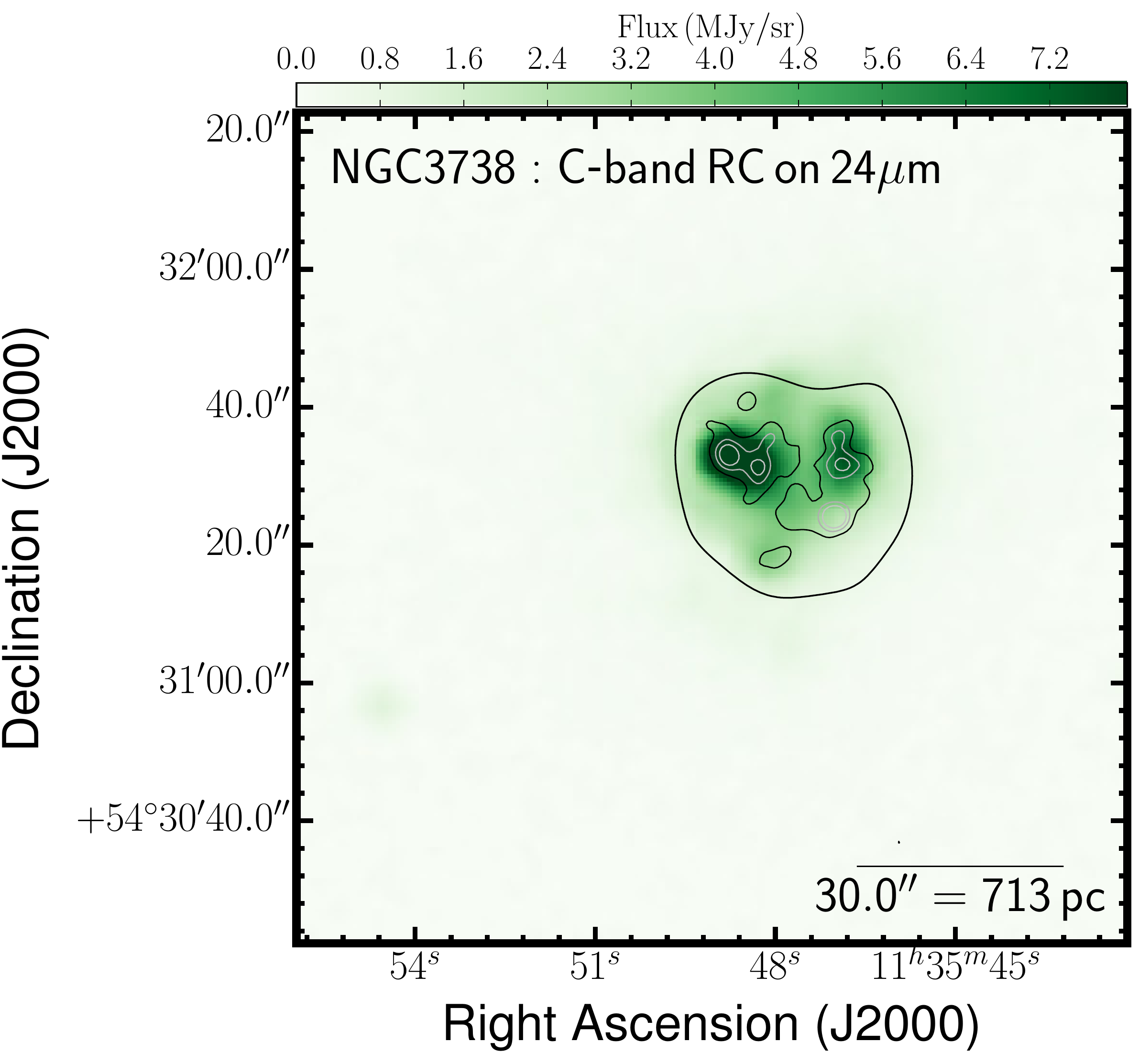} & \ 
    \includegraphics[width=0.31\linewidth,clip]{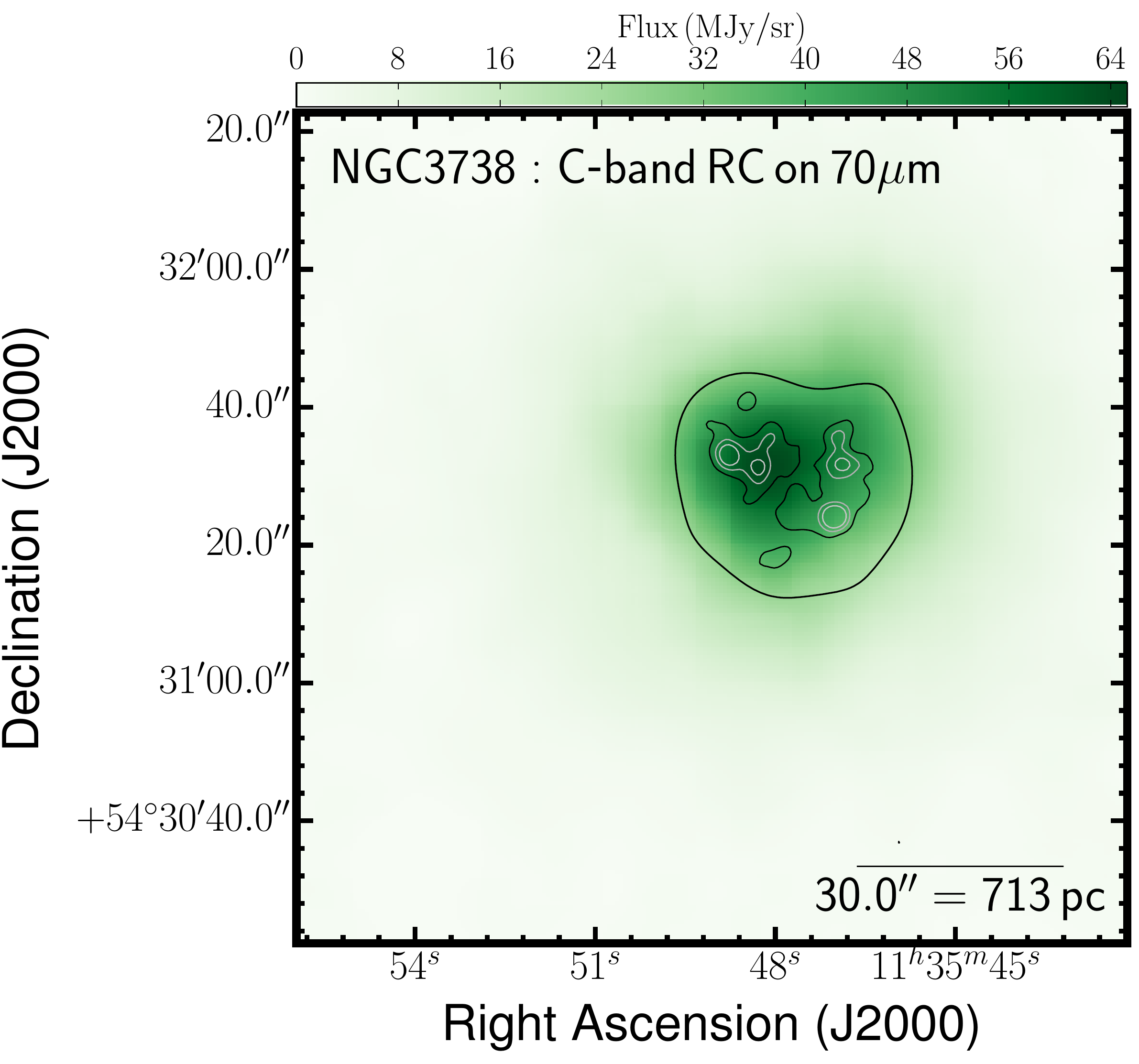} & \ 
    \includegraphics[width=0.31\linewidth,clip]{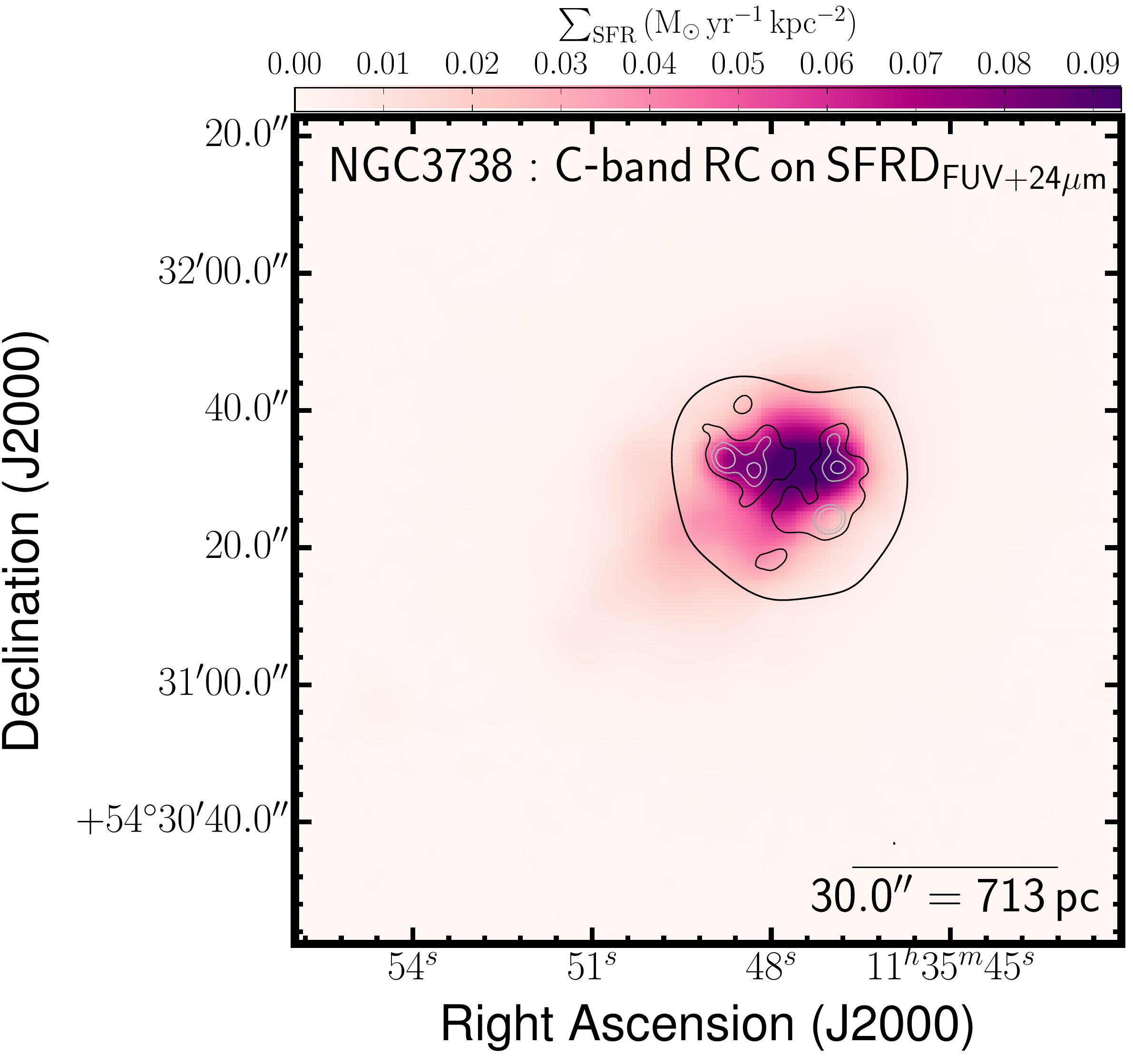} \\
  \end{tabular}
\caption[NGC\,3738 images: RC, IR, optical, and FUV]{Multi-wavelength coverage of NGC 3738 displaying a $2.0^\prime \times 2.0^\prime$ area. We show total RC flux density at the native resolution (top-left) and again with contours (top-centre). The RC contours are superposed on ancillary LITTLE THINGS images where possible: \halpha\ (middle-left); \RCNT\ obtained by subtracting the expected \RCT\ based on the \halpha-\RCT\ scaling factor of \cite{Deeg1997} from the total RC; {\em GALEX} FUV (middle-right); {\em Spitzer} 24\micron\ (bottom-left); {\em Spitzer} 70\micron\ (bottom-centre); FUV$+24{\rm \mu m}$--inferred SFRD from \citealp{Leroy2012} (bottom-right). We also show the RC that was isolated by the RC--based masking technique (top-right).}
  \label{figure:ngc3738Cc_maps}
\end{figure}

\begin{figure}
  \begin{tabular}{ccc}
    \includegraphics[width=0.31\linewidth,clip]{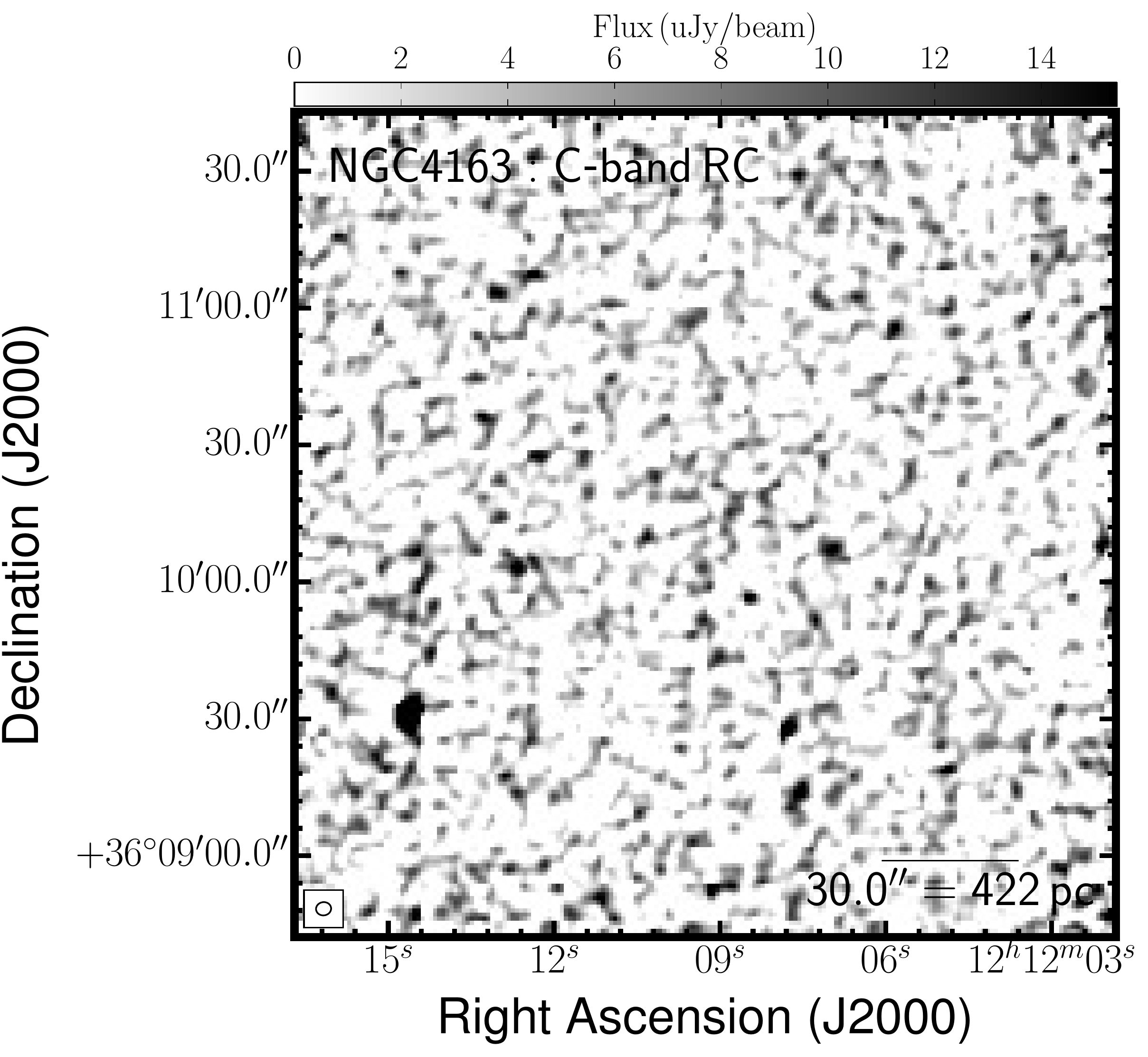} & \ 
    \includegraphics[width=0.31\linewidth,clip]{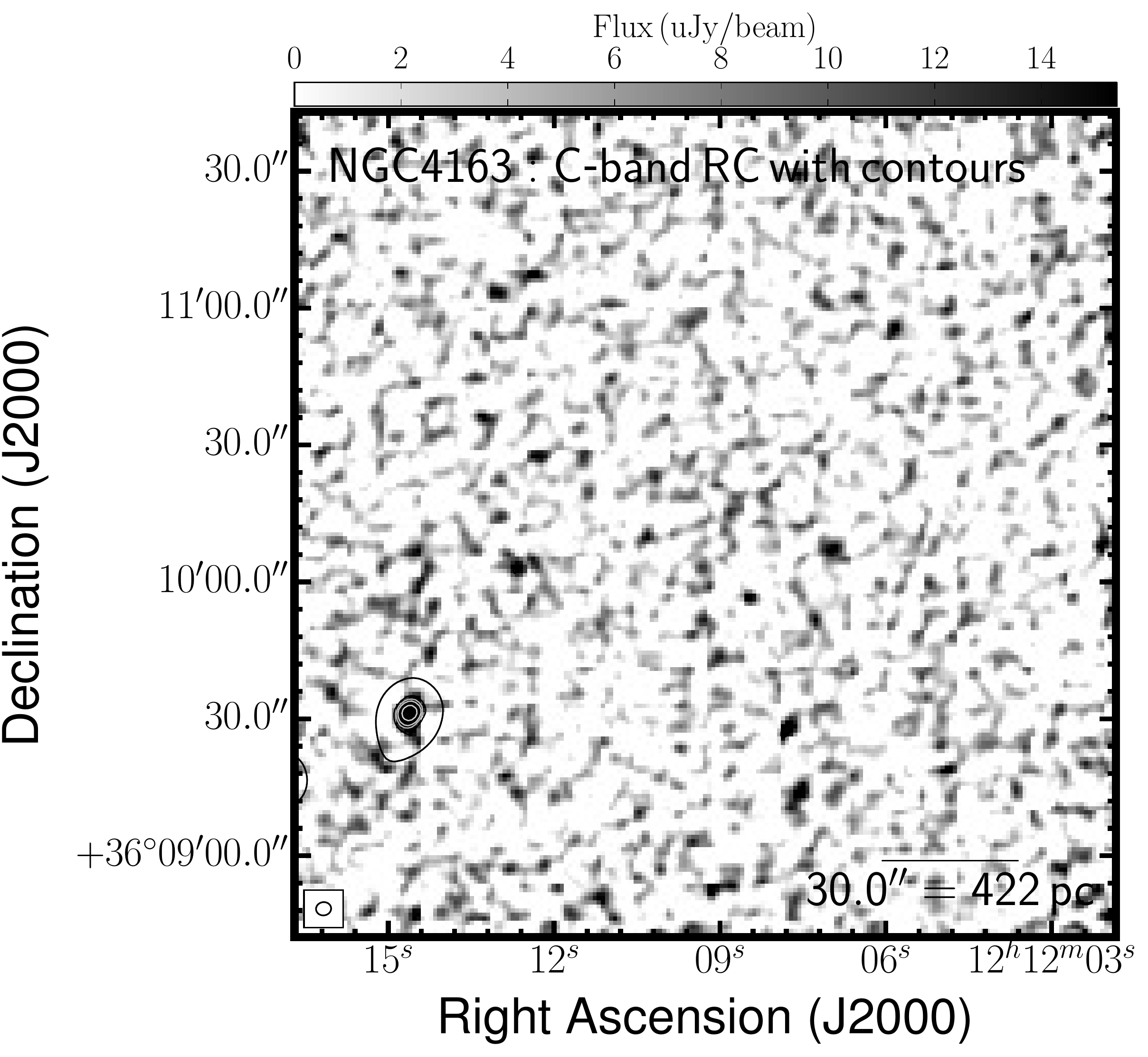} & \ 
    \includegraphics[width=0.31\linewidth,clip]{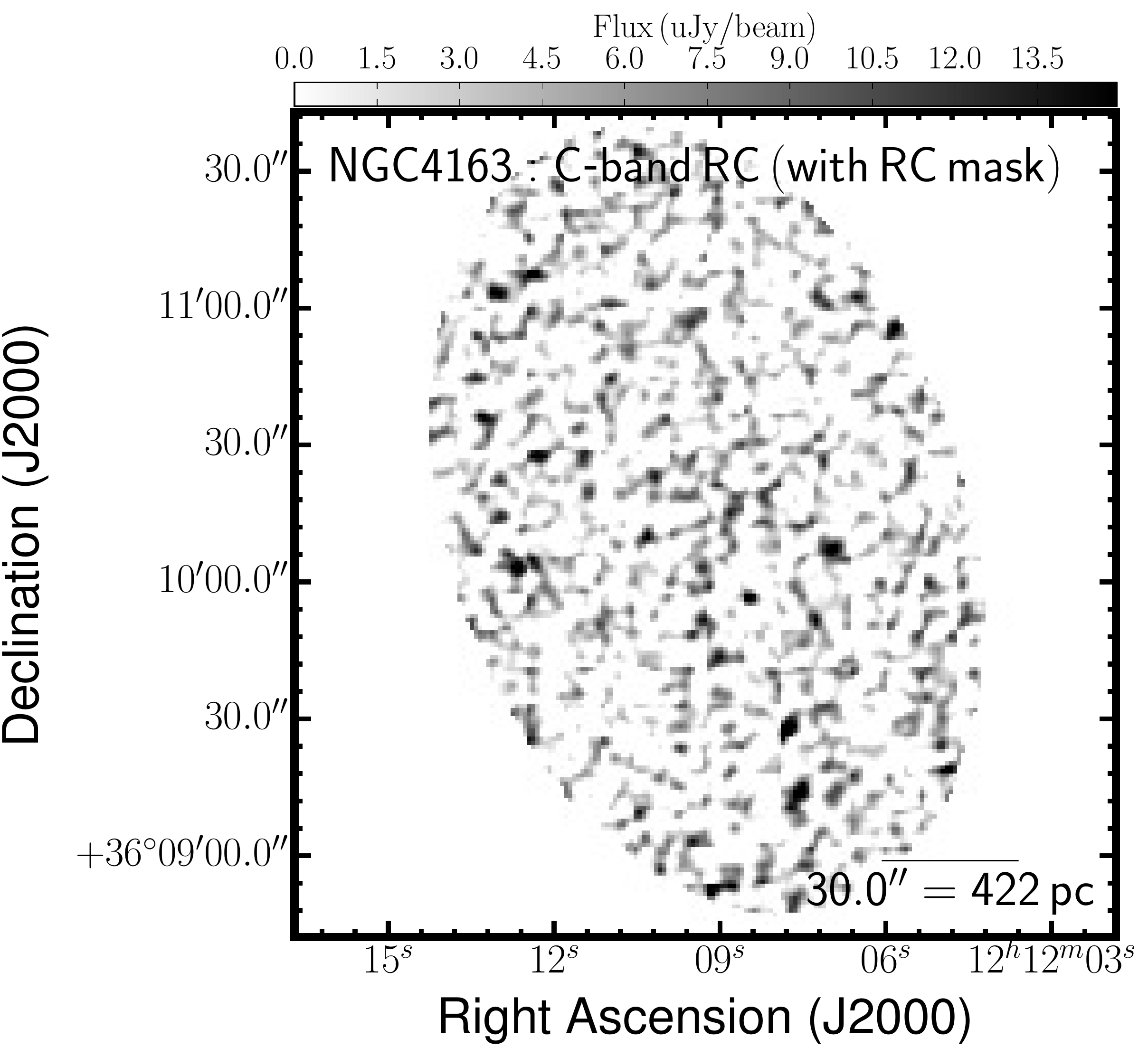} \\
    \includegraphics[width=0.31\linewidth,clip]{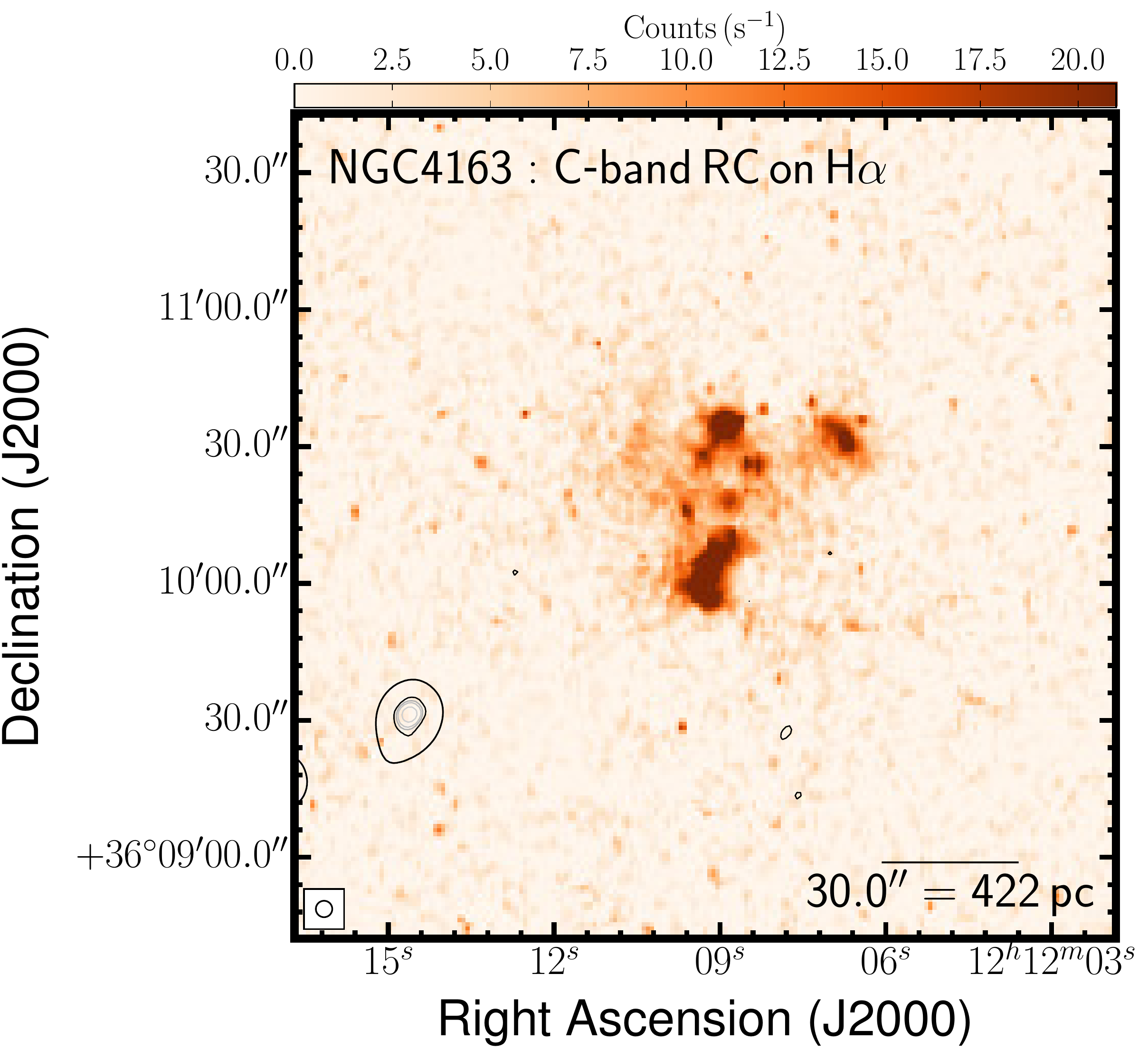} & \ 
    \includegraphics[width=0.31\linewidth,clip]{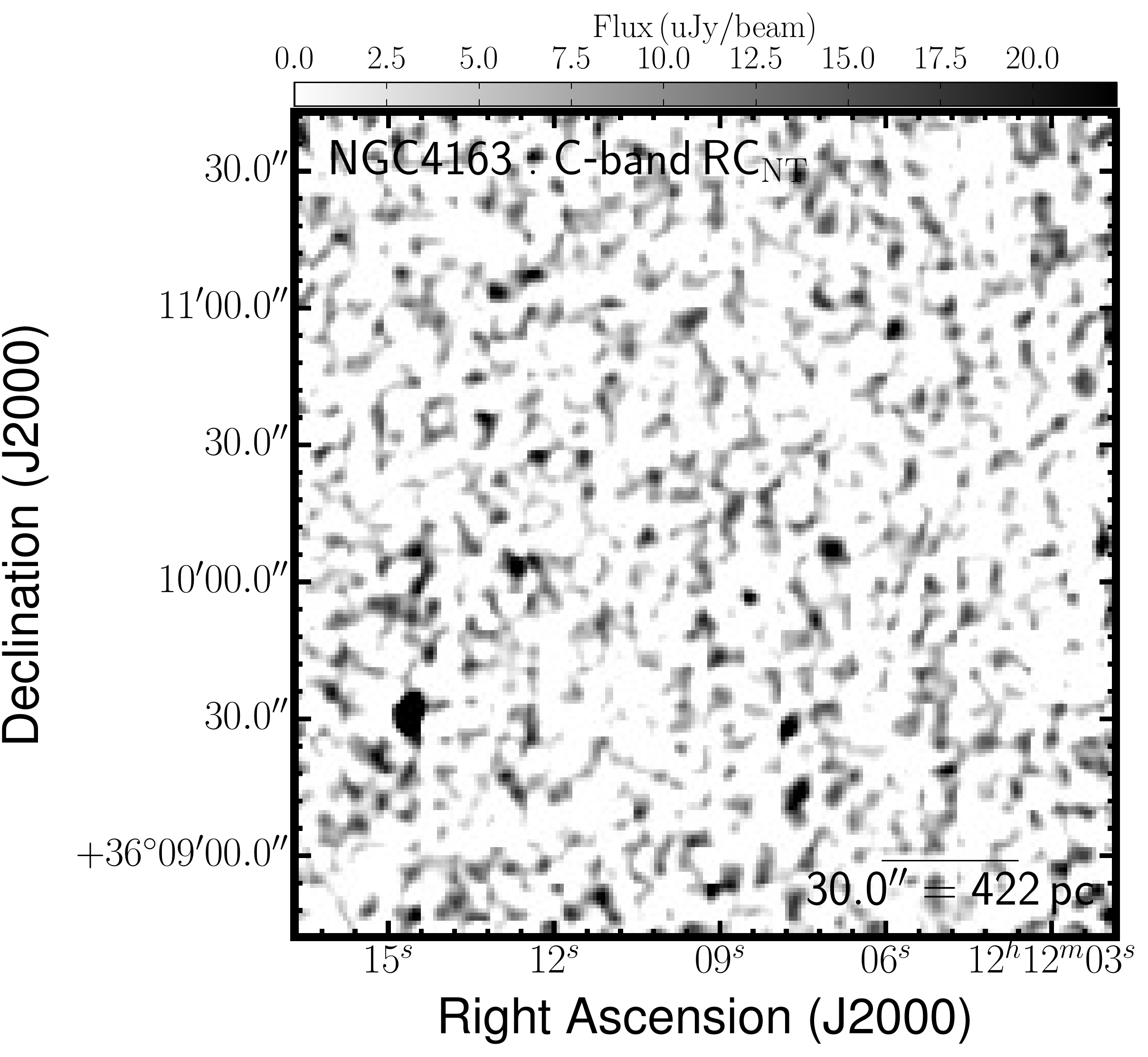} & \ 
    \includegraphics[width=0.31\linewidth,clip]{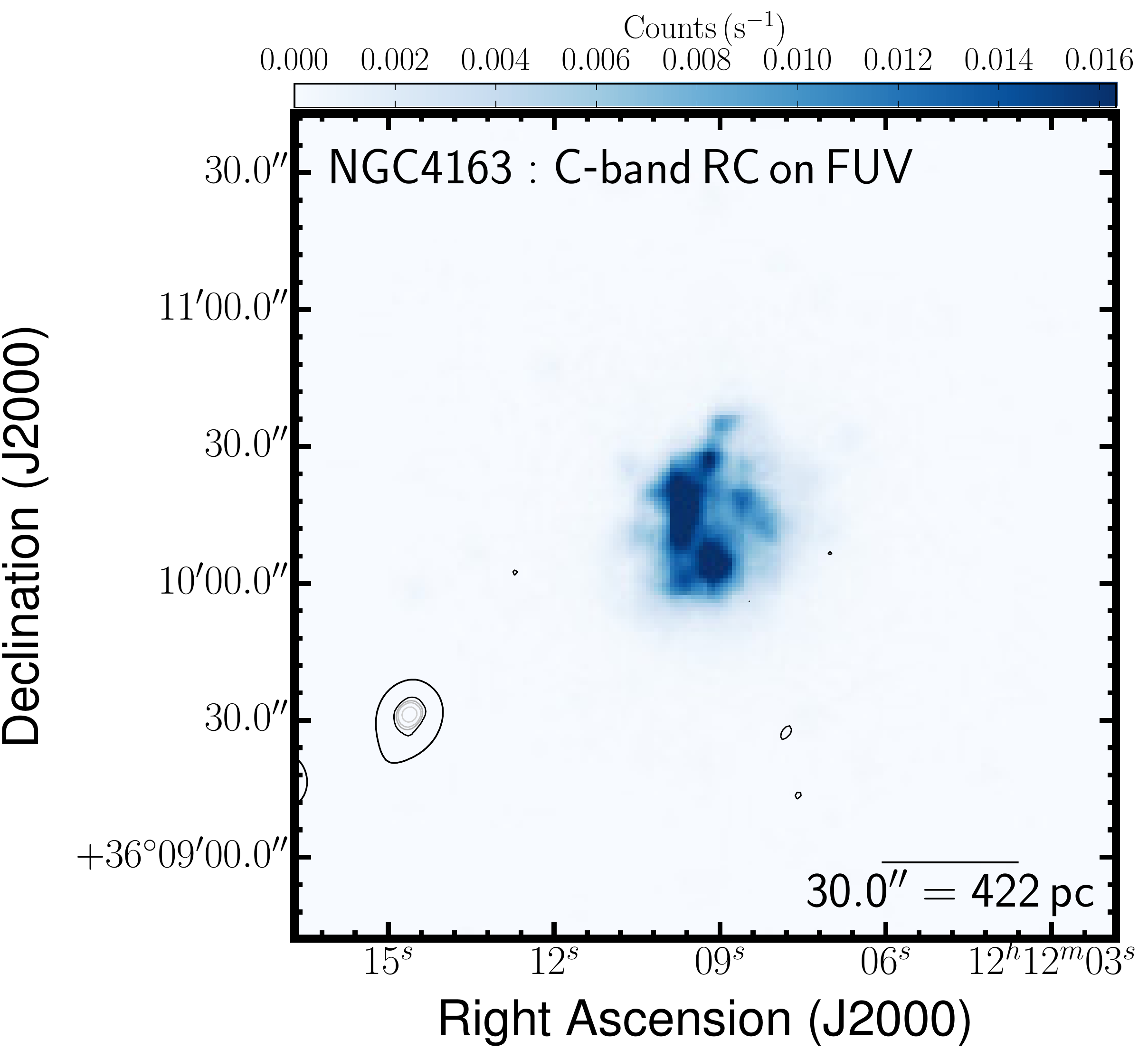} \\
    \includegraphics[width=0.31\linewidth,clip]{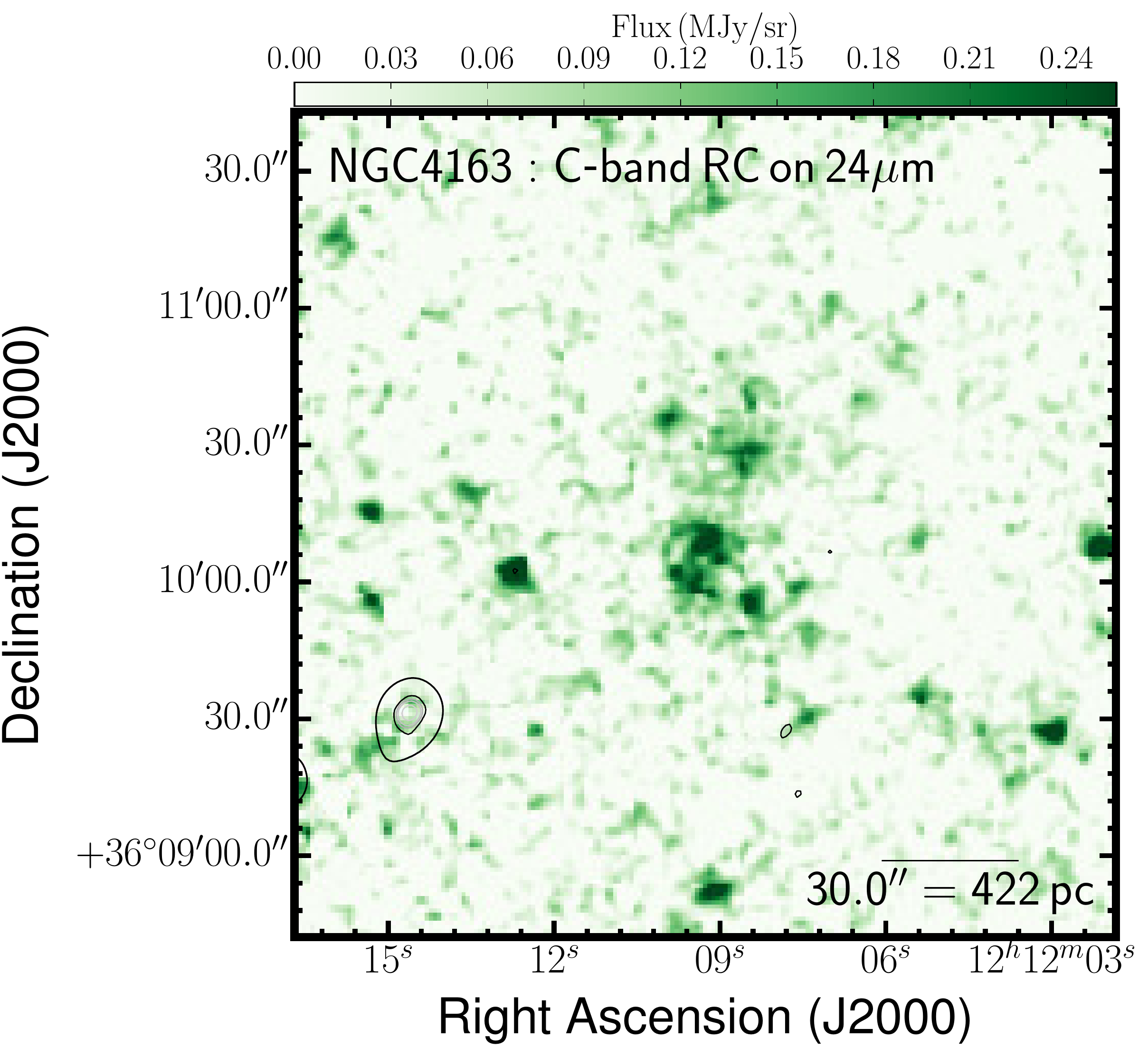} & \ 
    \includegraphics[width=0.31\linewidth,clip]{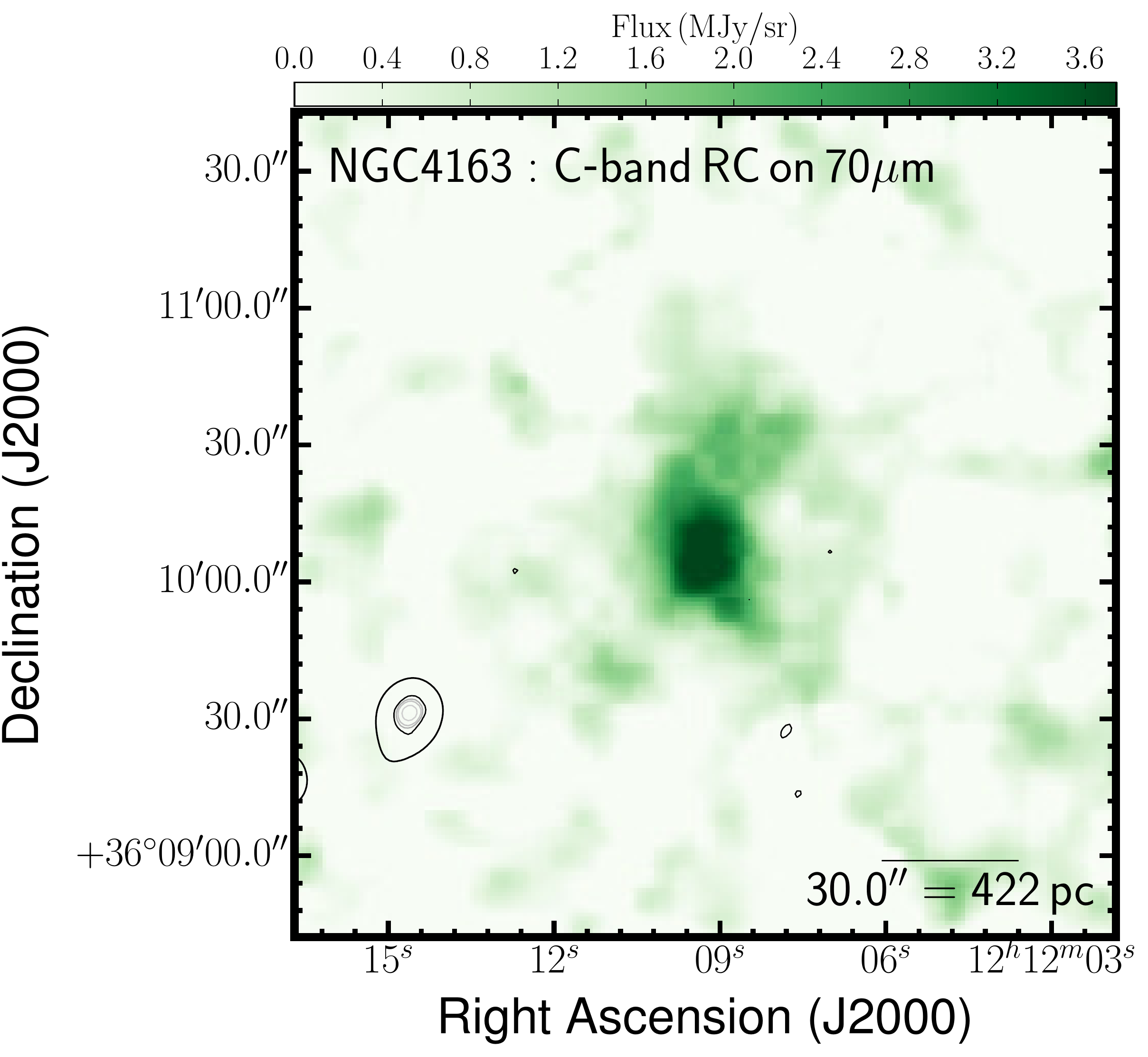} & \ 
    \includegraphics[width=0.31\linewidth,clip]{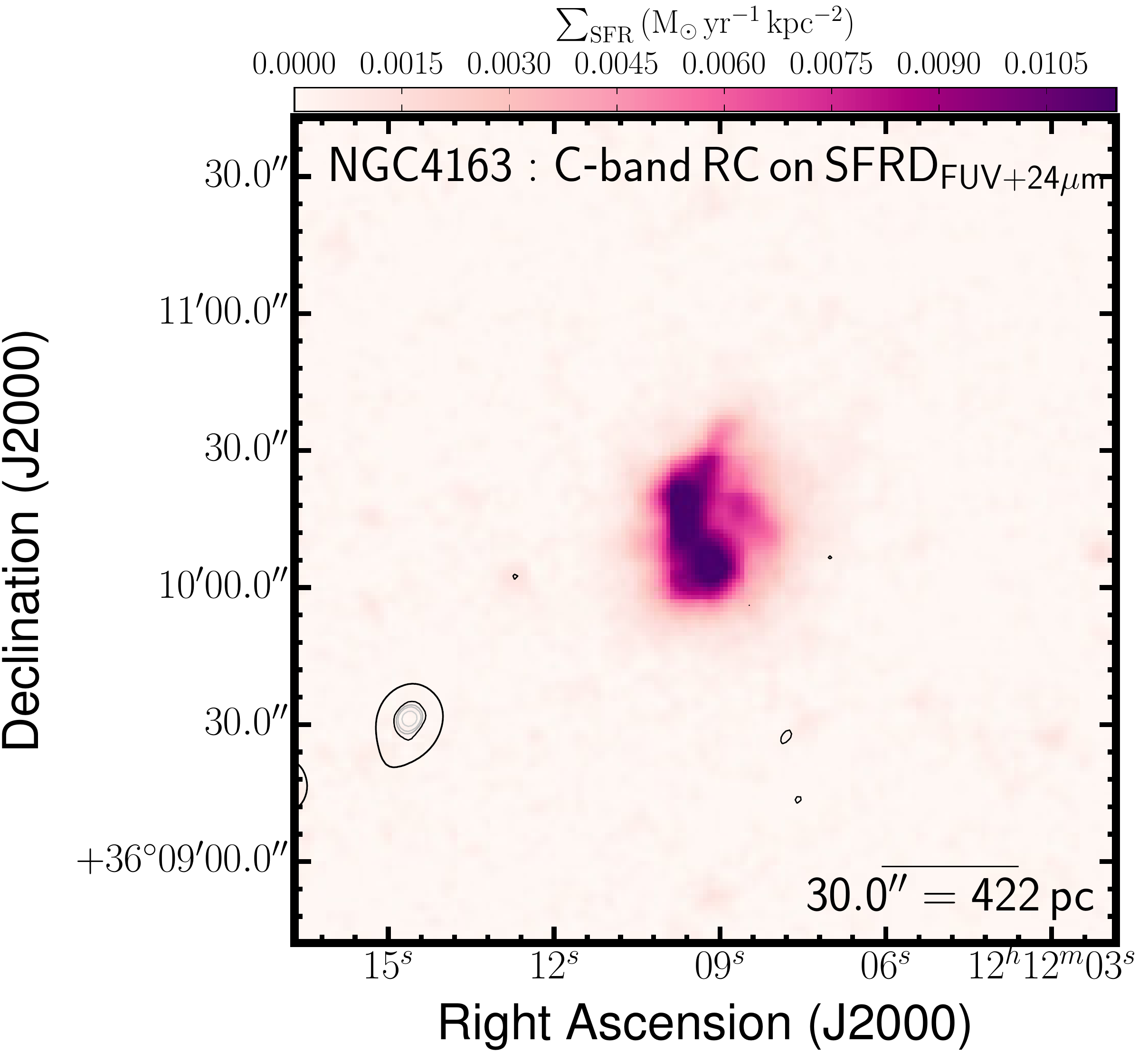} \\
  \end{tabular}
\caption[NGC\,4163 images: RC, IR, optical, and FUV]{Multi-wavelength coverage of NGC 4163 displaying a $3.0^\prime \times 3.0^\prime$ area. We show total RC flux density at the native resolution (top-left) and again with contours (top-centre). The RC contours are superposed on ancillary LITTLE THINGS images where possible: \halpha\ (middle-left); \RCNT\ obtained by subtracting the expected \RCT\ based on the \halpha-\RCT\ scaling factor of \cite{Deeg1997} from the total RC; {\em GALEX} FUV (middle-right); {\em Spitzer} 24\micron\ (bottom-left); {\em Spitzer} 70\micron\ (bottom-centre); FUV$+24{\rm \mu m}$--inferred SFRD from \citealp{Leroy2012} (bottom-right). We also show the RC that was isolated by the RC--based masking technique (top-right).}
  \label{figure:ngc4163Cc_maps}
\end{figure}

\begin{figure}
  \begin{tabular}{ccc}
    \includegraphics[width=0.31\linewidth,clip]{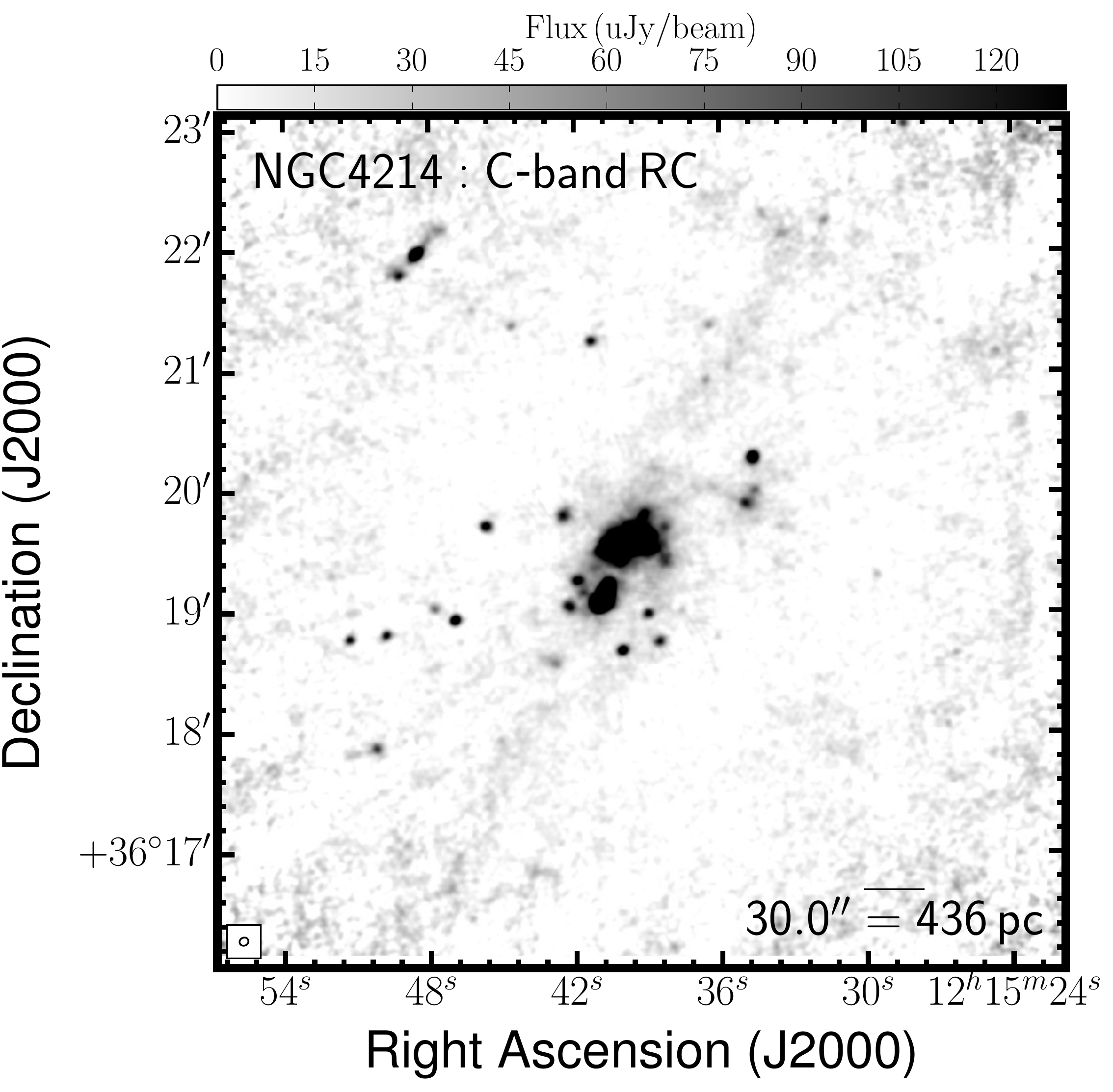} & \ 
    \includegraphics[width=0.31\linewidth,clip]{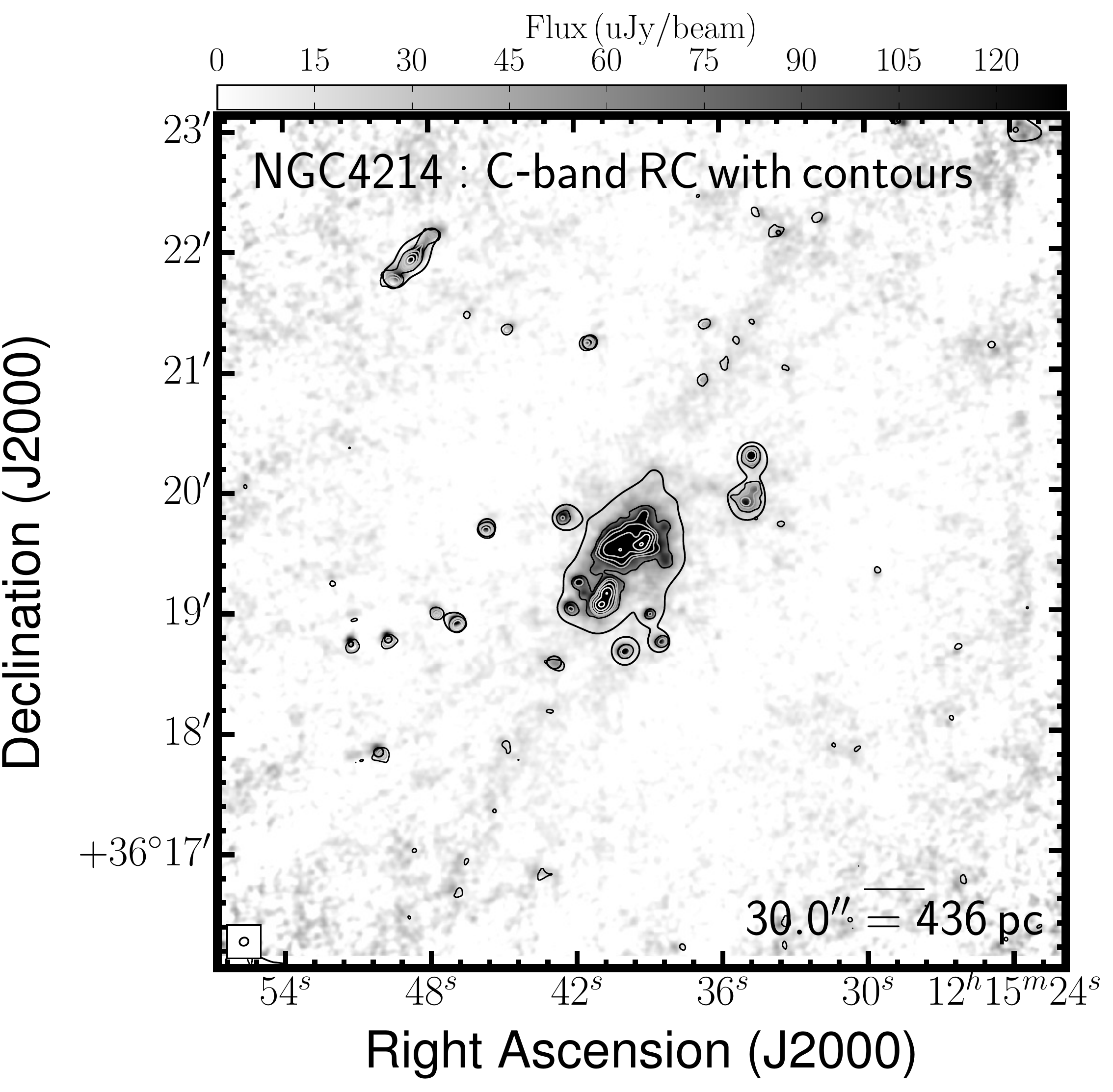} & \ 
    \includegraphics[width=0.31\linewidth,clip]{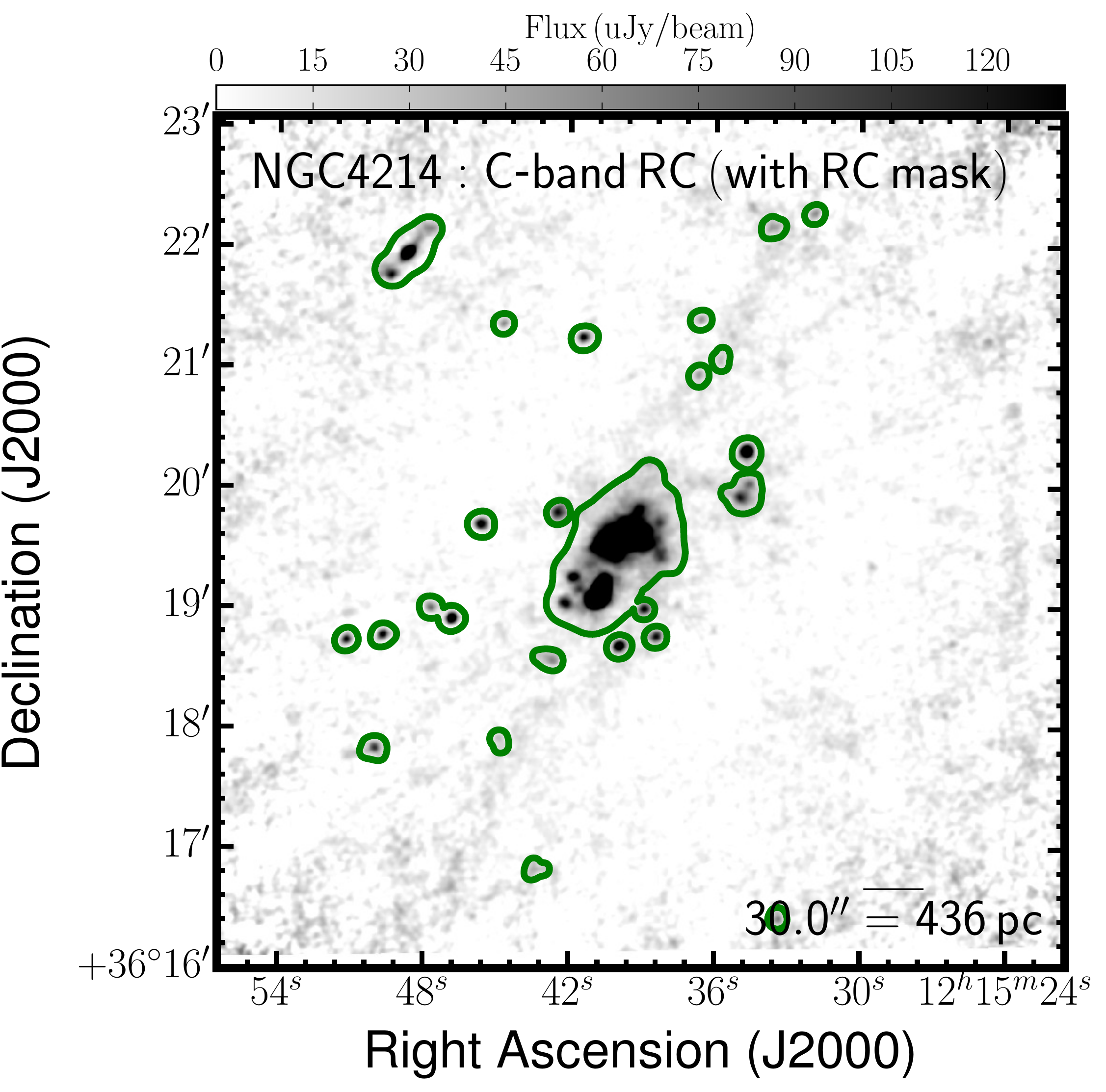} \\
    \includegraphics[width=0.31\linewidth,clip]{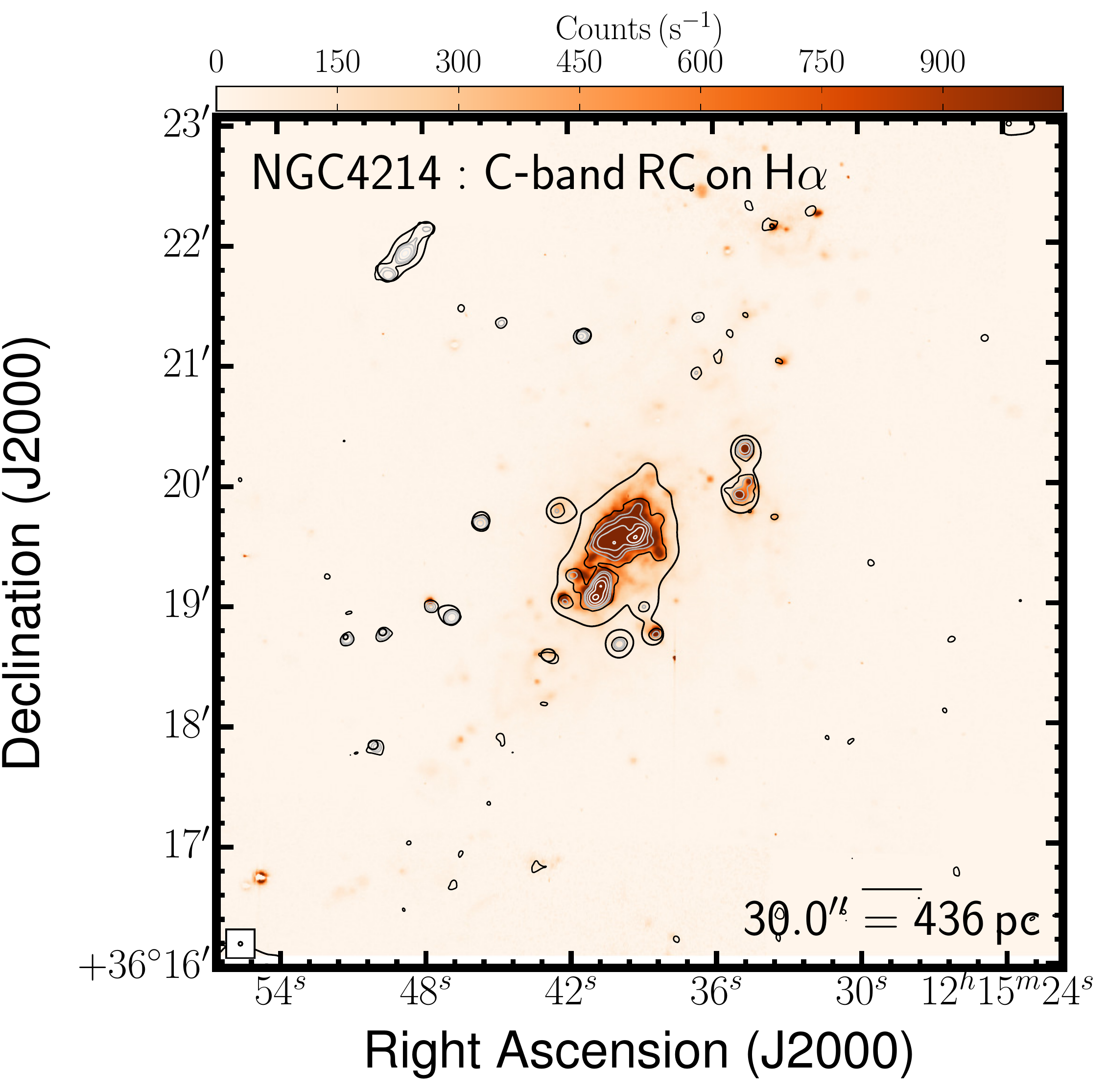} & \ 
    \includegraphics[width=0.31\linewidth,clip]{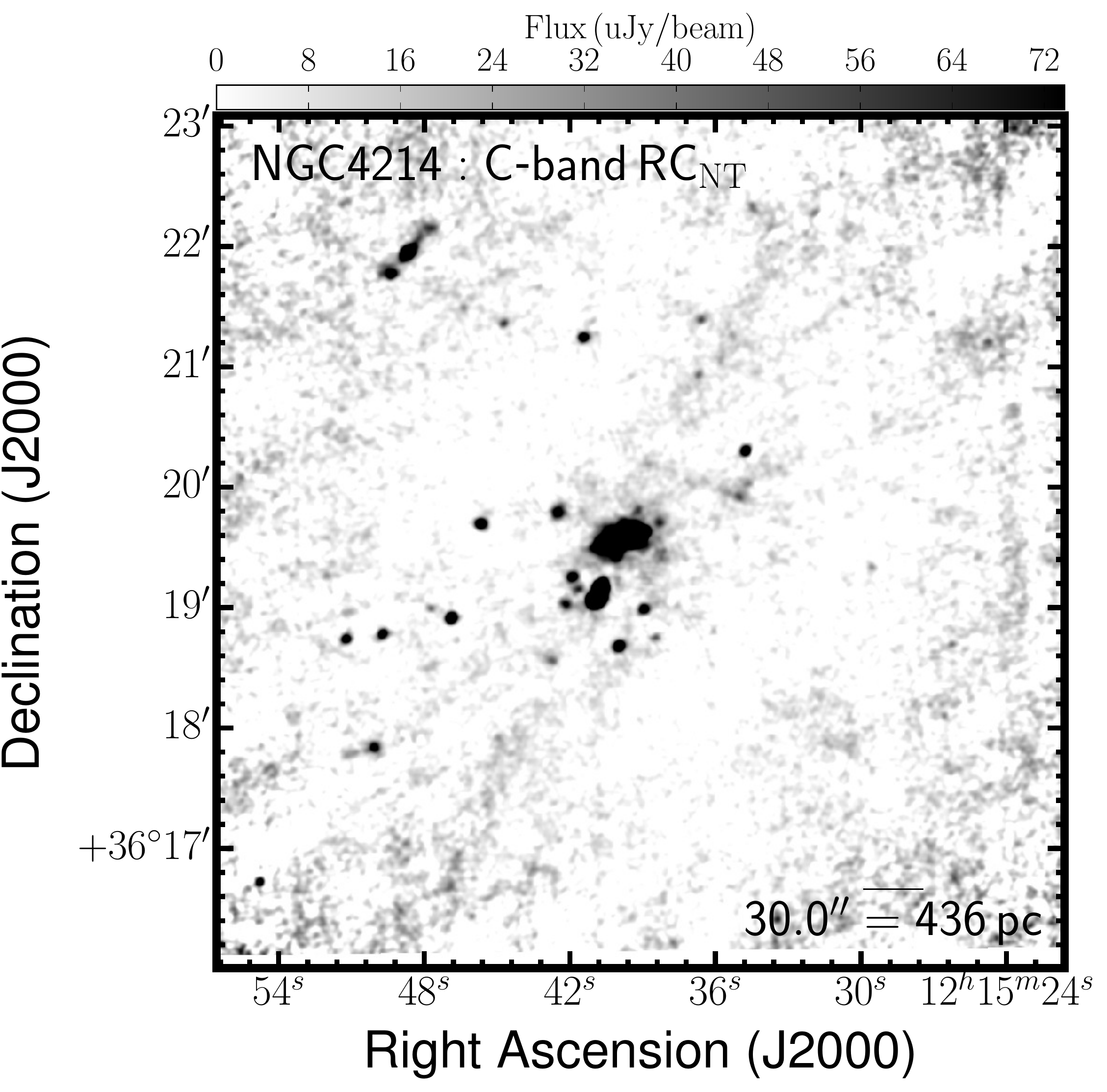} & \ 
    \includegraphics[width=0.31\linewidth,clip]{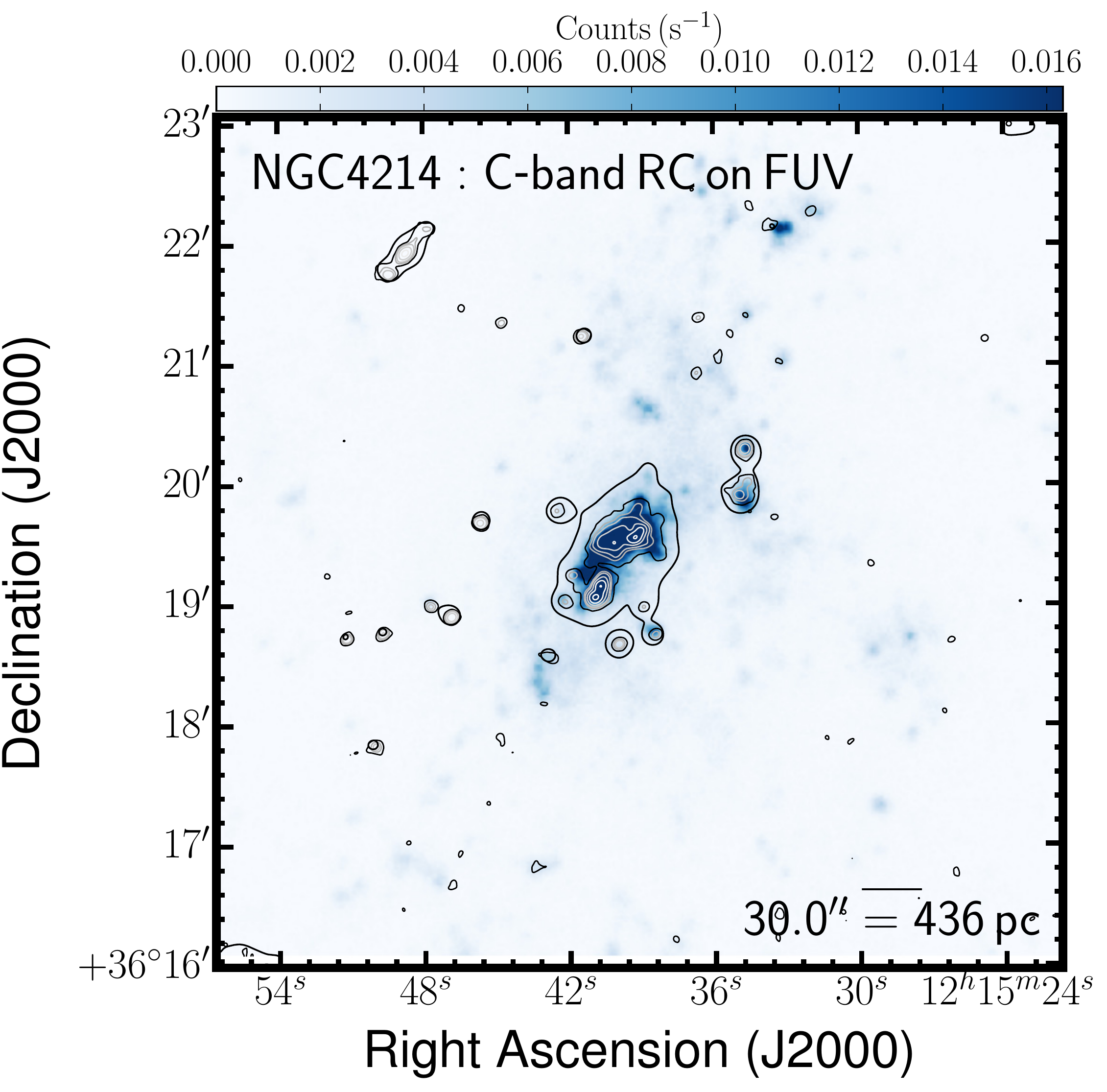} \\
    \includegraphics[width=0.31\linewidth,clip]{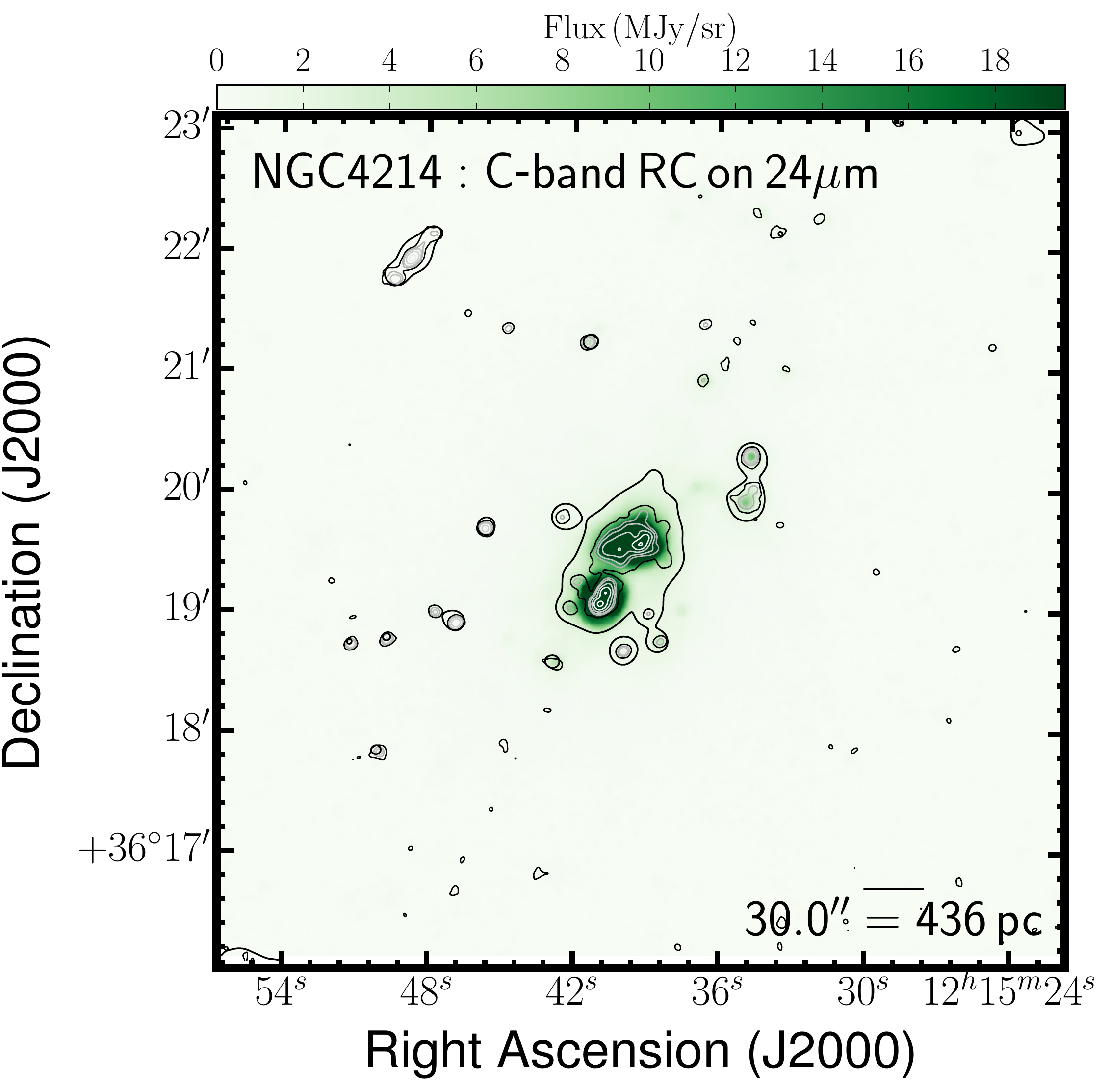} & \ 
    \includegraphics[width=0.31\linewidth,clip]{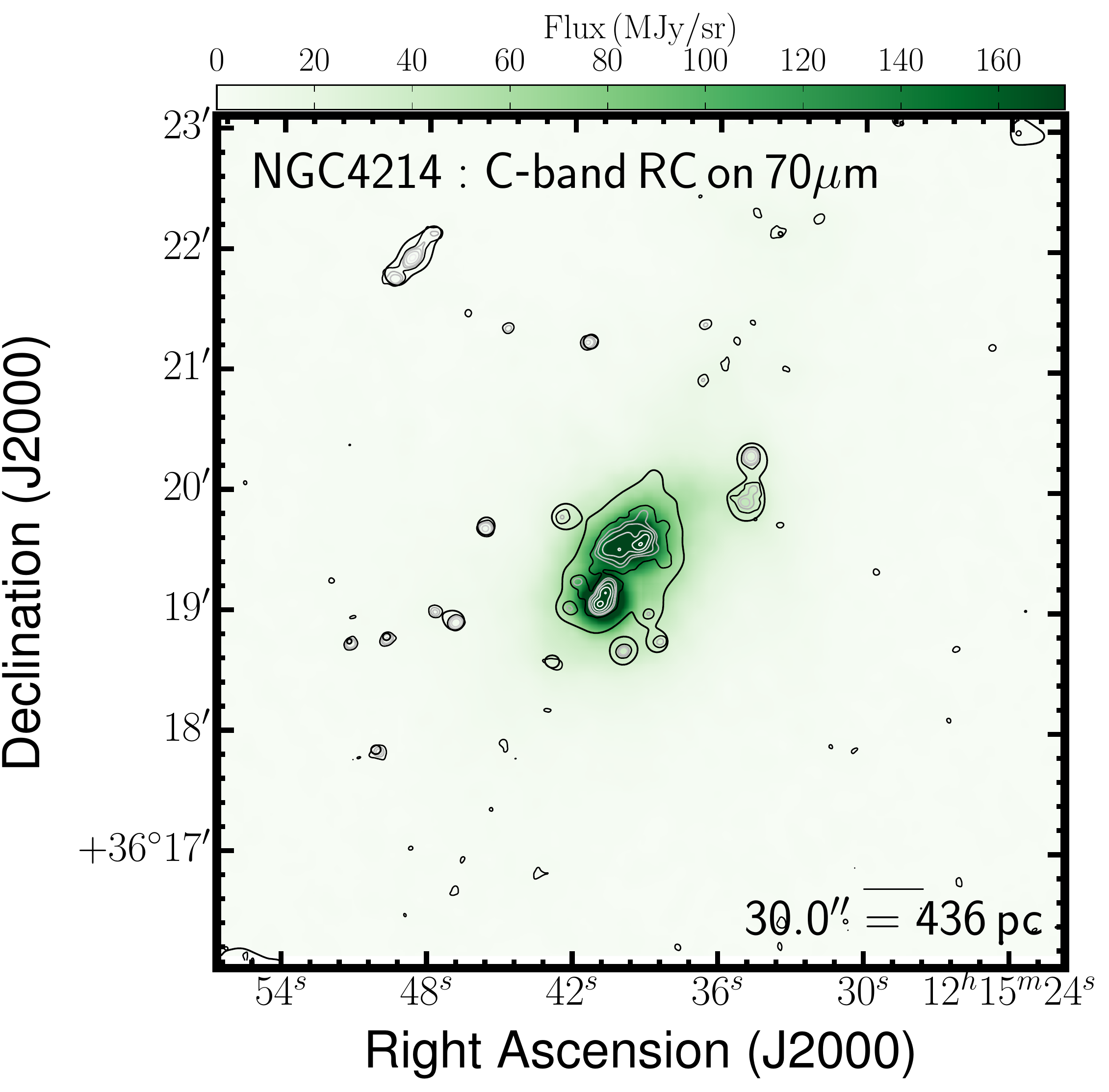} & \ 
    \includegraphics[width=0.31\linewidth,clip]{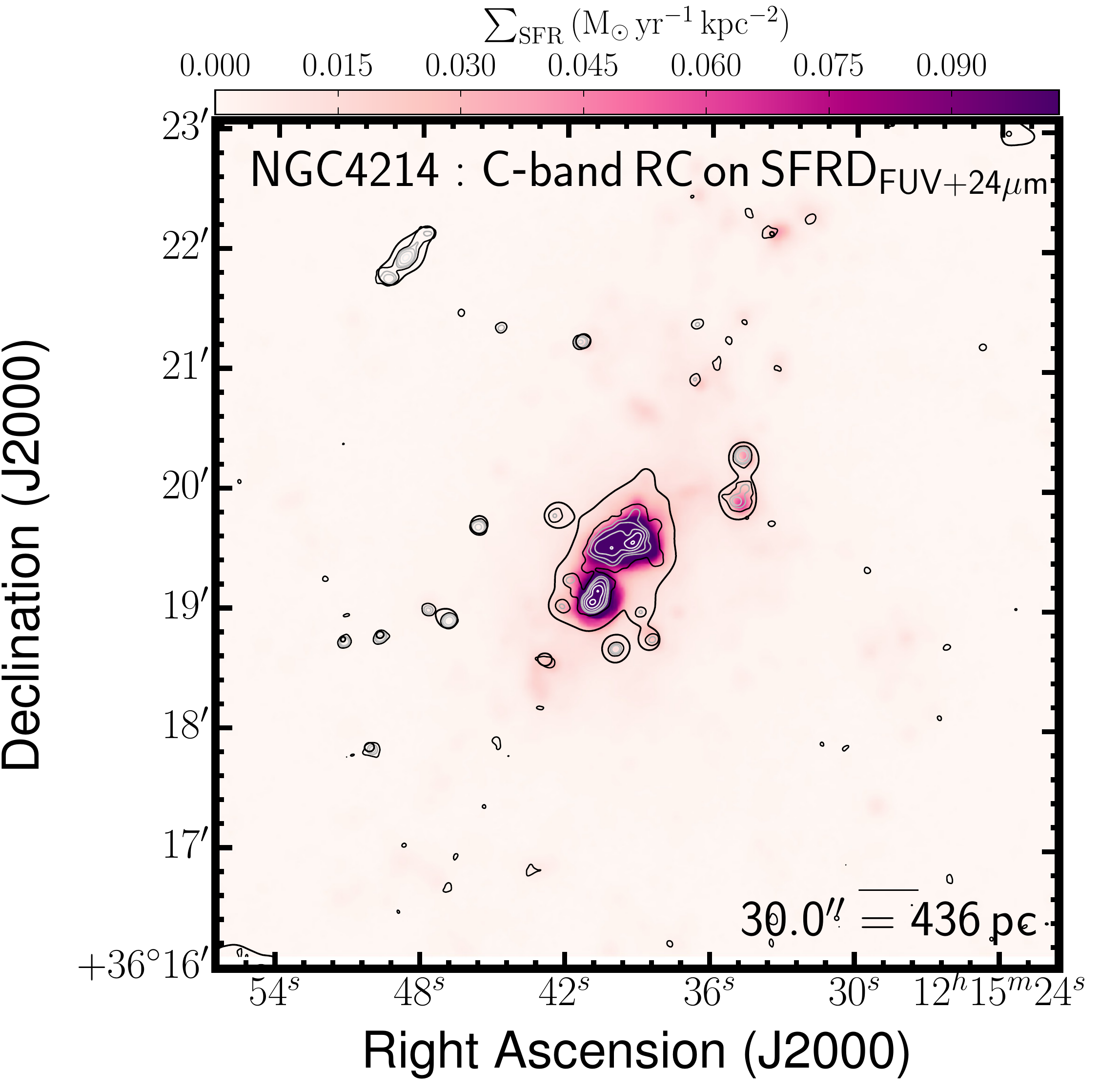} \\
  \end{tabular}
\caption[NGC\,4214 images: RC, IR, optical, and FUV]{Multi-wavelength coverage of NGC 4214 displaying a $7.0^\prime \times 7.0^\prime$ area. We show total RC flux density at the native resolution (top-left) and again with contours (top-centre). The RC contours are superposed on ancillary LITTLE THINGS images where possible: \halpha\ (middle-left); \RCNT\ obtained by subtracting the expected \RCT\ based on the \halpha-\RCT\ scaling factor of \cite{Deeg1997} from the total RC; {\em GALEX} FUV (middle-right); {\em Spitzer} 24\micron\ (bottom-left); {\em Spitzer} 70\micron\ (bottom-centre); FUV$+24{\rm \mu m}$--inferred SFRD from \citealp{Leroy2012} (bottom-right). We also show the RC that was isolated by the RC--based masking technique (top-right).}
  \label{figure:ngc4214Cc_maps}
\end{figure}

\begin{figure}
  \begin{tabular}{ccc}
    \includegraphics[width=0.31\linewidth,clip]{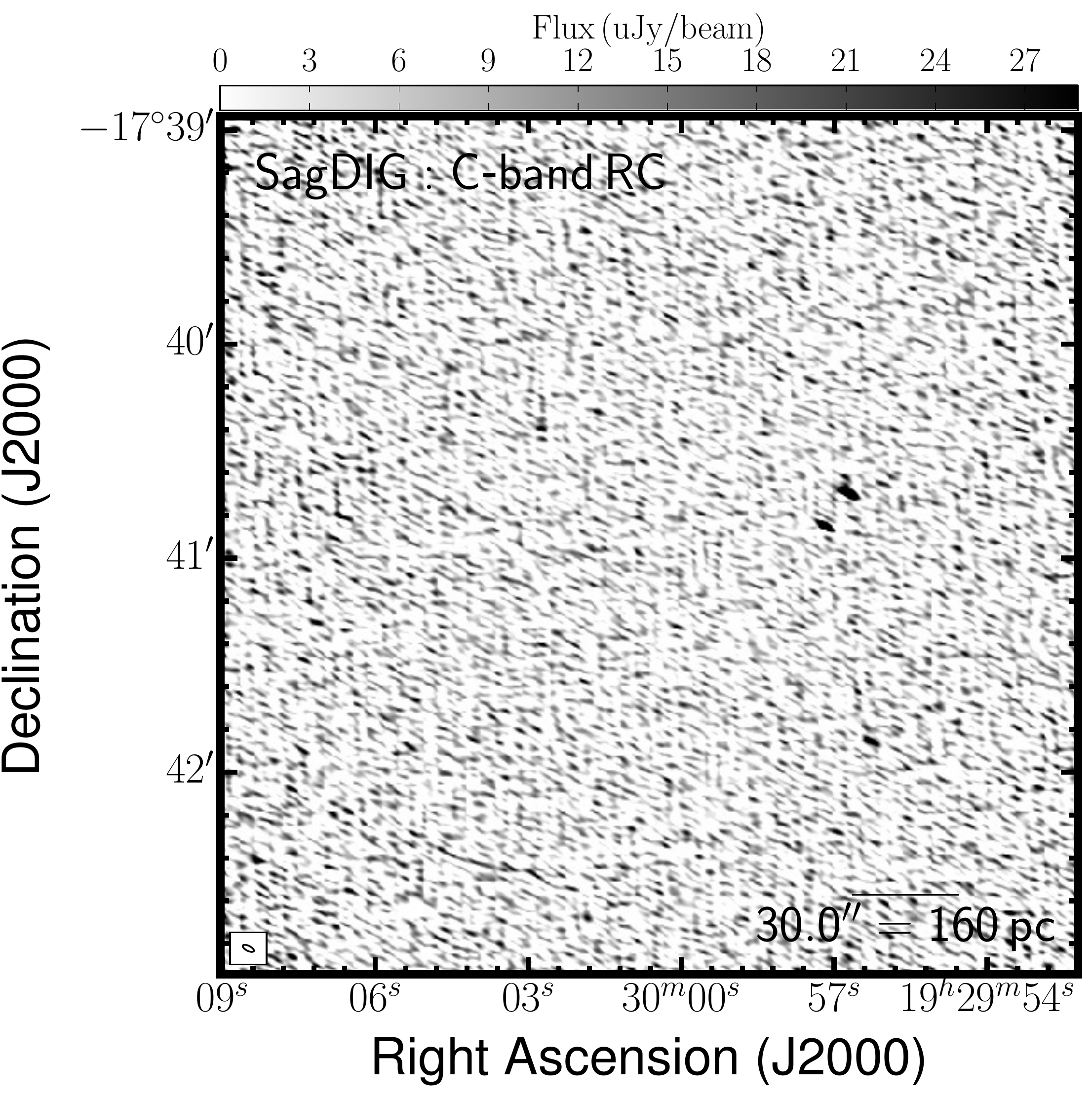} & \ 
    \includegraphics[width=0.31\linewidth,clip]{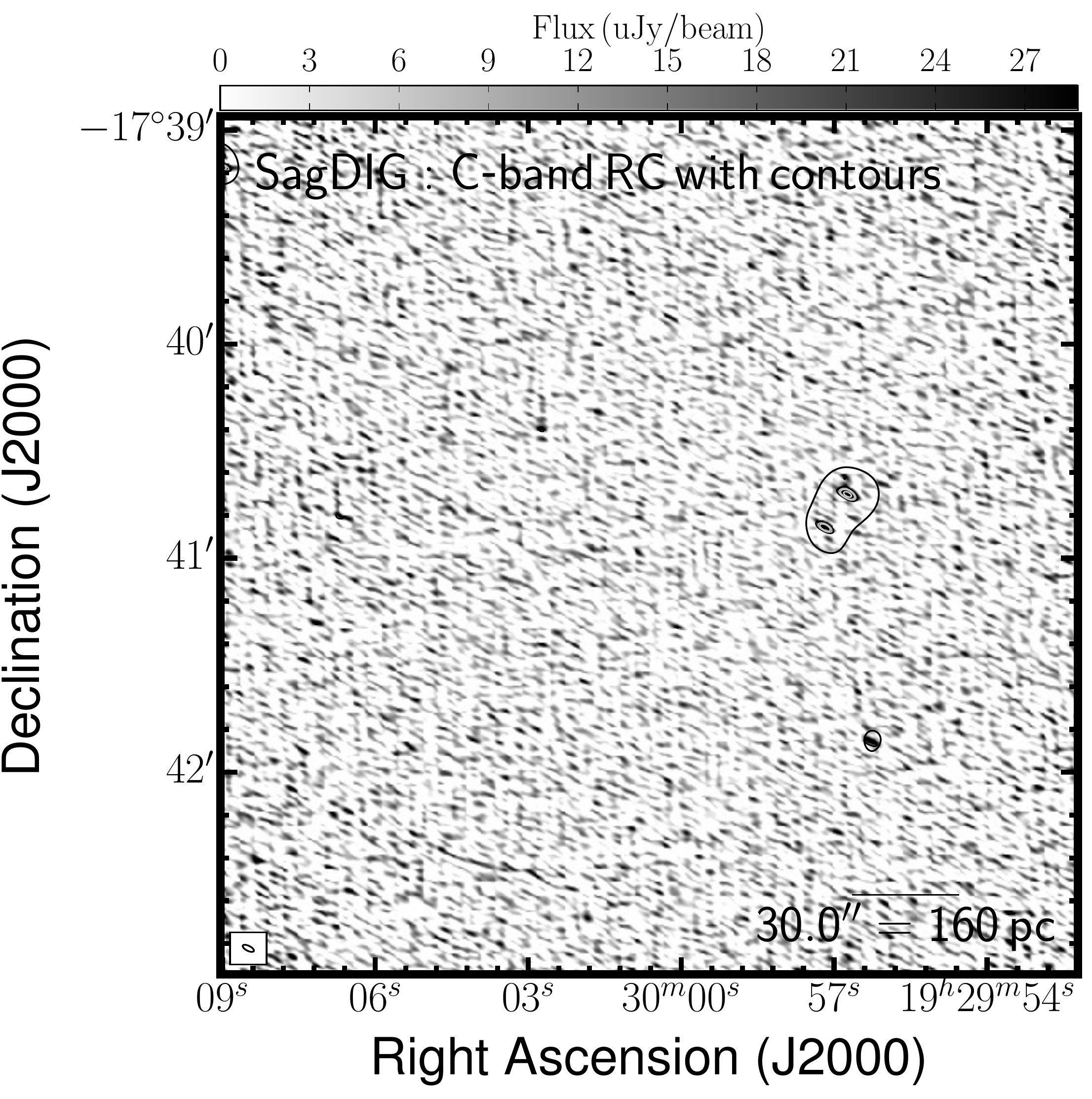} & \ 
    \includegraphics[width=0.31\linewidth,clip]{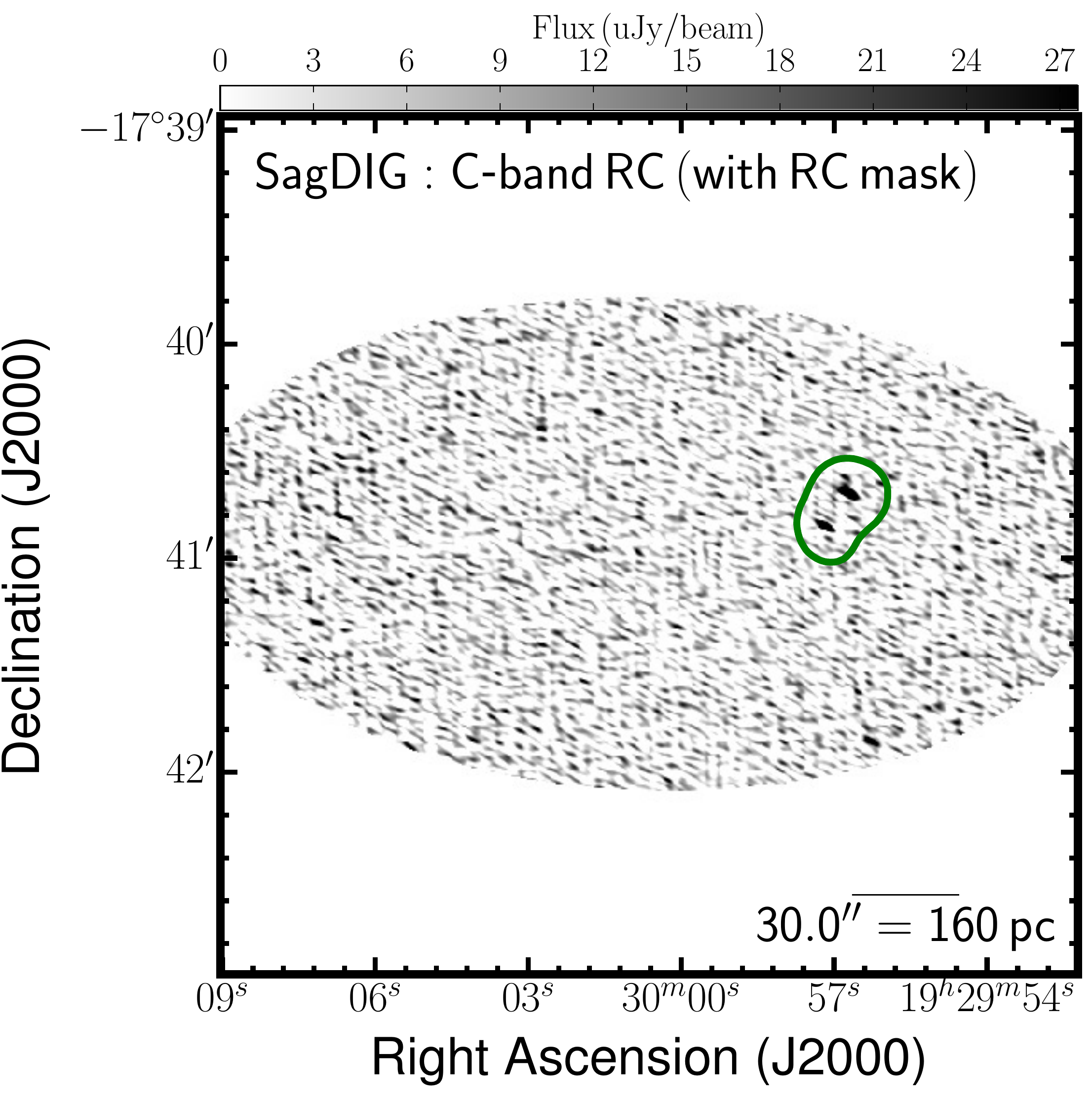} \\
    \includegraphics[width=0.31\linewidth,clip]{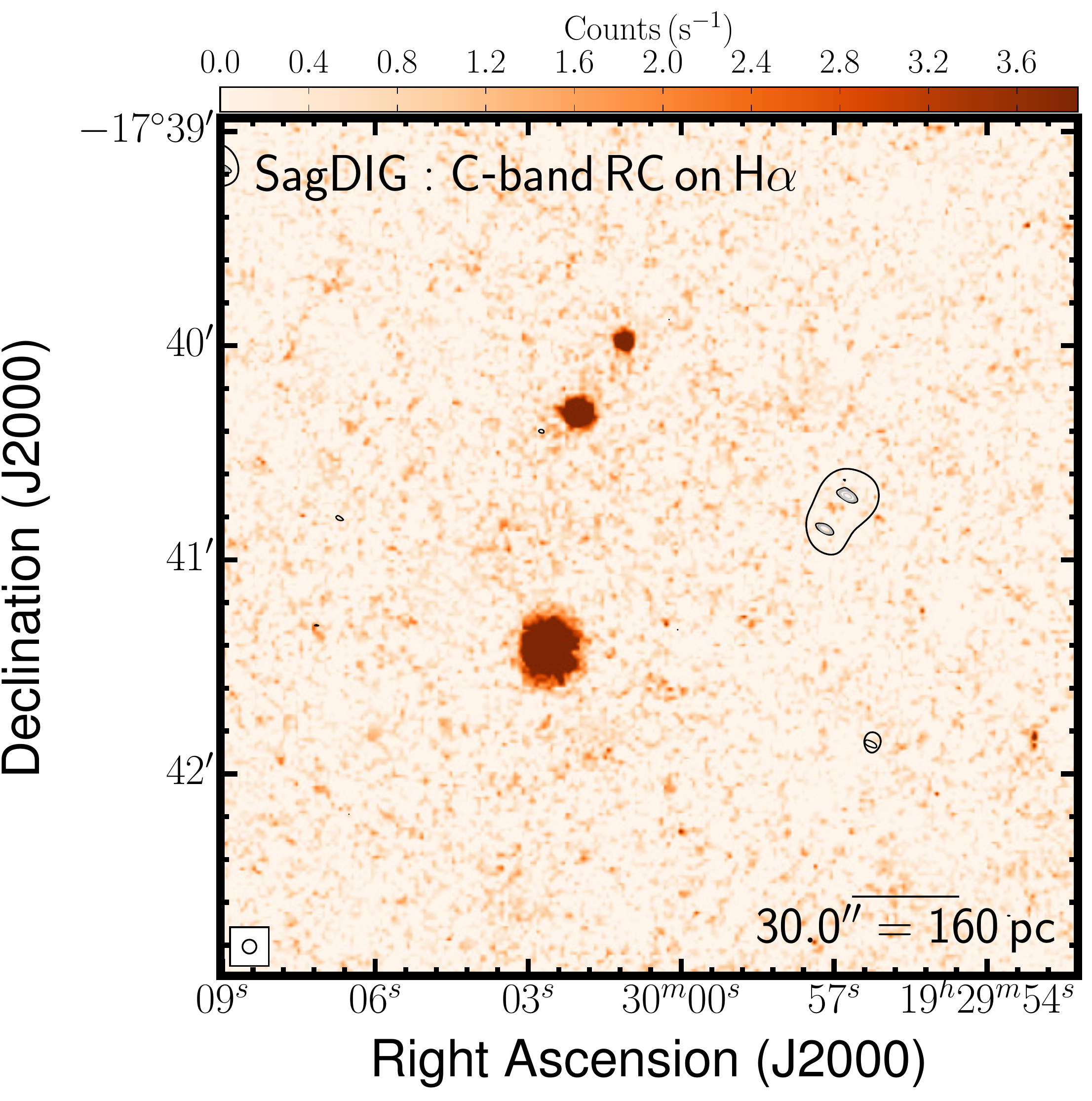} & \ 
    \includegraphics[width=0.31\linewidth,clip]{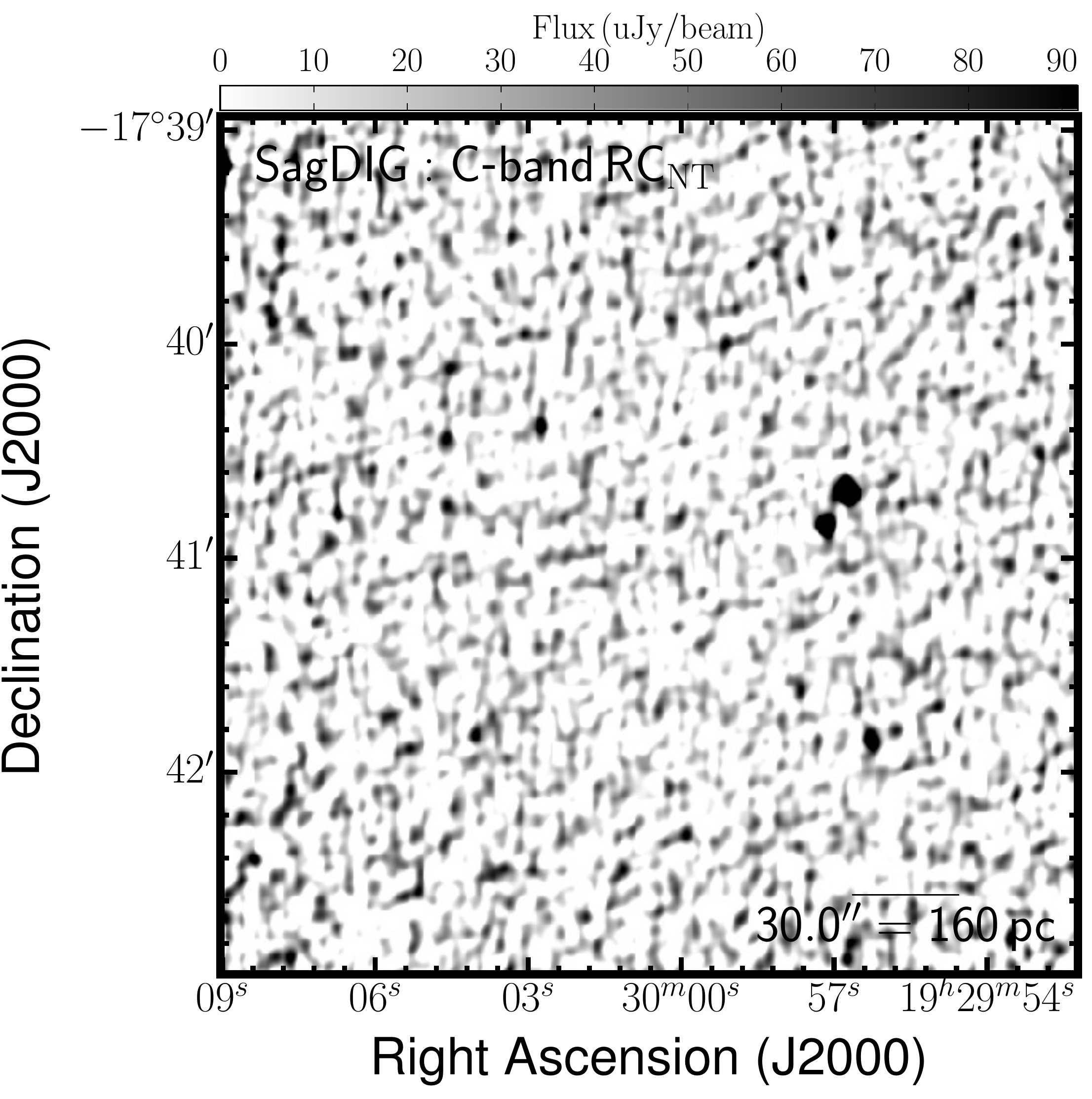} & \ 
    \includegraphics[width=0.31\linewidth,clip]{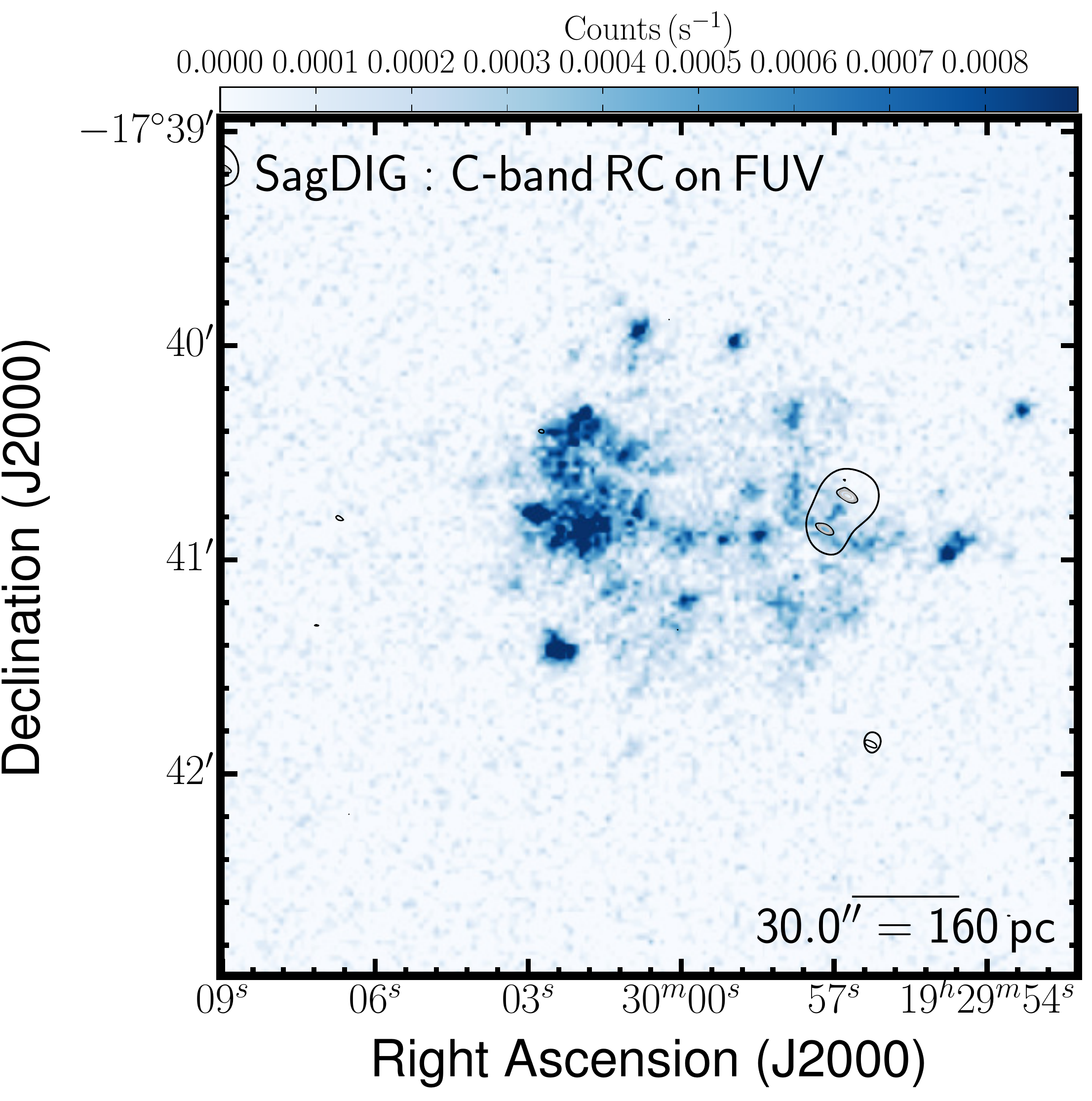} \\
    \includegraphics[width=0.31\linewidth,clip]{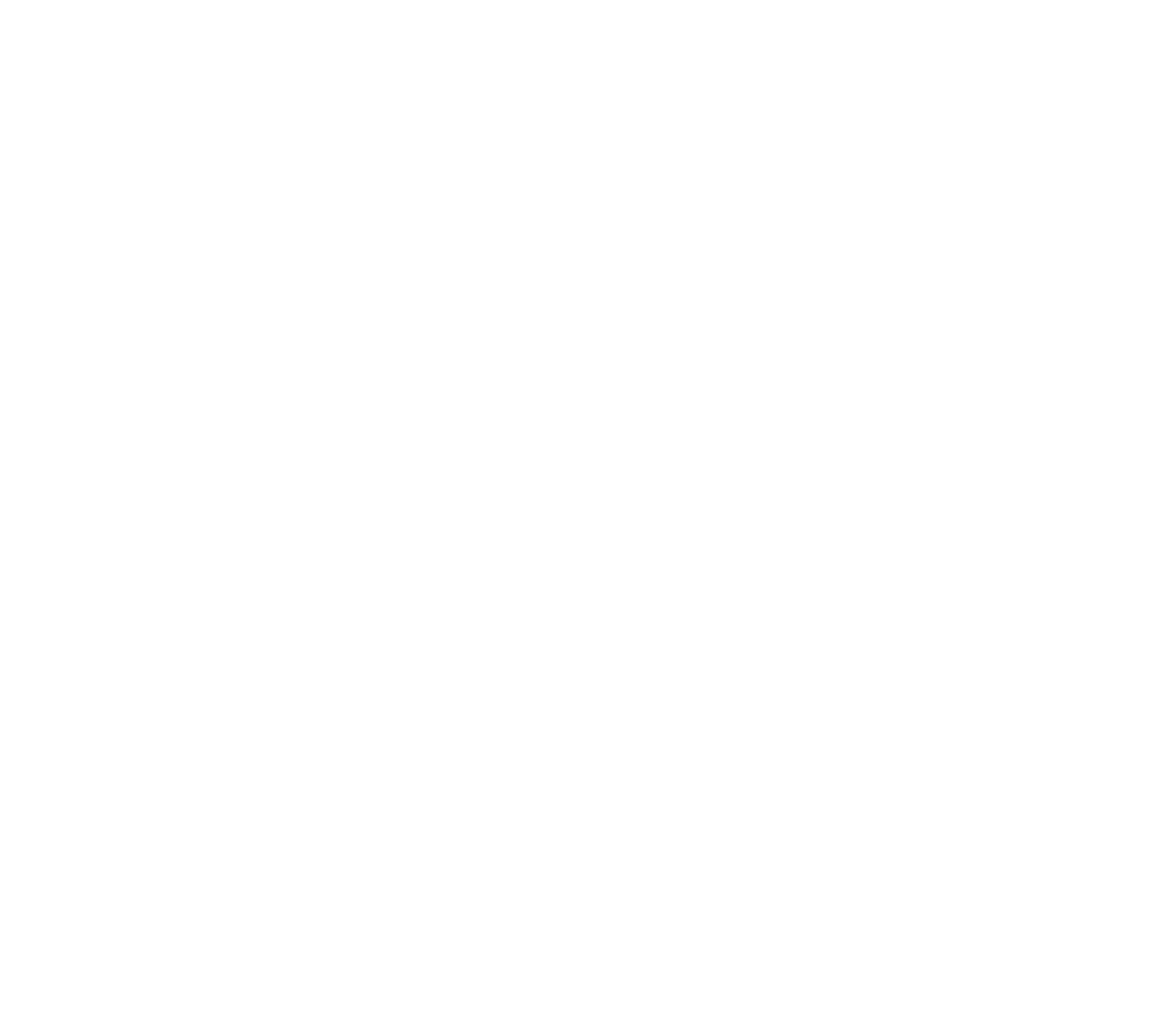} & \ 
    \includegraphics[width=0.31\linewidth,clip]{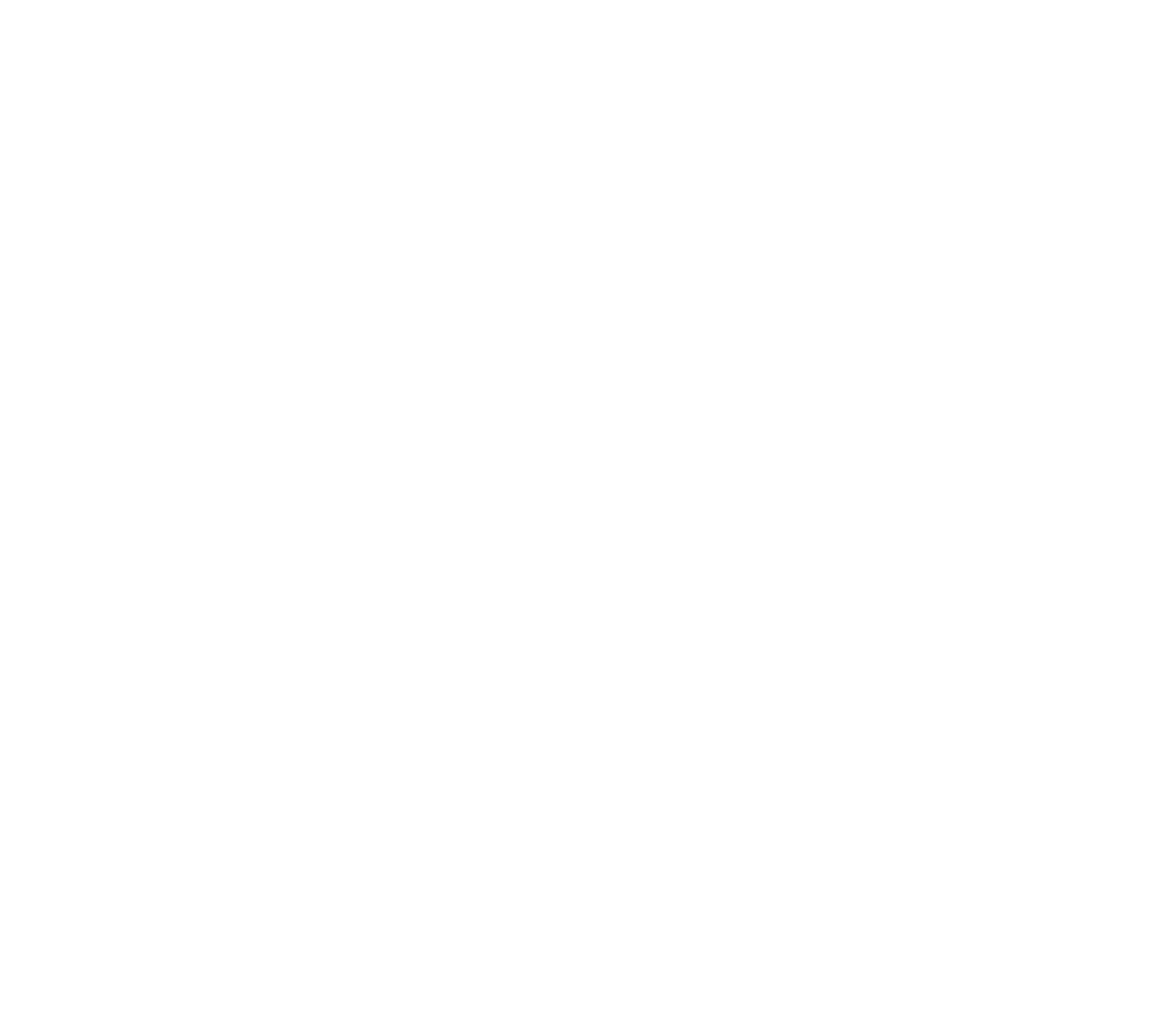} & \ 
    \includegraphics[width=0.31\linewidth,clip]{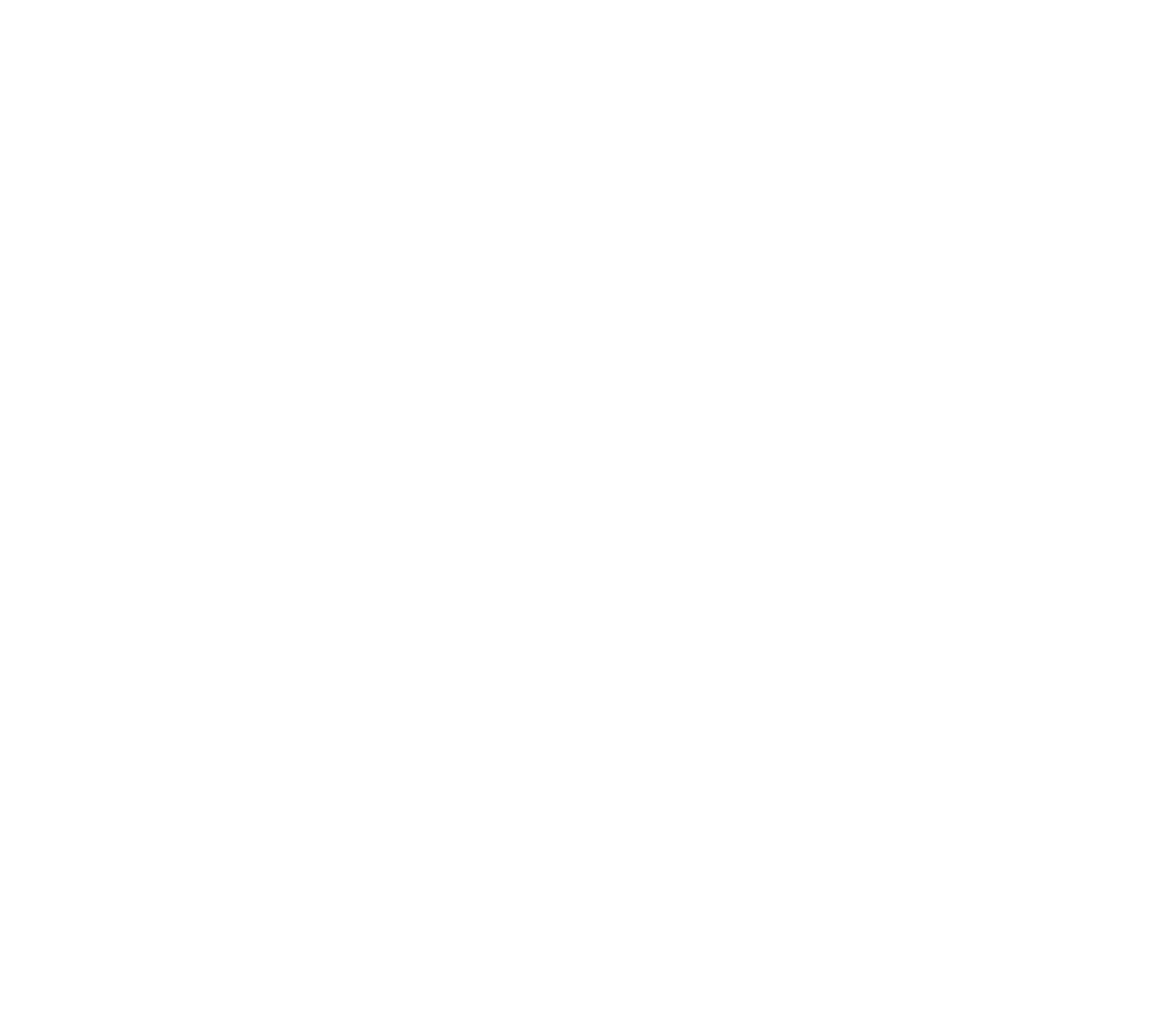} \\
  \end{tabular}
\caption[Sag\,DIG images: RC, IR, optical, and FUV]{Multi-wavelength coverage of Sag DIG displaying a $4.0^\prime \times 4.0^\prime$ area. We show total RC flux density at the native resolution (top-left) and again with contours (top-centre). The RC contours are superposed on ancillary LITTLE THINGS images where possible: \halpha\ (middle-left); \RCNT\ obtained by subtracting the expected \RCT\ based on the \halpha-\RCT\ scaling factor of \cite{Deeg1997} from the total RC; {\em GALEX} FUV (middle-right); {\em Spitzer} 24\micron\ (bottom-left); {\em Spitzer} 70\micron\ (bottom-centre); FUV$+24{\rm \mu m}$--inferred SFRD from \citealp{Leroy2012} (bottom-right). We also show the RC that was isolated by the RC--based masking technique (top-right).}
  \label{figure:sagdigCc_maps}
\end{figure}

\begin{figure}
  \begin{tabular}{ccc}
    \includegraphics[width=0.31\linewidth,clip]{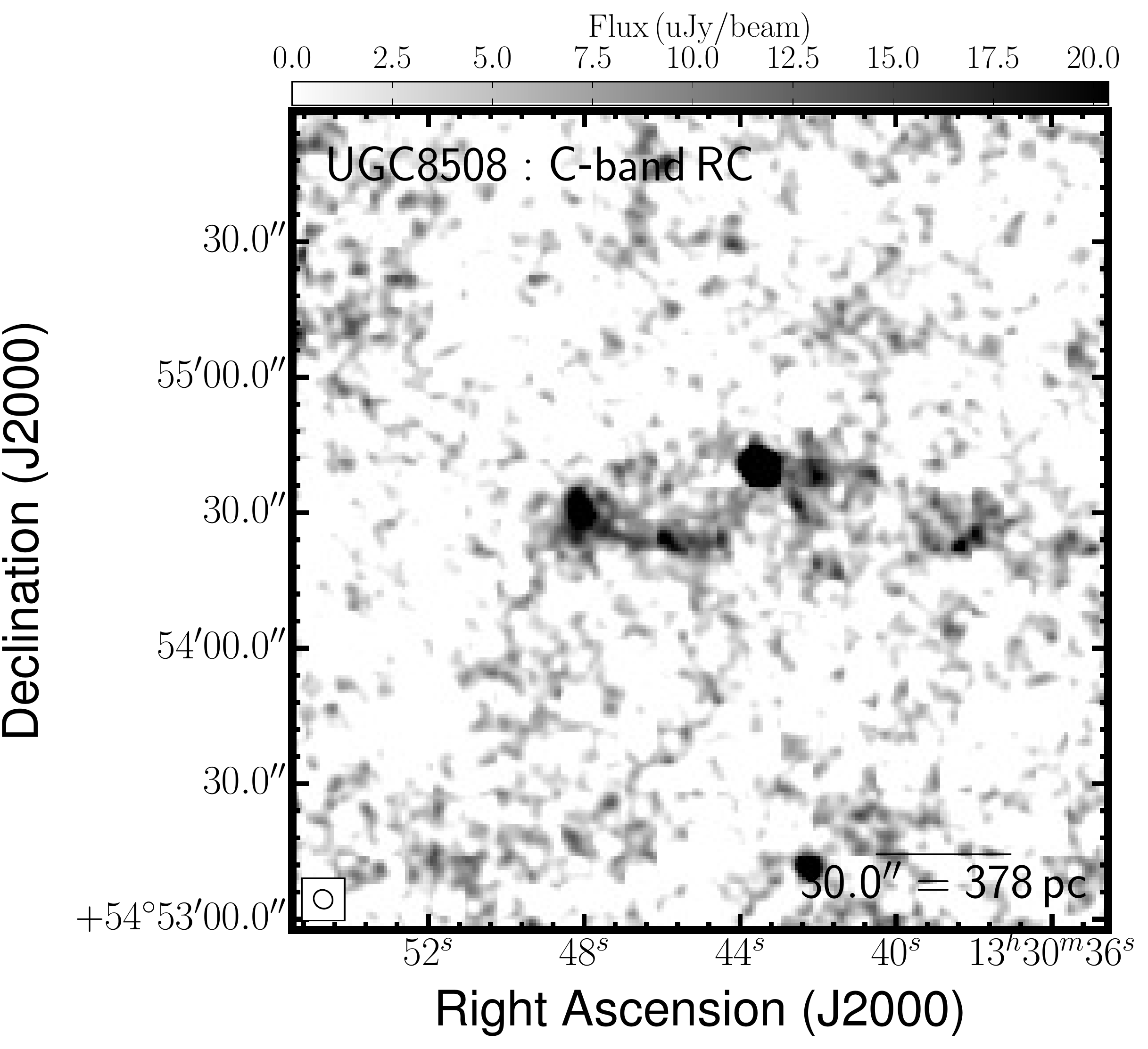} & \ 
    \includegraphics[width=0.31\linewidth,clip]{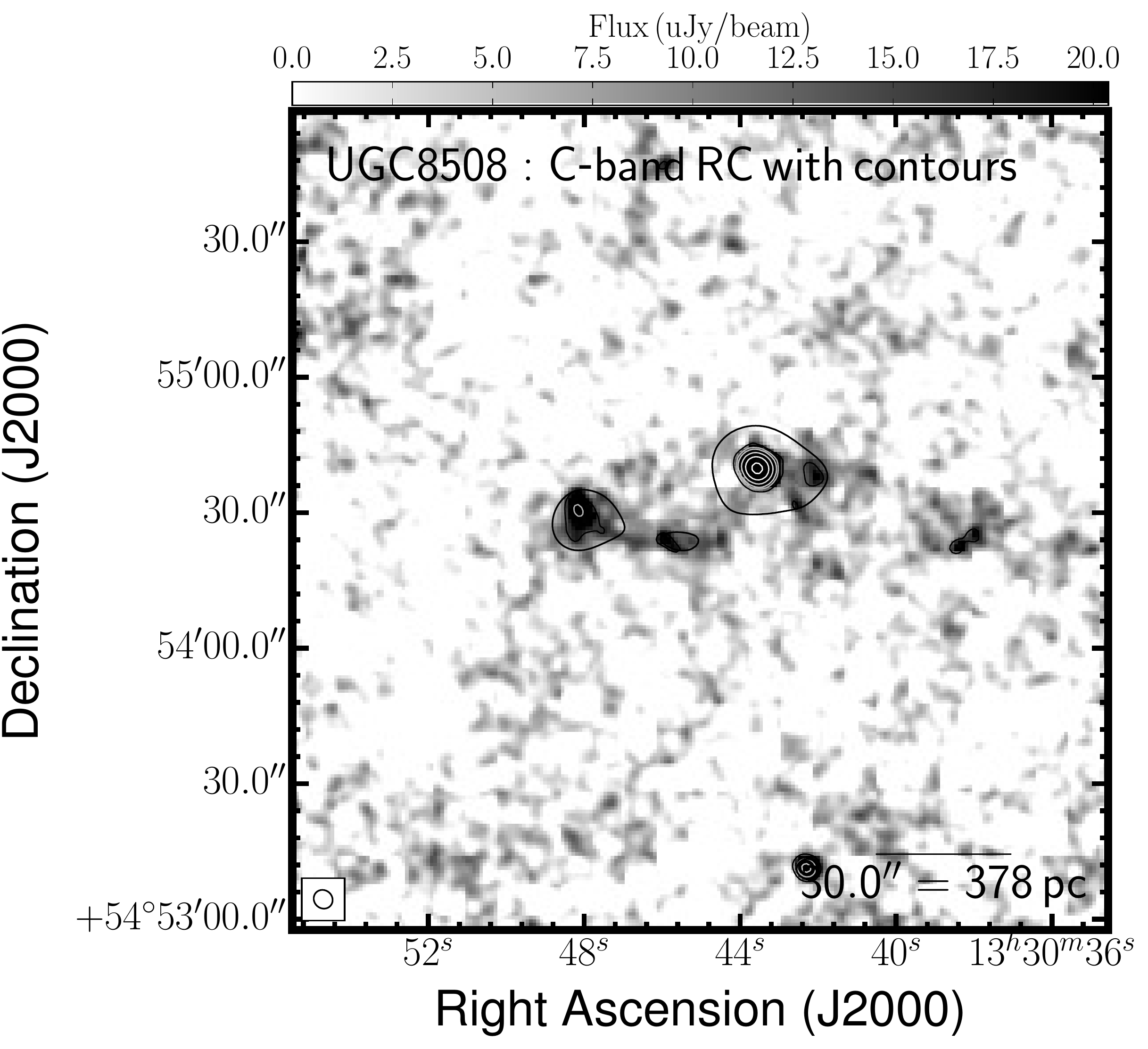} & \ 
    \includegraphics[width=0.31\linewidth,clip]{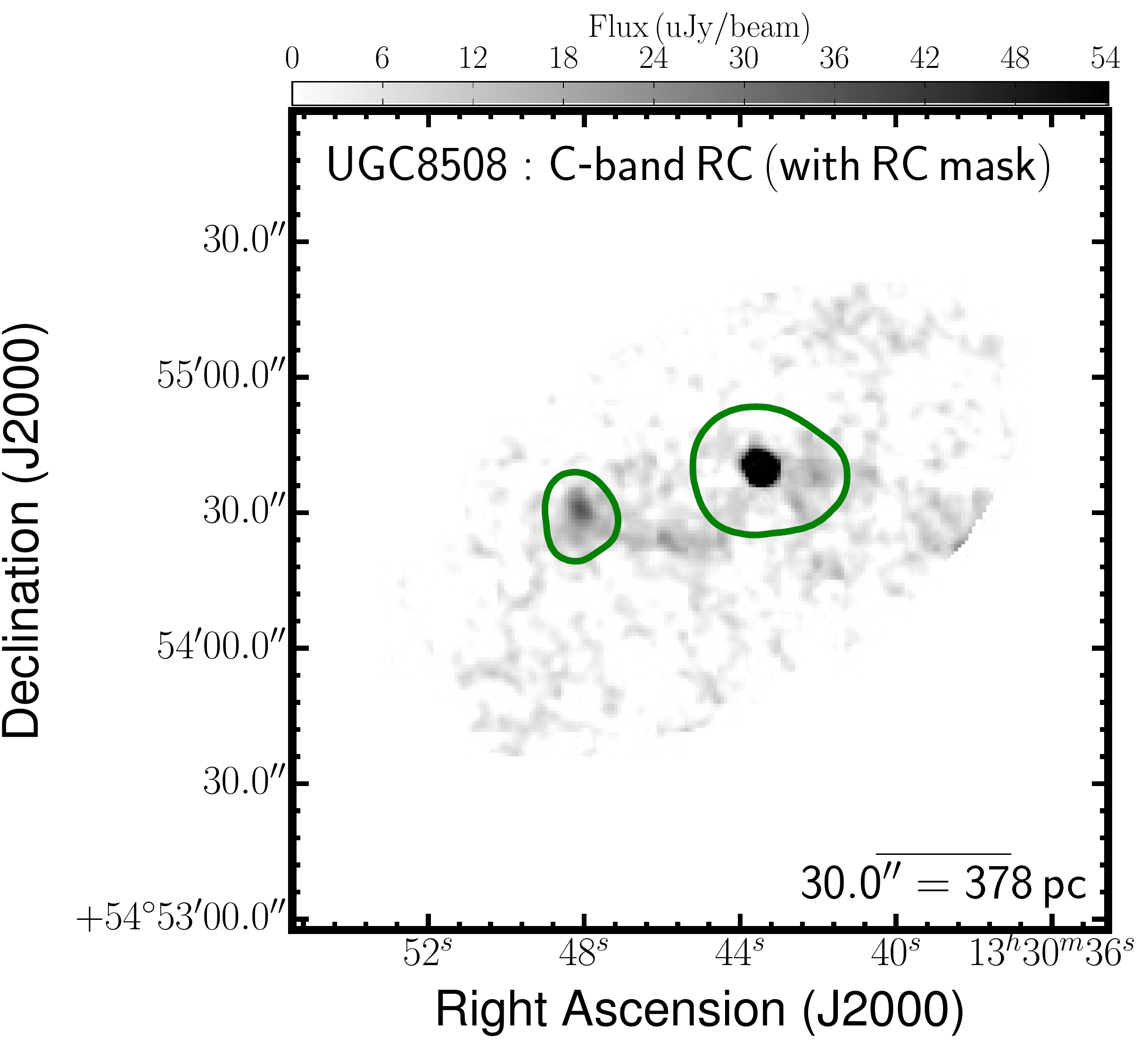} \\
    \includegraphics[width=0.31\linewidth,clip]{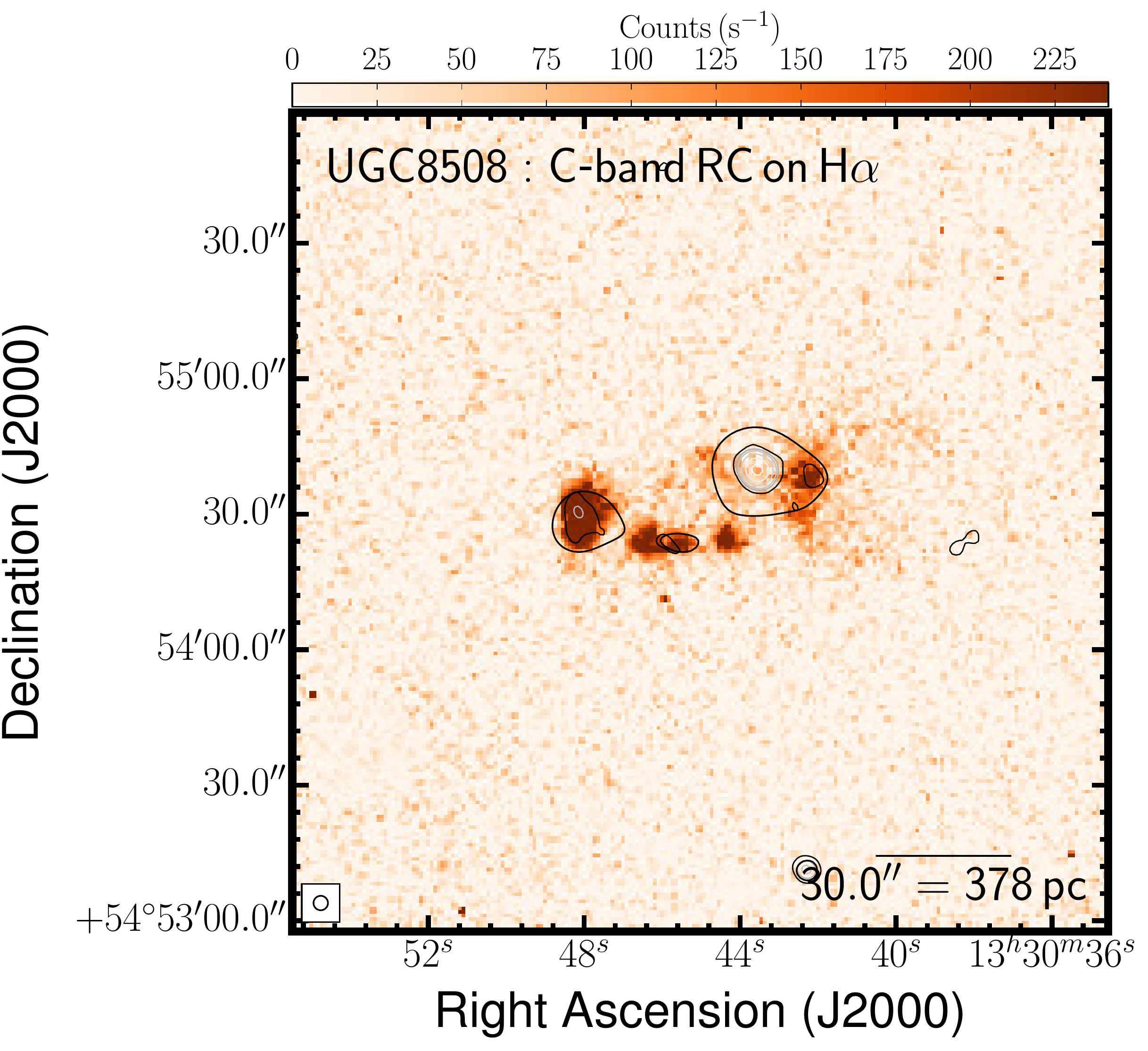} & \ 
    \includegraphics[width=0.31\linewidth,clip]{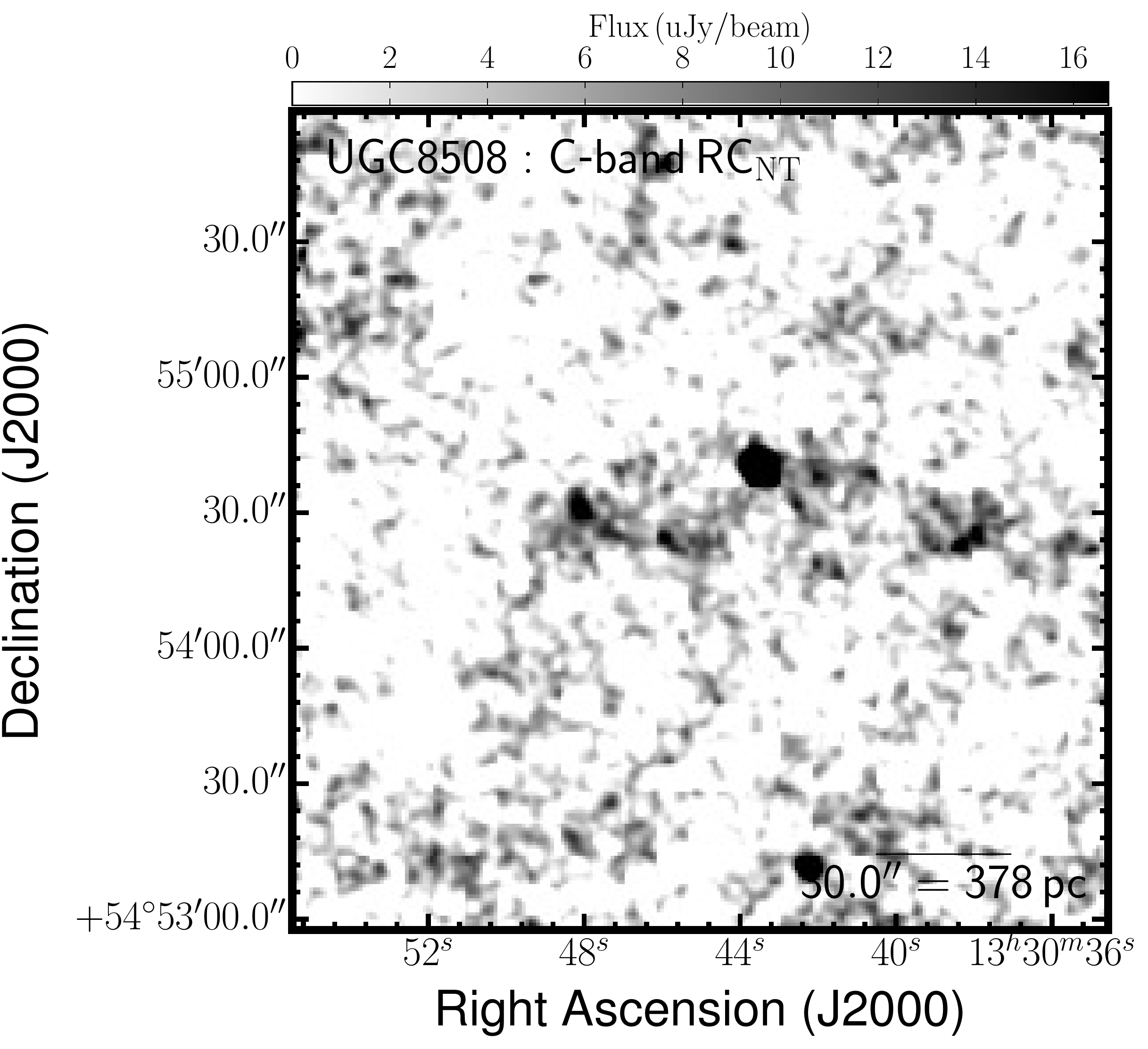} & \ 
    \includegraphics[width=0.31\linewidth,clip]{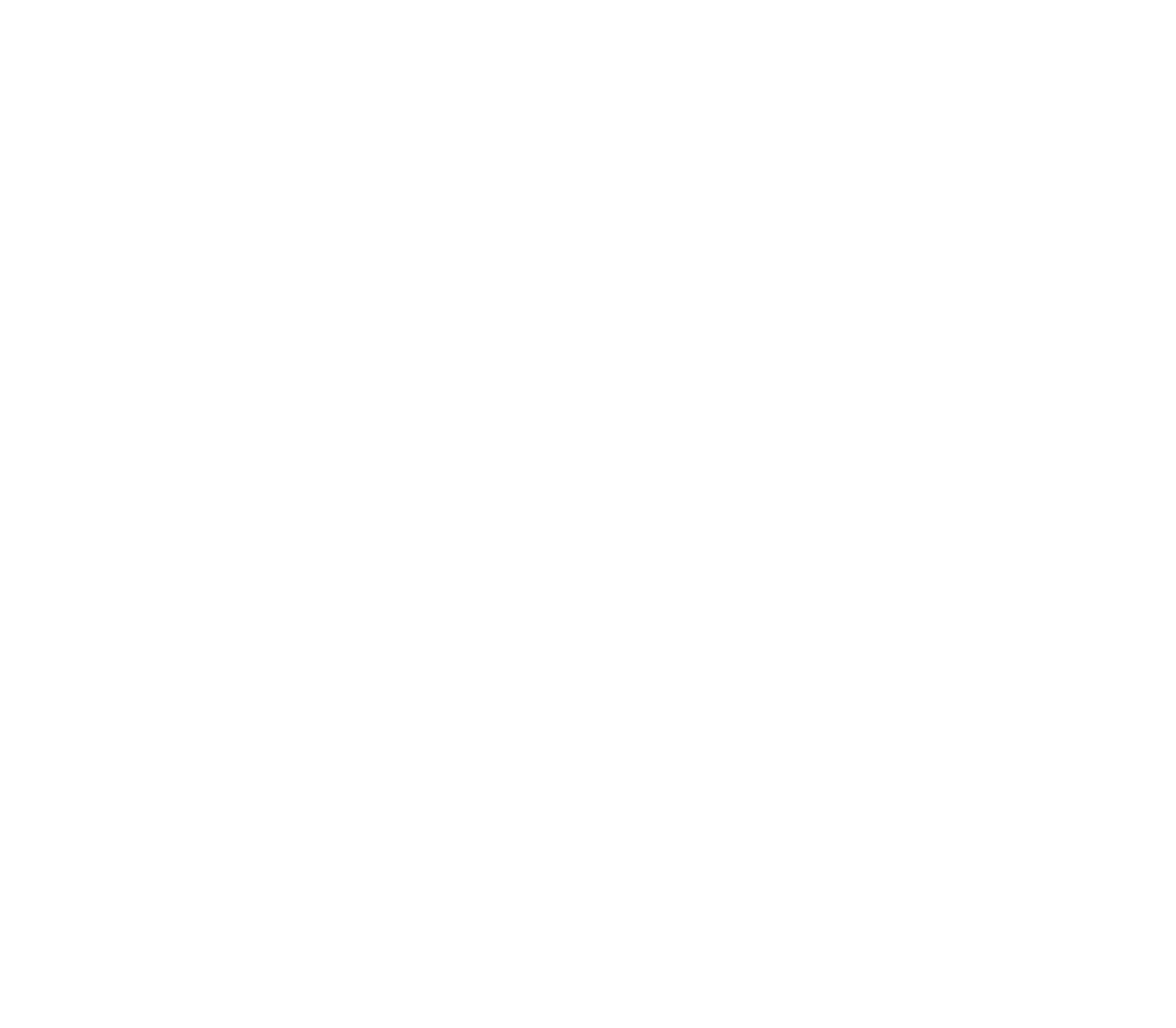} \\
    \includegraphics[width=0.31\linewidth,clip]{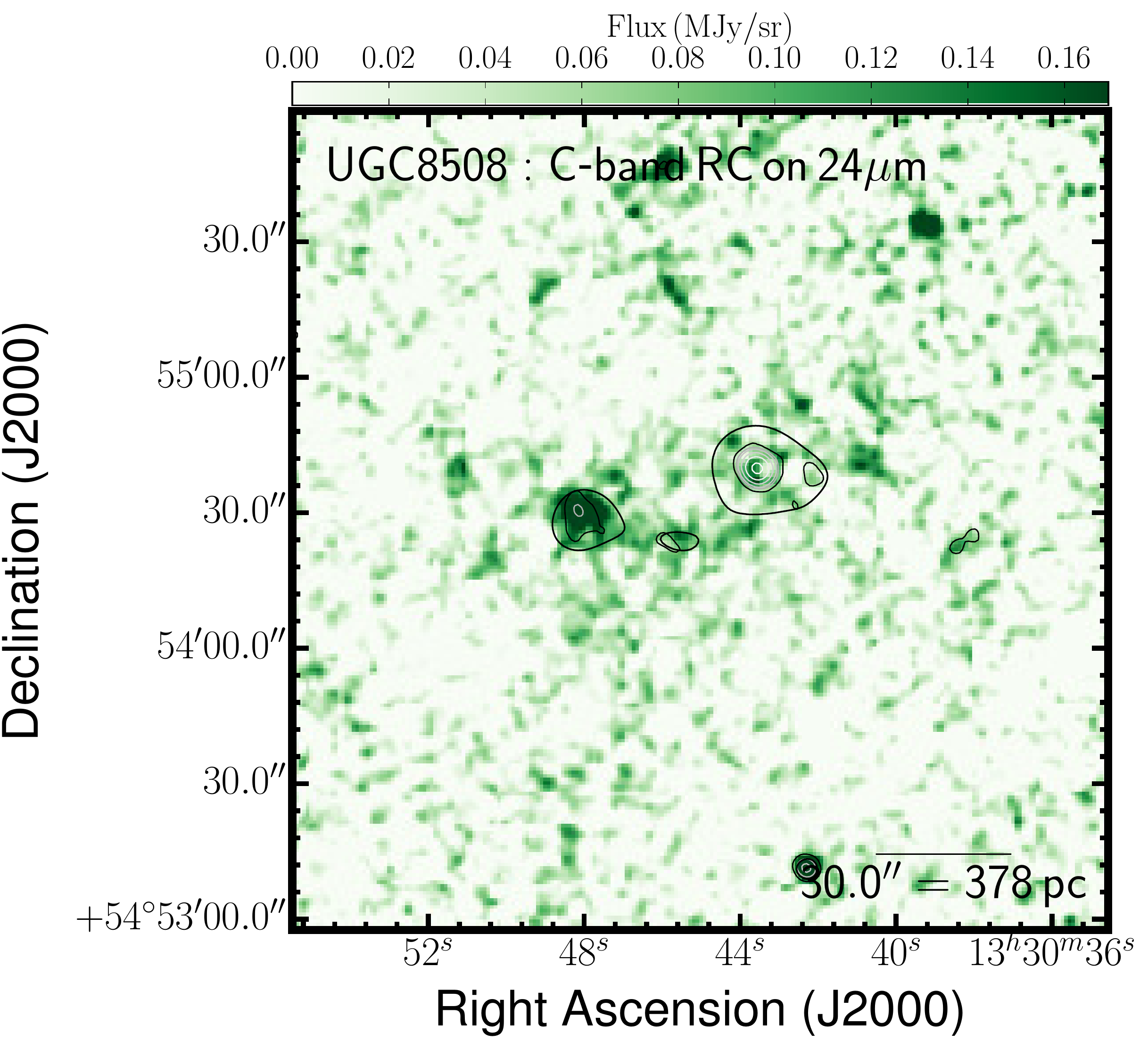} & \ 
    \includegraphics[width=0.31\linewidth,clip]{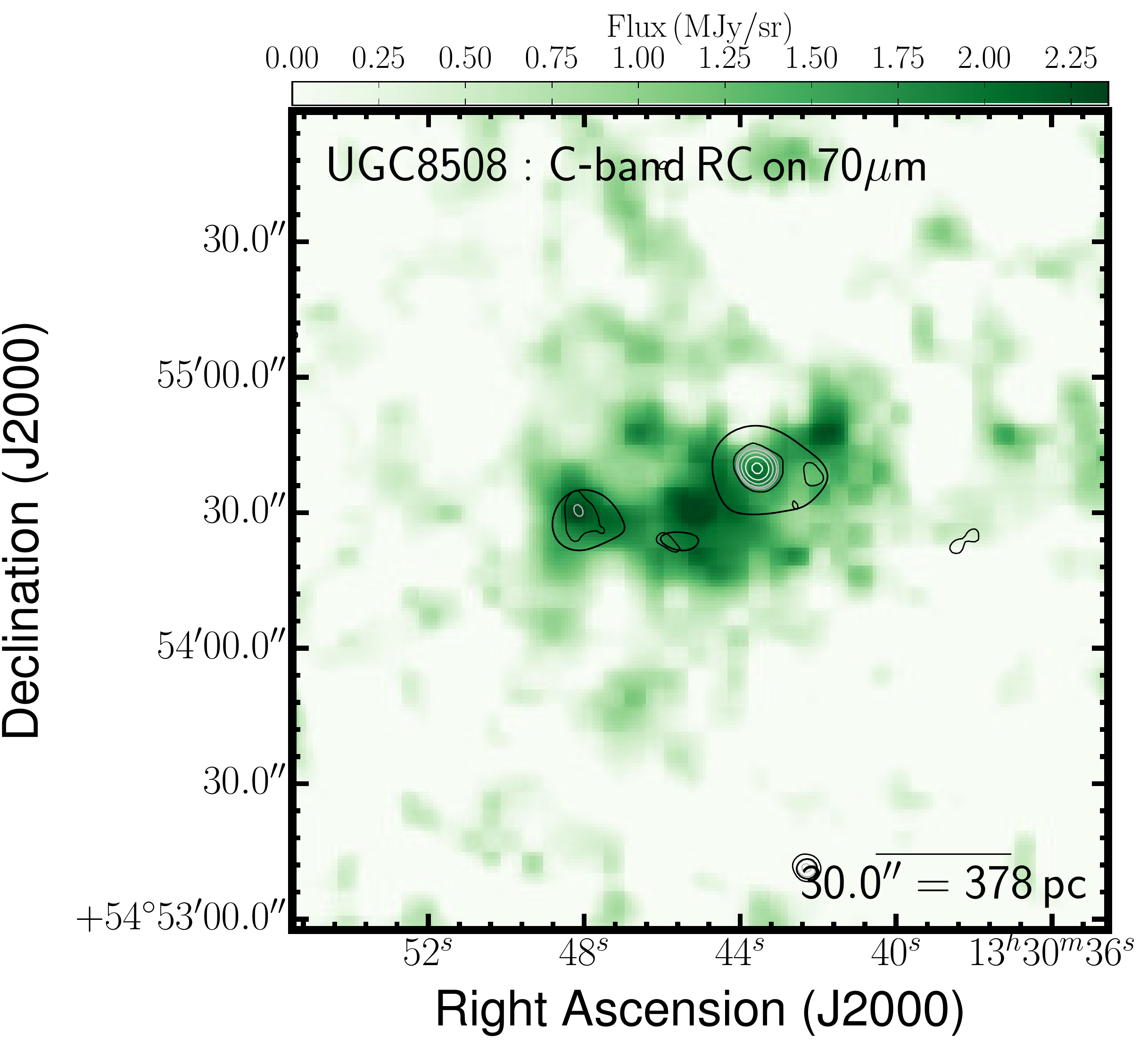} & \ 
    \includegraphics[width=0.31\linewidth,clip]{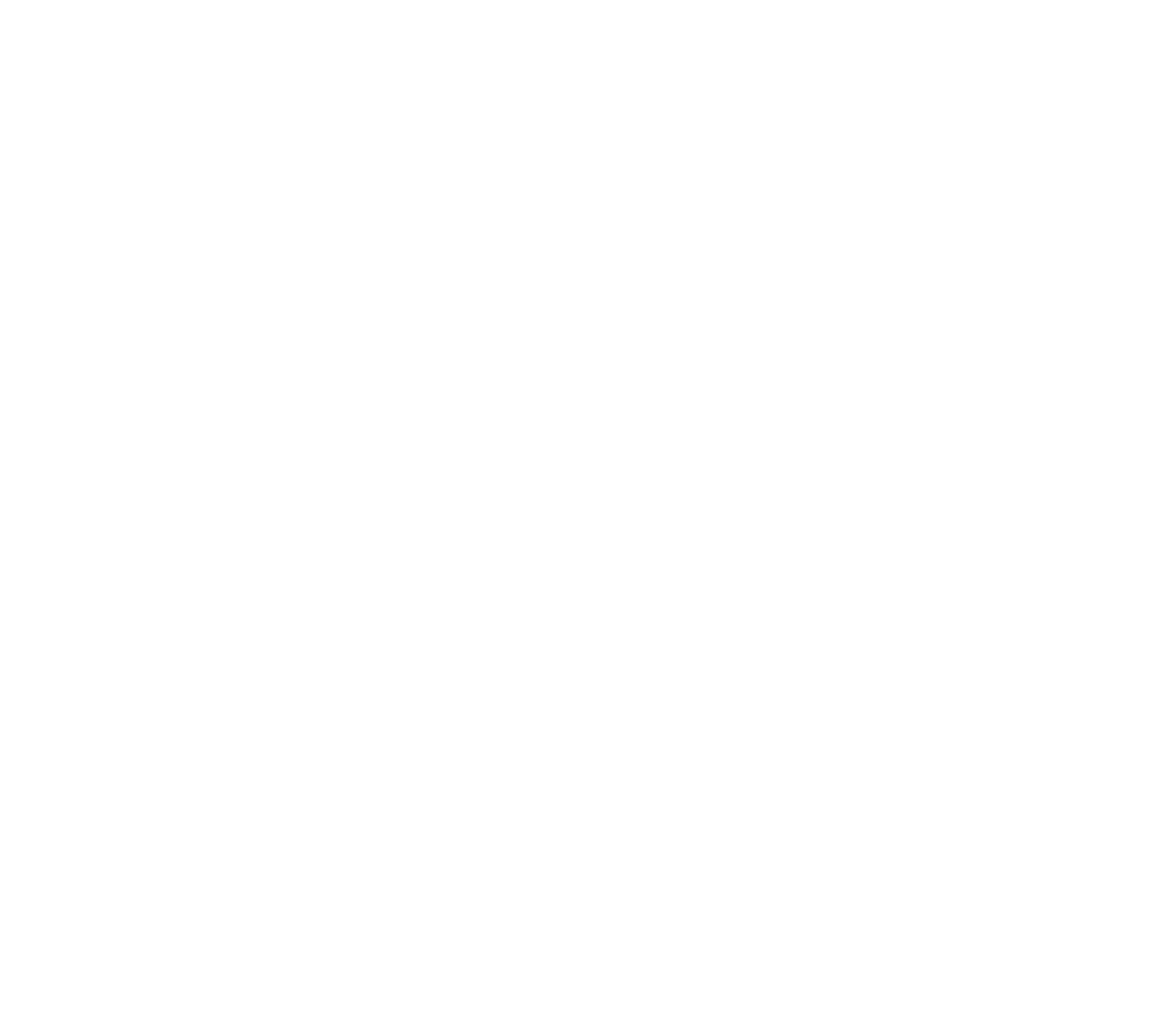} \\
  \end{tabular}
\caption[UGC\,8508 images: RC, IR, optical, and FUV]{Multi-wavelength coverage of UGC 8508 displaying a $3.0^\prime \times 3.0^\prime$ area. We show total RC flux density at the native resolution (top-left) and again with contours (top-centre). The RC contours are superposed on ancillary LITTLE THINGS images where possible: \halpha\ (middle-left); \RCNT\ obtained by subtracting the expected \RCT\ based on the \halpha-\RCT\ scaling factor of \cite{Deeg1997} from the total RC; {\em GALEX} FUV (middle-right); {\em Spitzer} 24\micron\ (bottom-left); {\em Spitzer} 70\micron\ (bottom-centre); FUV$+24{\rm \mu m}$--inferred SFRD from \citealp{Leroy2012} (bottom-right). We also show the RC that was isolated by the RC--based masking technique (top-right).}
  \label{figure:ugc8508Cc_maps}
\end{figure}

\begin{figure}
  \begin{tabular}{ccc}
    \includegraphics[width=0.31\linewidth,clip]{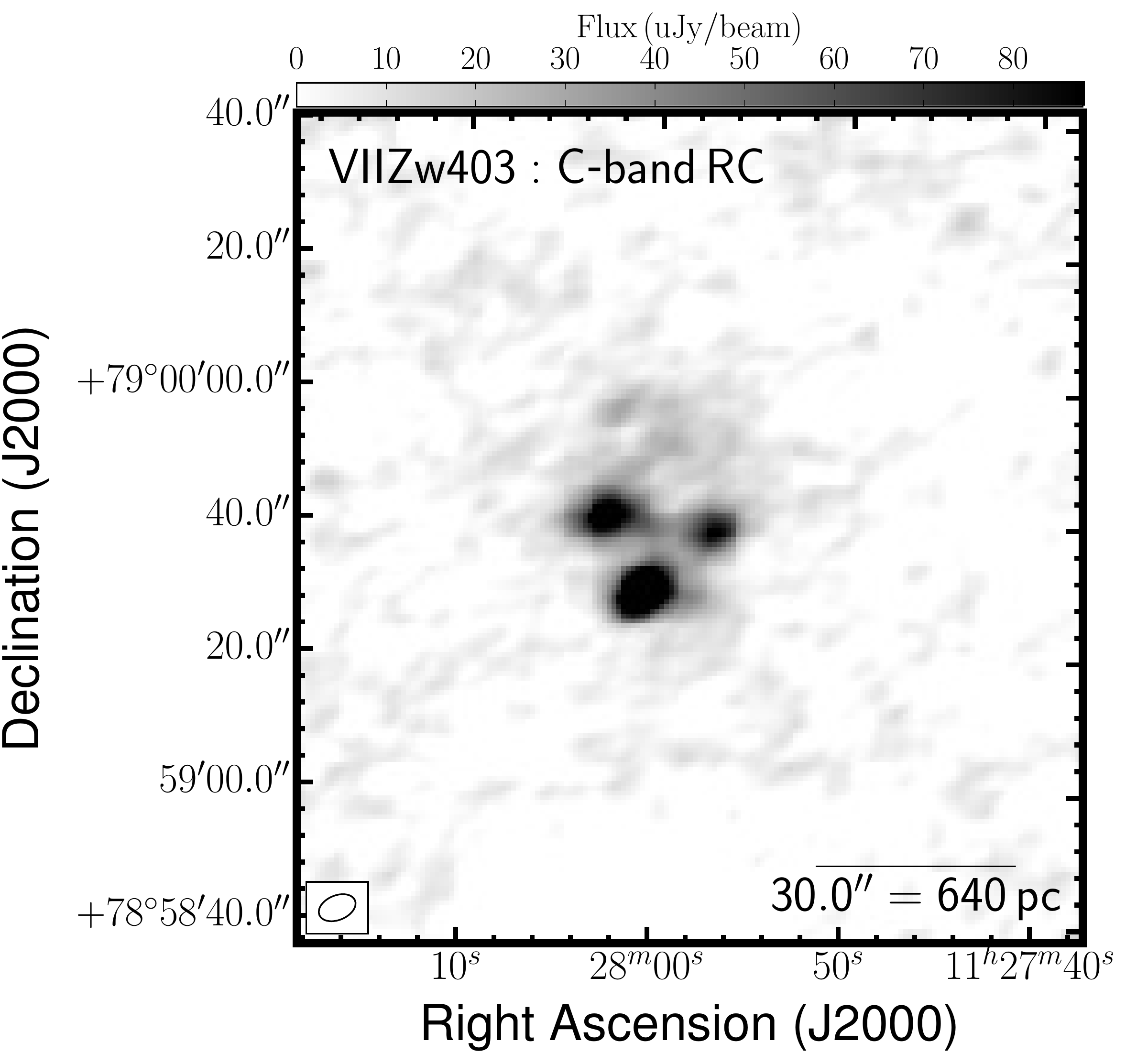} & \ 
    \includegraphics[width=0.31\linewidth,clip]{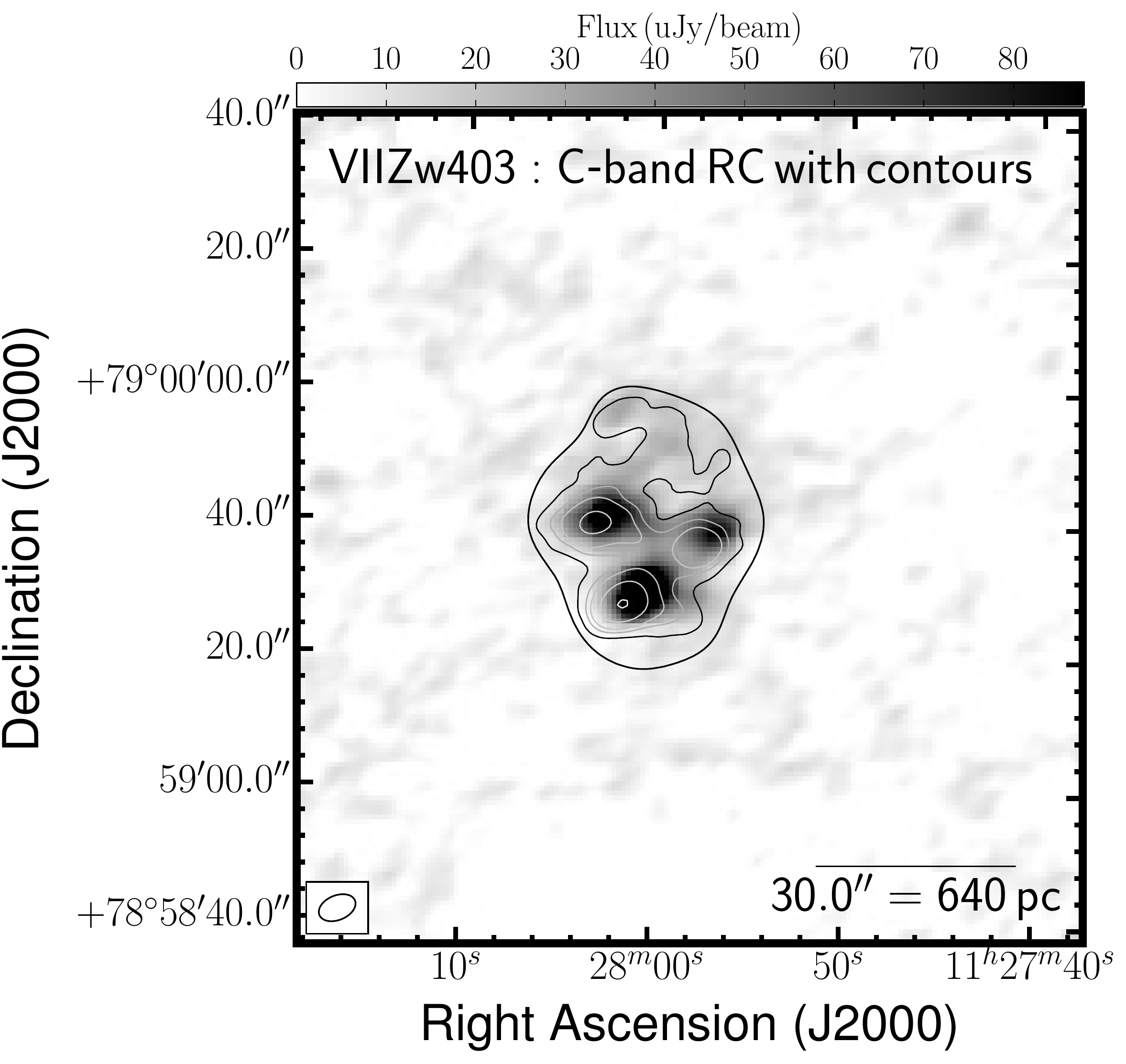} & \ 
    \includegraphics[width=0.31\linewidth,clip]{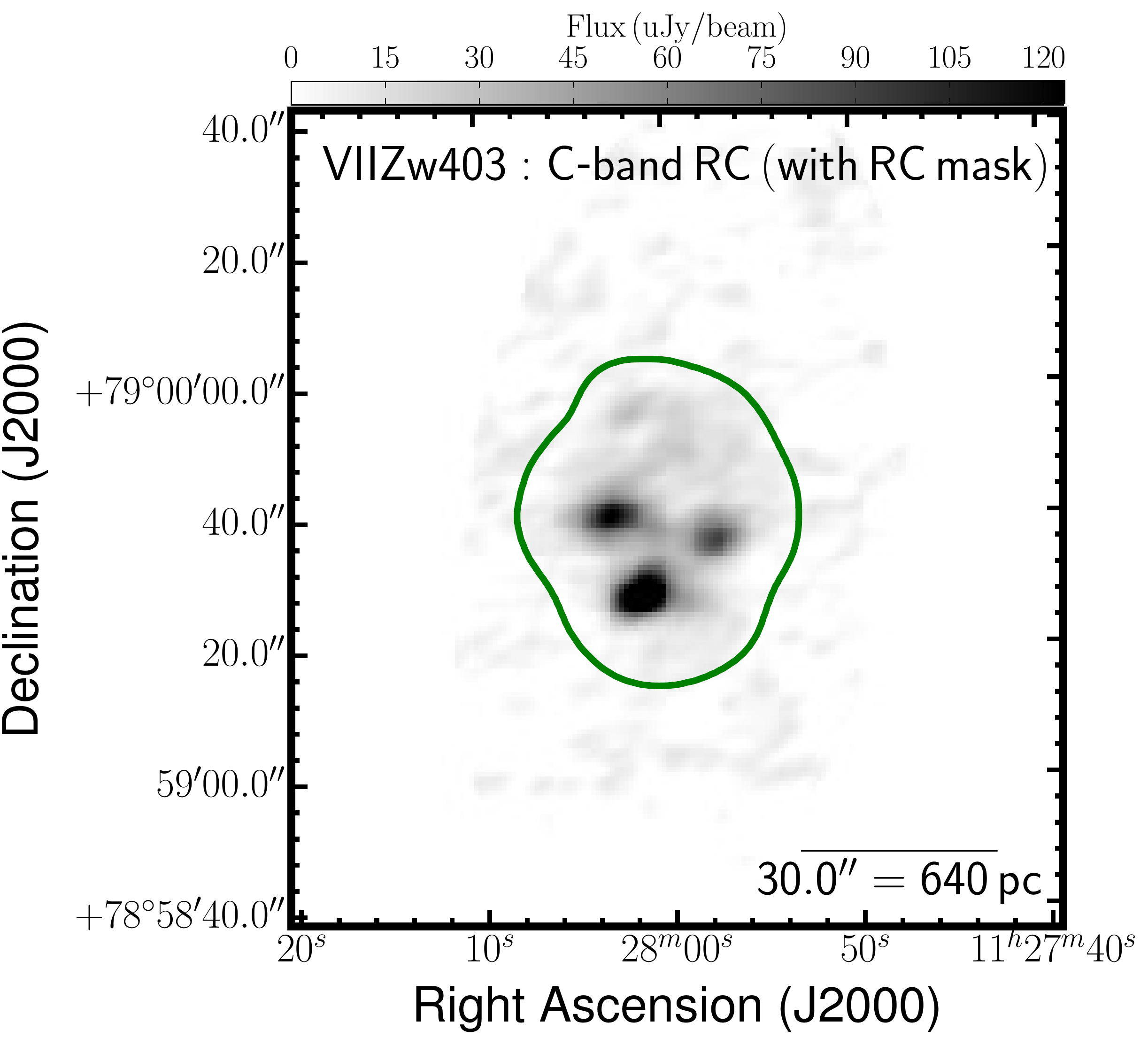} \\
    \includegraphics[width=0.31\linewidth,clip]{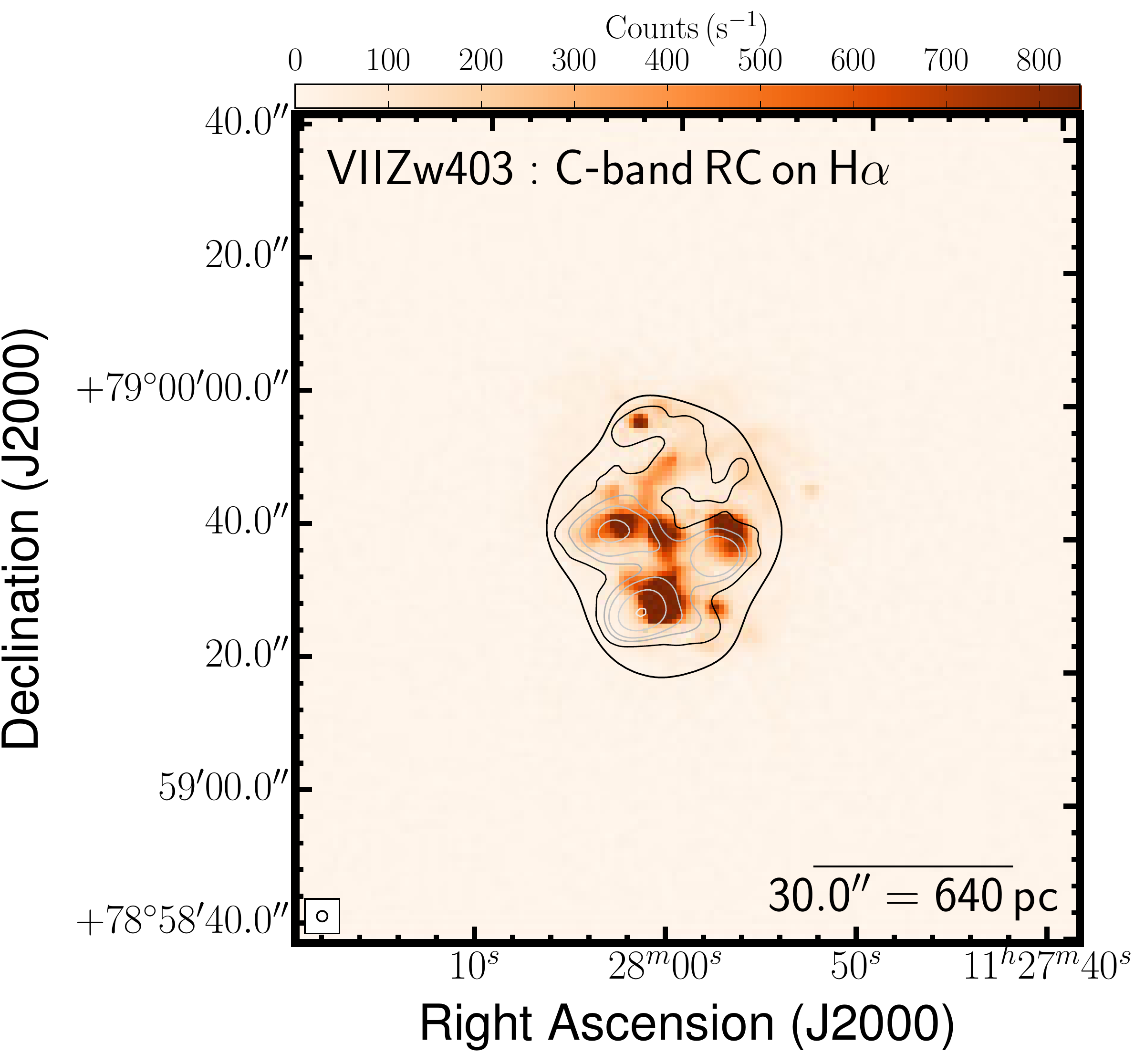} & \ 
    \includegraphics[width=0.31\linewidth,clip]{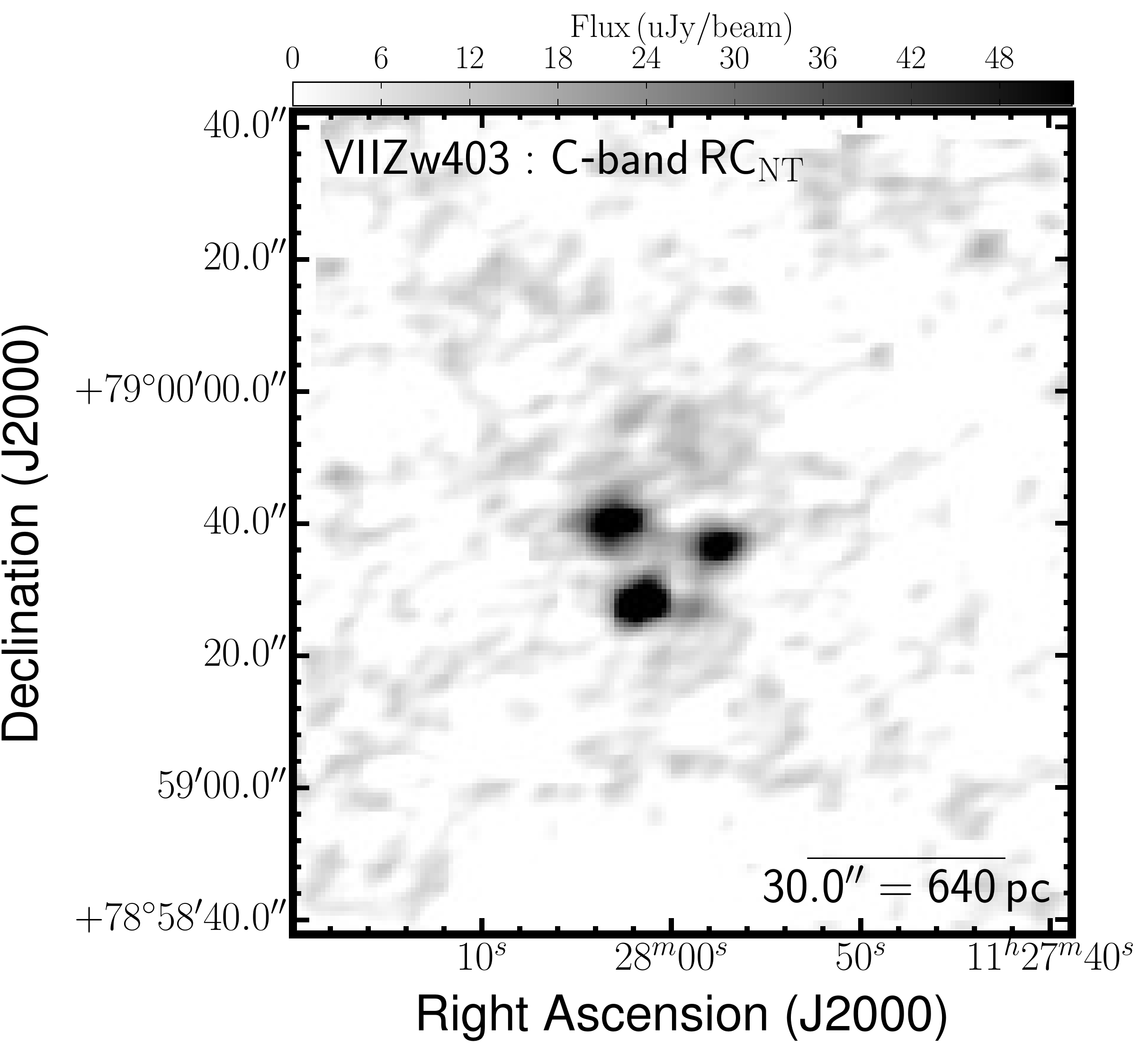} & \ 
    \includegraphics[width=0.31\linewidth,clip]{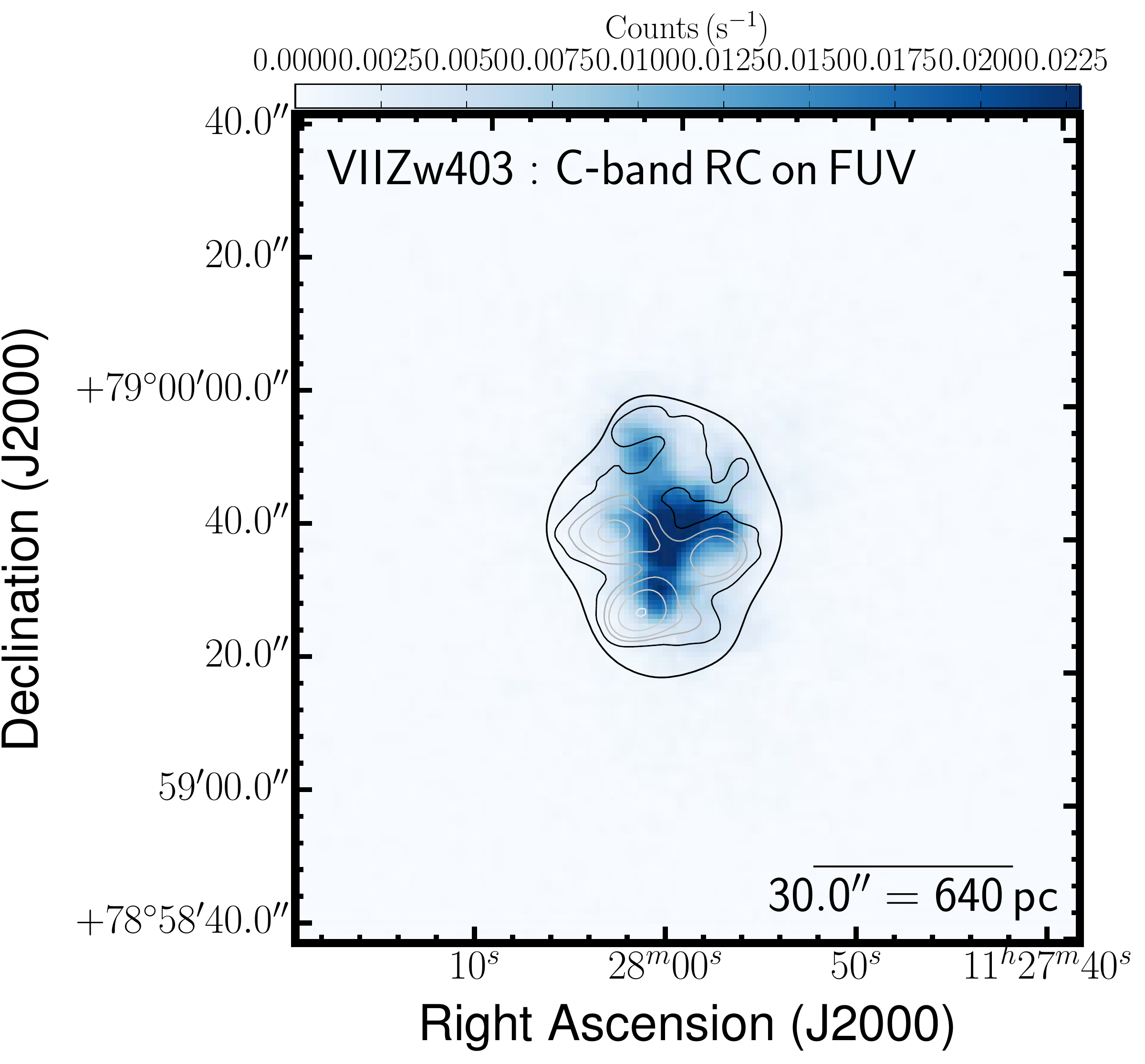} \\
    \includegraphics[width=0.31\linewidth,clip]{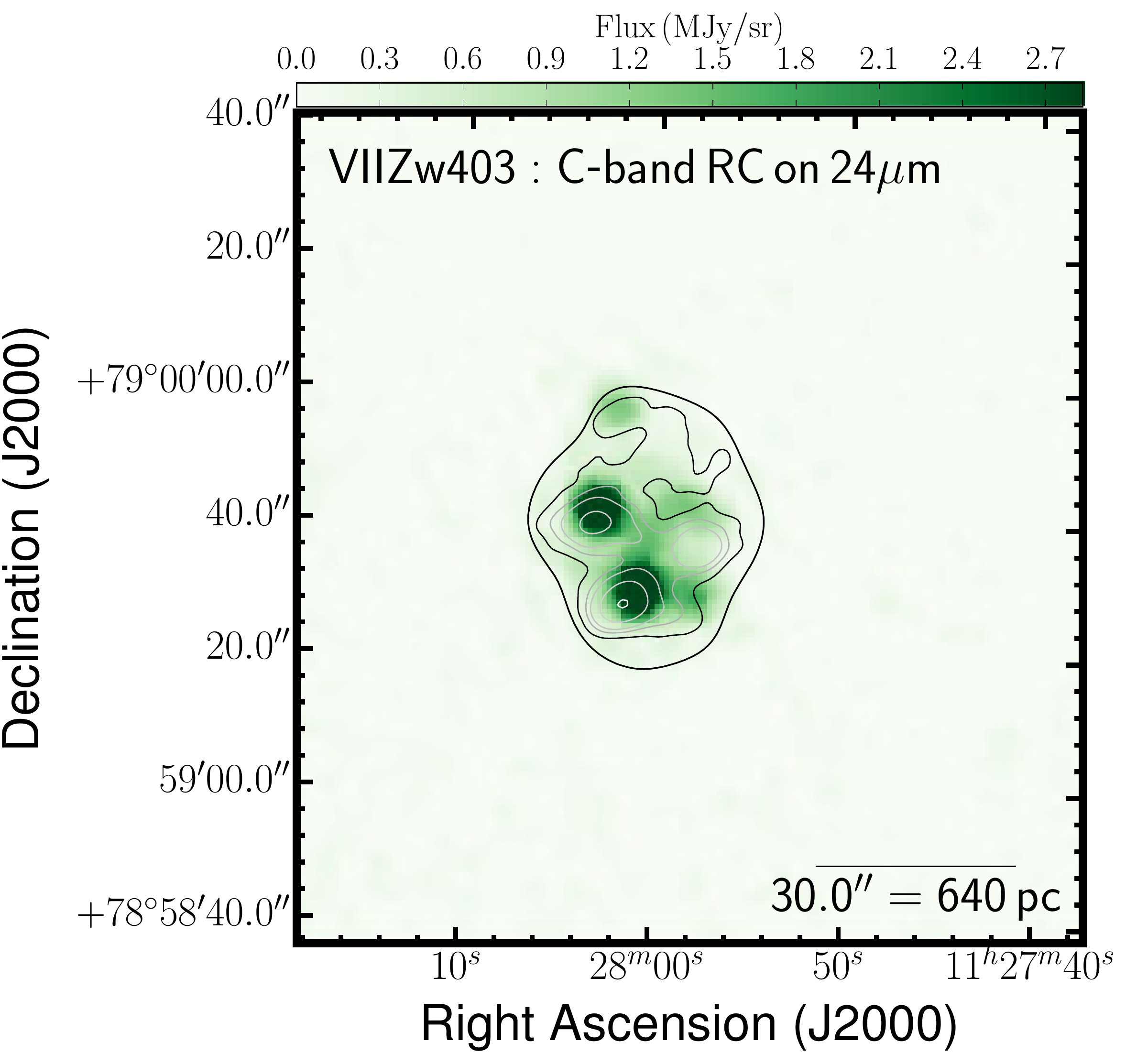} & \ 
    \includegraphics[width=0.31\linewidth,clip]{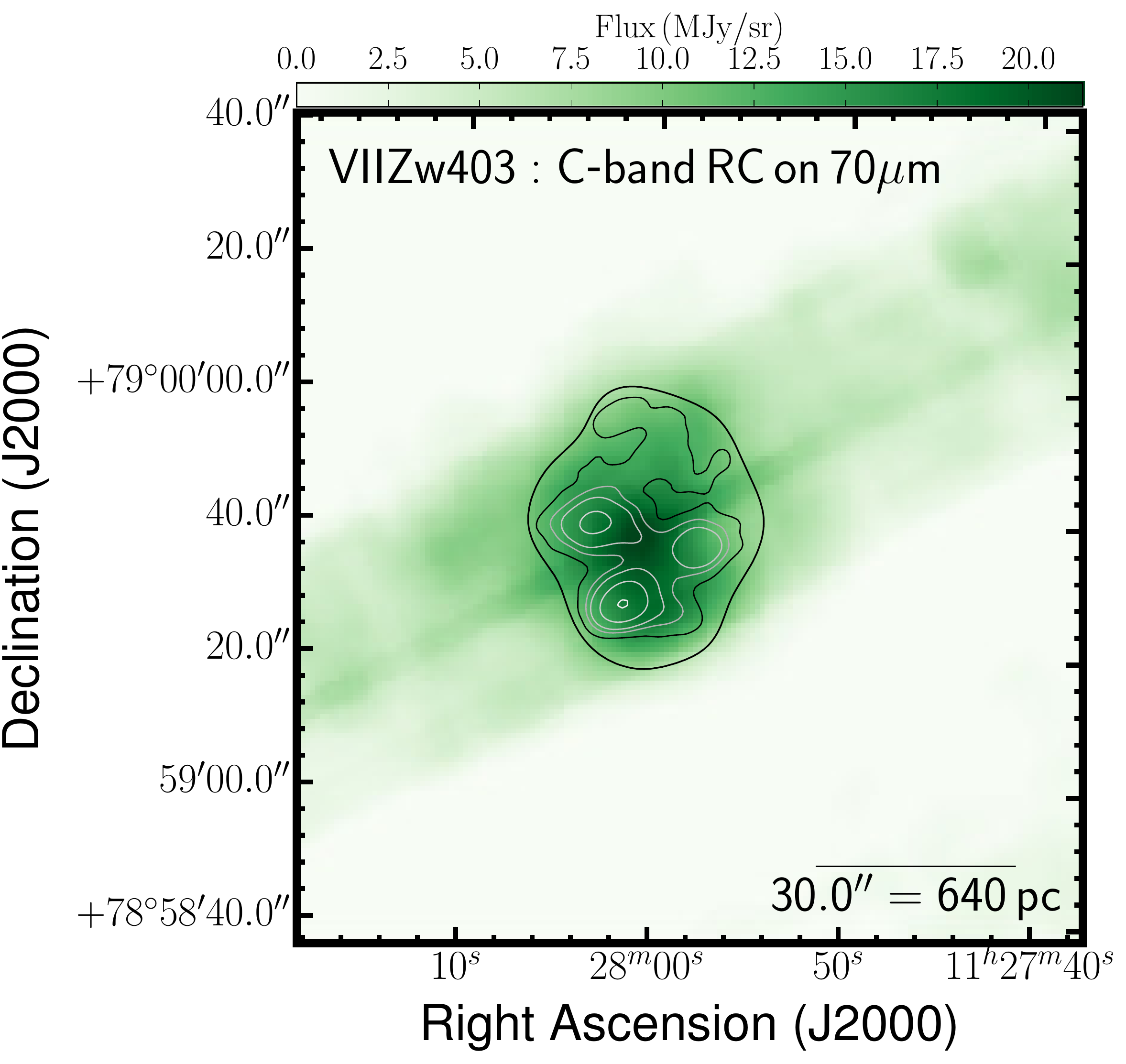} & \ 
    \includegraphics[width=0.31\linewidth,clip]{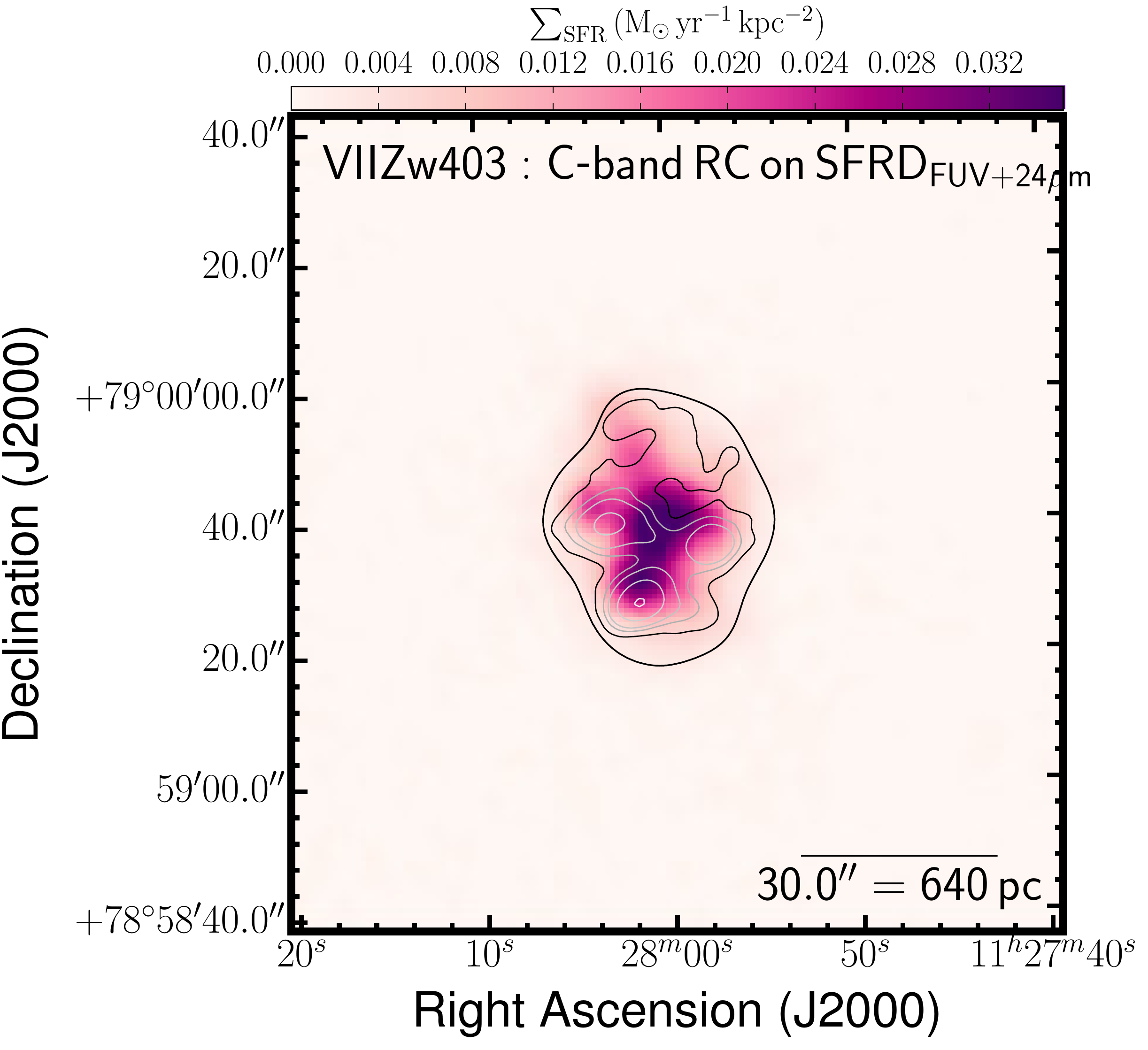} \\
  \end{tabular}
\caption[VIIZw\,403 images: RC, IR, optical, and FUV]{Multi-wavelength coverage of VIIZw 403 displaying a $2.6^\prime \times 2.6^\prime$ area. We show total RC flux density at the native resolution (top-left) and again with contours (top-centre). The RC contours are superposed on ancillary LITTLE THINGS images where possible: \halpha\ (middle-left); \RCNT\ obtained by subtracting the expected \RCT\ based on the \halpha-\RCT\ scaling factor of \cite{Deeg1997} from the total RC; {\em GALEX} FUV (middle-right); {\em Spitzer} 24\micron\ (bottom-left); {\em Spitzer} 70\micron\ (bottom-centre); FUV$+24{\rm \mu m}$--inferred SFRD from \citealp{Leroy2012} (bottom-right). We also show the RC that was isolated by the RC--based masking technique (top-right).}
  \label{figure:viizwCc_maps}
\end{figure}

\begin{figure}
  \begin{tabular}{ccc}
    \includegraphics[width=0.31\linewidth,clip]{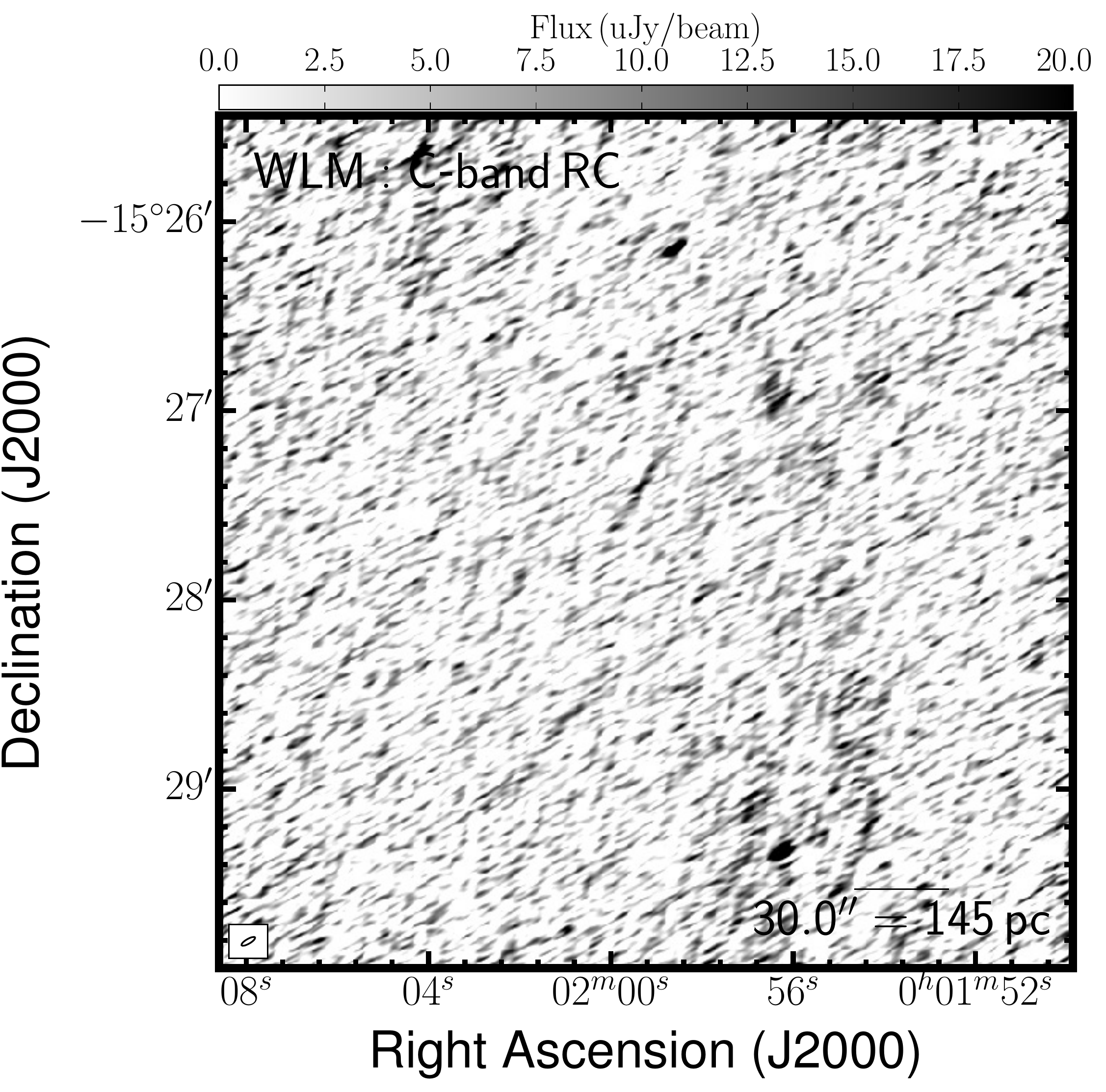} & \ 
    \includegraphics[width=0.31\linewidth,clip]{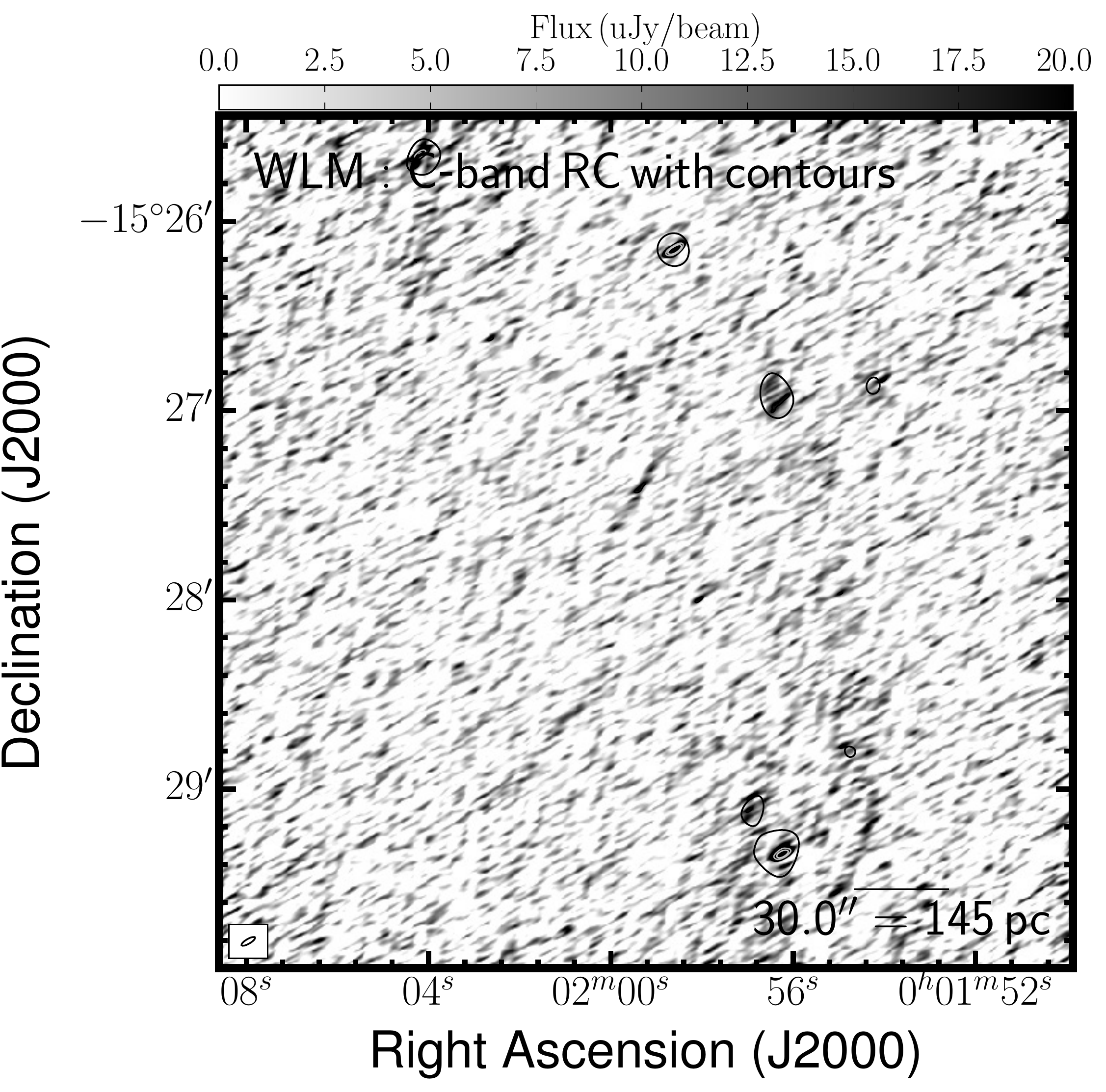} & \ 
    \includegraphics[width=0.31\linewidth,clip]{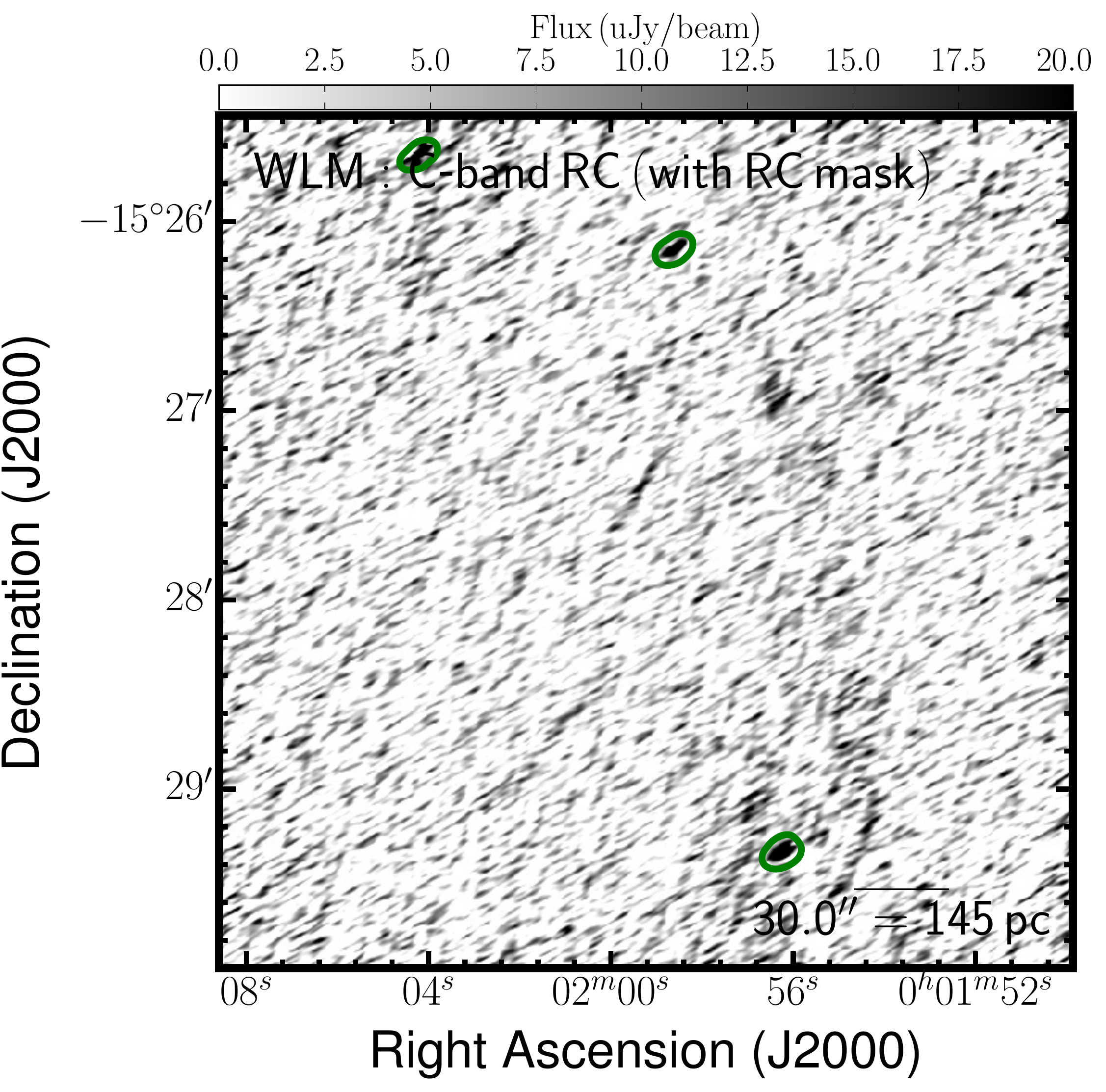} \\
    \includegraphics[width=0.31\linewidth,clip]{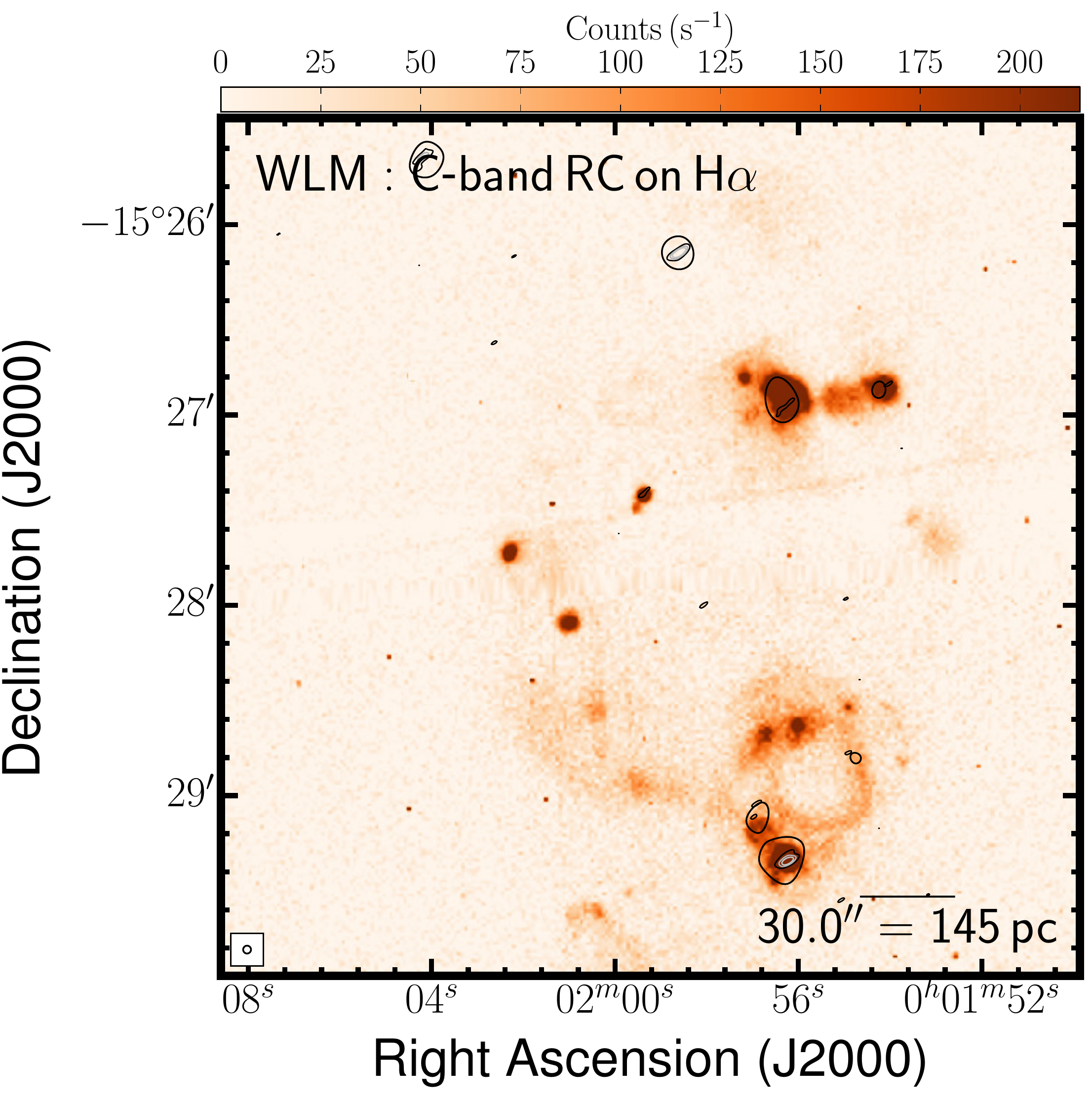} & \ 
    \includegraphics[width=0.31\linewidth,clip]{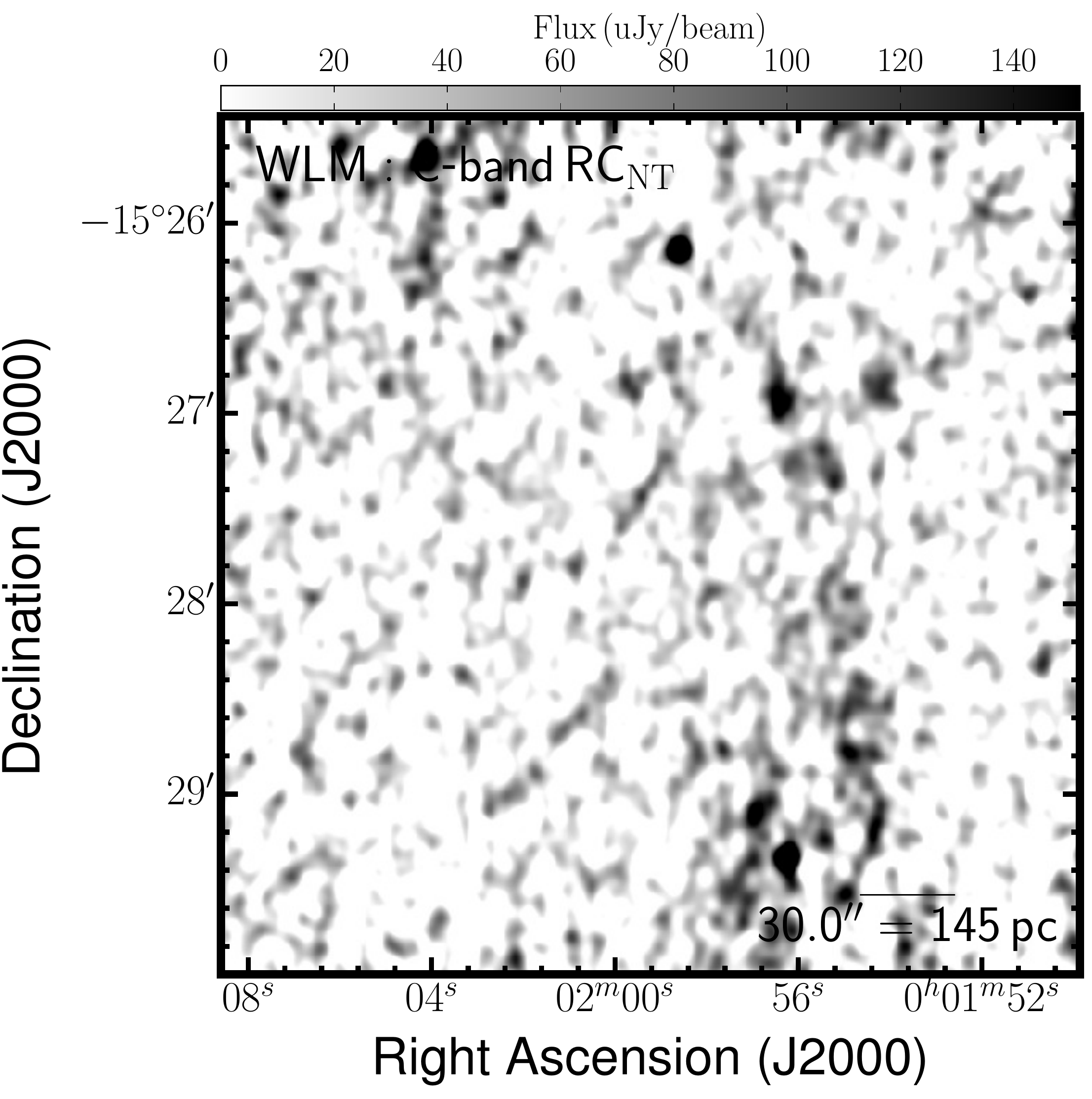} & \ 
    \includegraphics[width=0.31\linewidth,clip]{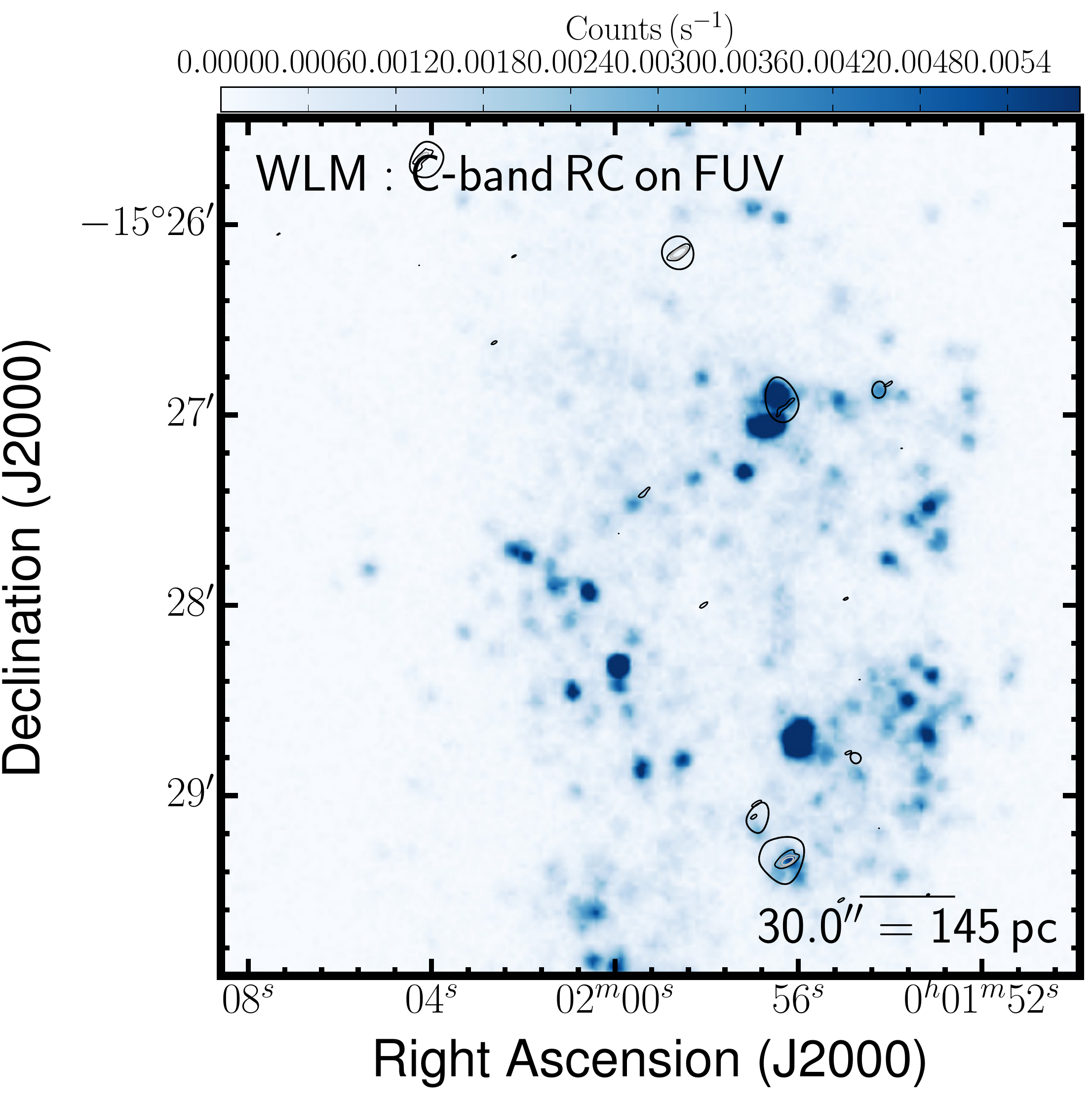} \\
    \includegraphics[width=0.31\linewidth,clip]{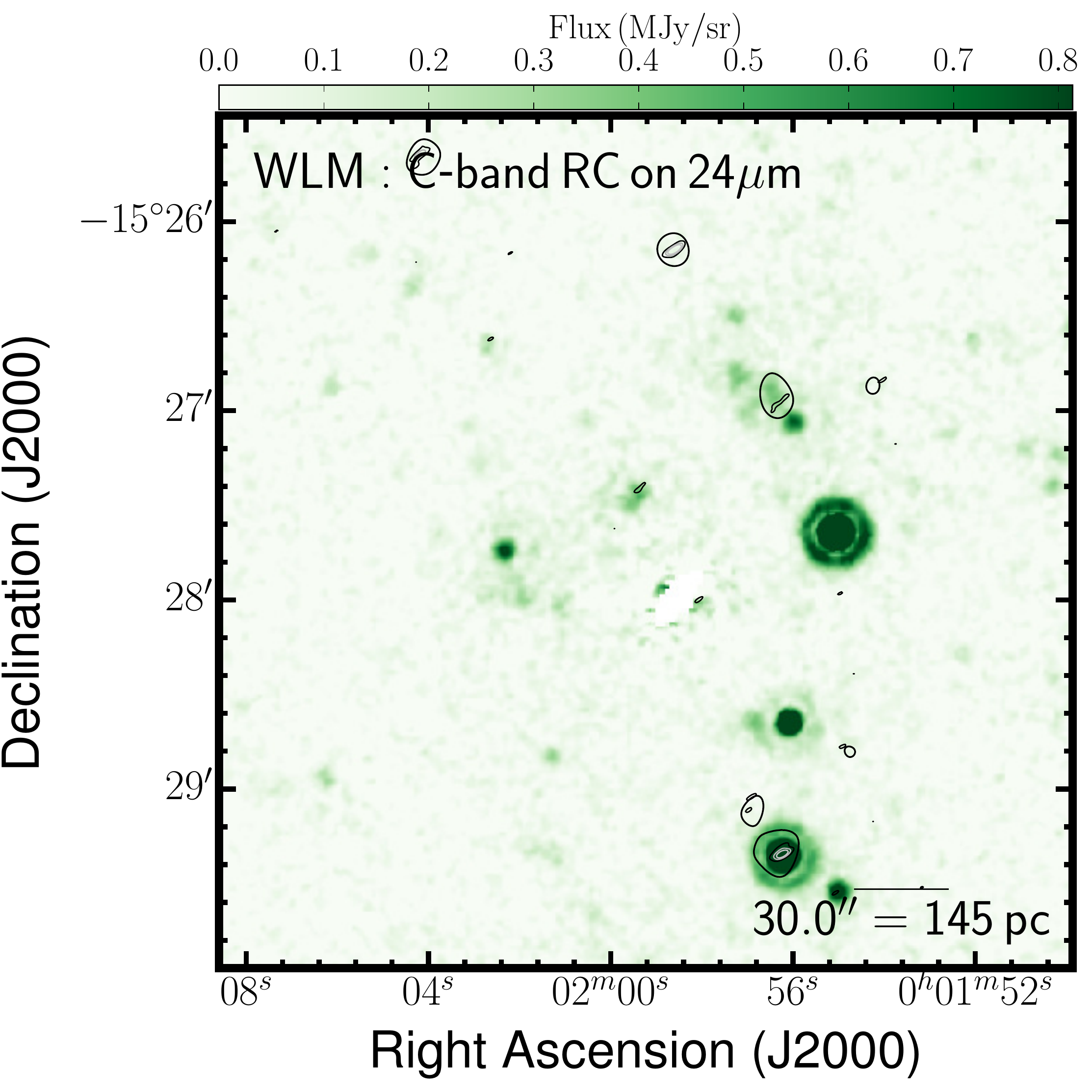} & \ 
    \includegraphics[width=0.31\linewidth,clip]{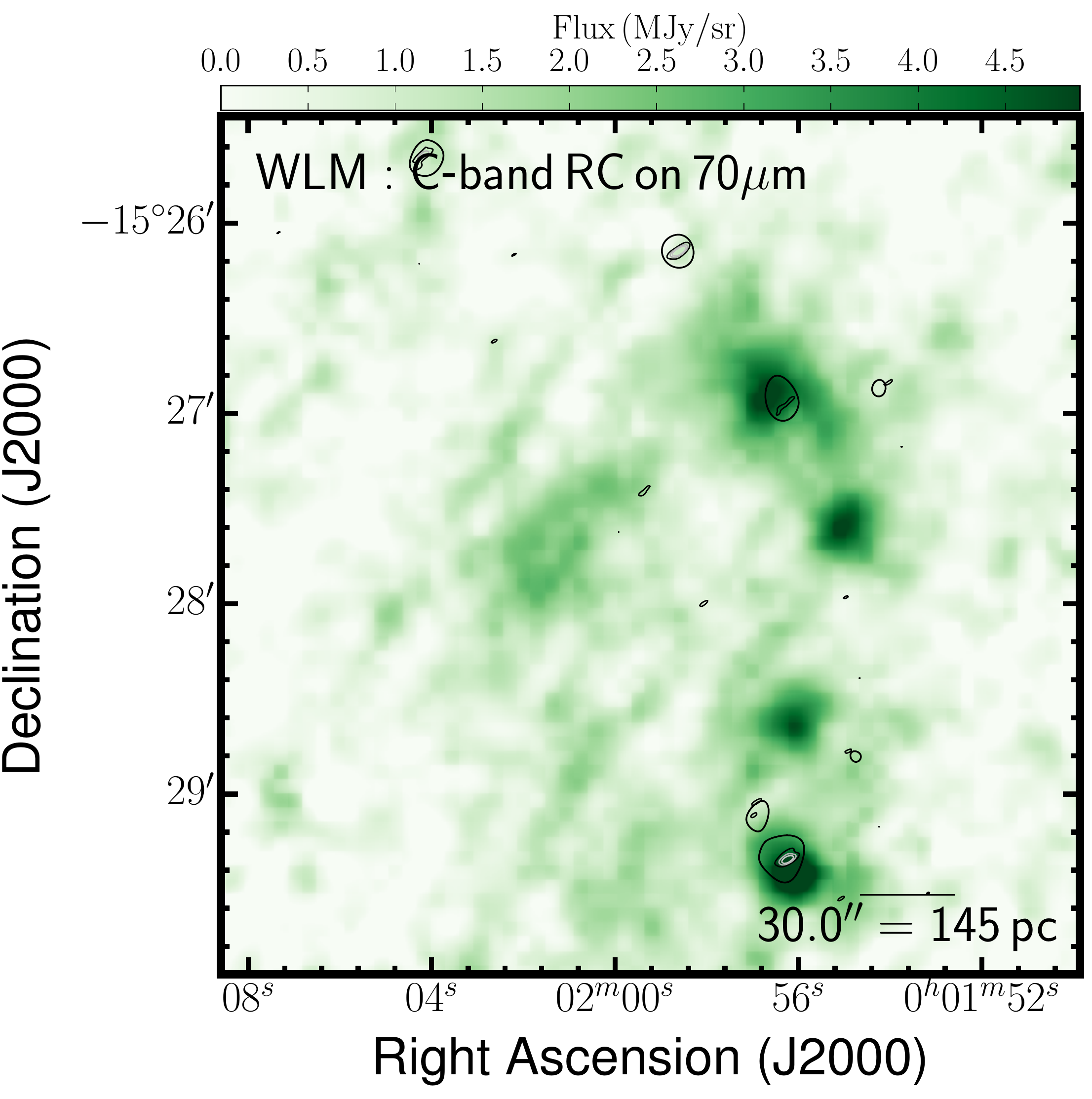} & \ 
    \includegraphics[width=0.31\linewidth,clip]{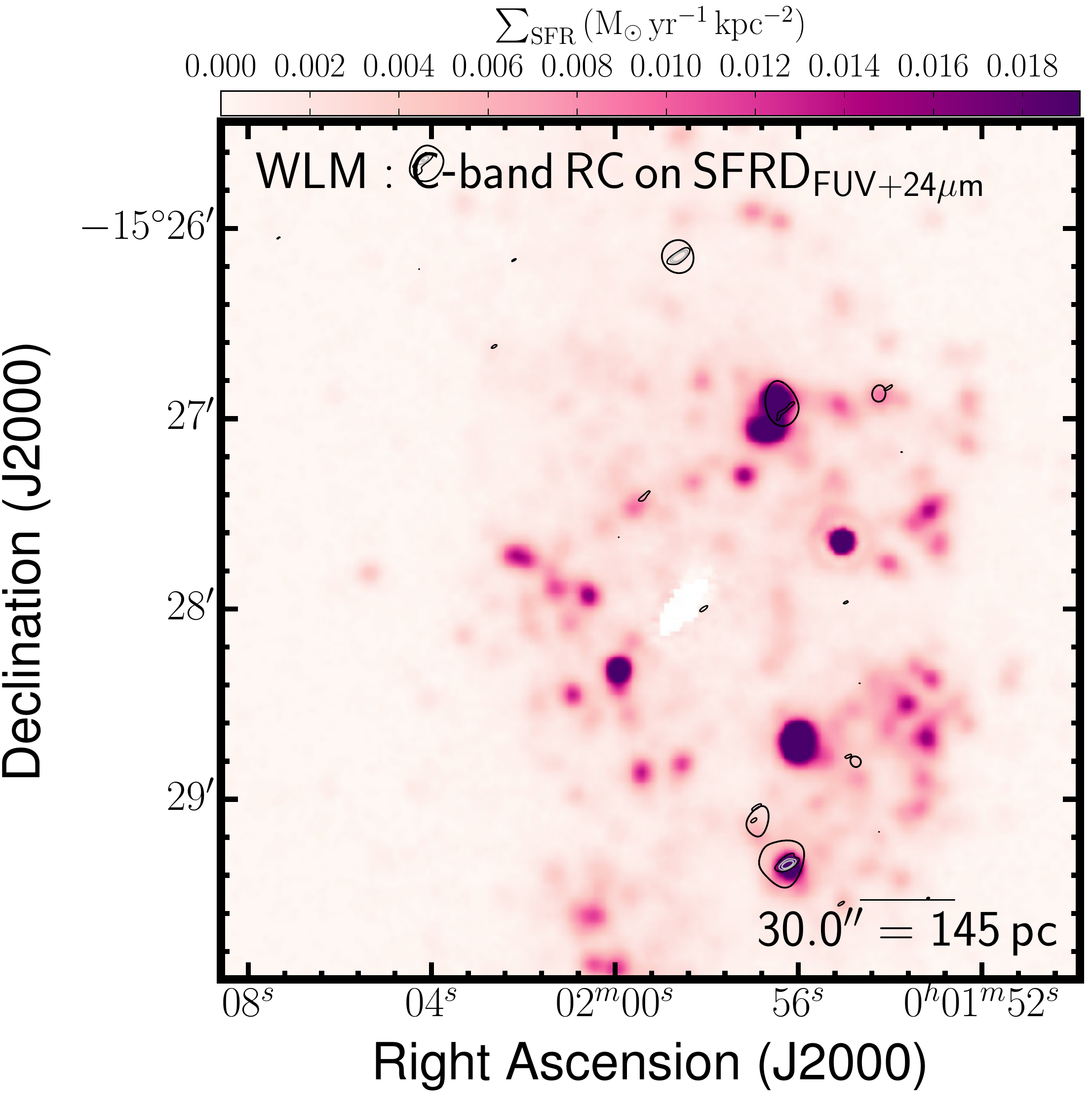} \\
  \end{tabular}
\caption[WLM images: RC, IR, optical, and FUV]{Multi-wavelength coverage of WLM displaying a $4.5^\prime \times 4.5^\prime$ area. We show total RC flux density at the native resolution (top-left) and again with contours (top-centre). The RC contours are superposed on ancillary LITTLE THINGS images where possible: \halpha\ (middle-left); \RCNT\ obtained by subtracting the expected \RCT\ based on the \halpha-\RCT\ scaling factor of \cite{Deeg1997} from the total RC; {\em GALEX} FUV (middle-right); {\em Spitzer} 24\micron\ (bottom-left); {\em Spitzer} 70\micron\ (bottom-centre); FUV$+24{\rm \mu m}$--inferred SFRD from \citealp{Leroy2012} (bottom-right). We also show the RC that was isolated by the RC--based masking technique (top-right).}
  \label{figure:wlmCc_maps}
\end{figure}

We provide images of the following dwarf galaxies: 
CVn\,I\,DwA (Fig.~\ref{figure:cvnidwaCc_maps}), 
DDO\,43 (Fig.~\ref{figure:ddo43Cc_maps}), 
DDO\,46 (Fig.~\ref{figure:ddo46Cc_maps}), 
DDO\,47 (Fig.~\ref{figure:ddo47Cc_maps}), 
DDO\,50 (Fig.~\ref{figure:ddo50Cc_maps}), 
DDO\,52 (Fig.~\ref{figure:ddo52Cc_maps}), 
DDO\,53 (Fig.~\ref{figure:ddo53Cc_maps}), 
DDO\,63 (Fig.~\ref{figure:ddo63Cc_maps}), 
DDO\,69 (Fig.~\ref{figure:ddo69Cc_maps}), 
DDO\,70 (Fig.~\ref{figure:ddo70Cc_maps}), 
DDO\,75 (Fig.~\ref{figure:ddo75Cc_maps}), 
DDO\,87 (Fig.~\ref{figure:ddo87Cc_maps}), 
DDO\,101 (Fig.~\ref{figure:ddo101Cc_maps}), 
DDO\,126 (Fig.~\ref{figure:ddo126Cc_maps}), 
DDO\,133 (Fig.~\ref{figure:ddo133Cc_maps}), 
DDO\,154 (Fig.~\ref{figure:ddo154Cc_maps}), 
DDO\,155 (Fig.~\ref{figure:ddo155Cc_maps}), 
DDO\,165 (Fig.~\ref{figure:ddo165Cc_maps}), 
DDO\,167 (Fig.~\ref{figure:ddo167Cc_maps}), 
DDO\,168 (Fig.~\ref{figure:ddo168Cc_maps}), 
DDO\,187 (Fig.~\ref{figure:ddo187Cc_maps}), 
DDO\,210 (Fig.~\ref{figure:ddo210Cc_maps}), 
DDO\,216 (Fig.~\ref{figure:ddo216Cc_maps}), 
F564\,V3 (Fig.~\ref{figure:f564v3Cc_maps}), 
Haro\,29 (Fig.~\ref{figure:haro29Cc_maps}), 
Haro\,36 (Fig.~\ref{figure:haro36Cc_maps}), 
IC\,10 (Fig.~\ref{figure:ic10Cc_maps}), 
IC\,1613 (Fig.~\ref{figure:ic1613C1d1_maps}), 
LGS\,3 (Fig.~\ref{figure:lgs3Cc_maps}), 
M81\,DwA (Fig.~\ref{figure:m81dwaCc_maps}), 
Mrk\,178 (Fig.~\ref{figure:mrk178Cc_maps}), 
NGC\,1569 (Fig.~\ref{figure:ngc1569Cc_maps}), 
NGC\,2366 (Fig.~\ref{figure:ngc2366Cc_maps}), 
NGC\,3738 (Fig.~\ref{figure:ngc3738Cc_maps}), 
NGC\,4163 (Fig.~\ref{figure:ngc4163Cc_maps}), 
NGC\,4214 (Fig.~\ref{figure:ngc4214Cc_maps}), 
Sag\,DIG (Fig.~\ref{figure:sagdigCc_maps}), 
UGC\,8508 (Fig.~\ref{figure:ugc8508Cc_maps}), 
VII\,Zw\,403 (Fig.~\ref{figure:viizwCc_maps}), and 
WLM (Figure~\ref{figure:wlmCc_maps}).
\clearpage

\clearpage

\clearpage

\clearpage

\clearpage

\clearpage

\clearpage

\clearpage

\clearpage

\clearpage

\clearpage

\clearpage

\clearpage

\clearpage

\clearpage

\clearpage

\clearpage

\clearpage

\clearpage

\clearpage

\clearpage

\clearpage

\clearpage

\clearpage

\clearpage

\clearpage

\clearpage

\clearpage

\clearpage

\clearpage

\clearpage

\clearpage

\clearpage

\clearpage

\clearpage

\clearpage

\clearpage

\clearpage

\clearpage

\clearpage
\clearpage

\section{Integrated Properties Including Ambiguous Sources}
\label{Section:Appendix_Tables}

\begin{deluxetable}{lccccccccccc}
\tablewidth{0pt}
\tabletypesize{\tiny}
\tablecolumns{13}
\rotate
\tablecaption{Integrated emission over the disk of the LITTLE THINGS Galaxies\label{table:12A-234_DiskQuantities_ambig}}
\tablehead{
\colhead{Galaxy}  & \colhead{Size} & \colhead{P.A.} & \colhead{6\,cm RC} & \colhead{\halpha}        & \colhead{FUV} & \colhead{24\micron\ MIR} & \colhead{70\micron\ FIR} & \colhead{6\,cm \RCNT} & \colhead{$B_{\rm eq}$} \\
  & ($^{\prime}$) & ($^\circ$)  & (mJy)  & ($10^{-13} $\,ergs\,s$^{-1}$\,cm$^{-2}$)   & (mJy) & ($10^{-2}$\,Jy) & ($10^{-2}$\,Jy) & (mJy) & (${\rm \mu G}$) \\
(1) & (2) & (3) & (4) & (5) & (6) & (7) & (8) & (9) & (10)}
\startdata
CVn I dwA	& $1.7 \times 1.4$ & $80$ &  $>0.29$ & $1.95\pm 0.03$ & $1.04\pm 0.11$ & $0.16\pm 0.06$ & $2.48\pm 0.04$ & $>0.29$ & $<2$ \\
DDO 43	& $1.8 \times 1.2$ & $6$ &  $>0.99$ & $1.28\pm 0.03$ & $1.07\pm 0.11$ & \nodata & \nodata & $>0.99$ & $<2$ \\
DDO 46$^\mathrm{V}$	& $3.8 \times 3.4$ & $84$ &  $>1.17$ & $1.09\pm 0.02$ & $1.76\pm 0.18$ & \nodata & \nodata & $>1.17$ & $<2$ \\
DDO 47	& $4.5 \times 2.3$ & $-79$ &  $>0.62$ & $3.01\pm 0.03$ & $3.02\pm 0.30$ & \nodata & \nodata & $>0.62$ & $<1$ \\
DDO 50	& $7.9 \times 5.7$ & $18$ &  $7.32 \pm 0.60$ & $60.49\pm 0.49$ & $42.13\pm 4.22$ & $17.89\pm 0.01$ & $322.60\pm 0.28$ & $1.56\pm 0.60$ & $<1$ \\
DDO 52	& $2.2 \times 1.4$ & $4$ &  $>1.28$ & $0.29\pm 0.01$ & $0.61\pm 0.06$ & $-0.04\pm 0.02$ & $1.81\pm 0.05$ & $>1.28$ & $<1$ \\
DDO 53	& $2.7 \times 1.4$ & $81$ &  $0.82 \pm 0.13$ & $4.46\pm 0.04$ & $2.68\pm 0.27$ & $2.52\pm 0.02$ & $25.74\pm 0.03$ & $0.40\pm 0.13$ & $<1$ \\
DDO 63	& $4.3 \times 4.3$ & $0$ &  $1.11 \pm 0.24$ & $4.46\pm 0.04$ & $5.16\pm 0.52$ & $1.83\pm 0.01$ & $3.88\pm 0.13$ & $0.69\pm 0.24$ & $<1$ \\
DDO 69	& $4.8 \times 2.7$ & $-64$ &  $>0.91$ & $1.71\pm 0.01$ & $4.87\pm 0.49$ & $-0.65\pm 0.01$ & $11.80\pm 0.07$ & $>0.91$ & $<1$ \\
DDO 70	& $7.4 \times 4.4$ & $88$ &  $>1.50$ & $6.38\pm 0.04$ & $11.68\pm 1.17$ & $0.58\pm 0.01$ & $63.30\pm 0.14$ & $>1.50$ & $<1$ \\
DDO 75	& $6.2 \times 5.2$ & $42$ &  $>1.95$ & $40.35\pm 0.10$ & $29.23\pm 2.92$ & $0.45\pm 0.01$ & $75.03\pm 0.19$ & $>1.95$ & $<1$ \\
DDO 87	& $2.3 \times 1.3$ & $76$ &  $>0.70$ & $0.68\pm 0.01$ & $0.65\pm 0.07$ & $0.09\pm 0.02$ & $7.60\pm 0.03$ & $>0.70$ & $<1$ \\
DDO 101	& $2.1 \times 1.5$ & $-69$ &  $>1.79$ & $0.82\pm 0.01$ & $0.39\pm 0.04$ & $0.24\pm 0.02$ & $-0.54\pm 0.04$ & $>1.79$ & $<1$ \\
DDO 126	& $3.5 \times 1.7$ & $-41$ &  $>0.57$ & $3.66\pm 0.08$ & $2.91\pm 0.29$ & $0.32\pm 0.03$ & $14.92\pm 0.10$ & $>0.57$ & $<1$ \\
DDO 133	& $4.7 \times 3.2$ & $-6$ &  $>1.19$ & $4.61\pm 0.03$ & $4.18\pm 0.42$ & $0.64\pm 0.01$ & $33.71\pm 0.13$ & $>1.19$ & $<2$ \\
DDO 154	& $3.1 \times 1.6$ & $46$ &  $>1.73$ & $2.21\pm 0.02$ & $3.77\pm 0.38$ & $0.28\pm 0.03$ & $3.66\pm 0.04$ & $>1.73$ & $<1$ \\
DDO 155	& $1.9 \times 1.3$ & $51$ &  $>0.47$ & $4.85\pm 0.05$ & \nodata & $0.22\pm 0.03$ & $16.15\pm 0.05$ & $>0.47$ & $<2$ \\
DDO 165	& $4.3 \times 2.3$ & $89$ &  $>1.19$ & $1.53\pm 0.01$ & \nodata & $0.03\pm 0.01$ & $10.69\pm 0.06$ & $>1.19$ & $<2$ \\
DDO 167	& $1.5 \times 1.0$ & $-24$ &  $>0.51$ & $0.80\pm 0.01$ & $1.05\pm 0.11$ & \nodata & \nodata & $>0.51$ & $<3$ \\
DDO 168	& $4.6 \times 2.9$ & $-25$ &  $>0.94$ & $5.95\pm 0.03$ & $5.57\pm 0.56$ & $0.67\pm 0.01$ & $42.28\pm 0.10$ & $>0.94$ & $<1$ \\
DDO 187	& $2.1 \times 1.7$ & $37$ &  $>1.17$ & $0.57\pm 0.01$ & $1.15\pm 0.12$ & $-0.02\pm 0.03$ & $-1.94\pm 0.09$ & $>1.17$ & $<1$ \\
DDO 210	& $2.6 \times 1.3$ & $-85$ &  $>0.88$ & \nodata & $0.80\pm 0.08$ & $-0.17\pm 0.02$ & $5.37\pm 0.04$ & $>0.87$ & $<2$ \\
DDO 216	& $8.0 \times 3.6$ & $-58$ &  $>1.29$ & $0.10\pm 0.01$ & $2.00\pm 0.20$ & $-0.09\pm 0.01$ & $9.96\pm 0.08$ & $>1.29$ & $<1$ \\
F564-V03$^\mathrm{V}$	& $1.3 \times 1.0$ & $7$ &  $>0.35$ & \nodata & $0.10\pm 0.01$ & \nodata & \nodata & $>0.35$ & $<3$ \\
Haro 29	& $1.7 \times 1.0$ & $85$ &  $2.18 \pm 0.11$ & $13.06\pm 0.45$ & $3.05\pm 0.33$ & $5.89\pm 0.05$ & $39.80\pm 0.05$ & $0.94\pm 0.12$ & $6$ \\
Haro 36$^\mathrm{V}$	& $1.5 \times 1.2$ & $90$ &  $0.94 \pm 0.09$ & $2.41\pm 0.03$ & $2.84\pm 0.29$ & $0.94\pm 0.04$ & $23.66\pm 0.06$ & $0.71\pm 0.09$ & $5$ \\
IC 1613	& $18.2 \times 14.7$ & $71$ &  $5.81 \pm 0.55$ & $56.21\pm 0.87$ & $93.29\pm 9.38$ & $6.86\pm 0.02$ & $417.00\pm 1.74$ & $0.51\pm 0.56$ & $5$ \\
IC 10$^\mathrm{V}$	& $11.6 \times 9.1$ & $-38$ &  $96.38 \pm 0.81$ & $1191.00\pm 5.73$ & \nodata & $3741.00\pm 4.83$ & $9547.00\pm 12.08$ & $-16.78\pm 0.97$ & $<1$ \\
LGS 3	& $1.9 \times 1.0$ & $-3$ &  $>0.57$ & \nodata & $0.08\pm 0.01$ & \nodata & \nodata & $>0.57$ & $<2$ \\
M81 dwA$^\mathrm{V}$	& $1.5 \times 1.1$ & $86$ &  $>1.28$ & \nodata & $0.38\pm 0.04$ & \nodata & \nodata & $>1.28$ & $<2$ \\
Mrk 178	& $2.0 \times 0.9$ & $-51$ &  $1.01 \pm 0.14$ & $5.38\pm 0.09$ & $2.56\pm 0.27$ & $0.45\pm 0.03$ & $0.45\pm 0.01$ & $0.50\pm 0.14$ & $5$ \\
NGC 1569$^\mathrm{V}$	& $2.3 \times 1.3$ & $-59$ &  $151.60 \pm 0.31$ & $489.50\pm 3.03$ & $750.50\pm 76.03$ & $709.10\pm 13.47$ & $3596.00\pm 2.69$ & $72.90\pm 0.58$ & $17$ \\
NGC 2366	& $9.4 \times 4.0$ & $33$ &  $9.82 \pm 0.59$ & $96.45\pm 1.11$ & $37.34\pm 3.74$ & $65.64\pm 0.01$ & $507.90\pm 0.31$ & $0.66\pm 0.60$ & $17$ \\
NGC 3738	& $4.8 \times 4.8$ & $0$ &  $2.62 \pm 0.48$ & $16.26\pm 0.17$ & $11.22\pm 1.13$ & $11.65\pm 0.03$ & $248.10\pm 0.41$ & $1.07\pm 0.48$ & $17$ \\
NGC 4163	& $2.9 \times 1.9$ & $18$ &  $>0.69$ & $1.48\pm 0.02$ & $2.68\pm 0.27$ & $0.44\pm 0.03$ & $10.16\pm 0.11$ & $>0.69$ & $<2$ \\
NGC 4214	& $9.3 \times 8.5$ & $16$ &  $27.74 \pm 0.55$ & $178.00\pm 0.92$ & $79.44\pm 7.96$ & $199.50\pm 0.01$ & $2383.00\pm 1.09$ & $10.83\pm 0.55$ & $6$ \\
Sag DIG$^\mathrm{V}$	& $4.3 \times 2.3$ & $88$ &  $>2.47$ & $1.28\pm 0.01$ & $4.55\pm 0.46$ & \nodata & \nodata & $>2.47$ & $6$ \\
UGC 8508	& $2.5 \times 1.4$ & $-60$ &  $0.96 \pm 0.13$ & $2.85\pm 0.04$ & \nodata & $0.40\pm 0.03$ & $13.04\pm 0.04$ & $0.69\pm 0.13$ & $4$ \\
VIIZw 403	& $2.2 \times 1.1$ & $-11$ &  $1.50 \pm 0.09$ & $7.44\pm 0.15$ & $3.66\pm 0.37$ & $2.05\pm 16.60$ & $62.33\pm 1.24$ & $0.80\pm 0.09$ & $6$ \\
WLM	& $11.6 \times 5.1$ & $-2$ &  $>2.51$ & $16.81\pm 0.06$ & $29.55\pm 2.96$ & $4.62\pm 0.01$ & $117.80\pm 0.18$ & $>2.51$ & $<1$ \\

\enddata
\tablecomments{ (Column 1) Name of dwarf galaxy. The superscript V means that disk properties (columns $2$--$5$) are taken from {\em V}--band data; otherwise, properties are taken from {\em B}--band; 
(Columns 2 \& 3) Size (major and minor axes) and position angle (P.A.) of the optical disk \cite[][]{Hunter2006}; 
(Column 4) 6\,cm ($\sim 6$\,GHz) radio continuum flux density. This and the following values are preseneted including ambiguous sources; 
(Column 5) \halpha\ flux (units of $10^{-13} $\,ergs\,s$^{-1}$\,cm$^{-2}$); 
(Column 6) {\em GALEX} FUV flux density; 
(Column 7) {\em Spitzer} 24\micron\ MIR flux density; 
(Column 8) {\em Spitzer} 70\micron\ FIR flux density; 
(Column 9) 6\,cm ($\sim 6$\,GHz) radio continuum non-thermal (synchrotron) flux density. All \RCNT\ emission is assumed to be synchrotron and is inferred by subtracting the \RCT\ component from the total RC following \cite{Deeg1997}. The quantity in parentheses is the amount that was regarded as ambiguous; 
(Column 13) Equipartition magnetic field strength in the plane of the sky \cite[see Equation 3 in][]{Beck2005}. }
\end{deluxetable}

\changetext{0.5cm}{}{}{1cm}{} 
\begin{deluxetable}{lcccccccccc}
\tablewidth{0pt}
\tabletypesize{\tiny}
\tablecolumns{13}
\rotate
\tablecaption{Integrated emission over the RC mask of the LITTLE THINGS Galaxies\label{table:12A-234_MaskQuantities_ambig}}
\tablehead{
\colhead{Galaxy}  & \colhead{R.A} & \colhead{Dec.} & \colhead{$f_{\rm disk}$} & \colhead{6\,cm RC} & \colhead{\halpha}        & \colhead{FUV} & \colhead{24\micron\ MIR} & \colhead{70\micron\ FIR} & \colhead{6\,cm \RCNT} & \colhead{$B_{\rm eq}$} \\
  &  hh\,mm\,ss.s    & dd\,mm\,ss.s     &  (\%) & (mJy)  & ($10^{-13} $\,ergs\,s$^{-1}$\,cm$^{-2}$)   & (mJy) & ($10^{-2}$\,Jy) & ($10^{-2}$\,Jy) & (mJy) & (${\rm \mu G}$) \\ 
(1) & (2) & (3) & (4) & (5) & (6) & (7) & (8) & (9) & (10) & (11)}
\startdata
DDO 46	& $07\,41\,26.6$ & $+40\,06\,39$ & $0.2$ 	& $1.29 \pm 0.02$ & $0.16\pm 0.01$ & $0.02\pm 0.01$ & \nodata & \nodata & $1.28\pm 0.02$ & $5$ \\
DDO 47	& $07\,41\,55.3$ & $+16\,48\,08$ & $0.8$ 	& $0.38 \pm 0.02$ & $0.16\pm 0.02$ & $0.06\pm 0.01$ & \nodata & \nodata & $0.37\pm 0.02$ & $4$ \\
DDO 50	& $08\,19\,08.7$ & $+70\,43\,25$ & $2.3$ 	& $6.81 \pm 0.09$ & $25.52\pm 0.42$ & $7.57\pm 0.77$ & $8.12\pm 0.06$ & $54.72\pm 0.04$ & $4.46\pm 0.10$ & $5$ \\
DDO 53	& $08\,34\,08.0$ & $+66\,10\,37$ & $3.7$ 	& $0.45 \pm 0.03$ & $1.91\pm 0.04$ & $0.74\pm 0.08$ & $1.41\pm 0.08$ & $5.66\pm 0.01$ & $0.27\pm 0.03$ & $4$ \\
DDO 63	& $09\,40\,30.4$ & $+71\,11\,02$ & $0.5$ 	& $1.56 \pm 0.02$ & $0.27\pm 0.01$ & $0.20\pm 0.02$ & $0.07\pm 0.17$ & $0.47\pm 0.01$ & $1.54\pm 0.02$ & $5$ \\
DDO 69	& $09\,59\,25.0$ & $+30\,44\,42$ & $1.8$ 	& $0.96 \pm 0.04$ & $0.02\pm 0.01$ & $0.08\pm 0.01$ & $0.12\pm 0.11$ & $0.74\pm 0.01$ & $0.96\pm 0.04$ & $4$ \\
DDO 70	& $10\,00\,00.9$ & $+05\,19\,50$ & $0.3$ 	& $1.53 \pm 0.03$ & $0.29\pm 0.02$ & $0.21\pm 0.03$ & $0.04\pm 0.19$ & $0.47\pm 0.01$ & $1.51\pm 0.03$ & $4$ \\
DDO 75	& $10\,10\,59.2$ & $-04\,41\,56$ & $0.4$ 	& $0.27 \pm 0.04$ & $2.89\pm 0.06$ & $0.90\pm 0.10$ & $0.06\pm 0.19$ & $1.71\pm 0.01$ & $0.01\pm 0.04$ & $4$ \\
DDO 87	& $10\,49\,34.7$ & $+65\,31\,46$ & $0.2$ 	& $0.13 \pm 0.01$ & $0.01\pm 0.01$ & $0.01\pm 0.01$ & $0.01\pm 0.45$ & $0.18\pm 0.01$ & $0.13\pm 0.01$ & $3$ \\
DDO 126	& $12\,27\,06.5$ & $+37\,08\,23$ & $4.2$ 	& $0.33 \pm 0.04$ & $1.45\pm 0.07$ & $0.54\pm 0.06$ & $0.15\pm 0.15$ & $2.14\pm 0.02$ & $0.21\pm 0.04$ & $3$ \\
DDO 133	& $12\,32\,55.4$ & $+31\,32\,14$ & $0.3$ 	& $0.39 \pm 0.02$ & $0.01\pm 0.01$ & $0.01\pm 0.01$ & $0.02\pm 0.21$ & $0.03\pm 0.01$ & $0.39\pm 0.02$ & $3$ \\
DDO 155	& $12\,58\,39.8$ & $+14\,13\,10$ & $5.2$ 	& $0.28 \pm 0.04$ & $2.23\pm 0.04$ & \nodata & $0.15\pm 0.12$ & $2.42\pm 0.01$ & $0.08\pm 0.04$ & $<0$ \\
DDO 168	& $13\,14\,27.2$ & $+45\,55\,46$ & $0.9$ 	& $0.24 \pm 0.03$ & $0.26\pm 0.01$ & $0.12\pm 0.01$ & $0.04\pm 0.09$ & $0.77\pm 0.01$ & $0.21\pm 0.03$ & $3$ \\
DDO 210	& $20\,46\,52.0$ & $-12\,50\,50$ & $0.6$ 	& $0.39 \pm 0.02$ & \nodata & $0.00\pm 0.01$ & $0.03\pm 0.28$ & $0.10\pm 0.01$ & $0.01\pm 0.01$ & $<2$ \\
Haro 29	& $12\,26\,16.7$ & $+48\,29\,38$ & $14.2$ 	& $2.04 \pm 0.04$ & $12.59\pm 0.45$ & $2.68\pm 0.29$ & $5.01\pm 0.13$ & $21.99\pm 0.02$ & $0.85\pm 0.06$ & $6$ \\
Haro 36	& $12\,46\,56.3$ & $+51\,36\,48$ & $9.2$ 	& $0.38 \pm 0.03$ & $1.17\pm 0.03$ & $1.94\pm 0.21$ & $0.41\pm 0.13$ & $6.88\pm 0.02$ & $0.26\pm 0.03$ & $4$ \\
IC 1613	& $01\,04\,49.2$ & $+02\,07\,48$ & $1.0$ 	& $3.69 \pm 0.06$ & $10.30\pm 0.43$ & $5.19\pm 0.71$ & $1.70\pm 0.19$ & $24.14\pm 0.15$ & $2.81\pm 0.07$ & $3$ \\
IC 10	& $00\,20\,17.5$ & $+59\,18\,14$ & $22.9$ 	& $99.33 \pm 0.39$ & $887.90\pm 5.68$ & \nodata & $1369.00\pm 10.10$ & $5482.00\pm 6.68$ & $14.96\pm 0.66$ & $8$ \\
M81 dwA	& $08\,23\,57.2$ & $+71\,01\,51$ & $0.2$ 	& $0.15 \pm 0.02$ & \nodata & $0.01\pm 0.01$ & \nodata & \nodata & $0.01\pm 0.01$ & $<2$ \\
Mrk 178	& $11\,33\,29.0$ & $+49\,14\,24$ & $3.8$ 	& $0.46 \pm 0.03$ & $2.33\pm 0.08$ & $0.97\pm 0.12$ & $0.16\pm 0.17$ & $0.16\pm 0.01$ & $0.25\pm 0.03$ & $4$ \\
NGC 1569	& $04\,30\,49.8$ & $+64\,50\,51$ & $125.5$ 	& $157.30 \pm 0.35$ & $506.70\pm 3.03$ & $759.10\pm 76.88$ & $719.60\pm 12.02$ & $3808.00\pm 3.01$ & $75.89\pm 0.60$ & $17$ \\
NGC 2366	& $07\,28\,48.8$ & $+69\,12\,22$ & $2.2$ 	& $12.05 \pm 0.09$ & $66.98\pm 1.10$ & $12.65\pm 1.28$ & $52.07\pm 0.04$ & $179.70\pm 0.05$ & $5.72\pm 0.14$ & $5$ \\
NGC 3738	& $11\,35\,49.0$ & $+54\,31\,23$ & $6.2$ 	& $2.98 \pm 0.12$ & $11.83\pm 0.17$ & $7.29\pm 0.75$ & $7.58\pm 0.13$ & $91.12\pm 0.10$ & $1.85\pm 0.12$ & $7$ \\
NGC 4163	& $12\,12\,09.2$ & $+36\,10\,13$ & $0.2$ 	& $0.23 \pm 0.01$ & $0.01\pm 0.01$ & $0.01\pm 0.01$ & $0.01\pm 0.61$ & $0.01\pm 0.01$ & $0.23\pm 0.01$ & $4$ \\
NGC 4214	& $12\,15\,39.2$ & $+36\,19\,38$ & $2.5$ 	& $23.16 \pm 0.09$ & $117.40\pm 0.91$ & $32.67\pm 3.29$ & $140.50\pm 0.09$ & $943.30\pm 0.17$ & $12.12\pm 0.12$ & $6$ \\
Sag DIG	& $19\,30\,00.6$ & $-17\,40\,56$ & $1.2$ 	& $0.56 \pm 0.09$ & $0.01\pm 0.01$ & $0.11\pm 0.01$ & \nodata & \nodata & $0.56\pm 0.09$ & $4$ \\
UGC 8508	& $13\,30\,44.9$ & $+54\,54\,29$ & $4.4$ 	& $0.71 \pm 0.03$ & $0.74\pm 0.02$ & \nodata & $0.08\pm 0.13$ & $1.66\pm 0.01$ & $0.64\pm 0.03$ & $3$ \\
VIIZw 403	& $11\,27\,58.2$ & $+78\,59\,39$ & $19.2$ 	& $1.29 \pm 0.04$ & $6.49\pm 0.15$ & $3.21\pm 0.33$ & $2.10\pm 37.85$ & $33.77\pm 0.54$ & $0.68\pm 0.04$ & $6$ \\
WLM	& $00\,01\,59.2$ & $-15\,27\,41$ & $0.1$ 	& $0.28 \pm 0.02$ & $0.79\pm 0.05$ & $0.11\pm 0.01$ & $0.27\pm 0.23$ & $0.36\pm 0.01$ & $0.23\pm 0.02$ & $2$ \\

\enddata
\tablecomments{ (Column 1) Name of dwarf galaxy;
(Columns 2 \& 3) Equatorial coordinates (J2000) of centre of the galaxy defined by the optical disk; 
(Column 4) Fraction of the disk (see table\,\ref{table:12A-234_DiskQuantities}) that has significant RC emission; 
(Column 5) 6\,cm ($\sim 6$\,GHz) radio continuum flux density. This and the following values are preseneted including ambiguous sources; 
(Column 6) \halpha\ flux (units of $10^{-13} $\,ergs\,s$^{-1}$\,cm$^{-2}$); 
(Column 7) {\em GALEX} FUV flux density; 
(Column 8) {\em Spitzer} 24\micron\ MIR flux density; 
(Column 9) {\em Spitzer} 70\micron\ FIR flux density; 
(Column 10) 6\,cm ($\sim 6$\,GHz) radio continuum non-thermal (synchrotron) flux density. All \RCNT\ emission is assumed to be synchrotron and is inferred by subtracting the \RCT\ component from the total RC following \cite{Deeg1997}. The quantity in parentheses is the amount that was regarded as ambiguous; 
(Column 11) Equipartition magnetic field strength in the plane of the sky \cite[see Equation 3 in][]{Beck2005}. }
\end{deluxetable}

\clearpage
\newpage
\notetoeditor{Table~\ref{tab-sample} should be able to fit on a landscape-orientated page with tabletypesize=tiny. There is something wrong with the formatting that pushes the Table off the page. Table~\ref{table:Temporary1} shows how Table~\ref{tab-sample} should be viewed.}
\changetext{2cm}{}{}{-2cm}{}
\begin{deluxetable}{lccccccccccc}
\tabletypesize{\tiny}
\rotate
\tablecolumns{12}
\tablewidth{689pt}
\tablecaption{The Galaxy Sample\label{Table:Temporary1} }
\tablehead{ &  & \colhead{D} &  & \colhead{M$_V$} & \colhead{$R_H$\tablenotemark{c}} & \colhead{$R_D$\tablenotemark{c}} &  
& \colhead{log$_{10}$ $\Sigma_\mathrm{SFR}\mathrm{(H\alpha)}$} & \colhead{log$_{10}$ $\Sigma_\mathrm{SFR}\mathrm{(FUV)}$} &  &  \\ \colhead{Galaxy}  &  \colhead{Other names\tablenotemark{a}} & \colhead{(Mpc)} & \colhead{Ref\tablenotemark{b}} & \colhead{(mag)} & \colhead{(arcmin)} & \colhead{(kpc)} & \colhead{E($B-V$)\tablenotemark{d}} & \colhead{(\msun yr$^{-1}$ kpc$^{-2}$)\tablenotemark{e}} & \colhead{(\msun yr$^{-1}$ kpc$^{-2}$)\tablenotemark{e}} & \colhead{$12+\log_{10} {\rm O/H}$\tablenotemark{f}}  & \colhead{Ref\tablenotemark{g}}  }
\startdata
\cutinhead{Im Galaxies}
\object[CVn I dwA]{CVnIdwA}   & UGCA 292                                                                          &  3.6 & 1             & -12.4 & 0.87 & $0.57 \pm 0.12$ & 0.01 & $-2.58 \pm 0.01$ & $-2.48 \pm 0.01$ & $7.3 \pm 0.06$ & 24 \\
\object{DDO 43}     &  PGC 21073, UGC 3860                                                  &  7.8 & 2              & -15.1 & 0.89 & $0.41\pm 0.03$ & 0.05 & $-1.78 \pm 0.01$ & $-1.55 \pm 0.01$ & $8.3 \pm 0.09$ & 25 \\
\object{DDO 46}     &  PGC 21585, UGC 3966                                                  &  6.1 & \nodata  & -14.7 & \nodata & $1.14 \pm 0.06$ & 0.05 & $-2.89 \pm 0.01$ & $-2.46 \pm 0.01$ & $8.1 \pm 0.1$  & 25 \\
\object{DDO 47}     &  PGC 21600, UGC 3974                                                  &  5.2  & 3             & -15.5 & 2.24 & $1.37 \pm 0.06$ & 0.02 & $-2.70 \pm 0.01$ & $-2.40 \pm 0.01$ & $7.8 \pm 0.2$ & 26 \\
\object{DDO 50}     & PGC 23324, UGC 4305, Holmberg II, VIIZw 223        &  3.4 & 1             & -16.6 & 3.97 & $1.10 \pm 0.05$ & 0.02 & $-1.67 \pm 0.01$ & $-1.55 \pm 0.01$ & $7.7 \pm 0.14$ & 27 \\
\object{DDO 52}     &  PGC 23769, UGC 4426                                                  & 10.3 & 4            & -15.4 & 1.08 & $1.30 \pm 0.13$ & 0.03 & $-3.20 \pm 0.01$ & $-2.43 \pm 0.01$ & (7.7)                   & 28 \\
\object{DDO 53}     &  PGC 24050, UGC 4459, VIIZw 238                              &  3.6 & 1             & -13.8 & 1.37 & $0.72 \pm 0.06$ & 0.03 & $-2.42 \pm 0.01$ &$-2.41 \pm 0.01$ & $7.6 \pm 0.11$ & 27 \\
\object{DDO 63}     &  PGC 27605, Holmberg I, UGC 5139,  Mailyan 044   &  3.9 & 1             & -14.8 & 2.17 & $0.68 \pm 0.01$ & 0.01 & $-2.32 \pm 0.01$ & $-1.95 \pm 0.00$ & $7.6 \pm 0.11$ & 27 \\
\object{DDO 69}     &  PGC 28868, UGC 5364, Leo A                                     &  0.8 & 5             & -11.7 & 2.40 & $0.19 \pm  0.01$ & 0.00 & $-2.83 \pm 0.01$ & $-2.22 \pm 0.01$ & $7.4 \pm 0.10$ & 29 \\
\object{DDO 70}     &  PGC 28913, UGC 5373, Sextans B                             &  1.3 & 6             & -14.1 & 3.71 & $0.48 \pm 0.01$ & 0.01 & $-2.85 \pm 0.01$ & $-2.16 \pm 0.00$ & $7.5 \pm 0.06$ & 30 \\
\object{DDO 75}     &  PGC 29653, UGCA 205, Sextans A                             &  1.3 & 7             & -13.9 & 3.09 & $0.22 \pm 0.01$ & 0.02 & $-1.28 \pm 0.01$ & $-1.07 \pm 0.01$ & $7.5 \pm 0.06$ & 30 \\
\object{DDO 87}     &  PGC 32405, UGC 5918, VIIZw 347                              &  7.7 & \nodata & -15.0 & 1.15 & $1.31 \pm 0.12$ & 0.00 & $-1.36 \pm 0.01$ & $-1.00 \pm 0.01$ & $7.8 \pm 0.04$ & 31 \\
\object{DDO 101}   &  PGC 37449, UGC 6900                                                 &  6.4 & \nodata & -15.0 & 1.05 & $0.94 \pm 0.03$ & 0.01 & $-2.85 \pm 0.01$ & $-2.81 \pm 0.01$ & $8.7 \pm 0.03$ & 25 \\
\object{DDO 126}   &  PGC 40791, UGC 7559                                                 &  4.9 & 8             & -14.9 & 1.76 & $0.87 \pm 0.03$ & 0.00 & $-2.37 \pm 0.01$ & $-2.10 \pm 0.01$ & (7.8)                    & 28 \\
\object{DDO 133}   &  PGC 41636, UGC 7698                                                 &  3.5 & \nodata  & -14.8 & 2.33 & $1.24 \pm 0.09$ & 0.00 & $-2.88 \pm 0.01$ & $-2.62 \pm 0.01$ & $8.2 \pm 0.09$  & 25 \\
\object{DDO 154}   &  PGC 43869, UGC 8024, NGC 4789A                         &  3.7 & \nodata  & -14.2 & 1.55 & $0.59 \pm 0.03$ & 0.01 & $-2.50 \pm 0.01$ & $-1.93 \pm 0.01$ & $7.5 \pm 0.09$ & 27 \\
\object{DDO 155}   &  PGC 44491, UGC 8091, GR 8, LSBC D646-07        &  2.2 & 9             & -12.5 & 0.95 & $0.15 \pm 0.01$ & 0.01 & $-1.44 \pm 0.01$ &\nodata & $7.7 \pm 0.06$ & 29 \\
\object{DDO 165}   &  PGC 45372, UGC 8201, IIZw 499, Mailyan 82          &  4.6 & 10           & -15.6 & 2.14 & $2.26 \pm 0.08$ & 0.01 & $-3.67 \pm 0.01$ & \nodata & $7.6 \pm0.08$ & 27 \\
\object{DDO 167}   &  PGC 45939, UGC 8308                                                 &  4.2 & 8             & -13.0 & 0.75 & $0.33 \pm 0.05$ & 0.00 & $-2.36 \pm 0.01$ & $-1.83 \pm 0.01$ & $7.7 \pm 0.2$ & 26 \\
\object{DDO 168}   &  PGC 46039, UGC 8320                                                 &  4.3 & 8             & -15.7 & 2.32 & $0.82 \pm 0.01$ & 0.00 & $-2.27 \pm 0.01$ &$-2.04 \pm 0.01$ & $8.3 \pm 0.07$ & 25 \\
\object{DDO 187}   &  PGC 50961, UGC 9128                                                  &  2.2 & 1            & -12.7 & 1.06 & $0.18 \pm 0.01$ & 0.00 & $-2.52 \pm 0.01$ & $-1.98 \pm 0.01$ & $7.7 \pm 0.09$ & 32 \\
\object{DDO 210}   &  PGC 65367, Aquarius Dwarf                                         &  0.9 & 10          & -10.9 & 1.31 & $0.17 \pm 0.01$ & 0.03 &  \nodata   & $-2.71 \pm 0.06$ & (7.2)                                      & 28 \\
\object{DDO 216}   &  PGC 71538, UGC 12613, Peg DIG, Pegasus Dwarf & 1.1  & 11         & -13.7 & 4.00 & $0.54 \pm 0.01$ & 0.02 & $-4.10 \pm 0.07$ & $-3.21 \pm 0.01$ & $7.9 \pm 0.15$ & 33 \\
\object[F564 V3]{F564-V3}    &  LSBC D564-08                                                                  &  8.7 & 4           & -14.0 & \nodata  & $0.53 \pm 0.03$ & 0.02 &  \nodata  & $-2.79 \pm 0.02$ & (7.6)                   & 28 \\
\object{IC 10}          &  PGC 1305, UGC 192                                                        &  0.7 & 12        & -16.3 & \nodata  & $0.40 \pm 0.01$ & 0.75 & $-1.11 \pm 0.01$ & \nodata & $8.2 \pm 0.12$ & 34 \\
\object{IC 1613}     &  PGC 3844, UGC 668, DDO 8                                          &  0.7 & 13         & -14.6 & 9.10 & $0.58 \pm 0.02$ & 0.00 & $-2.56 \pm 0.01$ & $-1.99 \pm 0.01$ & $7.6 \pm 0.05$ & 35 \\
\object{LGS 3}        &  PGC 3792, Pisces dwarf                                                  &  0.7 & 14         &    -9.7 & 0.96 & $0.23 \pm 0.02$ & 0.04 &  \nodata   & $-3.88 \pm 0.06$ & (7.0)                   & 28 \\
\object[M81 DwA]{M81dwA}    & PGC 23521                                                                         &  3.5 & 15         & -11.7 & \nodata  & $0.26 \pm 0.00$ & 0.02 &   \nodata   & $-2.26 \pm 0.01$ & (7.3)                   & 28 \\
\object{NGC 1569} & PGC 15345, UGC 3056, Arp 210, VIIZw 16                 &  3.4 & 16         & -18.2 & \nodata  & $0.38 \pm 0.02$& 0.51 & $0.19 \pm 0.01$ & $-0.01 \pm 0.01$ & $ 8.2 \pm 0.05$ & 36 \\
\object{NGC 2366} & PGC 21102, UGC 3851, DDO 42                                  &  3.4 & 17          & -16.8 & 4.72 & $1.36 \pm 0.04$ & 0.04 & $-1.67 \pm 0.01$ & $-1.66 \pm 0.01$ & $7.9 \pm 0.01$  & 37 \\
\object{NGC 3738} & PGC 35856, UGC 6565, Arp 234                                   &  4.9 & 3            & -17.1 & 2.40 & $0.78 \pm 0.01$ & 0.00 & $-1.66 \pm 0.01$ & $-1.53 \pm 0.01$ & $8.4 \pm 0.01$  & 25 \\
\object{NGC 4163} & PGC 38881, NGC 4167, UGC 7199                              &  2.9 & 1            & -14.4 & 1.47 & $0.27 \pm 0.03$ & 0.00 & $-2.28 \pm 0.13$ & $-1.74 \pm 0.01$ & $7.9 \pm 0.2$ & 38 \\
\object{NGC 4214} & PGC 39225, UGC 7278                                                   &  3.0 & 1            & -17.6 & 4.67 & $0.75 \pm 0.01$ & 0.00 & $-1.03 \pm 0.01$ & $-1.08 \pm 0.01$ & $8.2 \pm 0.06$ & 39 \\
\object[Sag DIG]{SagDIG}      & PGC 63287, Lowal's Object                                            &  1.1 & 19          & -12.5 & \nodata  & $0.23 \pm 0.03$ & 0.14 & $-2.97 \pm 0.04$ & $-2.11 \pm 0.01$ & $7.3 \pm 0.1$ & 35 \\
\object{UGC 8508} & PGC 47495, IZw 60                                                           &  2.6  & 1           & -13.6 & 1.28 & $0.27 \pm 0.01$ & 0.00 & $-2.03 \pm 0.01$ & \nodata & $7.9 \pm 0.2$ & 38 \\
\object{WLM}           & PGC 143, UGCA 444, DDO 221, Wolf-Lundmark-Melott & 1.0 & 20     & -14.4 & 5.81 & $0.57 \pm 0.03$ & 0.02 & $-2.77 \pm 0.01$ & $-2.05 \pm 0.01$ & $7.8 \pm 0.06$ & 40 \\
\cutinhead{BCD Galaxies}
\object{Haro 29}     & PGC 40665, UGCA 281, Mrk 209, I Zw 36                    &  5.8 & 21         & -14.6 & 0.84 & $0.29 \pm 0.01$ & 0.00 & $-0.77 \pm 0.01$ & $-1.07 \pm 0.01$ & $7.9 \pm 0.07$ & 41 \\
\object{Haro 36}     & PGC 43124, UGC 7950                                                  &  9.3 & \nodata & -15.9 & \nodata  & $0.69 \pm 0.01$ & 0.00 & $-1.86 \pm 0.01$ & $-1.55 \pm 0.01$ & $8.4 \pm 0.08$ & 25 \\
\object{Mrk 178}     & PGC 35684, UGC 6541                                                   &  3.9 & 8            & -14.1 & 1.01 & $0.33 \pm 0.01$ & 0.00 & $-1.60 \pm 0.01$ & $-1.66 \pm 0.01$ & $7.7 \pm 0.02 $ & 42 \\
\object[VII Zw 403]{VIIZw 403}  & PGC 35286, UGC 6456                                                   &  4.4 & 22,23    & -14.3 & 1.11 & $0.52 \pm 0.02$ & 0.02 & $-1.71 \pm 0.01$ & $-1.67 \pm 0.01$ & $7.7 \pm 0.01$ & 25 \\
\enddata
\tablenotetext{a}{Selected alternate identifications obtained from NED.}
\tablenotetext{b}{Reference for the distance to the galaxy. If no reference is given, the distance was determined from the galaxy's radial velocity, given by de Vaucouleurs et al. (1991), corrected for infall to the Virgo Cluster (Mould \et\ 2000) and a Hubble constant of 73 km s$^{-1}$ Mpc$^{-1}$.}
\tablenotetext{c}{$R_H$ is the Holmberg radius, the radius of the galaxy at a $B$-band isophote, corrected for reddening, of 26.7 mag arcsec$^{-2}$. $R_D$ is the disk scale length measured from $V$-band images. (Table from Hunter \& Elmegreen 2006).}
\tablenotetext{d}{Foreground reddening from Burstein \& Heiles (1984).}
\tablenotetext{e}{$\Sigma_\mathrm{SFR}\mathrm{(H\alpha)}$ is the Star Formation Rate Density (SFRD) measured from H$\alpha$, calculated over the area $\pi R_D^2$, where $R_D$ is the disk scale length (Hunter \& Elmegreen 2004). $\Sigma_\mathrm{SFR}\mathrm{(FUV)}$ is the SF rate density determined from {\it GALEX} FUV fluxes (Hunter et al.\ 2010, with an update of the  {\it GALEX} FUV photometry to the GR4/GR5 pipeline reduction).}
\tablenotetext{f}{Values in parentheses were determined from the empirical relationship between oxygen abundance and $M_B$ given by Richer \& McCall (1995) and are particularly uncertain.}
\tablenotetext{g}{Reference for the oxygen abundance.}
\tablerefs{
(1) Dalcanton et al. 2009;
(2) Karachentsev et al. 2004;
(3) Karachentsev et al. 2003a;
(4) Karachentsev et al. 2006;
(5) Dolphin et al. 2002;
(6) Sakai et al. 2004;
(7) Dolphin et al. 2003;
(8) Karachentsev et al. 2003b;
(9) Tolstoy et al. 1995a;
(10) Karachentsev et al. 2002;
(11) Meschin et al. 2009;
(12) Sakai et al. 1999;
(13) Pietrzynski et al. 2006;
(14) Miller et al. 2001;
(15) Freedman et al. 2001;
(16) Grocholski et al. 2008;
(17) Tolstoy et al. 1995b;
(18) Gieren et al. 2006;
(19) Momany et al. 2002;
(20) Gieren et al. 2008;
(21) Schulte-Ladbeck et al. 2001;
(22) Lynds et al. 1998;
(23) M\'endez et al. 2002;
(24) van Zee \& Haynes 2006;
(25) Hunter \& Hoffman 1999;
(26) Skillman, Kennicutt, \& Hodge 1989;
(27) Moustakas et al.\ 2010;
(28) Richer \& McCall 1995;
(29) van Zee et al. 2006;
(30) Kniazev et al. 2005;
(31) Croxall et al. 2009;
(32) Lee et al. 2003b;
(33) Skillman et al. 1997;
(34) Lequex et al.\ 1979;
(35) Lee et al. 2003a;
(36) Kobulnicky \& Skillman 1997;
(37) Gonz\'alez-Delgado et al.\ 1994;
(38) Moustakas \& Kennicutt (2006);
(39) Masegosa et al. 1991;
(40) Lee et al. 2005;
(41) Viallefond \& Thuan 1983;
(42) Gon\'zalez-Riestra \et\ 1988.
}
\end{deluxetable}
\end{document}